# Deep R Programming

**Marek Gagolewski**

Prof. **Marek Gagolewski**
Warsaw University of Technology, Poland
Systems Research Institute, Polish Academy of Sciences
https://www.gagolewski.com/

This project received no funding, administrative, technical, or editorial support from Warsaw University of Technology, Polish Academy of Sciences, Deakin University, nor any other source.

Any bug reports/corrections/feature requests are welcome. To make this textbook even better, please file them at https://github.com/gagolews/deepr.

Typeset with XeLaTeX. Please be understanding: it was an algorithmic process. Hence, the results are $\in$ [good enough, perfect).

Homepage: https://deepr.gagolewski.com/

Datasets: https://github.com/gagolews/teaching-data

*Deep R Programming* is a **comprehensive and in-depth introductory course** on one of the most popular languages for data science. It equips ambitious students, professionals, and researchers with the **knowledge and skills to become independent users** of this potent environment so that they can **tackle any problem** related to data wrangling and analytics, numerical computing, statistics, and machine learning.

For many students around the world, educational resources are hardly affordable. Therefore, I have decided that this book should **remain an independent, non-profit, open-access project** (available both in PDF[1] and HTML[2] forms). Whilst, for some people, the presence of a “designer tag” from a major publisher might still be a proxy for quality, it is my hope that this publication will prove useful to those seeking knowledge for knowledge’s sake.

Any bug/typo reports/fixes are appreciated. Please submit them via this project’s GitHub repository[3]. Thank you.

Make sure to check out *Minimalist Data Wrangling with Python*[5] [28], too.

[1] https://deepr.gagolewski.com/deepr.pdf
[2] https://deepr.gagolewski.com/
[3] https://github.com/gagolews/deepr/issues
[4] https://dx.doi.org/10.5281/zenodo.7490464
[5] https://datawranglingpy.gagolewski.com/

# 0 Preface

## 0.1 To R, or not to R

R has been named the eleventh most dreaded programming language in the 2022 StackOverflow Developer Survey[6].

Also, it is a free *app*, so there must be something wrong with it, right?

But whatever, R is deprecated anyway; the *modern* way is to use `tidyverse`.

Or we should all just switch to Python[7].

Yeah, nah.

## 0.2 R (GNU S) as a language and an environment

Let's get one[8] thing straight: R is *not* just a *statistical package*. It is a general-purpose, high-level programming language that happens to be very powerful for numerical, data-intense computing activities of any kind. It offers extensive support for statistical, machine learning, data analysis, data wrangling, and data visualisation applications, but there is much more.

As we detail below, R has a long history. It is an open-source version of the S environment, which was written for statisticians, by statisticians. Therefore, it is a free, yet often more capable alternative to other software (but without any strings attached). Unlike in some of them, in R, a spreadsheet-like GUI is not the main gateway for performing computations on data. Here, we must *write code* to get things done. Despite the beginning of the learning curve's being a little steeper for non-programmers, in the long run, R empowers us more because we are not limited to tackling the most common scenarios. If some functionality is missing or does not suit our needs, we can easily (re)implement it ourselves.

[6] https://survey.stackoverflow.co/2022

[7] https://datawranglingpy.gagolewski.com/

[8] Also, we must not confuse RStudio with R. The former is merely one of many development environments for our language. We program in R, not in RStudio.

R is thus very convenient for rapid prototyping. It helps turn our ideas into fully operational code that can be battle-tested, extended, polished, run in production, and otherwise enjoyed. As an interpreted language, it can be executed not only in an interactive read-eval-print loop (command–result, question–answer, …), but also in batch mode (running standalone scripts).

Therefore, we would rather position R amongst such environments for numerical or scientific computing as Python with `numpy` and `pandas`, Julia, GNU Octave, Scilab, and MATLAB. However, it is more *specialised* in data science applications than any of them. Hence, it provides a much smoother experience. This is why, over the years, R has become the de facto standard in statistics and related fields.

---

**Important** R is a whole ecosystem. Apart from the R language interpreter, it features advanced:

- graphics capabilities (see Chapter 13),
- a consistent, well-integrated help system (Section 1.4),
- ways for convenient interfacing with compiled code (Chapter 14),
- a package system and centralised package repositories (such as CRAN and Bioconductor; Section 7.3.1),
- a lively community of users and developers – curious and passionate people, like you and yours cordially.

---

**Note** R [71] is a dialect of the very popular S system designed in the mid-1970s by Rick A. Becker, John M. Chambers, and Allan R. Wilks at Bell Labs. For historical notes, see [3, 4, 5, 6]. For works on newer versions of S, refer to [7, 10, 14, 58]. Quoting from [4]:

> *The design goal for S is, most broadly stated, to enable and encourage good data analysis, that is, to provide users with specific facilities and a general environment that helps them quickly and conveniently look at many displays, summaries, and models for their data, and to follow the kind of iterative, exploratory path that most often leads to a thorough analysis. The system is designed for interactive use with simple but general expressions for the user to type, and immediate, informative feedback from the system, including graphic output on any of a variety of graphical devices.*

S became popular because it offered greater flexibility than the standalone statistical packages. It was praised for its high interactivity and array-centrism that was taken from APL, the familiar syntax of the C language involving {curly braces}, the ability to treat code as data known from Lisp (Chapter 15), the notion of lazy arguments (Chapter 17), and the ease of calling external C and Fortran routines (Chapter 14). Its newer versions were also somewhat object-orientated (Chapter 10).

However, S was a proprietary and closed-source system. To address this, Robert Gentleman and Ross Ihaka of the Statistics Department, University of Auckland developed R in the 1990s[9]. They were later joined by many contributors[10]. It has been decided that it will be distributed under the terms of the free GNU General Public License, version 2.

In essence, R was supposed to be backwards-compatible with S, but some design choices led to their evaluation models' being slightly different. In Chapter 16, we discuss that R's design was inspired by the Scheme language [1].

## 0.3 Aims, scope, and design philosophy

Many users were introduced to R by means of some very advanced operations involving data frames, formulae, and functions that rely on nonstandard evaluation (metaprogramming), like:

```
lm(
    Ozone~Solar.R+Temp,
    data=subset(airquality, Temp>60, select=-(Month:Day))
) |> summary()
```

This is horrible.

Another cohort was isolated from base R through a thick layer of popular third-party packages that introduce an overwhelming number of functions (every operation, regardless of its complexity, has a unique name). They often duplicate the core functionality, and might not be fully compatible with our traditional system.

Both user families ought to be fine, as long as they limit themselves to solving only the most common data processing problems.

But *we* yearn for more. We do not want hundreds of prefabricated *recipes* for popular dishes that we can mindlessly apply without much understanding.

Our aim is to learn the fundamentals of base R, which constitutes the lingua franca of all R users. We want to be able to indite code that everybody *should* understand; code that will work without modifications in the next decades, too.

[9] See [13, 38] for historical notes. R version 0.49 released in April 1997 (the first whose source code is available on CRAN; see https://cloud.r-project.org/src/base/R-0), was already fairly feature-rich. In particular, it implemented S3 methods, formulae, and data frames that were introduced in the 1991 version of S [14].

[10] The beauty of the employed open-source model is that all the contributors are real human beings, not anonymous contractors working for soulless corporations; see https://www.r-project.org/contributors.html.

We want to be able to tackle *any* data-intense problem. Furthermore, we want to develop *transferable* skills so that learning new tools such as Python with **numpy** and **pandas** (e.g., [28, 48]) or Julia will be much easier later. After all, R is not the only notable environment out there.

Anyway, enough preaching. This graduate[11]-level textbook is for readers who:

- would like to experience the *joy* of solving problems by programming,
- want to become *independent* users of the R environment,
- can appreciate a more cohesively and comprehensively[12] organised material,
- do not mind a slightly steeper learning curve at the beginning,
- do not want to be made obsolete by artificial "intelligence" in the future.

Some readers will benefit from this book's being their first introduction to R (yet, without all the pampering). For others[13], this will be a fine course from intermediate to advanced (do not skip the first chapters, though).

Either way, we should not forget to solve *all* the prescribed exercises.

Good luck!

## 0.4 Classification of R data types and book structure

The most commonly used R data types can be classified as follows; see also Figure 1.

1. *Basic types* are discussed in the first part of the book:
    - *atomic vectors* represent whole sequences of values, where every element is of the same type:
        - `logical` (Chapter 3) includes items that are `TRUE` ("yes", "present"), `FALSE` ("no", "absent"), or `NA` ("not available", "missing");

[11] The author taught similar courses for his wonderfully ambitious undergraduate data/computer science and mathematics students at the Warsaw University of Technology, where our approach has proven not difficult whatsoever.

[12] Yours truly has not chosen to play a role of a historian, a stenographer, nor a grammarian. Thus, he has made a few noninvasive idealisations for didactic purposes. Languages evolve over time, R is now different from what it used to be, and we can shape it (slowly; we value its stable API) to become something even better in the future.

[13] It might also happen that for certain readers, this will not be an appropriate course at all, either at this stage of their career (come back later) or in general (no dramas). This is a non-profit, open-access project, but it does not mean it is ideal for everyone. We recommend giving other sources a try, e.g., [9, 11, 16, 46, 59, 62, 63, 70], etc. Some of them are freely available.

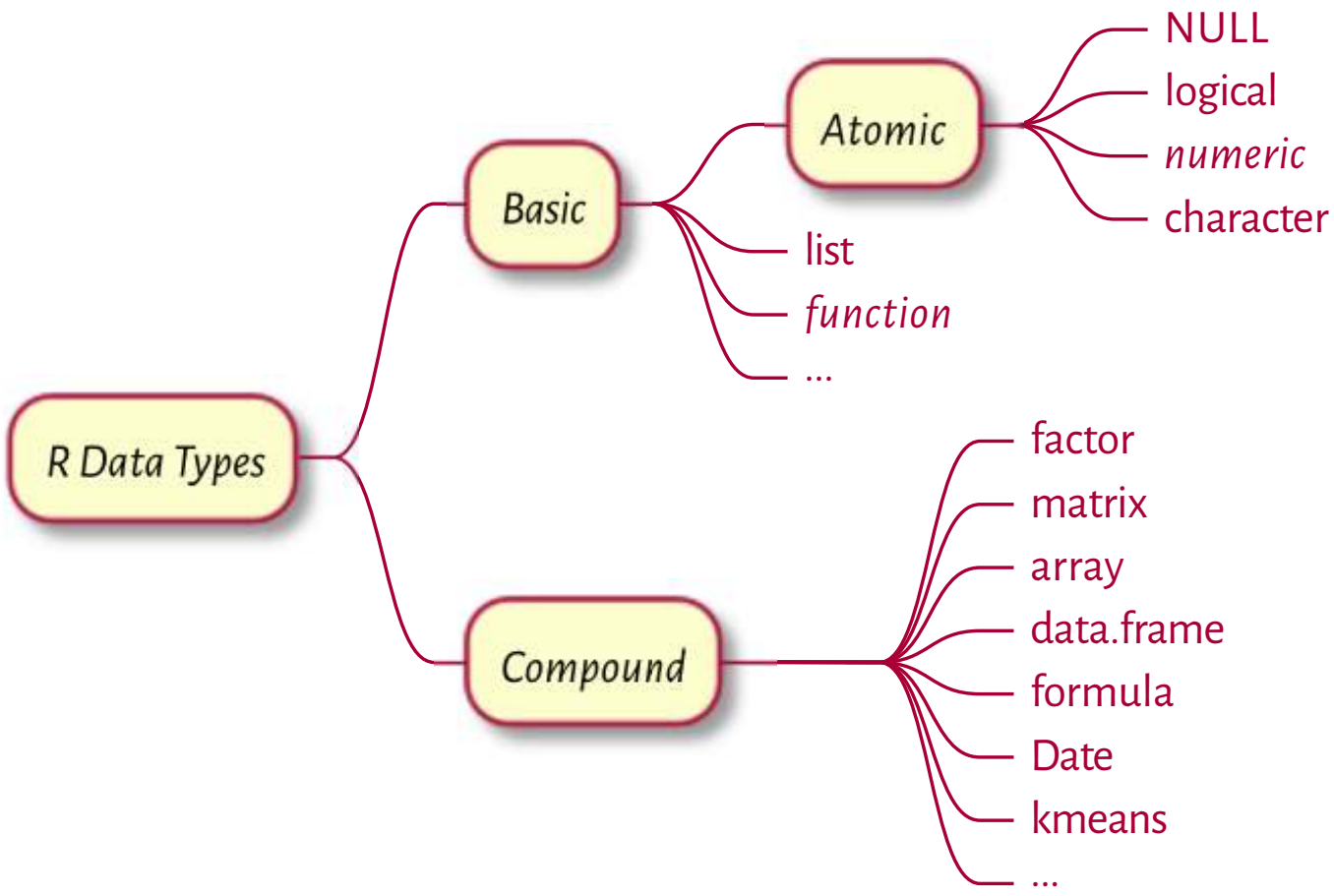

Figure 1. An overview of the most prevalent R data types; see Figure 17.2 for a more comprehensive list.

   – `numeric` (Chapter 2) represents real numbers, such as `1`, `3.14`, `-0.0000001`, etc.;

   – `character` (Chapter 6) contains strings of characters, e.g., `"groß"`, `"123"`, or "Добрий день";

 • `function` (Chapter 7) is used to group a series of expressions (code lines) so that they can be applied on miscellaneous input data to generate the (hopefully) desired outcomes, for instance, **`cat`**, **`print`**, **`plot`**, **`sample`**, and **`sum`**;

 • `list` (generic vector; Chapter 4) can store elements of mixed types.

The above will be complemented with a discussion on vector indexing (Chapter 5) and ways to control the program flow (Chapter 8).

2. *Compound types* are mostly discussed in the second part. They are wrappers around objects of basic types that might behave unlike the underlying primitives thanks to the dedicated operations *overloaded* for them. For instance:

 • `factor` (Section 10.3.2) is a vector-like object that represents qualitative data (on a nominal or an ordered scale);

 • `matrix` (Chapter 11) stores tabular data, i.e., arranged into rows and columns, where each cell is usually of the same type;

 • `data.frame` (Chapter 12) is also used for depositing tabular data, but this time such that each column can be of a different type;

 • `formula` (Section 17.6) is utilised by some functions to specify supervised learning models or define operations to be performed within data subgroups, amongst others;

- and many more, which we can arbitrarily define using the principles of S3-style object-orientated programming (Chapter 10).

In this part of the book, we also discuss the principles of sustainable coding (Chapter 9) as well as introduce ways to prepare publication-quality graphics (Chapter 13).

3. More advanced material is discussed in the third part. For most readers, it should be of theoretical interest only. However, it can help gain a complete understanding of and control over our environment. This includes the following data types:

   - `symbol` (`name`), `call`, `expression` (Chapter 15) are objects representing unevaluated R expressions that can be freely manipulated and executed if needed;
   - `environment` (Chapter 16) store named objects in hash maps and provides the basis for the environment model of evaluation;
   - `externalptr` (Section 14.2.8) provides the ability to maintain any dynamically allocated C/C++ objects between function calls.

We should not be surprised that we did not list any data types defined by a few trendy[14] third-party packages. We will later see that we can most often do without them. If that is not the case, we will become skilled enough to learn them quickly ourselves.

---

## 0.5 About the author

I, Marek Gagolewski[15] (pronounced like Maa'rek (Mark) Gong-o-leaf-ski), am currently an Associate Professor in Data Science at the Faculty of Mathematics and Information Science, Warsaw University of Technology.

My research interests are related to data science, in particular: modelling complex phenomena, developing usable, general-purpose algorithms, studying their analytical properties, and finding out how people use, misuse, understand, and misunderstand methods of data analysis in research, commercial, and decision-making settings. I am an author of ~100 publications, including journal papers in outlets such as *Proceedings of the National Academy of Sciences (PNAS)*, *Journal of Statistical Software*, *The R Journal*, *Journal of Classification*, *Information Fusion*, *International Journal of Forecasting*, *Statistical Modelling*, *Physica A: Statistical Mechanics and its Applications*, *Information Sciences*, *Knowledge-Based Systems*, *IEEE Transactions on Fuzzy Systems*, and *Journal of Informetrics*.

In my "spare" time, I write books for my students: check out my *Minimalist Data*

[14] Which does not automatically mean *good*. For instance, sugar, salt, and some drugs are very popular, but it does not make them healthy.

[15] https://www.gagolewski.com/

*Wrangling with Python*[16] [28]. I also develop[17] open-source software for data analysis, such as `stringi`[18] (one of the most often downloaded R packages) and `genieclust`[19] (a fast and robust clustering algorithm in both Python and R).

---

## 0.6 Acknowledgements

R, and its predecessor S, is the result of a collaborative effort of many programmers[20]. Without their generous intellectual contributions, the landscape of data analysis would not be as beautiful as it is now. R is distributed under the terms of the GNU General Public License version 2. We occasionally display fragments of its source code for didactic purposes.

We describe and use R version 4.6.1 (2026-06-24). However, we expect 99.9% of the material covered here to be valid in future releases (consider filing a bug report if you discover this is not the case).

*Deep R Programming* is based on the author's experience as an R user (since ~2003), developer of open-source packages, tutor/lecturer (since ~2008), and an author of a quite successful Polish textbook *Programowanie w języku R* [26] which was published by PWN (1st ed. 2014, 2nd ed. 2016). Even though the current book is an entirely different work, its predecessor served as an excellent test bed for many ideas conveyed here.

In particular, the teaching style exercised in this book has proven successful in many similar courses that yours truly was responsible for, including at Warsaw University of Technology, Data Science Retreat (Berlin), and Deakin University (Melbourne). I thank all my students and colleagues for the feedback given over the last 15-odd years.

This work received no funding, administrative, technical, or editorial support from Deakin University, Warsaw University of Technology, Polish Academy of Sciences, or any other source.

This book was prepared in a Markdown superset called MyST[21], `Sphinx`[22], and TeX (XeLaTeX). Code chunks were processed with the R package **`knitr`** [65]. All figures were plotted with the low-level **`graphics`** package using the author's own style template. A little help from Makefiles, custom shell scripts, and **`Sphinx`** plugins (`sphinxcontrib-bibtex`[23], `sphinxcontrib-proof`[24]) dotted the *j*'s and crossed the *f*'s.

[16] https://datawranglingpy.gagolewski.com/
[17] https://github.com/gagolews
[18] https://stringi.gagolewski.com/
[19] https://genieclust.gagolewski.com/
[20] https://www.r-project.org/contributors.html
[21] https://myst-parser.readthedocs.io/en/latest/index.html
[22] https://www.sphinx-doc.org/
[23] https://pypi.org/project/sphinxcontrib-bibtex
[24] https://pypi.org/project/sphinxcontrib-proof

The `Ubuntu Mono`[25] font is used for the display of code. The typesetting of the main text relies on the *Alegreya*[26] typeface.

## 0.7 You can make this book better

When it comes to quality assurance, open, non-profit projects have to resort to the generosity of the readers' community.

If you find a typo, a bug, or a passage that could be rewritten or extended for better readability/clarity, do not hesitate to report it via the *Issues* tracker available at https://github.com/gagolews/deepr. New feature requests are welcome as well.

[25] https://design.ubuntu.com/font
[26] https://www.huertatipografica.com/en

# Part I

# Deep

# 1

# *Introduction*

## 1.1 Hello, world!

Traditionally, every programming journey starts by printing a "Hello, world"-like greeting. Let's then get it over with asap:

```
cat("My hovercraft is full of eels.\n")  # `\n` == newline
## My hovercraft is full of eels.
```

By *calling (invoking)* the **cat** function, we printed out a given character string that we enclosed in double-quote characters.

Documenting code is a good development practice. It is thus worth knowing that any text following a hash sign (that is not part of a string) is a *comment*. It is ignored by the interpreter.

```
# This is a comment.
# This is another comment.
cat("I cannot wait", "till lunchtime.\n")  # two arguments (another comment)
## I cannot wait till lunchtime.
cat("# I will not buy this record.\n# It is scratched.\n")
## # I will not buy this record.
## # It is scratched.
```

By convention, in this book, R's textual output is always preceded by two hashes. This makes it easier to copy-paste all code chunks in case we would like to experiment with them (which is always highly encouraged).

Whenever a call to a function is to be made, *the round brackets are obligatory*. All objects within the parentheses (they are separated by commas) constitute the input data to be consumed by the operation. Thus, the syntax is: `a_function_to_call(argument1, argument2, etc.)`.

## 1.2 Setting up the development environment

### 1.2.1 Installing R

It is quite natural to pine for the ability to execute the foregoing code ourselves; to learn programming without getting our hands dirty is impossible.

The official precompiled binary distributions of R can be downloaded from https://cran.r-project.org/.

For serious programming work[1], we recommend, sooner rather than later, switching to[2] one of the UNIX-like operating systems. This includes the free, open-source (== good) variants of GNU/Linux, amongst others, or the proprietary (== not so good) m**OS. In such a case, we can employ our favourite package manager (e.g., **apt**, **dnf**, **pacman**, or **Homebrew**) to install R.

Other users (e.g., of Win***s) might consider installing Anaconda or Miniconda, especially if they would like to work with Jupyter (Section 1.2.5) or Python as well.

Below we review several ways in which we can write and execute R code. It is up to the benign readers to research, set up, and learn the development environment that suits their needs. As usual in real life, there is no single universal approach that always works best in all scenarios.

### 1.2.2 Interactive mode

Whenever we would like to compute something quickly, e.g., determine basic aggregates of a few numbers entered by hand or evaluate a mathematical expression like "2+2", R's *read-eval-print loop* (REPL) can give us instant gratification.

How to start the R console varies from system to system. For instance, the users of UNIX-like boxes can simply execute **R** from the terminal (shell, command line). Those on Win***s can activate **RGui** from the *Start* menu.

**Important** When working interactively, the default[3] command prompt, ">", means: *I am awaiting orders*. Moreover, "+" denotes: *Please continue*. In the latter case, we should either complete the unfinished expression or cancel the operation by pressing ESC or CTRL+C (depending on the operating system).

```
> cat("And now
+ for something
```

[1] For instance, when interoperability with other programming languages/environments is required or when we think about scheduling jobs on Linux-based computing/container clusters.

[2] Or at least trying out – by installing a copy of GNU/Linux on a virtual machine (VM).

[3] It can be changed; see **help**("options").

```
+ completely different
+
+
+ it is an unfinished expression...
+ awaiting another double quote character and then the closing bracket...
+
+ press ESC or CTRL+C to abort input
>
```

For readability, we never print out the command prompt characters in this book.

### 1.2.3 Batch mode: Working with R scripts (**)

The interactive mode of operation is unsuitable for more complicated tasks, though. The users of UNIX-like operating systems will be interested in another extreme, which involves writing standalone R scripts that can be executed line by line without any user intervention. To do so, in the terminal, we can invoke:

```
Rscript file.R
```

where `file.R` is the path to a source file; see Section 9.2.3 for more details

**Exercise 1.1** *(**) In your favourite text editor (e.g., **`Kate`**, **`vi`**, **`Emacs`**, **`Notepad++`**, **`RStudio`**, or **`VSCodium`**), create a file named `test.R`. Write a few calls to the **`cat`** function. Then, execute this script from the terminal through **`Rscript`**.*

### 1.2.4 Weaving: Automatic report generation (**)

Reproducible data analysis[4] requires us to keep the results (text, tables, plots, auxiliary files) synchronised with the code and data that generate them.

**`utils::Sweave`** (the **`Sweave`** function from the **`utils`** package) and **`knitr`** [65] are two example template processors that evaluate R code chunks within documents written in LaTeX, HTML, or other markup languages. The chunks are replaced by the outputs they yield.

This book is a showcase of such an approach: all the results, including Figure 2.3 and the message about busy hovercrafts, were generated programmatically. Thanks to its being written in the highly universal Markdown[5] language, it could be converted to a single PDF document[6] as well as the whole website[7]. This was facilitated by tools like **`pandoc`** and **`docutils`**.

[4] The idea dates back to Knuth's literate programming concept; see [41].
[5] https://daringfireball.net/projects/markdown
[6] https://deepr.gagolewski.com/deepr.pdf
[7] https://deepr.gagolewski.com/

**Exercise 1.2** *(**) Call* **`install.packages("knitr")`** *in R. Then, create a text file named* `test.Rmd` *with the following content:*

````
# Hello, Markdown!

This is my first automatically generated report,
where I print messages and stuff.

```{r}
print("G'day!")
print(2+2)
plot((1:10)^2)
```

Thank you for your attention.
````

*Assuming that the file is located in the current working directory (compare Section 7.3.2), call* **`knitr::knit("test.Rmd")`** *from the R console, or run in the terminal:*

```
Rscript -e 'knitr::knit("test.Rmd")'
```

*Inspect the generated Markdown file,* `test.md`.

*Furthermore, if you have the* **pandoc** *tool installed, to generate a standalone HTML file, execute in the terminal:*

```
pandoc test.md --standalone -o test.html
```

*Alternatively, see Section 7.3.2 for ways to call external programs from R.*

### 1.2.5 Semi-interactive modes (Jupyter Notebooks, sending code to the associated R console, etc.)

The nature of the most frequent use cases of R encourages a semi-interactive workflow, where we quickly progress with prototyping by trial and error. In this mode, we compose a series of short code fragments inside a standalone R script. Each fragment implements a simple, well-defined task, such as loading data files, data cleansing, feature visualisation, computations of information aggregates, etc. Importantly, any code chunk can be *sent* to the associated R console and executed there. This way, we can inspect the result it generates. If we are not happy with the outcome, we can apply the necessary corrections.

There are quite a few integrated development environments that enable such a workflow, including **JupyterLab**, **Emacs**, **RStudio**, and **VSCodium**. Some of them require additional plugins for R.

Executing an individual code line or a whole text selection is usually done by pressing (configurable) keyboard shortcuts such as `Ctrl+Enter` or `Shift+Enter`.

**Exercise 1.3** *(*) `JupyterLab`[8] is a development environment that runs in a web browser. It was programmed in Python, but supports many programming languages. Thanks to `IRkernel`[9], we can use it with R.*

1. *Install **`JupyterLab`** and **`IRkernel`** (for instance, if you use Anaconda, run `conda install -c r r-essentials`).*
2. *From the* File *menu, select* Create a new R source file *and save it as, e.g., `test.R`.*
3. *Click* File *and select* Create a new console for the editor running the R kernel.
4. *Input a few print "Hello, world"-like calls.*
5. *Press `Shift+Enter` (whilst working in the editor) to send different code fragments to the console and execute them. Inspect the results.*

*See Figure 1.1 for an illustration. Note that issuing **`options`**`(jupyter.rich_display=FALSE)` may be necessary to disable rich HTML outputs and make them look more like ones in this book.*

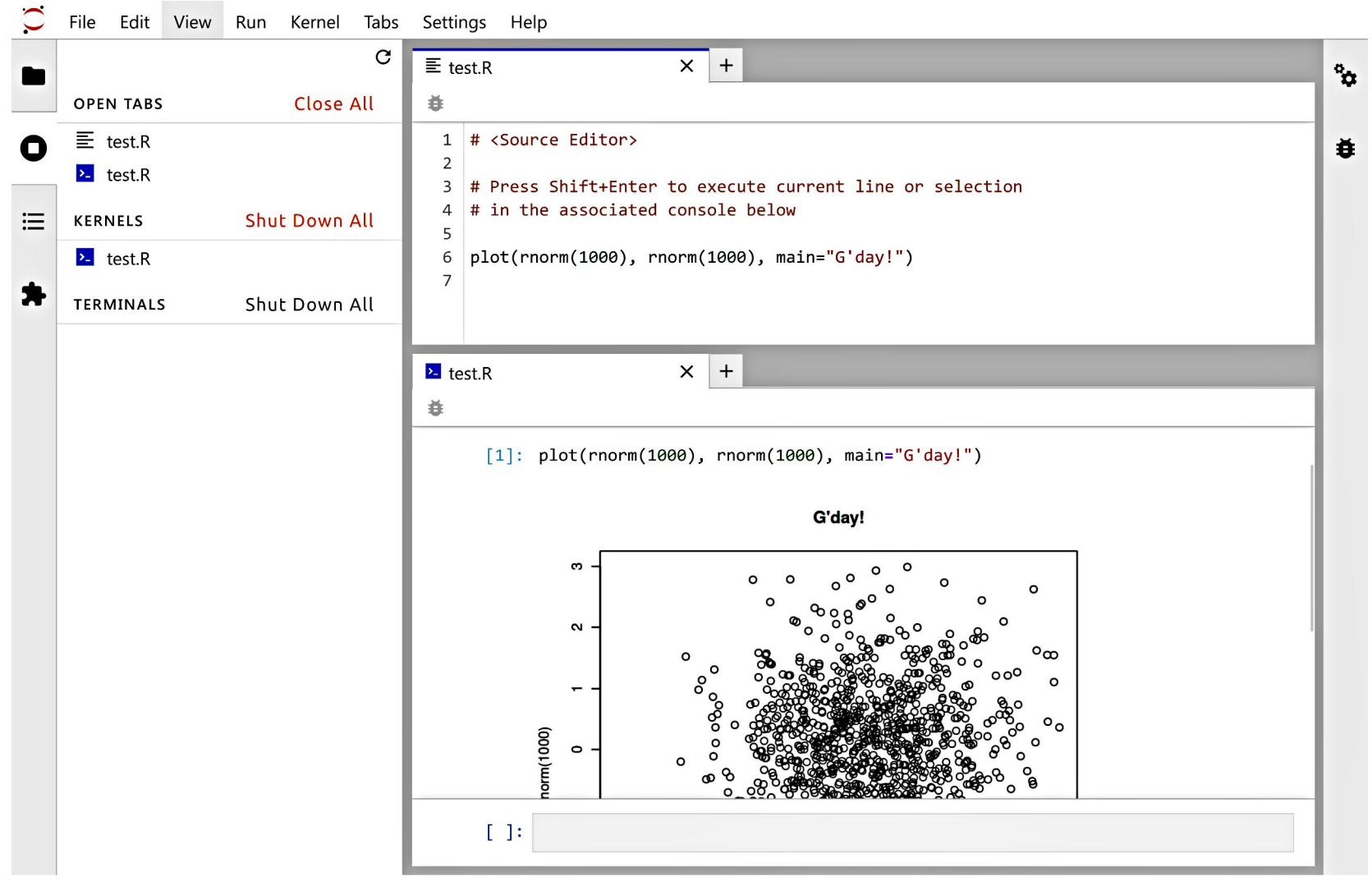

Figure 1.1. JupyterLab: A source file editor and the associated R console, where we can run arbitrary code fragments.

**Example 1.4** *(*) **`JupyterLab`** also handles dedicated Notebooks, where editable and executable code chunks and results they generate can be kept together in a single `.ipynb` (JSON) file; see Figure 1.2 for an illustration and Chapter 1 of [28] for a quick introduction (from the Python language kernel perspective).*

*This environment is convenient for live coding (e.g., for teachers) or performing exploratory data analyses. However, for more serious programming work, the code can get messy. Luckily, there is always an option to export a notebook to an executable, plain text R script.*

[8] https://jupyterlab.readthedocs.io/en/stable
[9] https://irkernel.github.io/

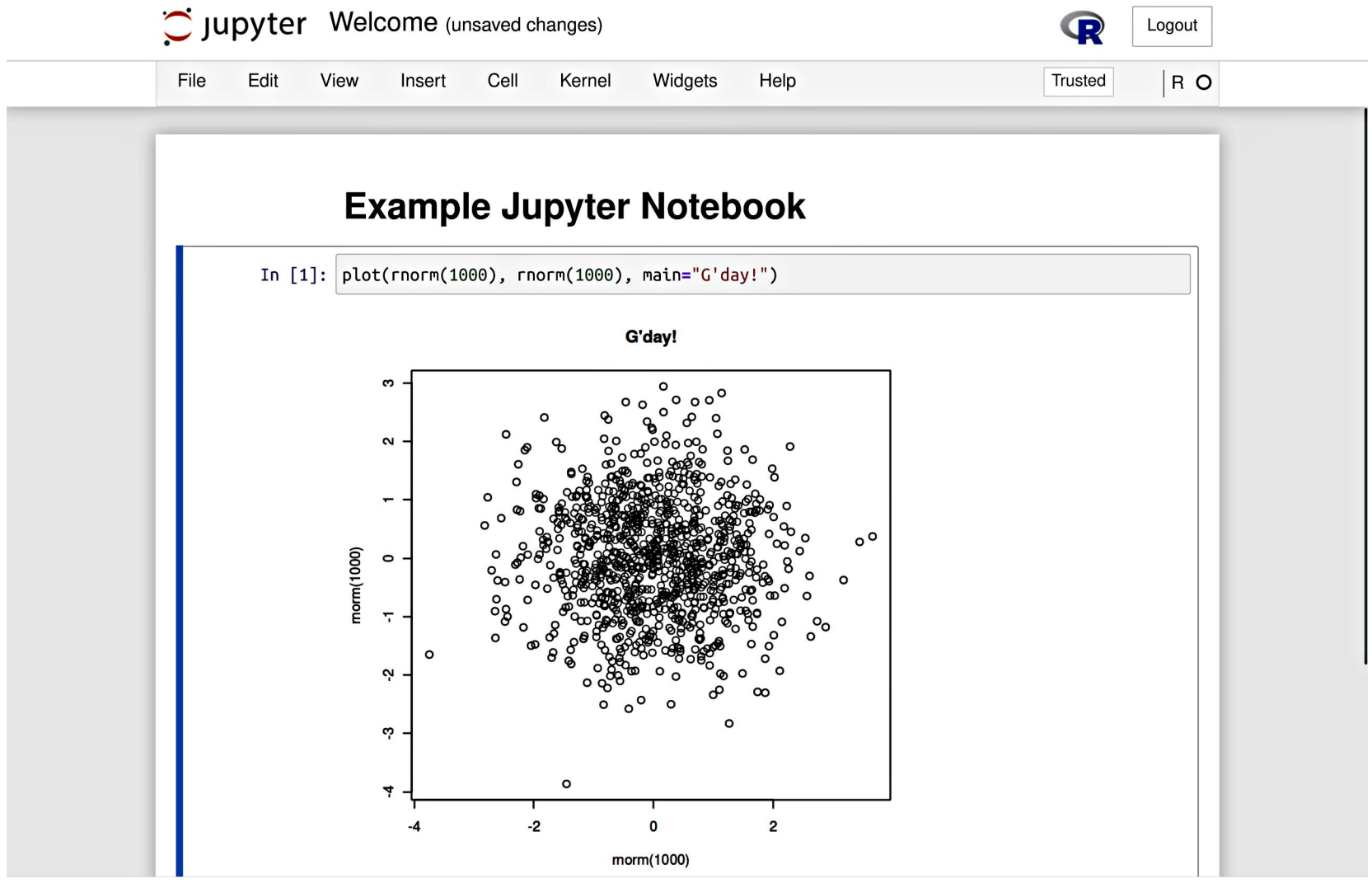

Figure 1.2. An example Jupyter Notebook, where we can keep code and results together.

---

## 1.3 Atomic vectors at a glance

After printing "Hello, world", a typical programming course would normally proceed with the discussion on basic data types for storing individual numeric or logical values. Next, we would be introduced to arithmetic and relational operations on such *scalars*, followed by the definition of whole arrays or other collections of values, complemented by the methods to iterate over them, one element after another.

In R, no separate types representing individual values have been defined. Instead, what seems to be a single datum, is already a *vector* (sequence, array) of length one.

```
2.71828          # input a number; here: the same as print(2.71828)
## [1] 2.7183
length(2.71828)  # it is a vector with one element
## [1] 1
```

To create a vector of any length, we can call the **c** function, which combines given arguments into a single sequence:

```
c(1, 2, 3)  # three values combined
## [1] 1 2 3
length(c(1, 2, 3))  # indeed, it is a vector of length three
## [1] 3
```

In Chapter 2, Chapter 3, and Chapter 6, we will discuss the most prevalent types of atomic vectors: numeric, logical, and character ones, respectively.

```
c(0, 1, -3.14159, 12345.6)           # four numbers
## [1]     0.0000     1.0000    -3.1416 12345.6000
c(TRUE, FALSE)                       # two logical values
## [1]  TRUE FALSE
c("spam", "bacon", "spam")           # three character strings
## [1] "spam"  "bacon" "spam"
```

We call them *atomic* for they can only group together values of the same type. Lists, which we will discuss in Chapter 4, are, on the other hand, referred to as *generic* vectors. They can be used for storing items of mixed types: other lists as well.

---

**Note** Not having separate scalar types greatly simplifies the programming of numerical computing tasks. Vectors are prevalent in our main areas of interest: statistics, simulations, data science, machine learning, and all other data-orientated computing. For example, columns and rows in tables (characteristics of clients, ratings of items given by users) or time series (stock market prices, readings from temperature sensors) are all best represented by means of such sequences.

The fact that vectors are the core part of the R language makes their use very natural, as opposed to the languages that require special add-ons for vector processing, e.g., `numpy` for Python [35]. By learning different ways to process them *as a whole* (instead of one element at a time), we will ensure that our ideas can quickly be turned into operational code. For instance, computing summary statistics such as, say, the mean absolute deviation of a sequence `x`, will be as effortless as writing `mean(abs(x-mean(x)))`. Such code is not only easy to read and maintain, but it is also fast to run.

---

## 1.4 Getting help

Our aim is to become independent, advanced R programmers.

Independent, however, does not mean omniscient. The *R help system* is the authoritative source of knowledge about specific functions or more general topics. To open a help page, we call:

```
help("topic")  # equivalently: ?"topic"
```

**Exercise 1.5** *Sight (without going into detail) the manual on the* `length` *function by calling* `help("length")`*. Note that most help pages are structured as follows:*

1. Header: *`package:base` means that the function is a base one (see Section 7.3.1 for more details on the R package system);*

2. Title*;*
3. Description*: a short description of what the function does;*
4. Usage*: the list of formal arguments (parameters) to the function;*
5. Arguments*: the meaning of each formal argument explained;*
6. Details*: technical information;*
7. Value*: return value explained;*
8. References*: further reading;*
9. See Also*: links to other help pages;*
10. Examples*: R code that is worth inspecting.*

We can also search within all the installed help pages by calling:

```
help.search("vague topic")  # equivalently: ??"vague topic"
```

This way, we will be able to find answers to our questions more reliably than when asking DuckDuckGo or G**gle, which commonly return many low-quality, irrelevant, or distracting results from splogs. We do not want to lose the sacred code writer's flow! It is a matter of personal hygiene and good self discipline.

---

**Important** All code chunks, including code comments and textual outputs, form an integral part of this book's text. They should not be skipped by the reader. On the contrary, they must become objects of our intense reflection and thorough investigation.

For instance, whenever we introduce a function, it may be a clever idea to look it up in the help system. Moreover, playing with the presented code (running, modifying, experimenting, etc.) is also very beneficial. We should develop the habit of asking ourselves questions like "What would happen if…", and then finding the answers on our own.

---

We are now ready to discuss the most significant operations on numeric vectors, which constitute the main theme of the next chapter. See you there.

## 1.5 Exercises

**Exercise 1.6** *What are the three most important types of atomic vectors?*

**Exercise 1.7** *According to the classification of the R data types we introduced in the previous chapter, are atomic vectors basic or compound types?*

# 2

# *Numeric vectors*

In this chapter, we discuss the uttermost common operations on numeric vectors. They are so fundamental that we will also find them in other scientific computing environments, including Python with **numpy** or **tensorflow**, Julia, MATLAB, GNU Octave, or Scilab.

At first blush, the number of functions we are going to explore may seem quite large. Still, the reader is kindly asked to place some trust (a rare thing these days) in yours truly. It is because our selection is comprised only of the most representative and educational amongst the plethora of possible choices. More complex building blocks can often be reduced to a creative combination of the former or be easily found in a number of additional packages or libraries (e.g., GNU GSL [29]).

A solid understanding of base R programming is crucial for dealing with popular packages (such as **data.table**, **dplyr**, or **caret**). Most importantly, base R's API is *stable*. Hence, the code we compose today will most likely work the same way in ten years. It is often not the case when we rely on external add-ons.

In the sequel, we will be advocating a minimalist, keep-it-simple approach to the art of programming data processing pipelines, one that is a healthy balance between "doing it all by ourselves", "minimising the information overload", "being lazy", and "standing on the shoulders of giants".

---

**Note** The exercises that we suggest in the sequel are all self-contained, unless explicitly stated otherwise. The use of language constructs that are yet to be formally introduced (in particular, `if`, `for`, and `while` explained in Chapter 8) is not just unnecessary: it is discouraged. Moreover, we recommend against taking shortcuts by looking up partial solutions on the internet. Rather, to get the most out of this course, we should be seeking relevant information within the current and preceding chapters as well as the R help system.

---

## 2.1 Creating numeric vectors

### 2.1.1 Numeric constants

The simplest numeric vectors are those of length one:

```
-3.14
## [1] -3.14
1.23e-4
## [1] 0.000123
```

The latter is in what we call *scientific notation*, which is a convenient means of entering numbers of very large or small orders of magnitude. Here, "e" stands for "... times 10 to the power of...". Therefore, `1.23e-4` is equal to $1.23 \times 10^{-4} = 0.000123$. In other words, given 1.23, we move the decimal separator by four digits towards the left, adding zeroes if necessary.

In real life, some information items may be inherently or temporarily missing, unknown, or Not Available. As R is orientated towards data processing, it was equipped with a special indicator:

```
NA_real_  # numeric NA (missing value)
## [1] NA
```

It is similar to the *Null* marker in database query languages such as SQL. Note that `NA_real_` is displayed simply as "NA", chiefly for readability.

Moreover, `Inf` denotes infinity, $\infty$, i.e., an element that is larger than the largest representable double-precision (64 bit) floating point value. Also, `NaN` stands for *not-a-number*, which is returned as the result of some illegal operations, e.g., $0/0$ or $\infty - \infty$.

Let's provide a few ways to create numeric vectors with possibly more than one element.

### 2.1.2 Concatenating vectors with `c`

First, the **c** function can be used to combine (concatenate) many numeric vectors, each of any length. It results in a single object:

```
c(1, 2, 3)  # three vectors of length one -> one vector of length three
## [1] 1 2 3
c(1, c(2, NA_real_, 4), 5, c(6, c(7, Inf)))
## [1]   1   2  NA   4   5   6   7 Inf
```

**Note** Running `help("c")`, we will see that its usage is like `c(...)`. In the current context, this means that the **c** function takes an arbitrary number of arguments. In Section 9.4.6, we will study the dot-dot-dot (ellipsis) parameter in more detail.

### 2.1.3 Repeating entries with `rep`

Second, **rep** *rep*licates the elements in a vector a given number of times.

```
rep(1, 5)
## [1] 1 1 1 1 1
rep(c(1, 2, 3), 4)
##  [1] 1 2 3 1 2 3 1 2 3 1 2 3
```

In the second case, the whole vector (1, 2, 3) has been *recycled* (tiled) four times. Interestingly, if the second argument is a vector of the same length as the first one, the behaviour will be different:

```
rep(c(1, 2, 3), c(2, 1, 4))
## [1] 1 1 2 3 3 3 3
rep(c(1, 2, 3), c(4, 4, 4))
##  [1] 1 1 1 1 2 2 2 2 3 3 3 3
```

Here, *each* element is repeated the *corresponding* number of times.

Calling `help("rep")`, we find that the function's usage is like `rep(x, ...)`. It is rather peculiar. However, reading further, we discover that the ellipsis (dot-dot-dot) may be fed with one of the following parameters:

- `times`,
- `length.out`[1],
- `each`.

So far, we have been playing with `times`, which is listed second in the parameter list (after `x`, the vector whose elements are to be repeated).

---

**Important** The undermentioned function calls are all equivalent:

```
rep(c(1, 2, 3), 4)   # positional matching of arguments: `x`, then `times`
rep(c(1, 2, 3), times=4)    # `times` is the second argument
rep(x=c(1, 2, 3), times=4)  # keyword arguments of the form name=value
rep(times=4, x=c(1, 2, 3))  # keyword arguments can be given in any order
rep(times=4, c(1, 2, 3))    # mixed positional and keyword arguments
```

---

We can also pass `each` or `length.out`, but their names must be mentioned explicitly:

```
rep(c(1, 2, 3), length.out=7)
## [1] 1 2 3 1 2 3 1
rep(c(1, 2, 3), each=3)
## [1] 1 1 1 2 2 2 3 3 3
rep(c(1, 2, 3), length.out=7, each=3)
## [1] 1 1 1 2 2 2 3
```

---

**Note** Whether we consider a good programming practice the implementation of a range of varied behaviours inside a single function is a question of taste. On the one hand, in all of the preceding examples, we *do* repeat the input elements somehow, so remembering just one function name is really convenient. Nevertheless, a drastic change in the repetition pattern depending, e.g., on the length of the `times` argument can be bug-prone. Anyway, we have been warned[2].

[1] A dot has no special meaning in R; see Section 2.2.

---

Zero-length vectors are possible too:

```
rep(c(1, 2, 3), 0)
## numeric(0)
```

Even though their handling might be a little tricky, we will later see that they are indispensable in contexts like "create an empty data frame with a specific column structure".

Also, note that R often allows for partial matching of named arguments, but its use is a bad programming practice; see Section 15.4.4 for more details.

```
rep(c(1, 2, 3), len=7)  # not recommended (see later)
## Warning in rep(c(1, 2, 3), len = 7): partial argument match of 'len' to
##     'length.out'
## [1] 1 2 3 1 2 3 1
```

We see the warning only because we have set **options**(warnPartialMatchArgs=TRUE) in our environment. It is not used by default.

### 2.1.4 Generating arithmetic progressions with seq and `:`

Third, we can call the **seq** function to create a sequence of equally-spaced numbers on a linear scale, i.e., an arithmetic progression.

```
seq(1, 15, 2)
## [1]  1  3  5  7  9 11 13 15
```

From the function's help page, we discover that **seq** accepts the `from`, `to`, `by`, and `length.out` arguments, amongst others. Thus, the preceding call is equivalent to:

```
seq(from=1, to=15, by=2)
## [1]  1  3  5  7  9 11 13 15
```

Note that `to` actually means "up to":

[2] Some "caring" R users might be tempted to introduce two new functions now, one for generating (1, 2, 3, 1, 2, 3, …) only and the other outputting patterns like (1, 1, 1, 2, 2, 2, …). They would most likely wrap them in a new package and announce that on social media. But this is nothing else than a multiplication of entities without actual necessity. This way, we would end up with three functions. First is the original one, **rep**, which everyone ought to know anyway because it is part of the standard library. Second and third are the two redundant procedures whose user-friendliness is only illusory. See also Chapter 9 for a discussion on the design of functions.

```
seq(from=1, to=16, by=2)
## [1]  1  3  5  7  9 11 13 15
```

We can also pass `length.out` instead of `by`. In such a case, the increments or decrements will be computed via the formula `((to - from)/(length.out - 1))`. This *default value* is reported in the *Usage* section of **help**("seq").

```
seq(1, 0, length.out=5)
## [1] 1.00 0.75 0.50 0.25 0.00
seq(length.out=5)  # default `from` is 1
## [1] 1 2 3 4 5
```

Arithmetic progressions with steps equal to 1 or -1 can also be generated via the `:` operator.

```
1:10    # seq(1, 10) or seq(1, 10, 1)
##  [1]  1  2  3  4  5  6  7  8  9 10
-1:10   # seq(-1, 10) or seq(-1, 10, 1)
##  [1] -1  0  1  2  3  4  5  6  7  8  9 10
-1:-10  # seq(-1, -10) or seq(-1, -10, -1)
##  [1]  -1  -2  -3  -4  -5  -6  -7  -8  -9 -10
```

Let's highlight the order of precedence of this operator: `-1:10` means `(-1):10`, and not `-(1:10)`; compare Section 2.4.3.

**Exercise 2.1** *Take a look at the manual page of **seq_along** and **seq_len** and determine whether we can do without them, having **seq**[3] at hand.*

### 2.1.5 Generating pseudorandom numbers

We can also generate sequences drawn independently from a range of univariate probability distributions.

```
runif(7)  # uniform U(0, 1)
## [1] 0.287578 0.788305 0.408977 0.883017 0.940467 0.045556 0.528105
rnorm(7)  # normal N(0, 1)
## [1]  1.23950 -0.10897 -0.11724  0.18308  1.28055 -1.72727  1.69018
```

These correspond to seven pseudorandom deviates from the uniform distribution on the unit interval (i.e., $(0, 1)$) and the standard normal distribution (i.e., with expectation 0 and standard deviation 1), respectively; compare Figure 2.3.

For more *named* distribution classes frequently occur in various real-world statistical modelling exercises, see Section 2.3.4.

Another worthwhile function picks items from a given vector, either with or without replacement:

[3] Certain configurations of **seq** and its variants might return vectors of the type `integer` instead of `double`, some of them in a compact (ALTREP) form; see Section 6.4.1.

```
sample(1:10, 20, replace=TRUE)  # 20 with replacement (allow repetitions)
##  [1]  3  3 10  2  6  5  4  6  9 10  5  3  9  9  9  3  8 10  7 10
sample(1:10, 5, replace=FALSE)  # 5 without replacement (do not repeat)
## [1] 9 3 4 6 1
```

The distribution of the sampled values does not need to be uniform; the `prob` argument may be fed with a vector of the corresponding probabilities. For example, here are 20 independent realisations of the random variable $X$ such that $\Pr(X = 0) = 0.9$ (the probability that we obtain 0 is equal to 90%) and $\Pr(X = 1) = 0.1$:

```
sample(0:1, 20, replace=TRUE, prob=c(0.9, 0.1))
##  [1] 0 0 0 0 1 0 0 0 0 0 1 0 0 0 0 0 0 0 0 1
```

---

**Note** If `n` is a single number (a numeric vector of length 1), then `sample(n, ...)` is equivalent to `sample(1:n, ...)`. Similarly, `seq(n)` is a synonym for `seq(1, n)` or `seq(1, length(n))`, depending on the length of `n`. This is a dangerous behaviour that can occasionally backfire and lead to bugs (check what happens when `n` is, e.g., 0). Nonetheless, we have been warned. From now on, we are going to be extra careful (but are we really?). Read more at `help("sample")` and `help("seq")`.

---

Let's stress that the numbers we obtain are merely *pseudo*random because they are generated algorithmically. R uses the Mersenne-Twister MT19937 method [47] by default; see `help("RNG")` and [22, 30, 43]. By setting the *seed* of the random number generator, i.e., resetting its state to a given one, we can obtain results that are *reproducible*.

```
set.seed(12345)  # seeds are specified with integers
sample(1:10, 5, replace=TRUE)  # a,b,c,d,e
## [1]  3 10  8 10  8
sample(1:10, 5, replace=TRUE)  # f,g,h,i,j
## [1]  2  6  6  7 10
```

Setting the seed to the one used previously gives:

```
set.seed(12345)
sample(1:10, 5, replace=TRUE)  # a,b,c,d,e
## [1]  3 10  8 10  8
```

We did not(?) expect that! And now for something completely different:

```
set.seed(12345)
sample(1:10, 10, replace=TRUE)  # a,b,c,d,e,f,g,h,i,j
##  [1]  3 10  8 10  8  2  6  6  7 10
```

Reproducibility is a crucial feature of each truly scientific experiment. The same initial condition (here: the same seed) leads to exactly the same outcomes.

---

**Note** Some claim that the only unsuspicious seed is 42 but in matters of taste, there can be no disputes. Everyone can use their favourite picks: yours truly savours 123, 1234, and 12345 as well.

When performing many runs of Monte Carlo experiments, it may also be a clever idea to call **set.seed**(i) in the *i*-th iteration of a simulation we are trying to program.

We should ensure that our seed settings are applied consistently across all our scripts. Otherwise, we might be accused of tampering with evidence. For instance, here is the ultimate proof that we are very lucky today:

```
set.seed(1679619)  # totally unsuspicious, right?
sample(0:1, 20, replace=TRUE)  # so random
##  [1] 1 1 1 1 1 1 1 1 1 1 1 1 1 1 1 1 1 1 1 1
```

This is exactly why reproducible scripts and auxiliary data should be published alongside all research reports or papers. Only open, transparent science can be fully trustworthy.

---

If **set.seed** is not called explicitly, and the random state is not restored from the previously saved R session (see Chapter 16), then the random generator is initialised based on the current wall time and the identifier of the running R instance (PID). This may justify the impression that the numbers we generate appear surprising.

To understand the "pseudo" part of the said randomness better, in Section 8.3, we will build a very simple random generator ourselves.

### 2.1.6 Reading data with scan

An example text file named `euraud-20200101-20200630.csv`[4] gives the EUR to AUD exchange rates (how many Australian Dollars can we buy for 1 Euro) from 1 January to 30 June 2020 (remember COVID-19?). Let's preview the first couple of lines:

```
# EUR/AUD Exchange Rates
# Source: Statistical Data Warehouse of the European Central Bank System
# https://www.ecb.europa.eu/stats/policy_and_exchange_rates/
# (provided free of charge)
NA
1.6006
1.6031
NA
```

The four header lines that begin with "#" merely serve as comments for us humans. They should be ignored by the interpreter. The first "real" value, `NA`, corresponds to 1 January (Wednesday, New Year's Day; Forex markets were closed, hence a missing observation).

We can invoke the **scan** function to read all the inputs and convert them to a single numeric vector:

[4] https://github.com/gagolews/teaching-data/raw/master/marek/euraud-20200101-20200630.csv

```
scan(paste0("https://github.com/gagolews/teaching-data/raw/",
    "master/marek/euraud-20200101-20200630.csv"), comment.char="#")
##  [1]     NA 1.6006 1.6031     NA     NA 1.6119 1.6251 1.6195 1.6193 1.6132
## [11]     NA     NA 1.6117 1.6110 1.6188 1.6115 1.6122     NA     NA 1.6154
## [21] 1.6177 1.6184 1.6149 1.6127     NA     NA 1.6291 1.6290 1.6299 1.6412
## [31] 1.6494     NA     NA 1.6521 1.6439 1.6299 1.6282 1.6417     NA     NA
## [41] 1.6373 1.6260 1.6175 1.6138 1.6151     NA     NA 1.6129 1.6195 1.6142
## [51] 1.6294 1.6363     NA     NA 1.6384 1.6442 1.6565 1.6672 1.6875     NA
## [61]     NA 1.6998 1.6911 1.6794 1.6917 1.7103     NA     NA 1.7330 1.7377
## [71] 1.7389 1.7674 1.7684     NA     NA 1.8198 1.8287 1.8568 1.8635 1.8226
## [81]     NA     NA 1.8586 1.8315 1.7993 1.8162 1.8209     NA     NA 1.8021
## [91] 1.7967 1.8053 1.7970 1.8004     NA     NA 1.7790 1.7578 1.7596
##  [ reached 'max' / getOption("max.print") -- omitted 83 entries ]
```

We used the **paste0** function (Section 6.1.3) to concatenate two long strings (too long to fit a single line of code) and form a single URL.

We can also read the files located on our computer. For example:

```
scan("~/Projects/teaching-data/marek/euraud-20200101-20200630.csv",
    comment.char="#")
```

It used an absolute file path that starts at the user's *home directory*, denoted "~". Yours truly's case is `/home/gagolews`.

---

**Note** For portability reasons, we suggest slashes, "/", as path separators; see also **help**("file.path") and **help**(".Platform"). They are recognised by all UNIX-like boxes as well as by other popular operating systems, including Win***s. Note that URLs, such as https://deepr.gagolewski.com/, consist of slashes, too.

---

Paths can also be relative to the *current working directory*, denoted ".", which can be read via a call to **getwd**. Usually, it is the location wherefrom the R session has been started. For instance, if the working directory was `/home/gagolews/Projects/teaching-data/marek`, we could write the file path equivalently as `./euraud-20200101-20200630.csv` or even `euraud-20200101-20200630.csv`.

On as side note, ".." marks the *parent directory* of the current working directory. In the above example, `../r/iris.csv` is equivalent to `/home/gagolews/Projects/teaching-data/r/iris.csv`.

**Exercise 2.2** *Read the help page about* **scan***. Take note of the following formal arguments and their meaning:* `dec`*,* `sep`*,* `what`*,* `comment.char`*, and* `na.strings`*.*

Later we will discuss the **read.table** and **read.csv** functions. They are wrappers around **scan** that reads *structured* data. Also, **write** exports an atomic vector's contents to a text file.

**Example 2.3** *Figure 2.1 shows the graph of the aforementioned exchange rates, which was generated by calling:*

```
plot(scan(paste0("https://github.com/gagolews/teaching-data/raw/",
    "master/marek/euraud-20200101-20200630.csv"), comment.char="#"),
    xlab="Day", ylab="EUR/AUD")
```

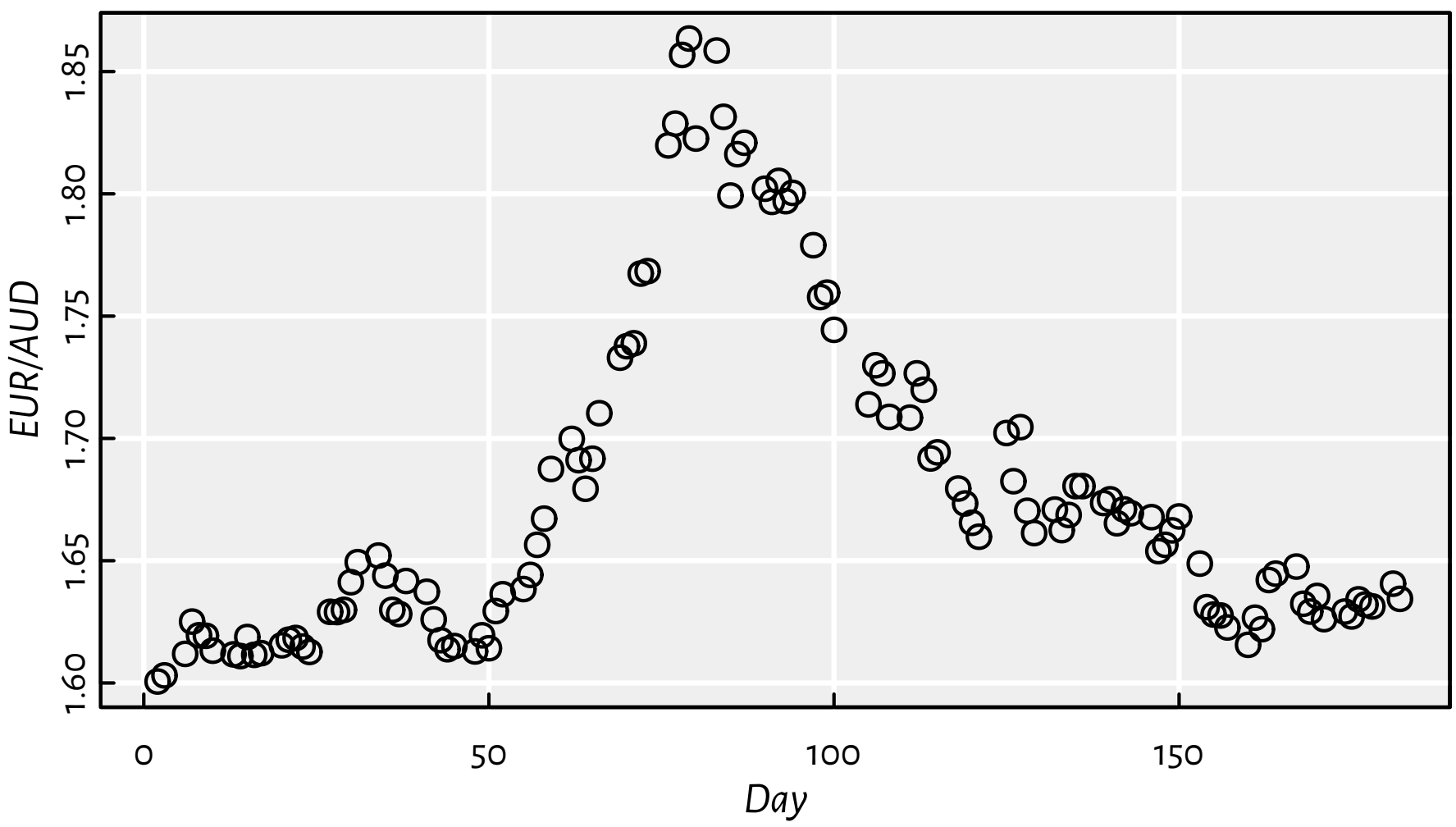

Figure 2.1. EUR/AUD exchange rates from 2020-01-01 (day 1) to 2020-06-30 (day 182).

*Somewhat misleadingly (and for reasons that will become apparent later), the documentation of* **`plot`** *can be accessed by calling* **`help`**`("plot.default")`*. Read about, and experiment with, different values of the* `main`, `xlab`, `ylab`, `type`, `col`, `pch`, `cex`, `lty`, *and* `lwd` *arguments. More plotting routines will be discussed in Chapter 13.*

## 2.2 Creating named objects

The objects we bring forth will often need to be memorised so that they can be referred to in further computations. The assignment operator, **`` `<-` ``**, can be used for this purpose:

```
x <- 1:3  # creates a numeric vector and binds the name `x` to it
```

The now-named object can be recalled[5] and dealt with as we please:

```
print(x)  # or just `x` in the R console
## [1] 1 2 3
```

[5] Name bindings are part of environment frames; see Chapter 16.

```
sum(x)    # example operation: compute the sum of all elements in `x`
## [1] 6
```

**Important** In R, all names are *case-sensitive*. Hence, `x` and `X` can coexist peacefully: when set, they refer to two different objects. If we tried calling `Print(x)`, `print(X)`, or `PRINT(x)`, we would get an error.

Typically, we will be using *syntactic names*. In `help("make.names")`, we read: *A syntactically valid name consists of letters, numbers and the dot or underline characters and starts with a letter or the dot not followed by a number. Names such as* `.2way` *are not valid, and neither are the reserved words* such as `if`, `for`, `function`, `next`, and `TRUE`, but see Section 9.3.1 for an exception.

A fine name is self-explanatory and thus reader-friendly: `patients`, `mean`, and `average_scores` are way better (if they are what they claim they are) than `xyz123`, `crap`, or `spam`. Also, it might not be such a bad idea to get used to denoting:

- vectors by `x`, `y`, `z`,
- matrices (and matrix-like objects) by `A`, `B`, …, `X`, `Y`, `Z`,
- integer indexes by letters `i`, `j`, `k`, `l`,
- object sizes by `n`, `m`, `d`, `p` or `nx`, `ny`, etc.,

especially when they are only of temporary nature (for storing auxiliary results, iterating over collections of objects, etc.).

There are numerous naming conventions that we can adopt, but most often they are a matter of taste; `snake_case`, `lowerCamelCase`, `UpperCamelCase`, `flatcase`, or `dot.case` are equally sound as long as they are used coherently (for instance, some use `snake_case` for vectors and `UpperCamelCase` for functions). Occasionally, we have little choice but to adhere to the naming conventions of the project we are about to contribute to.

**Note** Generally, a dot, "`.`", has no special meaning[6]; `na.omit` is as appropriate a name as `na_omit`, `naOmit`, `NAOMIT`, `naomit`, and `NaOmit`. Readers who know other programming languages will need to habituate themselves to this convention.

R, as a dynamic language, allows for introducing new variables at any time. Moreover, existing names can be bound to new values. For instance:

```
(y <- "spam")  # bracketed expression - printing not suppressed
## [1] "spam"
```

[6] See Section 10.2 and Section 16.2.1 for a few asterisks.

```
x <- y  # overwrites the previous `x`
print(x)
## [1] "spam"
```

Now x refers to a verbatim copy of y.

**Note** Objects are automatically destroyed when we cannot access them anymore. By now, the *garbage collector* is likely to have got rid of the foregoing 1:3 vector (to which the name x was bound previously).

## 2.3 Vectorised mathematical functions

Mathematically, we will be denoting a given vector $\boldsymbol{x}$ of length $n$ by $(x_1, x_2, \ldots, x_n)$. In other words, $x_i$ is its $i$-th element. Let's review a few operations that are ubiquitous in numerical computing.

### 2.3.1 abs and sqrt

R implements *vectorised* versions of the most popular mathematical functions, e.g., **abs** (absolute value, $|x|$) and **sqrt** (square root, $\sqrt{x}$).

```
abs(c(2, -1, 0, -3, NA_real_))
## [1]  2  1  0  3 NA
```

Here, *vectorised* means that instead of being defined to act on a single numeric value, they are applied on *each* element in a vector. The $i$-th resulting item is a transformed version of the $i$-th input:

$$|\boldsymbol{x}| = (|x_1|, |x_2|, \ldots, |x_n|).$$

Moreover, if an input is a missing value, the corresponding output will be marked as unknown as well.

Another example:

```
x <- c(4, 2, -1)
(y <- sqrt(x))
## Warning in sqrt(x): NaNs produced
## [1] 2.0000 1.4142    NaN
```

To attract our attention to the fact that computing the square root of a negative value is a reckless act, R generated an informative warning. However, *a warning is not an error*:

the result is being produced as usual. In this case, the ill value is marked as not-a-number.

Also, the fact that the irrational $\sqrt{2}$ is *displayed*[7] as `1.4142` does not mean that it is such a crude approximation to 1.4142135623730950488016887242096980.... It was rounded when printing purely for aesthetic reasons. In fact, in Section 3.2.3, we will point out that the computer's floating-point arithmetic has roughly 16 decimal digits precision (but we shall see that the devil is in the detail).

```
print(y, digits=16)  # display more significant figures
## [1] 2.000000000000000 1.414213562373095               NaN
```

### 2.3.2 Rounding

The following functions drop all or portions of fractional parts of numbers:

- **floor**(x) (rounds down to the nearest integer, denoted $\lfloor x \rfloor$),
- **ceiling**(x) (rounds up, denoted $\lceil x \rceil = -\lfloor -x \rfloor$),
- **trunc**(x) (rounds towards zero),
- **round**(x, digits=0) (rounds to the nearest number with `digits` decimal digits).

For instance:

```
x <- c(7.0001, 6.9999, -4.3149, -5.19999, 123.4567, -765.4321, 0.5, 1.5, 2.5)
floor(x)
## [1]    7    6   -5   -6  123 -766    0    1    2
ceiling(x)
## [1]    8    7   -4   -5  124 -765    1    2    3
trunc(x)
## [1]    7    6   -4   -5  123 -765    0    1    2
```

**Note** When we write that a function's usage is like **round**(x, digits=0), compare **help**("round"), we mean that the `digits` parameter is equipped with the *default value* of 0. In other words, if rounding to 0 decimal digits is what we need, the second argument can be omitted.

```
round(x)  # the same as round(x, 0); round half to even
## [1]    7    7   -4   -5  123 -765    0    2    2
round(x, 1)  # round to tenths (nearest 0.1s)
## [1]    7.0    7.0   -4.3   -5.2  123.5 -765.4    0.5    1.5    2.5
round(x, -2)  # round to hundreds (nearest 100s)
## [1]    0    0    0    0  100 -800    0    0    0
```

[7] There are a couple of settings in place that control the default behaviour of the **print** function; see `width`, `digits`, `max.print`, `OutDec`, `scipen`, etc. in **help**("options").

### 2.3.3 Natural exponential function and logarithm

Moreover:

- `exp(x)` outputs the natural exponential function, $e^x$, where Euler's number $e \simeq 2.718$,
- `log(x, base=exp(1))` computes, by default, the natural logarithm of $x$, $\log_e x$ (which is most often denoted simply by $\log x$).

Recall that if $x = e^y$, then $\log_e x = y$, i.e., one is the inverse of the other.

```
log(c(0, 1, 2.7183, 7.3891, 20.0855))  # grows slowly
## [1] -Inf    0    1    2    3
exp(c(0, 1, 2, 3))                      # grows fast
## [1]  1.0000  2.7183  7.3891 20.0855
```

These functions enjoy a number of very valuable identities and inequalities. In particular, we should know from school that $\log(x \cdot y) = \log x + \log y$, $\log(x^y) = y \log x$, and $e^{x+y} = e^x \cdot e^y$.

For the logarithm to a different base, say, $\log_{10} x$, we can call:

```
log(c(0, 1, 10, 100, 1000, 1e10), 10)  # or log(..., base=10)
## [1] -Inf    0    1    2    3   10
```

Recall that if $\log_b x = y$, then $x = b^y$, for any $1 \neq b > 0$.

**Example 2.4** *Commonly, a logarithmic scale is used for variables that grow rapidly when expressed as functions of each other; see Figure 2.2.*

```
x <- seq(0, 10, length.out=1001)
par(mfrow=c(1, 2))  # two plots in one figure (one row, two columns)
plot(x, exp(x), type="l")  # left subplot
plot(x, exp(x), type="l", log="y")  # log-scale on the y-axis; right subplot
```

*Let's highlight that $e^x$ on the log-scale is nothing more than a straight line. Such a transformation of the axes can only be applied in the case of values strictly greater than 0.*

### 2.3.4 Probability distributions (*)

It should come as no surprise that R offers extensive support for many univariate probability distribution families, including:

- continuous distributions, i.e., those whose support is comprised of uncountably many real numbers (e.g., some interval or the whole real line):
  - `*unif` (uniform),
  - `*norm` (normal),
  - `*exp` (exponential),

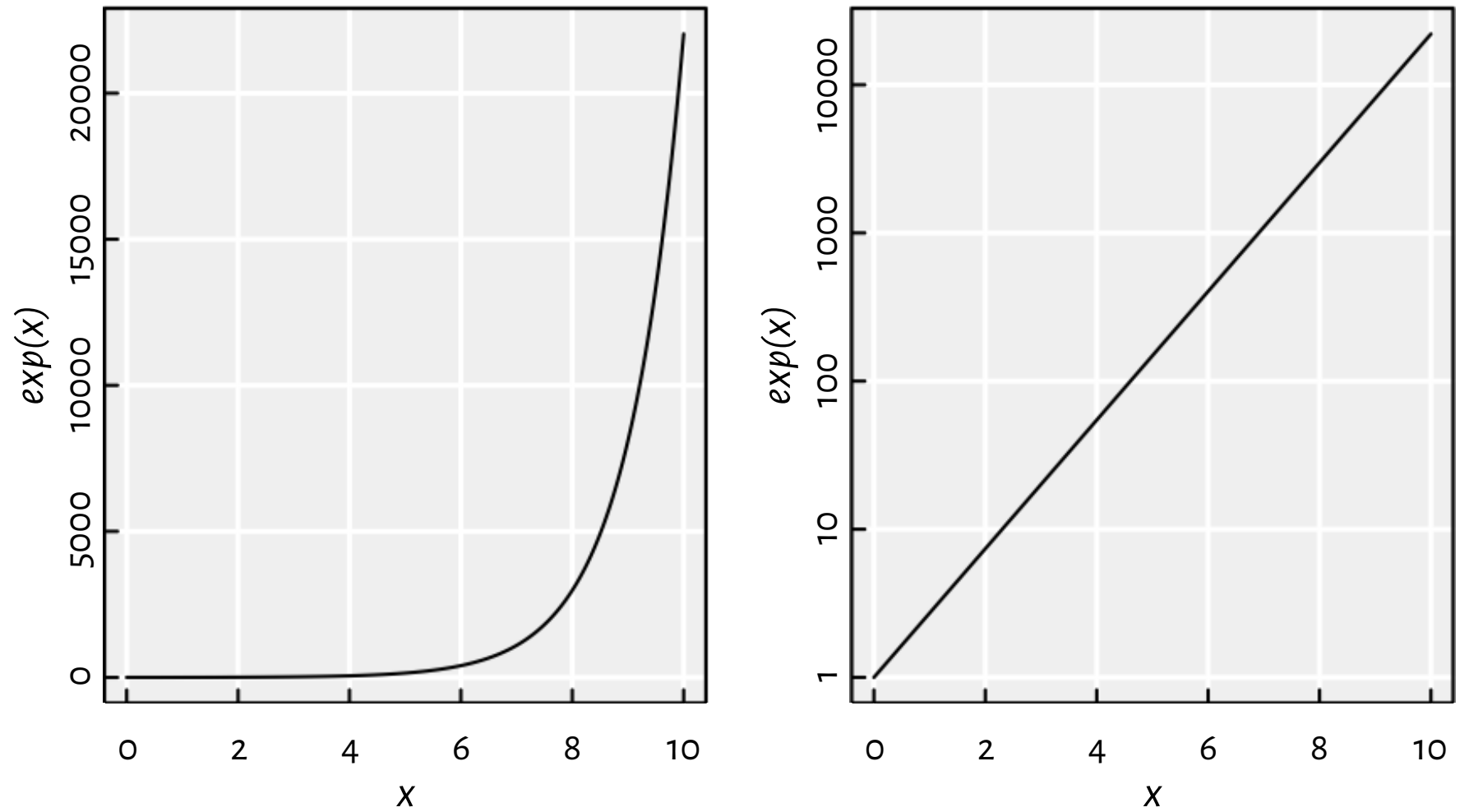

Figure 2.2. Linear- vs log-scale on the y-axis.

- `*gamma` (gamma, Γ),
- `*beta` (beta, B),
- `*lnorm` (log-normal),
- `*t` (Student),
- `*cauchy` (Cauchy–Lorentz),
- `*chisq` (chi-squared, $\chi^2$),
- `*f` (Snedecor–Fisher),
- `*weibull` (Weibull);

with the prefix "*" being one of:

- `d` (probability density function, PDF),
- `p` (cumulative distribution function, CDF; or survival function, SF),
- `q` (quantile function, being the inverse of the CDF),
- `r` (generation of random deviates; already mentioned above);

• discrete distributions, i.e., those whose possible outcomes can easily be enumerated (e.g., some integers):
  - `*binom` (binomial),
  - `*geom` (geometric),
  - `*pois` (Poisson),

- `*hyper` (hypergeometric),
- `*nbinom` (negative binomial);

prefixes "`p`" and "`r`" retain their meaning, however:

- `d` now gives the probability *mass* function (PMF),
- `q` brings about the quantile function, defined as a *generalised* inverse of the CDF.

Each distribution is characterised by a set of underlying parameters. For instance, a normal distribution $\mathrm{N}(\mu, \sigma)$ can be pinpointed by setting its expected value $\mu \in \mathbb{R}$ and standard deviation $\sigma > 0$. In R, these two have been named `mean` and `sd`, respectively; see `help("dnorm")`. Therefore, e.g., `dnorm(x, 1, 2)` computes the PDF of $\mathrm{N}(1, 2)$ at `x`.

---

**Note** The parametrisations assumed in R can be subtly different from what we know from statistical textbooks or probability courses. For example, the normal distribution can be identified based on either standard deviation or variance, and the exponential distribution can be defined via expected value or its reciprocal. We thus advise the reader to study carefully the documentation of `help("dnorm")`, `help("dunif")`, `help("dexp")`, `help("dbinom")`, and the like.

It is also worth knowing the typical use cases of each of the distributions listed, e.g., a Poisson distribution can describe the probability of observing the number of independent events in a fixed time interval (e.g., the number of users downloading a copy of R from CRAN per hour), and an exponential distribution can model the time between such events; compare [24].

---

**Exercise 2.5** *A call to* `hist(x)` *draws a histogram, which can serve as an estimator of the underlying continuous probability density function of a given sample; see Figure 2.3 for an illustration.*

```
par(mfrow=c(1, 2))  # two plots in one figure
# left subplot: uniform U(0, 1)
hist(runif(10000, 0, 1), col="white", probability=TRUE, main="")
x <- seq(0, 1, length.out=101)
lines(x, dunif(x, 0, 1), lwd=2)  # draw the true density function (PDF)
# right subplot: normal N(0, 1)
hist(rnorm(10000, 0, 1), col="white", probability=TRUE, main="")
x <- seq(-4, 4, length.out=101)
lines(x, dnorm(x, 0, 1), lwd=2)  # draw the PDF
```

*Draw a histogram of some random samples of different sizes* `n` *from the following distributions:*

- `rnorm(n, µ, σ)` – *normal* $\mathrm{N}(\mu, \sigma)$ *with expected values* $\mu \in \{-1, 0, 5\}$ *(i.e.,* $\mu$ *being equal to either* $-1$, $0$, *or* $5$*; read "*$\in$*" as "belongs to the given set" or "in") and standard deviations* $\sigma \in \{0.5, 1, 5\}$*;*

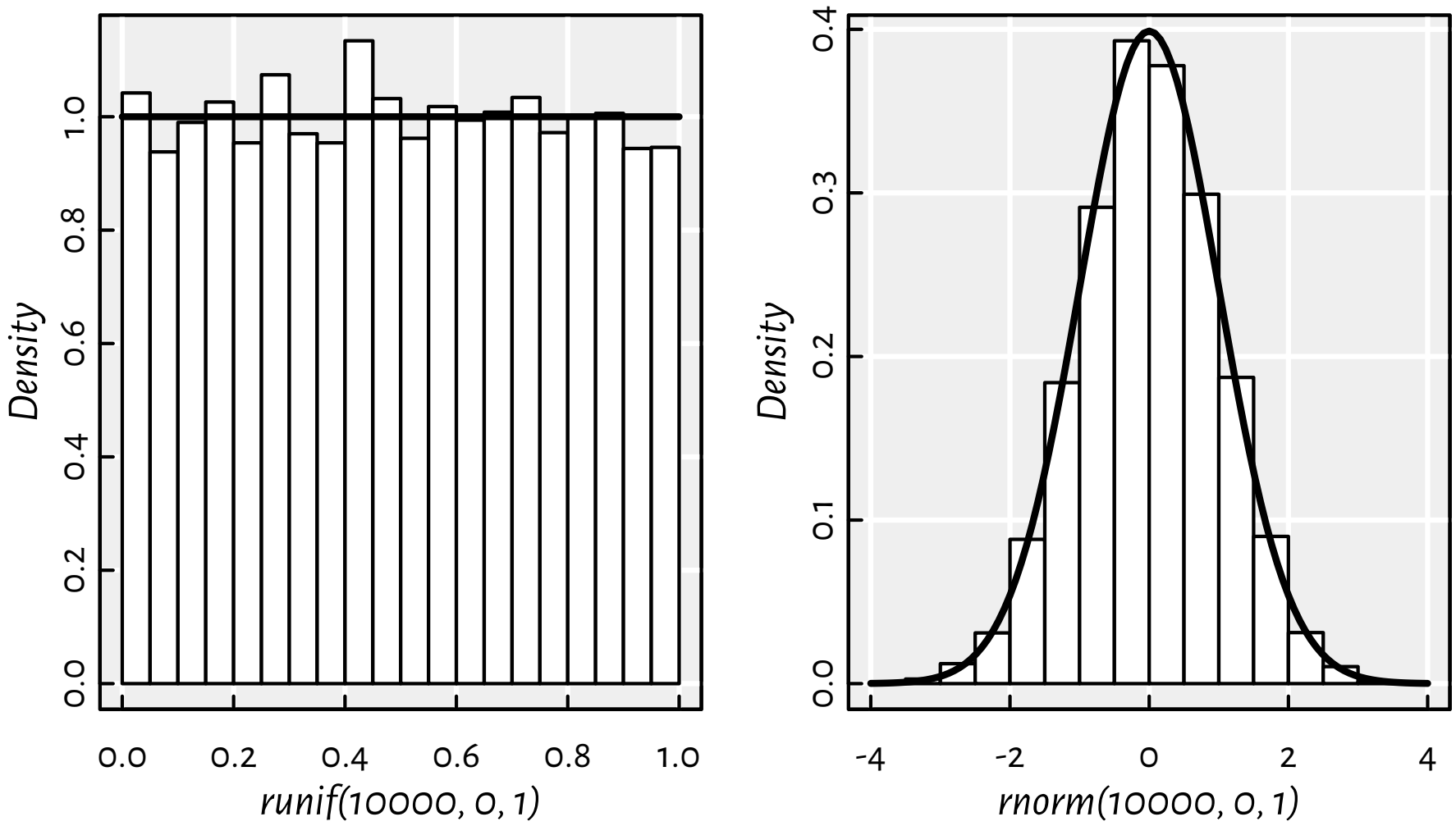

Figure 2.3. Example histograms of some pseudorandom samples and the true underlying probability density functions: the uniform distribution on the unit interval (left) and the standard normal distribution (right).

- ***`runif(n, a, b)`*** – *uniform* $\mathrm{U}(a,b)$ *on the interval* $(a,b)$ *with* $a = 0$ *and* $b = 1$ *as well as* $a = -1$ *and* $b = 1$;
- ***`rbeta(n, α, β)`*** – *beta* $\mathrm{B}(\alpha,\beta)$ *with* $\alpha,\beta \in \{0.5, 1, 2\}$;
- ***`rexp(n, λ)`*** – *exponential* $\mathrm{E}(\lambda)$ *with rates* $\lambda \in \{0.5, 1, 10\}$;

*Moreover, read about and play with the `breaks`, `main`, `xlab`, `ylab`, `xlim`, `ylim`, and `col` parameters; see **`help("hist")`**.*

**Example 2.6** *We roll a six-sided dice twelve times. Let C be a random variable describing the number of cases where the "1" face is thrown. C follows a binomial distribution* $\mathrm{Bin}(n,p)$ *with parameters* $n = 12$ *(the number of Bernoulli trials) and* $p = 1/6$ *(the probability of success in a single roll).*

*The probability mass function, **`dbinom`**, represents the probabilities that the number of "1"s rolled is equal to* $0, 1, \ldots,$ *or* $12$, *i.e.,* $P(C = 0), P(C = 1), \ldots,$ *or* $P(C = 12)$, *respectively:*

```
round(dbinom(0:12, 12, 1/6), 2)  # PMF at 13 different points
##  [1] 0.11 0.27 0.30 0.20 0.09 0.03 0.01 0.00 0.00 0.00 0.00 0.00 0.00
```

*On the other hand, the probability that we throw no more than three "1"s,* $P(C \le 3)$, *can be determined by means of the cumulative distribution function, **`pbinom`**:*

```
pbinom(3, 12, 1/6)  # pbinom(3, 12, 1/6, lower.tail=FALSE)
## [1] 0.87482
```

*The smallest c such that* $P(C \le c) \ge 0.95$ *can be computed based on the quantile function:*

```
qbinom(0.95, 12, 1/6)
## [1] 4
pbinom(3:4, 12, 1/6)   # for comparison: 0.95 is in-between
## [1] 0.87482 0.96365
```

*In other words, at least 95% of the time, we will be observing no more than four successes.*

*Also, here are 30 pseudorandom realisations (simulations) of the random variable C:*

```
rbinom(30, 12, 1/6)  # how many successes in 12 trials, repeated 30 times
##  [1] 1 3 2 4 4 0 2 4 2 2 4 2 3 2 0 4 1 0 1 4 4 3 2 6 2 3 2 2 1 1
```

### 2.3.5 Special functions (*)

Within mathematical formulae and across assorted application areas, certain functions appear more frequently than others. Hence, for the sake of notational brevity and computational precision, many of them have been assigned special names. For instance, the following functions are mentioned in the definitions related to a few probability distributions:

- **gamma**(x) for $x > 0$ computes $\Gamma(x) = \int_0^\infty t^{x-1} e^{-t}\, dt$,
- **beta**(a, b) for $a, b > 0$ yields $B(a, b) = \frac{\Gamma(a)\Gamma(b)}{\Gamma(a+b)} = \int_0^1 t^{a-1}(1-t)^{b-1}\, dt$.

Why do we have **beta** if it is merely a mix of **gamma**s? A specific, tailored function is expected to be faster and more precise than its DIY version; its underlying implementation does not have to involve any calls to **gamma**.

```
beta(0.25, 250)  # okay
## [1] 0.91213
gamma(0.25)*gamma(250)/gamma(250.25)  # not okay
## [1] NaN
```

The Γ function grows so rapidly that already **gamma**(172) gives rise to Inf. It is due to the fact that a computer's arithmetic is not infinitely precise; compare Section 3.2.3.

Special functions are plentiful; see the open-access *NIST Digital Library of Mathematical Functions* [52] for one of the most definitive references (and also [2] for its predecessor). R package **gsl** [34] provides a vectorised interface to the GNU GSL [29] library, which implements many of such routines.

**Exercise 2.7** *The Pochhammer symbol,* $(a)_x = \Gamma(a+x)/\Gamma(a)$, *can be computed via a call to* ***gsl::poch***(*a, x*), *i.e., the* ***poch*** *function from the* ***gsl*** *package:*

```
# call install.packages("gsl") first
library("gsl")  # load the package
poch(10, 3:6)   # calls gsl_sf_poch() from GNU GSL
## [1]    1320   17160  240240 3603600
```

*Read the documentation of the corresponding* **`gsl_sf_poch`** *function in the GNU GSL manual*[8]. *And when you are there, do not hesitate to go through the list of all functions, including those related to statistics, permutations, combinations, and so forth.*

Many functions also have their logarithm-of versions; see, e.g., **`lgamma`** and **`lbeta`**. Also, for instance, **`dnorm`** and **`dbeta`** have the `log` parameter. Their classical use case is the (numerical) maximum likelihood estimation, which involves the sums of the *logarithms* of densities.

## 2.4 Arithmetic operations

### 2.4.1 Vectorised arithmetic operators

R features the following binary arithmetic operators:

- **`` `+` ``** (addition) and **`` `-` ``** (subtraction),
- **`` `*` ``** (multiplication) and **`` `/` ``** (division),
- **`` `%/%` ``** (integer division) and **`` `%%` ``** (modulo, division remainder),
- **`` `^` ``** (exponentiation; synonym: **`` `**` ``**).

They are all *vectorised*: they take two vectors as input and produce another vector as output.

```
c(1, 2, 3) * c(10, 100, 1000)
## [1]   10  200 3000
```

The operation was performed in an *elementwise* fashion on the *corresponding* pairs of elements from both vectors. The first element in the left sequence was multiplied by the *corresponding* element in the right vector, and the result was stored in the first element of the output. Then, the second element in the left… all right, we get it.

Other operators behave similarly:

```
0:10 + seq(0, 1, 0.1)
##  [1]  0.0  1.1  2.2  3.3  4.4  5.5  6.6  7.7  8.8  9.9 11.0
0:7 / rep(3, length.out=8)   # division by 3
## [1] 0.00000 0.33333 0.66667 1.00000 1.33333 1.66667 2.00000 2.33333
0:7 %/% rep(3, length.out=8) # integer division
## [1] 0 0 0 1 1 1 2 2
0:7 %% rep(3, length.out=8)  # division remainder
## [1] 0 1 2 0 1 2 0 1
```

Operations involving missing values also yield `NA`s:

[8] https://www.gnu.org/software/gsl/doc/html

```
c(1, NA_real_, 3, NA_real_) + c(NA_real_, 2, 3, NA_real_)
## [1] NA NA  6 NA
```

### 2.4.2 Recycling rule

Some of the preceding statements can be written more concisely. When the operands are of different lengths, the shorter one is *recycled* as many times as necessary, as in `rep(y, length.out=length(x))`. For example:

```
0:7 / 3
## [1] 0.00000 0.33333 0.66667 1.00000 1.33333 1.66667 2.00000 2.33333
1:10 * c(-1, 1)
##  [1] -1  2 -3  4 -5  6 -7  8 -9 10
2 ^ (0:10)
##  [1]    1    2    4    8   16   32   64  128  256  512 1024
```

We call this the *recycling rule*.

If an operand cannot be recycled in its entirety, a warning[9] is generated, but the output is still available.

```
c(1, 10, 100) * 1:8
## Warning in c(1, 10, 100) * 1:8: longer object length is not a multiple of
##     shorter object length
## [1]   1  20 300   4  50 600   7  80
```

Vectorisation and the recycling rule are perhaps most fruitful when applying binary operators on sequences of identical lengths or when performing vector-scalar (i.e., a sequence vs a single value) operations. However, there is much more: schemes like "every $k$-th element" appear in Taylor series expansions (multiply by `c(-1, 1)`), $k$-fold cross-validation, etc.; see also Section 11.3.4 for use cases in matrix/tensor processing.

Also, `pmin` and `pmax` return the *parallel* minimum and maximum of the corresponding elements of the input vectors. Their behaviour is the same as the arithmetic operators, but we call them as ordinary functions:

```
pmin(c(1, 2, 3, 4), c(4, 2, 3, 1))
## [1] 1 2 3 1
pmin(3, 1:5)
## [1] 1 2 3 3 3
pmax(0, pmin(1, c(0.25, -2, 5, -0.5, 0, 1.3, 0.99)))  # clipping to [0, 1]
## [1] 0.25 0.00 1.00 0.00 0.00 1.00 0.99
```

**Note** Some functions can be very *deeply* vectorised, i.e., with respect to multiple arguments. For example:

[9] A few functions do not warn us whatsoever when they perform incomplete recycling (e.g., `paste`; see Section 6.1.3) or can even give an error (e.g., `as.data.frame.list`; compare Section 12.1.1). Consider this inconsistency as an annoying bug and hope it will be fixed, in the next decade or so.

```
runif(3, c(10, 20, 30), c(11, 22, 33))
## [1] 10.288 21.577 31.227
```

generates three random numbers uniformly distributed over the intervals $(10, 11)$, $(20, 22)$, and $(30, 33)$, respectively.

---

### 2.4.3 Operator precedence

Expressions involving multiple operators need a set of rules governing the order of computations (unless we enforce it using round brackets). We have said that `-1:10` means `(-1):10` rather than `-(1:10)`. But what about, say, `1+1+1+1+1*0` or `3*2^0:5+10`?

Let's list the operators mentioned so far *in their order of precedence*, from the least to the most binding (see also `help("Syntax")`):

1. `<-` (right to left),
2. `+` and `-` (binary),
3. `*` and `/`,
4. `%%` and `%/%`,
5. `:`,
6. `+` and `-` (unary),
7. `^` (right to left).

Hence, `-2^2/3+3*4` means `((-(2^2))/3)+(3*4)` and not, e.g., `-((2^(2/(3+3)))*4)`.

Notice that `+` and `-`, `*` and `/`, as well as `%%` and `%/%` have the same priority. Expressions involving a series of operations in the same group are evaluated left to right, with the exception of `^` and `<-`, which are performed the other way around. Therefore:

- `2*3/4*5` is equivalent to `((2*3)/4)*5`,
- `2^3^4` is `2^(3^4)` because, mathematically, we would write it as $2^{3^4} = 2^{81}$,
- "`x <- y <- 4*3%%8/2`" binds both `y` and `x` to 6, not `x` to the previous value of `y` and then `y` to 6.

When in doubt, we can always *bracket* a subexpression to ensure it is executed in the intended order. It can also increase the readability of our code.

### 2.4.4 Accumulating

The `+` and `*` operators, as well as the `pmin` and `pmax` functions, implement elementwise operations that are applied on the corresponding elements taken from two given

vectors. For instance:

$$\begin{pmatrix} x_1 \\ x_2 \\ x_3 \\ \vdots \\ x_n \end{pmatrix} + \begin{pmatrix} y_1 \\ y_2 \\ y_3 \\ \vdots \\ y_n \end{pmatrix} = \begin{pmatrix} x_1 + y_1 \\ x_2 + y_2 \\ x_3 + y_3 \\ \vdots \\ x_n + y_n \end{pmatrix}.$$

However, we can also scan through all the values in a *single* vector and combine the successive elements that we inspect using the corresponding operation:

- **cumsum**(x) gives the cumulative sum of the elements in a vector,
- **cumprod**(x) computes the cumulative product,
- **cummin**(x) yields the cumulative minimum,
- **cummax**(x) breeds the cumulative maximum.

The $i$-th element in the output vector will consist of the sum/product/min/max of the first $i$ inputs. For example:

$$\text{cumsum}\begin{pmatrix} x_1 \\ x_2 \\ x_3 \\ \vdots \\ x_n \end{pmatrix} = \begin{pmatrix} x_1 & & & & \\ x_1 + x_2 & & & & \\ x_1 + x_2 + x_3 & & & & \\ \vdots & & \ddots & & \\ x_1 + x_2 + x_3 + \cdots + x_n & & & & \end{pmatrix}.$$

```
cumsum(1:8)
## [1]  1  3  6 10 15 21 28 36
cumprod(1:8)
## [1]     1     2     6    24   120   720  5040 40320
cummin(c(3, 2, 4, 5, 1, 6, 0))
## [1] 3 2 2 2 1 1 0
cummax(c(3, 2, 4, 5, 1, 6, 0))
## [1] 3 3 4 5 5 6 6
```

**Example 2.8** *On a side note,* **diff** *can be considered an inverse to* **cumsum**. *It computes the iterated difference: subtracts the first two elements, then the second from the third one, the third from the fourth, and so on. In other words,* **diff**(x) *gives* **y** *such that* $y_i = x_{i+1} - x_i$.

```
x <- c(-2, 3, 6, 2, 15)
diff(x)
## [1]  5  3 -4 13
cumsum(diff(x))
## [1]  5  8  4 17
cumsum(c(-2, diff(x)))  # recreates x
## [1] -2  3  6  2 15
```

*Thanks to* **diff**, *we can compute the daily changes to the EUR/AUD forex rates studied earlier; see* Figure 2.4.

```
aud <- scan(paste0("https://github.com/gagolews/teaching-data/raw/",
    "master/marek/euraud-20200101-20200630.csv"), comment.char="#")
aud_all <- na.omit(aud)  # remove all missing values
plot(diff(aud_all), type="s", ylab="Daily change [EUR/AUD]")  # "steps"
abline(h=0, lty="dotted")  # draw a horizontal line at y=0
```

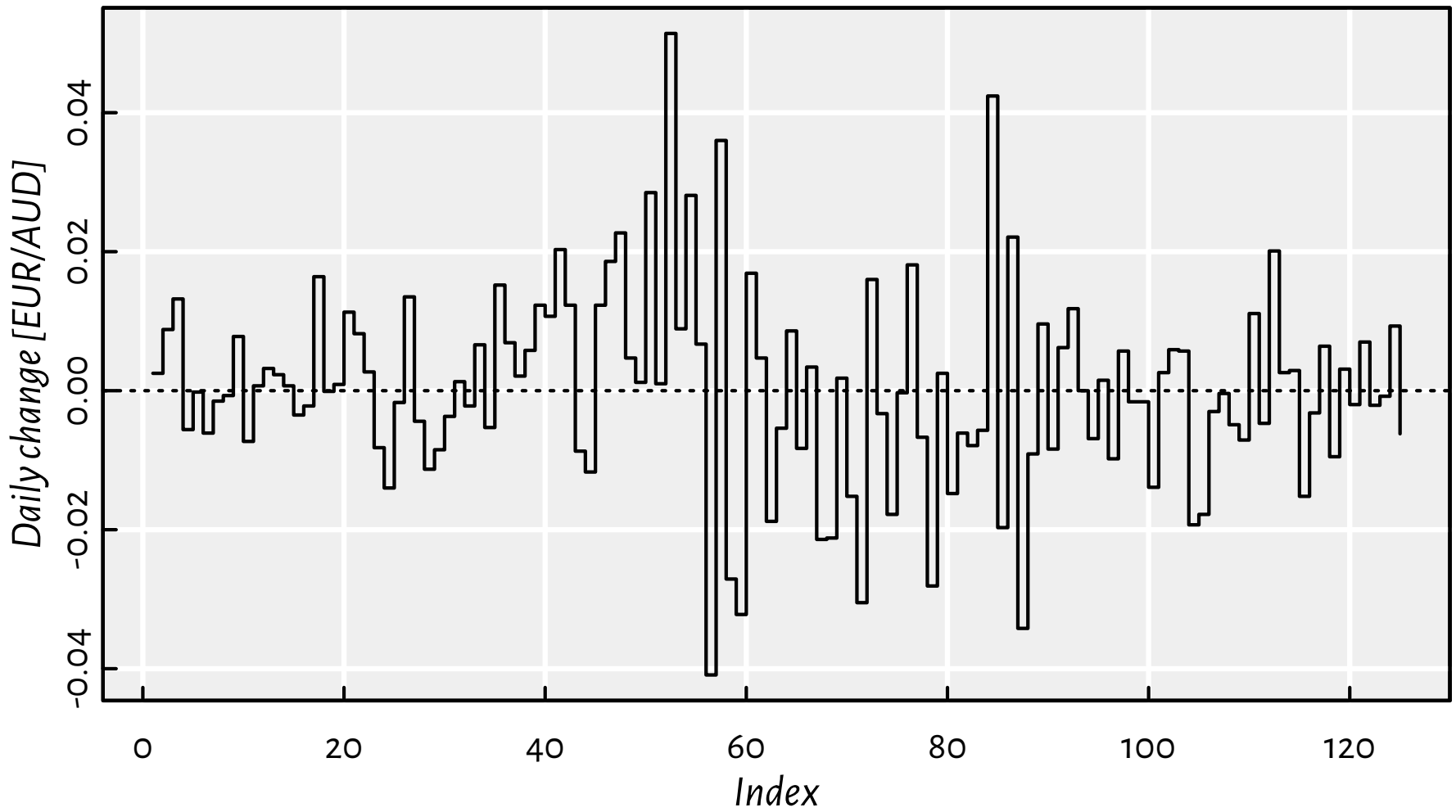

Figure 2.4. Iterated differences of the exchange rates (non-missing values only).

### 2.4.5 Aggregating

If we are only concerned with the last cumulant, which summarises *all* the inputs, we have the following[10] functions at our disposal:

- **sum**(x) computes the sum of elements in a vector, $\sum_{i=1}^{n} x_i = x_1 + x_2 + \cdots + x_n$,
- **prod**(x) outputs the product of all elements, $\prod_{i=1}^{n} x_i = x_1 x_2 \cdots x_n$,
- **min**(x) determines the minimum,
- **max**(x) reckons the greatest value.

```
sum(1:8)
## [1] 36
prod(1:8)
## [1] 40320
min(c(3, 2, 4, 5, 1, 6, 0))
## [1] 0
```

[10] Chapter 7 will discuss the **Reduce** function, which generalises the above by allowing any binary operation to be propagated over a given vector.

```
max(c(3, 2, 4, 5, 1, 6, 0))
## [1] 6
```

The foregoing functions form the basis for the popular summary statistics[11] (sample aggregates) such as:

- **mean**(x) yields the arithmetic mean, **sum**(x)/**length**(x),
- **var**(x) is the unbiased sample variance, **sum**((x-**mean**(x))^2)/(**length**(x)-1),
- **sd**(x) is the standard deviation, **sqrt**(**var**(x)).

Furthermore, **median**(x) computes the sample median, i.e., the middle value in the sorted[12] version of x.

For instance:

```
x <- runif(1000)
c(min(x), mean(x), median(x), max(x), sd(x))
## [1] 0.00046535 0.49727780 0.48995025 0.99940453 0.28748391
```

**Exercise 2.9** *Let **x** be any vector of length n with positive elements. Compute its geometric and harmonic mean, which are given by, respectively,*

$$\sqrt[n]{\prod_{i=1}^{n} x_i} = e^{\frac{1}{n}\sum_{i=1}^{n} \log x_i} \quad \text{and} \quad \frac{n}{\sum_{i=1}^{n} \frac{1}{x_i}}.$$

*When solving exercises like this one, it does not really matter what data you apply these functions on. We are being abstract in the sense that the **x** vector can be anything: from the one that features very accurate socioeconomic predictions that will help make this world less miserable, through the data you have been collecting for the last ten years in relation to your super important PhD research, whatever your company asked you to crunch today, to something related to the hobby project that you enjoy doing after hours. But you can also just test the above on something like "x <- **runif**(10)", and move on.*

All aggregation functions return a missing value if any of the input elements is unavailable. Luckily, they are equipped with the na.rm parameter, on behalf of which we can request the removal of NAs.

```
aud <- scan(paste0("https://github.com/gagolews/teaching-data/raw/",
    "master/marek/euraud-20200101-20200630.csv"), comment.char="#")
c(min(aud), mean(aud), max(aud))
## [1] NA NA NA
c(min(aud, na.rm=TRUE), mean(aud, na.rm=TRUE), max(aud, na.rm=TRUE))
## [1] 1.6006 1.6775 1.8635
```

[11] Actually, **var** and **median**, amongst others, are defined by the **stats** package. But this one is automatically loaded by default, so let's not make a fuss about it now.

[12] **min**, **median**, and **max** are generalised by the **quantile** function. We will discuss it much later (Section 4.4.3), because it returns a *named* vector.

Otherwise, we could have called, e.g., `mean(na.omit(x))`.

---

**Note** In the documentation, we read that the usage of `sum`, `prod`, `min`, and `max` is like `sum(..., na.rm=FALSE)`, etc. In this context, it means that they accept any number of input vectors, and each of them can be of arbitrary length. Therefore, `min(1, 2, 3)`, `min(c(1, 2, 3))` as well as `min(c(1, 2), 3)` all return the same result.

However, we also read that we have `mean(x, trim=0, na.rm=FALSE, ...)`. This time, only one vector can be aggregated, and any further arguments (except `trim` and `na.rm`) are ignored.

The extra flexibility (which we do not have to rely on, ever) of the former group is due to their being *associative* operations. We have, e.g., $(2+3)+4 = 2+(3+4)$. Hence, these operations can be performed in any order, in any group. They are primitive operations: it is `mean` that is based on `sum`, not vice versa.

---

## 2.5 Exercises

**Exercise 2.10** *Answer the following questions.*

- *What is the meaning of the dot-dot-dot parameter in the definition of the* `c` *function?*
- *We say that the* `round` *function is* vectorised. *What does that mean?*
- *What is wrong with a call to* `c(sqrt(1), sqrt(2), sqrt(3))`*?*
- *What do we mean by saying that multiplication operates element by element?*
- *How does the recycling rule work when applying* `` `+` ``*?*
- *How to (and why) set the seed of the pseudorandom number generator?*
- *What is the difference between* `NA_real_` *and* `NaN`*?*
- *How are default arguments specified in the manual of, e.g., the* `round` *function?*
- *Is a call to* `rep(times=4, x=1:5)` *equivalent to* `rep(4, 1:5)`*?*
- *List a few ways to generate a sequence like (-1, -0.75, -0.5, ..., 0.75, 1).*
- *Is* `-3:5` *the same as* `-(3:5)`*? What about the precedence of operators in expressions such as* `2^3/4*5^6`*,* `5*6+4/17%%8`*, and* `1+-2^3:4-1`*?*
- *If* `x` *is a numeric vector of length $n$ (for some $n \geq 0$), how many values will* `sample(x)` *output?*
- *Does* `scan` *support reading directly from compressed archives, e.g.,* `.csv.gz` *files?*

*When in doubt, refer back to the material discussed in this chapter or the R manual.*

**Exercise 2.11** *Thanks to vectorisation, implementing an example graph of arcsine and arccosine is straightforward.*

```
x <- seq(-1, 1, length.out=11)  # increase length.out for a smoother curve
plot(x, asin(x),                # asin() computed for 11 points
    type="l",                   # lines
    ylim=c(-pi/2, pi),          # y axis limits like c(y_min, y_max)
    ylab="asin(x), acos(x)")    # y axis label
lines(x, acos(x), col="red", lty="dashed")  # adds to the current plot
legend("topright", c("asin(x)", "acos(x)"),
    lty=c("solid", "dashed"), col=c("black", "red"), bg="white")
```

*Thusly inspired, plot the following functions:* $|\sin x^2|$, $|\sin |x||$, $\sqrt{\lfloor x \rfloor}$, *and* $1/(1+e^{-x})$. *Recall that the documentation of* `plot` *can be accessed by calling* `help("plot.default")`.

**Exercise 2.12** *The expression:*

$$4\sum_{i=1}^{n} \frac{(-1)^{i+1}}{2i-1} = 4\left(\frac{1}{1}-\frac{1}{3}+\frac{1}{5}-\frac{1}{7}+\cdots\right)$$

*slowly converges to* $\pi$ *as* $n$ *approaches* $\infty$. *Calculate it for* $n = 1\,000\,000$ *and* $n = 1\,000\,000\,000$ *using the vectorised functions and operators discussed in this chapter, making use of the recycling rule as much as possible.*

**Exercise 2.13** *Let x and y be two vectors of identical lengths n, say:*

```
x <- rnorm(100)
y <- 2*x+10+rnorm(100, 0, 0.5)
```

*Compute the Pearson linear correlation coefficient given by:*

$$r(\mathbf{x},\mathbf{y}) = \frac{\sum_{i=1}^{n}\left(x_i - \frac{1}{n}\sum_{j=1}^{n} x_j\right)\left(y_i - \frac{1}{n}\sum_{j=1}^{n} y_j\right)}{\sqrt{\sum_{i=1}^{n}\left(x_i - \frac{1}{n}\sum_{j=1}^{n} x_j\right)^2}\sqrt{\sum_{i=1}^{n}\left(y_i - \frac{1}{n}\sum_{j=1}^{n} y_j\right)^2}}.$$

*To make sure you have come up with a correct implementation, compare your result to a call to* `cor(x, y)`.

**Exercise 2.14** *(*) Find an R package providing a function to compute moving (rolling) averages and medians of a given vector. Apply them on the EUR/AUD currency exchange data. Draw thus obtained smoothened versions of the time series.*

**Exercise 2.15** *(**) Use a call to* `convolve(..., type="filter")` *to compute the k-moving average of a numeric vector.*

In the next chapter, we will study operations that involve logical values.

# 3

# *Logical vectors*

## 3.1 Creating logical vectors

R defines three(!) logical constants: `TRUE`, `FALSE`, and `NA` representing, respectively, "yes", "no", and "???". When instantiated, each of them is an atomic vector of length one. To generate logical vectors of arbitrary size, we can call some of the functions introduced in the previous chapter, for instance:

```
c(TRUE, FALSE, FALSE, NA, TRUE, FALSE)
## [1]  TRUE FALSE FALSE    NA  TRUE FALSE
rep(c(TRUE, FALSE, NA), each=2)
## [1]  TRUE  TRUE FALSE FALSE    NA    NA
sample(c(TRUE, FALSE), 10, replace=TRUE, prob=c(0.8, 0.2))
##  [1]  TRUE  TRUE  TRUE FALSE FALSE  TRUE  TRUE FALSE  TRUE  TRUE
```

**Note** By default, `T` is a synonym for `TRUE` and `F` stands for `FALSE`. However, these are not reserved keywords and can be reassigned to any other values. Therefore, we advise against relying on them: they are not used throughout the course of this course.

Also, notice that the logical missing value is spelled simply as `NA`, and not `NA_logical_`. Both the logical `NA` and the numeric `NA_real_` are, for the sake of our widely-conceived wellbeing, both *printed* as "`NA`" on the R console. This, however, does not mean they are identical; see Section 4.1 for discussion.

## 3.2 Comparing elements

### 3.2.1 Vectorised relational operators

Logical vectors frequently come into being as a result of various *testing* activities. In particular, the binary operators:

- `` `<` `` (less than),
- `` `<=` `` (less than or equal),

- `>` (greater than),
- `>=` (greater than or equal)
- `==` (equal),
- `!=` (not equal),

compare the *corresponding* elements of two numeric vectors and output a logical vector.

```
1 < 3
## [1] TRUE
c(1, 2, 3, 4) == c(2, 2, 3, 8)
## [1] FALSE  TRUE  TRUE FALSE
1:10 <= 10:1
##  [1]  TRUE  TRUE  TRUE  TRUE  TRUE FALSE FALSE FALSE FALSE FALSE
```

Thus, they operate in an elementwise manner. Moreover, the recycling rule is applied if necessary:

```
3 < 1:5  # c(3, 3, 3, 3, 3) < c(1, 2, 3, 4, 5)
## [1] FALSE FALSE FALSE  TRUE  TRUE
c(1, 4) == 1:4  # c(1, 4, 1, 4) == c(1, 2, 3, 4)
## [1]  TRUE FALSE FALSE  TRUE
```

Therefore, we can say that they are vectorised in the same manner as the arithmetic operators `+`, `*`, etc.; compare Section 2.4.1.

### 3.2.2 Testing for NA, NaN, and Inf

Comparisons against missing values and not-numbers yield NAs. Instead of the *incorrect* "x == NA_real_", testing for missingness should rather be performed via a call to the vectorised **is.na** function.

```
is.na(c(NA_real_, Inf, -Inf, NaN, -1, 0, 1))
## [1]  TRUE FALSE FALSE  TRUE FALSE FALSE FALSE
is.na(c(TRUE, FALSE, NA, TRUE))  # works for logical vectors, too
## [1] FALSE FALSE  TRUE FALSE
```

Moreover, **is.finite** is noteworthy since it returns FALSE on Infs, NA_real_s and NaNs.

```
is.finite(c(NA_real_, Inf, -Inf, NaN, -1, 0, 1))
## [1] FALSE FALSE FALSE FALSE  TRUE  TRUE  TRUE
```

See also the more specific **is.nan** and **is.infinite**.

### 3.2.3 Dealing with round-off errors (*)

In mathematics, real numbers are merely an idealisation. In practice, however, it is impossible to store them with infinite precision (think $\pi = 3.141592653589793...$): computer memory is limited, and our time is precious.

Therefore, a consensus had to be reached. In R, we rely on the *double-precision floating point format*. The *floating point* part means that the numbers can be both small (close to zero like $\pm 2.23 \times 10^{-308}$) and large (e.g., $\pm 1.79 \times 10^{308}$).

---

**Note**

```
2.23e-308 == 0.00000000000000000000000000000000000000000000000000
               00000000000000000000000000000000000000000000000000
               00000000000000000000000000000000000000000000000000
               00000000000000000000000000000000000000000000000000
               00000000000000000000000000000000000000000000000000
               00000000000000000000000000000000000000000000000000
               0000000223

1.79e308  ==                                            179000000
               00000000000000000000000000000000000000000000000000
               00000000000000000000000000000000000000000000000000
               00000000000000000000000000000000000000000000000000
               00000000000000000000000000000000000000000000000000
               00000000000000000000000000000000000000000000000000
               00000000000000000000000000000000000000000000000000
```

These two are pretty distant.

---

Every numeric value takes 8 bytes (or, equivalently, 64 bits) of memory. We are, however, able to store only *about* 15–17 decimal digits:

```
print(0.12345678901234567890123456789012345678901234, digits=22)  # 22 is max
## [1] 0.1234567890123456773699
```

which limits the precision of our computations. We wrote *about* because, unfortunately, the numbers are stored the computer-friendly *binary* base, not the human-aligned *decimal* one. This can lead to counterintuitive outcomes. In particular:

- `0.1` cannot be represented exactly for it cannot be written as a finite series of reciprocals of powers of 2 (we have $0.1 = 2^{-4} + 2^{-5} + 2^{-8} + 2^{-9} + ...$). This leads to surprising results such as:

  ```
  0.1 + 0.1 + 0.1 == 0.3
  ## [1] FALSE
  ```

  Quite strikingly, what follows does not *show* anything suspicious:

```
c(0.1, 0.1 + 0.1 + 0.1, 0.3)
## [1] 0.1 0.3 0.3
```

Printing involves rounding. In the above context, it is misleading. Actually, we experience something more like:

```
print(c(0.1, 0.1 + 0.1 + 0.1, 0.3), digits=22)
## [1] 0.1000000000000000055511 0.3000000000000000444089
## [3] 0.2999999999999999888978
```

- All integers between $-2^{53}$ and $2^{53}$ all stored *exactly*. This is good news. However, the next integer is beyond the representable range:

```
2^53 + 1 == 2^53
## [1] TRUE
```

- The order of operations may matter. In particular, the associativity property can be violated when dealing with numbers of contrasting orders of magnitude:

```
2^53 + 2^-53 - 2^53 - 2^-53  # should be == 0.0
## [1] -1.1102e-16
```

- Some numbers may just be too large, too small, or too close to zero to be represented exactly:

```
c(sum(2^((1023-52):1023)), sum(2^((1023-53):1023)))
## [1] 1.7977e+308         Inf
c(2^(-1022-52), 2^(-1022-53))
## [1] 4.9407e-324  0.0000e+00
```

---

**Important** The double-precision floating point format (IEEE 754) is not specific to R. It is used by most other computing environments, including Python and C++. For discussion, see [33, 36, 43], and the more statistically-orientated [31].

---

Can we do anything about these issues?

Firstly, dealing with *integers* of a *reasonable* order of magnitude (e.g., various resource or case IDs in our datasets) is *safe*. Their comparison, addition, subtraction, and multiplication are always precise.

In all other cases (including applying other operations on integers, e.g., division or **sqrt**), we need to be very careful with comparisons, especially involving testing for equality via `==`. The sole fact that $\sin\pi = 0$, mathematically speaking, does not mean that we should expect that:

```
sin(pi) == 0
## [1] FALSE
```

Instead, they are so close that we can *treat the difference between them as negligible*. Thus, in practice, instead of testing if $x = y$, we will be considering:

- $|x - y|$ (absolute error), or
- $\frac{|x-y|}{|y|}$ (relative error; which takes the order of magnitude of the numbers into account but obviously cannot be applied if $y$ is very close to 0),

and determining if these are less than an assumed error margin, $\varepsilon > 0$, say, $10^{-8}$ or $2^{-26}$. For example:

```
abs(sin(pi) - 0) < 2^-26
## [1] TRUE
```

**Note** Rounding can sometimes have a similar effect as testing for almost equality in terms of the absolute error.

```
round(sin(pi), 8) == 0
## [1] TRUE
```

**Important** The foregoing recommendations are valid for the most popular applications of R, i.e., statistical and, more generally, scientific computing[1]. Our datasets usually do not represent accurate measurements. Bah, the world itself is far from ideal! Therefore, we do not have to lose sleep over our not being able to precisely pinpoint the *exact* solutions.

## 3.3 Logical operations

### 3.3.1 Vectorised logical operators

The relational operators such as `` `==` `` and `` `>` `` accept only *two* arguments. Their chaining is forbidden. A test that we would mathematically write as $0 \leq x \leq 1$ (or $x \in [0, 1]$) *cannot* be expressed as "`0 <= x <= 1`" in R. Therefore, we need a way to combine two logical conditions so as to be able to state that "$x \geq 0$ *and, at the same time*, $x \leq 1$".

In such situations, the following logical operators and functions come in handy:

- `` `!` `` (not, negation; unary),

[1] However, in financial applications, we had rather rely on base-10 numbers (compare the aforementioned issue with 0.1). There are some libraries implementing higher precision floating-point numbers or even interval arithmetic that keeps track of error propagation in operation chains.

- `&` (and, conjunction; are both predicates true?),
- `|` (or, alternation; is at least one true?),
- `xor` (exclusive-or, exclusive disjunction, either-or; is one and only one of the predicates true?).

They again act elementwisely and implement the recycling rule if necessary (and applicable).

```
x <- c(-10, -1, -0.25, 0, 0.5, 1, 5, 100)
(x >= 0) & (x <= 1)
## [1] FALSE FALSE FALSE  TRUE  TRUE  TRUE FALSE FALSE
(x < 0) | (x > 1)
## [1]  TRUE  TRUE  TRUE FALSE FALSE FALSE  TRUE  TRUE
!((x < 0) | (x > 1))
## [1] FALSE FALSE FALSE  TRUE  TRUE  TRUE FALSE FALSE
xor(x >= -1, x <= 1)
## [1]  TRUE FALSE FALSE FALSE FALSE FALSE  TRUE  TRUE
```

**Important** The vectorised `&` and `|` operators should not be confused with their scalar, short-circuit counterparts, `&&` and `||`; see Section 8.1.4.

### 3.3.2 Operator precedence revisited

The operators introduced in this chapter have lower precedence than the arithmetic ones, including the binary `+` and `-`. Calling `help("Syntax")` reveals that we can extend our listing from Section 2.4.3 as follows:

1. `<-` *(right to left; least binding),*
2. `|`,
3. `&`,
4. `!` (unary),
5. `<`, `>`, `<=`, `>=`, `==`, and `!=`,
6. `+` *and* `-` *(binary),*
7. `*` *and* `/`,
8. ...

The order of precedence is intuitive, e.g., "`x+1 <= y & y <= z-1 | x <= z`" means "`(((x+1) <= y) & (y <= (z-1))) | (x <= z)`".

### 3.3.3 Dealing with missingness

Operations involving missing values follow the principles of Łukasiewicz's three-valued logic, which is based on common sense. For instance, "`NA | TRUE`" is `TRUE` because the alternative (*or*) needs *at least one* argument to be `TRUE` to generate a positive result. On the other hand, "`NA | FALSE`" is `NA` since the outcome would be different depending on what we substituted `NA` for.

Let's contemplate the logical operations' *truth tables* for all the possible combinations of inputs:

```
u <- c(TRUE, FALSE, NA,  TRUE,  FALSE, NA,    TRUE, FALSE, NA)
v <- c(TRUE, TRUE, TRUE, FALSE, FALSE, FALSE, NA,   NA,    NA)
!u
## [1] FALSE  TRUE    NA FALSE  TRUE    NA FALSE  TRUE    NA
u & v
## [1]  TRUE FALSE    NA FALSE FALSE FALSE    NA FALSE    NA
u | v
## [1]  TRUE  TRUE  TRUE  TRUE FALSE    NA  TRUE    NA    NA
xor(u, v)
## [1] FALSE  TRUE    NA  TRUE FALSE    NA    NA    NA    NA
```

### 3.3.4 Aggregating with `all`, `any`, and `sum`

Just like in the case of numeric vectors, we can summarise the contents of logical sequences. **`all`** tests whether *every* element in a logical vector is equal to `TRUE`. **`any`** determines if there *exists* an element that is `TRUE`.

```
x <- runif(10000)
all(x <= 0.2)  # are all values in x <= 0.2?
## [1] FALSE
any(x <= 0.2)  # is there at least one element in x that is <= 0.2?
## [1] TRUE
any(c(NA, FALSE, TRUE))
## [1] TRUE
all(c(TRUE, TRUE, NA))
## [1] NA
```

---

**Note** **`all`** will frequently be used in conjunction with **`` `==` ``** as the latter is itself *vectorised*: it does *not* test whether a vector *as a whole* is equal to another one.

```
z <- c(1, 2, 3)
z == 1:3  # elementwise equal
## [1] TRUE TRUE TRUE
all(z == 1:3)  # elementwise equal summarised
## [1] TRUE
```

However, let's keep in mind the warning about the testing for *exact* equality of floating-point numbers stated in Section 3.2.3. Sometimes, considering absolute or relative errors might be more appropriate.

```
z <- sin((0:10)*pi)  # sin(0), sin(pi), sin(2*pi), ..., sin(10*pi)
all(z == 0.0)  # danger zone! please don't...
## [1] FALSE
all(abs(z - 0.0) < 1e-8)  # are the absolute errors negligible?
## [1] TRUE
```

---

We can also call `sum` on a logical vector. Taking into account that it interprets `TRUE` as numeric 1 and `FALSE` as `0` (more on this in Section 4.1), it will give us the number of elements equal to `TRUE`.

```
sum(x <= 0.2)  # how many elements in x are <= 0.2?
## [1] 1998
```

Also, by computing `sum(x)/length(x)`, we can obtain the proportion (fraction) of values equal to `TRUE` in `x`. Equivalently:

```
mean(x <= 0.2)  # proportion of elements <= 0.2
## [1] 0.1998
```

Naturally, we *expect* `mean(runif(n) <= 0.2)` to be equal to 0.2 (20%), but with randomness, we can never be sure.

### 3.3.5 Simplifying predicates

Each aspiring programmer needs to become fluent with the rules governing the transformations of logical conditions, e.g., that the negation of "`(x >= 0) & (x < 1)`" is equivalent to "`(x < 0) | (x >= 1)`". Such rules are called *tautologies*. Here are a few of them:

- `!(!p)` is equivalent to `p` (double negation),
- `!(p & q)` holds if and only if `!p | !q` (De Morgan's law),
- `!(p | q)` is `!p & !q` (another De Morgan's law),
- `all(p)` is equivalent to `!any(!p)`.

Various combinations thereof are, of course, possible. Further simplifications are enabled by other properties of the binary operations:

- commutativity (symmetry), e.g., $a + b = b + a$, $a * b = b * a$,
- associativity, e.g., $(a + b) + c = a + (b + c)$, $\max(\max(a, b), c) = \max(a, \max(b, c))$,
- distributivity, e.g., $a * b + a * c = a * (b + c)$, $\min(\max(a, b), \max(a, c)) = \max(a, \min(b, c))$,

and relations, including:

- transitivity, e.g., if $a \leq b$ and $b \leq c$, then surely $a \leq c$.

**Exercise 3.1** *Assuming that a, b, and c are numeric vectors, simplify the following expressions:*

- `!(b>a & b<c)`,
- `!(a>=b & b>=c & a>=c)`,
- `a>b & a<c | a<c & a>b`,
- `a>b | a<=b`,
- `a<=b & a>c | a>b & a<=c`,
- `a<=b & (a>c | a>b) & a<=c`,
- `!all(a > b & b < c)`.

---

## 3.4 Choosing elements with `ifelse`

The **`ifelse`** function is a vectorised version of the scalar **`if`**…**`else`** conditional statement, which we will forgo for as long as until Chapter 8. It permits us to select an element from one of two vectors based on some logical condition. A call to `ifelse(l, t, f)`, where `l` is a logical vector, returns a vector `y` such that:

$$y_i = \begin{cases} t_i & \text{if } l_i \text{ is TRUE}, \\ f_i & \text{if } l_i \text{ is FALSE}. \end{cases}$$

In other words, the $i$-th element of the result vector is equal to $t_i$ if $l_i$ is `TRUE` and to $f_i$ otherwise. For example:

```
(z <- rnorm(6))  # example vector
## [1] -0.560476 -0.230177  1.558708  0.070508  0.129288  1.715065
ifelse(z >= 0, z, -z)  # like abs(z)
## [1] 0.560476 0.230177 1.558708 0.070508 0.129288 1.715065
```

or:

```
(x <- rnorm(6))  # example vector
## [1]  0.46092 -1.26506 -0.68685 -0.44566  1.22408  0.35981
(y <- rnorm(6))  # example vector
## [1]  0.40077  0.11068 -0.55584  1.78691  0.49785 -1.96662
ifelse(x >= y, x, y)   # like pmax(x, y)
## [1]  0.46092  0.11068 -0.55584  1.78691  1.22408  0.35981
```

We should not be surprised anymore that the recycling rule is fired up when necessary:

```
ifelse(x > 0, x^2, 0)  # squares of positive xs and 0 otherwise
## [1] 0.21244 0.00000 0.00000 0.00000 1.49838 0.12947
```

---

**Note** All arguments are evaluated in their entirety before deciding on which elements are selected. Therefore, the following call generates a warning:

```
ifelse(z >= 0, log(z), NA_real_)
## Warning in log(z): NaNs produced
## [1]       NA       NA  0.44386 -2.65202 -2.04571  0.53945
```

This is because, with `log(z)`, we compute the logarithms of negative values anyway. To fix this, we can write:

```
log(ifelse(z >= 0, z, NA_real_))
## [1]       NA       NA  0.44386 -2.65202 -2.04571  0.53945
```

---

In case we yearn for an **if…else if…else**-type expression, the calls to **ifelse** can naturally be nested.

**Example 3.2** *A version of* `pmax(pmax(x, y), z)` *can be written as:*

```
ifelse(x >= y,
    ifelse(z >= x, z, x),
    ifelse(z >= y, z, y)
)
## [1] 0.46092 0.11068 1.55871 1.78691 1.22408 1.71506
```

*However, determining three intermediate logical vectors is not necessary. We can save one call to* `` `>=` `` *by introducing an auxiliary variable:*

```
xy <- ifelse(x >= y, x, y)
ifelse(z >= xy, z, xy)
## [1] 0.46092 0.11068 1.55871 1.78691 1.22408 1.71506
```

**Exercise 3.3** *Figure 3.1 depicts a realisation of the mixture* $Z = 0.2X + 0.8Y$ *of two normal distributions* $X \sim \mathrm{N}(-2, 0.5)$ *and* $Y \sim \mathrm{N}(3, 1)$.

```
n <- 100000
z <- ifelse(runif(n) <= 0.2, rnorm(n, -2, 0.5), rnorm(n, 3, 1))
hist(z, breaks=101, probability=TRUE, main="", col="white")
```

*In other words, we generated a variate from the normal distribution that has the expected value of* $-2$ *with probability* $20\%$, *and from the one with the expectation of* $3$ *otherwise. Thus inspired, generate the Gaussian mixtures:*

- $\frac{2}{3}X + \frac{1}{3}Y$, *where* $X \sim \mathrm{N}(100, 16)$ *and* $Y \sim \mathrm{N}(116, 8)$,
- $0.3X + 0.4Y + 0.3Z$, *where* $X \sim \mathrm{N}(-10, 2)$, $Y \sim \mathrm{N}(0, 2)$, *and* $Z \sim \mathrm{N}(10, 2)$.

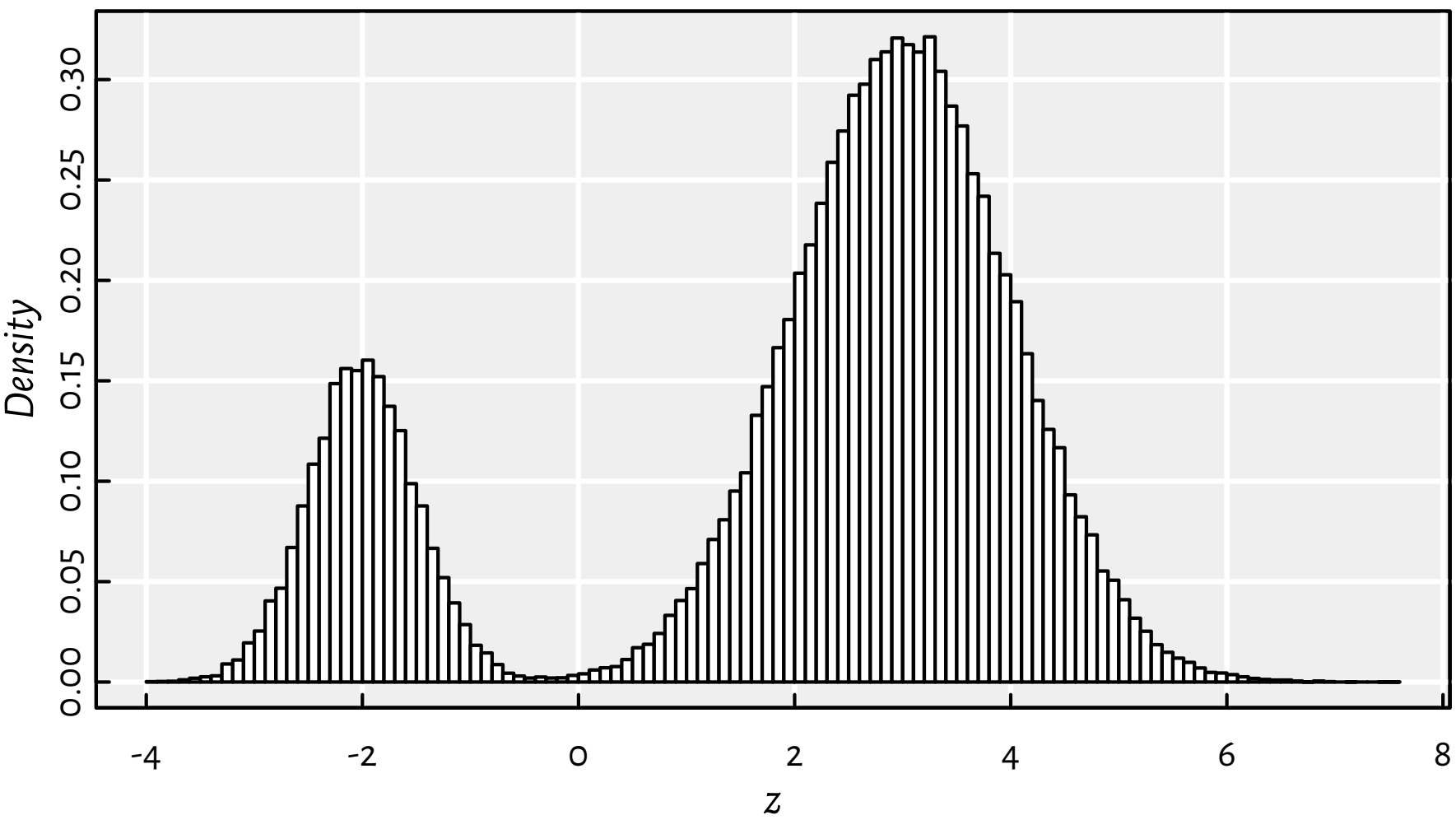

Figure 3.1. A mixture of two Gaussians generated with **`ifelse`**.

*(*) On a side note, knowing that if X follows* $\mathrm{N}(0, 1)$, *then the scaled-shifted* $\sigma X + \mu$ *is distributed* $\mathrm{N}(\mu, \sigma)$, *the above can be equivalently written as:*

```
w <- (runif(n) <= 0.2)
z <- rnorm(n, 0, 1)*ifelse(w, 0.5, 1) + ifelse(w, -2, 3)
```

## 3.5 Exercises

**Exercise 3.4** *Answer the following questions.*

- *Why the statement "The Earth is flat or the smallpox vaccine is proven effective" is obviously true?*
- *What is the difference between NA and* `NA_real_`*?*
- *Why is "*`FALSE & NA`*" equal to* `FALSE`*, but "*`TRUE & NA`*" is NA?*
- *Why has* `ifelse(x>=0, sqrt(x), NA_real_)` *a tendency to generate warnings and how to rewrite it so as to prevent that from happening?*
- *What is the interpretation of* `mean(x >= 0 & x <= 1)`*?*
- *For some integer x and y, how to verify whether* $0 < x < 100$, $0 < y < 100$, *and* $x < y$, *all at the same time?*
- *Mathematically, for all real* $x, y > 0$, *we have* $\log xy = \log x + \log y$. *Why then* `all(log(x*y) == log(x)+log(y))` *can sometimes return* `FALSE`*? How to fix this?*

- *Is `x/y/z` always equal to `x/(y/z)`? How to fix this?*
- *What is the purpose of very specific functions such as **`log1p`** and **`expm1`** (see their help page) and many others listed in, e.g., the GNU GSL library [29]? Is our referring to them a violation of the beloved "do not multiply entities without necessity" rule?*
- *If we know that $x$ may be subject to error, how to test whether $x > 0$ in a robust manner?*
- *Is "`y<-5`" the same as "`y <- 5`" or rather "`y < -5`"?*

**Exercise 3.5** *What is the difference between **`all`** and **`isTRUE`**? What about **`` `==` ``**, **`identical`**, and **`all.equal`**? Is the last one properly vectorised?*

**Exercise 3.6** *Compute the cross-entropy loss between a numeric vector **p** with values in the interval $(0,1)$ and a logical vector **y**, both of length $n$ (you can generate them randomly or manually, it does not matter, it is just an exercise):*

$$\mathcal{L}(\boldsymbol{p},\boldsymbol{y}) = \frac{1}{n}\sum_{i=1}^{n} \ell_i,$$

*where*

$$\ell_i = \begin{cases} -\log p_i & \text{if } y_i \text{ is TRUE}, \\ -\log(1-p_i) & \text{if } y_i \text{ is FALSE}. \end{cases}$$

*Interpretation: in classification problems, $y_i \in \{\mathit{FALSE}, \mathit{TRUE}\}$ denotes the true class of the i-th object (say, whether the i-th hospital patient is symptomatic) and $p_i \in (0,1)$ is a machine learning algorithm's* confidence *that i belongs to class TRUE (e.g., how sure a decision tree model is that the corresponding person is unwell). Ideally, if $y_i$ is TRUE, $p_i$ should be close to 1 and to $0$ otherwise. The cross-entropy loss quantifies by how much a classifier differs from the omniscient one. The use of the logarithm penalises strong beliefs in the wrong answer.*

By the way, if we have solved any of the exercises encountered so far by referring to **`if`** statements, **`for`** loops, vector indexing like `x[...]`, or any external R package, we recommend going back and rewrite our code. Let's keep things simple (effective, readable) by only using *base* R's vectorised operations that we have introduced.

# 4

# *Lists and attributes*

After two brain-teasing chapters, it is time to cool it down a little. In this more technical part, we will introduce *lists*, which serve as universal containers for R objects of any size and type. Moreover, we will also show that each R object can be equipped with a number of optional *attributes*. Thanks to them, we will be able to label elements in any vector, and, in Chapter 10, introduce new complex data types such as matrices and data frames.

## 4.1 Type hierarchy and conversion

So far, we have been playing with three types of atomic vectors:

1. `logical` (Chapter 3),
2. `numeric` (Chapter 2),
3. `character` (which we have barely touched upon yet, but rest assured that they will be covered in detail very soon; see Chapter 6).

To determine the type of an object programmatically, we can call the **`typeof`** function.

```
typeof(c(1, 2, 3))
## [1] "double"
typeof(c(TRUE, FALSE, TRUE, NA))
## [1] "logical"
typeof(c("spam", "spam", "bacon", "eggs", "spam"))
## [1] "character"
```

We can easily convert between these types, either on our explicit demand (*type casting*) or on-the-fly (*coercion*, when we perform an operation that expects something different from the kind of input it was fed with).

---

**Note** (*) Numeric vectors are reported as being either of the type `double` (double-precision floating-point numbers) or `integer` (32-bit; it is a subset of `double`); see Section 6.4.1. In most practical cases, this is a technical detail that we can risklessly ignore; compare also the **`mode`** function.

---

### 4.1.1 Explicit type casting

We can use functions such as **as.logical**, **as.numeric**[1], and **as.character** to *convert* given objects to the corresponding types.

```
as.numeric(c(TRUE, FALSE, NA, TRUE, NA, FALSE))  # synonym: as.double
## [1]  1  0 NA  1 NA  0
as.logical(c(-2, -1, 0, 1, 2, 3, NA_real_, -Inf, NaN))
## [1]  TRUE  TRUE FALSE  TRUE  TRUE  TRUE    NA  TRUE    NA
```

---

**Important** The rules are:

- TRUE → 1,
- FALSE → 0,
- NA → NA_real_,

and:

- 0 → FALSE,
- NA_real_ and NaN → NA,
- anything else → TRUE.

The distinction between zero and non-zero is commonly applied in other programming languages as well.

---

Moreover, in the case of the conversion involving character strings, we have:

```
as.character(c(TRUE, FALSE, NA, TRUE, NA, FALSE))
## [1] "TRUE"  "FALSE" NA      "TRUE"  NA      "FALSE"
as.character(c(-2, -1, 0, 1, 2, 3, NA_real_, -Inf, NaN))
## [1] "-2"   "-1"   "0"    "1"    "2"    "3"    NA     "-Inf" "NaN"
as.logical(c("TRUE", "True", "true", "T",
             "FALSE", "False", "false", "F",
             "anything other than these", NA_character_))
##  [1]  TRUE  TRUE  TRUE  TRUE FALSE FALSE FALSE FALSE    NA    NA
as.numeric(c("0", "-1.23e4", "pi", "2+2", "NaN", "-Inf", NA_character_))
## Warning: NAs introduced by coercion
## [1]      0 -12300     NA     NA    NaN   -Inf     NA
```

### 4.1.2 Implicit conversion (coercion)

Recall that we referred to the three vector types as *atomic* ones. They can only be used to store elements of the *same type*. If we make an attempt at composing an object of

[1] (*) **as.numeric** is a built-in generic function identical to (synonymous with) **as.double**; see Section 10.2.3. **is.numeric** is generic too, and is more universal than **is.double**, which only verifies whether **typeof** returns "double". For instance, vectors of the type *integer* which we mention later are considered numeric as well.

mixed types with `c`, the common type will be determined in such a way that data are stored without information loss:

```
c(-1, FALSE, TRUE, 2, "three", NA)
## [1] "-1"    "FALSE" "TRUE"  "2"     "three" NA
c("zero", TRUE, NA)
## [1] "zero" "TRUE" NA
c(-1, FALSE, TRUE, 2, NA)
## [1] -1  0  1  2 NA
```

Hence, we see that `logical` is the most specialised of the tree, whereas `character` is the most general.

---

**Note** The logical `NA` is converted to `NA_real_` and `NA_character_` in the preceding examples. R users tend to rely on implicit type conversion when they write `c(1, 2, NA, 4)` rather than `c(1, 2, NA_real_, 4)`. In most cases, this is fine, but it might make us less vigilant.

However, occasionally, it will be wiser to be more unequivocal. For instance, `rep(NA_real_, 1e9)` preallocates a long numeric vector instead of a logical one.

---

Some functions that expect vectors of specific types can apply coercion by themselves (or act as if they do so):

```
c(NA, FALSE, TRUE) + 10  # implicit conversion logical -> numeric
## [1] NA 10 11
c(-1, 0, 1) & TRUE  # implicit conversion numeric -> logical
## [1]  TRUE FALSE  TRUE
sum(c(TRUE, TRUE, FALSE, TRUE, FALSE))  # same as sum(as.numeric(...))
## [1] 3
cumsum(c(TRUE, TRUE, FALSE, TRUE, FALSE))
## [1] 1 2 2 3 3
cummin(c(TRUE, TRUE, FALSE, TRUE, FALSE))
## [1] 1 1 0 0 0
```

**Exercise 4.1** *In Exercise 3.6, we computed the cross-entropy loss between a logical vector* $\boldsymbol{y} \in \{0,1\}^n$ *and a numeric vector* $\boldsymbol{p} \in (0,1)^n$*. This measure can be equivalently defined as:*

$$\mathcal{L}(\boldsymbol{p},\boldsymbol{y}) = -\frac{1}{n}\left(\sum_{i=1}^{n} y_i \log(p_i) + (1-y_i)\log(1-p_i)\right).$$

*Implement this formula using vectorised operations, but not relying on* `ifelse` *this time. Then, compute the cross-entropy loss between, for instance, "*`y <- sample(c(FALSE, TRUE), n, replace=TRUE)`*" and "*`p <- runif(n)`*" for some* `n`*. Note how seamlessly we translate between* `FALSE`*/*`TRUE`*s and 0/1s in the above equation (in particular, where* $1-y_i$ *means the logical negation of* $y_i$*).*

## 4.2 Lists

*Lists* are *generalised* vectors. They can be comprised of R objects of any kind, also other lists. It is why we classify them as *recursive* (and not atomic) objects. They are especially useful wherever there is a need to handle some *multitude* as a single entity.

### 4.2.1 Creating lists

The most straightforward way to create a list is by means of the **list** function:

```
list(1, 2, 3)
## [[1]]
## [1] 1
##
## [[2]]
## [1] 2
##
## [[3]]
## [1] 3
```

Notice that it is not the same as `c(1, 2, 3)`. We got a sequence that wraps three numeric vectors, each of length one. More examples:

```
list(1:3, 4, c(TRUE, FALSE, NA, TRUE), "and so forth")  # different types
## [[1]]
## [1] 1 2 3
##
## [[2]]
## [1] 4
##
## [[3]]
## [1]  TRUE FALSE    NA  TRUE
##
## [[4]]
## [1] "and so forth"
list(list(c(TRUE, FALSE, NA, TRUE), letters), list(1:3))  # a list of lists
## [[1]]
## [[1]][[1]]
## [1]  TRUE FALSE    NA  TRUE
##
## [[1]][[2]]
##  [1] "a" "b" "c" "d" "e" "f" "g" "h" "i" "j" "k" "l" "m" "n" "o" "p" "q"
## [18] "r" "s" "t" "u" "v" "w" "x" "y" "z"
##
##
## [[2]]
```

```
## [[2]][[1]]
## [1] 1 2 3
```

The display of lists is (un)pretty bloated. However, the `str` function prints any R object in a more concise fashion:

```
str(list(list(c(TRUE, FALSE, NA, TRUE), letters), list(1:3)))
## List of 2
##  $ :List of 2
##   ..$ : logi [1:4] TRUE FALSE NA TRUE
##   ..$ : chr [1:26] "a" "b" "c" "d" ...
##  $ :List of 1
##   ..$ : int [1:3] 1 2 3
```

---

**Note** In Section 4.1, we said that the `c` function, when fed with arguments of mixed types, tries to determine the common type that retains the sense of data. If coercion to an atomic vector is not possible, the result will be a list.

```
c(1, "two", identity)  # `identity` is an object of the type "function"
## [[1]]
## [1] 1
##
## [[2]]
## [1] "two"
##
## [[3]]
## function (x)
## x
## <environment: namespace:base>
```

---

Thus, the `c` function can also be used to concatenate lists:

```
c(list(1), list(2), list(3))  # three lists -> one list
## [[1]]
## [1] 1
##
## [[2]]
## [1] 2
##
## [[3]]
## [1] 3
```

Lists can be repeated using `rep`:

```
rep(list(1:11, LETTERS), 2)
## [[1]]
```

```
##  [1]  1  2  3  4  5  6  7  8  9 10 11
##
## [[2]]
##  [1] "A" "B" "C" "D" "E" "F" "G" "H" "I" "J" "K" "L" "M" "N" "O" "P" "Q"
## [18] "R" "S" "T" "U" "V" "W" "X" "Y" "Z"
##
## [[3]]
##  [1]  1  2  3  4  5  6  7  8  9 10 11
##
## [[4]]
##  [1] "A" "B" "C" "D" "E" "F" "G" "H" "I" "J" "K" "L" "M" "N" "O" "P" "Q"
## [18] "R" "S" "T" "U" "V" "W" "X" "Y" "Z"
```

### 4.2.2 Converting to and from lists

The conversion of an atomic vector to a list of vectors of length one can be done via a call to **`as.list`**:

```
as.list(c(1, 2, 3))  # vector of length 3 -> list of 3 vectors of length 1
## [[1]]
## [1] 1
##
## [[2]]
## [1] 2
##
## [[3]]
## [1] 3
```

Unfortunately, calling, say, **`as.numeric`** on a list arouses an error, even if it is comprised of numeric vectors only. We can try flattening it to an atomic sequence by calling **`unlist`**:

```
unlist(list(list(1, 2), list(3, list(4:8)), 9))
## [1] 1 2 3 4 5 6 7 8 9
unlist(list(list(1, 2), list(3, list(4:8)), "spam"))
## [1] "1"    "2"    "3"    "4"    "5"    "6"    "7"    "8"    "spam"
```

**Note** (*) Chapter 11 will mention the **`simplify2array`** function, which generalises **`unlist`** in a way that can sometimes give rise to a matrix.

## 4.3 NULL

NULL, being the one and only instance of the eponymous type, can be used as a placeholder for an R object or designate the absence of any entities whatsoever.

```
list(NULL, NULL, month.name)
## [[1]]
## NULL
##
## [[2]]
## NULL
##
## [[3]]
##  [1] "January"   "February"  "March"     "April"     "May"
##  [6] "June"      "July"      "August"    "September" "October"
## [11] "November"  "December"
```

NULL is different from a vector of length zero because the latter has a type. However, NULL sometimes *behaves* like a zero-length vector. In particular, **length**(NULL) returns 0. Also, **c** called with no arguments returns NULL.

Testing for NULL-ness can be done with a call to **is.null**.

---

**Important** NULL is not the same as NA. The former cannot be emplaced in an atomic vector.

```
c(1, NA, 3, NULL, 5)  # here, NULL behaves like a zero-length vector
## [1]  1 NA  3  5
```

They have very distinct semantics (no value vs a missing value).

---

Later we will see that some functions return NULL invisibly when they have nothing interesting to report. This is the case of **print** or **plot**, which are called because of their side effects (printing and plotting).

Furthermore, in certain contexts, replacing content with NULL will actually result in its removal, e.g., when subsetting a list.

## 4.4 Object attributes

Lists can embrace many entities in the form of a single item sequence. *Attributes*, on the other hand, give means to inject *extra* data into an object. They are unordered

key=value pairs, where key is a single string, and value is any R object except NULL. We can introduce them by calling, amongst others[2], the **structure** function:

```
x_simple <- 1:10
x <- structure(
    x_simple,  # the object to be equipped with attributes
    attribute1="value1",
    attribute2=c(6, 100, 324)
)
```

### 4.4.1 Developing perceptual indifference to most attributes

Let's see how the foregoing x is reported on the console:

```
print(x)
##  [1]  1  2  3  4  5  6  7  8  9 10
## attr(,"attribute1")
## [1] "value1"
## attr(,"attribute2")
## [1]   6 100 324
```

The object of concern, 1:10, was displayed first. We need to get used to that. Most of the time, we suggest to treat the "attr…" parts of the display as if they were printed in tiny font.

Equipping an object with attributes *does not usually change its nature*; see, however, Chapter 10 for a few exceptions. The above x is still treated as an ordinary sequence of numbers by most functions:

```
sum(x)    # the same as sum(1:10); `sum` does not care about any attributes
## [1] 55
typeof(x)  # just a numeric vector, but with some perks
## [1] "integer"
```

---

**Important** Attributes are generally ignored by most functions unless they have specifically been programmed to pay attention to them.

---

### 4.4.2 But there are a few use cases

Some R functions add attributes to the return value to sneak extra information that *might* be useful, just in case. For instance, **na.omit**, whose main aim is to remove missing values from an atomic vector, yields:

[2] Other ways include the replacement versions of the **attr** and **attributes** functions; see Section 9.3.6.

```
y <- c(10, 20, NA, 40, 50, NA, 70)
(y_na_free <- na.omit(y))
## [1] 10 20 40 50 70
## attr(,"na.action")
## [1] 3 6
## attr(,"class")
## [1] "omit"
```

We can enjoy the `NA`-free version of `y` in further computations:

```
mean(y_na_free)
## [1] 38
```

Additionally, the `na.action` attribute indicates the former whereabouts of the missing observations:

```
attr(y_na_free, "na.action")  # read the attribute value
## [1] 3 6
## attr(,"class")
## [1] "omit"
```

We ignore the `class` part until Chapter 10.

As another example, **`gregexpr`** discussed in Chapter 6 searches for a given pattern in a character vector:

```
needle <- "spam|durian"  # pattern to search for: spam OR durian
haystack <- c("spam, bacon, and durian-flavoured spam", "spammer")  # text
(pos <- gregexpr(needle, haystack, perl=TRUE))
## [[1]]
## [1]  1 18 35
## attr(,"match.length")
## [1] 4 6 4
## attr(,"index.type")
## [1] "chars"
## attr(,"useBytes")
## [1] TRUE
##
## [[2]]
## [1] 1
## attr(,"match.length")
## [1] 4
## attr(,"index.type")
## [1] "chars"
## attr(,"useBytes")
## [1] TRUE
```

We sought all occurrences of the pattern within two character strings. As their number may vary from string to string, wrapping the results in a list was a good design choice. Each list element gives the starting positions where matches can be found: there are three and one match(es), respectively. Moreover, every vector of positions has a designated `match.length` attribute (amongst others), in case we need it.

**Exercise 4.2** *Create a list with EUR/AUD, EUR/GBP, and EUR/USD exchange rates read from the `euraud-*.csv`, `eurgbp-*.csv`, and `eurusd-*.csv` files in our data repository[3]. Each of its three elements should be a numeric vector storing the currency exchange rates. Furthermore, equip them with `currency_from`, `currency_to`, `date_from`, and `date_to` attributes. For example:*

```
##  [1]     NA 1.6006 1.6031     NA     NA 1.6119 1.6251 1.6195 1.6193 1.6132
## [11]     NA     NA 1.6117 1.6110 1.6188 1.6115 1.6122     NA
## attr(,"currency_from")
## [1] "EUR"
## attr(,"currency_to")
## [1] "AUD"
## attr(,"date_from")
## [1] "2020-01-01"
## attr(,"date_to")
## [1] "2020-06-30"
```

*Such an additional piece of information could be stored in a few separate variables (other vectors), but then it would not be as convenient to use as the above representation.*

### 4.4.3 Special attributes

Attributes have great potential which is somewhat wasted, for R users rarely know:

- that attributes exist (pessimistic scenario), or
- how to handle them (realistic scenario).

But *we* know now.

What is more, certain attributes have been predestined to play a unique role in R. Namely, the most prevalent amongst the *special attributes* are:

- `names`, `row.names`, and `dimnames` are used to label the elements of atomic and generic vectors (see below) as well as rows and columns in matrices (Chapter 11) and data frames (Chapter 12),
- `dim` turns flat vectors into matrices and other tensors (Chapter 11),
- `levels` labels the underlying integer codes in factor objects (Section 10.3.2),
- `class` can be used to bring forth new complex data structures based on basic types (Chapter 10).

We call them *special* because:

- they cannot be assigned arbitrary values; for instance, we will soon see that `names` can accept a character vector of a specific length,

[3] https://github.com/gagolews/teaching-data/tree/master/marek

- they can be accessed via designated functions, e.g., **names**, **class**, **dim**, **dimnames**, **levels**, etc.,
- they are widely recognised by many other functions.

However, in spite of the above, special attributes can still be managed as ordinary ones.

**Exercise 4.3** *comment is perhaps the most rarely used special attribute. Create an object (whatever) equipped with the comment attribute. Verify that assigning to it anything other than a character vector leads to an error. Read its value by calling the* ***comment*** *function. Display the object equipped with this attribute. Note that the* ***print*** *function ignores its existence whatsoever: this is how special it is.*

---

**Important** (*) The accessor functions such as **names** or **class** might return meaningful values, even if the corresponding attribute is not set explicitly; see, e.g., Section 11.1.5 for an example.

---

### 4.4.4 Labelling vector elements with the names attribute

The special attribute names labels atomic vectors' and lists' elements.

```
(x <- structure(c(13, 2, 6), names=c("spam", "sausage", "celery")))
##    spam sausage  celery
##      13       2       6
```

The labels may improve the expressivity and readability of our code and data.

**Exercise 4.4** *Verify that the above x is still an ordinary numeric vector by calling* ***typeof*** *and* ***sum*** *on it.*

Let's stress that we can ignore the names attribute whatsoever. If we apply any operation discussed in Chapter 2, we will garner the same result regardless whether such extra information is present or not.

It is just the **print** function that changed its behaviour slightly. After all, it is a special attribute. Instead of reporting:

```
## [1] 13  2  6
## attr(,"names")
## [1] "spam"    "sausage" "celery"
```

we got a nicely formatted table-like display. Non-special attributes are still printed in the standard way:

```
structure(x, additional_attribute=1:10)
##    spam sausage  celery
##      13       2       6
## attr(,"additional_attribute")
##  [1]  1  2  3  4  5  6  7  8  9 10
```

**Note** Chapter 5 will also mention that some operations (such as indexing) gain superpowers in the presence of the `names` attribute.

This attribute can be read by calling:

```
attr(x, "names")  # just like any other attribute
## [1] "spam"    "sausage" "celery"
names(x)  # because it is so special
## [1] "spam"    "sausage" "celery"
```

Named vectors can be easily created with the **c** and **list** functions as well:

```
c(a=1, b=2)
## a b
## 1 2
list(a=1, b=2)
## $a
## [1] 1
##
## $b
## [1] 2
c(a=c(x=1, y=2), b=3, c=c(z=4))  # this is smart
## a.x a.y   b c.z
##   1   2   3   4
```

Let's contemplate how a named list is printed on the console. Again, it is still a list, but with some extras.

**Exercise 4.5** *A whole lot of functions return named vectors. Evaluate the following expressions and read the corresponding pages in their documentation:*

- ***quantile**(**runif**(100)),*
- ***hist**(**runif**(100), plot=FALSE),*
- ***options**() (take note of `digits`, `scipen`, `max.print`, and `width`),*
- ***capabilities**().*

**Note** (*) Most of the time, lists are used merely as *containers* for other R objects. This is a dull yet essential role. However, let's just mention here that every data frame is, in fact, a generic vector (see Chapter 12). Each column corresponds to a named list element:

```
(df <- head(iris))  # some data frame
##   Sepal.Length Sepal.Width Petal.Length Petal.Width Species
## 1          5.1         3.5          1.4         0.2  setosa
## 2          4.9         3.0          1.4         0.2  setosa
## 3          4.7         3.2          1.3         0.2  setosa
```

```
## 4          4.6         3.1          1.5         0.2  setosa
## 5          5.0         3.6          1.4         0.2  setosa
## 6          5.4         3.9          1.7         0.4  setosa
typeof(df)  # it is just a list (with extras that will be discussed later)
## [1] "list"
unclass(df)  # how it is represented exactly (without the extras)
## $Sepal.Length
## [1] 5.1 4.9 4.7 4.6 5.0 5.4
##
## $Sepal.Width
## [1] 3.5 3.0 3.2 3.1 3.6 3.9
##
## $Petal.Length
## [1] 1.4 1.4 1.3 1.5 1.4 1.7
##
## $Petal.Width
## [1] 0.2 0.2 0.2 0.2 0.2 0.4
##
## $Species
## [1] setosa setosa setosa setosa setosa setosa
## Levels: setosa versicolor virginica
##
## attr(,"row.names")
## [1] 1 2 3 4 5 6
```

Therefore, the functions we discuss in this chapter are of use in processing such structured data, too.

### 4.4.5 Altering and removing attributes

We know that a single attribute can be read by calling **attr**. Their whole list is generated with a call to **attributes**.

```
(x <- structure(c("some", "object"), names=c("X", "Y"),
    attribute1="value1", attribute2="value2", attribute3="value3"))
##        X        Y
##   "some" "object"
## attr(,"attribute1")
## [1] "value1"
## attr(,"attribute2")
## [1] "value2"
## attr(,"attribute3")
## [1] "value3"
attr(x, "attribute1")  # reads a single attribute, returns NULL if unset
## [1] "value1"
attributes(x)  # returns a named list with all attributes of an object
## $names
```

```
## [1] "X" "Y"
##
## $attribute1
## [1] "value1"
##
## $attribute2
## [1] "value2"
##
## $attribute3
## [1] "value3"
```

We can alter an attribute's value or add further attributes by referring to the **structure** function once again. Moreover, setting an attribute's value to `NULL` gets rid of it completely.

```
structure(x, attribute1=NULL, attribute4="added", attribute3="modified")
##        X        Y
##   "some" "object"
## attr(,"attribute2")
## [1] "value2"
## attr(,"attribute3")
## [1] "modified"
## attr(,"attribute4")
## [1] "added"
```

As far as the `names` attribute is concerned, we may generate an unnamed copy of an object by calling:

```
unname(x)
## [1] "some"   "object"
## attr(,"attribute1")
## [1] "value1"
## attr(,"attribute2")
## [1] "value2"
## attr(,"attribute3")
## [1] "value3"
```

In Section 9.3.6, we will introduce *replacement functions*. They will enable us to modify or remove an object's attribute by calling "`attr(x, "some_attribute") <- new_value`".

Moreover, Section 5.5 highlights that certain operations (such as vector indexing, elementwise arithmetic operations, and coercion) might not preserve all attributes of the objects that were given as their inputs.

## 4.5 Exercises

**Exercise 4.6** *Provide an answer to the following questions.*

- *What is the meaning of* `c(TRUE, FALSE)*1:10`*?*
- *What does* `sum(as.logical(x))` *compute when* x *is a numeric vector?*
- *We said that atomic vectors of the type* `character` *are the most general ones. Therefore, is* `as.numeric(as.character(x))` *the same as* `as.numeric(x)`*, regardless of the type of* x*?*
- *What is the meaning of* `as.logical(x+y)` *if* x *and* y *are logical vectors? What about* `as.logical(x*y)`*,* `as.logical(1-x)`*, and* `as.logical(x!=y)`*?*
- *What is the meaning of the following when* x *is a logical vector?*
  - `cummin(x)` *and* `cummin(!x)`*,*
  - `cummax(x)` *and* `cummax(!x)`*,*
  - `cumsum(x)` *and* `cumsum(!x)`*,*
  - `cumprod(x)` *and* `cumprod(!x)`*.*
- *Let* x *be a* named *numeric vector, e.g., "*`x <- quantile(runif(100))`*". What is the result of* `2*x`*,* `mean(x)`*, and* `round(x, 2)`*?*
- *What is the meaning of* `x == NULL`*?*
- *Give two ways to create a named character vector.*
- *Give two ways (discussed above; there are more) to remove the* `names` *attribute from an object.*

**Exercise 4.7** *There are a few peculiarities when joining or coercing lists. Compare the results generated by the following pairs of expressions:*

```
# 1)
as.character(list(list(1, 2), list(3, list(4)), 5))
as.character(unlist(list(list(1, 2), list(3, list(4)), 5)))
# 2)
as.numeric(list(list(1, 2), list(3, list(4)), 5))
as.numeric(unlist(list(list(1, 2), list(3, list(4)), 5)))
# 3)
unlist(list(list(1, 2), sd))
list(1, 2, sd)
# 4)
c(list(c(1, 2), 3), 4, 5)
c(list(c(1, 2), 3), c(4, 5))
```

**Exercise 4.8** *Given numeric vectors* x*,* y*,* z*, and* w*, how to combine* x*,* y*, and* `list(z, w)` *so as to obtain* `list(x, y, z, w)`*? More generally, given a set of atomic vectors and lists of atomic* *vectors, how to combine them to obtain a single list of atomic vectors (not a list of atomic vectors and lists, not atomic vectors unwound, etc.)?*

**Exercise 4.9** *`saveRDS` serialises R objects and writes their snapshots to disk so that they can be restored via a call to `readRDS` at a later time. Verify that this function preserves object attributes. Also, check out `dput` and `dget` which work with objects' textual representations in the form of executable R code.*

**Exercise 4.10** *(*) Use `jsonlite::fromJSON` to read a JSON file in the form of a named list.*

In the extremely unlikely event of our finding the current chapter boring, let's rejoice: some of the exercises and remarks that we will encounter in the next part, which is devoted to vector indexing, will definitely be mouthwatering.

# 5

# *Vector indexing*

We now know plenty of ways to process vectors *in their entirety*, but how to extract and replace their specific *parts*? We will be collectively referring to such activities as *indexing*. This is because they are often performed through the *index operator*, `` `[` ``.

## 5.1 head and tail

Let's begin with something more lightweight, though. The **head** function fetches a few elements from the beginning of a vector.

```
x <- 1:10
head(x)  # head(x, 6)
## [1] 1 2 3 4 5 6
head(x, 3)   # get the first three
## [1] 1 2 3
head(x, -3)  # skip the last three
## [1] 1 2 3 4 5 6 7
```

Similarly, **tail** extracts a couple of items from the end of a sequence.

```
tail(x)  # tail(x, 6)
## [1]  5  6  7  8  9 10
tail(x, 3)   # get the last three
## [1]  8  9 10
tail(x, -3)  # skip the first three
## [1]  4  5  6  7  8  9 10
```

Both functions work on lists too[1]. They are useful for previewing the contents of *really big* objects. Also, they never complain about our trying to fetch supernumerary elements:

```
head(x, 100)  # no more than the first 100 elements
##  [1]  1  2  3  4  5  6  7  8  9 10
```

[1] **head** and **tail** are actually S3 generics defined in the **utils** package. We can call them on matrices and data frames, too; see Chapter 10.

## 5.2 Subsetting and extracting from vectors

Let x be a vector. Then x[i] returns its subset comprised of elements indicated by the indexer i, which can be a *single* vector of:

- nonnegative integers (gives the positions of elements to retrieve),
- negative integers (gives the positions to omit),
- logical values (states which items should be fetched or skipped),
- character strings (locates the elements with specific names).

### 5.2.1 Nonnegative indexes

Consider example vectors:

```
(x <- seq(10, 100, 10))
##  [1]  10  20  30  40  50  60  70  80  90 100
(y <- list(1, 11:12, 21:23))
## [[1]]
## [1] 1
##
## [[2]]
## [1] 11 12
##
## [[3]]
## [1] 21 22 23
```

The first element in a vector is at index 1. Hence:

```
x[1]          # the first element
## [1] 10
x[length(x)]  # the last element
## [1] 100
```

**Important** We might have wondered why "[1]" is displayed each time we print out an atomic vector on the console:

```
print((1:51)*10)
##  [1]  10  20  30  40  50  60  70  80  90 100 110 120 130 140 150 160 170
## [18] 180 190 200 210 220 230 240 250 260 270 280 290 300 310 320 330 340
## [35] 350 360 370 380 390 400 410 420 430 440 450 460 470 480 490 500 510
```

It is merely a visual hint indicating which vector element we output at the beginning of each line.

Vectorisation is a universal feature of R. It comes as no surprise that the indexer can also be of length greater than one.

```
x[c(1, length(x))]  # the first and the last
## [1]  10 100
x[1:3]  # the first three
## [1] 10 20 30
```

Take note of the boundary cases:

```
x[c(1, 2, 1, 0, 3, NA_real_, 1, 11)]  # repeated, 0, missing, out of bound
## [1] 10 20 10 30 NA 10 NA
x[c()]  # indexing by an empty vector
## numeric(0)
```

When applied on lists, the index operator always returns a list as well, even if we ask for a single element:

```
y[2]  # a list that includes the second element
## [[1]]
## [1] 11 12
y[c(1, 3)]  # not the same as x[1, 3] (a different story)
## [[1]]
## [1] 1
##
## [[2]]
## [1] 21 22 23
```

If we want to *extract* a component, i.e., to dig into what is inside a list at a specific location, we can refer to `` `[[` ``:

```
y[[2]]  # extract the second element
## [1] 11 12
```

This is exactly why R displays "`[[1]]`", "`[[2]]`", etc. when lists are printed.

On a side note, calling `x[[i]]` on an *atomic* vector, where `i` is a single value, has almost[2] the same effect as `x[i]`. However, `` `[[` `` generates an error if the subscript is out of bounds.

---

**Important** Let's reflect on the operators' behaviour in the case of nonexistent items:

```
c(1, 2, 3)[4]
## [1] NA
list(1, 2, 3)[4]
## [[1]]
## NULL
```

[2] See also Section 5.5 for the discussion on the preservation of object attributes.

```
c(1, 2, 3)[[4]]
## Error in `c(1, 2, 3)[[4]]`: ! subscript out of bounds
list(1, 2, 3)[[4]]
## Error in `list(1, 2, 3)[[4]]`: ! subscript out of bounds
```

**Note** (*) `[[` also supports multiple indexers.

```
y[[c(1, 3)]]
## Error in `y[[c(1, 3)]]`: ! subscript out of bounds
```

Its meaning is different from `y[c(1, 3)]`, though; we are about to extract a single value, remember? Here, indexing is applied *recursively*. Namely, the above is equivalent to `y[[1]][[3]]`. We got an error because `y[[1]]` is of a length smaller than three.

More examples:

```
y[[c(3, 1)]]  # y[[3]][[1]]
## [1] 21
list(list(7))[[c(1, 1)]]  # 7, not list(7)
## [1] 7
```

### 5.2.2 Negative indexes

The indexer can also be a vector of negative integers. This way, we can *exclude* the elements at given positions:

```
y[-1]  # all but the first
## [[1]]
## [1] 11 12
##
## [[2]]
## [1] 21 22 23
x[-(1:3)]  # all but the first three
## [1]  40  50  60  70  80  90 100
x[-c(1, 0, 2, 1, 1, 8:100)]  # 0, repeated, out of bound indexes
## [1] 30 40 50 60 70
```

**Note** Negative and positive indexes cannot be mixed.

```
x[-1:3]  # recall that -1:3 == (-1):3
## Error in `x[-1:3]`: ! only 0's may be mixed with negative subscripts
```

Also, `NA` indexes cannot be mixed with negative ones.

### 5.2.3 Logical indexer

A vector can also be subsetted by means of a logical vector. If they both are of identical lengths, the consecutive elements in the latter indicate whether the corresponding elements of the indexed vector are supposed to be selected (TRUE) or omitted (FALSE).

```
#   1***  2     3     4     5***  6***  7     8***  9?   10***
x[c(TRUE, FALSE, FALSE, FALSE, TRUE, TRUE, FALSE, TRUE, NA,  TRUE)]
## [1]  10  50  60  80  NA 100
```

In other words, x[l], where l is a logical vector, returns all x[i] with i such that l[i] is TRUE. We thus extracted the elements at indexes 1, 5, 6, 8, and 10.

---

**Important** Be careful: if the element selector is NA, we will get a missing value (for atomic vectors) or NULL (for lists).

```
c("one", "two", "three")[c(NA, TRUE, FALSE)]
## [1] NA    "two"
list("one", "two", "three")[c(NA, TRUE, FALSE)]
## [[1]]
## NULL
##
## [[2]]
## [1] "two"
```

This, lamentably, comes with no warning, which might be problematic when indexers are generated programmatically. As a remedy, we sometimes pass the logical indexer to the **which** function first. It returns the indexes of the elements equal to TRUE, ignoring the missing ones.

```
which(c(NA, TRUE, FALSE, TRUE, FALSE, NA, TRUE))
## [1] 2 4 7
c("one", "two", "three")[which(c(NA, TRUE, FALSE))]
## [1] "two"
```

---

Recall that in Chapter 3, we discussed ample vectorised operations that generate logical vectors. Anything that yields a logical vector of the same length as x can be passed as an indexer.

```
x > 60  # yes, it is a perfect indexer candidate
##  [1] FALSE FALSE FALSE FALSE FALSE FALSE  TRUE  TRUE  TRUE  TRUE
x[x > 60]  # select elements in `x` that are greater than 60
## [1]  70  80  90 100
x[x < 30 | 70 < x]  # elements not between 30 and 70
## [1]  10  20  80  90 100
x[x < mean(x)]  # elements smaller than the mean
## [1] 10 20 30 40 50
x[x^2 > 7777 | log10(x) <= 1.6]  # indexing via a transformed version of `x`
```

```
## [1]  10  20  30  90 100
(z <- round(runif(length(x)), 2))  # ten pseudorandom numbers
##  [1] 0.29 0.79 0.41 0.88 0.94 0.05 0.53 0.89 0.55 0.46
x[z <= 0.5]  # indexing based on `z`, not `x`: no problem
## [1]  10  30  60 100
```

The indexer is always evaluated first and then passed to the subsetting operation. The index operator does not care how an indexer is generated.

Furthermore, the recycling rule is applied when necessary:

```
x[c(FALSE, TRUE)]  # every second element
## [1]  20  40  60  80 100
y[c(TRUE, FALSE)]  # interestingly, there is no warning here
## [[1]]
## [1] 1
##
## [[2]]
## [1] 21 22 23
```

**Exercise 5.1** *Consider a simple database about six people, their favourite dishes, and birth years.*

```
name <- c("Graham", "John", "Terry", "Eric",  "Michael", "Terry")
food <- c("bacon",  "spam", "spam",  "eggs",  "spam",    "beans")
year <- c( 1941,     1939,   1942,    1943,    1943,      1940  )
```

*The consecutive elements in different vectors correspond to each other, e.g., Graham was born in 1941, and his go-to food was bacon.*

- *List the names of people born in 1941 or 1942.*
- *List the names of those who like spam.*
- *List the names of those who like spam and were born after 1940.*
- *Compute the average birth year of the lovers of spam.*
- *Give the average age, in 1969, of those who didn't find spam utmostly delicious.*

*The answers must be provided programmatically, i.e., do not just write* `"Eric"` *and* `"Graham"`*. Make the code generic enough so that it works in the case of any other database of this kind, no matter its size.*

**Exercise 5.2** *Remove missing values from a given vector without referring to* `na.omit`*.*

### 5.2.4 Character indexer

Let's consider a vector equipped with the `names` attribute:

```
x <- structure(x, names=letters[1:10])  # add names
print(x)
##   a   b   c   d   e   f   g   h   i   j
##  10  20  30  40  50  60  70  80  90 100
```

These labels can be referred to when extracting the elements. To do this, we use an indexer that is a character vector:

```
x[c("a", "f", "a", "g", "z")]
##    a    f    a    g <NA>
##   10   60   10   70   NA
```

---

**Important** We have said that special object attributes add *extra* functionality on top of the existing ones. Therefore, indexing by means of positive, negative, and logical vectors is still available:

```
x[1:3]
##  a  b  c
## 10 20 30
x[-(1:5)]
##   f   g   h   i   j
##  60  70  80  90 100
x[x > 70]
##   h   i   j
##  80  90 100
```

---

Lists can also be subsetted this way.

```
(y <- structure(y, names=c("first", "second", "third")))
## $first
## [1] 1
##
## $second
## [1] 11 12
##
## $third
## [1] 21 22 23
y[c("first", "second")]
## $first
## [1] 1
##
## $second
## [1] 11 12
y["third"]     # result is a list
## $third
## [1] 21 22 23
```

```
y[["third"]]  # result is the specific element unwrapped
## [1] 21 22 23
```

**Important** Labels do not have to be unique. When we have repeated names, the first matching element is extracted:

```
structure(c(1, 2, 3), names=c("a", "b", "a"))["a"]
## a
## 1
```

There is no direct way to select all *but* given names, just like with negative integer indexers. For a workaround, see Section 5.4.1.

**Exercise 5.3** *Rewrite the solution to Exercise 5.1 assuming that we now have three features wrapped inside a list.*

```
(people <- list(
    Name=c("Graham", "John", "Terry", "Eric",  "Michael", "Terry", "Steve"),
    Food=c("bacon",  "spam", "spam",  "eggs",  "spam",    "beans", "spam"),
    Year=c( 1941,     1939,   1942,    1943,    1943,      1940,   NA_real_)
))
## $Name
## [1] "Graham"  "John"    "Terry"   "Eric"    "Michael" "Terry"   "Steve"
##
## $Food
## [1] "bacon" "spam"  "spam"  "eggs"  "spam"  "beans" "spam"
##
## $Year
## [1] 1941 1939 1942 1943 1943 1940   NA
```

*Do not refer to `name`, `food`, and `year` directly. Instead, use the full `people[["Name"]]` etc. accessors. There is no need to pout: it is just a tiny bit of extra work. Also, notice that Steve has joined the group; hello, Steve.*

## 5.3 Replacing elements

### 5.3.1 Modifying atomic vectors

There are also *replacement* versions of the aforementioned indexing schemes. They allow us to substitute some new content for the old one.

```
(x <- 1:12)
##  [1]  1  2  3  4  5  6  7  8  9 10 11 12
```

```
x[length(x)] <- 42  # modify the last element
print(x)
##  [1]  1  2  3  4  5  6  7  8  9 10 11 42
```

The principles of vectorisation, recycling rule, and implicit coercion are all in place:

```
x[c(TRUE, FALSE)] <- c("a", "b", "c")
print(x)
##  [1] "a"  "2"  "b"  "4"  "c"  "6"  "a"  "8"  "b"  "10" "c"  "42"
```

Long story long: first, to ensure that the new content can be poured into the old wineskin, R coerced the numeric vector to a character one. Then, every second element therein, a total of six items, was replaced by a recycled version of the replacement sequence of length three. Finally, the name x was rebound to such a brought-forth object and the previous one became forgotten.

---

**Note** For more details on replacement functions in general, see Section 9.3.6. Such operations alter the state of the object they are called on (quite rare a behaviour in functional languages).

---

**Exercise 5.4** *Replace missing values in a given numeric vector with the arithmetic mean of its well-defined observations.*

### 5.3.2 Modifying lists

List contents can be altered as well. For modifying individual elements, the safest option is to use the replacement version of the `[[` operator:

```
y <- list(a=1, b=1:2, c=1:3)
y[[1]] <- 100:110
y[["c"]] <- -y[["c"]]
print(y)
## $a
##  [1] 100 101 102 103 104 105 106 107 108 109 110
##
## $b
## [1] 1 2
##
## $c
## [1] -1 -2 -3
```

The replacement version of `[` modifies a whole sub-list:

```
y[1:3] <- list(1, c("a", "b", "c"), c(TRUE, FALSE))
print(y)
## $a
```

```
## [1] 1
##
## $b
## [1] "a" "b" "c"
##
## $c
## [1]  TRUE FALSE
```

Moreover:

```
y[1] <- list(1:10)  # replace one element with one object
y[-1] <- 10:11      # replace two elements with two singletons
print(y)
## $a
##  [1]  1  2  3  4  5  6  7  8  9 10
##
## $b
## [1] 10
##
## $c
## [1] 11
```

---

**Note** Let `i` be a vector of positive indexes of elements to be modified. Overall, calling "`y[i] <- z`" behaves as if we wrote:

1. `y[[ i[1] ]] <- z[[1]]`,
2. `y[[ i[2] ]] <- z[[2]]`,
3. `y[[ i[3] ]] <- z[[3]]`,

and so forth.

Furthermore, `z` (but not `i`) will be recycled when necessary. In other words, we retrieve `z[[j %% length(z)]]` for consecutive `j`s from 1 to the length of `i`.

---

**Exercise 5.5** *Reflect on the results of the following expressions:*

- `y[1] <- c("a", "b", "c")`,
- `y[[1]] <- c("a", "b", "c")`,
- `y[[1]] <- list(c("a", "b", "c"))`,
- `y[1:3] <- c("a", "b", "c")`,
- `y[1:3] <- list(c("a", "b", "c"))`,
- `y[1:3] <- "a"`,
- `y[1:3] <- list("a")`,

- `y[c(1, 2, 1)] <- c("a", "b", "c")`.

---

**Important** Setting a list item to `NULL` removes it from the list completely.

```
y <- list(1, 2, 3, 4)
y[1] <- NULL        # removes the first element (i.e., 1)
y[[1]] <- NULL      # removes the first element (i.e., now 2)
y[1] <- list(NULL) # sets the first element (i.e., now 3) to NULL
print(y)
## [[1]]
## NULL
##
## [[2]]
## [1] 4
```

The same notation convention is used for dropping object attributes; see Section 9.3.6.

---

### 5.3.3 Inserting new elements

New elements can be pushed at the end of the vector easily[3].

```
(x <- 1:5)
## [1] 1 2 3 4 5
x[length(x)+1] <- 6  # insert at the end
print(x)
## [1] 1 2 3 4 5 6
x[10] <- 10  # insert at the end but add more items
print(x)
##  [1]  1  2  3  4  5  6 NA NA NA 10
```

The elements to be inserted can be named as well:

```
x["a"] <- 11  # still inserts at the end
x["z"] <- 12
x["c"] <- 13
x["z"] <- 14  # z is already there; replace
print(x)
##                               a  z  c
##  1  2  3  4  5  6 NA NA NA 10 11 14 13
```

Note that `x` was not equipped with the `names` attribute before. The unlabelled elements were assigned blank labels (empty strings).

---

**Note** It is not possible to insert new elements at the beginning or in the middle of a sequence, at least not with the index operator. By writing "`x[3:4] <- 1:5`" we do not

[3] And often cheaply; see Section 8.3.5 for performance notes. Still, a warning can be generated on each size extension if the `"check.bounds"` flag is set; see `help("options")`.

replace two elements in the middle with five other ones. However, we can always use the **c** function to slice parts of the vector and intertwine them with some new content:

```
x <- seq(10, 100, 10)
x <- c(x[1:2], 1:5, x[5:7])
print(x)
##  [1] 10 20  1  2  3  4  5 50 60 70
```

## 5.4 Functions related to indexing

Let's review a few operations which pinpoint interesting elements in a vector (or functions based on these).

### 5.4.1 Matching elements in another vector

We know that the **`==`** operator acts in an elementwise manner. It compares each element in a vector on its left side to the *corresponding* element in a vector on the right side. Thus, missing values and the recycling rule aside, if "`z <- (x == y)`", then `z[i]` is `TRUE` if and only if `x[i]` is equal to `y[i]`.

The **`%in%`** operator[4] is vectorised differently. It checks whether each element on the left-hand side matches *one of* the elements on the right. Given "`z <- (x %in% y)`", `z[i]` is `TRUE` whenever `x[i]` is equal to `y[j]` for some `j`.

```
c("spam", "bacon", "spam", "eggs", "spam") %in% c("eggs", "spam", "ham")
## [1]  TRUE FALSE  TRUE  TRUE  TRUE
```

**Example 5.6** *Here is how we can remove the elements of a vector that have been assigned specified labels.*

```
(x <- structure(1:12, names=month.abb))  # example vector
## Jan Feb Mar Apr May Jun Jul Aug Sep Oct Nov Dec
##   1   2   3   4   5   6   7   8   9  10  11  12
x[names(x) %notin% c("Jan", "May", "Sep", "Oct")]  # get rid of some elements
## Feb Mar Apr Jun Jul Aug Nov Dec
##   2   3   4   6   7   8  11  12
```

More generally, **match**(x, y) gives us the index of the element in `y` that matches each `x[i]`.

```
match(c("spam", "bacon", "spam", "eggs", "spam"), c("eggs", "spam", "ham"))
## [1]  2 NA  2  1  2
```

[4] A fantastic name; see Section 9.3.5.

```
match(month.abb, c("Jan", "May", "Sep", "Oct"))  # is the month on the list?
##  [1]  1 NA NA NA  2 NA NA NA  3  4 NA NA
match(c("Jan", "May", "Sep", "Oct"), month.abb)  # which month is it?
## [1]  1  5  9 10
```

By default, a missing value denotes a no-match.

**Exercise 5.7** *Check out the documentation of* `` `%in%` `` *to see how this operator is reduced to a call to* `match`*. Also, verify that it treats missing values as well-defined ones.*

If the elements in `y` are not unique, the smallest index `j` such that `x[i] == y[j]` is returned. Therefore, for example, `match(TRUE, l)` fetches the index of the first occurrence of a positive value in a logical vector `l`.

```
(x <- round(runif(10), 2))  # example vector
##  [1] 0.29 0.79 0.41 0.88 0.94 0.05 0.53 0.89 0.55 0.46
match(TRUE, x>0.8)  # index of the first value > 0.8 (from the left)
## [1] 4
```

### 5.4.2 Assigning numbers into intervals

`findInterval` can come in handy where the assigning of numeric values into real intervals is needed. Namely, `z <- findInterval(x, y)` for increasing `y` gives `z[i]` being the index `j` such that `x[i]` is between `y[j]` (by default, inclusive) and `y[j+1]` (by default, exclusive).

For example, a sequence of five knots $\boldsymbol{y} = (-\infty, 0.25, 0.5, 0.75, \infty)$ splits the real line into four intervals:

$$\underset{(1)}{[-\infty, 0.25)} \quad \underset{(2)}{[0.25, 0.5)} \quad \underset{(3)}{[0.5, 0.75)} \quad \underset{(4)}{[0.75, \infty)}$$

Hence, for instance:

```
findInterval(c(0, 0.2, 0.25, 0.4, 0.66, 1), c(-Inf, 0.25, 0.5, 0.75, Inf))
## [1] 1 1 2 2 3 4
```

**Exercise 5.8** *Refer to the manual of* `findInterval` *to verify the function's behaviour when we do not include* $\pm\infty$ *as endpoints and how to make* $\infty$ *classified as a member of the fourth interval.*

**Exercise 5.9** *Using a call to* `findInterval`*, compose a statement that generates a logical vector whose i-th element indicates whether* `x[i]` *is in the interval* $[0.25, 0.5]$*. Was this easier to write than an expression involving* `` `<=` `` *and* `` `>=` ``*?*

### 5.4.3 Splitting vectors into subgroups

`split(x, z)` can take the output of `match` or `findInterval` (and many other operations) and divide the elements in a vector `x` into subgroups corresponding to identical `z`s.

For instance, we can assign people into groups determined by their favourite dish:

```
name <- c("Graham", "John", "Terry", "Eric",  "Michael", "Terry")
food <- c("bacon",  "spam", "spam",  "eggs",  "spam",    "beans")
split(name, food)  # group names with respect to food
## $bacon
## [1] "Graham"
##
## $beans
## [1] "Terry"
##
## $eggs
## [1] "Eric"
##
## $spam
## [1] "John"    "Terry"   "Michael"
```

The result is a named list with labels determined by the unique elements in the second vector.

Here is another example, where we pigeonhole some numbers into the four previously mentioned intervals:

```
x <- c(0, 0.2, 0.25, 0.4, 0.66, 1)
split(x, findInterval(x, c(-Inf, 0.25, 0.5, 0.75, Inf)))
## $`1`
## [1] 0.0 0.2
##
## $`2`
## [1] 0.25 0.40
##
## $`3`
## [1] 0.66
##
## $`4`
## [1] 1
```

Items in the first argument that correspond to missing values in the grouping vector will be ignored. Also, unsurprisingly, the recycling rule is applied when necessary.

We can also split `x` into groups defined by a combination of levels of two or more variables `z1`, `z2`, etc., by calling `split(x, list(z1, z2, ...))`.

**Example 5.10** *The `ToothGrowth` dataset is a named list (more precisely, a data frame; see Chapter 12) that represents the results of an experimental study involving 60 guinea pigs. The experiment's aim was to measure the effect of different vitamin C `supp`lement types and `dose`s on the growth of the rodents' teeth `len`gths:*

```
ToothGrowth <- as.list(ToothGrowth)  # it is a list, but with extra attribs
ToothGrowth[["supp"]] <- as.character(ToothGrowth[["supp"]])  # was: factor
```

```
print(ToothGrowth)
## $len
##  [1]  4.2 11.5  7.3  5.8  6.4 10.0 11.2 11.2  5.2  7.0 16.5 16.5 15.2 17.3
## [15] 22.5 17.3 13.6 14.5 18.8 15.5 23.6 18.5 33.9 25.5 26.4 32.5 26.7 21.5
## [29] 23.3 29.5 15.2 21.5 17.6  9.7 14.5 10.0  8.2  9.4 16.5  9.7 19.7 23.3
## [43] 23.6 26.4 20.0 25.2 25.8 21.2 14.5 27.3 25.5 26.4 22.4 24.5 24.8 30.9
## [57] 26.4 27.3 29.4 23.0
##
## $supp
##  [1] "VC" "VC" "VC" "VC" "VC" "VC" "VC" "VC" "VC" "VC" "VC" "VC" "VC" "VC"
## [15] "VC" "VC" "VC" "VC" "VC" "VC" "VC" "VC" "VC" "VC" "VC" "VC" "VC" "VC"
## [29] "VC" "VC" "OJ" "OJ" "OJ" "OJ" "OJ" "OJ" "OJ" "OJ" "OJ" "OJ" "OJ" "OJ"
## [43] "OJ" "OJ" "OJ" "OJ" "OJ" "OJ" "OJ" "OJ" "OJ" "OJ" "OJ" "OJ" "OJ" "OJ"
## [57] "OJ" "OJ" "OJ" "OJ"
##
## $dose
##  [1] 0.5 0.5 0.5 0.5 0.5 0.5 0.5 0.5 0.5 0.5 1.0 1.0 1.0 1.0 1.0 1.0 1.0
## [18] 1.0 1.0 1.0 2.0 2.0 2.0 2.0 2.0 2.0 2.0 2.0 2.0 2.0 0.5 0.5 0.5 0.5
## [35] 0.5 0.5 0.5 0.5 0.5 0.5 1.0 1.0 1.0 1.0 1.0 1.0 1.0 1.0 1.0 1.0 2.0
## [52] 2.0 2.0 2.0 2.0 2.0 2.0 2.0 2.0 2.0
```

*We can split* `len` *with respect to the combinations of* `supp` *and* `dose` *(also called* interactions*) by calling:*

```
split(ToothGrowth[["len"]], ToothGrowth[c("supp", "dose")], sep="_")
## $OJ_0.5
##  [1] 15.2 21.5 17.6  9.7 14.5 10.0  8.2  9.4 16.5  9.7
##
## $VC_0.5
##  [1]  4.2 11.5  7.3  5.8  6.4 10.0 11.2 11.2  5.2  7.0
##
## $OJ_1
##  [1] 19.7 23.3 23.6 26.4 20.0 25.2 25.8 21.2 14.5 27.3
##
## $VC_1
##  [1] 16.5 16.5 15.2 17.3 22.5 17.3 13.6 14.5 18.8 15.5
##
## $OJ_2
##  [1] 25.5 26.4 22.4 24.5 24.8 30.9 26.4 27.3 29.4 23.0
##
## $VC_2
##  [1] 23.6 18.5 33.9 25.5 26.4 32.5 26.7 21.5 23.3 29.5
```

*Other synonyms are, of course, possible, e.g.,* `split(ToothGrowth[[1]], ToothGrowth[-1])` *and* `split(ToothGrowth[[1]], list(ToothGrowth[[2]], ToothGrowth[[3]]))`*. We recommend meditating upon our conscious use of double vs single square brackets here.*

*Functions such as* `Map` *(Section 7.2) will soon enable us to compute any summary statistics within groups, e.g., the group averages like those obtained by executing "*`SELECT AVG(len) FROM`

*`ToothGrowth GROUP BY supp, dose`" in SQL. As an appetiser, let's pass a list of vectors to the **`boxplot`** function; see Figure 5.1.*

```
boxplot(split(ToothGrowth[["len"]], ToothGrowth[c("supp", "dose")], sep="_"))
```

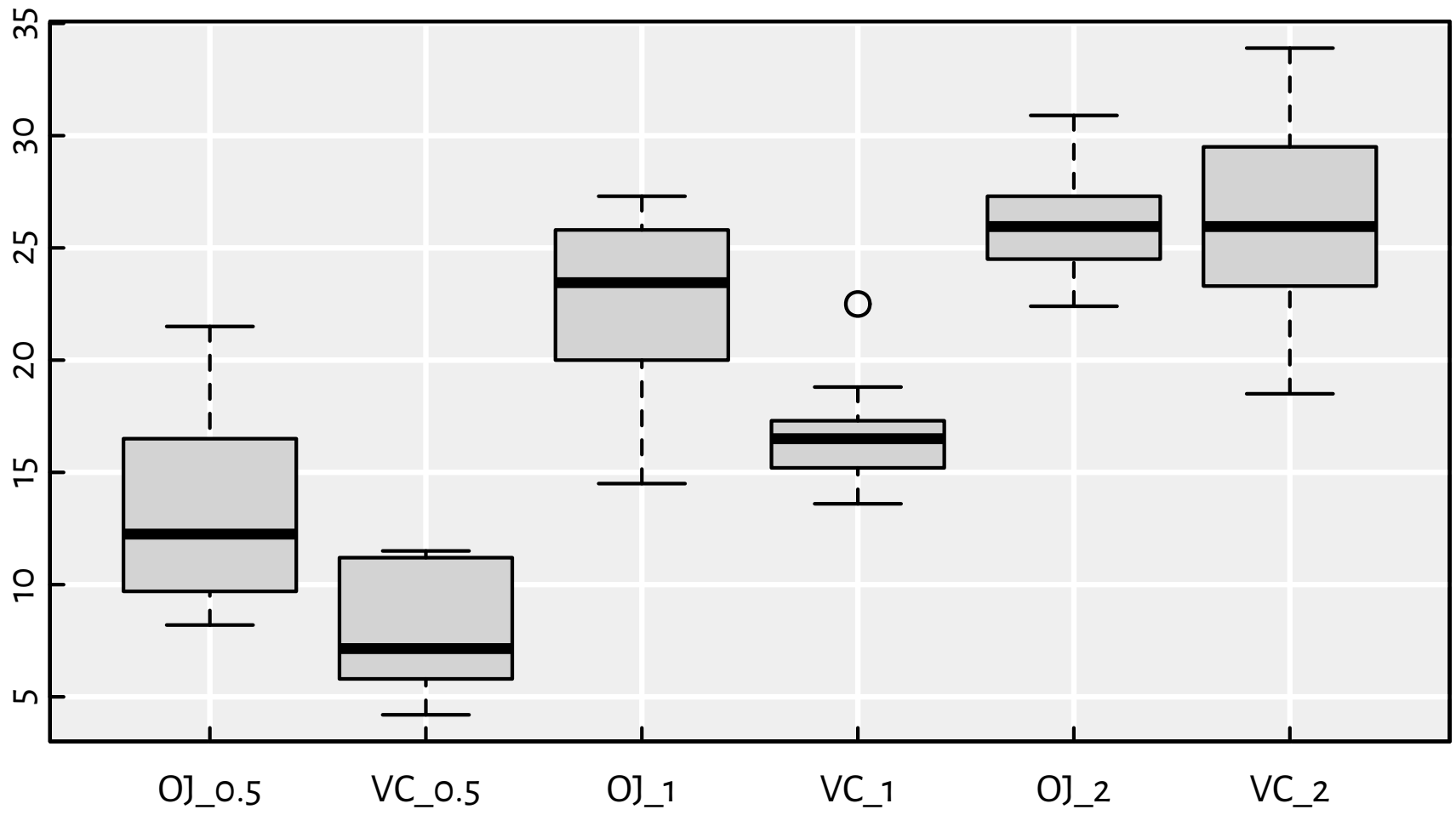

Figure 5.1. Box-and-whisker plots of `len` split by `supp` and `dose` in `ToothGrowth`.

**Note** **`unsplit`** revokes the effects of **`split`**. Later, we will get used to calling **`unsplit`**`(`**`Map`**`(some_transformation, `**`split`**`(x, z)), z)` to modify the values in `x` independently in each group defined by `z` (e.g., standardise the variables within each class separately).

## 5.4.4 Ordering elements

The **`order`** function finds the ordering permutation of a given vector, i.e., a sequence of indexes that leads to a sorted version thereof.

```
x <- c(1024, 7, 42, 666, 0, 32787)
(o <- order(x))  # the ordering permutation of x
## [1] 5 2 3 4 1 6
x[o]  # the ordered version of x
## [1]     0     7    42   666  1024 32787
```

Note that `o[1]` is the index of the smallest element in `x`, `o[2]` is the position of the second smallest, …, and `o[`**`length`**`(o)]` is the index of the greatest value. Hence, e.g., `x[o[1]]` is equivalent to **`min`**`(x)`.

Another example:

```
x <- c("b", "a", "abs", "bass", "aaargh", "aargh", "aaaargh")
(o <- order(x))
## [1] 2 7 5 6 3 1 4
x[o]
## [1] "a"       "aaaargh" "aaargh"  "aargh"   "abs"     "b"       "bass"
```

Here, as x is a character vector, the ordering is lexicographical (like in a dictionary). This is exactly how `` `<=` `` on strings works.

---

**Note** The ordering permutation that **order** returns is unique (that is why we call it *the* permutation), even for inputs containing duplicated elements. Owing to the use of a *stable* sorting algorithm, ties (repeated elements) will be listed in the order of occurrence.

```
order(c(10, 20, 40, 10, 10, 30, 20, 10, 10))
## [1] 1 4 5 8 9 2 7 6 3
```

We have, e.g., five 10s at positions 1, 4, 5, 8, and 9. These five indexes are guaranteed to be listed in this very order.

---

Ordering can also be performed in a nonincreasing manner:

```
x[order(x, decreasing=TRUE)]
## [1] "bass"    "b"       "abs"     "aargh"   "aaargh"  "aaaargh" "a"
```

---

**Note** A call to **sort**(x) is equivalent to x[**order**(x)], but the former function can be faster in certain scenarios. For instance, one of its arguments can induce a *partially* sorted vector which can be helpful if we only seek a few order statistics (e.g., the seven smallest values). Speed is rarely a bottleneck in the case of sorting (when it is, we have a problem!). This is why we will not bother ourselves with such topics until the last part of this pleasant book. Currently, we aim at expanding our skill repertoire so that we can implement anything we can think of.

---

**Exercise 5.11** *`is.unsorted(x)` determines if the elements in x are...* not *sorted with respect to `` `<=` ``. Write an R expression that generates the same result by referring to the **order** function. Also, assuming that x is numeric, do the same by means of a call to **diff**.*

**order** also accepts one or more arguments via the dot-dot-dot parameter, `` `...` ``. This way, we can sort a vector with respect to many criteria. If there are ties in the first variable, they will be resolved by the order of elements in the second variable. This is most useful for rearranging rows of a data frame, which we will exercise in Chapter 12.

```
x  <- c("a", "b", "a", "a", "b", "b")
y  <- c( 60,  40,  10,  30,  50,  20)
xy <- paste0(x, y)  # elementwise concatenate; see next chapter
```

```
xy[order(x)]  # ordered by x
## [1] "a60" "a10" "a30" "b40" "b50" "b20"
xy[order(y)]  # ordered by y
## [1] "a10" "b20" "a30" "b40" "b50" "a60"
xy[order(x, y)]  # ordered by x (primary) and y (secondary key)
## [1] "a10" "a30" "a60" "b20" "b40" "b50"
```

---

**Note** (*) We represent a permutation with a vector that is an arbitrary arrangement of $n$ consecutive natural numbers. A composition (product) of two permutations can be determined using simple vector indexing:

```
(o1 <- order(y))
## [1] 3 6 4 2 5 1
(o2 <- order(x[o1]))
## [1] 1 3 6 2 4 5
o1[o2]  # permutation composition (not the same as o2[o1])
## [1] 3 4 1 6 2 5
xy[ o1[o2] ]  # same as (xy[o1])[o2]
## [1] "a10" "a30" "a60" "b20" "b40" "b50"
```

---

---

**Note** (*) Calling **order** on a permutation determines its *inverse*.

```
z <- c(10, 30, 40, 20, 10, 10, 50, 30)
(o <- order(z))
## [1] 1 5 6 4 2 8 3 7
order(o)  # the inverse of the above permutation
## [1] 1 5 7 4 2 3 8 6
o[order(o)]  # the identity permutation
## [1] 1 2 3 4 5 6 7 8
order(o)[o]  # the identity permutation again
## [1] 1 2 3 4 5 6 7 8
(z[o])[order(o)]  # we get z again
## [1] 10 30 40 20 10 10 50 30
```

Note that **order**(**order**(z)) can be considered as a way to *rank* all the elements in z. For instance, the third value in z, 40, is assigned rank 7: it is the seventh smallest value in this vector. This breaks the ties on a first-come, first-served basis. But we can also write:

```
order(order(z, runif(length(z))))  # ranks with ties broken at random
## [1] 2 5 7 4 3 1 8 6
```

For different variations of these, see the **rank** function.

---

**Exercise 5.12** *Recall that **sample**(x) returns a pseudorandom permutation of elements of a*

*given vector unless* `x` *is a single positive number. Write an expression that always produces a proper rearrangement, regardless of the size of* `x`.

### 5.4.5 Identifying duplicates

Whether any element in a vector was already listed in the previous part of the sequence can be verified by calling:

```
x <- c(10, 20, 30, 20, 40, 50, 50, 50, 20, 20, 60)
duplicated(x)
##  [1] FALSE FALSE FALSE  TRUE FALSE FALSE  TRUE  TRUE  TRUE  TRUE FALSE
```

This function can be used to remove repeated observations; see also **unique**. This function returns a value that is not guaranteed to be sorted (unlike in some other languages/libraries).

**Exercise 5.13** *What can be the use case of a call to* `match(x, unique(x))`*?*

**Exercise 5.14** *Given two named lists* `x` *and* `y`*, which we treat as key-value pairs, determine their set-theoretic union (with respect to the keys). For example:*

```
x <- list(a=1, b=2)
y <- list(c=3, a=4)
z <- ...to.do...  # combine x and y
str(z)
## List of 3
##  $ a: num 4
##  $ b: num 2
##  $ c: num 3
```

### 5.4.6 Counting index occurrences

**tabulate** takes a vector of values from a set of small positive integers (e.g., indexes) and determines their number of occurrences:

```
x <- c(2, 4, 6, 2, 2, 2, 3, 6, 6, 3)
tabulate(x)
## [1] 0 4 2 1 0 3
```

In other words, there are 0 ones, 4 twos, …, and 3 sixes.

**Exercise 5.15** *Using a call to* `tabulate` *(amongst others), return a named vector with the number of occurrences of each unique element in a character vector. For example:*

```
y <- c("a", "b", "a", "c", "a", "d", "e", "e", "g", "g", "c", "c", "g")
result <- ...to.do...
print(result)
## a b c d e g
## 3 1 3 1 2 3
```

## 5.5 Preserving and losing attributes

Attributes are conceived of as *extra* data. It is thus up to a function's authors what they will decide to do with them. Generally, it is safe to assume that much thought has been put into the design of base R functions. Oftentimes, they behave fairly reasonably. This is why we are now going to spend some time now exploring their approaches to the handling of attributes.

Namely, for functions and operators that aim at transforming vectors passed as their inputs, the assumed strategy may be to:

- ignore the input attributes completely,
- equip the output object with the same set of attributes, or
- take care only of a few special attributes, such as `names`, if that makes sense.

Below we explore some common patterns; see also Section 1.3 of [69].

### 5.5.1 c

First, **c** drops[5] all attributes except `names`:

```
(x <- structure(1:4, names=c("a", "b", "c", "d"), attrib1="<3"))
## a b c d
## 1 2 3 4
## attr(,"attrib1")
## [1] "<3"
c(x)  # only `names` are preserved
## a b c d
## 1 2 3 4
```

We can therefore end up calling this function chiefly for this nice side effect. Also, recall that **unname** drops the labels.

```
unname(x)
## [1] 1 2 3 4
## attr(,"attrib1")
## [1] "<3"
```

### 5.5.2 as.something

**as.vector**, **as.numeric**, and similar drop all attributes in the case where the output is an atomic vector, but it might not necessarily do so in other cases (because they are S3 generics; see Chapter 10).

[5] To be precise, we mean the default S3 method of **c** here; compare Section 10.2.4.

```
as.vector(x)  # drops all attributes if x is atomic
## [1] 1 2 3 4
```

### 5.5.3 Subsetting

Subsetting with `[` (except where the indexer is not given) drops all attributes but `names` (as well as `dim` and `dimnames`; see Chapter 11), which is adjusted accordingly:

```
x[1]    # subset of labels
## a
## 1
x[[1]]  # this always drops the labels (makes sense, right?)
## [1] 1
```

The replacement version of the index operator modifies the values in an existing vector whilst preserving all the attributes. In particular, skipping the indexer replaces all the elements:

```
y <- x
y[] <- c("u", "v")  # note that c("u", "v") has no attributes
print(y)
##   a   b   c   d
## "u" "v" "u" "v"
## attr(,"attrib1")
## [1] "<3"
```

### 5.5.4 Vectorised functions

Vectorised unary functions tend to copy all the attributes.

```
round(x)
## a b c d
## 1 2 3 4
## attr(,"attrib1")
## [1] "<3"
```

Binary operations are expected to get the attributes from the longer input. If they are of equal sizes, the first argument is preferred to the second.

```
y <- structure(c(1, 10), names=c("f", "g"), attrib1=":|", attrib2=":O")
y * x  # x is longer
##  a  b  c  d
##  1 20  3 40
## attr(,"attrib1")
## [1] "<3"
y[c("h", "i")] <- c(100, 1000)  # add two new elements at the end
y * x
```

```
##     f     g     h     i
##     1    20   300  4000
## attr(,"attrib1")
## [1] ":|"
## attr(,"attrib2")
## [1] ":O"
x * y
##     a     b     c     d
##     1    20   300  4000
## attr(,"attrib1")
## [1] "<3"
## attr(,"attrib2")
## [1] ":O"
```

Also, Section 9.3.6 mentions a way to copy all attributes from one object to another.

**Important** Even in base R, the foregoing rules are not enforced strictly. We consider them inconsistencies that should be, for the time being, treated as features (with which we need to learn to live as they have not been fixed for years, but hope springs eternal).

As far as third-party extension packages are concerned, suffice it to say that a lot of R programmers do not know what attributes are whatsoever. It is always best to refer to the documentation, perform a few experiments, and/or manually ensure the preservation of the data we care about.

**Exercise 5.16** *Check what attributes are preserved by* `ifelse`*.*

## 5.6 Exercises

**Exercise 5.17** *Answer the following questions (contemplate first, then use R to find the answer).*

- *What is the result of* `x[c()]`*? Is it the same as* `x[]`*?*
- *Is* `x[c(1, 1, 1)]` *equivalent to* `x[1]`*?*
- *Is* `x[1]` *equivalent to* `x["1"]`*?*
- *Is* `x[c(-1, -1, -1)]` *equivalent to* `x[-1]`*?*
- *What does* `x[c(0, 1, 2, NA)]` *do?*
- *What does* `x[0]` *return?*
- *What does* `x[1, 2, 3]` *do?*
- *What about* `x[c(0, -1, -2)]` *and* `x[c(-1, -2, NA)]`*?*

- *Why `x[NA]` is so significantly different from `x[c(1, NA)]`?*
- *What is `x[c(FALSE, TRUE, 2)]`?*
- *What will we obtain by calling `x[x<min(x)]`?*
- *What about `x[length(x)+1]`?*
- *Why `x[min(y)]` is most probably a mistake? What could it mean? How can it be fixed?*
- *Why cannot we mix indexes of different types and write `x[c(1, "b", "c", 4)]`? Or can we?*
- *Why would we call `as.vector(na.omit(x))` instead of just `na.omit(x)`?*
- *What is the difference between `sort` and `order`?*
- *What is the type and the length of the object returned by a call to `split(a, u)`? What about `split(a, c(u, v))`?*
- *How to get rid of the seventh element from a list of ten elements?*
- *How to get rid of the seventh, eight, and ninth elements from a list with ten elements?*
- *How to get rid of the seventh element from an atomic vector of ten elements?*
- *If `y` is a list, by how many elements "`y[c(length(y)+1, length(y)+1, length(y)+1)] <- list(1, 2, 3)`" will extend it?*
- *What is the difference between `x[x>0]` and `x[which(x>0)]`?*

**Exercise 5.18** *If `x` is an atomic vector of length `n`, `x[5:n]` obviously extracts everything from the fifth element to the end. Does it, though? Check what happens when `x` is of length less than five, including 0. List different ways to correct this expression so that it makes (some) sense in the case of shorter vectors.*

**Exercise 5.19** *Similarly, `x[length(x) + 1 - 5:1]` is supposed to return the last five elements in `x`. Propose a few alternatives that are correct also for short `x`s.*

**Exercise 5.20** *Given a numeric vector, fetch its five largest elements. Ensure the code works for vectors of length less than five.*

**Exercise 5.21** *We can compute a* trimmed *mean of some `x` by setting the `trim` argument to the `mean` function. Compute a similar robust estimator of location – the p-winsorised mean, $p \in [0, 0.5]$ defined as the arithmetic mean of all elements in `x` clipped to the $[Q_p, Q_{1-p}]$ interval, where $Q_p$ is the vector's p-quantile; see `quantile`. For example, if `x` is (8, 5, 2, 9, 7, 4, 6, 1, 3), we have $Q_{0.25} = 3$ and $Q_{0.75} = 7$ and hence the 0.25-winsorised mean will be equal to the arithmetic mean of (7, 5, 3, 7, 7, 4, 6, 3, 3).*

**Exercise 5.22** *Let `x` and `y` be two vectors of the same length, $n$, and no ties. Implement the formula for the Spearman rank correlation coefficient:*

$$\varrho(\mathbf{x}, \mathbf{y}) = 1 - \frac{6 \sum_{i=1}^{n} d_i^2}{n(n^2 - 1)},$$

*where $d_i = r_i - s_i$, $i = 1, \ldots, n$, and $r_i$ and $s_i$ denote the rank of $x_i$ and $y_i$, respectively; see also* **`cor`**.

**Exercise 5.23** *(*) Given vectors x and y both of length n, a call to* **`approx`**`(x, y, ...)` *can be used to interpolate linearly between the points $(x_1, y_1), (x_2, y_2), \ldots, (x_n, y_n)$. We can use it to generate new ys for previously unobserved xs (somewhere "in-between" the data we already have). Moreover,* **`spline`**`(x, y, ...)` *can perform a cubic spline interpolation, which is smoother; see Figure 5.2.*

```
x <- c(1, 3,  5, 7, 10)
y <- c(1, 15, 25, 6,  0)
x_new <- seq(1, 10, by=0.25)
y_new1 <- approx(x, y, xout=x_new)[["y"]]
y_new2 <- spline(x, y, xout=x_new)[["y"]]
plot(x, y, ylim=c(-10, 30))  # the points to interpolate between
lines(x_new, y_new1, col="black", lty="solid")  # linear interpolation
lines(x_new, y_new2, col="darkred", lty="dashed")  # cubic interpolation
legend("topright", legend=c("linear", "cubic"),
    lty=c("solid", "dashed"), col=c("black", "darkred"), bg="white")
```

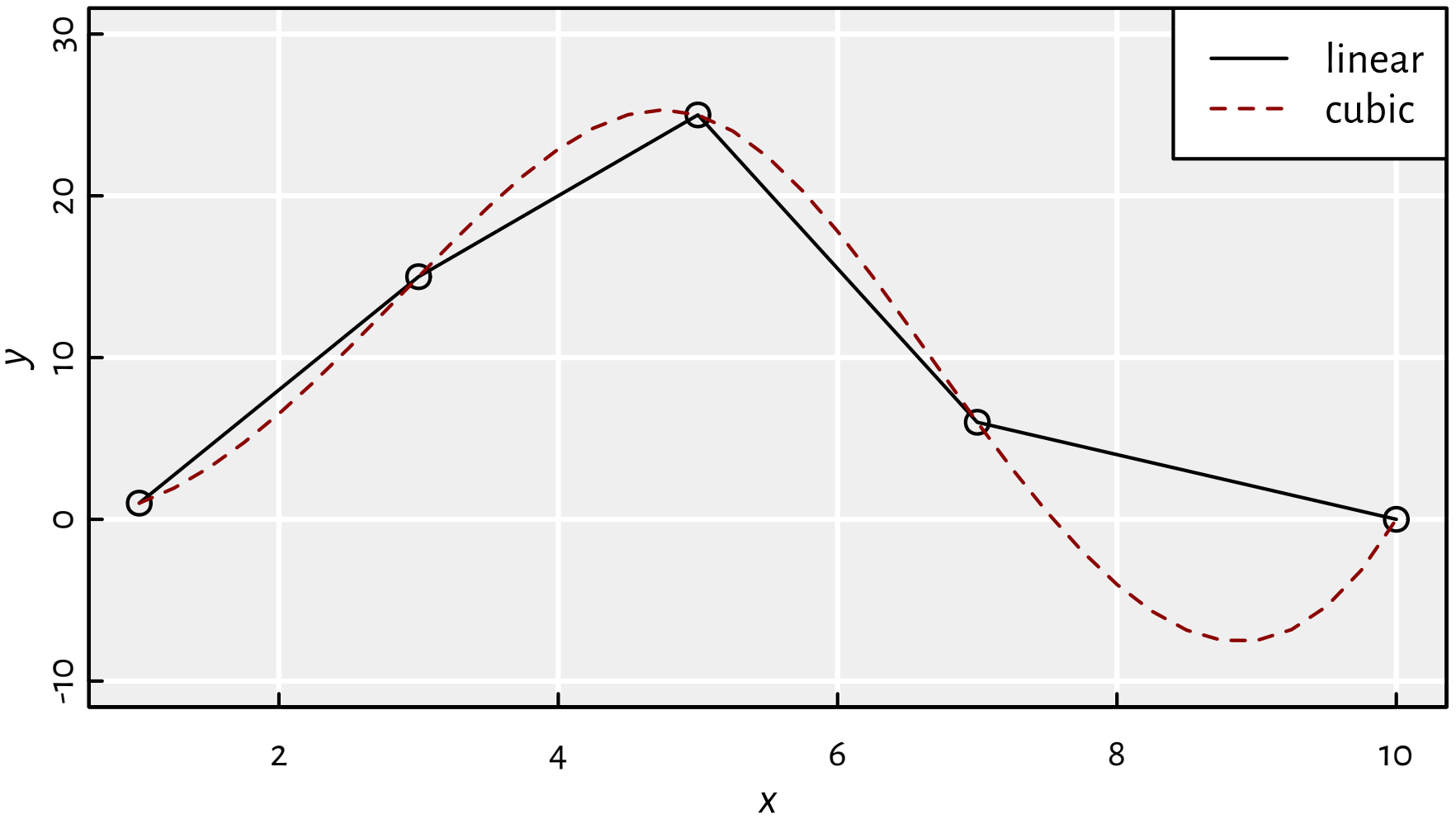

Figure 5.2. Piecewise linear and cubic spline interpolation.

*Using a call to one of the above, impute missing data in* `euraud-20200101-20200630.csv`[6]*, e.g., the blanks in* `(0.60, 0.62, NA, 0.64, NA, NA, 0.58)` *should be filled so as to obtain* `(0.60, 0.62, 0.63, 0.64, 0.62, 0.60, 0.58)`.

**Exercise 5.24** *Given some 1 ≤* `from` *≤* `to` *≤* `n`*, use* **`findInterval`** *to generate a logical vector of length* `n` *with* `TRUE` *elements only at indexes between* `from` *and* `to`*, inclusive.*

[6] https://github.com/gagolews/teaching-data/raw/master/marek/euraud-20200101-20200630.csv

**Exercise 5.25** *Implement expressions that give rise to the same results as calls to* **`which`**, **`which.min`**, **`which.max`**, *and* **`rev`** *functions. What is the difference between* `x[x>y]` *and* `x[which(x>y)]`*? What about* `which.min(x)` *vs* `which(x == min(x))`*?*

**Exercise 5.26** *Given two equal-length vectors* `x` *and* `y`*, fetch the value from the former that corresponds to the smallest value in the latter. Write three versions of such an expression, each dealing with potential ties in* `y` *differently. For example:*

```
x <- c("a", "b", "c", "d", "e", "f")
y <- c(  3,   1,   2,   1,   1,   4)
```

*It should choose the first (*`"b"`*), last (*`"e"`*), or random element from* `x` *fulfilling the above property (*`"b"`*,* `"d"`*, or* `"e"` *with equal probability). Make sure your code works for* `x` *being of the type* `character` *or* `numeric` *as well as an empty vector.*

**Exercise 5.27** *Implement an expression that yields the same result as* `duplicated(x)` *for a numeric vector* `x`*, but using* **`diff`** *and* **`order`**.

**Exercise 5.28** *Based on* **`match`** *and* **`unique`**, *implement your versions of* `union(x, y)`, `intersect(x, y)`, `setdiff(x, y)`, `is.element(x, y)`, *and* `setequal(x, y)` *for* `x` *and* `y` *being nonempty numeric vectors.*

# 6 Character vectors

Text is a universal, portable, economical, and efficient means of interacting between humans and computers as well as exchanging data between programs or APIs. This book is 99% made of text. And, wow, how much valuable knowledge is in it, innit?

## 6.1 Creating character vectors

### 6.1.1 Inputting individual strings

Specific character strings are delimited by a pair of either double or single quotes (apostrophes).

```
"a string"
## [1] "a string"
'another string'  # and, of course, neither 'like this" nor "like this'
## [1] "another string"
```

The only difference between these two is that we cannot directly include, e.g., an apostrophe in a single quote-delimited string. On the other hand, `"'tis good ol' spam"` and `'I "love" bacon'` are both okay.

To embrace characters whose inclusion might otherwise be difficult or impossible, we may always employ the so-called *escape sequences*. R uses the backslash, “\”, as the *escape character*. In particular:

- `\"` inputs a double quote,
- `\'` generates a single quote,
- `\\` includes a backslash,
- `\n` endows a new line.

```
(x <- "I \"love\" bacon\n\\\"/")
## [1] "I \"love\" bacon\n\\\"/"
```

The **`print`** function (which was implicitly called to display the above object) does not reveal the special meaning of the escape sequences. Instead, **`print`** outputs strings in the same way that we ourselves would follow when inputting them. The number of characters in x is 18, and not 23:

```
nchar(x)
## [1] 18
```

To display the string as-it-really-is, we call **cat**:

```
cat(x, sep="\n")
## I "love" bacon
## \"/
```

In *raw* character constants, the backslash character's special meaning is disabled. They can be entered using the notation like `r"(...)"`, `r"{...}"`, or `r"[...]"`; see **help**("Quotes"). These can be useful when inputting regular expressions (Section 6.2.4).

```
x <- r"(spam\n\\\"maps)"   # also: r"-(...)-", r"--(...)--", etc.
print(x)
## [1] "spam\\n\\\\\\\"maps"
cat(x, sep="\n")
## spam\n\\\"maps
```

Furthermore, the string version of the missing value marker is `NA_character_`.

---

**Note** (*) The Unicode standard 17.0 (version dated September 2025) defines 159 801 characters, i.a., letters from different scripts, mathematical symbols, and emojis. Each is assigned a unique numeric identifier; see the Unicode Character Code Charts[1]. For example, the *inverted exclamation mark* (see the Latin-1 Supplement section therein) has been mapped to the hexadecimal code 0xA1 (or 161 decimally). Knowing this magic number permits us to specify a Unicode code point using one of the following escape sequences:

- `\uxxxx` – codes using four hexadecimal digits,
- `\Uxxxxxxxx` – codes using eight hexadecimal digits.

For instance:

```
cat("!\u00a1!\U000000a1!", sep="\n")
## !¡!¡!
```

All R installations allow for working with Unicode strings. More precisely, they support dealing with *UTF-8*, being a super-encoding that is native to most UNIX-like boxes, including GNU/Linux and m**OS. Other operating systems may use some 8-bit encoding as the system one (e.g., `latin1` or `cp1252`), but they can be mixed with Unicode seamlessly; see **help**("Encoding"), **help**("iconv"), and [27] for discussion.

[1] https://www.unicode.org/charts

Nevertheless, certain output devices (web browsers, LaTeX renderers, text terminals) might fail to display some Unicode characters, e.g., because of missing fonts. However, as far as processing character data is concerned, this does not matter because R does it with its eyes closed. For example:

```
cat("\U0001f642\u2665\u0bb8\U0001f923\U0001f60d\u2307", sep="\n")
## □□□□□□
```

In the PDF version[2] of this adorable book, the Unicode glyphs are not rendered correctly for some reason. However, its HTML variant[3], generated from the same source files, should be displayed by most web browsers properly.

---

**Note** (*) Some output devices may support the following codes that control the position of the caret (text cursor):

- `\b` inserts a backspace (moves cursor one column to the left),
- `\t` implants a tabulator (advances to the next tab stop, e.g., a multiply of four or eight text columns),
- `\r` injects a carriage return (move to the beginning of the current line).

```
cat("abc\bd\tef\rg\nhij", sep="\n")
## gbd     ef
## hij
```

These can be used on unbuffered outputs like **`stderr`** to display the status of the current operation, for instance, an animated textual progress bar, the print-out of the ETA, or the percentage of work completed.

Further, certain terminals can also understand the ECMA-48/ANSI-X3.64 escape sequences[4] of the form `\u001b[...` to control the cursor's position, text colour, and even style. For example, `\u001b[1;31m` outputs red text in bold font and `\u001b[0m` resets the settings to default. We recommend giving, e.g., **`cat`**`("\u001b[1;31mspam\u001b[0m")` or **`cat`**`("\u001b[5;36m\u001b[Abacon\u001b[Espam\u001b[0m")` a try.

---

### 6.1.2 Many strings, one object

Less trivial character vectors (meaning, of length greater than one) can be constructed by means of, e.g., **`c`** or **`rep`**[5].

[2] https://deepr.gagolewski.com/deepr.pdf
[3] https://deepr.gagolewski.com/
[4] https://en.wikipedia.org/wiki/ANSI_escape_code
[5] Internally, there is a string cache (a hash table). Multiple clones of the same string do not occupy more RAM than necessary.

```
(x <- c(rep("spam", 3), "bacon", NA_character_, "spam"))
## [1] "spam"  "spam"  "spam"  "bacon" NA      "spam"
```

Thus, a character vector is, in fact, a sequence of sequences of characters[6]. As usual, the total number of strings can be fetched via the **length** function. However, the length of each string may be read with the vectorised **nchar**.

```
length(x)  # how many strings?
## [1] 6
nchar(x)   # the number of code points in each string
## [1]  4  4  4  5 NA  4
```

### 6.1.3 Concatenating character vectors

**paste** can be used to concatenate (join) the corresponding elements of two or more character vectors:

```
paste(c("a", "b", "c"), c("1", "2", "3"))  # sep=" " by default
## [1] "a 1" "b 2" "c 3"
paste(c("a", "b", "c"), c("1", "2", "3"), sep="")  # see also paste0
## [1] "a1" "b2" "c3"
```

The function is deeply vectorised (but note the lack of a warning about the partial recycling):

```
paste(c("a", "b", "c"), 1:5, c("!", "?"))  # coercion of numeric to character
## [1] "a 1 !" "b 2 ?" "c 3 !" "a 4 ?" "b 5 !"
```

We can also collapse (flatten, aggregate) a sequence of strings into a single string:

```
paste(c("a", "b", "c", "d"), collapse=",")
## [1] "a,b,c,d"
paste(c("a", "b", "c", "d"), 1:2, sep="", collapse="")
## [1] "a1b2c1d2"
```

Perhaps for convenience, alas, **paste** treats missing values differently from most other vectorised functions:

```
paste(c("A", NA_character_, "B"), "!", sep="")
## [1] "A!"  "NA!" "B!"
```

[6] (*) Chapter 14 will mention that objects of the type `character` are internally represented as objects with SEXPTYPE of STRSXP. They are arrays with elements whose SEXPTYPE is CHARSXP, each of which is a string of characters (`char*`).

### 6.1.4 Formatting objects

Strings can also arise by converting other-typed R objects into text. For example, the quite customisable (see Chapter 10) **format** function prepares data for display in dynamically generated reports.

```
x <- c(123456.789, -pi, NaN)
format(x)
## [1] "123456.7890" "     -3.1416" "         NaN"
cat(format(x, digits=8, scientific=FALSE, drop0trailing=TRUE), sep="\n")
## 123456.789
##     -3.1415927
##            NaN
```

Moreover, **sprintf** is a workhorse for turning possibly many atomic vectors into strings. Its first argument is a *format string*. Special escape sequences starting with the per cent sign, "%", serve as placeholders for the actual values. For instance, "%s" is replaced with a string and "%f" with a floating point value taken from further arguments.

```
sprintf("%s%s", "a", c("X", "Y", "Z"))  # like paste(...)
## [1] "aX" "aY" "aZ"
sprintf("key=%s, value=%f", c("spam", "eggs"), c(100000, 0))
## [1] "key=spam, value=100000.000000" "key=eggs, value=0.000000"
```

The numbers' precision, strings' widths and justification, etc., can be customised, e.g., "%6.2f" is a number that, when converted to text, will occupy six text columns[7], with two decimal digits of precision.

```
sprintf("%10s=%6.2f%%", "rate", 2/3*100)  # "%%" renders the per cent sign
## [1] "      rate= 66.67%"
sprintf("%.*f", 1:5, pi)  # variable precision
## [1] "3.1"     "3.14"     "3.142"     "3.1416"   "3.14159"
```

Also, e.g., "%1$s", "%2$s", … inserts the first, second, … argument as text.

```
sprintf("%1$s, %2$s, %1$s, and %1$s", "spam", "bacon")  # numbered argument
## [1] "spam, bacon, spam, and spam"
```

**Exercise 6.1** *Read* **help("sprintf")** *(highly recommended!).*

### 6.1.5 Reading text data from files

Given a raw text file, **readLines** loads it into memory and represents it as a character vector, with each line stored in a separate string.

[7] This is only true for 8-bit native encodings or ASCII; see also **sprintf** from the **stringx** package, which takes the text width and not the number of bytes into account.

```
head(readLines(
    "https://github.com/gagolews/teaching-data/raw/master/README.md"
))
## [1] "# Prof. [Marek](https://www.gagolewski.com)'s Data for Teaching"
## [2] ""
## [3] "> *See the comment lines within the files themselves for"
## [4] "> a detailed description of each dataset.*"
## [5] ""
## [6] "*Good* datasets are actually hard to find!"
```

**writeLines** is its counterpart. There is also an option to read or write parts of files at a time using file connections which we mention in Section 8.3.5. Moreover, **cat**(..., append=TRUE) can be used to create a text file incrementally.

## 6.2 Pattern searching

### 6.2.1 Comparing whole strings

We have already reviewed a couple of ways to compare strings as a whole. For instance, the `==` operator implements elementwise testing:

```
c("spam", "spam", "bacon", "eggs") == c("spam", "eggs")  # recycling rule
## [1]  TRUE FALSE FALSE  TRUE
```

In Section 5.4.1, we introduced the **match** function and its derivative, the `%in%` operator. They are vectorised in a different way:

```
match(c("spam", "spam", "bacon", "eggs"), c("spam", "eggs"))
## [1]  1  1 NA  2
c("spam", "spam", "bacon", "eggs") %in% c("spam", "eggs")
## [1]  TRUE  TRUE FALSE  TRUE
```

**Note** (*) **match** relies on a simple, bytewise comparison of the corresponding code points. It might not be valid in natural language processing activities, e.g., where the German word *groß* should be equivalent to *gross* [19]. Moreover, in the rare situations where we read Unicode-unnormalised data, canonically equivalent strings may be considered different; see [18].

### 6.2.2 Partial matching

When only a consideration of the initial part of each string is required, we can call:

```
startsWith(c("s", "spam", "spamtastic", "spontaneous", "spoon"), "spam")
## [1] FALSE  TRUE  TRUE FALSE FALSE
```

If we provide many prefixes, the above function will be applied elementwisely, just like the `==` operator.

On the other hand, **charmatch** performs a *partial matching* of strings. It is an each-vs-all version of **startsWith**:

```
charmatch(c("s", "sp", "spam", "spams", "eggs", "bacon"), c("spam", "eggs"))
## [1]  1  1  1 NA  2 NA
charmatch(c("s", "sp", "spam", "spoo", "spoof"), c("spam", "spoon"))
## [1]  0  0  1  2 NA
```

Note that 0 designates that there was an ambiguous match.

---

**Note** (*) In Section 9.4.7, we discuss **match.arg**, which a few R functions rely on when they need to select a value from a range of possible choices. Furthermore, Section 9.3.2 and Section 15.4.4 mention the (discouraged) partial matching of list labels and function argument names.

---

### 6.2.3 Matching anywhere within a string

Fixed patterns can also be searched for anywhere within character strings using **grepl**:

```
x <- c("spam", "y spammite spam", "yummy SPAM", "sram")
grepl("spam", x, fixed=TRUE)  # fixed patterns, as opposed to regexes below
## [1]  TRUE  TRUE FALSE FALSE
```

---

**Important** The order of arguments is like **grepl**(needle, haystack), not vice versa. Also, this function is not vectorised with respect to the first argument.

---

**Exercise 6.2** *How the calls to* ***grep****(y, x, value=FALSE) and* ***grep****(y, x, value=TRUE) can be implemented based on* ***grepl*** *and other operations we are already familiar with?*

---

**Note** (*) As a curiosity, **agrepl** performs *approximate* matching, which can account for a smöll nmber of tpyos.

```
agrepl("spam", x)
## [1]  TRUE  TRUE FALSE  TRUE
agrepl("ham", x, ignore.case=TRUE)
## [1] TRUE TRUE TRUE TRUE
```

It is based on *Levenshtein's edit distance* that measures the number of character insertions, deletions, or substitutions required to turn one string into another.

### 6.2.4 Using regular expressions (*)

Setting `perl=TRUE` allows for identifying occurrences of patterns specified by *regular expressions* (regexes).

```
grepl("^spam", x, perl=TRUE)  # strings that begin with `spam`
## [1]  TRUE FALSE FALSE FALSE
grepl("(?i)^spam|spam$", x, perl=TRUE)  # begin or end; case ignored
## [1]  TRUE  TRUE  TRUE FALSE
```

**Note** For more details on regular expressions in general, see, e.g., [25]. The ultimate reference on the PCRE2 pattern syntax is the Unix `man` page *pcre2pattern(3)*[8]. From now on, we assume that the reader is familiar with it.

Apart from the Perl-compatible regexes, R also gives access to the TRE library (ERE-like), which is the default one; see `help("regex")`. However, we discourage its use because it is feature-poorer.

**Exercise 6.3** *The `list.files` function generates the list of file names in a given directory that match a given regular expression. For instance, the following gives all CSV files in a folder:*

```
list.files("~/Projects/teaching-data/r/", "\\.csv$")
## [1] "air_quality_1973.csv" "anscombe.csv"         "iris.csv"
## [4] "titanic.csv"          "tooth_growth.csv"     "trees.csv"
## [7] "world_phones.csv"
```

*Write a single regular expression that matches file names ending with "`.csv`" or "`.csv.gz`". Also, scribble a regex that matches CSV files whose names do not begin with "`eurusd`".*

### 6.2.5 Locating pattern occurrences

`regexpr` finds the first occurrence of a pattern in each string:

```
regexpr("spam", x, fixed=TRUE)
## [1]  1  3 -1 -1
## attr(,"match.length")
## [1]  4  4 -1 -1
## attr(,"index.type")
## [1] "chars"
## attr(,"useBytes")
## [1] TRUE
```

[8] http://www.pcre.org/current/doc/html/pcre2pattern.html

In particular, there is a pattern occurrence starting at the third code point of the second string in `x`. Moreover, the last string has no pattern match, which is denoted by `-1`.

The `match.length` attribute is generally more informative when searching with regular expressions.

To locate all the matches, i.e., globally, we use **`gregexpr`**:

```
# `spam` followed by 0 or more letters, case insensitively
gregexpr("(?i)spam\\p{L}*", x, perl=TRUE)
## [[1]]
## [1] 1
## attr(,"match.length")
## [1] 4
## attr(,"index.type")
## [1] "chars"
## attr(,"useBytes")
## [1] TRUE
##
## [[2]]
## [1]  3 12
## attr(,"match.length")
## [1] 8 4
## attr(,"index.type")
## [1] "chars"
## attr(,"useBytes")
## [1] TRUE
##
## [[3]]
## [1] 7
## attr(,"match.length")
## [1] 4
## attr(,"index.type")
## [1] "chars"
## attr(,"useBytes")
## [1] TRUE
##
## [[4]]
## [1] -1
## attr(,"match.length")
## [1] -1
## attr(,"index.type")
## [1] "chars"
## attr(,"useBytes")
## [1] TRUE
```

As we noted in Section 4.4.2, wrapping the results in a list was a clever choice for the number of matches can obviously vary between strings.

In Section 7.2, we will look at the **`Map`** function, which, along with **`substring`** intro-duced below, can aid in getting the most out of such data. Meanwhile, let's just mention that **regmatches** extracts the matching substrings:

```
regmatches(x, gregexpr("(?i)spam\\p{L}*", x, perl=TRUE))
## [[1]]
## [1] "spam"
##
## [[2]]
## [1] "spammite" "spam"
##
## [[3]]
## [1] "SPAM"
##
## [[4]]
## character(0)
```

---

**Note** (*) Consider what happens when a regular expression contains *parenthesised subexpressions (capture groups)*.

```
r <- "(?<basename>[^. ]+)\\.(?<extension>[^ ]*)"
```

This regex consists of two capture groups separated by a dot. The first one is labelled "`basename`". It comprises several arbitrary characters except for spaces and dots. The second group, named "`extension`", is a substring consisting of anything but spaces.

Such a pattern can be used for unpacking space-delimited lists of file names.

```
z <- "dataset.csv.gz something_else.txt spam"
regexpr(r, z, perl=TRUE)
## [1] 1
## attr(,"match.length")
## [1] 14
## attr(,"index.type")
## [1] "chars"
## attr(,"useBytes")
## [1] TRUE
## attr(,"capture.start")
##      basename extension
## [1,]        1         9
## attr(,"capture.length")
##      basename extension
## [1,]        7         6
## attr(,"capture.names")
## [1] "basename"  "extension"
gregexpr(r, z, perl=TRUE)
## [[1]]
## [1]  1 16
## attr(,"match.length")
## [1] 14 18
```

```
## attr(,"index.type")
## [1] "chars"
## attr(,"useBytes")
## [1] TRUE
## attr(,"capture.start")
##      basename extension
## [1,]        1         9
## [2,]       16        31
## attr(,"capture.length")
##      basename extension
## [1,]        7         6
## [2,]       14         3
## attr(,"capture.names")
## [1] "basename"  "extension"
```

The `capture.*` attributes give us access to the matches to the individual capture groups, i.e., the `basename` and the `extension`.

---

**Exercise 6.4** (*) *Check out the difference between the results generated by **`regexec`** and **`regexpr`** as well as between the outputs of **`gregexec`** and **`gregexpr`**.*

### 6.2.6 Replacing pattern occurrences

**`sub`** and **`gsub`** can replace the first and all, respectively, matches to a pattern:

```
x <- c("spam", "y spammite spam", "yummy SPAM", "sram")
sub("spam", "ham", x, fixed=TRUE)
## [1] "ham"             "y hammite spam" "yummy SPAM"     "sram"
gsub("spam", "ham", x, fixed=TRUE)
## [1] "ham"            "y hammite ham" "yummy SPAM"    "sram"
```

---

**Note** (*) If a regex defines capture groups, matches thereto can be mentioned not only in the pattern itself but also in the replacement string:

```
gsub("(\\p{L})\\p{L}\\1", "\\1", "aha egg gag NaN spam", perl=TRUE)
## [1] "a egg g N spam"
```

Matched are, in the following order: a letter (it is a capture group), another letter, and the former letter again. Each such palindrome of length three is replaced with just the repeated letter.

---

**Exercise 6.5** (*) *Display the source code of **`glob2rx`** by calling **`print`**`(glob2rx)` and study how this function converts wildcards such as `file???.*` or `*.csv` to regular expressions that can be passed to, e.g., **`list.files`**.*

### 6.2.7 Splitting strings into tokens

`strsplit` divides each string in a character vector into chunks.

```
strsplit(c("spam;spam;eggs;;bacon", "spam"), ";", fixed=TRUE)
## [[1]]
## [1] "spam"  "spam"  "eggs"  ""      "bacon"
##
## [[2]]
## [1] "spam"
```

Note that this time the search pattern specifying the token delimiter is given as the *second* argument (an inconsistency).

## 6.3 Other string operations

### 6.3.1 Extracting substrings

`substring` extracts parts of strings between given character position ranges.

```
substring("spammity spam", 1, 4)  # from the first to the fourth character
## [1] "spam"
substring("spammity spam", 10)  # from the tenth to end
## [1] "spam"
substring("spammity spam", c(1, 10), c(4, 14))  # vectorisation
## [1] "spam" "spam"
substring(c("spammity spam", "bacon and eggs"), 1, c(4, 5))
## [1] "spam"  "bacon"
```

**Note** There is also a replacement (compare Section 9.3.6) version of the foregoing:

```
x <- "spam, spam, bacon, and spam"
substring(x, 7, 11) <- "eggs"
print(x)
## [1] "spam, eggs, bacon, and spam"
```

Unfortunately, the number of characters in the replacement string should not exceed the length of the part being substituted (try `"chickpeas"` instead of `"eggs"`). However, substring replacement can be written as a composition of substring extraction and concatenation:

```
paste(substring(x, 1, 6), "chickpeas", substring(x, 11), sep="")
## [1] "spam, chickpeas, bacon, and spam"
```

**Exercise 6.6** *Take the output generated by* **`regexpr`** *and apply* **`substring`** *to extract the pattern occurrences. If there is no match in a string, the corresponding output should be* `NA`.

### 6.3.2 Translating characters

**`tolower`** and **`toupper`** converts between lower and upper case:

```
toupper("spam")
## [1] "SPAM"
```

---

**Note** Like many other string operations in base R, these functions perform very simple character substitutions. They might not be valid in natural language processing tasks. For instance, *groß* is *not* converted to *GROSS*, being the correct case folding in German.

---

Moreover, **`chartr`** translates individual characters:

```
chartr("\\", "/", "c:\\windows\\system\\cmd.exe")  # chartr(old, new, x)
## [1] "c:/windows/system/cmd.exe"
chartr("([S", ")]*", ":( :S :[")
## [1] ":) :* :]"
```

In the first line, we replace each backslash with a slash. The second example replaces "(", "[", and "S" with ")", "]", and "*", respectively.

### 6.3.3 Ordering strings

We have previously mentioned that operators and functions such as **`` `<` ``**, **`` `>=` ``**, **`sort`**, **`order`**, **`rank`**, and **`xtfrm`**[9] are based on the lexicographic ordering of strings.

```
sort(c("chłodny", "hardy", "chladný", "hladný"))
## [1] "chladný" "chłodny" "hardy"   "hladný"
```

It is worth noting that the ordering depends on the currently selected locale; see **`Sys.getlocale`**`("LC_COLLATE")`. For instance, in the Slovak language setting, we would obtain `"hardy"` < `"hladný"` < `"chladný"` < `"chłodny"`.

---

**Note** Many "structured" data items can be displayed or transmitted as human-readable strings. In particular, we know that **`as.numeric`** can convert a string to a number. Moreover, Section 10.3.1 will discuss date-time objects such as `"1970-01-01 00:00:00 GMT"`. We will be processing them with specialised functions such as **`strptime`** and **`strftime`**.

---

[9] See Section 12.3.1 for a use case.

**Important** (*) Many string operations in base R are not necessarily portable. The **stringx** package defines drop-in, "fixed" replacements therefor. They are based on the International Components for Unicode (ICU[10]) library, a de facto standard for processing Unicode text, and the R package **stringi**; see [27].

```
# call install.packages("stringx") first
suppressPackageStartupMessages(library("stringx"))  # load the package
sort(c("chłodny", "hardy", "chladný", "hladný"), locale="sk_SK")
## [1] "hardy"   "hladný"  "chladný" "chłodny"
toupper("gro\u00DF")  # compare base::toupper("gro\u00DF")
## [1] "GROSS"
detach("package:stringx")  # remove the package from the search path
```

## 6.4 Other atomic vector types (*)

We have discussed four vector types: `logical`, `double`, `character`, and `list`. To get a more complete picture of the sequence-like types in R, let's briefly mention `integer`, `complex`, and `raw` atomic types so that we are not surprised when we encounter them.

### 6.4.1 Integer vectors (*)

Integer scalars can be input manually by using the `L` suffix:

```
(x <- c(1L, 2L, -1L, NA_integer_))  # looks like numeric
## [1]  1  2 -1 NA
typeof(x)  # but is integer
## [1] "integer"
```

Some functions return them in a few contexts[11]:

```
typeof(1:10)  # seq(1, 10) as well, but not seq(1, 10, 1)
## [1] "integer"
as.integer(c(-1.1, 0, 1.9, 2.1))  # truncate/round towards 0
## [1] -1  0  1  2
```

In most expressions, integer vectors behave like numeric ones. They are silently co-

[10] https://icu.unicode.org/

[11] Actually, `1:10` returns an integer vector in a compact (ALTREP; see [56]) form; compare the results of the call to `.Internal(inspect(1:10))` and `.Internal(inspect(seq(1, 10, 1)))`. This way, the whole vector does not have to be allocated. This saves memory and time. At the R level, though, it behaves as any other integer (numeric) sequence.

erced to double if need be. Usually, there is no practical[12] reason to distinguish between them. For example:

```
1L/2L  # like 1/2 == 1.0/2.0
## [1] 0.5
```

---

**Note** (*) R `integer`s are 32-bit signed types. In the `double` type, we can store more of them. The maximal contiguously representable integer is $2^{31} - 1$ and $2^{53}$, respectively; see Section 3.2.3:

```
as.integer(2^31-1) + 1L  # 32-bit integer overflow
## Warning in as.integer(2^31 - 1) + 1L: NAs produced by integer overflow
## [1] NA
as.integer(2^31-1) + 1 == 2^31 # integer+double == double - OK
## [1] TRUE
(2^53 - 1) + 1 == 2^53  # OK
## [1] TRUE
(2^53 + 1) - 1 == 2^53  # lost due to FP rounding; left side equals 2^53 - 1
## [1] FALSE
```

---

---

**Note** Since R 3.0, there is support for vectors longer than $2^{31} - 1$ elements. As there are no 64-bit integers in R, long vectors are indexed by `double`s (we have been doing all this time). In particular, `x[1.9]` is the same as `x[1]`, and `x[-1.9]` means `x[-1]`, i.e., the fractional part is truncated. It is why the notation like `x[length(x)*0.2]` works, whether the length of `x` is a multiple of five or not.

---

### 6.4.2 Raw vectors (*)

Vectors of the type `raw` can store bytes, i.e., unsigned 8-bit integers, whose range is 0–255. For example:

```
as.raw(c(-1, 0, 1, 2, 0xc0, 254, 255, 256, NA))
## Warning: out-of-range values treated as 0 in coercion to raw
## [1] 00 00 01 02 c0 fe ff 00 00
```

They are displayed as two-digit hexadecimal (base-16) numbers. There are no raw `NA`s.

Only a few functions deal with such vectors: e.g., **`readBin`**, **`charToRaw`**, and **`rawToChar`**.

Interestingly, the meaning of the logical operators differs for raw vectors; they denote bitwise operations. See also **`bitwAnd`**, **`bitwOr`** etc. that work on integer vectors.

[12] They are of *internal* interest, e.g., when writing C/C++ extensions; see Chapter 14.

```
xor(as.raw(0xf0), as.raw(0x0f))
## [1] ff
bitwXor(0x0fff0f00, 0x0f00f0ff)
## [1] 16777215
```

**Example 6.7** *(*) One use case of bitwise operations is for representing a selection of items in a small set of possible values. This can be useful for communicating with routines implemented in C/C++. For instance, let's define three flags:*

```
HAS_SPAM  <- 0x01  # binary 00000001
HAS_BACON <- 0x02  # binary 00000010
HAS_EGGS  <- 0x04  # binary 00000100
```

*Now a particular subset can be created using the bitwise OR:*

```
dish <- bitwOr(HAS_SPAM, HAS_EGGS)  # {spam, eggs}
```

*Testing for inclusion is done via the bitwise AND:*

```
as.logical(bitwAnd(dish, c(HAS_SPAM, HAS_BACON, HAS_EGGS)))
## [1]  TRUE FALSE  TRUE
```

### 6.4.3 Complex vectors (*)

We can also play with vectors of the type `complex`, with "`1i`" representing the imaginary unit, $\sqrt{-1}$. Complex numbers appear in quite a few engineering or scientific applications, e.g., in physics, electronics, or signal processing. They are (at least: ought to be) part of introductory subjects or textbooks in university-level mathematics, including the statistics- and machine learning-orientated ones because of their heavy use of numerical computing; see e.g., [20, 32].

```
c(0, 1i, pi+pi*1i, NA_complex_)
## [1] 0.0000+0.0000i 0.0000+1.0000i 3.1416+3.1416i             NA
```

Apart from the basic operators, mathematical and aggregation functions, procedures like **fft**, **solve**, **qr**, or **svd** can be fed with or produce such data. For more details, see **help**("complex") and some matrix examples in Chapter 11.

## 6.5 Exercises

Exercises marked with (*) might require tinkering with regular expressions or third-party R packages.

**Exercise 6.8** *Answer the following questions.*

- *How many characters are there in the string* `"ab\n\\\t\\\\\""`*? What about* `r"-{ab\n\\\t\\\\\"-)}-"`*?*
- *What is the result of a call to* `paste(NA, 1:5, collapse="")`*?*
- *What is the meaning of the following* `sprintf` *format strings: "*`%s`*", "*`%20s`*", "*`%-20s`*", "*`%f`*", "*`%g`*", "*`%e`*", "*`%5f`*", "*`%5.2f%%`*", "*`%.2f`*", "*`%0+5f`*", and "*`[%+-5.2f]`*"?*
- *What is the difference between* `regexpr` *and* `gregexpr`*? What does "g" in the latter function name stand for?*
- *What is the result of a call to* `grepl(c("spam", "spammity spam", "aubergines"), "spam")`*?*
- *Is it always the case that "*`"Aaron" < "Zorro"`*"?*
- *Why "*`x < "10"`*" and "*`x < 10`*" may return different results?*
- *If* `x` *is a character vector, is "*`x == x`*" always equal to* `TRUE`*?*
- *If* `x` *and* `y` *are character vectors of lengths $n$ and $m$, respectively, what is the length of the output of* `match(x, y)`*?*
- *If* `x` *is a named vector, why is there a difference between* `x[NA]` *and* `x[NA_character_]`*?*
- *What is the difference between "*`x == y`*" and "*`x %in% y`*"?*

**Exercise 6.9** *Let* `x`*,* `y`*, and* `z` *be atomic vectors and* `a` *and* `b` *be single strings. Generate the same results as* `pastena(x, collapse=b)`*,* `pastena(x, y, sep=a)`*,* `pastena(x, y, sep=a, collapse=b)`*,* `pastena(x, y, z, sep=a)`*,* `pastena(x, y, z, sep=a, collapse=b)`*, assuming that* `pastena` *is a version of* `paste` *(which we do not have) that handles missing data in a way consistent with most other functions.*

**Exercise 6.10** *Based on* `list.files` *and* `glob2rx`*, generate the list of all PDFs on your computer. Then, use* `file.size` *to filter out the files smaller than 10 MiB.*

**Exercise 6.11** *Read a text file that stores a long paragraph of some banal prose. Concatenate all the lines to form a single, long string. Using* `strwrap` *and* `cat`*, output the paragraph on the console, nicely formatted to fit a block of text of an aesthetic width, say, 60 columns.*

**Exercise 6.12** *(*) Implement a simplified version of* `basename` *and* `dirname`*.*

**Exercise 6.13** *(*) Implement an operation similar to* `trimws` *using the functions introduced in this chapter.*

**Exercise 6.14** *(*) Write a regex that extracts all* words *from each string in a given character vector.*

**Exercise 6.15** *(*) Write a regex that extracts, from each string in a character vector, all:*

- *integers numbers (signed or unsigned),*
- *floating-point numbers,*
- *numbers of any kind (including those in scientific notation),*
- *#hashtags,*

- *email@address.es,*
- *hyperlinks of the form http://... and https://....*

**Exercise 6.16** *(*) What do `42i`, `42L`, and `0x42` stand for?*

**Exercise 6.17** *(*) Check out **`stri_sort`** in the **`stringi`** package (or **`sort.character`** in **`stringx`**) for a way to obtain an ordering like `"a1"` < `"a2"` < `"a10"` < `"a11"` < `"a100"`.*

**Exercise 6.18** *(*) In **`sprintf`**, the formatter `"%20s"` means that if a string is less than 20 bytes long, the remaining bytes will be replaced with spaces. Only for ASCII characters (English letters, digits, some punctuation marks, etc.), it is true that one character is represented by one byte. Other Unicode code points can take up between two and four bytes.*

```
cat(sprintf("..%6s..", c("abc", "1!<", "aßc", "qß©")), sep="\n")  # aligned?
## ..   abc..
## ..   1!<..
## ..  aßc..
## ..qß©..
```

*Use the **`stri_pad`** function from the **`stringi`** package to align the strings aesthetically. Alternatively, check out **`sprintf`** from **`stringx`**.*

**Exercise 6.19** *(*) Implement an operation similar to **`stri_pad`** from **`stringi`** using the functions introduced in this chapter.*

# 7

# *Functions*

R is a *functional* language, i.e., one where functions play first fiddle. Each action we perform reduces itself to a call to some function or a combination thereof.

So far, we have been tinkering with dozens of available functions which were mostly part of base R. They constitute the essential vocabulary that everyone *must* be able to speak fluently.

Any operation, be it **sum**, **sqrt**, or **paste**, when fed with a number of arguments, generates a (hopefully fruitful) return value.

```
sum(1:10)  # invoking `sum` on a specific argument
## [1] 55
```

From a user's perspective, each function is merely a tool. To achieve a goal at hand, we do not have to care about what is going on under its bonnet, i.e., how the inputs are being transformed so that, after a couple of nanoseconds or hours, we can relish what has been bred. This is very convenient: all we need to know is the function's specification which can be stated, for example, informally, in plain Polish or Malay, on its help page.

In this chapter, we will learn how to write our *own* functions. Using this skill is a good development practice when we expect that *the same operations* will need to be *executed many times* but perhaps on *different data*.

Also, some functions invoke other procedures, for instance, on every element in a list or every section of a data frame grouped by a qualitative variable. Thus, it is advisable to learn how we can specify a custom operation to be propagated thereover.

**Example 7.1** *Given some objects (whatever):*

```
x1 <- runif(16)
x2 <- runif(32)
x3 <- runif(64)
```

*assume we want to apply the same action on different data, say, compute the root mean square. Then, instead of retyping almost identical expressions (or a bunch of them) over and over again:*

```
sqrt(mean(x1^2))  # very fresh
## [1] 0.6545
sqrt(mean(x2^2))  # the same second time; borderline okay
```

```
## [1] 0.56203
sqrt(mean(x3^2))  # third time the same; tedious, barbarous, and error-prone
## [1] 0.57206
```

*we can* generalise *the operation to any object like x:*

```
rms <-                  # bind the name `rms` to...
    function(x)         # a function that takes one parameter, `x`
        sqrt(mean(x^2)) # transforming the input to yield output this way
```

*and then reuse it on different* concrete *data instances:*

```
rms(x1)
## [1] 0.6545
rms(x2)
## [1] 0.56203
rms(x3)
## [1] 0.57206
```

*or even combine it with other function calls:*

```
rms(sqrt(c(x1, x2, x3)))^2
## [1] 0.50824
```

*Thus, custom functions are very useful.*

---

**Important** Does writing own functions equal reinventing the wheel? Can everything be found online these days (including on Stack Overflow, GitHub, or CRAN)? Luckily, it is not the case. Otherwise, data analysts', researchers', and developers' lives would be monotonous, dreary, and uninspiring. What is more, we might be able to compose a function from scratch much more quickly than to get through the whole garbage dump called the internet from where, only occasionally, we can dig out some pearls. Let's remember that we advocate for minimalism in this book. We will reflect on such issues in Chapter 9. There is also the *personal growth* side: we become more skilled programmers by crunching those exercises.

---

## 7.1 Creating and invoking functions

### 7.1.1 Anonymous functions

Functions are usually created through the following notation:

```
function(args) body
```

First, `args` is a (possibly empty) list of comma-separated parameter names which act as *input* variables. Second, `body` is a *single* R expression that is evaluated when the function is called. The value this expression yields will constitute the function's output.

For example, here is a definition of a function that takes no inputs and generates a constant output:

```
function() 1
## function ()
## 1
```

We thus created a *function* object. However, as we have not used it at all, it disappeared immediately thereafter.

Any function **f** can be *invoked*, i.e., evaluated on *concrete* data, using the syntax `f(arg1, ..., argn)`. Here, `arg1`, …, `argn` are expressions passed as *arguments* to **f**.

```
(function() 1)()  # invoking f like f(); here, no arguments are expected
## [1] 1
```

Only now have we obtained a return value.

---

**Note** (*) Calling **typeof** on a function object will report `"closure"` (user-defined functions), `"builtin"`, or `"primitive"` (built-in, base ones) for the reasons that we explain in more detail in Section 9.4.3 and Section 16.3.2. In our case:

```
typeof(function() 1)
## [1] "closure"
```

---

### 7.1.2 Named functions

Names can be bound to function objects. This way, we can refer to them multiple times:

```
one <- function() 1  # one <- (function() 1)
```

We created an object named **one** (we use bold font to indicate that it is of the type function for functions are so crucial in R). We are very familiar with such a notation, as not since yesterday we are used to writing "`x <- 1`", etc.

Invoking **one**, which can be done by writing **one()**, will generate a return value:

```
one()  # (function() 1)()
## [1] 1
```

This output can be used in further computations. For instance:

```
0:2 - one()  # 0:2 - (function() 1)(), i.e., 0:2 - 1
## [1] -1  0  1
```

### 7.1.3 Passing arguments to functions

Functions with no arguments are kind of boring. Thus, let's distil a more highbrowed operation:

```
concat <- function(x, y) paste(x, y, sep="")
```

We created a mapping whose aim is to concatenate two objects using a specialised call to **paste**. Yours faithfully pleads guilty to multiplying entities needlessly: it *should* not be a problem for anyone to write **paste**(x, y, sep="") each time. Yet, 'tis merely an illustration.

The **concat** function has two *parameters*, x and y. Hence, calling it will require the provision of two *arguments*, which we put within round brackets and separate from each other by commas.

```
u <- 1:5
concat("spam", u)  # i.e., concat(x="spam", y=1:5)
## [1] "spam1" "spam2" "spam3" "spam4" "spam5"
```

---

**Important** Notice the distinction: *parameters* (*formal arguments*) are abstract, general, or symbolic; "something, anything that will be put in place of x when the function is invoked". Contrastingly, *arguments* (*actual parameters*) are concrete, specific, and real.

---

During the above call, x in the function's body is precisely "spam" and nothing else. Also, the u object from the caller's environment can be accessed via y in **concat**. Most of the time (yet, see Section 16.3), it is best to think of the function as being fed not with u per se but the value that u is bound to, i.e., 1:5.

Also:

```
x <- 1:5
y <- "spam"
concat(y, x)  # concat(x="spam", y=1:5)
## [1] "spam1" "spam2" "spam3" "spam4" "spam5"
```

This call is equivalent to **concat**(x=y, y=x). The argument x is assigned the value of y from the calling environment, "spam". Let's stress that one x is not the same as the other x; which is which is unambiguously defined by the context.

**Exercise 7.2** *Write a function* **standardise** *that takes a numeric vector x as argument and returns its standardised version, i.e., from each element in x, subtract the sample arithmetic mean and then divide it by the standard deviation.*

**Note** Section 2.1.3 mentioned that, syntactically speaking, the following are perfectly valid alternatives to the positionally-matched call `concat("spam", u)`:

```
concat(x="spam", y=u)
concat(y=u, x="spam")
concat("spam", y=u)
concat(u, x="spam")
concat(x="spam", u)
concat(y=u, "spam")
```

However, we recommend avoiding the last two for the sake of the readers' sanity. It is best to provide positionally-matched arguments before the keyword-based ones; see Section 15.4.4 for more details.

Also, Section 10.4 mentions the (overused) forward pipe operator, `` `|>` ``, which will enable us to rewrite the above as "`"spam" |> concat(u)`".

### 7.1.4 Grouping expressions with curly braces, `` `{` ``

We have been informed that a function's body is a *single* R expression whose evaluated value is passed to the user as its output. This may sound restrictive and in contrast with what we have experienced so far. Seldom are we faced with such simple computing tasks, and we have already seen R functions performing quite sophisticated operations.

Grammatically, a single R expression can be arbitrarily complex (Chapter 15). We can use curly braces to group many calls that are to be evaluated one after another. For instance:

```
{
    cat("first expression\n")
    cat("second expression\n")
    # ...
    cat("last expression\n")
}
## first expression
## second expression
## last expression
```

We used four spaces to visually indent the constituents for greater readability (some developers prefer tabs over spaces, others find two or three spaces more urbane, but we do not). This single (compound) expression can now play a role of a function's body.

**Important** The last expression evaluated in a curly-braces delimited block will be considered its output value.

```
x <- {
    1
    2
    3  # <--- last expression: will be taken as the output value
}
print(x)
## [1] 3
```

This code block can also be written more concisely by replacing newlines with semi-colons, albeit with perhaps some loss in readability:

```
{1; 2; 3}
## [1] 3
```

Section 9.3 will give a few more details about `` `{` ``.

---

**Example 7.3** *Here is a version of our* **`concat`** *function that guarantees a more Chapter 2-style missing values' propagation:*

```
concat <- function(a, b)
{
    z <- paste(a, b, sep="")
    z[is.na(a) | is.na(b)] <- NA_character_
    z  # last expression in the block - return value
}
```

*Example calls:*

```
concat("a", 1:3)
## [1] "a1" "a2" "a3"
concat(NA_character_, 1:3)
## [1] NA NA NA
concat(1:6, c("a", NA_character_, "c"))
## [1] "1a" NA   "3c" "4a" NA   "6c"
```

*Let's appreciate the fact that we could keep the code brief thanks to* **`paste`***'s and* `` `|` ``*'s implementing the recycling rule.*

**Exercise 7.4** *Write a function* **`normalise`** *that takes a numeric vector x and returns its version shifted and scaled to the [0, 1] interval. To do so, subtract the sample minimum from each element, and then divide it by the range, i.e., the difference between the maximum and the minimum. Avoid computing* **`min`**`(x)` *twice.*

**Exercise 7.5** *Write a function that applies the robust standardisation of a numeric vector: subtract the median and divide it by the median absolute deviation, 1.4826 times the median of the absolute differences between the values and their median.*

---

**Note** R is an open-source (free, libre) project distributed under the terms of the GNU

General Public License version 2. Therefore, we are not only encouraged to run the software for whatever purpose, but also study and modify its source code without restrictions. To facilitate this, we can display all function definitions:

```
print(concat)  # the code of the above procedure
## function (a, b)
## {
##     z <- paste(a, b, sep = "")
##     z[is.na(a) | is.na(b)] <- NA_character_
##     z
## }
print(union)  # a built-in function
## function (x, y)
## {
##     if (.set_ops_need_as_vector(x, y)) {
##         x <- as.vector(x)
##         y <- as.vector(y)
##     }
##     else if (!isa(y, class(x)))
##         x <- c(y[0L], x)
##     x <- unique(x)
##     names(x) <- NULL
##     y <- unique(y)
##     names(y) <- NULL
##     c(x, y[match(y, x, 0L) == 0L])
## }
## <environment: namespace:base>
```

Nevertheless, some functionality might be implemented in compiled programming languages such as C, C++, or Fortran; notice a call to `.Internal` in the source code of `paste`, `.Primitive` in `list`, or `.Call` in `runif`. Therefore, we will sometimes have to dig a bit deeper to access the underlying definition; see Chapter 14 for more details.

## 7.2 Functional programming

R is a *functional* programming language. As such, it shares several features with other languages that emphasise the role of function manipulation in software development (e.g., Common Lisp, Scheme, OCaml, Haskell, Clojure, F#). Let's explore these commonalities now.

### 7.2.1 Functions are objects

R functions were given the right to a *fair go*; they are what we refer to as *first-class citizens*. In other words, our interaction with them is not limited to their invocation; we treat them as any other language object.

- They can be stored inside list objects, which can embrace R objects of *any* kind:

```
list(identity, NROW, sum)  # a list storing three functions
## [[1]]
## function (x)
## x
## <environment: namespace:base>
##
## [[2]]
## function (x)
## if (length(d <- dim(x))) d[1L] else length(x)
## <environment: namespace:base>
##
## [[3]]
## function (..., na.rm = FALSE)  .Primitive("sum")
```

- They can be created and then called inside another function's body:

```
euclidean_distance <- function(x, y)
{
    square <- function(z) z^2  # auxiliary/internal/helper function
    sqrt(sum(square(x-y)))     # square root of the sum of squares
}

euclidean_distance(c(0, 1), c(1, 0))  # example call
## [1] 1.4142
```

  This is why we tend to classify functions as representatives of *recursive* types (compare **is.recursive**).

- They can be passed as arguments to other operations:

```
# Replaces missing values with a given aggregate
# of all non-missing elements:
fill_na <- function(x, filler_fun)
{
    missing_ones <- is.na(x)  # otherwise, we'd have to call is.na twice
    replacement_value <- filler_fun(x[!missing_ones])
    x[missing_ones] <- replacement_value
    x
}

fill_na(c(0, NA_real_, NA_real_, 2, 3, 7, NA_real_), mean)
## [1] 0 3 3 2 3 7 3
fill_na(c(0, NA_real_, NA_real_, 2, 3, 7, NA_real_), median)
## [1] 0.0 2.5 2.5 2.0 3.0 7.0 2.5
```

  Procedures like this are called *higher-order functions*.

**Note** More advanced techniques, which we discuss in the third part of the book, will let the functions be:

- returned as other functions' outputs,
- equipped with auxiliary data,
- generated programmatically on the fly,
- modified at runtime.

Let's review the most essential higher-order functions, including `do.call` and `Map`.

### 7.2.2 Calling on precomputed arguments with `do.call`

Notation like `f(arg1, ..., argn)` has no monopoly over how we call a function on a specific sequence of arguments. The list of actual parameters does not have to be hardcoded.

Here is an alternative. We can first prepare a number of objects to be passed as `f`'s inputs, wrap them in a list `l`, and then invoke `do.call(f, l)` to get the same result.

```
words <- list(
    c("spam",      "bacon",  "eggs"),
    c("buckwheat", "quinoa", "barley"),
    c("ham",       "spam",   "spam")
)
do.call(paste, words)  # paste(words[[1]], words[[2]], words[[3]])
## [1] "spam buckwheat ham" "bacon quinoa spam"  "eggs barley spam"
do.call(cbind, words)  # column-bind; returns a matrix (explained later)
##      [,1]    [,2]        [,3]
## [1,] "spam"  "buckwheat" "ham"
## [2,] "bacon" "quinoa"    "spam"
## [3,] "eggs"  "barley"    "spam"
do.call(rbind, words)  # row-bind (explained later)
##      [,1]        [,2]     [,3]
## [1,] "spam"      "bacon"  "eggs"
## [2,] "buckwheat" "quinoa" "barley"
## [3,] "ham"       "spam"   "spam"
```

The length and content of the list passed as the second argument of `do.call` can be arbitrary (possibly unknown at the time of writing the code). See Section 12.1.2 for more use cases, e.g., ways to concatenate a list of data frames (perhaps produced by some complex chain of commands) into a single data frame.

If elements of the list are named, they will be matched to the corresponding keyword arguments.

```
x <- 2^(seq(-2, 2, length.out=101))
```

```
plot_opts <- list(col="red", lty="dashed", type="l")
do.call(plot, c(list(x, log2(x), xlab="x", ylab="log2(x)"), plot_opts))
## (plot display suppressed)
```

Notice that our favourite `plot_opts` can now be reused in further calls to graphics functions. This is very convenient as it avoids repetitions.

### 7.2.3 Common higher-order functions

There is an important class of higher-order functions that permit us to apply custom operations on consecutive elements of sequences without relying on loop-like statements, at least explicitly. They can be found in all functional programming languages (e.g., Lisp, Haskell, Scala) and have been ported to various add-on libraries (**functools** in Python, more recent versions of the C++ Standard Library, etc.) or frameworks (Apache Spark and the like). Their presence reflects the obvious truth that certain operations occur more frequently than others. In particular:

- **Map** calls a function on each element of a sequence in order to transform:
  - their individual components (just like **sqrt**, **round**, or the unary `` `!` `` operator in R), or
  - the corresponding elements of many sequences so as to vectorise a given operation elementwisely (compare the binary `` `+` `` or **paste**),
- **Reduce** (also called accumulate) applies a binary operation to combine consecutive elements in a sequence, e.g., to generate the aggregates, like, totally (compare **sum**, **prod**, **all**, **max**) or cumulatively (compare **cumsum**, **cummin**),
- **Filter** creates a subset of a sequence that is comprised of elements that enjoy a given property (which we typically achieve in R by means of the `` `[` `` operator),
- **Find** locates the first element that fulfils some logical condition (compare **which**).

Below we will only focus on the **Map** function. The inspection of the remaining ones is left as an exercise. This is because, oftentimes, we can be better off with their more R-ish versions (e.g., using the subsetting operator, `` `[` ``).

### 7.2.4 Vectorising functions with Map

In data-centric computing, we are frequently faced with tasks that involve processing each vector element independently, one after another. Such use cases can benefit from vectorised operations like those discussed in Chapter 2, Chapter 3, and Chapter 6.

Unfortunately, most of the functions that we introduced so far cannot be applied on lists. For instance, if we try calling **sqrt** on a generic vector, we will get an error, even if it is a list of numeric sequences only. One way to compute the square root of all elements would be to invoke `sqrt(unlist(...))`. It is a go-to approach if we want to treat all the list's elements as one sequence. However, this comes at the price of losing the list's structure.

We have also discussed a few operations that are not vectorised with respect to all their arguments, even though they could have been designed this way, e.g., **grepl**.

The **Map** function[1] applies an operation on each element in a vector or the corresponding elements in a number of vectors. In many situations, it may be used as a more elegant alternative to **for** loops that we will introduce in the next chapter.

First[2], a call to `Map(f, x)` yields a list whose $i$-th element is equal to `f(x[[i]])`. For example:

```
x <- list(  # an example named list
    x1=1:3,
    x2=seq(0, 1, by=0.25),
    x3=c(1, 0, NA_real_, 0, 0, 1, NA_real_)
)
Map(sqrt, x)  # x is named, hence the result will be named as well
## $x1
## [1] 1.0000 1.4142 1.7321
##
## $x2
## [1] 0.00000 0.50000 0.70711 0.86603 1.00000
##
## $x3
## [1]  1  0 NA  0  0  1 NA
Map(length, x)
## $x1
## [1] 3
##
## $x2
## [1] 5
##
## $x3
## [1] 7
unlist(Map(mean, x))  # compute three aggregates, convert to an atomic vector
##  x1  x2  x3
## 2.0 0.5  NA
```

**Exercise 7.6** *Given a list of numeric vectors, fetch their last elements in the form of an atomic vector.*

Recall that `` `[[` `` works on atomic vectors, too. Thus, we can call, e.g.:

[1] Yes, the author is aware that **Map** was implemented using the slightly more primitive **mapply** but we are not fond of the latter function's having the `SIMPLIFY` argument set to `TRUE` by default.

[2] This use case scenario can also be programmed using **lapply**; `lapply(x, f, ...)` is equivalent to `Map(f, x, MoreArgs=list(...))`.

```
x <- c(2, 4, 6)
Map(function(n) round(runif(n, -1, 1), 1), x)
## [[1]]
## [1] 0.4 0.8
##
## [[2]]
## [1]  0.5  0.8 -0.1 -0.7
##
## [[3]]
## [1] -0.3  0.0  0.5  1.0 -0.9 -0.7
```

Next, we can vectorise a given function over several parameters. A call to, e.g., `Map(f, x, y, z)` breeds a list whose $i$-th element is equal to `f(x[[i]], y[[i]], z[[i]])`. Like in the case of, e.g., **`paste`**, the recycling rule will be applied if necessary.

For example, the following generates `list(seq(1, 6), seq(11, 13), seq(21, 29))`:

```
Map(seq, c(1, 11, 21), c(6, 13, 29))
## [[1]]
## [1] 1 2 3 4 5 6
##
## [[2]]
## [1] 11 12 13
##
## [[3]]
## [1] 21 22 23 24 25 26 27 28 29
```

Moreover, we can get `list(seq(1, 40, length.out=10), seq(11, 40, length.out=5), seq(21, 40, length.out=10), seq(31, 40, length.out=5))` by calling:

```
Map(seq, c(1, 11, 21, 31), 40, length.out=c(10, 5))
## [[1]]
##  [1]  1.0000  5.3333  9.6667 14.0000 18.3333 22.6667 27.0000 31.3333
##  [9] 35.6667 40.0000
##
## [[2]]
## [1] 11.00 18.25 25.50 32.75 40.00
##
## [[3]]
##  [1] 21.000 23.111 25.222 27.333 29.444 31.556 33.667 35.778 37.889 40.000
##
## [[4]]
## [1] 31.00 33.25 35.50 37.75 40.00
```

**Note** If we have some additional arguments to be passed to the function applied (which it does not have to be vectorised over), we can wrap them inside a separate list and toss it via the `MoreArgs` argument (à la **`do.call`**).

```
unlist(Map(mean, x, MoreArgs=list(na.rm=TRUE)))  # mean(..., na.rm=TRUE)
## [1] 2 4 6
```

Alternatively, we can always construct a custom anonymous function:

```
unlist(Map(function(xi) mean(xi, na.rm=TRUE), x))
## [1] 2 4 6
```

---

**Exercise 7.7** *Here is an example list of files (see our teaching data repository[3]) with daily Forex rates:*

```
file_names <- c(
    "euraud-20200101-20200630.csv",
    "eurgbp-20200101-20200630.csv",
    "eurusd-20200101-20200630.csv"
)
```

*Call `Map` to read them with `scan`. Determine each series' minimal, mean, and maximal value.*

**Exercise 7.8** *Implement your version of the `Filter` function based on a call to `Map`.*

---

## 7.3 Accessing third-party functions

When we indulge in the writing of a software piece, a few questions naturally arise. Is the problem we are facing fairly complex? Has it already been successfully addressed in its entirety? If not, can it, or its parts, be split into manageable chunks? Can it be constructed based on some readily available nontrivial components?

A smart developer is independent but knows when to stand on the shoulders to cry on. Let's explore a few ways to reuse the existing function libraries.

### 7.3.1 Using R packages

Most contributed R extensions come in the form of *add-on packages*, which can include:

- reusable code (e.g., new functions),
- data (which we can exercise on),
- documentation (manuals, vignettes, etc.);

see Section 9.2.2 for more and *Writing R Extensions* [66] for all the details.

Most packages are published in the moderated repository that is part of the *Compre-*

[3] https://github.com/gagolews/teaching-data/tree/master/marek

*hensive R Archive Network* (CRAN[4]). However, there are also other popular sources such as Bioconductor[5] which specialises in bioinformatics.

We call **install.packages**("pkg") to fetch a package **pkg** from a repository (CRAN by default; see, however, the `repos` argument).

A call to **library**("pkg") loads an indicated package and makes the exported objects available to the user (i.e., attaches it to the search path; see Section 16.2.6).

For instance, in one of the previous chapters, we have mentioned the **gsl** package:

```
# call install.packages("gsl") first
library("gsl")  # load the package
poch(10, 3:6)   # calls gsl_sf_poch() from GNU GSL
## [1]    1320   17160  240240 3603600
```

Here, **poch** is an object exported by package **gsl**. If we did not call **library**("gsl"), trying to access the former would raise an error.

We could have also accessed the preceding function without attaching it to the search path using the **pkg::fun** syntax, namely, **gsl::poch**.

---

**Note** For more information about any R extension, call **help**(package="pkg"). Also, it is advisable to visit the package's CRAN entry at an address like *https://CRAN.R-project.org/package=pkg* to access additional information, e.g., vignettes. Why waste our time and energy by querying a web search engine that will likely lead us to a dodgy middleman when we can acquire authoritative knowledge directly from the source?

Moreover, it is worth exploring various CRAN Task Views[6] that group the packages into topics such as *Genetics*, *Graphics*, and *Optimisation*. They are curated by experts in their relevant fields.

---

**Important** Frequently, R packages are written in their respective authors' free time, many of whom are volunteers. Neither get they paid for this, nor do it as part of the so-called *their job*. Yes, not everyone is driven by money or fame.

Someday, when we come up with something valuable for the community, we will become one of them. Before this happens, we can show appreciation for their generosity by, e.g., spreading the word about their software by citing it in publications (see **citation**(package="pkg")), talking about them during lunchtime, or mentioning them in (un)social media. We can also help them improve the existing code base by reporting bugs, polishing documentation, proposing new features, or cleaning up the redundant fragments of their APIs.

---

[4] https://cloud.r-project.org/
[5] https://bioconductor.org/
[6] https://cloud.r-project.org/web/views

### Default packages

The **base** package is omnipresent. It provides us with the most crucial functions such as the vector addition, **c**, **Map**, and **library**. Certain other extensions are also loaded by default:

```
getOption("defaultPackages")
## [1] "datasets"  "utils"     "grDevices" "graphics"  "stats"
## [6] "methods"
```

In this book, we assume that they are always attached (even though this list can, theoretically, be changed[7]). Due to this, in Section 2.4.5, there was no need to call, for example, **library**("stats") before referring to the **var** and **sd** functions.

On a side note, **grDevices** and **graphics** will be discussed in Chapter 13. **methods** will be mentioned in Section 10.5. **datasets** brings a few example R objects on which we can exercise our skills. The functions from **utils**, **graphics**, and **stats** already appeared here and there.

**Exercise 7.9** *Use the **find** function to determine which packages define **mean**, **var**, **find**, and **Map**. Recall from Section 1.4 where such information can be found in these objects' manual pages.*

### Source vs binary packages (*)

R is an open environment. Therefore, its packages are published primarily in the source form. This way, anyone can study how they work and improve them or reuse parts thereof in different projects.

If we call **install.packages**("path", repos=NULL, type="source"), we should be able to install a package from sources: path can be pinpointing either a directory or a source tarball (most often as a compressed pkg_version.tar.gz file; see **help**("untar")).

Note that type="source" is the default unless one is on a Win***s or m**OS box; see **getOption**("pkgType"). This is because these two operating systems require additional build tools, especially if a package relies on C or C++ code; see Chapter 14 and Section C.3 of [68]:

- RTools[8] on Win***s,
- Xcode Command Line Tools[9] on m**OS.

These systems are less developer-orientated. Thus, as a courtesy to their users, CRAN also distributes the platform-specific binary versions of the packages (.zip or .tgz files). **install.packages** will try to fetch them by default.

**Example 7.10** *It is very easy to retrieve a package's source directly from GitLab and GitHub, which are popular hosting platforms. The relevant links are, respectively:*

[7] (*) R is greatly configurable: we can have custom ~/.Renviron and ~/.Rprofile files that are processed on R's startup; see **help**("Startup").

[8] https://cran.r-project.org/bin/windows/Rtools

[9] https://developer.apple.com/xcode/resources

- *https://gitlab.com/*user/repo/*-/archive/*branch/repo-branch.*zip*,
- *https://github.com/*user/repo/*archive/*branch.*zip*.

*For example, to download the contents of the* master *branch in the GitHub repository* rpackagedemo *owned by* gagolews*, we can call:*

```
f <- tempfile()  # download destination: a temporary file name
download.file("https://github.com/gagolews/rpackagedemo/archive/master.zip",
    destfile=f)
```

*Next, the contents can be extracted with* **unzip***:*

```
t <- tempdir()  # temporary directory for extracted files
(d <- unzip(f, exdir=t))  # returns extracted file paths
```

*The path where the files were extracted can be passed to* **install.packages***:*

```
install.packages(dirname(d)[1], repos=NULL, type="source")
file.remove(c(f, d))  # clean up
```

**Exercise 7.11** *Use the* **git2r** *package to clone the* **git** *repository located at https://github.com/gagolews/rpackagedemo.git and install the package published therein.*

### Managing dependencies (*)

By calling **update.packages**, all installed add-on packages may be upgraded to their most recent versions available on CRAN or other indicated repository.

As a general rule, the more experienced we become, the less excited we get about the *new*. Sure, bug fixes and well-thought-out additional features are usually welcome. Still, just we wait until someone updates a package's API for the $n$-th time, $n \geq 2$, breaking our so-far flawless program.

Hence, when designing software projects (see Chapter 9 for more details), we must ask ourselves the ultimate question: do we really need to import that package with lots of dependencies from which we will just use only about 3–5 functions? Wouldn't it be better to write our own version of some functionality (and learn something new, exercise our brain, etc.), or call a mature terminal-based tool?

Otherwise, as all the historical versions of the packages are archived on CRAN[10], simple software dependency management can be conducted by storing different releases of packages in different directories. This way, we can create an isolated environment for the add-ons. To fetch the locations where packages are sought (in this very order), we call:

```
.libPaths()
## [1] "/home/gagolews/R/x86_64-suse-linux-gnu-library/4.6"
## [2] "/usr/lib64/R/library"
```

[10] https://cran.r-project.org/src/contrib/Archive

The same function can add new folders to the search path; see also the environment variable `R_LIBS_USER` that we can set using **`Sys.setenv`**. The **`install.packages`** function will honour them as target directories; see its `lib` parameter for more details. Note that only one version of a package can be loaded at a time, though.

Moreover, the packages may deposit auxiliary data on the user's machine. Therefore, it might be worthwhile to set the following directories (via the corresponding environment variables) relative to the current project:

```
tools::R_user_dir("pkg", "data")    # R_USER_DATA_DIR
## [1] "/home/gagolews/.local/share/R/pkg"
tools::R_user_dir("pkg", "config")  # R_USER_CONFIG_DIR
## [1] "/home/gagolews/.config/R/pkg"
tools::R_user_dir("pkg", "cache")   # R_USER_CACHE_DIR
## [1] "/home/gagolews/.cache/R/pkg"
```

### 7.3.2 Calling external programs

Many tasks can be accomplished by calling external programs. Such an approach is particularly natural on UNIX-like systems, which classically follow modular, minimalist design patterns. There are many tools at a developer's hand and each of them is specialised at solving a single, well-defined problem. Apart from the many standard UNIX commands[11], we may consider:

- **`pandoc`**[12] converts documents between markup formats, e.g., Markdown, HTML, reStructuredText, and LaTeX and can generate LibreOffice Writer documents, EPUB or PDF files, or slides;
- **`jupyter-nbconvert`** converts Jupyter[13] notebooks (see Section 1.2.5) to other formats such as LaTeX, HTML, Markdown, etc.;
- **`convert`** (from **`ImageMagick`**[14]) applies various operations on bitmap graphics (scaling, cropping, conversion between formats);
- **`graphviz`**[15] and **`PlantUML`**[16] draws graphs and diagrams;
- **`python`**, **`perl`**, … can be called to perform tasks that can be expressed more easily in languages other than R.

The good news is that we are not limited to calling R from the system shell in the interactive or batch mode; see Section 1.2. Our environment serves particularly well as a glue language, too.

The **`system2`** function invokes an external command. The communication between different programs may be done using, e.g., intermediate text, JSON, CSV, XML, or any other files. The `stdin`, `stdout`, and `stderr` arguments control the redirection of the standard I/O streams.

[11] https://en.wikipedia.org/wiki/List_of_Unix_commands
[12] https://pandoc.org/
[13] https://jupyter.org/
[14] https://imagemagick.org/
[15] https://graphviz.org/
[16] https://plantuml.com/

```
system2("pandoc", "-s input.md -o output.html")
system2("bash", "-c 'for i in `seq 1 2 10`; do echo $i; done'", stdout=TRUE)
## [1] "1" "3" "5" "7" "9"
system2("python3", "-", stdout=TRUE,
    input=c(
    "import numpy as np",
    "print(repr(np.arange(5)))"
    ))
## [1] "array([0, 1, 2, 3, 4])"
```

On a side note, the current working directory can be read and changed through a call to **getwd** and **setwd**, respectively. By default, it is the directory where the current R session was started.

---

**Important** Relying on **system2** assumes that the commands referred to are available on the target platform. Hence, it might not be portable unless additional assumptions are made, e.g., that a user runs a UNIX-like system and that certain libraries are available. We strongly recommend GNU/Linux or FreeBSD for both software development and production use, as they are free, open, developer-friendly, user-loving, reliable, ethical, and sustainable. Users of other operating systems are missing out on so many good features.

---

### 7.3.3 Interfacing C, C++, Fortran, Python, Java, etc. (**)

Most standalone data processing algorithms are implemented in compiled, slightly lower-level programming languages. This usually makes them faster and more reusable in other environments. For instance, an industry-standard library might be written in very portable C, C++, or Fortran and define bindings for easier access from within R, Python, Julia, etc. It is the case with FFTW, LIBSVM, mlpack, OpenBLAS, ICU, and GNU GSL, amongst many others. Chapter 14 explains basic ways to refer to such compiled code.

Also, the **rJava** package can dynamically create JVM objects and access their fields and methods. Similarly, **reticulate** can be used to access Python objects, including **numpy** arrays and **pandas** data frames (but see also the **rpy2** package for Python).

---

**Important** We should not feel obliged to use R in all parts of a data processing pipeline. Some activities can be expressed more naturally in other languages or environments (e.g., parse raw data and create a SQL database in Python but visualise it in R).

---

## 7.4 Exercises

**Exercise 7.12** *Answer the following questions.*

- *What is the result of "*`{x <- "x"; x <- function(x) x; x(x)}`*"?*
- *How to compose a function that returns two objects?*
- *What is a higher-order function?*
- *What are the use cases of* `do.call`*?*
- *Why a call to* `Map` *is redundant in the expression* `Map(paste, x, y, z)`*?*
- *What is the difference between* `Map(mean, x, na.rm=TRUE)` *and* `Map(mean, x, MoreArgs=list(na.rm=TRUE))`*?*
- *What do we mean when we write* `stringx::sprintf`*?*
- *How to get access to the vignettes (tutorials, FAQs, etc.) of the* `data.table` *and* `dplyr` *packages? Why perhaps 95% of R users would just googleit, and what is suboptimal about this strategy?*
- *What is the difference between a source and a binary package?*
- *How to update the* `base` *package?*
- *How to ensure that we will always run an R session with only specific versions of a set of packages?*

**Exercise 7.13** *Write a function that computes the Gini index of a vector of positive integers* `x`*, which, assuming $x_1 \le x_2 \le \ldots \le x_n$, is equal to:*

$$G(x_1, \ldots, x_n) = \frac{\sum_{i=1}^{n} (n - 2i + 1) x_i}{(n-1) \sum_{i=1}^{n} x_i}.$$

**Exercise 7.14** *Implement a function* `between(x, a, b)` *that verifies whether each element in* `x` *is in the* `[a, b]` *interval. Return a logical vector of the same length as* `x`*. Ensure the function is correctly vectorised with respect to all the arguments and handles missing data correctly.*

**Exercise 7.15** *Write your version of the* `strrep` *function called* `dup`*.*

```
dup <- ...to.do...
dup(c("a", "b", "c"), c(1, 3, 5))
## [1] "a"     "bbb"   "ccccc"
dup("a", 1:3)
## [1] "a"   "aa"  "aaa"
dup(c("a", "b", "c"), 4)
## [1] "aaaa" "bbbb" "cccc"
```

**Exercise 7.16** *Given a list* `x`*, generate its sublist with all the elements equal to* `NULL` *removed.*

**Exercise 7.17** *Implement your version of the* **`sequence`** *function.*

**Exercise 7.18** *Using* **`Map`**, *how can we generate window indexes like below?*

```
## [[1]]
## [1] 1 2 3
##
## [[2]]
## [1] 2 3 4
##
## [[3]]
## [1] 3 4 5
##
## [[4]]
## [1] 4 5 6
```

*Write a function* **`windows(k, n)`** *that yields index windows of length $k$ with elements between $1$ and $n$ (the above example is for $k = 3$ and $k = 6$).*

**Exercise 7.19** *Write a function to extract all q-grams, $q \geq 1$, from a given character vector. Return a list of character vectors. For example, bigrams (2-grams) in* `"abcd"` *are:* `"ab"`, `"bc"`, *"cd"`.*

**Exercise 7.20** *Implement a function* **`movstat(f, x, k)`** *that computes, using* **`Map`**, *a given aggregate* **`f`** *of each $k$ consecutive elements in* `x`. *For instance:*

```
movstat <- ...to.do...
x <- c(1, 3, 5, 10, 25, -25)  # example data
movstat(mean, x, 3)           # 3-moving mean
## [1]  3.0000  6.0000 13.3333  3.3333
movstat(median, x, 3)         # 3-moving median
## [1]  3.0000  6.0000 13.3333  3.3333
```

**Exercise 7.21** *Recode a character vector with a small number of distinct values to a vector where each unique code is assigned a positive integer from $1$ to $k$. Here are example calls and the corresponding expected results:*

```
recode <- ...to.do...
recode(c("a", "a", "a", "b", "b"))
## [1] 1 1 1 2 2
recode(c("x", "z", "y", "x", "y", "x"))
## [1] 1 3 2 1 2 1
```

**Exercise 7.22** *Implement a function that returns the number of occurrences of each unique element in a given atomic vector. The return value should be a numeric vector equipped with the* `names` *attribute. Hint: use* **`match`** *and* **`tabulate`**.

```
count <- ...to.do...
count(c(5, 5, 5, 5, 42, 42, 954))
##   5  42 954
```

```
##    4    2    1
count(c("x", "z", "y", "x", "y", "x", "w", "x", "x", "y", NA_character_))
##     w     x     y     z <NA>
##     1     5     3     1    1
```

**Exercise 7.23** *Extend the built-in* **`duplicated`** *function. For each vector element, indicate which occurrence of a repeated value is it (starting from the beginning of the vector).*

```
duplicatedn <- ...to.do...
duplicatedn(c("a", "a", "a", "b", "b"))
## [1] 1 2 3 1 2
duplicatedn(c("x", "z", "y", "x", "y", "x", "w", "x", "x", "y", "z"))
##  [1] 1 1 1 2 2 3 1 4 5 3 2
```

**Exercise 7.24** *Based on a call to* **`Map`***, implement your version of* **`split`** *that takes two atomic vectors as arguments. Then, extend it to handle the second argument being a list of the form* **`list`**`(y1, y2, ...)` *representing the product of many levels. If the* `y`*s are of different lengths, apply the recycling rule.*

**Exercise 7.25** *Implement* **`my_unsplit`** *being your version of* **`unsplit`***. For any* `x` *and* `g` *of the same lengths, ensure that* **`my_unsplit`**`(`**`split`**`(x, g), g)` *is equal to* `x`*.*

**Exercise 7.26** *Write a function that takes as arguments: (a) an integer $n$, (b) a numeric vector* `x` *of length $k$ and no duplicated elements, (c) a vector of probabilities* `p` *of length $k$. Verify that $p_i \geq 0$ for all $i$ and $\sum_{i=1}^{k} p_i \simeq 1$. Based on a random number generator from the uniform distribution on the unit interval, generate $n$ independent realisations of a random variable $X$ such that $\Pr(X = x_i) = p_i$ for $i = 1, \ldots, k$. To obtain a single value:*

1. *generate $u \in [0, 1]$,*
2. *find $m \in \{1, \ldots, k\}$ such that $u \in \left(\sum_{j=1}^{m-1} p_j, \sum_{j=1}^{m} p_j\right]$,*
3. *the result is then $x_m$.*

**Exercise 7.27** *Write a function that takes as arguments: (a) an increasingly sorted vector* `x` *of length $n$, (b) any vector* `y` *of length $n$, (c) a vector* `z` *of length* `k` *and elements in $[x_1, x_n)$. Let $f$ be the piecewise linear spline that interpolates the points $(x_1, y_1), \ldots, (x_n, y_n)$. Return a vector* `w` *of length* `k` *such that $w_i = f(z_i)$.*

**Exercise 7.28** *(*) Write functions* **`dpareto`***,* **`ppareto`***,* **`qpareto`***, and* **`rpareto`** *that implement the functions related to the Pareto distribution; compare Section 2.3.4.*

# 8

# *Flow of execution*

The **ifelse** and **Map** functions are potent. However, they allow us to process only the *consecutive* elements in a vector. Below we will (finally!) discuss different ways to alter a program's control flow manually, based on some criterion, and to evaluate the same expression many times, but perhaps on different data. Nevertheless, before proceeding any further, let's meditate on the fact that we have managed *without* them for such a long time, even though the data processing exercises we solved were far from trivial.

## 8.1 Conditional evaluation

Life is full of surprises, so it would be nice if we were able to adapt to any future challenges. The following evaluates a given `expression` *if and only if* a logical `condition` is true.

```
if (condition) expression
```

When performing some `other_expression` is preferred rather than doing nothing in the case of the `condition`'s being false, we can write:

```
if (condition) expression else other_expression
```

For instance:

```
(x <- runif(1))  # to spice things up
## [1] 0.28758
if (x > 0.5) cat("head\n") else cat("tail\n")
## tail
```

Many expressions can, of course, be grouped with curly braces, **`{`**.

```
if (x > 0.5) {
    cat("head\n")
    x <- 1
} else {  # do not put newline before else!
    cat("tail\n")
    x <- 0
}
```

```
## tail
print(x)
## [1] 0
```

**Important** At the top level, we should not put a new line before `else`. Otherwise, we will get an error like `Error: unexpected 'else' in "else"`. This is because the interpreter enthusiastically executes the statements read line by line as soon as it regards them as standalone expressions. In this case, we first get an `if` without `else`, and then, separately, a *dangling* `else` without the preceding `if`.

This is not an issue when a conditional statement is part of an expression group as the latter is read in its entirety.

```
function (x)
{  # opening bracket - start
    if (x > 0.5)
        cat("head\n")
    else              # not dandling because {...} is read as a whole
        cat("tail\n")
}  # closing bracket - expression ends
```

As an exercise, try removing the curly braces and see what happens.

### 8.1.1 Return value

`` `if` `` is a function (compare Section 9.3). Hence, it has a return value: the result of evaluating the conditional expression.

```
(x <- runif(1))
## [1] 0.28758
y <- if (x > 0.5) "head"  # no else
print(y)
## NULL
y <- if (x > 0.5) "head" else "tail"
print(y)
## [1] "tail"
```

This is particularly useful when a call to `` `if` `` is the last expression in a curly brace-delimited code block that constitutes a function's body.

```
mint <- function(x)
{
    cond <- (x > 0.5)  # could be something more sophisticated
    if (cond)  # the last expression in the code block
        "head"        # this can be the return value...
    else
```

```
        "tail"        # or this one, depending on the condition
}

mint(x)
## [1] "tail"
unlist(Map(mint, runif(5)))
## [1] "tail" "head" "tail" "head" "head"
```

**Example 8.1** *Add-on packages can be loaded using* **`requireNamespace`**. *Contrary to* **`library`**, *the former does not fail when a package is not available. Also, it does not attach it to the search path; see Section 16.2.6. Instead, it returns a logical value indicating if the package is available for use. This can be helpful in situations where the availability of some features depends on the user environment's configuration:*

```
process_data <- function(x)
{
    if (requireNamespace("some_extension_package", quietly=TRUE))
        some_extension_package::very_fast_method(x)
    else
        normal_method(x)
}
```

### 8.1.2 Nested `if`s

If more than two test cases are possible, i.e., when we need to go beyond either `condition` or `!condition`, then we can use the following construct:

```
if (a) {
    expression_a
} else if (b) {
    expression_b
} else if (c) {
    expression_c
} else {
    expression_else
}
```

This evaluates all conditions `a`, `b`, … (in this order) until the first positive case is found and then executes the corresponding `expression`. It is worth stressing that the foregoing is nothing else than a series of nested **`if`** statements but written in a more readable[1] manner:

```
if (a) {
    expression_a
```

[1] (*) Somewhat related is the **`switch`** function which relies on the lazy evaluation of its arguments (Chapter 17). However, it can always be replaced by a series of **`if`**s.

```
} else {
    if (b) {
    expression_b
    } else {
        if (c) {
            expression_c
        } else {
            expression_else
        }
    }
}
```

**Exercise 8.2** *Write a function named* `sign` *that determines if a given numeric value is* `"positive"`*,* `"negative"`*, or* `"zero"`*.*

### 8.1.3 Condition: Either TRUE or FALSE

`if` expects a `condition` that is a single, well-defined logical value, either `TRUE` or `FALSE`. Thence, problems may arise when this is not the case. First, if the `condition` is not of length one, we get an error:

```
if (c(TRUE, FALSE)) cat("spam\n")
## Error in `if (c(TRUE, FALSE)) ...`: ! the condition has length > 1
if (logical(0)) cat("bacon\n")
## Error in `if (logical(0)) ...`: ! argument is of length zero
```

We cannot pass a missing value either:

```
if (NA) cat("ham\n")
## Error in `if (NA) ...`: ! missing value where TRUE/FALSE needed
```

---

**Important** If we think that we are immune to writing code violating the preceding constraints, just we wait until the `condition` becomes a function of data for which there is no sanity-checking in place.

```
mint <- function(x)
    if (x > 0.5) "head" else "tail"

mint(0.25)
## [1] "tail"
mint(runif(5))
## Error in `if (x > 0.5) ...`: ! the condition has length > 1
mint(log(rnorm(1)))  # not obvious, error raised occasionally
## Warning in log(rnorm(1)): NaNs produced
## Error in `if (x > 0.5) ...`: ! missing value where TRUE/FALSE needed
```

Chapter 9 will be concerned with ensuring input data integrity so that such cases will either fail gracefully or succeed bombastically. In the above example, we should probably verify that x is a single finite numeric value. Alternatively, we might need to apply **ifelse**, **all**, or **any**.

---

Interestingly, conditions other that logical are coerced:

```
x <- 1:5
if (length(x))  # i.e., length(x) != 0, but way less readable
    cat("length is not zero\n")
## length is not zero
```

Recall that coercion of numeric to logical yields FALSE if and only if the original value is zero.

### 8.1.4 Short-circuit evaluation

Especially for formulating logical conditions in **if** and **while** (see below), we have the *scalar* `||` (alternative) and `&&` (conjunction) operators.

```
FALSE || TRUE
## [1] TRUE
NA || TRUE
## [1] TRUE
```

Contrary to their vectorised counterparts (`|` and `&`), the scalar operators are lazy (Chapter 17) in the sense that they evaluate the first operand and then determine if the computing of the second one is necessary (because, e.g., FALSE && whatever is always FALSE anyway). Therefore,

```
if (a && b)
    expression
```

is equivalent to:

```
if (a) {
    if (b) {  # compute b only if a is TRUE
        expression
    }
}
```

and

```
if (a || b)
    expression
```

corresponds to:

```
if (a) {
```

```
    expression
} else if (b) {  # compute b only if a is FALSE
    expression
}
```

For instance, "`is.vector(x) && length(x) > 0 && x[[1]] > 0`" is a risk-free test. It takes into account that `x[[1]]` has the desired meaning only for objects that are nonempty vectors.

Some more examples:

```
{cat("spam"); FALSE} || {cat("ham"); TRUE} || {cat("cherries"); FALSE}
## spamham
## [1] TRUE
{cat("spam"); TRUE} && {cat("ham"); FALSE} && {cat("cherries"); TRUE}
## spamham
## [1] FALSE
```

Recall that the expressions within the curly braces are evaluated one after another and that the result is determined by the last value in the series.

**Exercise 8.3** *Study the source code of* **`isTRUE`** *and* **`isFALSE`** *and determine if these functions could be useful in formulating the conditions within the* **`if`** *expressions.*

## 8.2 Exception handling

Exceptions are exceptional, but they may happen and break stuff. For instance, we are in deep skit when the internet connection drops while we try to download a file, an optimisation algorithm fails to converge, or:

```
read.csv("/path/to/a/file/that/does/not/exist")
## Warning in file(file, "rt"): cannot open file '/path/to/a/file/that/does/
##     not/exist': No such file or directory
## Error in `file()`: ! cannot open the connection
```

Three types of *conditions* are frequently observed:

- *errors* stop the flow of execution,
- *warnings* are not critical, but can be turned into errors (see `warn` in **`option`**),
- *messages* transmit diagnostic information.

They can be manually triggered using the **`stop`**, **`warning`**, and **`message`** functions.

Errors (but warnings too) can be handled by means of the **`tryCatch`** function, amongst others.

```
tryCatch({          # block of expressions to execute, until an error occurs
        cat("a...\n")
        stop("b!")    # error - breaks the linear control flow
        cat("c?\n")
    },
    error = function(e) {  # executed immediately on an error
        cat(sprintf("[error] %s\n", e[["message"]]))
    },
    finally = {  # always executed at the end, regardless of error occurrence
        cat("d.\n")
    }
)
## a...
## [error] b!
## d.
```

The two other conditions can be ignored by calling `suppressWarnings` and `suppressMessages`.

```
log(-1)
## Warning in log(-1): NaNs produced
## [1] NaN
suppressWarnings(log(-1))  # yeah, yeah, we know what we're doing
## [1] NaN
```

**Exercise 8.4** *At the time of writing this book, when the **`data.table`** package is attached, it emits a message. Call **`suppressMessages`** to silence it. Note that consecutive calls to **`library`** do not reload an already loaded package. Therefore, the message will only be seen once per R session.*

Related functions include `stopifnot` discussed in Section 9.1 and `on.exit` mentioned in Section 17.4; see Section 9.2.4 for some code debugging tips.

---

## 8.3 Repeated evaluation

And now for something completely different… time for the elephant in the room!

We have been able to manage without loops so far (and will be quite all right in the second part of the book too). This is because many data processing tasks can be written in terms of vectorised operations such as `` `+` ``, `sqrt`, `sum`, `` `[` ``, `Map`, and `Reduce`. Oftentimes, compared to their loop-based counterparts, they are more readable and efficient. We will explore this in the coming exercises.

However, at times, using an explicit `while` or `for` loop might be the only natural way to solve a problem, for instance, when processing chunks of data streams. Also, an explicitly “looped” algorithm may occasionally have better[2] time or memory complexity.

### 8.3.1 `while`

`if` considers a logical condition provided and determines whether to execute a given statement. On the other hand:

```
while (condition)  # single TRUE or FALSE, as in `if`
    expression
```

evaluates a given `expression` *as long as* the logical `condition` is true. Therefore, it is advisable to make the `condition` dependent on some variable that the `expression` can modify.

```
i <- 1
while (i <= 3) {
    cat(sprintf("%d, ", i))
    i <- i + 1
}
## 1, 2, 3,
```

Nested loops are possible too:

```
i <- 1
while (i <= 2) {
    j <- 1
    while (j <= 3) {
        cat(sprintf("%d %d, ", i, j))
        j <- j + 1
    }
    cat("\n")
    i <- i + 1
}
## 1 1, 1 2, 1 3,
## 2 1, 2 2, 2 3,
```

**Exercise 8.5** *Implement a simple linear congruential pseudorandom number generator that, given some seed $X_0 \in [0, m)$, outputs a sequence $(X_1, X_2, \dots)$ defined by:*

$$X_i = (aX_{i-1} + c) \mod m,$$

*with, e.g., $a = 75$, $c = 74$, and $m = 2^{16} + 1$ (here,* mod *is the division remainder, `%%`). This generator has poor statistical properties and its use in practice is discouraged. In particular, after a rather small number of iterations $k$, we will find a cycle such that $X_k = X_1, X_{k+1} = X_2, \dots$.*

[2] In such a case, rewriting it in C or C++ might be beneficial; see Chapter 14.

### 8.3.2 for

The for-each loop:

```
for (name in vector)
    expression
```

takes each element, from the beginning to the end, in a given `vector`, assigns it some `name`, and evaluates the `expression`. For example:

```
fridge <- c("spam", "spam", "bacon", "eggs")
for (food in fridge)
    cat(sprintf("%s, ", food))
## spam, spam, bacon, eggs,
```

Another example:

```
for (i in 1:length(fridge))  # better: seq_along(fridge); see below
    cat(sprintf("%s, ", fridge[i]))
## spam, spam, bacon, eggs,
```

One more:

```
for (i in 1:2) {
    for (j in 1:3)
        cat(sprintf("%d %d, ", i, j))
    cat("\n")
}
## 1 1, 1 2, 1 3,
## 2 1, 2 2, 2 3,
```

The iterator still exists after the loop's watch has ended:

```
print(i)
## [1] 2
print(j)
## [1] 3
```

---

**Important** Writing:

```
for (i in 1:length(x))
    print(x[i])
```

is reckless. If `x` is an empty vector, then we will observe undesired behaviour because we ask to iterate over `1:0`:

```
x <- logical(0)
for (i in 1:length(x))
    print(x[i])
```

```
## [1] NA
## logical(0)
```

Recall from Chapter 5 that x[1] tries to access an out-of-bounds element here, and x[0] returns nothing. We generally suggest replacing 1:**length**(x) with **seq_along**(x) or **seq_len**(**length**(x)) wherever possible.

---

**Note** The preceding model **for** loop is roughly equivalent to:

```
name <- NULL
tmp_vector <- vector
tmp_iter <- 1
while (tmp_iter <= length(tmp_vector)) {
    name <- tmp_vector[[tmp_iter]]
    expression
    tmp_iter <- tmp_iter + 1
}
```

Note that the tmp_vector is determined before the loop itself. Hence, any changes to the vector will not influence the execution flow. Furthermore, due to the use of `` `[[` ``, the loop can also be applied on lists.

---

**Example 8.6** *Let x be a list and **f** be a function. The following code generates the same result as **Map**(**f**, x):*

```
n <- length(x)
ret <- vector("list", n)  # a new list of length `n`
for (i in seq_len(n))
    ret[[i]] <- f(x[[i]])
```

**Example 8.7** *Let x and y be two lists and **f** be a function. Here is the most basic version of **Map**(**f**, x, y).*

```
nx <- length(x)
ny <- length(y)
n <- max(nx, ny)
ret <- vector("list", n)
for (i in seq_len(n))
    ret[[i]] <- f(x[[((i-1)%%nx)+1]], y[[((i-1)%%ny)+1]])
```

*Note that x and y might be of different lengths. Feel free to upgrade this code by adding a warning like* the longer argument is not a multiple of the length of the shorter one. *Also, rewrite it without using the modulo operator, `` `%%` ``.*

### 8.3.3 break and next

**break** can be used to escape the current loop. **next** skips the remaining expressions and advances to the next iteration (where the testing of the logical condition occurs). Here is a rather random example:

```
x <- c(10, 0.03, 0.04, 1, 0.001, 0.05)
s <- 0
for (e in x) {
    if (e > 0.1)  # skip the current element if it is greater than 0.1
        next

    print(e)
    if (e < 0.01)  # stop at the first element less than 0.01
        break

    s <- s + e
}
## [1] 0.03
## [1] 0.04
## [1] 0.001
print(s)
## [1] 0.07
```

We have used a frequently occurring design pattern:

```
for (e in x) {
    if (condition)
        next

    many_statements...
}
```

which is equivalent to:

```
for (e in x) {
    if (!condition) {
        many_statements...
    }
}
```

but which avoids introducing a nested block of expressions.

---

**Note** (*) There is a third loop type,

```
repeat
    expression
```

which is a shorthand for:

```
while (TRUE)
    expression
```

i.e., it is a possibly infinite loop. Such constructs are invaluable when expressing situations like *repeat*-something-*until*-success, e.g., when we want to execute a command at least once.

```
i <- 1
repeat {  # while (TRUE)
    # simulate dice casting until we throw "1"
    if (runif(1) < 1/6) break  # repeat until this
    i <- i+1  # how many times until success
}
print(i)
## [1] 6
```

---

**Exercise 8.8** *What is wrong with the following code?*

```
j <- 1
while (j <= 10) {
    if (j %% 2 == 0) next
    print(j)
    j <- j + 1
}
```

**Exercise 8.9** *What about this one?*

```
j <- 1
while (j <= 10);
    j <- j + 1
```

### 8.3.4 `return`

`return`, when called from within a function, immediately yields a specified value and goes back to the caller. For example, here is a simple recursive function that flattens a given list:

```
my_unlist <- function(x)
{
    if (is.atomic(x))
        return(x)

    # so if we are here, x is definitely not atomic
    out <- NULL
    for (e in x)
        out <- c(out, my_unlist(e))
```

```
    out  # or return(out); not necessary as it's the last expression
}

my_unlist(list(list(list(1, 2), 3), list(4, list(5, list(6, 7:10)))))
##  [1]  1  2  3  4  5  6  7  8  9 10
```

`return` is a function: the round brackets are obligatory.

### 8.3.5 Time and space complexity of algorithms (*)

Analysis of algorithms can give us a rough estimate of their run time or memory consumption as a function of the input problem size, especially for *big* data (e.g., [15, 44]). In scientific computing and data science, we often deal with vectors (sequences) or matrices/data frames (tabular data). Therefore, we might be interested in determining how many *primitive operations* need to be performed as a function of their length $n$ or the number of rows $n$ and columns $m$, respectively.

The $O$ (Big-Oh) notation can express the upper bounds for time/resource consumption in asymptotic cases. For instance, we say that the time complexity is $O(n^2)$, if for large $n$, the number of operations to perform or memory cells to use will be proportional to *at most* the square of the vector size (more precisely, there exists $m$ and $C > 0$ such that for all $n > m$, the number of operations is $\leq Cn^2$).

Therefore, if we have two algorithms that solve the same task, one that has $O(n^2)$ time complexity, and other of $O(n^3)$, it is better to choose the former. For large problem sizes, we expect it to be faster. Moreover, whether time grows proportionally to $\log n$, $\sqrt{n}$, $n$, $n \log n$, $n^2$, $n^3$, or $2^n$, can be informative in predicting how big the data can be if we have a fixed deadline or not enough space left on the disk.

**Exercise 8.10** *The `hclust` function determines a hierarchical clustering of a dataset. It is fed with an object that stores the distance between all the pairs of input points. There are $n(n-1)/2$ (i.e., $O(n^2)$) unique point pairs for any given $n$. One numeric scalar (`double` type) takes 8 bytes of storage. If you have 16 GiB of RAM, what is the largest dataset that you can process on your machine using this function?*

Oftentimes, we can learn about the time or memory complexity of the functions we use from their documentation; see, e.g., `help("findInterval")`.

**Example 8.11** *A course in data structures in algorithms, which this one is not, will give us plenty of opportunities to implement many algorithms ourselves. This way, we can gain a lot of insights and intuitions. For instance, here is an $O(n)$-time algorithm:*

```
for (i in seq_len(n))
    expression
```

*and this one runs in $O(n^2)$ time:*

```
for (i in seq_len(n))
```

```
    for (j in seq_len(n))
        expression
```

*as long as, of course, the* `expression` *is rather primitive (e.g., operations on scalar variables).*

*R is a very expressive language. Hence, complex and lengthy operations can look pretty innocent. After all, it is a glue language for rapid prototyping. For example:*

```
for (i in seq_len(n))
    for (j in seq_len(n))
        z <- z + x[[i]] + y[[j]]
```

*can be seen as running in $O(n^3)$ time if each element in the lists x and y as well as z itself are atomic vectors of length $n$. Similarly,*

```
Map(mean, x)
```

*is $O(n^2)$ if x is a list of n atomic vectors, each of length $n$.*

---

**Note** A quite common statistical scenario involves generating a data buffer of a fixed size:

```
ret <- c()  # start with an empty vector
for (i in seq_len(n))
    ret[[i]] <- generate_data(i)  # here: ret[[length(ret)+1]] <- ...
```

This notation, however, involves growing the `ret` array in each iteration. Luckily, since R version 3.4.0, each such size extension has amortised $O(1)$ time as some more memory is internally reserved for its prospective growth (dynamic arrays; see, e.g., Chapter 17 of [15]).

However, it is better to preallocate the output vector of the desired final size. We can construct vectors of specific lengths and types in an efficient way (more efficient than with **rep**) by calling:

```
numeric(3)
## [1] 0 0 0
numeric(0)
## numeric(0)
logical(5)
## [1] FALSE FALSE FALSE FALSE FALSE
character(2)
## [1] "" ""
vector("numeric", 8)
## [1] 0 0 0 0 0 0 0 0
vector("list", 2)
## [[1]]
## NULL
```

```
##
## [[2]]
## NULL
```

**Note** Not all data fit into memory, but it does not mean that we should start installing Apache Hadoop and Spark immediately. Some datasets can be processed chunk by chunk. R enables data stream handling (some can be of infinite length) through file connections. For example:

```
f <- file("https://github.com/gagolews/teaching-data/raw/master/README.md",
    open="r")  # a big file, the biggest file ever
i <- 0
while (TRUE) {
    few_lines <- readLines(f, n=4)  # reads only four lines at a time
    if (length(few_lines) == 0) break
    i <- i + length(few_lines)
}
close(f)
print(i)  # the number of lines
## [1] 90
```

Many functions support reading from/writing to already established connections of different types, e.g., **file**, **gzfile**, **textConnection**, batch by batch. A common scenario involves reading a very large CSV, JSON, or XML file only by thousands of lines/records at a time, parsing and cleansing them, and exporting them to SQL databases (which we will exercise in Chapter 12).

## 8.4 Exercises

From now on, we must stay alert. Many, if not all, of the undermentioned tasks, can still be implemented without the explicit use of the R loops but based only on the operations covered in the previous chapters. If this is the case, try composing both the looped and loop-free versions. Use **proc.time** to compare their run times[3].

**Exercise 8.12** *Answer the following questions.*

- *Let x be a numeric vector. When does "`if(x > 0) ...`" yield a warning? When does it give an error? How to guard ourselves against them?*
- *What is a dangling **else**?*

[3] It might be the case that a **for**-based solution is faster (e.g., for larger objects) because of the use of a more efficient algorithm. Such cases will benefit from a rewrite in C or C++ (Chapter 14).

- *What happens if you put* `if` *as the last expression in a curly braces block within a function's body?*
- *Why do we say that* `` `&&` `` *and* `` `||` `` *are lazy? What are their use cases?*
- *What is the difference between* `` `&&` `` *and* `` `&` ``*?*
- *Can* `while` *always be replaced with* `for`*? What about the other way around?*
- *What is wrong with "*`return (1+2)*3`*"?*

**Exercise 8.13** *Verify which of the following can be safely used as logical conditions in* `if` *statements. If that is not the case for all* `x`*,* `y`*, …, determine the additional conditions that must be imposed to make them valid.*

- `x == 0`,
- `x[y] > 0`,
- `any(x>0)`,
- `match(x, y)`,
- `any(x %in% y)`.

**Exercise 8.14** *What can go wrong in the following code chunk, depending on the type and form of* `x`*? Consider as many scenarios as possible.*

```
count <- 0
for (i in 1:length(x))
    if (x[i] > 0)
        count <- count + 1
```

**Exercise 8.15** *Implement* `shift_left(x, n)` *and* `shift_right(x, n)`*. The former function gets rid of the first* `n` *observations in* `x` *and adds* `n` *missing values at the end of the resulting vector, e.g.,* `shift_left(c(1, 2, 3, 4, 5), 2)` *is* `c(3, 4, 5, NA, NA)`*. On the other hand,* `shift_right(c(1, 2, 3, 4, 5), 2)` *is* `c(NA, NA, 1, 2, 3)`*.*

**Exercise 8.16** *Implement your version of* `diff`*.*

**Exercise 8.17** *Write a function that determines the longest ascending trend in a given numeric vector, i.e., the length of the longest subsequence of consecutive increasing elements. For example, the input* `c(1, 2, 3, 2, 1, 2, 3, 4, 3)` *should yield* 4.

**Exercise 8.18** *Implement the functions that round down and round up each element in a numeric vector to a number of decimal digits.*

This concludes the first part of this magnificent book.

# Part II

# Deeper

# 9

# *Designing functions*

In Chapter 7, we learnt how to compose simple functions. This skill is vital to enforcing the good development practice of avoiding code repetition: running the same command sequence on different data.

The current chapter is devoted to designing reusable methods so that they are easier to use, test, and maintain. We also provide more technical details about functions. They were not of the highest importance during our first exposure to this topic but are crucial to our better understanding of how R works.

## 9.1 Managing data flow

A function, most of the time, can and should be treated as a black box. Its callers do not have to care what it hides inside. After all, they are supposed to *use* it. Given some *inputs*, they expect well-defined *outputs* that are explained in detail in the function's manual.

### 9.1.1 Checking input data integrity and argument handling

A function takes R objects of any kind as arguments, but it does not mean feeding it with everything is healthy for its guts. When designing functions, it is best to handle the inputs in a manner similar to base R's behaviour. This will make our contributions easier to work with. Lamentably, base functions frequently do not process arguments of a similar *kind* fully consistently. Such variability might be due to many reasons and, in essence, is not necessarily bad. Usually, there might be many possible behaviours and choosing one over another would make a few users unhappy anyway. Some choices might not be optimal, but they are for historical compatibility (e.g., with S). Of course, it might also happen that something is poorly designed or there is a bug (but the likelihood is low). This is why we should rather keep our vocabulary restricted. Even if there are exceptions to the general rules, with fewer functions, they are easier to remember. We advocate for such minimalism in this book.

Consider the following case study, illustrating that even the extremely simple scenario dealing with a single positive integer is not necessarily straightforward.

**Exercise 9.1** *In mathematical notation, we usually denote the number of objects in a collection by the famous "n". It is implicitly assumed that such n is a single natural number (albeit whether* *this includes 0 or not should be specified at some point). The functions* `runif`, `sample`, `seq`, `rep`, `strrep`, *and* `class::knn` *take it as arguments. Nonetheless, nothing stops us from trying to challenge them by passing:*

- `2.5`, `-1`, `0`, `1-1e-16` *(non-positive numbers, non-integers);*
- `NA_real_`, `Inf` *(not finite);*
- `1:5` *(not of length 1; after all, there are no scalars in R);*
- `numeric(0)` *(an empty vector);*
- `TRUE`, `NA`, `c(TRUE, FALSE, NA)`, `"1"`, `c("1", "2", "3")` *(non-numeric, but coercible to);*
- `list(1)`, `list(1, 2, 3)`, `list(1:3, 4)` *(non-atomic);*
- `"Spanish Inquisition"` *(unexpected nonsense);*
- `as.matrix(1)`, `factor(7)`, `factor(c(3, 4, 2, 3))`, *etc. (compound types;* *Chapter 10**).*

*Read the aforementioned functions' reference manuals and call them on different inputs. Notice how differently they handle such atypical arguments.*

Sometimes we will rely on other functions to check data integrity for us.

**Example 9.2** *Consider a function that generates n pseudorandom numbers from the unit interval rounded to d decimal digits. We strongly believe, or at least hope (the good faith and high competence assumption), that its author knew what he was doing when he wrote:*

```
round_rand <- function(n, d)
{
    x <- runif(n)  # runif will check if `n` makes sense
    round(x, d)    # round will determine the appropriateness of `d`
}
```

*What constitutes correct n and d and how the function behaves when not provided with positive integers is determined by the two underlying functions,* `runif` *and* `round`*:*

```
round_rand(4, 1)  # the expected use case
## [1] 0.3 0.8 0.4 0.9
round_rand(4.8, 1.9)  # 4, 2
## [1] 0.94 0.05 0.53 0.89
round_rand(4, NA)
## [1] NA NA NA NA
round_rand(0, 1)
## numeric(0)
```

Some design choices can be defended if they are well thought out and adequately documented. Certain programmers will opt for high uniformity/compatibility across numerous tools, as there are cases where diversity does more good than harm.

Our functions might become part of a more complicated data flow pipeline. Let's consider what happens when another procedure generates a value that we did not expect (due to a bug or because we did not study its manual). The problem arises when this *unthinkable* value is passed to our function. In our case, this would correspond to the said *n*'s or *d*'s being determined programmatically.

**Example 9.3** *Continuing the previous example, the following might be somewhat challenging with regard to our being flexible and open-minded:*

```
round_rand(c(100, 42, 63, 30), 1)  # n=length(c(...))
## [1] 0.7 0.6 0.1 0.9
round_rand("4", 1)  # n=as.numeric("4")
## [1] 0.2 0.0 0.3 1.0
```

*Sure, it is convenient. Nevertheless, it might lead to problems that are hard to diagnose.*

*Also, note the not so informative error messages in cases like:*

```
round_rand(NA, 1)
## Error in `runif()`: ! invalid arguments
round_rand(4, "1")
## Error in `round()`: ! non-numeric argument to mathematical function
```

*Defensive design* strategies are always welcome, especially if they lead to constructive error messages.

---

**Important** **`stopifnot`** gives a convenient means to assert the enjoyment of our expectations about a function's arguments (or intermediate values). A call to **`stopifnot`**`(cond1, cond2, ...)` is more or less equivalent to:

```
if (!(is.logical(cond1) && !any(is.na(cond1)) && all(cond1)))
    stop("`cond1` are not all TRUE")
if (!(is.logical(cond2) && !any(is.na(cond2)) && all(cond2)))
    stop("`cond2` are not all TRUE")
...
```

Thus, if all the elements in the given logical vectors are `TRUE`, nothing happens. We can move on with certainty.

---

**Example 9.4** *We can rewrite the preceding function as:*

```
round_rand2 <- function(n, d)
{
    stopifnot(
        is.numeric(n), length(n) == 1,
        is.finite(n), n > 0, n == floor(n),
        is.numeric(d), length(d) == 1,
        is.finite(d), d > 0, d == floor(d)
```

```
    )
    x <- runif(n)
    round(x, d)
}

round_rand2(5, 1)
## [1] 0.7 0.7 0.5 0.6 0.3
round_rand2(5.4, 1)
## Error in `round_rand2()`: ! n == floor(n) is not TRUE
round_rand2(5, "1")
## Error in `round_rand2()`: ! is.numeric(d) is not TRUE
```

*It is the strictest test for "a single positive integer" possible. In the case of any violation of the underlying condition, we get a very informative error message.*

**Example 9.5** *At other times, we might be interested in a more liberal yet still foolproof argument checking like:*

```
if (!is.numeric(n))
    n <- as.numeric(n)
if (length(n) > 1) {
    warning("only the first element will be used")
    n <- n[1]
}
n <- floor(n)
stopifnot(is.finite(n), n > 0)
```

*This way, `"4"` and `c(4.9, 100)` will all be accepted as 4*[1].

We see that there is always a tension between being generous/flexible and precise/restrictive. Also, because of their particular use cases, for certain functions, it will be better to behave differently from the others. Excessive uniformity is as bad as chaos. We are always expected to rely on common sense. Let's not be boring bureaucrats.

Still, it is our duty to be explicit about all the assumptions we make or exceptions we tolerate (by writing comprehensive documentation; see Section 9.2.2).

---

**Note** (*) Example exercises related to improving the consistency of base R's argument handling in different domains include the **vctrs** and **stringx** packages. Can these contributions be justified?

---

**Exercise 9.6** *Reflect on how you would respond to miscellaneous boundary cases in the following scenarios (and how base R and other packages or languages you know deal with them):*

- *a vectorised mathematical function (empty vector? non-numeric input? what if it is equipped with the `names` attribute? what if it has other ones?);*

[1] We rely on the S3 generics `is.numeric` and `as.numeric` here; see Section 10.2.3.

- *an aggregation function (what about missing values? empty vector?);*
- *a function vectorised with regard to two arguments (elementwise vectorisation? recycling rule? only scalar vs vector, or vector vs vector of the same length allowed? what if one argument is a row vector and the other is a column vector?);*
- *a function vectorised with respect to all arguments (really all? maybe some exceptions are necessary?);*
- *a function vectorised with respect to the first argument but not the second (why such a restriction? when?).*

*Find a few functions that match each case.*

### 9.1.2 Putting outputs into context

Our functions do not exist in a vacuum. We should put them into a much broader context: how can they be combined with other tools? As a general rule, we ought to generate outputs of a *predictable* kind. This way, we can easily deduce what will happen in the code chunks that utilise them.

**Example 9.7** *Some base R functions do not adhere to this rule for the sake of (questionable) users' convenience. We will meet a few of them in Chapter 11 and Chapter 12. In particular,* **`sapply`** *and the underlying* **`simplify2array`***, can return a list, an atomic vector, or a matrix.*

```
simplify2array(list(1, 3:4))    # list
## [[1]]
## [1] 1
##
## [[2]]
## [1] 3 4
simplify2array(list(1, 3))      # vector
## [1] 1 3
simplify2array(list(1:2, 3:4))  # matrix
##      [,1] [,2]
## [1,]    1    3
## [2,]    2    4
```

*Further, the index operator with `drop=TRUE`, which is the default, may output an atomic vector. However, it may as well yield a matrix or a data frame.*

```
(A <- matrix(1:6, nrow=3))  # an example matrix
##      [,1] [,2]
## [1,]    1    4
## [2,]    2    5
## [3,]    3    6
A[1, ]    # vector
## [1] 1 4
A[1:2, ]  # matrix
##      [,1] [,2]
```

```
## [1,]    1    4
## [2,]    2    5
A[1, , drop=FALSE]  # matrix with 1 row
##      [,1] [,2]
## [1,]    1    4
```

We proclaim that, if there are many options, the default behaviour should be to return an object of the most generic kind possible, even when it is not the most convenient form. Then, either:

- we equip the function with a further argument which must be explicitly set if we *really* want to simplify the output, or
- we ask the user to call a simplifier explicitly after the function call; in this case, if the simplifier cannot neaten the object, it should probably fail by issuing an error or at least try to apply some brute force solution (e.g., "fill the gaps" somehow itself, preferably with a warning).

For instance:

```
as.numeric(A[1:2, ])  # always returns a vector
## [1] 1 2 4 5
stringi::stri_list2matrix(list(1, 3:4))  # fills the gaps with NAs
##      [,1] [,2]
## [1,] "1"  "3"
## [2,] NA   "4"
```

Ideally, a function is expected to perform one (and only one) well-defined task. If it tends to generate objects of different kinds, depending on the arguments provided, it might be better to compose two or more separate procedures instead.

**Exercise 9.8** *Functions such as* **`rep`**, **`seq`**, *and* **`sample`** *do not perform a single task. Or do they?*

---

**Note** (*) In a purely functional programming language, we can assume the so-called *referential transparency*: a call to a *pure function* can always be replaced with the *value* it generates. If this is true, then for the same set of argument values, the output is always the same. Furthermore, there are no side effects. In R, it is not exactly the case:

- a call can introduce/modify/delete variables in other environments (see Chapter 16), e.g., the state of the random number generator,
- due to lazy evaluation, functions are free to interpret the argument *forms* (passed *expressions*, i.e., not only: *values*) however they like; see Section 9.4.7, Section 12.3.9, and Section 17.5,
- printing, plotting, file writing, and database access have apparent consequences with regard to the state of certain external devices or resources.

---

**Important** Each function *must* return a value. However, in several instances (e.g., plotting, printing) this does not necessarily make sense. In such a case, we may consider returning `invisible(NULL)`, a `NULL` whose *first* printing will be suppressed. Compare the following:

```
f <- function() invisible(NULL)
f()  # printing suppressed
x <- f()  # by the way, assignment also returns an invisible value
print(x)  # no longer invisible
## NULL
```

## 9.2 Organising and maintaining functions

### 9.2.1 Function libraries

Definitions of frequently-used functions or datasets can be emplaced in separate source files (.R extension) for further reference. Such code banks can be executed by calling:

```
source("path_to_file.R")
```

**Exercise 9.9** *Create a source file (script) named `mylib.R`, where you define a function called **`nlargest`** which returns a few largest elements in a given atomic vector. From within another script, call **`source("mylib.R")`**; note that relative paths refer to the current working directory (Section 2.1.6). Then, write a few lines of code where you test **`nlargest`** on some example inputs.*

### 9.2.2 Writing R packages (*)

When a function library grows substantially, there is a need for equipping its contents with the relevant help pages, or we wish to rely on compiled code, turning it into an R package might be worth considering.

**Important** Packages can be written only for ourselves or a small team's purpose. We *do not have to* publish them on CRAN[2]. Have mercy on the busy CRAN maintainers and do not contribute to the information overload unless we have come up with something potentially of service[3] for other R users. Packages can always be hosted on and installed from GitLab or GitHub.

[2] Always consult the CRAN Repository Policy at https://cran.r-project.org/web/packages/policies.html.

---

### Package structure (*)

A *source package* is a directory containing the following special files and subdirectories:

- `DESCRIPTION` – a text file that gives the name of the project, its version, authors, dependencies on other packages, license, etc.;
- `NAMESPACE` – a text file containing directives stating which objects are available to the package users and which names are imported from other packages;
- `R` – a directory with R scripts (.R files), which define, e.g., functions, example datasets, etc.;
- `man` – a directory with R documentation files (.Rd), describing at least all the exported objects (Section 9.2.2);
- `src` – optional; compiled code (Chapter 14);
- `tests` – optional; tests to run on the package check (Section 9.2.4).

See Section 1 of *Writing R Extensions* [66] for more details and other options. We do not need to repeat the information from the official manual as all readers can read it themselves.

**Exercise 9.10** *Inspect the source code of the example package available for download from https://github.com/gagolews/rpackagedemo.*

### Building and installing (*)

Recall from Section 7.3.1 that a source package can be built and installed by calling:

```
install.packages("pkg_directory", repos=NULL, type="source")
```

Then it can be used as any other R package (Section 7.3.1). In particular, it can be loaded and attached to the search path (Section 16.2.6) via a call to:

```
library("pkg")
```

All the exported objects mentioned in its `NAMESPACE` file are now available to the user; see also Section 16.3.5.

**Exercise 9.11** *Create a package* **`mypkg`** *with the solutions to the exercises listed in the previous chapter. When in doubt, refer to the official manual [66].*

---

[3] Let's make it less about ourselves and more about the community. Developing expertise in any complex area takes years of hard work. In the meantime, we can help open-source projects by spreading the good word about them, submitting bug fixes, extending documentation, supporting other users through their journey, etc.

**Note** (*) The building and installing of packages also be done from the command line:

```
R CMD build pkg_directory  # creates a distributable source tarball (.tar.gz)
R CMD INSTALL pkg-version.tar.gz
R CMD INSTALL --build pkg_directory
```

Also, some users may benefit from authoring Makefiles that help automate the processes of building, testing, checking, etc.

### Documenting (*)

Documenting functions and commenting code thoroughly is critical, even if we just write for ourselves. Most programmers sooner or later will notice that they find it hard to determine what a piece of code is doing after they took a break from it. In some sense, we always communicate with external audiences, which includes our future selves.

The help system is one of the stronger assets of the R environment. By far, we most likely have interacted with many documentation pages and got a general idea of what constitutes an informative documentation piece.

From the technical side, documentation (.Rd) files are located in the `man` subdirectory of a source package. All exported objects (e.g., functions) should be described clearly. Additional topics can be covered, too. Documentation files use a LaTeX-like syntax, which looks obscure to an untrained eye. The relevant commands are explained in very detail in Section 2 of [66]. During the package installation, the .Rd files are converted to various output formats, e.g., HTML or plain text, and displayed on a call to the well-known **`help`** function.

**Note** The process of writing .Rd files by hand might be tedious, especially keeping track of the changes to the `\usage` and `\arguments` commands. Rarely do we recommend using external packages for base R facilities are usually sufficient. But **`roxygen2`** might be worth a try because it makes the developers' lives easier. Most importantly, it allows the documentation to be specified alongside the functions' definitions, which is much more natural.

**Exercise 9.12** *Add a few manual pages to your example R package.*

## 9.2.3 Writing standalone programs (**)

Section 7.3.2 mentioned how to call external programs using **`system2`**. On UNIX-like operating systems, it is easy to turn our R scripts into standalone tools that can be run from the terminal: we have already touched upon this topic in Section 1.2.3. As an addition, the **`commandArgs`** function returns the list of arguments passed from the command line to our script in the form of a character vector. What we do with them is up to us. Moreover, **q** can terminate a script, yielding any integer return code. By convention, anything other than 0 indicates an error.

**Example 9.13** *Say we have the following script named* `testfile` *in the current directory:*

```
#!/bin/env -S Rscript --vanilla

argv <- commandArgs(trailingOnly=TRUE)
cat("commandArgs:\n")
print(argv)

if (length(argv) == 0) {
    cat("Usage: testfiles file1 file2 ...\n")
    q(save="no", status=1)  # exit with code 1
}

if (!all(file.exists(argv))) {
    cat("Some files do not exist.\n")
    q(save="no", status=2)  # exit with code 2
}

cat("All files exist.\n")

# exits with code 0 (success)
```

*Example interactions with this program from a UNIX-like shell (***bash***):*

```
chmod u+x testfiles  # add permission to execute
./testfiles
## commandArgs:
## character(0)
## Usage: testfiles file1 file2 ...
./testfiles spanish_inquisition
## commandArgs:
## [1] "spanish_inquisition"
## Some files do not exist.
./testfiles spam bacon eggs spam
## commandArgs:
## [1] "spam"  "bacon" "eggs"  "spam"
## All files exist.
```

**stdin**, **stdout**, and **stderr** represent the always-open connections mapped to the standard input ("keyboard"), as well as the normal and error output. They can be read from or written to using functions such as **scan** or **cat**.

During run time, we can redirect **stdout** and **stderr** to different files or even strings using **sink**.

### 9.2.4 Assuring quality code

Below we mention some good development practices related to maintaining quality code. This is an important topic, but writing about them is tedious to the same extent that reading about them is dull. It is the more *artistic* part of software engineering as such heuristics are learnt best by observing and mimicking what more skilled programmers are doing (the coming exercises aim to make up for our not having them at hand at the moment).

#### Managing changes and working collaboratively

We recommend employing a source code version control system, such as **`git`**, to keep track of the changes made to the software.

---

**Note** It is worth investing time and effort to learn how to use **`git`** from the command line; see https://git-scm.com/doc.

---

There are a few hosting providers for **`git`** repositories, with GitLab and GitHub being particularly popular among open-source software developers. They support working collaboratively on the projects and are equipped with additional tools for reporting bugs, suggesting feature requests, etc.

**Exercise 9.14** *Find source code of your favourite R packages or other projects. Explore the corresponding repositories, feature trackers, wikis, discussion boards, etc. Each community is different and is governed by varied, sometimes contrasting guidelines; after all, we come from all corners of the world.*

#### Test-driven development and continuous integration

It is often hygienic to include some principles of *test-driven development*.

**Exercise 9.15** *Assume that, for some reason, we were asked to compose a function to compute the root mean square (quadratic mean) of a given numeric vector. Before implementing the actual routine, we need to reflect upon what we want to achieve, especially how we want our function to behave in certain boundary cases.*

***`stopifnot`*** *gives simple means to ensure that a given assertion is fulfilled. If that is the case, it will move forward without fuss. Say we have come up with the following set of* expectations:

```
stopifnot(all.equal(rms(1), 1))
stopifnot(all.equal(rms(1:100), 58.16786054171151931769))
stopifnot(all.equal(rms(rep(pi, 10)), pi))
stopifnot(all.equal(rms(numeric(0)), 0))
```

*Write a function* **`rms`** *that fulfils these assertions.*

**Exercise 9.16** *Implement your version of the* **`sample`** *function (assuming* `replace=TRUE`*), using calls to* **`runif`**. *Start by writing a few unit tests.*

A couple of R packages support writing and executing unit tests, including **`testthat`**,

`tinytest`, `RUnit`, or `realtest`. However, in the most typical use cases, relying on `stopifnot` is powerful enough.

**Exercise 9.17** *(*) Consult the* Writing R Extensions *manual [66] about where and how to include unit tests in your example package.*

---

**Note** (*) R can check a couple of code quality areas: running `R CMD check pkg_directory` from the command line (preferably using the most recent version of the environment) will suggest several improvements. Also, it is possible to use various continuous integration techniques that are automatically triggered when pushing changes to our software repositories; see GitLab CI or GitHub Actions. For instance, we can run a package build, install, and check process on every `git` commit. Also, CRAN deploys continuous integration services, including checking the package on various platforms.

---

### Debugging

R has an interactive debugger; see the `browser` function and Section 9 of [70] for more details. Some IDEs (e.g., `RStudio`) also support this feature; see their corresponding documentation.

However, for all his life, the current author has been debugging his programs primarily by manually printing the state of the suspicious variables (`printf` and the like) in different code areas. This is old-school but uncannily efficient.

### Profiling

Typically, a program spends relatively long time executing only a small portion of code. The `Rprof` function can be a helpful tool to identify which chunks might need a rewrite, for instance, using a compiled language (Chapter 14). Please remember, though, that bottlenecks are not only formed by using algorithms with high computational complexity, but also data input and output (such as reading files from disk, printing messages on the console, querying Web APIs, etc.).

## 9.3 Special functions: Syntactic sugar

Some functions, such as `` `*` ``, are somewhat special. They can be referred to using *infix* syntax which, for obvious reasons, most of us accepted as the default one. However, we will later reveal, amongst others, that "`5 * 9`" reduces to an ordinary function *call*:

```
`*`(5, 9)  # a call to `*` with two arguments, equivalent to 5 * 9
## [1] 45
```

### 9.3.1 Backticks

In Section 2.2, we mentioned that via `` `<-` `` we can assign *syntactically valid names* to our objects. Most identifiers comprised of letters, digits, dots, and underscores can be used directly in R code. Nevertheless, it is possible to label our objects however we like. Not syntactically valid (nonstandard) identifiers just need to be enclosed in backticks (back quotes, grave accents):

```
`42 a quite peculiar name :O` <- c(a=1, `b c`=2, `42`=3, `!`=4)
1/(1+exp(-`42 a quite peculiar name :O`))
##       a       b c      42       !
## 0.73106 0.88080 0.95257 0.98201
```

Such names are less convenient but backticks allow us to refer to them in *any* setting.

### 9.3.2 Dollar, `$` (*)

The dollar operator, `$`, can be an alternative accessor to a single element in a named list[4]. If a `label` is a syntactically valid name, then `x$label` does the same job as `x[["label"]]` (saving five keystrokes: such a burden!).

```
x <- list(spam="a", eggs="b", `eggs and spam`="c", best.spam.ever="d")
x$eggs
## [1] "b"
x$best.spam.ever  # recall that a dot has no special meaning in most contexts
## [1] "d"
```

Nonstandard names must still be enclosed in backticks (or quotes):

```
x$`eggs and spam`  # x[["eggs and spam"]] is okay as usual
## [1] "c"
```

We are minimalist by design here. Thence, we will avoid this operator for it does not increase the expressive power of our function repertoire. Also, it does not work on atomic vectors nor matrices. Furthermore, it does not support names that are generated programmatically:

```
what <- "spam"
x$what  # the same as x[["what"]]; we do not want this
## NULL
x[[what]]  # works fine
## [1] "a"
```

The support for the *partial matching* of element names has been added to provide users working in interactive programming sessions with some relief in the case where they find typing the whole label daunting:

[4] And hence also in data frames.

```
x$s
## Warning in x$s: partial match of 's' to 'spam'
## [1] "a"
```

Compare:

```
x[["s"]]  # no warning here...
## NULL
x[["s", exact=FALSE]]
## [1] "a"
```

Partial matching is generally a rubbishy programming practice. The result depends on the names of other items in `x` (which might change later) and can decrease code readability. The only reason why we obtained a warning message was because this book enforces the `options(warnPartialMatchDollar=TRUE)` setting, which, sadly, is not the default.

Note the behaviour on an ambiguous partial match:

```
x$egg  # ambiguous resolution
## NULL
```

as well as on an element assignment:

```
x$s <- "e"
str(x)
## List of 5
##  $ spam          : chr "a"
##  $ eggs          : chr "b"
##  $ eggs and spam : chr "c"
##  $ best.spam.ever: chr "d"
##  $ s             : chr "e"
```

It did not modify `spam` but added a new element, `s`. Confusing? Just let's not use the dollar operator and we will have one less thing to worry about.

### 9.3.3 Curly braces, `{`

A block of statements grouped with curly braces, `{`, corresponds to a function call. When we write:

```
{
    print(TRUE)
    cat("two")
    3
}
## [1] TRUE
## two
## [1] 3
```

The parser translates it to a call to:

```
`{`(print(TRUE), cat("two"), 3)
## [1] TRUE
## two
## [1] 3
```

When it is executed, every argument to `` `{` `` is evaluated one by one. Then, the last value is returned as the result of that call.

### 9.3.4 `` `if` ``

**`if`** is a function too. As mentioned in Section 8.1, it returns the value corresponding to the expression that is evaluated conditionally. Hence, we may write:

```
if (runif(1) < 0.5) "head" else "tail"
## [1] "head"
```

but also:

```
`if`(runif(1) < 0.5, "head", "tail")
## [1] "head"
```

---

**Note** A call like `` `if`(test, what_if_true, what_if_false) `` can only work correctly because of the lazy evaluation of function arguments; see Chapter 17.

---

On a side note, **`while`**, **`for`**, **`repeat`** can also be called that way, but they return **`invisible`**`(NULL)`.

### 9.3.5 Operators are functions

#### Calling built-in operators as functions

Every arithmetic, logical, and relational operator is translated to a call to the corresponding function. For instance:

```
`<`(`+`(`*`(`-`(3), 4)), 5)  # 2+(-3)*4 < 5
## [1] TRUE
```

Also, `x[i]` is equivalent to `` `[`(x, i) `` and `x[[i]]` maps to `` `[[`(x, i) ``.

Knowing the above will not only enable us to manipulate unevaluated R code (Chapter 15) or access the corresponding manual pages (see, e.g., **`help`**`("[")`), but also verbalise certain operations more concisely. For instance:

```
x <- list(1:5, 11:17, 21:23)
unlist(Map(`[`, x, 1))  # 1 is a further argument passed to `[`
## [1]  1 11 21
```

is equivalent to a call to Map(function(e) e[1], x).

---

**Note** Unsurprisingly, the assignment operator, `<-`, is also a function. It returns the assigned value invisibly. `<-` binds right to left (compare help("Syntax")). Thus, the expression “a <- b <- 1” assigns 1 to both b and a. It is equivalent to `<-`("a", `<-`("b", 1)) and `<-`("b", 1) returns 1.

Owing to the pass-by-value-like semantics (Section 9.4.1), we can also expect that we will be assigning a *copy*[5] of the value on the right side of the operator (with the exception of environments; Chapter 16).

```
x <- 1:6
y <- x  # makes a copy (but delayed, on demand, for performance reasons)
y[c(TRUE, FALSE)] <- NA_real_  # modify every second element
print(y)
## [1] NA  2 NA  4 NA  6
print(x)  # state of x has not changed; x and y are different objects
## [1] 1 2 3 4 5 6
```

However, with no harm to the semantics, the copying of x is postponed until absolutely necessary (Section 16.1.4). This is efficient both time- and memory-wisely.

---

#### Defining binary operators

We can also introduce custom binary operators named like `%myopname%`:

```
`%:)%` <- function(e1, e2) (e1+e2)/2
5 %:)% 1:10
##  [1] 3.0 3.5 4.0 4.5 5.0 5.5 6.0 6.5 7.0 7.5
```

Recall that `%%`, `%/%`, and `%in%` are built-in operators denoting division remainder, integer division, and testing for set inclusion. Also, in Chapter 11, we will learn about `%*%`, which implements matrix multiplication.

---

**Note** Chapter 10 notes that most existing operators can be overloaded for objects of custom types.

---

### 9.3.6 Replacement functions

Functions generally do not change the state of their arguments. However, there is some syntactic sugar that permits us to replace objects or their parts with new content. We call them *replacement functions*.

For instance, three of the following calls *replace* the input x with its modified version:

[5] This is especially worth pointing out to Python (amongst others) programmers, where the example assignment below would mean that x and y both refer to the same (shared) object in the computer’s memory.

```
x <- 1:5  # example input
x[3] <- 0  # replace the third element with 0
length(x) <- 7  # "replace" length
names(x) <- LETTERS[seq_along(x)]  # replace the names attribute
print(x)  # `x` is now different
##  A  B  C  D  E  F  G
##  1  2  0  4  5 NA NA
```

### Creating replacement functions

A *replacement function* is a mapping named like `` `f<-` `` with at least two parameters:

- `x` (the object to be modified),
- `...` (possible further arguments),
- `value` (as the last parameter; the object on the right-hand side of the `` `<-` `` operator).

We will most often interact with *existing* replacement functions, not create our own ones. But knowing how to do the latter is vital to understanding this language feature. For example:

```
`add<-` <- function(x, where=TRUE, value)
{
    x[where] <- x[where] + value
    x  # the modified object that will replace the original one
}
```

This function aims to add a `value` to a subset of the input vector `x` (by default, to each element therein). Then, it returns its altered version.

```
y <- 1:5           # example vector
add(y) <- 10       # calls y <- `add<-`(y, value=10)
print(y)           # y has changed
## [1] 11 12 13 14 15
add(y, 3) <- 1000  # calls y <- `add<-`(y, 3, value=1000)
print(y)           # y has changed again
## [1]   11   12 1013   14   15
```

Thus, invoking "`add(y, w) <- v`" is equivalent to "``y <- `add<-`(y, w, value=v)``".

---

**Note** (*) According to [70], a call "`add(y, 3) <- 1000`" is a syntactic sugar precisely for:

```
`*tmp*` <- y  # temporary substitution
y <- `add<-`(`*tmp*`, 3, value=1000)
rm("*tmp*")  # remove the named object from the current scope
```

This has at least two implications. First, in the unlikely event that a variable `` `*tmp*` ``

existed before the call to the replacement function, it will be no more, it will cease to be. It will be an ex-variable. Second, the temporary substitution guarantees that `y` must exist before the call (due to lazy evaluation, a function's body does not have to refer to all the arguments passed).

---

### Substituting parts of vectors

The replacement versions of the index-like operators are named as follows:

- **`` `[<-` ``** is used in substitutions like "`x[i] <- value`",
- **`` `[[<-` ``** is called when we perform "`x[[i]] <- value`",
- **`` `$<-` ``** is used whilst calling "`x$i <- value`".

```
x <- 1:5
`[<-`(x, c(3, 5), NA_real_)  # returns a new object
## [1]  1  2 NA  4 NA
print(x)  # does not change the original input
## [1] 1 2 3 4 5
```

**Exercise 9.18** *Write a function* **`` `extend<-` ``***, which pushes new elements at the end of a given vector, modifying it in place.*

```
`extend<-` <- function(x, value) ...to.do...
```

*Example use:*

```
x <- 1
extend(x) <- 2     # push 2 at the back
extend(x) <- 3:10  # add 3, 4, ..., 10
print(x)
##  [1]  1  2  3  4  5  6  7  8  9 10
```

### Replacing attributes

There are many replacement functions to reset object attributes (Section 4.4). In particular, each special attribute has its replacement procedure, e.g., **`` `names<-` ``**, **`` `class<-` ``**, **`` `dim<-` ``**, **`` `levels<-` ``**, etc.

```
x <- 1:3
names(x) <- c("a", "b", "c")  # change the `names` attribute
print(x)  # x has been altered
## a b c
## 1 2 3
```

Individual (arbitrary, including non-special ones) attributes can be set using **`` `attr<-` ``**, and all of them can be established via a single call to **`` `attributes<-` ``**.

```
x <- "spam"
attributes(x) <- list(shape="oval", smell="meaty")
attributes(x) <- c(attributes(x), taste="umami")
attr(x, "colour") <- "rose"
print(x)
## [1] "spam"
## attr(,"shape")
## [1] "oval"
## attr(,"smell")
## [1] "meaty"
## attr(,"taste")
## [1] "umami"
## attr(,"colour")
## [1] "rose"
```

Also, setting an attribute to `NULL` results, by convention, in its removal:

```
attr(x, "taste") <- NULL  # it is tasteless now
print(x)
## [1] "spam"
## attr(,"shape")
## [1] "oval"
## attr(,"smell")
## [1] "meaty"
## attr(,"colour")
## [1] "rose"
attributes(x) <- NULL  # remove all
print(x)
## [1] "spam"
```

Which can be worthwhile in contexts such as:

```
x <- structure(c(a=1, b=2, c=3), some_attrib="value")
y <- `attributes<-`(x, NULL)
```

`y` is a version of `x` with metadata removed. The latter remains unchanged.

#### Compositions of replacement functions (*)

Updating only selected names like:

```
x <- c(a=1, b=2, c=3)
names(x)[2] <- "spam"
print(x)
##    a spam    c
##    1    2    3
```

is possible due to the fact that "`names(x)[i] <- v`" is equivalent to:

```r
old_names <- names(x)
new_names <- `[<-`(old_names, i, value=v)
x <- `names<-`(x, value=new_names)
```

---

**Important** More generally, a composition of replacement calls "`g(f(x, a), b) <- y`" yields a result equivalent to "`x <- `f<-`(x, a, value=`g<-`(f(x, a), b, value=y))`". Both **`f`** and **`` `f<-` ``** need to be defined, but having **`g`** is not necessary.

---

**Exercise 9.19** *(*) What is "`h(g(f(x, a), b), c) <- y`" equivalent to?*

**Exercise 9.20** *Write a (convenient!) function **`` `recode<-` ``** which replaces specific elements in a character vector with other ones, allowing the following interface:*

```r
`recode<-` <- function(x, value) ...to.do...
x <- c("spam", "bacon", "eggs", "spam", "eggs")
recode(x) <- c(eggs="best spam", bacon="yummy spam")
print(x)
## [1] "spam"       "yummy spam" "best spam"  "spam"       "best spam"
```

*We see that the named character vector gives a few `from="to"` pairs, e.g., all `eggs` are to be replaced by `best spam`. Determine which calls are equivalent to the following:*

```r
x <- c(a=1, b=2, c=3)
recode(names(x)) <- c(c="z", b="y")  # or equivalently = ... ?
print(x)
## a y z
## 1 2 3
y <- list(c("spam", "bacon", "spam"), c("spam", "eggs", "cauliflower"))
recode(y[[2]]) <- c(cauliflower="broccoli")  # or = ... ?
print(y)
## [[1]]
## [1] "spam"  "bacon" "spam"
##
## [[2]]
## [1] "spam"     "eggs"     "broccoli"
```

**Exercise 9.21** *(*) Consider an example matrix with the `dimnames` attribute whose `names` attribute is set (more details in Chapter 11):*

```r
(x <- Titanic["Crew", , "Adult", ])
##         Survived
## Sex       No Yes
##   Male   670 192
##   Female   3  20
```

*Applying the **`` `recode<-` ``** function from Exercise 9.20, we can change the x object in place:*

```
recode(names(dimnames(x))) <- c(Sex="sex", Survived="survived")
print(x)
##         survived
## sex       No Yes
##   Male   670 192
##   Female   3  20
```

*Compose a single call that alters* **names**(**dimnames**(x)) *without modifying* x *in place but returning a recoded copy of the following:*

- **names**(**dimnames**(x)),
- **dimnames**(x),
- x.

**Exercise 9.22** *(*) Consider the* `` `recode<-` `` *function again but now let an example object be a list with an element of the* factor *class:*

```
x <- as.list(iris[c(1, 2, 51, 101), c(1, 5)])
recode(levels(x[["Species"]])) <- c(
    setosa="SET", versicolor="VER", virginica="VIR"
)
print(x)
## $Sepal.Length
## [1] 5.1 4.9 7.0 6.3
##
## $Species
## [1] SET SET VER VIR
## Levels: SET VER VIR
```

*How to change* **levels**(x[["Species"]]) *and return an altered copy of:*

- **levels**(x[["Species"]]),
- x[["Species"]],
- x

*without modifying* x *in place?*

## 9.4 Arguments and local variables

### 9.4.1 Call by "value"

As a general rule, functions cannot change the state of their arguments[6]. We can think of them as being passed by *value*, i.e., as if their copy was made.

```
test_change <- function(y)
{
    y[1] <- 7
    y
}

x <- 1:5
test_change(x)
## [1] 7 2 3 4 5
print(x)  # same
## [1] 1 2 3 4 5
```

If the preceding statement was not true, the state of `x` would change after the call.

### 9.4.2 Variable scope

Function arguments and any other variables we create inside a function's body are *relative* to each call to that function.

```
test_change <- function(x)
{
    x <- x+1
    z <- -x
    z
}

x <- 1:5
test_change(x*10)
## [1] -11 -21 -31 -41 -51
print(x)  # x in the function's body was a different x
## [1] 1 2 3 4 5
print(z)  # z was local
## Error: ! object 'z' not found
```

Both `x` and `z` are local variables. They only live whilst our function is being executed. The former temporarily *masks*[7] the object of the same name from the caller's context.

[6] With the exception of objects of the type `environment`, which are passed by reference; see Chapter 16. Also, the fact that we have access to unevaluated R expressions can cause further deviations to this rule because, actually, R implements the *call-by-need* strategy; see Chapter 17.

[7] Chapter 16 discusses this topic in-depth: names are bound to objects within environment frames.

**Important** It is a good development practice to refrain from referring to objects not created within the current function, especially to "global" variables. We can always pass an object as an argument explicitly.

**Note** It is a function call as such, not curly braces per se that form a local scope. When we run "`x <- { y <- 1; y + 1 }`", y is not a temporary variable. It is an ordinary named object created alongside x.

On the other hand, in "`x <- (function() { z <- 1; z + 1 })()`", z will not be available thereafter.

### 9.4.3 Closures (*)

Most user-defined functions are, in fact, instances of the so-called *closures*; see Section 16.3.2 and [1]. They not only consist of an R expression to evaluate but also can carry auxiliary data.

For instance, given two numeric vectors x and y of the same length, a call to **approxfun**(x, y) returns a *function* that linearly interpolates between the consecutive points $(x_1, y_1)$, $(x_2, y_2)$, etc., so that a corresponding $y$ can be determined for any $x$.

```
x <- seq(0, 1, length.out=11)
f1 <- approxfun(x, x^2)
f2 <- approxfun(x, x^3)
f1(0.75)  # check that it is close to the true 0.75^2
## [1] 0.565
f2(0.75)  # compare with 0.75^3
## [1] 0.4275
```

Let's inspect the source code of the above functions:

```
print(f1)
## function (v)
## .approxfun(x, y, v, method, yleft, yright, f, na.rm)
## <environment: 0x5620adb098b8>
print(f2)
## function (v)
## .approxfun(x, y, v, method, yleft, yright, f, na.rm)
## <environment: 0x5620adb4c268>
```

We might wonder why they produce different results: after all, they are identical. It turns out, however, that they internally store additional data that are referred to when they are called:

Moreover, R uses lexical (static) scoping, which is not necessarily intuitive, especially taking into account that a function's environment can always be changed.

```
environment(f1)[["y"]]
##  [1] 0.00 0.01 0.04 0.09 0.16 0.25 0.36 0.49 0.64 0.81 1.00
environment(f2)[["y"]]
##  [1] 0.000 0.001 0.008 0.027 0.064 0.125 0.216 0.343 0.512 0.729 1.000
```

We will explore these concepts in detail in the third part of this book.

### 9.4.4 Default arguments

We often need to find a sweet spot between being generous, mindful of the diverse needs of our users, and making the API neither overwhelming nor oversimplistic. We have established that it is best if a function performs a single, well-specified task. However, we are always delighted when it also lets us tweak its behaviour should we wish to do so. The use of *default arguments* can facilitate this principle.

For instance, **log** computes logarithms, by default, the natural ones.

```
log(2.718)  # the same as log(2.718, base=exp(1)), i.e., default base, e
## [1] 0.9999
log(4, base=2)  # different base
## [1] 2
```

**Exercise 9.23** *Study the documentation of the following functions and note the default values they define:* ***round****,* ***hist****,* ***grep****, and* ***download.file****.*

Let's create a function equipped with such *recommended* settings:

```
test_default <- function(x=1) x

test_default()   # use default
## [1] 1
test_default(2)  # use something else
## [1] 2
```

Most often, default arguments are just constants, e.g., `1`. Generally, though, they can be any R expressions, also ones that include a reference to other arguments passed to the same function; see Section 17.2.

Default arguments usually appear at the end of the parameter list, but see Section 9.3.6 (on replacement functions) for a well-justified exception.

### 9.4.5 Lazy vs eager evaluation

In some languages, function arguments are always evaluated prior to a call. In R, though, they are only computed when actually needed. We call it *lazy* or *delayed* evaluation. Recall that in Section 8.1.4, we introduced the short-circuit evaluation operators **`||`** (or) and **`&&`** (and). They can do their job precisely thanks to this mechanism.

**Example 9.24** *In the following example, we do not use the function's argument at all:*

```
lazy_test1 <- function(x) 1  # x is not used

lazy_test1({cat("and now for something completely different!"); 7})
## [1] 1
```

*Otherwise, we would see a message being printed out on the console.*

**Example 9.25** *Next, let's use x amidst other expressions in a function's body:*

```
lazy_test2 <- function(x)
{
    cat("it's... ")
    y <- x+x  # using x twice
    cat(" a man with two noses")
    y
}

lazy_test2({cat("and now for something completely different!"); 7})
## it's... and now for something completely different! a man with two noses
## [1] 14
```

*An argument is evaluated once, and its value is stored for further reference. If that was not the case, we would see two messages like "`and now...`". We will elaborate on this in Chapter 17.*

### 9.4.6 Ellipsis, `...`

We will start with an exercise.

**Exercise 9.26** *Notice the presence of `` `...` `` in the parameter list of **`c`**, **`list`**, **`structure`**, **`cbind`**, **`rbind`**, **`cat`**, **`Map`** (and the underlying **`mapply`**), **`lapply`** (a specialised version of **`Map`**), **`optimise`**, **`optim`**, **`uniroot`**, **`integrate`**, **`outer`**, **`aggregate`**. What purpose does it serve, according to these functions' documentation pages?*

We can create a *variadic function* by including `` `...` `` (dot-dot-dot, ellipsis; see **`help`**`("dots")`) somewhere in its parameter list. The ellipsis serves as a placeholder for all objects passed to the function but not matched by any formal (named) parameters.

The easiest way to process arguments passed via `` `...` `` programmatically (see also Section 17.3) is by redirecting them to **`list`**.

```
test_dots <- function(...)
    list(...)

test_dots(1, a=2)
## [[1]]
## [1] 1
##
## $a
## [1] 2
```

Such a list can be processed just like… any other generic vector. What we can do with these arguments is only limited by our creativity (in particular, recall from Section 7.2.2 the very powerful **do.call** function). There are two primary use cases of the ellipsis[8]:

- create a new object by combining an arbitrary number of other objects:

```
c(1, 2, 3)   # three arguments
## [1] 1 2 3
c(1:5, 6:7)  # two arguments
## [1] 1 2 3 4 5 6 7
structure("spam")  # no additional arguments
## [1] "spam"
structure("spam", color="rose", taste="umami")  # two further arguments
## [1] "spam"
## attr(,"color")
## [1] "rose"
## attr(,"taste")
## [1] "umami"
cbind(1:2, 3:4)  # two
##      [,1] [,2]
## [1,]    1    3
## [2,]    2    4
cbind(1:2, 3:4, 5:6, 7:8)  # four
##      [,1] [,2] [,3] [,4]
## [1,]    1    3    5    7
## [2,]    2    4    6    8
sum(1, 2, 3, 4, 5, 6, 7, 8, 9, 10, 11, 42)  # twelve
## [1] 108
```

- pass further arguments (as-is) to other methods:

```
lapply(list(c(1, NA, 3), 4:9), mean, na.rm=TRUE)  # mean(x, na.rm=TRUE)
## [[1]]
## [1] 2
##
## [[2]]
## [1] 6.5
integrate(dbeta, 0, 1,
    shape1=2.5, shape2=0.5)  # dbeta(x, shape1=2.5, shape2=0.5)
## 1 with absolute error < 1.2e-05
```

**Example 9.27** *The documentation of* **lapply** *states that this function is defined like* **lapply**`(X, FUN, ...)`*. Here, the ellipsis is a placeholder for a number of optional arguments that can be passed to* **FUN***. Hence, if we denote the i-th element of a vector* `X` *by* `X[[i]]`*, calling* **lapply**`(X, FUN, ...)` *will return a list whose i-th element will be equal to* **FUN**`(X[[i]], ...)`*.*

**Exercise 9.28** *Using a single call to* **lapply***, generate a list with three numeric vectors of*

[8] Which is somewhat similar to Python's `*args` and `**kwargs` in a function's parameter list.

*lengths 3, 9, and 7, respectively, drawn from the uniform distribution on the unit interval. Then, upgrade your code to get numbers sampled from the interval* $[-1, 1]$.

**Example 9.29** *Chapter 4 mentioned that concatenating a mix of lists and atomic vectors with* **c**, *unfortunately, unrolls the latter:*

```
str(c(u=list(1:2), v=list(a=3:4, b=5:6), w=7:8))
## List of 5
##  $ u  : int [1:2] 1 2
##  $ v.a: int [1:2] 3 4
##  $ v.b: int [1:2] 5 6
##  $ w1 : int 7
##  $ w2 : int 8
```

*Let's implement a fix:*

```
as.list2 <- function(x) if (is.list(x)) x else list(x)
clist <- function(...) do.call(c, lapply(list(...), as.list2))
str(clist(u=list(1:2), v=list(a=3:4, b=5:6), w=7:8))
## List of 4
##  $ u  : int [1:2] 1 2
##  $ v.a: int [1:2] 3 4
##  $ v.b: int [1:2] 5 6
##  $ w  : int [1:2] 7 8
```

### 9.4.7 Metaprogramming (*)

We can access *expressions* passed as a function's arguments *without evaluating them*. In particular, a call to the composition of **deparse** and **substitute** converts them to a character vector.

```
test_deparse_substitute <- function(x)
    deparse(substitute(x))  # does not evaluate whatever is behind `x`

test_deparse_substitute(testing+1+2+3)
## [1] "testing + 1 + 2 + 3"
test_deparse_substitute(spam & spam^2 & bacon | grilled(spam))
## [1] "spam & spam^2 & bacon | grilled(spam)"
```

**Exercise 9.30** *Check out the y-axis label generated by* ***plot.default**((1:100)^2)*. *Inspect its source code. Notice a call to the two aforementioned functions. Similarly, call* ***shapiro.test**(**log**(**rlnorm**(100)))* *and take note of the "*`data:`*" field.*

A function is free to do with such an expression whatever it likes. For instance, it can modify the expression and then evaluate it in a very different context. Such a language feature allows certain operations to be expressed much more compactly. In theory, it is a potent tool. Alas, it is easy to find many practical examples where it was over/misused and made learning or using R confusing.

**Example 9.31** *(*) In Section 12.3.9 and Section 17.5, we explain that* **`subset`** *and* **`transform`** *use metaprogramming techniques to specify basic data frame transformations. For instance:*

```
transform(
    subset(
        iris,
        Sepal.Length>=7.7 & Sepal.Width >= 3.0,  # huh?
        select=c(Species, Sepal.Length:Sepal.Width)  # le what?
    ),
    Sepal.Length.mm=Sepal.Length/10  # pardon my French, but pardon?
)
##       Species Sepal.Length Sepal.Width Sepal.Length.mm
## 118 virginica          7.7         3.8            0.77
## 132 virginica          7.9         3.8            0.79
## 136 virginica          7.7         3.0            0.77
```

*None of the arguments (except `iris`) makes sense outside of the function's call. In particular, neither `Sepal.Length` nor `Sepal.Width` exists as a standalone variable.*

*The authors of the two functions took the liberty to interpret the arguments passed how they wanted. They created virtual realities within our well-defined world. The reader must refer to the documentation to understand the meaning of the new syntax.*

---

**Note** (*) Some functions have rather bizarre default arguments. For instance, in the manual page of **`prop.test`**, we read that the `alternative` parameter defaults to `c("two.sided", "less", "greater")`. However, if a user does not set this argument explicitly, `alternative="two.sided"` (the first element in the above vector), will actually be assumed.

If we call `print(prop.test)`, we will find the code line responsible for this odd behaviour: "`alternative <- match.arg(alternative)`". Consider the following example:

```
test_match_arg <- function(x=c("a", "b", "c")) match.arg(x)

test_match_arg()  # missing argument; choose first
## [1] "a"
test_match_arg("c")  # one of the predefined options
## [1] "c"
test_match_arg("d")  # unexpected setting
## Error in `match.arg()`: ! 'arg' should be one of "a", "b", "c"
```

In the current context, **`match.arg`** only allows an actual parameter from a given set of choices. However, if the argument is missing, it selects the first option.

Unfortunately, we have to learn this behaviour by heart, because the above source code is far from self-explanatory. If such an expression was normally evaluated, we would use either the default argument or whatever the user passed as `x` (but then the function would not know the range of possible choices). A call to `match.arg(x, c("a", "b", "c"))` could guarantee the desired functionality and would be much more read-able. Instead, metaprogramming techniques enabled `match.arg` to access the enclosing function's default argument list without explicitly referring to them.

One may ask: why is it so? The only sensible answer to this will be "because its programmer decided it must be this way". Let's contemplate this for a while. In cases like these, we are not dealing with some base R language design choice that we might like or dislike, but which we should just accept as an inherent feature. Instead, we are struggling intellectually because of some programmers' (mis)use (in good faith...) of R's flexibility itself. They have introduced a slang/dialect on top of our mother tongue, whose meaning is valid only within this function. Blame the middleman, not the environment, please.

This is why we generally advocate for avoiding metaprogramming-based techniques wherever possible. We shall elaborate on this topic in the third part of this book.

## 9.5 Principles of sustainable design (*)

Fine design is more art than science. As usual in real life, we will need to make many compromises. This is because improving things with regard to one criterion sometimes makes them worse with respect to other aspects[9] (also those that we are not aware of). Moreover, not everything that counts can nor will be counted.

We do not want to be considered heedless enablers who say that if anything is possible, it should be done. Therefore, below we serve some food for thought. However, as there is no accounting for taste, the kind readers might as well decide to skip this spicy meal.

### 9.5.1 To write or abstain

Our functions can often be considered merely creative combinations of the building blocks available in base R or a few high-quality add-on packages. Some are simpler than others. Thus, there is a question if a new operation should be introduced at all: whether we are faced with the case of multiplying entities without necessity.

On the one hand, the DRY (don't repeat yourself) principle tells us that the most frequently used code chunks (say, called at least thrice) should be generalised in the form of a new function. As far as *complex* operations are concerned, this is definitely a correct approach.

On the other hand, not every generalisation is necessarily welcome. Let's say we are tired of writing `g(f(x))` for the $n$-th time, $n \geq 2$. Why not introduce `h` defined as a combination of `g` and `f`? This might *seem* like a clever idea, but let's not take it for granted. Being tired might be an indication that we need a rest. Being lazy can be a call

[9] Compare the notion of Pareto efficiency.

for more self-discipline (not an overly popular word these days, but still, an endearing trait).

**Example 9.32** *`paste0` is a specialised version of `paste`, but has the `sep` argument hardcoded to an empty string.*

- *Even if this might be the most often applied use case, is the introduction of a new function justifiable? Is it so hard to write `sep=""` each time?*
- *Would changing `paste`'s default argument be better? That, of course, would harm backward compatibility, but what strategies could we apply to make the transition as smooth as possible?*
- *What about introducing a new version of `paste` with `sep` defaulting to `""`, and informing the users that the old version is deprecated and will be removed in, say, two years? (or maybe one month is preferable? or five?)*

**Example 9.33** *R 4.0 defined a new function called `deparse1`. It is nothing but a combination of `deparse` and `paste`:*

```
print(deparse1)
## function (expr, collapse = " ", width.cutoff = 500L, ...)
## paste(deparse(expr, width.cutoff, ...), collapse = collapse)
## <environment: namespace:base>
```

*Let's say this covers 90% of use cases: was introducing it a justified idea then? What if that number was 99%? Might it lead to new users' not knowing that the more primitive operations are available?*

Overall, more functions contribute to information overload. We do not want our users to be overwhelmed by unreasonably many choices. Luckily, nothing is cemented once and for all. Had we made bad design choices resulting in our API's being bloated, we could always cancel those that no longer spark joy.

### 9.5.2 To pamper or challenge

We should think about the kind of audience we would like to serve: is it our team only, students, professionals, certain client groups, etc.? Do they have mathematical, programming, engineering, or scientific background? Not everything appropriate for one cohort will be valuable for another. Not everything pleasing some *now* will benefit them in the long run: people (their skills, attitudes, etc.) change.

**Example 9.34** *Assume we are writing a friendly package for novices who would like to grasp the rudiments of data analysis as quickly as possible. Without much effort, it could enable them to solve 80–95% of the most common, easy problems.*

*Think of introducing the students to a function that returns the five largest observations in a given vector. Let's call it `nlargest`. So pleasant. It makes the students feel* empowered *and improves their retention*[10].

---

[10] Brought to the extreme, this strategy is employed by certain companies (and drug dealers): make the in-

*However, when faced with the remaining 5–20% of tasks, they will have to learn another, more advanced, generic, and capable tool anyway (in our case, the base R itself). Are they determined and skilled enough to do that? Some might, unfortunately, say: "it is not my problem, I made sure everyone was happy at that time". Due to this shortsightedness, it is* our *problem now.*

*Recall that it took us some time to arrive at* `order` *and subsetting via* `` `[` ``. *Assuming that we read this book from the beginning to the end and solve all the exercises, which we should, we are now able to author the said* `nlargest` *(and lots of other functions) ourselves, using a single line of code. This will also pay off in many scenarios that we will be facing in the future, e.g., when we consider matrices and data frames.*

*Yes, everyone will be reinventing their own* `nlargest` *this way. But this constitutes a great exercise: by our being immoderately* nice *(spoonfeeding), some might have lost an opportunity to learn a new, more universal skill.*

Although most users would love to minimise the effort put into all their activities, ultimately, they sometimes need to learn new things. Let's thus not be afraid to teach them stuff.

Furthermore, we do not want to discourage experts (or experts to-be) by presenting them with overly simplified solutions that keep their hands tied when something more ambitious needs to be done.

### 9.5.3 To build or reuse

The *fail-fast* philosophy encourages us to build applications using prefabricated components. This is fantastic at the early stage of their life cycles. Nonetheless, if we construct something uncomplicated or whose only purpose is to illustrate an idea, educate, or show off, let's be explicit about it so that other users do not feel obliged to treat our product (exercise) seriously.

In the (not so likely, probabilistically speaking) event of its becoming successful, we are expected to start thinking about the project's long-term stability and sustainability. After all, relying on third-party functions, packages, or programs makes our software projects less… independent. This may be problematic because:

- the dependencies might not be available on every platform or may behave differently across various system configurations,
- they may be huge (and can depend on other external software too),
- their APIs may be altered, which can cause our code to break,
- their functionality can change, which can lead to unexpected behaviour.

Hence, it might be better to rewrite some parts from scratch on our own.

**Exercise 9.35** *Identify a few R packages on CRAN with many dependencies. See what functions they import from other packages. How often do they only borrow a few lines of code?*

---

troductory experience smooth and fun. At the same time, do not permit your users to become independent too easily. Instead, make them rely on your product lines/proprietary solutions/payable services, etc.

The UNIX philosophy emphasises building and using minimalist yet nontrivial, single-purpose, high-quality pieces of software that can work as parts of more complex pipelines. R serves as a glue language very well.

In the long run, our software project might mature to become such a tool. Thus, at some point, we might have to standardise its API (e.g., make it available from the command line; Section 1.2) so that the users of other languages can benefit from our work.

---

**Important** If our project is merely a modified interface/front-end to a standalone program developed by others, we should be humble about it. We should strive to ensure we are not the ones who get all the credit for other people's work. Also, we must clearly state how the original tools can be used to achieve the same goals, e.g., when working from the command line. In other words, let's not be selfish jerks.

---

### 9.5.4 To revolt or evolve

The wise, gradual improving of things is generally welcome. It gives everyone time to adjust. Some projects, however, are developed in a compulsive way, reinforced by neurotic thinking that "stakeholders need to be kept engaged or we're going to lose popularity". It is not a sustainable strategy. Less is better, even though slightly more challenging. Put good engineering first.

Someday we might realise that "everything so far was wrong and we need a global reset". But if we become very successful, we will cause a *divide* in the community. Especially when we decide to duplicate the existing, base functionality, we should note that some users will be introduced to the system through the supplementary interface and they will not be familiar with the classic one. Others will have to learn the added syntax to be able to communicate with the former group. This gives rise to a whole new set of issues (how to make all the functions interoperable with each other seamlessly, etc.). Such moves are sometimes necessary, but let's not treat them lightly; it is a great responsibility.

## 9.6 Exercises

**Exercise 9.36** *Answer the following questions.*

- *Will* `stopifnot(1)` *stop? What about* `stopifnot(NA)`, `stopifnot(TRUE, FALSE)`, *and* `stopifnot(c(TRUE, TRUE, NA))`*?*
- *What does the* `` `if` `` *function return?*
- *Does* `` `attributes<-`(x, NULL) `` *modify* x*?*
- *When can we be interested in calling* `` `[` `` *and* `` `[<-` `` *as functions (and not as operators) directly?*

- *How to define a new binary operator? Can it be equipped with default arguments?*
- *What are the main use cases of the ellipsis?*
- *What is wrong with* `transform`, `subset`, *and* `match.arg`*?*
- *When a call like* `f(-1, do_something_that_takes_a_million_years())` *does not necessarily have to be a regrettable action?*
- *What is the difference between "*`names(x)[1] <- new_name`*" and "*`names(x[1]) <- new_name`*"?*
- *What might be the form of* `x` *if it is legit to call it like* `x[[c(1, 2)]]()()()[[1]]()()`*?*

**Exercise 9.37** *Consider a function:*

```
f <- function(x)
    for (e in x)
        print(e)
```

*What is the return value of a call to* `f(list(1, 2, 3))`*? Is it* `NULL`, `invisible(NULL)`, `x[[length(x)]]`*, or* `invisible(x[[length(x)]])`*? Does it change relative to whether* `x` *is empty or not?*

**Exercise 9.38** *The* `split` *function also has its replacement version. Study its documentation to learn how it works.*

**Exercise 9.39** *A call to* `ls(envir=baseenv())` *returns all objects defined in the* **base** *package (see Chapter 16). List the names corresponding to replacement functions.*

---

**Important** Apply the principle of test-driven development when solving the remaining exercises.

---

**Exercise 9.40** *Implement your version of the* `Position` *and* `Find` *functions. Evaluation should stop as soon as the first element fulfilling a given predicate has been found.*

**Exercise 9.41** *Implement your version of the* `Reduce` *function.*

**Exercise 9.42** *Write a function* `slide(f, x, k, ...)` *which returns a list* `y` *with* `length(x)-k+1` *elements such that* `y[[i]] = f(x[i:(i+k-1)], ...)`

```
unlist(slide(sum, 1:5, 1))
## [1] 1 2 3 4 5
unlist(slide(sum, 1:5, 3))
## [1]  6  9 12
unlist(slide(sum, 1:5, 5))
## [1] 15
```

**Exercise 9.43** *Using* `slide` *defined above, write another function that counts how many increasing pairs of numbers are in a given numeric vector. For instance, in (0, 2, 1, 1, 0, 1, 6, 0), there are three such pairs: (0, 2), (0, 1), (1, 6).*

**Exercise 9.44** *(*) Implement* **`tools::package_dependencies`** *with* `reverse=TRUE` *based on information extracted by calling* **`utils::available.packages`**.

**Exercise 9.45** *(**) Write a standalone program which can be run from the system shell and which computes the total size of all the files in directories given as the script's arguments (via* **`commandArgs`***).*

# 10

# *S3 classes*

Let x be a randomly generated matrix with 1 000 000 rows and 1 000 columns, y be a data frame with results from the latest survey indicating that things are way more complicated than what most people think, and z be another matrix, this time with many zeroes.

The human brain is not capable of dealing with excessive amounts of data that are immoderately specific. This is why we have a natural tendency to *group* different entities based on their similarities. This way, we form more abstract classes of objects.

Also, many of us are inherently lazy. Oftentimes we take shortcuts to minimise energy (at a price to be paid later).

Printing out a matrix, a data frame, and a time series are all instances of the displaying of things, although they undoubtedly differ in detail. By now, we have probably forgotten which objects are hidden behind the aforementioned x, y, and z. Being able to simply call `print(y)` without having to recall that, yes, y is a data frame, seems pretty appealing.

This chapter introduces *S3 classes* [14]. They provide a lightweight object-orientated programming (OOP) approach for automated dispatching calls to *generics* of the type `print(y)` to concrete *methods* such as `print.data.frame(y)`, based on the *class* of the object they are invoked on.

We shall see that S3 classes in their essence are beautifully simple[1]. Ultimately, *generics* and *methods* are ordinary R functions (Chapter 7) and *classes* are merely additional object attributes (Section 4.4).

Of course, this does not mean that wrapping our heads around them will be effortless. However, unlike other "class systems"[2], S3 is ubiquitous in most R programming projects. Suffice it to say that factors, matrices, and data frames discussed in the coming chapters are straightforward, S3-based extensions of the concepts we are about to introduce.

[1] They were built on top of the ordinary ("old R") S so they have inherent limitations that we discuss in the sequel: classes cannot be formally defined (often we will use named lists for representing objects, and we know we cannot be any more flexible than this), and method dispatching can only be based on the class of one of the arguments (usually the first one, but, e.g., binary operators take both types into account).

[2] Other class systems may give an impression that they are alien implants which were forcefully added to our language to solve a specific, rather narrow class of problems; e.g., S4 (Section 10.5), reference classes (Section 16.1.5), and other ones proposed by third-party packages.

## 10.1 Object type vs class

Recall that **typeof** (introduced in Section 4.1) returns the *internal* type of an object. So far, we were mostly focused on atomic and generic vectors; compare Figure 1 in the Preface.

```
typeof(NULL)
## [1] "NULL"
typeof(c(TRUE, FALSE, NA))
## [1] "logical"
typeof(c(1, 2, 3, NA_real_))
## [1] "double"
typeof(c("a", "b", NA_character_))
## [1] "character"
typeof(list(list(1, 2, 3), LETTERS))
## [1] "list"
typeof(function(x) x)
## [1] "closure"
```

The number of admissible types is small[3], but they open the world of endless possibilities[4]. They provide a basis for more complex data structures. This is thanks to the fact that they can be equipped with arbitrary attributes (Section 4.4).

Most *compound types* constructed using the mechanisms discussed in this chapter only *pretend* they are something different from what they actually are. Still, they often do their job very well. By looking under their bonnet, we will be able to manipulate their state outside of the prescribed use cases.

---

**Important** Setting the `class` attribute might make some objects behave differently in certain scenarios.

---

**Example 10.1** *Let's equip two identical objects with different* `class` *attributes.*

```
xt <- structure(123, class="POSIXct")  # POSIX calendar time
xd <- structure(123, class="Date")
```

*Both objects are* represented *using numeric vectors:*

```
c(typeof(xt), typeof(xd))
## [1] "double" "double"
```

*However, when printed, they are* decoded *differently:*

[3] Their list is hardcoded at the C language level; see the list of `SEXPTYPE`s in Table 14.1 and [69].

[4] In particular, Section 14.2.8 mentions `externalptr`s which are simple pointers to memory blocks that can be instances of any C `struct`s or C++ `class`es. This makes R a very extensible language.

```
print(xt)
## [1] "1970-01-01 01:02:03 CET"
print(xd)
## [1] "1970-05-04"
```

*In the former case, 123 is understood as the number of* seconds *since the* UNIX epoch, *1970-01-01T00:00:00+0000. The latter is deciphered as the number of* days *since the said timestamp. Therefore, we expect that there must exist a mechanism that calls a* version of **`print`** *dependent on an object's virtual class. That it only relies on the* `class` *attribute, which might be set, unset, or reset freely, is emphasised below.*

```
attr(xt, "class") <- "Date"  # change class from POSIXct to Date
print(xt)  # same 123, but now interpreted as Date
## [1] "1970-05-04"
as.numeric(xt)  # drops all attributes
## [1] 123
unclass(xd)  # drops the class attribute; `attr<-`(xd, "class", NULL)
## [1] 123
```

We are having so much fun that one more illustration can only grow our joy.

**Example 10.2** *Consider an example data frame:*

```
x <- iris[1:3, 1:2]  # a subset of an example data frame
print(x)
##   Sepal.Length Sepal.Width
## 1          5.1         3.5
## 2          4.9         3.0
## 3          4.7         3.2
```

*It is an object of the class (an object whose* `class` *attribute is set to):*

```
attr(x, "class")
## [1] "data.frame"
```

*Some may say, and they are absolutely right, that we have not covered data frames yet. After all, they are the topic of Chapter 12, which is still ahead of us. However, from the current perspective, we should know that R data frames are nothing but lists of vectors of the same lengths equipped with the* `names` *and* `row.names` *attributes.*

```
typeof(x)
## [1] "list"
`attr<-`(x, "class", NULL)  # or unclass(x)
## $Sepal.Length
## [1] 5.1 4.9 4.7
##
## $Sepal.Width
## [1] 3.5 3.0 3.2
##
```

```
## attr(,"row.names")
## [1] 1 2 3
print(x)
##   Sepal.Length Sepal.Width
## 1          5.1         3.5
## 2          4.9         3.0
## 3          4.7         3.2
```

**Important** Revealing how x is *actually* represented enables us to process it using the extensive skill set that we have already[5] developed by studying the material covered in the previous part of our book (including all the exercises). This fact is noteworthy because some built-in and third-party data types are not particularly well-designed.

Let's underline again that attributes are simple additions to R objects. However, as we said in Section 4.4.3, certain attributes are special, and `class` is one of them. In particular, we can only set `class` to be a character vector (possibly of length greater than one; see Section 10.2.5).

```
x <- 12345
attr(x, "class") <- 1  # character vectors only
## Error in `attr(x, "class") <- 1`: ! attempt to set invalid 'class'
##     attribute
```

Furthermore, the **class** function can read the value of the `class` attribute. Its replacement version is also available.

```
class(x) <- "Date"  # set; the same as attr(x, "class") <- "Date"
class(x)  # get; here, it is the same as attr(x, "class")
## [1] "Date"
```

**Important** The **class** function always yields a value, even if the corresponding attribute is not set. We call it an *implicit* class. Compare the following results:

```
class(NULL)  # no `class` set because NULL cannot have any attributes
## [1] "NULL"
class(c(TRUE, FALSE, NA))  # no attributes so class is implicit (= typeof)
## [1] "logical"
class(c(1, 2, 3, NA_real_))  # typeof returns "double"
## [1] "numeric"
class(c("a", "b", NA_character_))
## [1] "character"
class(list(list(1, 2, 3), LETTERS))
```

[5] For instance, consider once again the example from Section 5.4.3 that applies the **split** function on a data frame reduced to a list.

```
## [1] "list"
class(function(x) x)  # typeof gives "closure"
## [1] "function"
```

Also, Chapter 11 will explain that any object equipped with the `dim` attribute also has an implicit class:

```
(x <- as.matrix(c(1, 2, 3)))
##      [,1]
## [1,]    1
## [2,]    2
## [3,]    3
attributes(x)  # `class` is not amongst the attributes
## $dim
## [1] 3 1
class(x)  # implicit class
## [1] "matrix" "array"
typeof(x)  # it is still a numeric vector (under the bonnet)
## [1] "double"
```

## 10.2 Generics and method dispatching

### 10.2.1 Generics, default, and custom methods

Let's inspect the source code of the **`print`** function:

```
print(print)  # sic!
## function (x, ...)
## UseMethod("print")
## <environment: namespace:base>
```

Any function like the above[6] we will call from now on a *generic* (an *S3 generic*, from S version 3 [14]). Its only job is to invoke **`UseMethod`**`("print")`. It dispatches the control flow to another function, referred to as a *method*, based on the class of the first argument.

[6] Some functions can have a version of **`UseMethod`** hidden at the C language level (internally); see Section 10.2.3.

**Important** All arguments passed to the generic will also be available[7] in the method dispatched to.

For example, let's define an object of the class `categorical` (a name that we have just come up with; we could have called it `cat`, `CATEGORICAL`, or `SpanishInquisition` as well). It will be our version of the `factor` type that we discuss later.

```
x <- structure(
    c(1, 3, 2, 1, 1, 1, 3),
    levels=c("a", "b", "c"),
    class="categorical"
)
```

We assume that such an object is a sequence of small positive integers (codes). It is equipped with the `levels` attribute, which is a character vector of length not less than the maximum of the said integers. In particular, the first level deciphers the meaning of the code `1`. Hence, the above vector represents a sequence *a, c, b, a, a, a, c*.

There is no special method for displaying objects of the `categorical` class. Hence, when we call **`print`**, the *default* (fallback) method is invoked:

```
print(x)
## [1] 1 3 2 1 1 1 3
## attr(,"levels")
## [1] "a" "b" "c"
## attr(,"class")
## [1] "categorical"
```

This is the standard function for displaying numeric vectors. We are well familiar with it. Its name is **`print.default`**, and we can always call it directly:

```
print.default(x)  # the default print method
## [1] 1 3 2 1 1 1 3
## attr(,"levels")
## [1] "a" "b" "c"
## attr(,"class")
## [1] "categorical"
```

However, we can introduce a designated method for printing `categorical` objects. Its name must precisely be **`print.categorical`**:

```
print.categorical <- function(x, ...)
{
```

[7] However, it cannot be implied by reading the preceding source code. **`UseMethod`** heavily relies on some obscure hacks. We may only call it inside a function's body. Once invoked, it does not return to the generic. Before dispatching to a particular method, it creates a couple of hidden variables which give more detail on the operation conveyed, e.g., `` `.Generic` `` or `` `.Class` ``; see **`help`**`("UseMethod")` and Section 5 of [70].

```
    x_character <- attr(x, "levels")[unclass(x)]
    print(x_character)  # calls `print.default`
    cat(sprintf("Categories: %s\n",
        paste(attr(x, "levels"), collapse=", ")))
    invisible(x)  # this is what all print methods do; see help("print")
}
```

Calling **print** automatically dispatches the control flow to this method:

```
print(x)
## [1] "a" "c" "b" "a" "a" "a" "c"
## Categories: a, b, c
```

Of course, the default method can still be called. Referring to **print.default**(x) directly will output the same result as the one a few chunks above.

---

**Note** **print.categorical** has been equipped with the dot-dot-dot attribute since the generic **print** had one too[8].

---

### 10.2.2 Creating generics

Introducing new S3 generics is as straightforward as defining a function that calls **UseMethod**. For instance, here is a dispatcher which creates new objects of the categorical class based on other objects:

```
as.categorical <- function(x, ...)
    UseMethod("as.categorical")  # synonym: UseMethod("as.categorical", x)
```

We always need to define the default method:

```
as.categorical.default <- function(x, ...)
{
    if (!is.character(x))
        x <- as.character(x)
    xu <- unique(sort(x))  # drops NAs
    structure(
        match(x, xu),
        class="categorical",
        levels=xu
    )
}
```

Testing:

[8] (*) Ensuring S3 generic/method consistency is part of R package check.

```
as.categorical(c("a", "c", "a", "a", "d", "c"))
## [1] "a" "c" "a" "a" "d" "c"
## Categories: a, c, d
as.categorical(c(3, 6, 4, NA, 9, 9, 6, NA, 3))
## [1] "3" "6" "4" NA  "9" "9" "6" NA  "3"
## Categories: 3, 4, 6, 9
```

This method is already quite flexible. It handles a wide variety of data types because it relies on the *built-in generic* **as.character** (Section 10.2.3).

**Example 10.3** *We might want to forbid the conversion from lists because it does not necessarily make sense:*

```
as.categorical.list <- function(x, ...)
    stop("conversion of lists to categorical is not supported")
```

*The users can always be instructed in the method's documentation that they are responsible for converting lists to another type prior to a call to* ***as.categorical****.*

**Example 10.4** *The default method deals with logical vectors perfectly fine:*

```
as.categorical(c(TRUE, FALSE, NA, NA, FALSE))  # as.categorical.default
## [1] "TRUE"  "FALSE" NA      NA      "FALSE"
## Categories: FALSE, TRUE
```

*However, we might still want to introduce its specialised version. This is because we know a slightly more efficient algorithm (and we have nothing better to do) based on the fact that FALSE and TRUE converted to numeric yield 0 and 1, respectively:*

```
as.categorical.logical <- function(x, ...)
{
    if (!is.logical(x))
        x <- as.logical(x)  # or maybe stopifnot(is.logical(x))?
    structure(
        x + 1,  # only 1, 2, and NAs will be generated
        class="categorical",
        levels=c("FALSE", "TRUE")
    )
}
```

*It spawns the same result as the default method but is slightly faster.*

```
as.categorical(c(TRUE, FALSE, NA, NA, FALSE))  # as.categorical.logical
## [1] "TRUE"  "FALSE" NA      NA      "FALSE"
## Categories: FALSE, TRUE
```

*We performed some argument consolidation at the beginning because a user is always able to call a method directly on an R object of any kind (which is a good thing; see Section 10.2.4). In other words, there is no guarantee that the argument x must be of type logical.*

### 10.2.3 Built-in generics

Many[9] functions and operators we have introduced so far are, in fact, S3 generics: `print`, `head`, `` `[` ``, `` `[[` ``, `` `[<-` ``, `` `[[<-` ``, `length`, `` `+` ``, `` `<=` ``, `is.numeric`, `as.numeric`, `is.character`, `as.character`, `as.list`, `round`, `log`, `sum`, `rep`, `c`, and `na.omit`, to name a few.

**Example 10.5** *Let's overload the* `as.character` *method. The default one does not make much sense for the objects of our custom type:*

```
as.character(x)
## [1] "1" "3" "2" "1" "1" "1" "3"
```

*So:*

```
as.character.categorical <- function(x, ...)
    attr(x, "levels")[unclass(x)]
```

*And now:*

```
as.character(x)
## [1] "a" "c" "b" "a" "a" "a" "c"
```

**Exercise 10.6** *Overload the* `unique` *and* `rep` *methods for objects of the class* `categorical`.

**Example 10.7** *New types ought to be designed carefully. For instance, if we forget to overload the to-numeric converter, some users might be puzzled*[10] *when they see:*

```
(x <- as.categorical(c(4, 9, 100, 9, 9, 100, 42, 666, 4)))
## [1] "4"   "9"   "100" "9"   "9"   "100" "42"  "666" "4"
## Categories: 100, 4, 42, 666, 9
as.double(x)  # synonym: as.numeric(x); here, it calls as.double.default(x)
## [1] 2 5 1 5 5 1 3 4 2
```

*Hence, we might want to introduce a new method:*

---

[9] Generating the list of all S3 generics is somewhat tricky, but at least the internal ones are enumerated in `help("InternalMethods")` and `help("groupGeneric")`; compare `` `.S3PrimitiveGenerics` ``, `` `.internalGenerics` ``, `` `.knownS3Generics` ``, and `` `.S3_methods_table` ``. Some of them do not even call `UseMethod` explicitly; they dispatch at the C language level. This is unfortunate as it decreases transparency. Instead of simply inspecting a function's source code (compare, e.g., `cbind`), we need to look this information up in the documentation. Also, methods may be hardcoded internally, and thus be unoverloadable. However, sometimes these design choices can be defended because they improve execution speed or memory consumption.

[10] It is a different story if we *really* want this behaviour. Provided that we document it *thoroughly* (see how `help("factor")` discusses the behaviour of a to-numeric conversion), we can start holding the users responsible for their feeling confused (those who have experience in teaching others will certainly agree how complex this matter is). Remember that we can never make an API fully foolproof and that there will always be someone to challenge/stress-test our ideas. Bad design is always wrong, but being overprotective or too defensive also has its cons. We should maintain our audience wisely. Users of open-source software are not our *clients*. We do not work *for* them. We are in this *together*.

```
as.double.categorical <- function(x, ...)  # not: as.numeric.categorical
{
    # actually: as.double.default(as.character.categorical(x))
    as.double(as.character(x))
}
```

*It now yields:*

```
as.double(x)  # or as.numeric(x); calls as.double.categorical(x)
## [1]   4   9 100   9   9 100  42 666   4
```

---

**Note** We can still use **unclass** to fetch the codes:

```
unclass(x)
## [1] 2 5 1 5 5 1 3 4 2
## attr(,"levels")
## [1] "100" "4"   "42"  "666" "9"
```

It is because the foregoing returns a class-free object, which is now guaranteed to be processed by the default methods (**print**, subsetting, **as.character**, etc.).

---

**Exercise 10.8** *What would happen if we used* **as.numeric** *instead of* **unclass** *in* **print.categorical** *and* **as.character.categorical***?*

**Exercise 10.9** *Update the preceding methods so that we can also create* named *objects of the class* `categorical` *(i.e., equipped with the* `names` *attribute).*

**Exercise 10.10** *The levels of* x *are sorted lexicographically, not numerically. Introduce a single method that would let the above code (when rerun without any alterations) generate a more natural result.*

### 10.2.4 First-argument dispatch and calling S3 methods directly

With S3, dispatching is most often done based on the class of only one[11] argument: by default, the first one from the parameter list.

For example, the **c** function is a generic that dispatches on the first argument's class. Let's overload it for `categorical` objects. In other words, we will create a function to be called by the generic when it is invoked on a series of objects whose first element is of the said class.

[11] There are many exceptions to this rule. They were made for the (debatable) sake of the R users' convenience. In particular, in Section 12.1.2 we mention that **cbind** and **rbind** will dispatch to the `data.frame` method if at least one argument is a data frame (and others are unclassed). Binary operators consider the type of both operands; see Section 10.2.6. Furthermore, it is worth noting that the S4 class system (Section 10.5) allows for dispatching based on the classes many arguments.

```
c.categorical <- function(...)
    as.categorical(
        unlist(
            lapply(list(...), as.character)
        )
    )
```

It converts each argument to a character vector, relying on the generic **as.character** to take care of the details. It works because **unlist** converts a list of such atomic vectors to a single sequence of strings.

Calling **c** with the first argument of the class `categorical` dispatches to the above method:

```
x <- c(9, 5, 7, 7, 2)
xc <- as.categorical(x)
c(xc, x)  # c.categorical
##  [1] "9" "5" "7" "7" "2" "9" "5" "7" "7" "2"
## Categories: 2, 5, 7, 9
```

However, if the first argument is, say, unclassed, the default method will be consulted:

```
c(x, xc)  # default c
##  [1] 9 5 7 7 2 4 2 3 3 1
```

It ignored the `class` attribute and saw xc *as it is*, a bareboned numeric vector:

```
`attributes<-`(xc, NULL)  # the underlying codes
## [1] 4 2 3 3 1
```

It is not a bug. It is a well-documented (and now explained) behaviour. After all, compound types (classed objects) are emulated through the basic ones.

---

**Important** In most cases, S3 methods can be called directly to get the desired outcome:

```
c.categorical(x, xc)  # force a call to the specific method
##  [1] "9" "5" "7" "7" "2" "9" "5" "7" "7" "2"
## Categories: 2, 5, 7, 9
```

We said *in most cases* because methods can be:

- hardcoded at the C language level (e.g., there is no **c.default** defined at all[12]),
- hidden (defined in a package's namespace but not exported; Section 16.3.6),

[12] Dispatching to internal methods can also be done… internally. For instance, overloading **`<.character`** (or **Compare.character**; see below) will have no effect unless the base **`<`** is replaced with a custom one that makes an explicit call to **UseMethod**. Most often, we can expect that the built-in types (e.g., atomic vectors), factors, data frames, and matrices and other arrays might be treated specially.

- overloaded as a group; see Section 10.2.6 and **help**("groupGeneric").

---

**Example 10.11** *Purely for jollity, let's find a partition of the* `iris` *dataset into three clusters using the k-means algorithm:*

```
res <- kmeans(iris[-5], centers=3, nstart=10)
print(res)
## K-means clustering with 3 clusters of sizes 50, 62, 38
##
## Cluster means:
##   Sepal.Length Sepal.Width Petal.Length Petal.Width
## 1       5.0060      3.4280       1.4620      0.2460
## 2       5.9016      2.7484       4.3935      1.4339
## 3       6.8500      3.0737       5.7421      2.0711
##
## Clustering vector:
##  [1] 1 1 1 1 1 1 1 1 1 1 1 1 1 1 1 1 1 1 1 1 1 1 1 1 1 1 1 1 1 1 1 1 1 1 1
## [36] 1 1 1 1 1 1 1 1 1 1 1 1 1 1 1 2 2 3 2 2 2 2 2 2 2 2 2 2 2 2 2 2 2 2 2
## [71] 2 2 2 2 2 2 2 3 2 2 2 2 2 2 2 2 2 2 2 2 2 2 2 2 2 2 2 2 2 2
##  [ reached 'max' / getOption("max.print") -- omitted 51 entries ]
##
## Within cluster sum of squares by cluster:
## [1] 15.151 39.821 23.879
##  (between_SS / total_SS =  88.4 %)
##
## Available components:
##
## [1] "cluster"      "centers"      "totss"        "withinss"
## [5] "tot.withinss" "betweenss"    "size"         "iter"
## [9] "ifault"
```

*It is an object of the class:*

```
class(res)
## [1] "kmeans"
```

*which, in fact, is a:*

```
typeof(res)
## [1] "list"
```

*The underlying list looks like:*

```
unclass(res)
## $cluster
##  [1] 1 1 1 1 1 1 1 1 1 1 1 1 1 1 1 1 1 1 1 1 1 1 1 1 1 1 1 1 1 1 1 1 1 1 1
## [36] 1 1 1 1 1 1 1 1 1 1 1 1 1 1 1 2 2 3 2 2 2 2 2 2 2 2 2 2 2 2 2 2 2 2 2
## [71] 2 2 2 2 2 2 2 3 2 2 2 2 2 2 2 2 2 2 2 2 2 2 2 2 2 2 2 2 2 2
##  [ reached 'max' / getOption("max.print") -- omitted 51 entries ]
```

```
##
## $centers
##   Sepal.Length Sepal.Width Petal.Length Petal.Width
## 1       5.0060      3.4280       1.4620      0.2460
## 2       5.9016      2.7484       4.3935      1.4339
## 3       6.8500      3.0737       5.7421      2.0711
##
## $totss
## [1] 681.37
##
## $withinss
## [1] 15.151 39.821 23.879
##
## $tot.withinss
## [1] 78.851
##
## $betweenss
## [1] 602.52
##
## $size
## [1] 50 62 38
##
## $iter
## [1] 2
##
## $ifault
## [1] 0
```

*We already know that `res` was displayed in a fancy way only because there is a **`print`** method overloaded for objects of the `kmeans` class.*

*But is there?*

```
print.kmeans
## Error: ! object 'print.kmeans' not found
```

*Even though the method is hidden (internal) in the **stats** package's namespace, from Section 16.3.6 we will learn that it can be accessed by calling **`getS3method`**`("print", "kmeans")` or referring to **`stats:::print.kmeans`** (note the triple colon).*

### 10.2.5 Multi-class-ness

The `class` attribute can be instantiated as a character vector of any length. For example:

```
(t1 <- Sys.time())
## [1] "2026-07-30 14:11:33 CEST"
```

```
(t2 <- strptime("2021-08-15T12:59:59+1000", "%Y-%m-%dT%H:%M:%S%z"))
## [1] "2021-08-15 04:59:59"
```

Let's inspect the classes of these two objects:

```
class(t1)
## [1] "POSIXct" "POSIXt"
class(t2)
## [1] "POSIXlt" "POSIXt"
```

Section 10.3.1 will discuss date-time classes in more detail. It will highlight that the former is represented as a numeric vector, while the latter is a list. Thus, these two should primarily be seen as instances of two distinct types. However, as both of them have a lot in common, it was a wise design choice to allow them to be seen also as the representatives of the same generic category of *POSIX time* objects.

---

**Important** When calling a generic function[13] `f` on an object `x` of the classes[14] `class1`, `class2`, …, `classK` (in this order), `UseMethod(f, x)` dispatches to the method determined as follows:

1. if **`f.class1`** is available[15], call it;
2. otherwise, if **`f.class2`** is available, call this one;
3. …;
4. otherwise, if **`f.classK`** is available, invoke it;
5. otherwise, refer to the fallback **`f.default`**.

---

**Example 10.12** *There is a method* ***`diff`*** *for objects of the class* `POSIXt` *that carries a statement:*

```
r <- if (inherits(x, "POSIXlt")) as.POSIXct(x) else x
```

*This way, we can process both* `POSIXct` *and* `POSIXlt` *instances using the same procedure.*

We should see no magic in this simple scheme. It is nothing more than a way to determine the method to be called for a particular R object. It can be used as a mechanism to mimic the idea of inheritance in object-orientated programming languages. However, the S3 system does not allow for defining classes in any formal manner.

For example, we cannot say that objects of the class `POSIXct` inherit from `POSIXt`. Neither can we say that each object of the class `POSIXct` is also an instance of `POSIXt`. The class attribute can still be set arbitrarily on a per-object basis. We can create ones whose class is simply `POSIXct` (without the `POSIXt` part) or even `c("POSIXt", "POSIXct")` (in this order).

---

[13] The case of binary operators is handled differently; see Section 10.2.6.

[14] **`UseMethod`** dispatches on the implicit class as determined by the **`class`** function. Note that the `class` attribute does not necessarily have to be set in order for **`class`** to return a sensible answer.

[15] For more details on S3 method lookup, see Section 16.3.6.

**Note** In any method, it is possible to call the method corresponding to the next class by calling **NextMethod**.

For instance, if we are in **f.class1**, a call to **NextMethod**(f) will try invoking **f.class2**. If such a method does not exist, further methods in the search chain will be probed, falling back to the default method if necessary. We will give an illustration later.

### 10.2.6 Operator overloading

Operators are ordinary functions (Section 9.3.5). Even though what follows can partially be implied by what we have already said, as usual in R, some oddities are to be expected.

For example, let's overload the index operator for objects of the class `categorical`. Looking at **help**("["), we see that the default method has two arguments: `x` (the `categorical` object being sliced) and `i` (the indexer). Ours will have the same interface then:

```
`[.categorical` <- function(x, i)
{
    structure(
        unclass(x)[i],  # `[`(unclass(x), i)
        class="categorical",
        levels=attr(x, "levels")  # the same levels as input
    )
}
```

The default S3 method, **`[.default`**, is hardcoded at the C language level and we cannot refer to it directly. This is why we called **unclass** instead. Alternatively, we can also invoke **NextMethod**:

```
`[.categorical` <- function(x, i)
{
    structure(
        NextMethod("["),  # call default method, passing `x` and `i`
        class="categorical",
        levels=attr(x, "levels")  # the same levels as input
    )
}
```

We can also introduce the replacement version of this operator:

```
`[<-.categorical` <- function(x, i, value)
{
```

```
    levels <- attr(x, "levels")
    value <- match(value, levels)  # integer codes corresponding to levels
    structure(
        NextMethod("[<-"),  # call default method, passing `x`, `i`, `values`
        class="categorical",
        levels=levels  # same levels as input
    )

    # # or, equivalently:
    # structure(
    #     `[<-`(unclass(x), i, value=match(value, attr(x, "levels"))),
    #     class="categorical",
    #     levels=attr(x, "levels")
    # )
}
```

Testing:

```
x <- as.categorical(c(3, 6, 4, NA, 9, 9, 6, NA, 3))
x[1:4]
## [1] "3" "6" "4" NA
## Categories: 3, 4, 6, 9
x[1:4] <- c("6", "7")
print(x)
## [1] "6" NA  "6" NA  "9" "9" "6" NA  "3"
## Categories: 3, 4, 6, 9
```

Notice how we handled the case of nonexistent levels and that the recycling rule has been automagically inherited (amongst other features) from the default index operator.

**Exercise 10.13** *Do these two operators preserve the* `names` *attribute of* `x`*? Is indexing with negative integers or logical vectors supported as well? Why is that/is that not the case?*

Furthermore, let's overload the `` `==` `` operator. Assume[16] that we would like two `categorical` objects to be compared based on the actual labels they encode, in an elementwise manner:

```
`==.categorical` <- function(e1, e2)
    as.character(e1) == as.character(e2)
```

We are feeling lucky: by not performing any type checking, we rely on the particular **as.character** methods corresponding to the types of `e1` and `e2`. Also, assuming that

[16] There are, of course, many possible ways to implement the `` `==` `` operator for the discussed objects. For instance, it may return either a single `TRUE` or `FALSE` depending on if two objects are identical (although probably overloading **all.equal** would be a better idea). We could also compare the corresponding underlying integer codes instead of the labels, etc.

`as.character` always[17] returns a `character` object, we dispatch to the default method for `` `==` `` (which handles atomic vectors).

Some examples:

```
as.categorical(c(1, 3, 5, 1)) == as.categorical(c(1, 3, 1, 1))
## [1]  TRUE  TRUE FALSE  TRUE
as.categorical(c(1, 3, 5, 1)) == c(1, 3, 1, 1)
## [1]  TRUE  TRUE FALSE  TRUE
c(1, 3, 5, 1) == as.categorical(c(1, 3, 1, 1))
## [1]  TRUE  TRUE FALSE  TRUE
```

---

**Important** In the case of binary operators, dispatching is done based on the classes of both arguments. In all three preceding calls, we call `` `==.categorical` ``, regardless of whether the classed object is the first or the second operand.

If two operands are classed, and different methods are overloaded for both, a warning will be generated, and the default internal method will be called.

```
`==.A` <- function(e1, e2) "A"
`==.B` <- function(e1, e2) "B"
structure(c(1, 2, 3), class="A") == structure(c(2, NA, 3), class="B")
## Warning: Incompatible methods ("==.A", "==.B") for "=="
## [1] FALSE    NA  TRUE
```

---

---

**Note** (*) By creating a single `Ops` method, we can define the meaning of all binary operators at once.

```
Ops.categorical <- function(e1, e2)
{
    if (.Generic %notin% c("<", ">", "<=", ">=", "==", "!="))
        stop(sprintf("%s not defined for 'categorical' objects", .Generic))
    e1 <- as.character(e1)
    e2 <- as.character(e2)
    NextMethod(.Generic)  # dispatch to the default method (for character)
}

as.categorical(c(1, 3, 5, 1)) > c(1, 2, 4, 2)
## [1] FALSE  TRUE  TRUE FALSE
```

Here, `` `.Generic` `` is a variable representing the name of the operator (generic) being invoked; see Section 16.3.6.

Other *group generics* are: `Summary` (including functions such as `min`, `sum`, and `all`),

---

[17] Which, of course, does not have to be the case; it is merely an assumption based on our belief in the common sense of other developers.

**Math** (**abs**, **log**, **round**, etc.), **matrixOps** (`` `%*%` ``), and **Complex** (e.g., **Re**, **Im**); see **help**("groupGeneric") for more details.

Sometimes we must rely on the `` `.S3method` `` function to let R recognise a custom method related to such generics.

## 10.3 Common built-in S3 classes

Below we discuss a few noteworthy classes, including those representing date-time information and factors (ordered or not). Note that classes representing tabular data will be dealt with in separate parts, owing to their importance and ubiquity. Namely, matrices and other arrays are covered in Chapter 11, and data frames are discussed in Chapter 12. Inspecting other[18] interesting compound types is left as a simple exercise for the studious reader.

### 10.3.1 Date, time, etc.

The Date class represents… dates (calendar ones, not fruits).

```
(curd <- c(Sys.Date(), as.Date(c("1969-12-31", "1970-01-01", "2026-07-30"))))
## [1] "2026-07-30" "1969-12-31" "1970-01-01" "2026-07-30"
class(curd)
## [1] "Date"
```

Complex types are built on basic ones. Underneath, what we deal with here is:

```
typeof(curd)
## [1] "double"
unclass(curd)
## [1] 20664    -1     0 20664
```

which is the number of days since the *UNIX epoch*, 1970-01-01T00:00:00+0000 (midnight GMT/UTC).

The POSIXct (calendar time) class represents date-time objects:

```
(curt <- Sys.time())
## [1] "2026-07-30 14:11:33 CEST"
class(curt)
## [1] "POSIXct" "POSIXt"
typeof(curt)
## [1] "double"
```

[18] **unique**(.S3_methods_table[, 2]) approximates the list of available classes.

```
unclass(curt)
## [1] 1785413493
```

Underneath, it is the number of seconds since the UNIX epoch. By default, whilst printing, the current default time zone is used (see **Sys.timezone**). However, such objects can be equipped with the tzone attribute.

```
structure(1, class=c("POSIXct", "POSIXt"))  # using current default time zone
## [1] "1970-01-01 01:00:01 CET"
structure(1, class=c("POSIXct", "POSIXt"), tzone="UTC")
## [1] "1970-01-01 00:00:01 UTC"
```

In both cases, the time is 1 second after the beginning of the UNIX epoch. On the author's PC, the former is displayed in the current local time zone, though.

**Exercise 10.14** *Inspect how **ISOdatetime** displays midnights in different time zones.*

The POSIXlt (local time) class is represented using a list of atomic vectors[19].

```
(x <- as.POSIXlt(c(a="1970-01-01 00:00:00", b="2030-12-31 23:59:59")))
##                         a                         b
## "1970-01-01 00:00:00 CET" "2030-12-31 23:59:59 CET"
class(x)
## [1] "POSIXlt" "POSIXt"
typeof(x)
## [1] "list"
str(unclass(x))  # calling str instead of print to make display more compact
## List of 11
##  $ sec   : num [1:2] 0 59
##  $ min   : int [1:2] 0 59
##  $ hour  : int [1:2] 0 23
##  $ mday  : int [1:2] 1 31
##  $ mon   : int [1:2] 0 11
##  $ year  : Named int [1:2] 70 130
##   ..- attr(*, "names")= chr [1:2] "a" "b"
##  $ wday  : int [1:2] 4 2
##  $ yday  : int [1:2] 0 364
##  $ isdst : int [1:2] 0 0
##  $ zone  : chr [1:2] "CET" "CET"
##  $ gmtoff: int [1:2] NA NA
##  - attr(*, "tzone")= chr [1:3] "" "CET" "CEST"
##  - attr(*, "balanced")= logi TRUE
```

**Exercise 10.15** *Read about the meaning of each named element, especially mon and year; see **help**("DateTimeClasses").*

The manual states that POSIXlt is supposedly *closer to human-readable forms* than

[19] Which was inspired by struct tm in C's <time.h>.

`POSIXct`, but it is a matter of taste. Some R functions return the former, and other output the latter type.

**Exercise 10.16** *The two main functions for date formatting and parsing, **`strftime`** and **`strptime`**, use special field formatters (similar to **`sprintf`**). Read about them in the R manual. What type of inputs do they accept? What outputs do they produce?*

There are several methods overloaded for objects of the said classes. In fact, the first call in this section already involved the use of **`c.Date`**.

**Exercise 10.17** *Play around with the overloaded versions of **`seq`**, **`rep`**, and **`as.character`**.*

A specific number of days or seconds can be added to or subtracted from a date or time, respectively. However, **`` `-` ``** and **`diff`** can also be applied on date-time objects; they yield an object of the class `difftime`.

```
curt - (curt-1)
## Time difference of 1 secs
diff(c(curt, curt+1, curt+7, curt+67))
## Time differences in secs
## [1]  1  6 60
diff(c(curt, curt+1, curt+7, curt+67), units="m")
## Time differences in mins
## [1] 0.016667 0.100000 1.000000
```

**Exercise 10.18** *Check out how objects of the class `difftime` are internally represented.*

Applying other arithmetic operations on date-time objects raises an error. Because date-time objects are just numbers, they can be compared to each other using binary operators[20]. Also, methods such as **`sort`** and **`order`**[21] could be applied on them.

**Exercise 10.19** *Check out the **`stringx`** package, which replaces the base R date-time processing functions with their more portable counterparts.*

**Exercise 10.20** ***`proc.time`** can be used to measure the time to execute a given expression:*

```
t0 <- proc.time()  # timer start
# ... to do - something time-consuming ...
sum(runif(1e7))  # whatever, just testing
## [1] 4999488
print(proc.time() - t0)  # elapsed time
##    user  system elapsed
##   0.104   0.006   0.111
```

*The function returns an object of the class `proc_time`. Inspect how it is represented internally.*

---

[20] The overloaded group generic **`Ops`** prevents us from adding or multiplying two dates and defines the meaning of the relational operators. As an exercise, check out its source code.

[21] See an exercise below on the use of `xtfrm`.

### 10.3.2 Factors

The `factor` class is often used to represent qualitative data, e.g., species, groups, types. In fact, `categorical` (our example class) was inspired by the built-in `factor`.

```
(x <- c("spam", "spam", "bacon", "sausage", "spam", "bacon"))
## [1] "spam"    "spam"    "bacon"   "sausage" "spam"    "bacon"
(f <- factor(x))
## [1] spam    spam    bacon   sausage spam    bacon
## Levels: bacon sausage spam
```

Note how factors are printed. There are no double quote characters around the labels. The list of levels is given at the end.

Internally, such objects are represented as integer vectors (Section 6.4.1) with elements between 1 and $k$. They are equipped with the special (as in Section 4.4.3) `levels` attribute, which is a character vector of length $k$[22].

```
class(f)
## [1] "factor"
typeof(f)
## [1] "integer"
unclass(f)
## [1] 3 3 1 2 3 1
## attr(,"levels")
## [1] "bacon"   "sausage" "spam"
attr(f, "levels")  # also: levels(f)
## [1] "bacon"   "sausage" "spam"
```

Factors are often used instead of character vectors defined over a small number of unique labels[23], where there is a need to manipulate their levels conveniently.

```
attr(f, "levels") <- c("a", "b", "c")  # also levels(f) <- c(....new...)
print(f)
## [1] c c a b c a
## Levels: a b c
```

The underlying integer codes remain the same.

Certain operations on vectors of small integers are relatively easy to express, especially those concerning element grouping: splitting, counting, and plotting (e.g., Figure 13.17). It is because the integer codes can naturally be used whilst indexing other vectors. Section 5.4 mentioned a few functions related to this, such as **`match`**, **`split`**, **`findInterval`**, and **`tabulate`**. Specifically, the latter can be implemented like "for each `i`, increase `count[factor_codes[i]]` by one".

---

[22] [70] states: *Factors are currently implemented using an integer array to specify the actual levels and a second array of names that are mapped to the integers. Rather unfortunately users often make use of the implementation in order to make some calculations easier. This, however, is an implementation issue and is not guaranteed to hold in all implementations of R.* Still, *fortunately*, this has been a de facto standard for factors for a very long time.

[23] Recall that there is a global (internal) string cache. Hence, having many duplicated strings is not a burden on the computer's memory.

**Exercise 10.21** *Study the source code of the **`factor`** function. Note the use of **`as.character`**, **`unique`**, **`order`**, and **`match`**.*

**Exercise 10.22** *Implement a simplified version of **`table`** based on **`tabulate`**. It should work for objects of the class `factor` and return a named numeric vector.*

**Exercise 10.23** *Implement a version of **`cut`** based on **`findInterval`**.*

---

**Important** The **`as.numeric`** method has not been overloaded for factors. Therefore, when we call the generic, the default method is used: it returns the underlying integer codes as-is. This can surprise unaware users when they play with factors representing integer numbers:

```
(g <- factor(c(11, 15, 16, 11, 13, 4, 15)))  # converts numbers to strings
## [1] 11 15 16 11 13 4  15
## Levels: 4 11 13 15 16
as.numeric(g)  # the underlying codes
## [1] 2 4 5 2 3 1 4
as.numeric(as.character(g))  # to get the numbers encoded
## [1] 11 15 16 11 13  4 15
```

Alas, support for factors is often hardcoded at the C language level. From the end user perspective, it makes this class behave less predictably. In particular, the manual overloading of certain methods for factor objects might have no effect.

---

---

**Important** If `f` is a factor, then `x[f]` does not behave like `x[`**`as.character`**`(f)]`, i.e., it is not indexing by labels using the `names` attribute. Instead, we get `x[`**`as.numeric`**`(f)]`; the underlying codes determine the positions.

```
h <- factor(c("a", "b", "a", "c", "a", "c"))
levels(h)[h]  # the same as c("a", "b", "c")[c(1, 2, 1, 3, 1, 3)]
## [1] "a" "b" "a" "c" "a" "c"
c(b="x", c="y", a="z")[h]  # names are not used whilst indexing
##   b   c   b   a   b   a
## "x" "y" "x" "z" "x" "z"
c(b="x", c="y", a="z")[as.character(h)]  # names are used now
##   a   b   a   c   a   c
## "z" "x" "z" "y" "z" "y"
```

More often than not, indexing by factors will happen "accidentally"[24], leaving us slightly puzzled. In particular, factors look much like character vectors when they are carried in data frames:

---

[24] (*) Up until R 4.0, many functions (including **`data.frame`** and **`read.csv`**) had the `stringsAsFactors` option set to `TRUE`; see **`help`**`("options")`. It resulted in all character vectors' being automatically converted to factors, e.g., when creating data frames (compare Section 12.1.5). Luckily, this is no longer the case. However, factor objects can still be encountered; for instance, check the class of `iris[["Species"]]`.

```
(df <- data.frame(A=c("x", "y", "z"), B=factor(c("x", "y", "z"))))
##   A B
## 1 x x
## 2 y y
## 3 z z
class(df[["A"]])
## [1] "character"
class(df[["B"]])
## [1] "factor"
```

**Important** Be careful when combining factors and not-factors:

```
x <- factor(c("A", "B", "A"))
c(x, "C")
## [1] "1" "2" "1" "C"
c(x, factor("C"))
## [1] A B A C
## Levels: A B C
```

**Exercise 10.24** *When subsetting a factor object, the result will inherit the* `levels` *attribute in its entirety:*

```
f[c(1, 2)]  # drop=FALSE
## [1] c c
## Levels: a b c
```

*However:*

```
f[c(1, 2), drop=TRUE]
## [1] c c
## Levels: c
```

*Implement your version of the* **`droplevels`** *function, which removes the unused attributes.*

**Exercise 10.25** *The replacement version of the index operator does not automatically add new levels to the modified object:*

```
x <- factor(c("A", "B", "A"))
`[<-`(x, 4, value="C")  # like in x[4] <- "C"
## Warning in `[<-.factor`(x, 4, value = "C"): invalid factor level, NA
##     generated
## [1] A    B    A    <NA>
## Levels: A B
```

*Implement a version of* **`` `[<-.factor` ``** *that has such a capability.*

### 10.3.3 Ordered factors

When creating factors, we can enforce a particular ordering and the number of levels:

```
x <- c("spam", "spam", "bacon", "sausage", "spam", "bacon")
factor(x, levels=c("eggs", "bacon", "sausage", "spam"))
## [1] spam    spam    bacon   sausage spam    bacon
## Levels: eggs bacon sausage spam
```

If we want the arrangement of the levels to define a linear ordering relation over the set of labels, we can call:

```
(f <- factor(x, levels=c("eggs", "bacon", "sausage", "spam"), ordered=TRUE))
## [1] spam    spam    bacon   sausage spam    bacon
## Levels: eggs < bacon < sausage < spam
class(f)
## [1] "ordered" "factor"
```

It yields an ordered factor, which enables comparisons like:

```
f[f >= "bacon"]  # what's not worse than bacon?
## [1] spam    spam    bacon   sausage spam    bacon
## Levels: eggs < bacon < sausage < spam
```

How is that possible? Well, based on information provided in this chapter, it will come as no surprise that it is because… someone has created a relational operator for objects of the class `ordered`.

### 10.3.4 Formulae (*)

Formulae are created using the `` `~` `` operator. Some R users employ them to specify widely-conceived *statistical models* in functions such as **`lm`** (e.g., linear regression), **`glm`** (generalised linear models like logistic regression etc.), **`aov`** (analysis of variance), **`wilcox.test`** (the two-sample Mann–Whitney–Wilcoxon test), **`aggregate`** (computing aggregates within data groups), **`boxplot`** (box-and-whisker plots for a variable split by a combination of factors), or **`plot`** (scatter plots); see also Chapter 11 of [59]. For instance, formulae can be used to describe symbolic relationships such as:

- "y as a linear combination of `x1`, `x2`, and `x3`",
- "y grouped/split by a combination of `x1` and `x2`",

where `y`, `x1`, etc., are, for example, column names in a data frame.

Formulae are interpreted by the corresponding functions, and not the R language itself. Thus, programmers are free to assign them any meaning. As their syntax is quite esoteric, beginners might find them confusing. Hence, we will postpone discussing them until Section 17.6. Luckily, the use of formulae can usually easily be avoided[25].

[25] For example, `lm.fit` can be used instead of `lm`. It is slightly more difficult to learn, but it has the added

## 10.4 (Over)using the forward pipe operator, `|>` (*)

The OOP paradigm is utile when we wish to define a new data type, perhaps even a hierarchy of types. Many development teams find it an efficient tool to organise larger pieces of software. However, in the data science and numerical computing domains, more often than not, we are the *consumers* of object orientation rather than class *designers*.

Thanks to the S3 method dispatch mechanism, our language is easily extensible. Something that mimics a new data type can easily be introduced. Most importantly, methods can be added[26] or removed during runtime, e.g., when importing external packages.

However, R is still a functional programming language, where functions are not just first-class citizens: they are privileged. In functional programming, the emphasis is on operations (verbs), not data (nouns). It leads to a very readable syntax. For example, assuming that **square**, x, and y are sensibly defined, the mean squared difference can be written as:

```
mean(square(x-y))  # read: mean of squares of differences
```

**Example 10.26** *Base R is extremely flexible. We can introduce new vocabulary as we please. In Section 12.3.7, we will study an example where we define:*

- ***group_by*** *(a function that splits a data frame with respect to a combination of levels in given named columns and returns a list of data frames with class* `list_dfs`*),*
- ***aggregate.list_dfs*** *(which applies an aggregation function on every column of all data frames in a given list), and*
- ***mean.list_dfs*** *(a specialised version of the former that calls* ***mean****).*

*The specifics do not matter now. Let's just consider the notation we use when the operations are chained:*

```
# select a few rows and columns from the `iris` data frame:
iris_subset <- iris[51:150, c("Sepal.Width", "Petal.Length", "Species")]
# compute the averages of all variables grouped by Species:
mean(group_by(iris_subset, "Species"))
##      Species            x  Mean
```

benefit of ensuring the user knows that the emergence of all model variables is not magical (especially the nonlinear/mixed effect terms).

[26] (*) In more traditional object-orientated programming languages, the method list is often sealed inside the class' definition (like in C++) or cumbersome patches must be applied to inject a method (like in Python; see also the concept of extension methods in C# or Kotlin and, to some extent, of class inheritance). When methods are parts of particular classes, there can be a lot of duplicated code. Functional OOP can be more developer-friendly as we can provide all methods related to roughly the same functionality in one spot.

```
## 1 versicolor  Sepal.Width 2.770
## 2 versicolor Petal.Length 4.260
## 3  virginica  Sepal.Width 2.974
## 4  virginica Petal.Length 5.552
```

*The functional syntax is very* reader-*centric (by the way, self-explanatory variable names and rich comments are priceless). We compute the mean in groups defined by `Species` in a subset of the `iris` data frame. All* verbs *appear on the left side of the expression, with the final (the most important?) operation being listed first.*

Nonetheless, when implementing more complex data processing pipelines, *programmers* think in different categories: "first, we need to do this, then we need to do that, and afterwards…". When they write their ideas down, they have to press Home and End or arrow keys a few times to move the caret to the right places:

```
finally(thereafter(then(first(x))))
```

As people are inherently lazy, they might want to "optimise" their workflow to save a bit of energy. Thus, many popular languages rely on message-passing syntax, where operations are propagated (and written) left-to-right instead of inside-out. For instance, `object.method1().method2()` might mean "call **method1** on the `object` and then call **method2** on the result". Here, the `object` is *told* what to do.

Since R 4.1.0, we have the *pipe operator*[27], `` `|>` ``. It is merely syntactic sugar for translating between the message-passing and function-centric notion. In a nutshell, instead of writing:

```
h(g(f(x), y))
mean(square(x-y))
mean(group_by(iris_subset, "Species"))
```

we have the following equivalent forms:

```
x |> f() |> g(y) |> h()
(x-y) |> square() |> mean()
iris_subset |> group_by("Species") |> mean()
```

Such syntax is developer-centric. It might be faster to write. It emphasises the order in which the operations are executed. However, we must stress that there is nothing that cannot be achieved through the function-centric form and perhaps a few auxiliary variables. As this book is minimalist by design, we refrain ourselves from using it. Those unconvinced should take note of the following.

First, expressions on the right side of the pipe operator must always be proper calls. Therefore, the use of round brackets is obligatory. Thus, when passing anonymous functions, we must write:

[27] It was inspired by `` `|` `` in Bash and `` `|>@` `` in F# and Julia (which are part of the language specification). Also, there is a `` `%>%` `` operator (and related ones) in the R package **magrittr**.

```
runif(10) |> (function(x) mean((x-mean(x))^2))()  # note the "()" at the end
## [1] 0.078184
```

Peculiarly, in R 4.1.0, a “shorthand” notation for creating functions was introduced. We can save seven keystrokes and scribble “`\(...) expr`” instead of “`function(...) expr`”.

```
runif(10) |> (\(x) mean((x-mean(x))^2))()  # again: "()" at the end
## [1] 0.078184
```

Also, the placeholder `` `_` `` can be used on the right side of the pipe operator (only once) to indicate that the left side must be matched with a specific *named* argument of the function to be called. Otherwise, the left side is always passed as the first argument. Therefore, the two following expressions are equivalent:

```
x |> median() |> `-`(e1=x, e2=_) |> abs() |> median()
median(abs(x-median(x)))
```

---

**Note** When writing code interactively, we may sometimes benefit from using the rightward `` `->` `` operator. Suffice it to say that “`name <- value`” and “`value -> name`” are synonymous. This way, we can type some lengthy code, store the result in an intermediate variable, and then continue in the next line (possibly referring to that auxiliary value more than once). For instance:

```
runif(10) -> .
mean((.-mean(.))^2)
## [1] 0.078184
```

Recall that `` `.` `` is as valid a variable name as any other one. Another example:

```
iris[, c("Sepal.Width", "Petal.Length", "Species")] -> .
.[ .[, "Species"] %in% c("versicolor", "virginica"), ] -> .
mean(group_by(., "Species"))
##      Species            x  Mean
## 1 versicolor  Sepal.Width 2.770
## 2 versicolor Petal.Length 4.260
## 3  virginica  Sepal.Width 2.974
## 4  virginica Petal.Length 5.552
```

---

## 10.5 S4 classes (*)

The S3-style OOP is based on a brilliantly simple idea: calling a generic `f(x)` dispatches automatically to a method `f.class_of_x(x)` or `f.default(x)` in the case where the former does not exist. Naturally, S3 has some inherent limitations:

- classes cannot be formally defined; the `class` attribute may be assigned arbitrarily to any object[28],
- argument dispatch is performed only[29] with regard to one data type[30].

In most cases, and with an appropriate level of mindfulness, they are not a problem at all. However, it is a typical condition of programmers who come to our world from more mainstream languages (e.g., C++ or Java; yours truly included) until they appreciate the true beauty of R's being somewhat different. Before they fully develop such an acquired taste, though, they grow restless as "R is has no real OOP because it lacks polymorphism, encapsulation, formal inheritance, and so on, and something must be done about it!". The truth is that it had not had to, but with high probability, it would have anyway in one way or another.

And so the fourth version of the S language was introduced in 1998 (see [10]). It brought a new object-orientated system, which we are used to referring to as S4. Its R version is defined by the **methods** package. Below we discuss it briefly. For more details, see `help("Classes_Details")` and `help("Methods_Details")` as well as [11] and [12].

**Note** (*) S4 was loosely inspired by the Common Lisp Object System (with its `defclass`, `defmethod`, etc.; see, e.g., [21]). In the current author's opinion, the S4 system is somewhat of an afterthought. Due to appendages like this, R seems like a patchwork language. Suffice it to say that it was not the last attempt to introduce a "real" OOP in the overall functional R: the story will resume in Section 16.1.5.

The main issue with all the supplementary OOP approaches is that each of them is *parallel* to S3 which never lost its popularity and is still in the very core of our language. We are thus covering them only for the sake of completeness for the readers might come across such objects. In particular, we shall explain the meaning of a notation like `x@slot`. Moreover, in Section 11.4.7 we mention the `Matrix` class which is perhaps the most prominent showcase of S4.

Nonetheless, the current author advises taking with a pinch of salt statements such as "for new projects, it is recommended to use the more flexible and robust S4 scheme provided in the **methods** package" mentioned in `help("UseMethod")`.

[28] A partial solution to this could involve defining a method like `validate.class_name`, which is called frequently and which checks whether a given object enjoys a few desirable constraints.

[29] Certain functions implement ad hoc workarounds (see, e.g., `cbind`, which dispatches to `cbind.data.frame` if one argument is a data frame and the remaining ones are vectors or matrices). Also, we said in the previous chapter that binary operators consider the classes of both operands.

[30] Hypothetically, we can imagine an OOP system relying on methods named like `method.class_name1.class_name2` where dispatching is based on two argument types. This would be beautiful, but it is not the case in R.

### 10.5.1 Defining S4 classes

An S4 class can be formally registered through a call to **setClass**. For instance:

```
library("methods")  # in the case where it is not auto-loaded
setClass("qualitative", slots=c(data="integer", levels="character"))
```

We defined a class named `qualitative` (similarity to our own `categorical` and the built-in `factor` S3 classes is intended). It has two slots: `data` and `levels` being integer and character vectors, respectively. This notation is already outlandish. There is no assignment suggesting that we have introduced something novel.

An object of the above class can be instantiated by calling **new**:

```
z <- new("qualitative", data=c(1L, 2L, 2L, 1L, 1L), levels=c("a", "b"))
print(z)
## An object of class "qualitative"
## Slot "data":
## [1] 1 2 2 1 1
##
## Slot "levels":
## [1] "a" "b"
```

That z is of this class can be verified by calling **is**.

```
is(z, "qualitative")
## [1] TRUE
class(z)  # also: attr(z, "class")
## [1] "qualitative"
## attr(,"package")
## [1] ".GlobalEnv"
```

---

**Important** A few R packages import the **methods** package only to get access the handy **is** function. It does not mean they are defining new S4 classes.

---

**Note** S4 objects are marked as being of the following basic type:

```
typeof(z)
## [1] "S4"
```

See Section 1.12 of [69] for technical details on how they are internally represented. In particular, in our case, all the slots are simply stored as object attributes:

```
attributes(z)
## $data
## [1] 1 2 2 1 1
##
## $levels
## [1] "a" "b"
##
## $class
## [1] "qualitative"
## attr(,"package")
## [1] ".GlobalEnv"
```

### 10.5.2 Accessing slots

Reading or writing slot contents can be done via the `` `@` `` operator and the **slot** function or their replacement versions.

```
z@data  # or slot(z, "data")
## [1] 1 2 2 1 1
z@levels <- c("A", "B")
```

**Note** The `` `@` `` operator can only be used on S4 objects, and some sanity checks are automatically performed:

```
z@unknown <- "spam"
## Error: ! 'unknown' is not a slot in class "qualitative"
z@data <- "spam"
## Error: ! assignment of an object of class "character" is not valid for
##     @'data' in an object of class "qualitative"; is(value, "integer") is
##     not TRUE
```

### 10.5.3 Defining methods

For the S4 counterparts of the S3 generics (Section 10.2), see **help**("setGeneric"). Luckily, there is a reasonable degree of interoperability between the S3 and S4 systems.

Let's introduce a new method for the well-known **as.character** generic. Instead of defining **as.character.qualitative**, we need to register a new routine with **setMethod**.

```
setMethod(
    "as.character",      # name of the generic
    "qualitative",       # class of 1st arg; or: signature=c(x="qualitative")
    function(x, ...)     # method definition
```

```
        x@levels[x@data]
)
```

Testing:

```
as.character(z)
## [1] "A" "B" "B" "A" "A"
```

**show** is the S4 counterpart of **print**:

```
setMethod(
    "show",
    "qualitative",
    function(object)
    {
        x <- as.character(object)
        print(x)  # calls `print.default`
        cat(sprintf("Categories: %s\n",
            paste(object@levels, collapse=", ")))
    }
)
```

Interestingly, it is involved automatically on a call to **print**:

```
print(z)  # calls `show` for `qualitative`
## [1] "A" "B" "B" "A" "A"
## Categories: A, B
```

Methods that dispatch on the type of multiple arguments are also possible. For example:

```
setMethod(
    "split",
    c(x="ANY", f="qualitative"),
    function (x, f, drop=FALSE, ...)
        split(x, as.character(f), drop=drop, ...)
)
```

It permits the first argument to be of any type (like a default method). Moreover, here is its version tailored for matrices (see Chapter 11).

```
setMethod(
    "split",
    c(x="matrix", f="qualitative"),
    function (x, f, drop=FALSE, ...)
        lapply(
            split(seq_len(NROW(x)), f, drop=drop, ...),  # calls the above
            function(i) x[i, , drop=FALSE])
)
```

Some tests:

```
A <- matrix(1:35, nrow=5)  # whatever
split(A, z)    # matrix, qualitative
## $A
##      [,1] [,2] [,3] [,4] [,5] [,6] [,7]
## [1,]    1    6   11   16   21   26   31
## [2,]    4    9   14   19   24   29   34
## [3,]    5   10   15   20   25   30   35
##
## $B
##      [,1] [,2] [,3] [,4] [,5] [,6] [,7]
## [1,]    2    7   12   17   22   27   32
## [2,]    3    8   13   18   23   28   33
split(1:5, z)  # ANY, qualitative
## $A
## [1] 1 4 5
##
## $B
## [1] 2 3
```

**Exercise 10.27** *Overload the `` `[` `` operator for the `qualitative` class.*

### 10.5.4 Defining constructors

We can also overload the **`initialize`** method, which is automatically called by **`new`**:

```
setMethod(
    "initialize",   # note the American spelling
    "qualitative",
    function(.Object, x)
    {  # the method itself
        if (!is.character(x))
            x <- as.character(x)  # see above
        xu <- unique(sort(x))  # drops NAs

        .Object@data <- match(x, xu)
        .Object@levels <- xu

        .Object  # return value - a modified object
    }
)
```

This constructor yields instances of the class `qualitative` based on an object coercible to a character vector. For example:

```
w <- new("qualitative", c("a", "c", "a", "a", "d", "c"))
print(w)
## [1] "a" "c" "a" "a" "d" "c"
## Categories: a, c, d
```

**Exercise 10.28** *Set up a validating method for our class; see* **help**("setValidity").

### 10.5.5 Inheritance

New S4 classes can be derived from existing ones. For instance:

```
setClass("binary", contains="qualitative")
```

It is a child class that inherits all slots from its parent. We can overload its initialisation method:

```
setMethod(
    "initialize",
    "binary",
    function(.Object, x)
    {
        if (!is.logical(x))
            x <- as.logical(x)
        x <- as.character(as.integer(x))
        xu <- c("0", "1")
        .Object@data <- match(x, xu)
        .Object@levels <- xu
        .Object
    }
)
```

Testing:

```
new("binary", c(TRUE, FALSE, TRUE, FALSE, NA, TRUE))
## [1] "1" "0" "1" "0" NA  "1"
## Categories: 0, 1
```

We can still use the **show** method of the parent class.

## 10.6 Exercises

**Exercise 10.29** *Answer the following questions.*

- *How to display the source code of the default methods for* **head** *and* **tail**?
- *Can there be, at the same time, one object of the class* **c**("A", "B") *and another one of the class* **c**("B", "A")?
- *If* f *is a factor, what are the relationships between* **as.character**(f), **as.numeric**(f), **as.character**(**as.numeric**(f)), *and* **as.numeric**(**as.character**(f))?
- *If* x *is a named vector and* f *is a factor, is* x[f] *equivalent to* x[**as.character**(f)] *or rather* x[**as.numeric**(f)]?

**Exercise 10.30** *A user calls:*

```
plot(x, y, col="red", ylim=c(1, max(x)), log="y")
```

*where* x *and* y *are numeric vectors. Consult* `help("plot")` *for the meaning of the* `ylim` *and* `log` *arguments. Was that straightforward?*

**Exercise 10.31** *Explain why the two following calls return significantly different results.*

```
c(Sys.Date(), "1970-01-01")
## [1] "2026-07-30" "1970-01-01"
c("1970-01-01", Sys.Date())
## [1] "1970-01-01" "20664"
```

*Propose a workaround.*

**Exercise 10.32** *Write methods* `head` *and* `tail` *for our example* `categorical` *class.*

**Exercise 10.33** *(*) Write an R package that defines S3 class* `categorical`*. Add a few methods for this class. Note the need to use the* `S3method` *directive in the* NAMESPACE *file; see [66].*

**Exercise 10.34** *Inspect the result of a call to* `binom.test(79, 100)` *and to* `rle(c(1, 1, 1, 4, 3, 3, 3, 3, 3, 2, 2))`*. Find the methods responsible for such objects' pretty-printing.*

**Exercise 10.35** *Read more about the* `connection` *class. In particular, see the* Value *section of* `help("connections")`*.*

**Exercise 10.36** *Read about the subsetting operators overloaded for the* `package_version` *class; see* `help("numeric_version")`*.*

**Exercise 10.37** *There are* `xtfrm` *methods overloaded for classes such as* `numeric_version`*,* `difftime`*,* `Date`*, and* `factor`*. Find out how they work and where they might be of service (especially in relation to* `order` *and* `sort`*; see also Section 12.3.1).*

**Exercise 10.38** *Give an example where* `split(x, list(y1, y2))` *(with default arguments) will fail to generate the correct result.*

**Exercise 10.39** *Write a function that determines the mode, i.e., the most frequently occurring value in a given object of the class* `factor`*. If the mode is not unique, return a randomly chosen one (each with the same probability).*

**Exercise 10.40** *Implement your version of the* `gl` *function.*

# 11

# *Matrices and other arrays*

When we equip an atomic or generic vector with the `dim` attribute, it automatically becomes an object of the S3 class `array`. In particular, two-dimensional arrays (of the primary S3 class `matrix`) allow us to represent *tabular* data where items are aligned into rows and columns:

```
structure(1:6, dim=c(2, 3))  # a matrix with two rows and three columns
##      [,1] [,2] [,3]
## [1,]    1    3    5
## [2,]    2    4    6
```

Combined with the fact that there are many functions overloaded for the `matrix` class, we have just opened up a whole world of new possibilities, which we explore in this chapter. Namely, we will discuss how to perform the algebraic operations such as matrix multiplication, transpose, finding eigenvalues, and performing various decompositions. We will also cover data wrangling operations such as array subsetting and column- and rowwise aggregation. Furthermore, the next chapter will present data frames: matrix-like objects whose columns can be of any (not necessarily the same) type.

---

**Important** Oftentimes, a numeric matrix with $n$ rows and $m$ columns is used to represent $n$ points (samples, observations) in an $m$-dimensional space (with $m$ numeric features or variables), $\mathbb{R}^m$.

---

## 11.1 Creating arrays

### 11.1.1 `matrix` and `array`

A matrix can be conveniently created using the following function.

```
(A <- matrix(1:6, byrow=TRUE, nrow=2))
##      [,1] [,2] [,3]
## [1,]    1    2    3
## [2,]    4    5    6
```

It converted an atomic vector of length six to a matrix with two rows. The number of columns was determined automatically (`ncol=3` could have been passed, additionally or instead, to get the same result).

---

**Important** By default, the elements of the input vector are read column by column:

```
matrix(1:6, ncol=3)  # byrow=FALSE
##      [,1] [,2] [,3]
## [1,]    1    3    5
## [2,]    2    4    6
```

---

A matrix can be equipped with an attribute that defines dimension names, being a list of two character vectors of appropriate sizes which label each row and column:

```
matrix(1:6, byrow=TRUE, nrow=2, dimnames=list(c("x", "y"), c("a", "b", "c")))
##   a b c
## x 1 2 3
## y 4 5 6
```

Alternatively, to create a matrix, we can use the **array** function. It requires the number of rows and columns to be specified explicitly.

```
array(1:6, dim=c(2, 3))
##      [,1] [,2] [,3]
## [1,]    1    3    5
## [2,]    2    4    6
```

The elements were consumed in the column-major order.

Arrays of other dimensionalities are also possible. Here is a one-dimensional array:

```
array(1:6, dim=6)
## [1] 1 2 3 4 5 6
```

When printed, it is indistinguishable from an atomic vector (but the `class` attribute is still set to `array`). And now for something completely different: a three-dimensional array of size $3 \times 4 \times 2$:

```
array(1:24, dim=c(3, 4, 2))
## , , 1
##
##      [,1] [,2] [,3] [,4]
## [1,]    1    4    7   10
## [2,]    2    5    8   11
## [3,]    3    6    9   12
##
## , , 2
##
##      [,1] [,2] [,3] [,4]
```

```
## [1,]   13   16   19   22
## [2,]   14   17   20   23
## [3,]   15   18   21   24
```

It can be thought of as two matrices of size $3 \times 4$ (because how else can we print out a 3D object on a 2D console?).

The **array** function can be fed with the `dimnames` argument, too. For instance, the above three-dimensional hypertable would require a list of three character vectors of sizes 3, 4, and 2, respectively.

**Exercise 11.1** *Verify that 5-dimensional arrays can also be created.*

### 11.1.2 Promoting and stacking vectors

We can promote an ordinary vector to a *column vector*, i.e., a matrix with one column, by calling:

```
as.matrix(1:2)
##      [,1]
## [1,]    1
## [2,]    2
cbind(1:2)
##      [,1]
## [1,]    1
## [2,]    2
```

and to a row vector:

```
t(1:3)  # transpose
##      [,1] [,2] [,3]
## [1,]    1    2    3
rbind(1:3)
##      [,1] [,2] [,3]
## [1,]    1    2    3
```

Actually, **cbind** and **rbind** stand for column- and row-bind. They permit multiple vectors and matrices to be stacked one after/below another:

```
rbind(1:4, 5:8, 9:10, 11)  # row-bind
##      [,1] [,2] [,3] [,4]
## [1,]    1    2    3    4
## [2,]    5    6    7    8
## [3,]    9   10    9   10
## [4,]   11   11   11   11
cbind(1:4, 5:8, 9:10, 11)  # column-bind
##      [,1] [,2] [,3] [,4]
## [1,]    1    5    9   11
```

```
## [2,]    2    6   10   11
## [3,]    3    7    9   11
## [4,]    4    8   10   11
cbind(1:2, 3:4, rbind(11:13, 21:23))  # vector, vector, 2x3 matrix
##      [,1] [,2] [,3] [,4] [,5]
## [1,]    1    3   11   12   13
## [2,]    2    4   21   22   23
```

and so forth. Unfortunately, the *generalised* recycling rule is not implemented in full:

```
cbind(1:4, 5:8, cbind(9:10, 11))  # different from cbind(1:4, 5:8, 9:10, 11)
## Warning in cbind(1:4, 5:8, cbind(9:10, 11)): number of rows of result is
##     not a multiple of vector length (arg 1)
##      [,1] [,2] [,3] [,4]
## [1,]    1    5    9   11
## [2,]    2    6   10   11
```

Note that the first two arguments were of length four.

### 11.1.3 Simplifying lists

**`simplify2array`** is an extension of the **`unlist`** function. Given a list of atomic vectors, each of length one, it will return a flat atomic vector. However, if longer vectors of the same lengths are given, they will be converted to a matrix.

```
simplify2array(list(1, 11, 21))  # each of length one
## [1]  1 11 21
simplify2array(list(1:3, 11:13, 21:23, 31:33))  # each of length three
##      [,1] [,2] [,3] [,4]
## [1,]    1   11   21   31
## [2,]    2   12   22   32
## [3,]    3   13   23   33
simplify2array(list(1, 11:12, 21:23))  # no can do (without warning!)
## [[1]]
## [1] 1
##
## [[2]]
## [1] 11 12
##
## [[3]]
## [1] 21 22 23
```

In the second example, each vector becomes a separate column of the resulting matrix, which can easily be justified by the fact that matrix elements are stored in a columnwise order.

**Example 11.2** *Quite a few functions call the foregoing automatically; compare the `simplify` argument to **`apply`**, **`sapply`**, **`tapply`**, or **`replicate`**, and the `SIMPLIFY` (sic!) argument to **`mapply`**. For instance, **`sapply`** combines **`lapply`** with **`simplify2array`**:*

```
min_mean_max <- function(x) c(Min=min(x), Mean=mean(x), Max=max(x))
sapply(split(iris[["Sepal.Length"]], iris[["Species"]]), min_mean_max)
##      setosa versicolor virginica
## Min   4.300      4.900     4.900
## Mean  5.006      5.936     6.588
## Max   5.800      7.000     7.900
```

*Take note of what constitutes the columns of the return matrix.*

**Exercise 11.3** *Inspect the behaviour of **`as.matrix`** on list arguments. Write your version of **`simplify2array`** named **`as.matrix.list`** that* always *returns a matrix. If a list of non-equisized vectors is given, fill the missing cells with NAs and generate a warning.*

---

**Important** Sometimes a call to **`do.call(cbind, x)`** might be a better idea than a referral to **`simplify2array`**. Provided that x is a list of atomic vectors, it *always* returns a matrix: shorter vectors are recycled (which might be welcome, but not necessarily).

```
do.call(cbind, list(a=c(u=1), b=c(v=2, w=3), c=c(i=4, j=5, k=6)))
## Warning in (function (..., deparse.level = 1) : number of rows of result
##     is not a multiple of vector length (arg 2)
##   a b c
## i 1 2 4
## j 1 3 5
## k 1 2 6
```

---

**Example 11.4** *Consider a toy named list of numeric vectors:*

```
x <- list(a=runif(10), b=rnorm(15))
```

*Compare the results generated by **`sapply`** (which calls **`simplify2array`**):*

```
sapply(x, function(e) c(Mean=mean(e)))
##  a.Mean  b.Mean
## 0.57825 0.12431
sapply(x, function(e) c(Min=min(e), Max=max(e)))
##             a       b
## Min 0.045556 -1.9666
## Max 0.940467  1.7869
```

*with its version based on **`do.call`** and **`cbind`**:*

```
sapply2 <- function(...)
    do.call(cbind, lapply(...))

sapply2(x, function(e) c(Mean=mean(e)))
##            a       b
## Mean 0.57825 0.12431
```

```
sapply2(x, function(e) c(Min=min(e), Max=max(e)))
##            a       b
## Min 0.045556 -1.9666
## Max 0.940467  1.7869
```

*Notice that **`sapply`** may return an atomic vector with somewhat surprising `names`.*

More examples appear in Section 12.3.7.

### 11.1.4 Beyond numeric arrays

Arrays based on non-numeric vectors are also possible. For instance, we will later stress that matrix comparisons are performed elementwisely. They spawn logical matrices:

```
A >= 3
##       [,1]  [,2] [,3]
## [1,] FALSE FALSE TRUE
## [2,]  TRUE  TRUE TRUE
```

Matrices of character strings can be useful too:

```
matrix(strrep(LETTERS[1:6], 1:6), ncol=3)
##      [,1] [,2]   [,3]
## [1,] "A"  "CCC"  "EEEEE"
## [2,] "BB" "DDDD" "FFFFFF"
```

And, of course, complex matrices:

```
A + 1i
##      [,1] [,2] [,3]
## [1,] 1+1i 2+1i 3+1i
## [2,] 4+1i 5+1i 6+1i
```

We are not limited to *atomic* vectors. Lists can be a basis for arrays as well:

```
matrix(list(1, 11:21, "A", list(1, 2, 3)), nrow=2)
##      [,1]       [,2]
## [1,] 1          "A"
## [2,] integer,11 list,3
```

Certain elements are not *displayed* correctly, but they *are* still there.

### 11.1.5 Internal representation

An object of the S3 class `array` is an atomic vector or a list equipped with the `dim` attribute being a vector of nonnegative integers. Interestingly, we do not have to set the class attribute explicitly: the accessor function **class** will return an implicit[1] class anyway.

```
class(1)  # atomic vector
## [1] "numeric"
class(structure(1, dim=rep(1, 1)))  # 1D array (vector)
## [1] "array"
class(structure(1, dim=rep(1, 2)))  # 2D array (matrix)
## [1] "matrix" "array"
class(structure(1, dim=rep(1, 3)))  # 3D array
## [1] "array"
```

Note that a two-dimensional array is additionally of the matrix class.

Optional dimension names are represented by means of the dimnames attribute, which is a list of $d$ character vectors, where $d$ is the array's dimensionality.

```
(A <- structure(1:6, dim=c(2, 3), dimnames=list(letters[1:2], LETTERS[1:3])))
##   A B C
## a 1 3 5
## b 2 4 6
dim(A)  # or attr(A, "dim")
## [1] 2 3
dimnames(A)  # or attr(A, "dimnames")
## [[1]]
## [1] "a" "b"
##
## [[2]]
## [1] "A" "B" "C"
```

---

**Important** Internally, elements in an array are stored in the column-major (Fortran) order:

```
as.numeric(A)  # drop all attributes to reveal the underlying numeric vector
## [1] 1 2 3 4 5 6
```

Setting byrow=TRUE in a call to the **matrix** function only affects the order in which this constructor *reads* a given source vector, not the resulting column/row-majorness.

```
(B <- matrix(1:6, ncol=3, byrow=TRUE))
##      [,1] [,2] [,3]
## [1,]    1    2    3
## [2,]    4    5    6
as.numeric(B)
## [1] 1 4 2 5 3 6
```

---

The two said special attributes can be modified through the replacement functions

[1] See Section 10.1. Interestingly, calling **unclass** on a matrix has no effect.

`` `dim<-` `` and `` `dimnames<-` `` (and, of course, `` `attr<-` `` as well). In particular, changing `dim` does not alter the underlying atomic vector. It only affects how other functions, including the corresponding **`print`** method, interpret their placement on a virtual grid:

```
`dim<-`(A, c(3, 2))  # not the same as the transpose of `A`
##      [,1] [,2]
## [1,]    1    4
## [2,]    2    5
## [3,]    3    6
```

We obtained a different *view* of the same *flat* data vector. Also, the `dimnames` attribute was dropped because its size became incompatible with the newly requested dimensionality.

**Exercise 11.5** *Study the source code of the* ***`nrow`****,* ***`NROW`****,* ***`ncol`****,* ***`NCOL`****,* ***`rownames`****,* ***`row.names`****, and* ***`colnames`*** *functions.*

Interestingly, for one-dimensional arrays, the **`names`** function returns a reasonable value (based on the `dimnames` attribute, which is a list with one character vector), despite the `names` attribute's not being set.

What is more, the `dimnames` attribute itself can be named:

```
names(dimnames(A)) <- c("ROWS", "COLUMNS")
print(A)
##     COLUMNS
## ROWS A B C
##    a 1 3 5
##    b 2 4 6
```

It is still a numeric matrix, but its presentation has been slightly prettified.

**Exercise 11.6** ***`outer`*** *applies an elementwisely vectorised function on each pair of elements from two vectors, forming a two-dimensional result grid. Implement it yourself based on two calls to* ***`rep`****. Some examples:*

```
outer(c(x=1, y=10, z=100), c(a=1, b=2, c=3, d=4), "*")  # multiplication
##     a   b   c   d
## x   1   2   3   4
## y  10  20  30  40
## z 100 200 300 400
outer(c("A", "B"), 1:8, paste, sep="-")  # concatenate strings
##      [,1]  [,2]  [,3]  [,4]  [,5]  [,6]  [,7]  [,8]
## [1,] "A-1" "A-2" "A-3" "A-4" "A-5" "A-6" "A-7" "A-8"
## [2,] "B-1" "B-2" "B-3" "B-4" "B-5" "B-6" "B-7" "B-8"
```

**Exercise 11.7** *Show how* ***`match(y, z)`*** *can be implemented using* ***`outer`****. Is its time and memory complexity optimal, though?*

**Exercise 11.8** ***`table`*** *creates a contingency matrix/array that counts the number of unique elements or unique pairs of corresponding items from one or more vectors of equal lengths. Write its one- and two-argument version based on* ***`tabulate`****. For example:*

```
tips <- read.csv(paste0("https://github.com/gagolews/teaching-data/raw/",
    "master/other/tips.csv"), comment.char="#")  # a data.frame (list)
table(tips[["day"]])
##
##  Fri  Sat  Sun Thur
##   19   87   76   62
table(tips[["smoker"]], tips[["day"]])
##
##        Fri Sat Sun Thur
##   No     4  45  57   45
##   Yes   15  42  19   17
```

## 11.2 Array indexing

Array subsetting can be performed by means of the overloaded[2] `` `[` `` method.

### 11.2.1 Arrays are built on basic vectors

Consider two example matrices:

```
(A <- matrix(1:12, byrow=TRUE, nrow=3))
##      [,1] [,2] [,3] [,4]
## [1,]    1    2    3    4
## [2,]    5    6    7    8
## [3,]    9   10   11   12
(B <- `dimnames<-`(A, list(  # copy of `A` with `dimnames` set
    c("a", "b", "c"),       # row labels
    c("x", "y", "z", "w")  # column labels
)))
##   x  y  z  w
## a 1  2  3  4
## b 5  6  7  8
## c 9 10 11 12
```

Subsetting based on one indexer (as in Chapter 5) will refer to the underlying flat vector. For instance:

```
A[6]
## [1] 10
```

It is the element in the third row, second column. Recall that values are stored in the column-major order.

[2] Hidden deeply at the C language level; see `help("[")`.

### 11.2.2 Selecting individual elements

Our example $3 \times 4$ real matrix $\mathbf{A} \in \mathbb{R}^{3\times 4}$ is like:

$$\mathbf{A} = \left[\begin{array}{cccc} a_{1,1} & a_{1,2} & a_{1,3} & a_{1,4} \\ a_{2,1} & a_{2,2} & a_{2,3} & a_{2,4} \\ a_{3,1} & a_{3,2} & a_{3,3} & a_{3,4} \end{array}\right] = \left[\begin{array}{cccc} 1 & 2 & 3 & 4 \\ 5 & 6 & 7 & 8 \\ 9 & 10 & 11 & 12 \end{array}\right].$$

Matrix elements are aligned in a two-dimensional grid. Hence, we can pinpoint a cell using two indexes. In mathematical notation, $a_{i,j}$ refers to the $i$-th row and the $j$-th column. Similarly in R:

```
A[3, 2]  # the third row, the second column
## [1] 10
B["c", "y"]  # using dimnames == B[3, 2]
## [1] 10
```

### 11.2.3 Selecting rows and columns

Some textbooks, and we are fond of this notation here as well, mark with $\mathbf{a}_{i,\cdot}$ a vector that consists of all the elements in the $i$-th row and with $\mathbf{a}_{\cdot,j}$ all items in the $j$-th column. In R, this corresponds to one of the indexers being left out.

```
A[3, ]  # the third row
## [1]  9 10 11 12
A[, 2]  # the second column
## [1]  2  6 10
B["c", ]  # or B[3, ]
##  x  y  z  w
##  9 10 11 12
B[, "y"]  # or B[, 2]
##  a  b  c
##  2  6 10
```

Let's stress that `A[1]`, `A[1, ]`, and `A[, 1]` have different meanings. Also, we see that the results' `dimnames` are adjusted accordingly; see also **`unname`**, which can take care of them once and for all.

**Exercise 11.9** *Use **`duplicated`** to remove repeating rows in a given numeric matrix (see also **`unique`**).*

### 11.2.4 Dropping dimensions

Extracting an individual element or a single row/column from a matrix brings about an atomic vector. If the resulting object's `dim` attribute consists of 1s only, it will be removed whatsoever; see also the **`drop`** function which removes the dimensions with only one level.

In order to obtain proper row and column vectors, we can request the preservation of the dimensionality of the output object (and, more precisely, the length of `dim`). This can be done by passing `drop=FALSE` to `` `[` ``.

```
A[1, 2, drop=FALSE]  # the first row, second column
##      [,1]
## [1,]    2
A[1,  , drop=FALSE]  # the first row
##      [,1] [,2] [,3] [,4]
## [1,]    1    2    3    4
A[ , 2, drop=FALSE]  # the second column
##      [,1]
## [1,]    2
## [2,]    6
## [3,]   10
```

**Important** Unfortunately, the `drop` argument defaults to `TRUE`. Many bugs could be avoided otherwise, primarily when the indexers are generated programmatically.

**Note** For list-based matrices, we can also use a multi-argument version of `` `[[` `` to extract the individual elements.

```
C <- matrix(list(1, 11:12, 21:23, 31:34), nrow=2)
C[1, 2]  # for `[`, input type is the same as the output type, hence a list
## [[1]]
## [1] 21 22 23
C[1, 2, drop=FALSE]
##      [,1]
## [1,] integer,3
C[[1, 2]]  # extract
## [1] 21 22 23
```

### 11.2.5 Selecting submatrices

Indexing based on two vectors, both of length two or more, extracts a sub-block of a given matrix.

```
A[1:2, c(1, 2, 4)]  # rows 1 and 2, columns 1, 2, and 4
##      [,1] [,2] [,3]
## [1,]    1    2    4
## [2,]    5    6    8
B[c("a", "b"), -3]  # some rows, omit the third column
##   x y w
## a 1 2 4
## b 5 6 8
```

Note again that we have `drop=TRUE` by default, which affects the operator's behaviour if one of the indexers is a scalar.

```
A[c(1, 3), 3]
## [1]  3 11
A[c(1, 3), 3, drop=FALSE]
##      [,1]
## [1,]    3
## [2,]   11
```

**Exercise 11.10** *Define the* **`split`** *method for the* `matrix` *class that returns a list of n matrices when given a matrix with n rows and an object of the class* `factor` *of length n (or a list of such objects). For example:*

```
split.matrix <- ...to.do...
A <- matrix(1:12, nrow=3)  # matrix whose rows are to be split
s <- factor(c("a", "b", "a"))  # determines a grouping of rows
split(A, s)
## $a
##      [,1] [,2] [,3] [,4]
## [1,]    1    4    7   10
## [2,]    3    6    9   12
##
## $b
##      [,1] [,2] [,3] [,4]
## [1,]    2    5    8   11
```

### 11.2.6 Selecting elements based on logical vectors

Logical vectors can also be used as indexers, with consequences that are not hard to guess:

```
A[c(TRUE, FALSE, TRUE), -1]  # select 1st and 3rd row, omit 1st column
##      [,1] [,2] [,3]
## [1,]    4    7   10
## [2,]    6    9   12
B[B[, "x"]>1 & B[, "x"]<=9, ]  # all rows where x's contents are in (1, 9]
##   x  y  z  w
## b 5  6  7  8
## c 9 10 11 12
A[2, colMeans(A)>6, drop=FALSE]  # 2nd row and the columns whose means > 6
##      [,1] [,2]
## [1,]    8   11
```

**Note** Section 11.3 notes that comparisons involving matrices are performed in an elementwise manner. For example:

```
A>7
##       [,1]  [,2]  [,3] [,4]
## [1,] FALSE FALSE FALSE TRUE
## [2,] FALSE FALSE  TRUE TRUE
## [3,] FALSE FALSE  TRUE TRUE
```

Such logical matrices can be used to subset other matrices of the same size. This kind of indexing always gives rise to a (flat) vector:

```
A[A>7]
## [1]  8  9 10 11 12
```

It is nothing else than the single-indexer subsetting involving two flat vectors (a numeric and a logical one). The `dim` attributes are not considered here.

---

**Exercise 11.11** *Implement your versions of* ***`max.col`***, ***`lower.tri`***, *and* ***`upper.tri`***.

### 11.2.7 Selecting based on two-column numeric matrices

We can also index a matrix `A` by a two-column matrix of positive integers `I`. For instance:

```
(I <- cbind(
    c(1, 3, 2, 1, 2),
    c(2, 3, 2, 2, 4)
))
##      [,1] [,2]
## [1,]    1    2
## [2,]    3    3
## [3,]    2    2
## [4,]    1    2
## [5,]    2    4
```

Now `A[I]` gives easy access to:

- `A[ I[1, 1], I[1, 2] ]`,
- `A[ I[2, 1], I[2, 2] ]`,
- `A[ I[3, 1], I[3, 2] ]`,
- …

and so forth. In other words, each row of `I` gives the coordinates of the elements to extract. The result is always a flat vector.

```
A[I]
## [1]  4  9  5  4 11
```

This is exactly `A[1, 2]`, `A[3, 3]`, `A[2, 2]`, `A[1, 2]`, `A[2, 4]`.

**Note** `which` can also return a list of index matrices:

```
which(A>7, arr.ind=TRUE)
##      row col
## [1,]   2   3
## [2,]   3   3
## [3,]   1   4
## [4,]   2   4
## [5,]   3   4
```

Moreover, `arrayInd` converts flat indexes to multidimensional ones.

**Exercise 11.12** *Implement your version of* `arrayInd` *and a function performing the inverse operation.*

**Exercise 11.13** *Write your version of* `diag`*.*

### 11.2.8 Higher-dimensional arrays

For $d$-dimensional arrays, indexing can involve up to $d$ indexes. It is particularly valuable for arrays with the `dimnames` attribute set representing contingency tables over a Cartesian product of multiple factors. The `datasets::Titanic` object is an exemplary four-dimensional table:

```
str(dimnames(Titanic))  # for reference (note that dimnames are named)
## List of 4
##  $ Class   : chr [1:4] "1st" "2nd" "3rd" "Crew"
##  $ Sex     : chr [1:2] "Male" "Female"
##  $ Age     : chr [1:2] "Child" "Adult"
##  $ Survived: chr [1:2] "No" "Yes"
```

Here is the number of adult male crew members who survived the accident:

```
Titanic["Crew", "Male", "Adult", "Yes"]
## [1] 192
```

Moreover, let's fetch a slice corresponding to adults travelling in the first class:

```
Titanic["1st", , "Adult", ]
##         Survived
## Sex       No Yes
##   Male   118  57
##   Female   4 140
```

**Exercise 11.14** *Check if the above four-dimensional array can be indexed using matrices with four columns.*

### 11.2.9 Replacing elements

Generally, subsetting drops all attributes except `names`, `dim`, and `dimnames` (unless it does not make sense otherwise). The replacement variant of the index operator modifies vector values but generally preserves all the attributes. This enables transforming matrix elements like:

```
B[B<10] <- A[B<10]^2  # `A` has no `dimnames` set
print(B)
##   x  y  z   w
## a 1 16 49 100
## b 4 25 64 121
## c 9 10 11  12
B[] <- rep(seq_len(NROW(B)), NCOL(B))  # NOT the same as B <- ...
print(B)  # `dim` and `dimnames` were preserved
##   x y z w
## a 1 1 1 1
## b 2 2 2 2
## c 3 3 3 3
```

**Exercise 11.15** *Given a character matrix with entities that can be interpreted as numbers like:*

```
(X <- rbind(x=c(a="1", b="2"), y=c("3", "4")))
##   a   b
## x "1" "2"
## y "3" "4"
```

*convert it to a numeric matrix with a single line of code. Preserve all attributes.*

---

## 11.3 Common operations

### 11.3.1 Matrix transpose

The matrix *transpose*, mathematically denoted by $\mathbf{A}^T$, is available via a call to `t`:

```
(A <- matrix(1:6, byrow=TRUE, nrow=2))
##      [,1] [,2] [,3]
## [1,]    1    2    3
## [2,]    4    5    6
t(A)
##      [,1] [,2]
## [1,]    1    4
## [2,]    2    5
## [3,]    3    6
```

Hence, if $\mathbf{B} = \mathbf{A}^T$, then it is a matrix such that $b_{i,j} = a_{j,i}$. In other words, in the transposed matrix, rows become columns, and columns become rows.

For higher-dimensional arrays, a generalised transpose can be obtained through **aperm** (try permuting the dimensions of `Titanic`). Also, the conjugate transpose of a complex matrix **A** is done via `Conj(t(A))`.

### 11.3.2 Vectorised mathematical functions

Vectorised functions such as **sqrt**, **abs**, **round**, **log**, **exp**, **cos**, **sin**, etc., operate on each array element[3].

```
A <- matrix(1/(1:6), nrow=2)
round(A, 2)  # rounds every element in A
##      [,1] [,2] [,3]
## [1,]  1.0 0.33 0.20
## [2,]  0.5 0.25 0.17
```

**Exercise 11.16** *Using a single call to **matplot**, which allows the y argument to be a matrix, draw a plot of* $\sin(x)$, $\cos(x)$, $|\sin(x)|$, *and* $|\cos(x)|$ *for* $x \in [-2\pi, 6\pi]$*; see Section 13.3 for more details.*

### 11.3.3 Aggregating rows and columns

When we call an aggregation function on an array, it will reduce all elements to a single number:

```
(A <- matrix(1:12, byrow=TRUE, nrow=3))
##      [,1] [,2] [,3] [,4]
## [1,]    1    2    3    4
## [2,]    5    6    7    8
## [3,]    9   10   11   12
mean(A)
## [1] 6.5
```

The **apply** function may be used to summarise individual rows or columns in a matrix:

- `apply(A, 1, f)` applies a given function **f** on each *row* of a matrix `A` (over the first axis),
- `apply(A, 2, f)` applies **f** on each *column* of `A` (over the second axis).

For instance:

```
apply(A, 1, mean)  # synonym: rowMeans(A)
## [1]  2.5  6.5 10.5
apply(A, 2, mean)  # synonym: colMeans(A)
## [1] 5 6 7 8
```

The function being applied does not have to return a single number:

[3] They are simply applied on each element of the underlying flat vector. Section 5.5 mentioned that unary functions preserve *all* attributes of their inputs, hence also `dim` and `dimnames`.

```
apply(A, 2, range)  # min and max
##      [,1] [,2] [,3] [,4]
## [1,]    1    2    3    4
## [2,]    9   10   11   12
apply(A, 1, function(row) c(Min=min(row), Mean=mean(row), Max=max(row)))
##      [,1] [,2] [,3]
## Min   1.0  5.0  9.0
## Mean  2.5  6.5 10.5
## Max   4.0  8.0 12.0
```

Take note of the columnwise order of the output values. Moreover, **apply** also works on higher-dimensional arrays:

```
apply(Titanic, 1, mean)  # over the first axis, "Class" (dimnames works too)
##     1st     2nd     3rd    Crew
##  40.625  35.625  88.250 110.625
apply(Titanic, c(1, 3), mean)  # over c("Class", "Age")
##       Age
## Class  Child  Adult
##   1st   1.50  79.75
##   2nd   6.00  65.25
##   3rd  19.75 156.75
##   Crew  0.00 221.25
```

### 11.3.4 Binary operators

In Section 5.5, we stated that binary elementwise operations, such as addition or multiplication, preserve the attributes of the longer input or both (with the first argument preferred to the second) if they are of equal sizes. Taking into account that:

- an array is simply a flat vector equipped with the `dim` attribute, and
- we refer to the respective *default* methods when applying binary operators,

we can deduce how **`+`**, **`<=`**, **`&`**, etc. behave in several different contexts.

**Array-array.** First, let's note what happens when we operate on two arrays of identical dimensionalities.

```
(A <- rbind(c(1, 10, 100), c(-1, -10, -100)))
##      [,1] [,2] [,3]
## [1,]    1   10  100
## [2,]   -1  -10 -100
(B <- matrix(1:6, byrow=TRUE, nrow=2))
##      [,1] [,2] [,3]
## [1,]    1    2    3
## [2,]    4    5    6
A + B  # elementwise addition
##      [,1] [,2] [,3]
```

```
## [1,]    2   12  103
## [2,]    3   -5  -94
A * B  # elementwise multiplication (not: algebraic matrix multiply)
##      [,1] [,2] [,3]
## [1,]    1   20  300
## [2,]   -4  -50 -600
```

They are simply the addition and multiplication of the corresponding elements of two given matrices.

**Array-scalar**. Second, we can apply matrix-scalar operations:

```
(-1)*B
##      [,1] [,2] [,3]
## [1,]   -1   -2   -3
## [2,]   -4   -5   -6
A^2
##      [,1] [,2]  [,3]
## [1,]    1  100 10000
## [2,]    1  100 10000
```

They multiplied each element in `B` by -1 and squared every element in `A`, respectively. The behaviour of relational operators is of course similar:

```
A >= 1 & A <= 100
##       [,1]  [,2]  [,3]
## [1,]  TRUE  TRUE  TRUE
## [2,] FALSE FALSE FALSE
```

**Array-vector**. Next, based on the recycling rule and the fact that matrix elements are ordered columnwisely, we have that:

```
B * c(10, 100)
##      [,1] [,2] [,3]
## [1,]   10   20   30
## [2,]  400  500  600
```

It multiplied every element in the first *row* by 10 and each element in the second row by 100. If we wish to multiply each element in the first, second, …, etc. *column* by the first, second, …, etc. value in a vector, we should *not* call:

```
B * c(1, 100, 1000)
##      [,1] [,2] [,3]
## [1,]    1 2000  300
## [2,]  400    5 6000
```

but rather:

```
t(t(B) * c(1, 100, 1000))
##      [,1] [,2] [,3]
## [1,]    1  200 3000
## [2,]    4  500 6000
```

or:

```
t(apply(B, 1, `*`, c(1, 100, 1000)))
##      [,1] [,2] [,3]
## [1,]    1  200 3000
## [2,]    4  500 6000
```

**Exercise 11.17** *Write a function that standardises the values in each column of a given matrix: for all elements in each column, subtract their mean and then divide them by the standard deviation. Try to implement it in a few different ways, including via a call to* **`apply`**, **`sweep`**, **`scale`**, *or based solely on arithmetic operators.*

---

**Note** Some sanity checks are done on the `dim` attributes, so not every configuration is possible. Notice some peculiarities:

```
A + t(B)  # `dim` equal to c(2, 3) vs c(3, 2)
## Error in `A + t(B)`: ! non-conformable arrays
A * cbind(1, 10, 100)  # this is too good to be true
## Error in `A * cbind(1, 10, 100)`: ! non-conformable arrays
A * rbind(1, 10)  # but A * c(1, 10) works...
## Error in `A * rbind(1, 10)`: ! non-conformable arrays
A + 1:12  # `A` has six elements
## Error: ! dims [product 6] do not match the length of object [12]
A + 1:5  # partial recycling is okay
## Warning in A + 1:5: longer object length is not a multiple of shorter
##     object length
##      [,1] [,2] [,3]
## [1,]    2   13  105
## [2,]    1   -6  -99
```

---

## 11.4 Numerical matrix algebra (*)

Many data analysis and machine learning algorithms, in their essence, involve rather straightforward matrix algebra and numerical mathematics. Suffice it to say that anyone serious about data science and scientific computing should learn the necessary theory; see, for example, [32] and [31].

R is a convenient interface to the stable and well-tested algorithms from, amongst others, BLAS[4] and LAPACK. Below we mention a few of them. External packages implement hundreds of algorithms tackling differential equations, constrained and unconstrained optimisation, etc.; CRAN Task Views[5] provide a good overview.

### 11.4.1 Matrix multiplication

`*` performs *elementwise* multiplication. For what we call the (algebraic) *matrix* multiplication, we use the `%*%` operator. It can only be performed on two matrices of *compatible sizes*: the number of columns in the left matrix must match the number of rows in the right operand. Given $\mathbf{A} \in \mathbb{R}^{n\times p}$ and $\mathbf{B} \in \mathbb{R}^{p\times m}$, their multiply is a matrix $\mathbf{C} = \mathbf{A}\mathbf{B} \in \mathbb{R}^{n\times m}$ such that $c_{i,j}$ is the dot product of the $i$-th row in $\mathbf{A}$ and the $j$-th column in $\mathbf{B}$:

$$c_{i,j} = \mathbf{a}_{i,\cdot} \cdot \mathbf{b}_{\cdot,j} = \sum_{k=1}^{p} a_{i,k} b_{k,j},$$

for $i = 1, \dots, n$ and $j = 1, \dots, m$. For instance:

```
(A <- rbind(c(0, 1, 3), c(-1, 1, -2)))
##      [,1] [,2] [,3]
## [1,]    0    1    3
## [2,]   -1    1   -2
(B <- rbind(c(3, -1), c(1, 2), c(6, 1)))
##      [,1] [,2]
## [1,]    3   -1
## [2,]    1    2
## [3,]    6    1
A %*% B
##      [,1] [,2]
## [1,]   19    5
## [2,]  -14    1
```

**Note** When applying `%*%` on one or more flat vectors, their dimensionality will be promoted automatically to make the operation possible. However, `c(a, b) %*% c(c, d)` gives a scalar $ac + bd$, and not a $2 \times 2$ matrix.

Further, `crossprod(A, B)` yields $\mathbf{A}^T\mathbf{B}$ and `tcrossprod(A, B)` determines $\mathbf{A}\mathbf{B}^T$ more efficiently than relying on `%*%`. We can omit the second argument and get $\mathbf{A}^T\mathbf{A}$ and $\mathbf{A}\mathbf{A}^T$, respectively.

```
crossprod(c(2, 1))  # Euclidean norm squared
##      [,1]
## [1,]    5
```

[4] (*) We can select the underlying implementation of BLAS at R's compile time; see Section A.3 of [68]. Some of them are faster than others.

[5] https://cran.r-project.org/web/views

```
crossprod(c(2, 1), c(-1, 2))  # dot product of two vectors
##      [,1]
## [1,]    0
crossprod(A)  # same as t(A) %*% A, i.e., dot products of all column pairs
##      [,1] [,2] [,3]
## [1,]    1   -1    2
## [2,]   -1    2    1
## [3,]    2    1   13
```

Recall that if the dot product of two vectors equals 0, we say that they are orthogonal (perpendicular).

**Exercise 11.18** *(*) Write your versions of* **cov** *and* **cor***: functions to compute the covariance and correlation matrices. Make use of the fact that the former can be determined with* **crossprod** *based on a centred version of an input matrix.*

### 11.4.2 Solving systems of linear equations

The **solve** function can be used to determine the solution to $m$ systems of $n$ linear equations of the form $\mathbf{AX} = \mathbf{B}$, where $\mathbf{A} \in \mathbb{R}^{n\times n}$ and $\mathbf{X}, \mathbf{B} \in \mathbb{R}^{n\times m}$ (via the LU decomposition with partial pivoting and row interchanges).

### 11.4.3 Norms and metrics

Given an $n \times m$ matrix $\mathbf{A}$, calling `norm(A, "1")`, `norm(A, "2")`, and `norm(A, "I")`, we can compute the operator norms:

$$
\begin{aligned}
\|\mathbf{A}\|_1 &= \max_{j=1,\dots,m} \textstyle\sum_{i=1}^{n} |a_{i,j}|, \\
\|\mathbf{A}\|_2 &= \sigma_1(\mathbf{A}) = \sup_{\mathbf{0}\neq\mathbf{x}\in\mathbb{R}^m} \frac{\|\mathbf{A}\mathbf{x}\|_2}{\|\mathbf{x}\|_2}, \\
\|\mathbf{A}\|_I &= \max_{i=1,\dots,n} \textstyle\sum_{j=1}^{m} |a_{i,j}|,
\end{aligned}
$$

where $\sigma_1$ gives the largest singular value (see below). Also, passing `"F"` as the second argument yields the Frobenius norm, $\|\mathbf{A}\|_F = \sqrt{\sum_{i=1}^{n}\sum_{j=1}^{m} a_{i,j}^2}$, and `"M"` computes the maximum norm, $\|\mathbf{A}\|_M = \max_{\substack{i=1,\dots,n\\ j=1,\dots,m}} |a_{i,j}|$.

If $\mathbf{A}$ is a column vector, then $\|\mathbf{A}\|_F$ and $\|\mathbf{A}\|_2$ are equivalent. They are referred to as the Euclidean norm. Moreover, $\|\mathbf{A}\|_M = \|\mathbf{A}\|_I$ gives the supremum norm and $\|\mathbf{A}\|_1$ outputs the Manhattan (taxicab) one.

**Exercise 11.19** *Given an $n\times m$ matrix $\mathbf{A}$, normalise each column so that it becomes a unit vector, i.e., whose Euclidean norm equals 1.*

Further, **dist** determines all pairwise distances between a set of $n$ vectors in $\mathbb{R}^m$, written as an $n \times m$ matrix. For example, let's consider three vectors in $\mathbb{R}^2$:

```
(X <- rbind(c(1, 1), c(1, -2), c(0, 0)))
```

```
##      [,1] [,2]
## [1,]    1    1
## [2,]    1   -2
## [3,]    0    0
as.matrix(dist(X, "euclidean"))
##        1      2      3
## 1 0.0000 3.0000 1.4142
## 2 3.0000 0.0000 2.2361
## 3 1.4142 2.2361 0.0000
```

Thus, the Euclidean distance between the first and the third vector, $\|\mathbf{x}_{1,\cdot} - \mathbf{x}_{3,\cdot}\|_2 = \sqrt{(x_{1,1} - x_{3,1})^2 + (x_{1,2} - x_{3,2})^2}$, is roughly 1.4142. The maximum, Manhattan, and Canberra distances/metrics are also available, amongst others.

**Exercise 11.20** *`dist` returns an object of the S3 class `dist`. Inspect how it is represented.*

**Example 11.21** *`adist` implements a couple of string metrics. For example:*

```
x <- c("spam", "bacon", "eggs", "spa", "spams", "legs")
names(x) <- x
(d <- adist(x))
##       spam bacon eggs spa spams legs
## spam     0     5    4   1     1    4
## bacon    5     0    5   5     5    5
## eggs     4     5    0   4     4    2
## spa      1     5    4   0     2    4
## spams    1     5    4   2     0    4
## legs     4     5    2   4     4    0
```

*It gave the Levenshtein distances between each pair of strings. In particular, we need two edit operations (character insertions, deletions, or replacements) to turn `"eggs"` into `"legs"` (add `l` and remove `g`).*

**Example 11.22** *Objects of the class `dist` can be used to find a hierarchical clustering of a dataset. For example:*

```
h <- hclust(as.dist(d), method="average")  # see also: plot(h, labels=x)
cutree(h, 3)
##  spam bacon  eggs   spa spams  legs
##     1     2     3     1     1     3
```

*It determined three clusters using the average linkage strategy (`"legs"` and `"eggs"` are grouped together, `"spam"`, `"spa"`, `"spams"` form another cluster, and `"bacon"` is a singleton).*

### 11.4.4 Eigenvalues and eigenvectors

**eigen** returns a sequence of eigenvalues $(\lambda_1, \dots, \lambda_n)$ ordered nondecreasingly w.r.t. $|\lambda_i|$, and a matrix $\mathbf{V}$ whose columns define the corresponding eigenvectors (scaled to the unit length) of a given matrix $\mathbf{X}$. By definition, for all $j$, we have $\mathbf{X}\mathbf{v}_{\cdot,j} = \lambda_j \mathbf{v}_{\cdot,j}$.

**Example 11.23** *(*) Here are the eigenvalues and the corresponding eigenvectors of the matrix defining the rotation in the xy-plane about the origin $(0,0)$ by the counterclockwise angle $\pi/6$:*

```
(R <- rbind(c( cos(pi/6), sin(pi/6)),
            c(-sin(pi/6), cos(pi/6))))
##          [,1]    [,2]
## [1,]  0.86603 0.50000
## [2,] -0.50000 0.86603
eigen(R)
## eigen() decomposition
## $values
## [1] 0.86603+0.5i 0.86603-0.5i
##
## $vectors
##                  [,1]             [,2]
## [1,] 0.70711+0.00000i 0.70711+0.00000i
## [2,] 0.00000+0.70711i 0.00000-0.70711i
```

*The complex eigenvalues are $e^{-\pi/6i}$ and $e^{\pi/6i}$ and we have $|e^{-\pi/6i}| = |e^{\pi/6i}| = 1$.*

**Example 11.24** *(*) Consider a pseudorandom sample that we depict in Figure 11.1:*

```
S <- rbind(c(sqrt(5),       0 ),
           c(      0 , sqrt(2)))
mu <- c(10, -3)
Z <- matrix(rnorm(2000), ncol=2)  # each row is a standard normal 2-vector
X <- t(t(Z %*% S %*% R)+mu)  # scale, rotate, shift
plot(X, asp=1)  # scatter plot
# draw principal axes:
A <- t(t(matrix(c(0,0, 1,0, 0,1), ncol=2, byrow=TRUE) %*% S %*% R)+mu)
arrows(A[1, 1], A[1, 2], A[-1, 1], A[-1, 2], col="red", lwd=1, length=0.1)
```

*$\mathbf{X}$ was created by generating a realisation of a two-dimensional standard normal vector $\mathbf{Z}$, scaling it by $\left(\sqrt{5}, \sqrt{2}\right)$, rotating by the counterclockwise angle $\pi/6$, and shifting by $(10,-3)$, which we denote by $\mathbf{X} = \mathbf{ZSR} + \boldsymbol{\mu}^T$. It follows a bivariate*[6] *normal distribution centred at $\boldsymbol{\mu} = (10,-3)$ and with the covariance matrix $\boldsymbol{\Sigma} = (\mathbf{SR})^T(\mathbf{SR})$:*

```
crossprod(S %*% R)  # covariance matrix
##       [,1]  [,2]
## [1,] 4.250 1.299
## [2,] 1.299 2.750
cov(X)  # compare: sample covariance matrix (estimator)
##        [,1]   [,2]
```

[6] For drawing random samples from any multivariate distribution, refer to the theory of copulas, e.g., [50]. There are a few R packages on CRAN that implement the most popular models.

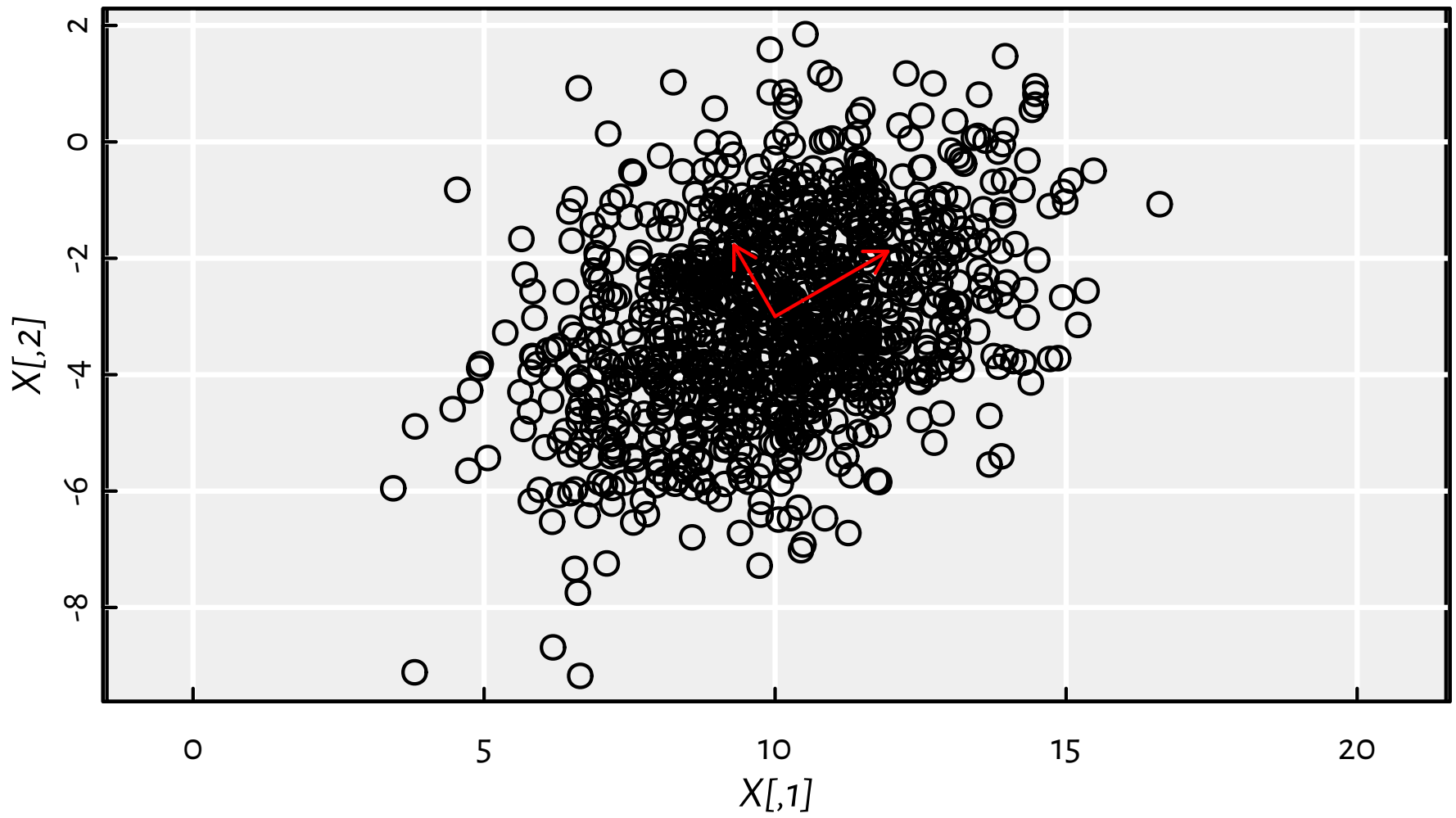

Figure 11.1. A sample from a bivariate normal distribution and its principal axes.

```
## [1,] 4.1965 1.2386
## [2,] 1.2386 2.7973
```

*It is known that eigenvectors of the covariance matrix correspond to the principal components of the original dataset. Furthermore, its eigenvalues give the variances explained by each of them.*

```
eigen(cov(X))
## eigen() decomposition
## $values
## [1] 4.9195 2.0744
##
## $vectors
##          [,1]     [,2]
## [1,] -0.86366  0.50408
## [2,] -0.50408 -0.86366
```

*It roughly corresponds to the principal directions* $(\cos\pi/6, \sin\pi/6) \simeq (0.866, 0.5)$ *and the thereto-orthogonal* $(-\sin\pi/6, \cos\pi/6) \simeq (-0.5, 0.866)$ *(up to an orientation inverse) with the corresponding variances of 5 and 2, respectively (i.e., standard deviations of* $\sqrt{5}$ *and* $\sqrt{2}$*). Note that this method of performing principal component analysis, i.e., recreating the scale and rotation transformation applied on* $\mathbf{Z}$ *based only on* $\mathbf{X}$*, is not particularly numerically stable; see Exercise 11.26 for an alternative.*

### 11.4.5 QR decomposition

Let $n \geq m$. We say that a real $n \times m$ matrix $\mathbf{Q}$ is *orthogonal*, whenever $\mathbf{Q}^T\mathbf{Q} = \mathbf{I}$ (identity matrix). This is equivalent to $\mathbf{Q}$'s columns' being orthogonal unit vectors. Also, if $\mathbf{Q}$ is a square matrix, then $\mathbf{Q}^T = \mathbf{Q}^{-1}$ if and only if $\mathbf{Q}^T\mathbf{Q} = \mathbf{Q}\mathbf{Q}^T = \mathbf{I}$.

Let $\mathbf{A}$ be a real[7] $n \times m$ matrix with $n \geq m$. Then $\mathbf{A} = \mathbf{QR}$ is its QR decomposition, if $\mathbf{Q}$ is an orthogonal $n \times m$ matrix and $\mathbf{R}$ is an upper triangular $m \times m$ one. Note that such a decomposition is not necessarily unique (without imposing additional requirements), and that we speak here of a QR factorisation in the so-called narrow form.

The `qr` function returns an object of the S3 class `qr` from which we can extract the two components of interest; see the `qr.Q` and `qr.R` functions.

**Example 11.25** *Let $\mathbf{X}$ be an $n \times m$ data matrix, representing $n$ points in $\mathbb{R}^m$, and a vector $\mathbf{y} \in \mathbb{R}^n$, where $y_i$ gives the* desired *output corresponding to the input $\mathbf{x}_{i,\cdot}$.*

*Let $\boldsymbol\theta$ be a vector of $m$ parameters. For fitting a linear model $y = \mathbf{x}^T\boldsymbol\theta = \theta_1 x_1 + \cdots + \theta_m x_m$, we can use the method of least squares, which minimises the quadratic loss function:*

$$\mathcal{L}(\boldsymbol\theta) = \sum_{i=1}^{n} \left(\mathbf{x}_{i,\cdot}^T\boldsymbol\theta - y_i\right)^2 = \|\mathbf{X}\boldsymbol\theta - \mathbf{y}\|_2^2.$$

*It might be shown that if we have the QR factorisation $\mathbf{X} = \mathbf{QR}$, then the optimal $\boldsymbol\theta$ is given by:*

$$\boldsymbol\theta = \left(\mathbf{X}^T\mathbf{X}\right)^{-1}\mathbf{X}^T\mathbf{y} = \mathbf{R}^{-1}\mathbf{Q}^T\mathbf{y},$$

*which can conveniently be determined via a call to* `qr.coef`.

*In particular, we can fit a simple linear regression model $y = ax + b = \mathbf{x}^T\boldsymbol\theta$ by considering $\mathbf{x} = (x, 1)$ and $\boldsymbol\theta = (a, b)$. For instance:*

```
x <- cars[["speed"]]
y <- cars[["dist"]]
X1 <- cbind(x, 1)  # the model is theta[1]*x + theta[2]*1
qrX1 <- qr(X1)
(theta <- qr.coef(qrX1, y))
##        x
##   3.9324 -17.5791
plot(x, y, xlab="speed", ylab="dist")  # scatter plot
abline(theta[2], theta[1], lty=2)  # add the regression line
```

*Thus, the fitted model is $y = 3.9324x - 17.5791$, see Figure 11.2.*

*The same approach is used by* `lm.fit`*, the workhorse behind the* `lm` *method that allows for specifying regression models using an R formula (which some readers might be familiar with; compare Section 17.6).*

[7] If $\mathbf{A}$ is a complex matrix, its QR decomposition spawns $\mathbf{Q}$ that is a unitary matrix.

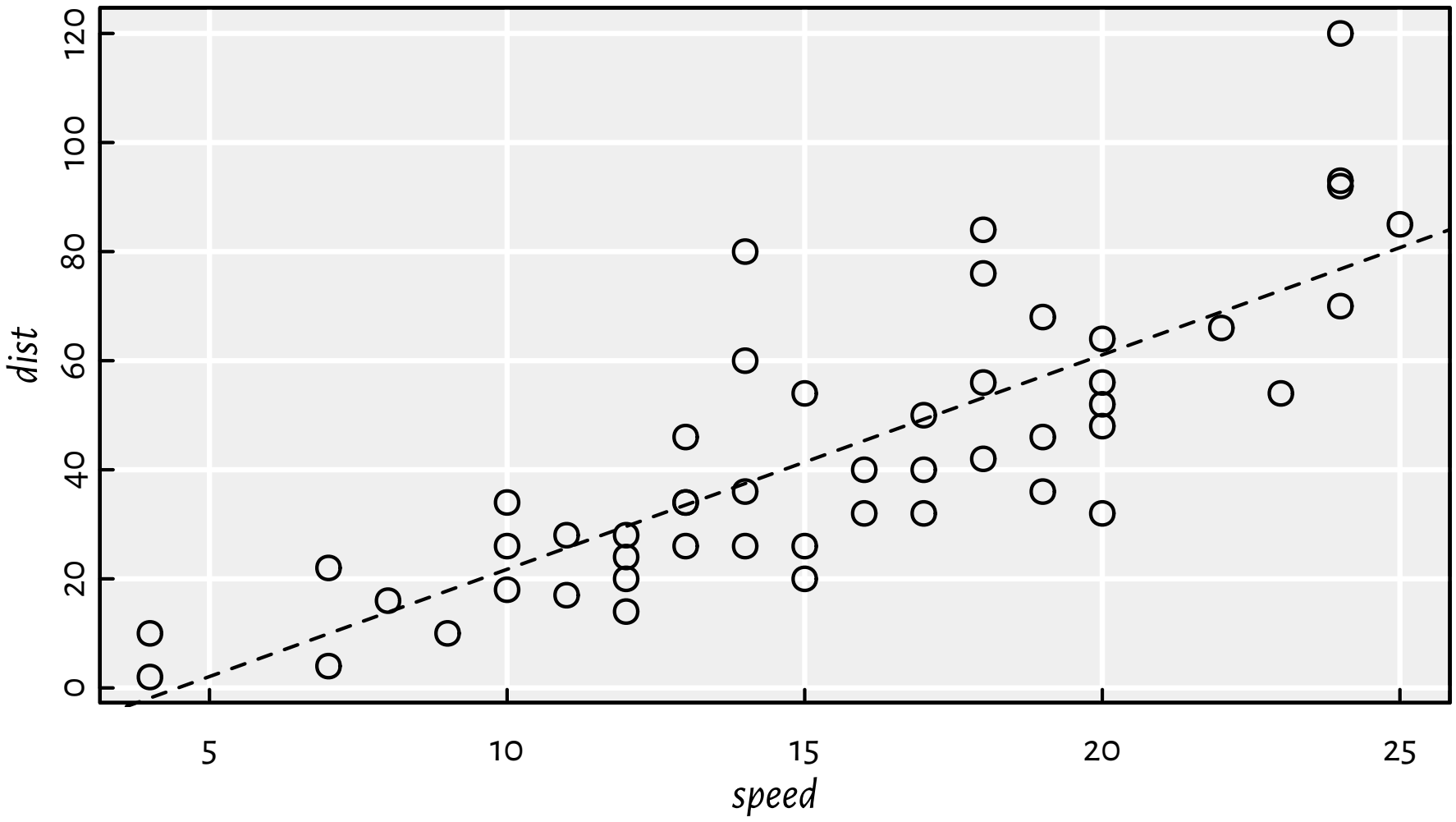

Figure 11.2. The `cars` dataset and the fitted regression line.

```
lm.fit(cbind(x, 1), y)[["coefficients"]]  # also: lm(dist~speed, data=cars)
##        x
##   3.9324 -17.5791
```

*As an exercise, let us compute $\boldsymbol{\theta} = \mathbf{R}^{-1}\mathbf{Q}^T\mathbf{y}$ manually. We note that the one-argument version of* **`solve`** *determines the inverse of a given matrix. Thus:*

```
Q <- qr.Q(qrX1)
R <- qr.R(qrX1)
solve(R) %*% t(Q) %*% y  # or solve(R) %*% crossprod(Q, y)
##        [,1]
## x    3.9324
##   -17.5791
```

*However, from the perspective of numerical stability, computing a matrix inverse is rarely a good idea. Multiplying both sides of the equation $\boldsymbol{\theta} = \mathbf{R}^{-1}\mathbf{Q}^T\mathbf{y}$ by $\mathbf{R}$, we get that it holds $\mathbf{R}\boldsymbol{\theta} = \mathbf{b}$ with $\mathbf{b} = \mathbf{Q}^T\mathbf{y}$. This is a triangular system of linear equations, which can be efficiently solved using a designated routine:*

```
backsolve(R, crossprod(Q, y))
##          [,1]
## [1,]   3.9324
## [2,] -17.5791
```

### 11.4.6 SVD decomposition

Given a real $n \times m$ matrix $\mathbf{X}$, its singular value decomposition (SVD) is given by $\mathbf{X} = \mathbf{U}\mathbf{D}\mathbf{V}^T$, where $\mathbf{U}$ and $\mathbf{V}$ are orthogonal matrices of dimensions $n \times p$ and $m \times p$, respectively, and $\mathbf{D}$ is a $p \times p$ diagonal matrix with the singular values of $\mathbf{X}$. It is usually assumed that $d_{1,1} \geq \ldots \geq d_{p,p} \geq 0$, $p = \min\{n, m\}$, and then the SVD factorisation is unique.

The corresponding **svd** function may be used to perform the principal component analysis[8]. Namely, if $\mathbf{X} = \mathbf{U}\mathbf{D}\mathbf{V}^T$, then the columns of $\mathbf{V}$ give the eigenvectors of $\mathbf{X}^T\mathbf{X}$. The latter is precisely its scaled covariance matrix if we assume that $\mathbf{X}$ is centred at 0.

**Example 11.26** *(*) Continuing Exercise 11.24 that features a rotated bivariate normal sample, we can determine the principal directions by referring to the* $\mathbf{V}$ *component of the SVD decomposition of a centred version of the data matrix:*

```
Xc <- t(t(X)-colMeans(X))  # centred version of X
svd(Xc)[["v"]]
##          [,1]     [,2]
## [1,] -0.86366 -0.50408
## [2,] -0.50408  0.86366
```

The SVD factorisation can also aid in determining the solution to linear regression. In the previous section, we mentioned that the parameter vector $\boldsymbol{\theta}$ minimising $\|\mathbf{X}\boldsymbol{\theta} - \mathbf{y}\|_2^2$ is given by $\boldsymbol{\theta} = \left(\mathbf{X}^T\mathbf{X}\right)^{-1}\mathbf{X}^T\mathbf{y}$. The component $\mathbf{X}^+ = \left(\mathbf{X}^T\mathbf{X}\right)^{-1}\mathbf{X}^T$ is called the pseudoinverse of $\mathbf{X}$ for it holds that $\mathbf{X}^+\mathbf{X} = \mathbf{I}$. If the SVD decomposition is $\mathbf{X} = \mathbf{U}\mathbf{D}\mathbf{V}^T$, then $\mathbf{X}^+ = \mathbf{V}\mathbf{D}^+\mathbf{U}^T$, where $\mathbf{D}^+$ is the transposed version of $\mathbf{D}$ carrying the reciprocals of its non-zero elements.

**Example 11.27** *Let us go back to the simple linear regression model discussed in Exercise 11.25. The same solution can be obtained by computing:*

```
svdX1 <- svd(X1)
V <- svdX1[["v"]]
d <- svdX1[["d"]]  # only the elements on the diagonal
U <- svdX1[["u"]]
V %*% diag(1/d) %*% t(U) %*% y
##          [,1]
## [1,]   3.9324
## [2,] -17.5791
```

The book [8] gives many more applications of the SVD factorisation in data science.

### 11.4.7 A note on the `Matrix` package

The **Matrix** package is perhaps the most widely known showcase of the S4 object orientation (Section 10.5). It defines classes and methods for dense and sparse matrices, including rectangular, symmetric, triangular, band, and diagonal ones. In particular,

[8] See the source code of **getS3method**("prcomp", "default").

large graph (e.g., in network sciences) or preference (e.g., in recommender systems) data can be represented using sparse matrices, i.e., those with many zeroes. After all, it is much more likely for two vertices in a network *not* to be joined by an edge than to be connected. For example:

```
library("Matrix")
(D <- Diagonal(x=1:5))
## 5 x 5 diagonal matrix of class "ddiMatrix"
##      [,1] [,2] [,3] [,4] [,5]
## [1,]    1    .    .    .    .
## [2,]    .    2    .    .    .
## [3,]    .    .    3    .    .
## [4,]    .    .    .    4    .
## [5,]    .    .    .    .    5
```

We created a real diagonal matrix of size $5 \times 5$; 20 elements equal to zero are specially marked. Moreover:

```
S <- as(D, "sparseMatrix")
S[1, 2] <- 7
S[4, 1] <- 42
print(S)
## 5 x 5 sparse Matrix of class "dgCMatrix"
##
## [1,]  1 7 . . .
## [2,]  . 2 . . .
## [3,]  . . 3 . .
## [4,] 42 . . 4 .
## [5,]  . . . . 5
```

It yielded a general sparse real matrix in the CSC (compressed, sparse, column-orientated) format.

For more information on this package, see **vignette**(package="Matrix").

## 11.5 Exercises

**Exercise 11.28** *Let X be a matrix with* `dimnames` *set. For instance:*

```
X <- matrix(1:12, byrow=TRUE, nrow=3)      # example matrix
dimnames(X)[[2]] <- c("a", "b", "c", "d")  # set column names
print(X)
##      a  b  c  d
## [1,] 1  2  3  4
## [2,] 5  6  7  8
## [3,] 9 10 11 12
```

*Explain the meaning of the following expressions involving matrix subsetting. Note that a few of them are invalid.*

- `X[1, ]`,
- `X[, 3]`,
- `X[, 3, drop=FALSE]`,
- `X[3]`,
- `X[, "a"]`,
- `X[, c("a", "b", "c")]`,
- `X[, -2]`,
- `X[X[,1] > 5, ]`,
- `X[X[,1]>5, c("a", "b", "c")]`,
- `X[X[,1]>=5 & X[,1]<=10, ]`,
- `X[X[,1]>=5 & X[,1]<=10, c("a", "b", "c")]`,
- `X[, c(1, "b", "d")]`.

**Exercise 11.29** *Assuming that X is an array, what is the difference between the following operations involving indexing?*

- `X["1", ]` vs `X[1, ]`,
- `X[, "a", "b", "c"]` vs `X["a", "b", "c"]` vs `X[, c("a", "b", "c")]` vs `X[c("a", "b", "c")]`,
- `X[1]` vs `X[, 1]` vs `X[1, ]`,
- `X[X>0]` vs `X[X>0, ]` vs `X[, X>0]`,
- `X[X[, 1]>0]` vs `X[X[, 1]>0,]` vs `X[,X[,1]>0]`,
- `X[X[, 1]>5, X[1, ]<10]` vs `X[X[1, ]>5, X[, 1]<10]`.

**Exercise 11.30** *Give a few ways to create a matrix like:*

```
##      [,1] [,2]
## [1,]    1    1
## [2,]    1    2
## [3,]    1    3
## [4,]    2    1
## [5,]    2    2
## [6,]    2    3
```

*and one like:*

```
##       [,1] [,2] [,3]
##  [1,]    1    1    1
##  [2,]    1    1    2
##  [3,]    1    2    1
##  [4,]    1    2    2
##  [5,]    1    3    1
##  [6,]    1    3    2
##  [7,]    2    1    1
##  [8,]    2    1    2
##  [9,]    2    2    1
## [10,]    2    2    2
## [11,]    2    3    1
## [12,]    2    3    2
```

**Exercise 11.31** *For a given real $n \times m$ matrix $\mathbf{X}$, encoding $n$ input points in an $m$-dimensional space, determine their bounding hyperrectangle, i.e., return a $2 \times m$ matrix $\mathbf{B}$ with $b_{1,j} = \min_i x_{i,j}$ and $b_{2,j} = \max_i x_{i,j}$.*

**Exercise 11.32** *Let $\mathbf{t}$ be a vector of $n$ integers in $\{1, \dots, k\}$. Write a function to one-hot encode each $t_i$, i.e., return a 0–1 matrix $\mathbf{R}$ of size $n \times k$ such that $r_{i,j} = 1$ if and only if $j = t_i$ (such a representation is beneficial when solving, e.g., a multiclass classification problem by means of $k$ binary classifiers). For example, if $\mathbf{t} = [1, 2, 3, 2, 4]$ and $k = 4$, then:*

$$\mathbf{R} = \begin{bmatrix} 1 & 0 & 0 & 0 \\ 0 & 1 & 0 & 0 \\ 0 & 0 & 1 & 0 \\ 0 & 1 & 0 & 0 \\ 0 & 0 & 0 & 1 \end{bmatrix}.$$

*Then, compose another function, but this time setting $r_{i,j} = 1$ if and only if $j \ge t_i$, e.g.:*

$$R = \begin{bmatrix} 1 & 1 & 1 & 1 \\ 0 & 1 & 1 & 1 \\ 0 & 0 & 1 & 1 \\ 0 & 1 & 1 & 1 \\ 0 & 0 & 0 & 1 \end{bmatrix}.$$

---

**Important** As usual, try to solve all the exercises without using explicit `for` and `while` loops (provided that it is possible).

---

**Exercise 11.33** *Given an $n \times k$ real matrix, apply the softmax function on each row, i.e., map $x_{i,j}$ to $\frac{\exp(x_{i,j})}{\sum_{l=1}^{k} \exp(x_{i,l})}$. Then, one-hot decode the values in each row, i.e., find the column number with the greatest value. Return a vector of size $n$ with elements in $\{1, \dots, k\}$.*

**Exercise 11.34** *Assume that an $n \times m$ real matrix $\mathbf{X}$ represents $n$ points in $\mathbb{R}^m$. Write a function (but do not refer to `dist`) that determines the pairwise Euclidean distances between all the $n$ points and a given $\mathbf{y} \in \mathbb{R}^m$. Return a vector $\mathbf{d}$ of length $n$ with $d_i = \|\mathbf{x}_{i,\cdot} - \mathbf{y}\|_2$.*

**Exercise 11.35** *Let* $\mathbf{X}$ *and* $\mathbf{Y}$ *be two real-valued matrices of sizes* $n \times m$ *and* $k \times m$*, respectively, representing two sets of points in* $\mathbb{R}^m$*. Return an integer vector* $\mathbf{r}$ *of length* $k$ *such that* $r_i$ *indicates the index of the point in* $\mathbf{X}$ *with the least distance to (the closest to) the* $i$*-th point in* $\mathbf{Y}$*, i.e.,* $r_i = \arg\min_j \|\mathbf{x}_{j,\cdot} - \mathbf{y}_{i,\cdot}\|_2$.

**Exercise 11.36** *Write your version of* **`utils::combn`**.

**Exercise 11.37** *Time series are vectors or matrices of the class* `ts` *equipped with the* `tsp` *attribute, amongst others. Refer to* **`help("ts")`** *for more information about how they are represented and what S3 methods have been overloaded for them.*

**Exercise 11.38** *(*) Numeric matrices can be stored in a CSV file, amongst others. Usually, we will be loading them via* **`read.csv`**, *which returns a data frame (see Chapter 12). For example:*

```
X <- as.matrix(read.csv(
    paste0(
        "https://github.com/gagolews/teaching-data/",
        "raw/master/marek/eurxxx-20200101-20200630.csv"
    ),
    comment.char="#",
    sep=","
))
```

*Write a function* **`read_numeric_matrix(file_name, comment, sep)`** *which is based on a few calls to* **`scan`** *instead. Use* **`file`** *to establish a file connection so that you can ignore the comment lines and fetch the column names before reading the actual numeric values.*

**Exercise 11.39** *(*) Using* **`readBin`**, *read the* `t10k-images-idx3-ubyte.gz` *from the MNIST database homepage*[9]*. The output object should be a three-dimensional,* $10000 \times 28 \times 28$ *array with real elements between 0 and 255. Refer to the* File Formats *section therein for more details.*

**Exercise 11.40** *(**) Circular convolution of discrete-valued multidimensional signals can be performed by means of* **`fft`** *and matrix multiplication, whereas affine transformations require only the latter. Apply various image transformations such as sharpening, shearing, and rotating on the MNIST digits and plot the results using the* **`image`** *function.*

**Exercise 11.41** *(*) Using* **`constrOptim`**, *find the minimum of the Constrained Betts Function* $f(x_1, x_2) = 0.01x_1^2 + x_2^2 - 100$ *with linear constraints* $2 \le x_1 \le 50$, $-50 \le x_2 \le 50$, *and* $10x_1 \ge 10 + x_2$. *(**) Also, use* **`solve.QP`** *from the* **quadprog** *package to find the minimum.*

[9] https://web.archive.org/web/20211107114045/http://yann.lecun.com/exdb/mnist

# 12

# *Data frames*

Most matrices are built on top of atomic vectors. Hence, only items of the same type can be arranged into rows and columns. On the other hand, *data frames* (objects of the S3 class `data.frame`, first introduced in [14]) are *collections* of vectors of the same lengths or matrices with identical row counts. Hence, they represent structured[1] data of possibly *heterogeneous* types. For instance:

```
class(iris)  # `iris` is an example data frame
## [1] "data.frame"
iris[c(1, 51, 101), ]  # three chosen rows from `iris`
##     Sepal.Length Sepal.Width Petal.Length Petal.Width    Species
## 1            5.1         3.5          1.4         0.2     setosa
## 51           7.0         3.2          4.7         1.4 versicolor
## 101          6.3         3.3          6.0         2.5  virginica
```

It is a mix of numeric and factor-type data.

The good news is that not only are data frames built on *named lists* (e.g., to extract a column, we can refer to `` `[[` ``), but also many functions consider them *matrix-like* (e.g., to select specific rows and columns, two indexes can be passed to `` `[` `` like in the preceding example). Hence, it will soon turn out that we already know a lot about performing basic data wrangling activities, even if we do not fully realise it now.

## 12.1 Creating data frames

### 12.1.1 `data.frame` and `as.data.frame`

Most frequently, we create data frames based on a series of logical, numeric, or character vectors of identical lengths. In such a scenario, the **`data.frame`** function is particularly worthwhile.

```
(x <- data.frame(
    a=c(TRUE, FALSE),
```

[1] We are already highly skilled in dealing with unstructured data and turning them into something that is much more regular. The numerous functions, which we have covered in the first part of this book, allow us to extract meaningful data from text, handle missing values, engineer features, and so forth.

```
    b=1:6,
    c=runif(6),
    d=c("spam", "spam", "eggs")
))
##       a b       c    d
## 1  TRUE 1 0.77437 spam
## 2 FALSE 2 0.19722 spam
## 3  TRUE 3 0.97801 eggs
## 4 FALSE 4 0.20133 spam
## 5  TRUE 5 0.36124 spam
## 6 FALSE 6 0.74261 eggs
```

The shorter vectors were recycled. We can verify that the diverse column types were retained and no coercion was made by calling:

```
str(x)
## 'data.frame':        6 obs. of  4 variables:
##  $ a: logi  TRUE FALSE TRUE FALSE TRUE FALSE
##  $ b: int  1 2 3 4 5 6
##  $ c: num  0.774 0.197 0.978 0.201 0.361 ...
##  $ d: chr  "spam" "spam" "eggs" "spam" ...
```

---

**Important** For many reasons (see, e.g., Section 12.1.5 and Section 12.1.6), we recommend having the type of each column always checked, e.g., by calling the **str** function.

---

Many objects, such as matrices, can easily be coerced to data frames using particular **as.data.frame** methods. Here is an example matrix:

```
(A <- matrix(1:6, nrow=3,
    dimnames=list(
        NULL,        # no row labels
        c("u", "v")  # some column labels
)))
##      u v
## [1,] 1 4
## [2,] 2 5
## [3,] 3 6
```

Let's convert it to a data frame:

```
as.data.frame(A)  # as.data.frame.matrix
##   u v
## 1 1 4
## 2 2 5
## 3 3 6
```

Note that a matrix with no row labels is printed slightly differently than a data frame with (as it will soon turn out) the default `row.names`.

Named lists are amongst other aspirants to a meaningful conversion. Consider an example list where all elements are vectors of the same length:

```
(l <- Map(
    function(x) {
        c(Min=min(x), Median=median(x), Mean=mean(x), Max=max(x))
    },
    split(iris[["Sepal.Length"]], iris[["Species"]])
))
## $setosa
##    Min Median   Mean    Max
##  4.300  5.000  5.006  5.800
##
## $versicolor
##    Min Median   Mean    Max
##  4.900  5.900  5.936  7.000
##
## $virginica
##    Min Median   Mean    Max
##  4.900  6.500  6.588  7.900
```

Each list element will be turned into a separate column:

```
as.data.frame(l)  # as.data.frame.list
##        setosa versicolor virginica
## Min     4.300      4.900     4.900
## Median  5.000      5.900     6.500
## Mean    5.006      5.936     6.588
## Max     5.800      7.000     7.900
```

Sadly, **`as.data.frame`** is not particularly fond of lists of vectors of incompatible lengths:

```
as.data.frame(list(a=1, b=11:12, c=21:23))
## Error: ! arguments imply differing number of rows: 1, 2, 3
```

These vectors could have been recycled with a warning. But they were not.

```
as.data.frame(list(a=1:4, b=11:12, c=21))  # recycling rule okay
##   a  b  c
## 1 1 11 21
## 2 2 12 21
## 3 3 11 21
## 4 4 12 21
```

The method for the S3 class `table` (mentioned in Chapter 11) can be helpful as well. Here is an example contingency table together with its *unstacked* (wide) version.

```
(t <- table(mtcars[["vs"]], mtcars[["cyl"]]))
##
##       4  6  8
##   0   1  3 14
##   1  10  4  0
as.data.frame(t)  # as.data.frame.table; see the stringsAsFactors note below!
##   Var1 Var2 Freq
## 1    0    4    1
## 2    1    4   10
## 3    0    6    3
## 4    1    6    4
## 5    0    8   14
## 6    1    8    0
```

**`as.data.frame.table`** is so handy that we might want to call it directly on any array. This way, we can convert it from the wide format to the long (tall) one; see Section 12.3.6 for more details.

---

**Note** The aforementioned method is based on **`expand.grid`**, which determines all combinations of a given series of vectors.

```
expand.grid(1:2, c("a", "b", "c"))  # see the stringsAsFactors note below!
##   Var1 Var2
## 1    1    a
## 2    2    a
## 3    1    b
## 4    2    b
## 5    1    c
## 6    2    c
```

---

Overall, many classes of objects can be included[2] in a data frame. The popular choices include `Date`, `POSIXct`, and `factor`.

**Example 12.1** *It is worth noting that **`format`** is used whilst printing the columns. Here is its custom method for what we would like to call from now on the S3 class `spam`:*

```
format.spam <- function(x, ...)
    paste0("<", x, ">")
```

*Testing data frame printing:*

```
data.frame(
    a=structure(c("lovely", "yummy", "delicious"), class="spam"),
    b=factor(c("spam", "bacon", "spam")),
    c=Sys.Date()+1:3
```

[2] The attributes of objects stored as columns will generally be preserved (even if they are not displayed by **`print`**; see **`str`** though).

```
)
##              a     b          c
## 1    <lovely>  spam 2026-07-31
## 2     <yummy> bacon 2026-08-01
## 3 <delicious>  spam 2026-08-02
```

### 12.1.2 cbind.data.frame and rbind.data.frame

There are data frame-specific versions of **cbind** or **rbind** (which we discussed in the context of stacking matrices; Section 11.1.2). They are used quite eagerly: **help**("cbind") states that they will be referred to if at least[3] one of its arguments is a data frame, and the other arguments are atomic vectors or lists (possibly with the dim attribute). For example:

```
x <- iris[c(1, 51, 101), c("Sepal.Length", "Species")]  # whatever
cbind(Yummy=c(TRUE, FALSE, TRUE), x)
##     Yummy Sepal.Length    Species
## 1    TRUE          5.1     setosa
## 51  FALSE          7.0 versicolor
## 101  TRUE          6.3  virginica
```

It added a new column to a data frame x. Moreover:

```
rbind(x, list(42, "virginica"))
##     Sepal.Length    Species
## 1            5.1     setosa
## 51           7.0 versicolor
## 101          6.3  virginica
## 11          42.0  virginica
```

It added a new row. Note that columns are of different types. Hence, the values to row-bind had to be provided as a list.

The generic vector used as a new row specifier can also be named. It can consist of sequences of length greater than one that are given in any order:

```
rbind(x, list(
    Species=c("virginica", "setosa"),
    Sepal.Length=c(42, 7)
))
##     Sepal.Length    Species
## 1            5.1     setosa
## 51           7.0 versicolor
## 101          6.3  virginica
```

[3] This is a clear violation of the rule that an S3 generic dispatches on the type of only one argument (usually: the first). It is an exception made for the sake of the questionable user *convenience*. Also, note that there is no **cbind.default** method available: it is hardcoded at the C language level.

```
## 11          42.0  virginica
## 2            7.0     setosa
```

A direct referral to **cbind.data.frame** and **rbind.data.frame** will sometimes be necessary. Consider an example list of atomic vectors:

```
x <- list(a=1:3, b=11:13, c=21:23)
```

First, we call the generic, which dispatches to the default method:

```
do.call(cbind, x)
##      a  b  c
## [1,] 1 11 21
## [2,] 2 12 22
## [3,] 3 13 23
```

It created a matrix. If we want to ensure we garner a data frame, we need to write:

```
do.call(cbind.data.frame, x)
##   a  b  c
## 1 1 11 21
## 2 2 12 22
## 3 3 13 23
```

This is useful for fetching outputs from **Map** et al., as they are wrapped inside a list. Here is a fancy way to obtain an illustrative list:

```
l <- unname(Map(
    function(x) list(  # objects are of different types, hence a list
        Sepal.Length=mean(x[["Sepal.Length"]]),
        Sepal.Width=mean(x[["Sepal.Width"]]),
        Species=x[["Species"]][1]  # all are the same, so the first will do
    ),
    split(iris, iris[["Species"]])  # split.data.frame; see below
))
str(l)
## List of 3
##  $ :List of 3
##   ..$ Sepal.Length: num 5.01
##   ..$ Sepal.Width : num 3.43
##   ..$ Species     : Factor w/ 3 levels "setosa","versicolor",..: 1
##  $ :List of 3
##   ..$ Sepal.Length: num 5.94
##   ..$ Sepal.Width : num 2.77
##   ..$ Species     : Factor w/ 3 levels "setosa","versicolor",..: 2
##  $ :List of 3
##   ..$ Sepal.Length: num 6.59
##   ..$ Sepal.Width : num 2.97
##   ..$ Species     : Factor w/ 3 levels "setosa","versicolor",..: 3
```

We may now turn it into a data frame by calling:

```
do.call(rbind.data.frame, l)
##   Sepal.Length Sepal.Width    Species
## 1        5.006       3.428     setosa
## 2        5.936       2.770 versicolor
## 3        6.588       2.974  virginica
```

On the other hand, **do.call**(rbind, l) does not return an amiable object type:

```
do.call(rbind, l)
##      Sepal.Length Sepal.Width Species
## [1,] 5.006        3.428       setosa
## [2,] 5.936        2.77        versicolor
## [3,] 6.588        2.974       virginica
```

Despite the pretty face, it is a matrix... over a list:

```
str(do.call(rbind, l))
## List of 9
##  $ : num 5.01
##  $ : num 5.94
##  $ : num 6.59
##  $ : num 3.43
##  $ : num 2.77
##  $ : num 2.97
##  $ : Factor w/ 3 levels "setosa","versicolor",..: 1
##  $ : Factor w/ 3 levels "setosa","versicolor",..: 2
##  $ : Factor w/ 3 levels "setosa","versicolor",..: 3
##  - attr(*, "dim")= int [1:2] 3 3
##  - attr(*, "dimnames")=List of 2
##   ..$ : NULL
##   ..$ : chr [1:3] "Sepal.Length" "Sepal.Width" "Species"
```

### 12.1.3 Reading data frames

Structured data can be imported from external sources, such as CSV/TSV (comma/tab-separated values) or HDF5 files, relational databases supporting SQL (see Section 12.1.4), web APIs (e.g., through the **curl** and **jsonlite** packages), spreadsheets [67], and so on. In particular, **read.csv** and the like fetch data from plain text files consisting of records, where commas, semicolons, tabs, etc. separate the fields. For instance:

```
x <- data.frame(a=runif(3), b=c(TRUE, FALSE, TRUE))  # example data frame
f <- tempfile()  # temporary file name
write.csv(x, f, row.names=FALSE)  # export
```

It created a CSV file that looks like:

```r
cat(readLines(f), sep="\n")  # print file contents
## "a","b"
## 0.287577520124614,TRUE
## 0.788305135443807,FALSE
## 0.4089769218117,TRUE
```

which can be read by calling:

```r
read.csv(f)
##         a     b
## 1 0.28758  TRUE
## 2 0.78831 FALSE
## 3 0.40898  TRUE
```

**Exercise 12.2** *Check out* **`help`**`("read.table")` *for a long list of tunable parameters, especially:* `sep`, `dec`, `quote`, `header`, `comment.char`, *and* `row.names`. *Further, note that reading from compressed files and interned URLs is supported directly.*

---

**Important** CSV is the most portable and user-friendly format for exchanging matrix-like objects between different programs and computing languages (Python, Julia, LibreOffice Calc, etc.). Such files can be opened in any text editor.

Also, as mentioned in Section 8.3.5, we can process data frames chunk by chunk. This is beneficial especially when data do not fit into memory (compare the `nrows` argument to **`read.csv`**).

---

### 12.1.4 Interfacing relational databases and querying with SQL (*)

The **`DBI`** package provides a universal interface for many database management systems whose drivers are implemented in add-ons such as **`RSQLite`**, **`RMariaDB`**, **`RPostgreSQL`**, etc., or, more generally, **`RODBC`** or **`odbc`**. For more details, see Section 4 of [67].

**Example 12.3** *Let's play with an in-memory (volatile) instance of an SQLite database.*

```r
library("DBI")
con <- dbConnect(RSQLite::SQLite(), ":memory:")
```

*It returned an object representing a database connection which we can refer to in further communication. An easy way to create a database table is to call:*

```r
dbWriteTable(con, "mtcars", mtcars)  # `mtcars` is a toy data frame
```

*Alternatively, we could have called* **`dbExecute`** *to send SQL statements such as "*`CREATE TABLE ...`*" followed by a series of "*`INSERT INTO ...`*". We can now retrieve some data:*

```r
dbGetQuery(con, "
    SELECT cyl, vs, AVG(mpg) AS mpg_ave, AVG(hp) AS hp_ave
```

```
    FROM mtcars
    GROUP BY cyl, vs
")
##   cyl vs mpg_ave hp_ave
## 1   4  0  26.000  91.00
## 2   4  1  26.730  81.80
## 3   6  0  20.567 131.67
## 4   6  1  19.125 115.25
## 5   8  0  15.100 209.21
```

*It gave us an ordinary R data frame. We can process it in the same fashion as any other object of this kind.*

*At the end, the database connection must be closed.*

```
dbDisconnect(con)
```

**Exercise 12.4** *Database passwords must never be stored in plain text files, let alone in R scripts in version-controlled repositories. Consider a few ways to fetch credentials programmatically:*

- *using environment variables (see* `help("Sys.getenv")`*),*
- *using the* **`keyring`** *package,*
- *calling* **`system2`** *(Section 7.3.2) to retrieve it from the system keyring (e.g., the* **`keyring`** *package for Python provides a platform-independent command-line utility).*

### 12.1.5 Strings as factors?

Some functions related to data frames automatically convert character vectors to factors. This behaviour is frequently controlled by an argument named `stringsAsFactors`. It can be particularly problematic because, when printed, factor and character columns look identical:

```
(x <- data.frame(a=factor(c("U", "V")), b=c("U", "V")))
##   a b
## 1 U U
## 2 V V
```

We recall from Section 10.3.2 that factors can be nasty. For example, passing factors as indexers in `` `[` `` or converting them with **`as.numeric`** might give counterintuitive results. Also, when we want to extend factors by previously unobserved data, new levels must be added manually. This can cause unexpected behaviour in contexts such as:

```
rbind(x, c("W", "W"))
## Warning in `[<-.factor`(`*tmp*`, ri, value = "W"): invalid factor level,
##    NA generated
##      a b
## 1    U U
```

```
## 2    V V
## 3 <NA> W
```

Therefore, always having the data types checked is a praiseworthy habit. For instance:

```
str(x)
## 'data.frame':	2 obs. of  2 variables:
##  $ a: Factor w/ 2 levels "U","V": 1 2
##  $ b: chr  "U" "V"
```

Before R 4.0, certain functions, including **data.frame** and **read.csv** had the `stringsAsFactors` argument defaulting to `TRUE`. It is no longer the case. However, exceptions to this rule still exist, e.g., including **as.data.frame.table** and **expand.grid**. Besides, some example data frames continue to enjoy factor-typed columns, e.g.:

```
class(iris[["Species"]])
## [1] "factor"
```

In particular, adding a new flower variety might be oblique:

```
iris2 <- iris[c(1, 101), ]  # example subset
rbind(iris2, c(6, 3, 3, 2, "croatica"))
## Warning in `[<-.factor`(`*tmp*`, ri, value = "croatica"): invalid factor
##     level, NA generated
##     Sepal.Length Sepal.Width Petal.Length Petal.Width   Species
## 1            5.1         3.5          1.4         0.2    setosa
## 101          6.3         3.3            6         2.5 virginica
## 3              6           3            3           2      <NA>
```

Compare it to:

```
levels(iris2[["Species"]])[nlevels(iris2[["Species"]])+1] <- "croatica"
rbind(iris2, c(6, 3, 3, 2, "croatica"))
##     Sepal.Length Sepal.Width Petal.Length Petal.Width   Species
## 1            5.1         3.5          1.4         0.2    setosa
## 101          6.3         3.3            6         2.5 virginica
## 3              6           3            3           2  croatica
```

### 12.1.6 Internal representation

Objects of the S3 class `data.frame` are erected on lists of vectors of the same length or matrices with identical row counts. Each list element defines a column or column group. Apart from `class`, data frames must be equipped with the following special attributes:

- `names` – a character vector (as usual in any named list) that gives the column labels,
- `row.names` – a character or integer vector with no duplicates nor missing values, doing what is advertised.

Therefore, a data frame can be created from scratch by calling, for example:

```
structure(
    list(a=11:13, b=21:23),  # sets the `names` attribute
    row.names=1:3,
    class="data.frame"
)
##    a  b
## 1 11 21
## 2 12 22
## 3 13 23
```

Here is a data frame based on a list of length five, a matrix with five rows, and a numeric vector with five items. We added some fancy row names on top:

```
structure(
    list(
        a=list(1, 1:2, 1:3, numeric(0), -(4:1)),
        b=cbind(u=11:15, v=21:25),
        c=runif(5)
    ),
    row.names=c("spam", "bacon", "eggs", "ham", "aubergine"),
    class="data.frame"
)
##                        a b.u b.v       c
## spam                   1  11  21 0.28758
## bacon               1, 2  12  22 0.78831
## eggs             1, 2, 3  13  23 0.40898
## ham                       14  24 0.88302
## aubergine -4, -3, -2, -1  15  25 0.94047
```

In general, the columns of the type `list` can contain anything, e.g., other lists or R functions. Including atomic vectors of varying lengths, just like above, permits us to create something à la *ragged arrays*.

The issue with matrix entries, on the other hand, is that they appear as if they were many columns. Still, as it will turn out in the sequel, they are often treated as a single complex column, e.g., by the index operator (see Section 12.2). Therefore, from this perspective, the aforementioned data frame has three columns, not four. Such compound columns can be output by **aggregate** (see Section 12.3), amongst others. They are valuable in certain contexts: the *column groups* can be easily accessed as a whole and batch-processed in the same way.

---

**Important** Alas, data frames with list or matrix columns cannot be created with the **data.frame** nor **cbind** functions. This might explain why they are less popular. This behaviour is dictated by the underlying **as.data.frame** methods, which they both call. As a curiosity, see **help**("I"), though.

---

**Exercise 12.5** *Verify that if a data frame carries a matrix column, this matrix does not need to have any column names (the second element of `dimnames`).*

The `names` and `row.names` attributes are *special* in the sense of Section 4.4.3. In particular, they can be accessed or modified via the dedicated functions.

It is worth noting that `row.names(df)` always returns a character vector, even when `attr(df, "row.names")` is integer. Further, calling "`row.names(df) <- NULL`" will reset[4] this attribute to the most commonly desired case of consecutive natural numbers. For example:

```
(x <- iris[c(1, 51, 101), ])  # comes with some sad row names
##     Sepal.Length Sepal.Width Petal.Length Petal.Width    Species
## 1            5.1         3.5          1.4         0.2     setosa
## 51           7.0         3.2          4.7         1.4 versicolor
## 101          6.3         3.3          6.0         2.5  virginica
`row.names<-`(x, NULL)  # reset to seq_len(NROW(x))
##   Sepal.Length Sepal.Width Petal.Length Petal.Width    Species
## 1          5.1         3.5          1.4         0.2     setosa
## 2          7.0         3.2          4.7         1.4 versicolor
## 3          6.3         3.3          6.0         2.5  virginica
```

**Exercise 12.6** *Implement your version of `expand.grid`.*

**Exercise 12.7** *Write a version of `xtabs` that does not rely on a formula interface (compare Section 10.3.4). Allow three parameters: a data frame, the name of the "counts" column, and the names of the cross-classifying factors. Hence, `my_xtabs(x, "Freq", c("Var1", "Var2"))` should be equivalent to `xtabs(Freq~Var1+Var2, x)`.*

## 12.2 Data frame subsetting

### 12.2.1 Data frames are lists

A data frame is a named list whose elements represent individual columns. Therefore[5], `length` yields the number of columns and `names` gives their respective labels.

Let's play around with this data frame:

```
(x <- data.frame(
    a=runif(6),
```

[4] `` `attr<-`(df, "row.names", value) `` does not run the same sanity checks as `` `row.names<-`(df, value) ``. For instance, it is easy to corrupt a data frame by setting too short a `row.names` attribute.

[5] This is a strong word. This implication relies on an implicit assumption that the primitive functions `length` and `names` have not been contaminated by treating data frames differently from named lists. Luckily, that is indeed not the case. Even though we have the index operators specially overloaded for the `data.frame` class, they behave reasonably. As we will see, they support a mix of list- and matrix-like behaviours.

```
    b=rnorm(6),
    c=LETTERS[1:6],
    d1=c(FALSE, TRUE, FALSE, NA, FALSE, NA),
    d2=c(FALSE, TRUE, FALSE, TRUE, FALSE, TRUE)
))
##          a         b c    d1    d2
## 1 0.287578  0.070508 A FALSE FALSE
## 2 0.788305  0.129288 B  TRUE  TRUE
## 3 0.408977  1.715065 C FALSE FALSE
## 4 0.883017  0.460916 D    NA  TRUE
## 5 0.940467 -1.265061 E FALSE FALSE
## 6 0.045556 -0.686853 F    NA  TRUE
typeof(x)  # each data frame is a list
## [1] "list"
length(x)  # the number of columns
## [1] 5
names(x)   # column labels
## [1] "a"  "b"  "c"  "d1" "d2"
```

The one-argument versions of extract and index operators behave as expected. `` `[[` `` fetches (looks inside) the contents of a given column:

```
x[["a"]]  # or x[[1]]
## [1] 0.287578 0.788305 0.408977 0.883017 0.940467 0.045556
```

`` `[` `` returns a data frame (a list with extras):

```
x["a"]  # or x[1]; a data frame with one column
##          a
## 1 0.287578
## 2 0.788305
## 3 0.408977
## 4 0.883017
## 5 0.940467
## 6 0.045556
x[c(TRUE, TRUE, FALSE, TRUE, FALSE)]
##          a         b    d1
## 1 0.287578  0.070508 FALSE
## 2 0.788305  0.129288  TRUE
## 3 0.408977  1.715065 FALSE
## 4 0.883017  0.460916    NA
## 5 0.940467 -1.265061 FALSE
## 6 0.045556 -0.686853    NA
```

Just like with lists, the replacement versions of these operators can add new columns or modify existing ones.

```
(y <- head(x, 1))  # example data frame
##         a        b c    d1    d2
## 1 0.28758 0.070508 A FALSE FALSE
y[["a"]] <- round(y[["a"]], 1)  # replaces the column with new content
y[["b"]] <- NULL  # removes the column, like, totally
y[["e"]] <- 10*y[["a"]]^2  # adds a new column at the end
print(y)
##     a c    d1    d2   e
## 1 0.3 A FALSE FALSE 0.9
```

**Example 12.8** *Some spam for thought to show how much we already know. Here are a few common scenarios involving indexing.*

```
(y <- head(x, 1))  # example data frame
##         a        b c    d1    d2
## 1 0.28758 0.070508 A FALSE FALSE
```

*Move the column a to the end:*

```
y[unique(c(names(y), "a"), fromLast=TRUE)]
##          b c    d1    d2       a
## 1 0.070508 A FALSE FALSE 0.28758
```

*Remove the columns a and c:*

```
y[-match(c("a", "c"), names(y))]  # or y[setdiff(names(y), c("a", "c"))]
##          b    d1    d2
## 1 0.070508 FALSE FALSE
```

*Select all columns between a and c:*

```
y[match("a", names(y)):match("c", names(y))]
##         a        b c
## 1 0.28758 0.070508 A
```

*Fetch the columns with names starting with d:*

```
y[grep("^d", names(y), perl=TRUE)]
##      d1    d2
## 1 FALSE FALSE
```

*Change the name of column c to z:*

```
names(y)[names(y) == "c"] <- "z"
print(y)  # `names<-`(y, `[<-`(names(y), names(y) == "c", "z"))
##         a        b z    d1    d2
## 1 0.28758 0.070508 A FALSE FALSE
```

*Change names: d2 to u and d1 to v:*

```
names(y)[match(c("d2", "d1"), names(y))] <- c("v", "u")
print(y)
##         a        b z     u     v
## 1 0.28758 0.070508 A FALSE FALSE
```

**Note** Some users prefer the `` `$` `` operator over `` `[[` ``, but we do not. By default, the former supports partial matching of column names which might be appealing when R is used interactively. Nonetheless, it does not work on matrices nor it allows for programmatically generated names. It is also trickier to use on not syntactically valid labels; compare Section 9.3.1.

**Exercise 12.9** *Write a function* `rename` *that changes the names of columns based on a translation table given in a* `from=to` *fashion (we have already solved a similar exercise in* Chapter 9*). For instance:*

```
rename <- function(x, ...) ...to.do...
rename(head(x, 1), c="new_c", a="new_a")
##     new_a        b new_c    d1    d2
## 1 0.28758 0.070508     A FALSE FALSE
```

### 12.2.2 Data frames are matrix-like

Data frames can be considered "generalised" matrices. They store data of any kind (possibly mixed) organised in a *tabular* fashion. A few functions mentioned in the previous chapter are overloaded for the data frame case. They include: **`dim`** (despite the lack of the `dim` attribute), **`NROW`**, **`NCOL`**, and **`dimnames`** (which is, of course, based on `row.names` and `names`). For example:

```
(x <- data.frame(
    a=runif(6),
    b=rnorm(6),
    c=LETTERS[1:6],
    d1=c(FALSE, TRUE, FALSE, NA, FALSE, NA),
    d2=c(FALSE, TRUE, FALSE, TRUE, FALSE, TRUE)
))
##          a         b c    d1    d2
## 1 0.287578  0.070508 A FALSE FALSE
## 2 0.788305  0.129288 B  TRUE  TRUE
## 3 0.408977  1.715065 C FALSE FALSE
## 4 0.883017  0.460916 D    NA  TRUE
## 5 0.940467 -1.265061 E FALSE FALSE
## 6 0.045556 -0.686853 F    NA  TRUE
dim(x)  # the number of rows and columns
## [1] 6 5
dimnames(x)  # row and column labels
## [[1]]
```

```
## [1] "1" "2" "3" "4" "5" "6"
##
## [[2]]
## [1] "a"  "b"  "c"  "d1" "d2"
```

In addition to the list-like behaviour, which only allows for dealing with particular columns or their groups, the `[` operator can also take two indexers:

```
x[1:2, ]  # first two rows
##         a        b c    d1    d2
## 1 0.28758 0.070508 A FALSE FALSE
## 2 0.78831 0.129288 B  TRUE  TRUE
x[x[["a"]] >= 0.3  &  x[["a"]] <= 0.8, -2]  # or use x[, "a"]
##         a c    d1    d2
## 2 0.78831 B  TRUE  TRUE
## 3 0.40898 C FALSE FALSE
```

Recall the drop argument to `[` and its effects on matrix indexing (Section 11.2.4). In the current case, its behaviour will be similar with regard to the operations on individual columns:

```
x[, 1]             # synonym: x[[1]] because drop=TRUE
## [1] 0.287578 0.788305 0.408977 0.883017 0.940467 0.045556
x[, 1, drop=FALSE]  # synonym: x[1]
##          a
## 1 0.287578
## 2 0.788305
## 3 0.408977
## 4 0.883017
## 5 0.940467
## 6 0.045556
```

When we extract a single row and more than one column, drop does not apply. It is because columns (unlike in matrices) can potentially be of different types:

```
x[1, 1:2]  # two numeric columns but the result is still a numeric
##         a        b
## 1 0.28758 0.070508
```

However:

```
x[1, 1]  # a single value
## [1] 0.28758
x[1, 1, drop=FALSE]  # a data frame with one row and one column
##         a
## 1 0.28758
```

**Note** Once again, let's take note of logical indexing in the presence of missing values:

```
x[x[["d1"]], ]  # `d1` is of the type logical
##              a       b    c   d1   d2
## 2      0.78831 0.12929    B TRUE TRUE
## NA          NA      NA <NA>   NA   NA
## NA.1        NA      NA <NA>   NA   NA
x[which(x[["d1"]]), ]  # `which` drops missing values
##         a       b c   d1   d2
## 2 0.78831 0.12929 B TRUE TRUE
```

The default behaviour is consistent with many other R functions. It explicitly indicates that something is missing. After all, when we select a "don't know", the result is unknown as well. Regretfully, this comes with no warning. As we seldom check missing values in the outputs manually, our absent-mindedness can lead to code bugs.

By far, we might have already noted that the index operator adjusts (not: resets) the `row.names` attribute. For instance:

```
(xs <- x[order(x[["a"]], decreasing=TRUE)[1:3], ])
##         a        b c    d1    d2
## 5 0.94047 -1.26506 E FALSE FALSE
## 4 0.88302  0.46092 D    NA  TRUE
## 2 0.78831  0.12929 B  TRUE  TRUE
```

It is a version of `x` comprised of the top three values in the `a` column. Indexing by means of character vectors will refer to `row.names` and `names`:

```
xs["5", c("a", "b")]
##         a       b
## 5 0.94047 -1.2651
```

It is not the same as `xs[5, c("a", "b")]`, even though `row.names` is formally an integer vector here.

Regarding the replacement version of the two-indexer variant of the `` `[` `` operator, it is a flexible tool. It permits the new content to be a vector, a data frame, a list, or even a matrix. Verifying this is left as an exercise.

**Note** If a data frame carries a matrix, to access a specific sub-column, we need to use the index/extract operator twice:

```
(x <- aggregate(iris[1], iris[5], function(x) c(Min=min(x), Max=max(x))))
##      Species Sepal.Length.Min Sepal.Length.Max
## 1     setosa              4.3              5.8
## 2 versicolor              4.9              7.0
## 3  virginica              4.9              7.9
```

```
x[["Sepal.Length"]][, "Min"]
## [1] 4.3 4.9 4.9
```

In other words, neither `x[["Sepal.Length.Min"]]` nor `x[, "Sepal.Length.Min"]` works.

---

**Exercise 12.10** *Write two replacement functions*[6]*. First, author* **`set_row_names`** *which replaces the* `row.names` *of a data frame with the contents of a specific column. For example:*

```
(x <- aggregate(iris[1], iris[5], mean))  # an example data frame
##      Species Sepal.Length
## 1     setosa        5.006
## 2 versicolor        5.936
## 3  virginica        6.588
set_row_names(x) <- "Species"
print(x)
##            Sepal.Length
## setosa            5.006
## versicolor        5.936
## virginica         6.588
```

*Second, implement* **`reset_row_names`** *which converts* `row.names` *to a standalone column of a given name. For instance:*

```
reset_row_names(x) <- "Type"
print(x)
##   Sepal.Length       Type
## 1        5.006     setosa
## 2        5.936 versicolor
## 3        6.588  virginica
```

*These two functions may be handy for they enable writing* `x[something, ]` *instead of* `x[x[["column"]] %in% something, ]`.

---

## 12.3 Common operations

Below we review the most commonly applied operations related to data frame wrangling. We have a few dedicated functions or methods overloaded for the `data.frame` class. However, we have already mastered most skills to deal with such objects effectively. Let's repeat: data frames are just lists exhibiting matrix-like behaviour.

[6] (*) Compare `pandas.DataFrame.set_index` and `pandas.DataFrame.reset_index` in Python.

### 12.3.1 Ordering rows

Ordering rows in a data frame with respect to different criteria can be easily achieved through the **order** function and the two-indexer version of `` `[` ``. For instance, here are the six fastest cars from `mtcars` in terms of the time (in seconds) to complete a 402-metre race:

```
mtcars6 <- mtcars[order(mtcars[["qsec"]])[1:6], c("qsec", "cyl", "gear")]
(mtcars6 <- `row.names<-`(cbind(model=row.names(mtcars6), mtcars6), NULL))
##             model  qsec cyl gear
## 1 Ford Pantera L 14.50   8    5
## 2  Maserati Bora 14.60   8    5
## 3     Camaro Z28 15.41   8    3
## 4   Ferrari Dino 15.50   6    5
## 5     Duster 360 15.84   8    3
## 6      Mazda RX4 16.46   6    4
```

**order** uses a stable sorting algorithm. Therefore, any sorting with respect to a different criterion will not break the *relative* ordering of `qsec` in row groups with ties:

```
mtcars6[order(mtcars6[["cyl"]]), ]
##             model  qsec cyl gear
## 4   Ferrari Dino 15.50   6    5
## 6      Mazda RX4 16.46   6    4
## 1 Ford Pantera L 14.50   8    5
## 2  Maserati Bora 14.60   8    5
## 3     Camaro Z28 15.41   8    3
## 5     Duster 360 15.84   8    3
```

`qsec` is still increasing in each of the two `cyl` groups.

**Example 12.11** *Notice the difference between ordering by* `cyl` *and* `gear`*:*

```
mtcars6[order(mtcars6[["cyl"]], mtcars6[["gear"]]), ]
##             model  qsec cyl gear
## 6      Mazda RX4 16.46   6    4
## 4   Ferrari Dino 15.50   6    5
## 3     Camaro Z28 15.41   8    3
## 5     Duster 360 15.84   8    3
## 1 Ford Pantera L 14.50   8    5
## 2  Maserati Bora 14.60   8    5
```

*vs* `gear` *and* `cyl`*:*

```
mtcars6[order(mtcars6[["gear"]], mtcars6[["cyl"]]), ]
##             model  qsec cyl gear
## 3     Camaro Z28 15.41   8    3
## 5     Duster 360 15.84   8    3
## 6      Mazda RX4 16.46   6    4
## 4   Ferrari Dino 15.50   6    5
```

```
## 1 Ford Pantera L 14.50   8    5
## 2  Maserati Bora 14.60   8    5
```

**Note** Mixing increasing and decreasing ordering is tricky as the `decreasing` argument to **`order`** currently does not accept multiple flags in all the contexts. Perhaps the easiest way to change the ordering direction is to use the unary minus operator on the column(s) to be sorted decreasingly.

```
mtcars6[order(mtcars6[["gear"]], -mtcars6[["cyl"]]), ]
##            model  qsec cyl gear
## 3     Camaro Z28 15.41   8    3
## 5     Duster 360 15.84   8    3
## 6      Mazda RX4 16.46   6    4
## 1 Ford Pantera L 14.50   8    5
## 2  Maserati Bora 14.60   8    5
## 4   Ferrari Dino 15.50   6    5
```

For factor and character columns, **`xtfrm`** can convert them to sort keys first.

```
mtcars6[order(mtcars6[["cyl"]], -xtfrm(mtcars6[["model"]])), ]
##            model  qsec cyl gear
## 6      Mazda RX4 16.46   6    4
## 4   Ferrari Dino 15.50   6    5
## 2  Maserati Bora 14.60   8    5
## 1 Ford Pantera L 14.50   8    5
## 5     Duster 360 15.84   8    3
## 3     Camaro Z28 15.41   8    3
```

Both statements act *like* the unsupported `decreasing=c(FALSE, TRUE)`.

**Exercise 12.12** *Write a method* ***`sort.data.frame`*** *that orders a data frame with respect to a given set of columns.*

```
sort.data.frame <- function(x, decreasing=FALSE, cols) ...to.do...
sort(mtcars6, cols=c("cyl", "model"))
##            model  qsec cyl gear
## 4   Ferrari Dino 15.50   6    5
## 6      Mazda RX4 16.46   6    4
## 3     Camaro Z28 15.41   8    3
## 5     Duster 360 15.84   8    3
## 1 Ford Pantera L 14.50   8    5
## 2  Maserati Bora 14.60   8    5
```

*Unfortunately, that `decreasing` must be of length one and be placed as the second argument is imposed by the* ***`sort`*** *S3 generic.*

### 12.3.2 Handling duplicated rows

`duplicated`, `anyDuplicated`, and `unique` have methods overloaded for the `data.frame` class. They can be used to indicate, get rid of, or replace the repeating rows.

```
sum(duplicated(iris))  # how many duplicated rows are there?
## [1] 1
iris[duplicated(iris), ]  # show the duplicated rows
##     Sepal.Length Sepal.Width Petal.Length Petal.Width   Species
## 143          5.8         2.7          5.1         1.9 virginica
```

### 12.3.3 Joining (merging) data frames

The `merge` function can perform the *join* operation that some readers might know from SQL[7]. It matches the items in the columns that two given data frames somewhat share. Then, it returns the combination of the corresponding rows.

**Example 12.13** *Two calls to* `merge` *could be used to match data on programmers (each identified by* `developer_id` *and giving such details as their name, location, main skill, etc.) with the information about the open-source projects (each identified by* `project_id` *and informing us about its title, scope, website, and so forth) they are engaged in (based on a third data frame defining* `developer_id` *and* `project_id` *pairs).*

As a simple illustration, consider two objects:

```
A <- data.frame(
    u=c("b0", "b1", "b2", "b3"),
    v=c("a0", "a1", "a2", "a3")
)

B <- data.frame(
    v=c("a0", "a2", "a2", "a4"),
    w=c("c0", "c1", "c2", "c3")
)
```

The two *common* columns, i.e., storing data of similar nature (a-something strings), are both named `v`.

First is the *inner (natural) join*, where we list only the matching pairs:

```
merge(A, B)  # x=A, y=B, by="v", all.x=FALSE, all.y=FALSE
##    v  u  w
## 1 a0 b0 c0
## 2 a2 b2 c1
## 3 a2 b2 c2
```

[7] Join is the reverse operation to data normalisation from relational database theory. It reduces data redundancy and increases their integrity. What data scientists need in data analysis, visualisation, and processing activities is sometimes the opposite of what the art of data management focuses on, i.e., efficient collection and storage of information. The readers are encouraged to learn about various normalisation forms from, e.g., [17] or any other course covering this topic.

The common column is included in the result only once. Next, the *left join* guarantees that all elements in the first data frame will be included in the result:

```
merge(A, B, all.x=TRUE)  # by="v", all.y=FALSE
##    v  u    w
## 1 a0 b0   c0
## 2 a1 b1 <NA>
## 3 a2 b2   c1
## 4 a2 b2   c2
## 5 a3 b3 <NA>
```

The *right join* includes all records in the second argument:

```
merge(A, B, all.y=TRUE)  # by="v", all.x=FALSE
##    v    u  w
## 1 a0   b0 c0
## 2 a2   b2 c1
## 3 a2   b2 c2
## 4 a4 <NA> c3
```

Lastly, the *full outer join* is their set-theoretic union:

```
merge(A, B, all.x=TRUE, all.y=TRUE)  # by="v"
##    v    u    w
## 1 a0   b0   c0
## 2 a1   b1 <NA>
## 3 a2   b2   c1
## 4 a2   b2   c2
## 5 a3   b3 <NA>
## 6 a4 <NA>   c3
```

Joining on more than one common column is also supported.

**Exercise 12.14** *Show how* **match** *(Section 5.4.1) can help author a very basic version of* **merge***.*

**Exercise 12.15** *Implement a version of* **match** *that allows the* `x` *and* `table` *arguments to be data frames with the same number of columns so that also the matching of pairs, triples, etc. is possible.*

### 12.3.4 Aggregating and transforming columns

It might be tempting to try aggregating data frames with **apply**. Sadly, currently, this function coerces its argument to a matrix. Hence, we should refrain from applying it on data frames whose columns are of mixed types. However, taking into account that data frames are special lists, we can always call **Map** and its relatives.

**Example 12.16** *Consider an example data frame:*

```
(iris_sample <- iris[sample(NROW(iris), 6), ])
##     Sepal.Length Sepal.Width Petal.Length Petal.Width    Species
```

```
## 28          5.2         3.5          1.5         0.2     setosa
## 80          5.7         2.6          3.5         1.0 versicolor
## 101         6.3         3.3          6.0         2.5  virginica
## 111         6.5         3.2          5.1         2.0  virginica
## 137         6.3         3.4          5.6         2.4  virginica
## 133         6.4         2.8          5.6         2.2  virginica
```

*To get the class of each column, we can call:*

```
sapply(iris_sample, class)  # or unlist(Map(class, iris))
## Sepal.Length  Sepal.Width Petal.Length  Petal.Width      Species
##    "numeric"    "numeric"    "numeric"    "numeric"     "factor"
```

*Next, here is a way to compute some aggregates of the numeric columns:*

```
unlist(Map(mean, Filter(is.numeric, iris_sample)))
## Sepal.Length  Sepal.Width Petal.Length  Petal.Width
##       6.0667       3.1333       4.5500       1.7167
```

*or:*

```
sapply(iris_sample[sapply(iris_sample, is.numeric)], mean)
## Sepal.Length  Sepal.Width Petal.Length  Petal.Width
##       6.0667       3.1333       4.5500       1.7167
```

*We can also fetch more than a single summary of each column:*

```
as.data.frame(Map(
    function(x) c(Min=min(x), Max=max(x)),
    Filter(is.numeric, iris_sample)
))
##     Sepal.Length Sepal.Width Petal.Length Petal.Width
## Min          5.2         2.6          1.5         0.2
## Max          6.5         3.5          6.0         2.5
```

*or:*

```
sapply(iris_sample[sapply(iris_sample, is.numeric)], quantile, c(0, 1))
##      Sepal.Length Sepal.Width Petal.Length Petal.Width
## 0%            5.2         2.6          1.5         0.2
## 100%          6.5         3.5          6.0         2.5
```

*The latter called* **`simplify2array`** *automatically. Thus, the result is a matrix.*

*On the other hand, the standardisation of all numeric features can be performed, e.g., via a call:*

```
iris_sample[] <- Map(function(x) {
    if (!is.numeric(x)) x else (x-mean(x))/sd(x)
}, iris_sample)
```

```
print(iris_sample)
##     Sepal.Length Sepal.Width Petal.Length Petal.Width    Species
## 28      -1.70405     1.03024     -1.76004    -1.65318     setosa
## 80      -0.72094    -1.49854     -0.60591    -0.78117 versicolor
## 101      0.45878     0.46829      0.83674     0.85384  virginica
## 111      0.85202     0.18732      0.31738     0.30884  virginica
## 137      0.45878     0.74927      0.60591     0.74484  virginica
## 133      0.65540    -0.93659      0.60591     0.52684  virginica
```

### 12.3.5 Handling missing values

The **`is.na`** method for objects of the class `data.frame` returns a logical matrix of the same dimensionality[8], indicating whether the corresponding items are missing or not. Of course, the default method can still be called on individual columns. Further, **`na.omit`** gets rid of *rows* with missing values.

**Exercise 12.17** *Given a data frame, use* **`is.na`** *and other functions like* **`apply`** *or* **`approx`** *to:*

1. *remove all rows that bear at least one missing value,*
2. *remove all rows that only consist of missing values,*
3. *remove all columns that carry at least one missing value,*
4. *for each column, replace all missing values with the column averages,*
5. *for each column, replace all missing values with values that linearly interpolate between the preceding and succeeding well-defined observations (which is of use in time series processing), e.g., the blanks in* **`c(0.60, 0.62, NA, 0.64, NA, NA, 0.58)`** *should be filled to obtain* **`c(0.60, 0.62, 0.63, 0.64, 0.62, 0.60, 0.58)`**.

### 12.3.6 Reshaping data frames

Consider an example matrix:

```
A <- matrix(round(runif(6), 2), nrow=3,
    dimnames=list(
        c("X", "Y", "Z"),  # row labels
        c("u", "v")        # column labels
))
names(dimnames(A)) <- c("Row", "Col")
print(A)
##    Col
## Row    u    v
##   X 0.29 0.88
##   Y 0.79 0.94
##   Z 0.41 0.05
```

[8] Provided that a data frame does not carry a matrix column.

The **as.data.frame** method for the table class can be called directly on any array-like object:

```
as.data.frame.table(A, responseName="Val", stringsAsFactors=FALSE)
##   Row Col  Val
## 1   X   u 0.29
## 2   Y   u 0.79
## 3   Z   u 0.41
## 4   X   v 0.88
## 5   Y   v 0.94
## 6   Z   v 0.05
```

It is an instance of array *reshaping*. More precisely, we call it *stacking*. We converted from a *wide* (okay, in this example, not so wide, as we only have two columns) to a *long* (tall) format.

The above can also be achieved by means of the **reshape** function which is more flexible and operates directly on data frames (but is harder to use):

```
(df <- `names<-`(
    data.frame(row.names(A), A, row.names=NULL),
    c("Row", "Col.u", "Col.v")))
##   Row Col.u Col.v
## 1   X  0.29  0.88
## 2   Y  0.79  0.94
## 3   Z  0.41  0.05
(stacked <- reshape(df, varying=2:3, direction="long"))
##     Row time  Col id
## 1.u   X    u 0.29  1
## 2.u   Y    u 0.79  2
## 3.u   Z    u 0.41  3
## 1.v   X    v 0.88  1
## 2.v   Y    v 0.94  2
## 3.v   Z    v 0.05  3
```

Maybe the default column names are not superb, but we can adjust them manually afterwards. The reverse operation is called *unstacking*:

```
reshape(stacked, idvar="Row", timevar="time", drop="id", direction="wide")
##     Row Col.u Col.v
## 1.u   X  0.29  0.88
## 2.u   Y  0.79  0.94
## 3.u   Z  0.41  0.05
```

**Exercise 12.18** *Given a named numeric vector, convert it to a data frame with two columns. For instance:*

```
convert <- function(x) ...to.do...
x <- c(spam=42, eggs=7, bacon=3)
convert(x)
```

```
##     key value
## 1   spam    42
## 2   eggs     7
## 3  bacon     3
```

**Exercise 12.19** *Stack the* `WorldPhones` *dataset. Then, unstack it back. Furthermore, unstack the stacked set after removing*[9] *five random rows from it and randomly permuting all the remaining rows. Fill in the missing entries with* `NA`*s.*

**Exercise 12.20** *Implement a basic version of* **`as.data.frame.table`** *manually (using* **`rep`** *etc.). Also, write a function* **`as.table.data.frame`** *that computes its reverse. Make sure both functions are compatible with each other.*

**Exercise 12.21** `Titanic` *is a four-dimensional array. Convert it to a long data frame.*

**Exercise 12.22** *Perform what follows on the undermentioned data frame:*

1. *convert the second column to a list of character vectors (split at* `","`*);*
2. *extract the first elements from each of such vectors;*
3. *extract the last elements;*
4. *(*) unstack the split data frame;*
5. *(*) stack it back to a data frame that carries a list;*
6. *convert the list back to a character column (concatenate with* `","` *as separator).*

```
(x <- data.frame(
    name=c("Kat", "Ron", "Jo", "Mary"),
    food=c("buckwheat", "spam,bacon,spam", "", "eggs,spam,spam,lollipops")
))
##   name                     food
## 1  Kat                buckwheat
## 2  Ron          spam,bacon,spam
## 3   Jo
## 4 Mary eggs,spam,spam,lollipops
```

**Exercise 12.23** *Write a function that converts all matrix-based columns in a given data frame to separate atomic columns. Furthermore, author a function that does the opposite, i.e., groups all columns with similar prefixes and turns them into matrices.*

### 12.3.7 Aggregating data in groups

We can straightforwardly apply various transforms on data groups determined by a factor-like variable or their combination thanks to the **`split.data.frame`** method, which returns a list of data frames. For example:

[9] The original dataset can be thought of as representing a fully crossed design experiment (all combinations of two grouping variables are present). Its truncated version is like an incomplete crossed design.

```
x <- data.frame(
    a=c(    10,     20,     30,     40,     50),
    u=c("spam", "spam", "eggs", "spam", "eggs"),
    v=c(     1,      2,      1,      1,      1)
)
split(x, x["u"])  # i.e., split.data.frame(x, x["u"]) or x[["u"]]
## $eggs
##    a    u v
## 3 30 eggs 1
## 5 50 eggs 1
##
## $spam
##    a    u v
## 1 10 spam 1
## 2 20 spam 2
## 4 40 spam 1
```

It split x with respect to the u column, which served as the grouping variable. On the other hand:

```
split(x, x[c("u", "v")])  # sep="."
## $eggs.1
##    a    u v
## 3 30 eggs 1
## 5 50 eggs 1
##
## $spam.1
##    a    u v
## 1 10 spam 1
## 4 40 spam 1
##
## $eggs.2
## [1] a u v
## <0 rows> (or 0-length row.names)
##
## $spam.2
##    a    u v
## 2 20 spam 2
```

It partitioned with respect to a combination of two factor-like sequences. A nonexistent level pair (eggs, 2) resulted in an empty data frame.

**Exercise 12.24** ***split.data.frame*** *(when called directly) can also be used to break a matrix into a list of matrices (rowwisely). Given a matrix, perform its train-test split: allocate, say, 70% of the rows at random into one matrix and the remaining 30% into another.*

**sapply** is convenient if we need to aggregate grouped numeric data. To recall, it is a combination of **lapply** (one-vector version of **Map**) and **simplify2array** (Section 11.1.3).

```
sapply(split(iris[1:2], iris[5]), sapply, mean)
##                setosa versicolor virginica
## Sepal.Length  5.006      5.936     6.588
## Sepal.Width   3.428      2.770     2.974
```

If the function to apply returns more than a single value, **sapply** will not return too informative a result. The list of matrices converted to a matrix will not have the `row.names` argument set:

```
MinMax <- function(x) c(Min=min(x), Max=max(x))
sapply(split(iris[1:2], iris[5]), sapply, MinMax)
##      setosa versicolor virginica
## [1,]    4.3        4.9       4.9
## [2,]    5.8        7.0       7.9
## [3,]    2.3        2.0       2.2
## [4,]    4.4        3.4       3.8
```

As a workaround, we either call **simplify2array** explicitly, or pass `simplify="array"` to **sapply**:

```
(res <- sapply(
    split(iris[1:2], iris[5]),
    sapply,
    MinMax,
    simplify="array"
))  # or simplify2array(lapply(...) or Map(...) etc.)
## , , setosa
##
##     Sepal.Length Sepal.Width
## Min          4.3         2.3
## Max          5.8         4.4
##
## , , versicolor
##
##     Sepal.Length Sepal.Width
## Min          4.9         2.0
## Max          7.0         3.4
##
## , , virginica
##
##     Sepal.Length Sepal.Width
## Min          4.9         2.2
## Max          7.9         3.8
```

It produced a three-dimensional array, which is particularly handy if we now wish to access specific results by name:

```
res[, "Sepal.Length", "setosa"]
```

```
## Min Max
## 4.3 5.8
```

The previously mentioned **as.data.frame.table** method will work on it like a charm (up to the column names, which we can change):

```
as.data.frame.table(res, stringsAsFactors=FALSE)
##     Var1         Var2       Var3 Freq
## 1    Min Sepal.Length     setosa  4.3
## 2    Max Sepal.Length     setosa  5.8
## 3    Min  Sepal.Width     setosa  2.3
## 4    Max  Sepal.Width     setosa  4.4
## 5    Min Sepal.Length versicolor  4.9
## 6    Max Sepal.Length versicolor  7.0
## 7    Min  Sepal.Width versicolor  2.0
## 8    Max  Sepal.Width versicolor  3.4
## 9    Min Sepal.Length  virginica  4.9
## 10   Max Sepal.Length  virginica  7.9
## 11   Min  Sepal.Width  virginica  2.2
## 12   Max  Sepal.Width  virginica  3.8
```

---

**Note** If the grouping (by) variable is a list of two or more factors, the combined levels will be concatenated to a single string. This behaviour yields a result that may be deemed convenient in some contexts but not necessarily so in other ones.

```
as.data.frame.table(as.array(sapply(
    split(ToothGrowth["len"], ToothGrowth[c("supp", "dose")], sep="_"),
    sapply,  # but check also: function(...) as.matrix(sapply(...)),
    mean
)), stringsAsFactors=FALSE)
##          Var1  Freq
## 1 OJ_0.5.len 13.23
## 2 VC_0.5.len  7.98
## 3   OJ_1.len 22.70
## 4   VC_1.len 16.77
## 5   OJ_2.len 26.06
## 6   VC_2.len 26.14
```

The name of the aggregated column (`len`) has been included, because **sapply** simplifies the result to a flat vector too eagerly.

---

**aggregate** can assist us when a single function is to be applied on all columns in a data frame:

```
aggregate(iris[-5], iris[5], mean)  # neither iris[[5]] nor iris[, 5]
##      Species Sepal.Length Sepal.Width Petal.Length Petal.Width
```

```
## 1     setosa        5.006       3.428        1.462       0.246
## 2 versicolor        5.936       2.770        4.260       1.326
## 3  virginica        6.588       2.974        5.552       2.026
aggregate(ToothGrowth["len"], ToothGrowth[c("supp", "dose")], mean)
##   supp dose   len
## 1   OJ  0.5 13.23
## 2   VC  0.5  7.98
## 3   OJ  1.0 22.70
## 4   VC  1.0 16.77
## 5   OJ  2.0 26.06
## 6   VC  2.0 26.14
```

The second argument, by, must be list-like (this includes data frames). Neither a factor nor an atomic vector is acceptable. Also, if the function being applied returns many values, they will be wrapped into a matrix column:

```
(x <- aggregate(iris[2], iris[5], function(x) c(Min=min(x), Max=max(x))))
##      Species Sepal.Width.Min Sepal.Width.Max
## 1     setosa             2.3             4.4
## 2 versicolor             2.0             3.4
## 3  virginica             2.2             3.8
class(x[["Sepal.Width"]])
## [1] "matrix" "array"
x[["Sepal.Width"]]  # not: Sepal.Width.Max, etc.
##      Min Max
## [1,] 2.3 4.4
## [2,] 2.0 3.4
## [3,] 2.2 3.8
```

It is actually handy: by referring to x[["Sepal.Width"]], we access all the stats for this column. Further, if many columns are being aggregated simultaneously, we can process all the summaries in the same way.

**Exercise 12.25** *Check out the **by** function, which supports basic split-apply-bind use cases. Note the particularly odd behaviour of the **print** method for the by class.*

The most flexible scenario involves applying a custom function returning any set of aggregates in the form of a list and then row-binding the results to obtain a data frame.

**Example 12.26** *The following implements an R version of what we would express in SQL as:*

```
SELECT supp, dose, AVG(len) AS ave_len, COUNT(*) AS count
FROM ToothGrowth
GROUP BY supp, dose
```

*Ad rem:*

```
do.call(rbind.data.frame, lapply(
    split(ToothGrowth, ToothGrowth[c("supp", "dose")]),
```

```
    function(df) list(
        supp=df[1, "supp"],
        dose=df[1, "dose"],
        ave_len=mean(df[["len"]]),
        count=NROW(df)
    )
))
##        supp dose ave_len count
## OJ.0.5   OJ  0.5   13.23    10
## VC.0.5   VC  0.5    7.98    10
## OJ.1     OJ  1.0   22.70    10
## VC.1     VC  1.0   16.77    10
## OJ.2     OJ  2.0   26.06    10
## VC.2     VC  2.0   26.14    10
```

**Exercise 12.27** *Many aggregation functions are idempotent, which means that when they are fed with a vector with all the elements being identical, the result is exactly that unique element:* `min`*,* `mean`*,* `median`*, and* `max` *behave this way. Overload the* `mean` *and* `median` *methods for character vectors and factors. They should return* NA *and give a warning for sequences where not all elements are the same. Otherwise, they are expected to output the unique value.*

```
mean.character <- function(x, na.rm=FALSE, ...) ...to.do...
mean.factor <- function(x, na.rm=FALSE, ...) ...to.do...
```

*This way, we can also aggregate the grouping variables conveniently:*

```
do.call(rbind.data.frame,
    lapply(split(ToothGrowth, ToothGrowth[c("supp", "dose")]), lapply, mean))
##          len supp dose
## OJ.0.5 13.23   OJ  0.5
## VC.0.5  7.98   VC  0.5
## OJ.1   22.70   OJ  1.0
## VC.1   16.77   VC  1.0
## OJ.2   26.06   OJ  2.0
## VC.2   26.14   VC  2.0
```

**Example 12.28** *Let's study a function that takes a named list* `x` *(can be a data frame) and a sequence of* `col=f` *pairs, and applies the function* `f` *(or each function from a list of functions* `f`*) on the element named* `col` *in* `x`*:*

```
napply <- function(x, ...)
{
    fs <- list(...)
    cols <- names(fs)
    stopifnot(is.list(x), !is.null(names(x)))
    stopifnot(all(cols %in% names(x)))
    do.call(
        c,  # concatenates lists
```

```
        lapply(
            structure(seq_along(fs), names=cols),
            function(i)
            {   # always returns a list
                y <- x[[ cols[i] ]]
                if (is.function(fs[[i]]))
                    list(fs[[i]](y))
                else
                    lapply(fs[[i]], function(f) f(y))
            }
        )
    )
}
```

For example:

```
first <- function(x, ...) head(x, n=1L, ...)  # helper function
napply(ToothGrowth,
    supp=first, dose=first, len=list(ave=mean, count=length)
)
## $supp
## [1] VC
## Levels: OJ VC
##
## $dose
## [1] 0.5
##
## $len.ave
## [1] 18.813
##
## $len.count
## [1] 60
```

It applied **`first`** on both `ToothGrowth[["supp"]]` and `ToothGrowth[["dose"]]` as well as **`mean`** and **`length`** on `ToothGrowth[["len"]]`. We included list names for a more dramatic effect. And now:

```
do.call(
    rbind.data.frame,
    lapply(
        split(ToothGrowth, ToothGrowth[c("supp", "dose")]),
        napply,
        supp=first, dose=first, len=list(ave=mean, count=length)
    )
)
##         supp dose len.ave len.count
## OJ.0.5    OJ  0.5   13.23        10
## VC.0.5    VC  0.5    7.98        10
```

```
## OJ.1      OJ  1.0   22.70         10
## VC.1      VC  1.0   16.77         10
## OJ.2      OJ  2.0   26.06         10
## VC.2      VC  2.0   26.14         10
```

*or even:*

```
gapply <- function(x, by, ...)
    do.call(rbind.data.frame, lapply(
        split(x, x[by]),
        function(x, ...)
            do.call(napply, c(  # add all by=first calls
                x=list(x),
                `names<-`(rep(list(first), length(by)), by),
                list(...)
            )),
        ...
    ))
```

*And now:*

```
gapply(iris, "Species", Sepal.Length=mean, Sepal.Width=list(min, max))
##               Species Sepal.Length Sepal.Width1 Sepal.Width2
## setosa         setosa        5.006          2.3          4.4
## versicolor versicolor        5.936          2.0          3.4
## virginica   virginica        6.588          2.2          3.8
gapply(ToothGrowth, c("supp", "dose"), len=list(ave=mean, count=length))
##        supp dose len.ave len.count
## OJ.0.5   OJ  0.5   13.23        10
## VC.0.5   VC  0.5    7.98        10
## OJ.1     OJ  1.0   22.70        10
## VC.1     VC  1.0   16.77        10
## OJ.2     OJ  2.0   26.06        10
## VC.2     VC  2.0   26.14        10
```

*This brings fun back to R programming in the sad times when many things are given to us on a plate (the thorough testing of the above is left as an exercise).*

**Example 12.29** *In Section 10.4, we mentioned (without giving the implementation) the* **group_by** *function returning a list of the class* `list_dfs`*. It splits a data frame into a list of data frames with respect to a combination of levels in given named columns:*

```
group_by <- function(df, by)
{
    stopifnot(is.character(by), is.data.frame(df))
    df <- droplevels(df)  # factors may have unused levels
    structure(
        split(df, df[names(df) %in% by]),
        class="list_dfs",
```

```
        by=by
    )
}
```

*The next function applies a set of aggregates on every column of each data frame in a given list (two nested* **`lapply`***s plus cosmetic additions):*

```
aggregate.list_dfs <- function(x, FUN, ...)
{
    aggregates <- lapply(x, function(df) {
        is_by <- names(df) %in% attr(x, "by")
        res <- lapply(df[!is_by], FUN, ...)
        res_mat <- do.call(rbind, res)
        if (is.null(dimnames(res_mat)[[2]]))
            dimnames(res_mat)[[2]] <- paste0("f", seq_len(NCOL(res_mat)))
        cbind(
            `row.names<-`(df[1, is_by, drop=FALSE], NULL),
            x=row.names(res_mat),
            `row.names<-`(res_mat, NULL)
        )
    })
    combined_aggregates <- do.call(rbind.data.frame, aggregates)
    `row.names<-`(combined_aggregates, NULL)
}
aggregate(group_by(ToothGrowth, c("supp", "dose")), range)
##   supp dose   x   f1   f2
## 1   OJ  0.5 len  8.2 21.5
## 2   VC  0.5 len  4.2 11.5
## 3   OJ  1.0 len 14.5 27.3
## 4   VC  1.0 len 13.6 22.5
## 5   OJ  2.0 len 22.4 30.9
## 6   VC  2.0 len 18.5 33.9
```

*We really want our API to be bloated, so let's introduce a* convenience *function, which is a specialised version of the above:*

```
mean.list_dfs <- function(x, ...)
    aggregate.list_dfs(x, function(y) c(Mean=mean(y, ...)))
mean(group_by(iris[51:150, c(2, 3, 5)], "Species"))
##      Species            x  Mean
## 1 versicolor  Sepal.Width 2.770
## 2 versicolor Petal.Length 4.260
## 3  virginica  Sepal.Width 2.974
## 4  virginica Petal.Length 5.552
```

### 12.3.8 Transforming data in groups

Variables will sometimes need to be transformed relative to what is happening in a dataset's subsets. This is the case, e.g., where we decide that missing values should be replaced by the corresponding within-group averages or want to compute the relative ranks or z-scores.

If the loss of the original ordering of rows is not an issue, the standard split-apply-bind will suffice. Here is an example data frame:

```
(x <- data.frame(
    a=c( 10,   1,  NA,  NA,  NA,   4),
    b=c( -1,  10,  40,  30,   1,  20),
    c=runif(6),
    d=c("v", "u", "u", "u", "v", "u")
))
##    a  b       c d
## 1 10 -1 0.52811 v
## 2  1 10 0.89242 u
## 3 NA 40 0.55144 u
## 4 NA 30 0.45661 u
## 5 NA  1 0.95683 v
## 6  4 20 0.45333 u
```

Some operations:

```
fill_na <- function(x) `[<-`(x, is.na(x), value=mean(x[!is.na(x)]))
standardise <- function(x) (x-mean(x))/sd(x)
```

And now:

```
x_groups <- lapply(
    split(x, x["d"]),
    function(df) {
        df[["a"]] <- fill_na(df[["a"]])
        df[["b"]] <- rank(df[["b"]])
        df[["c"]] <- standardise(df[["c"]])
        df
    }
)
do.call(rbind.data.frame, x_groups)
##        a b        c d
## u.2  1.0 1  1.46357 u
## u.3  2.5 4 -0.17823 u
## u.4  2.5 3 -0.63478 u
## u.6  4.0 2 -0.65057 u
## v.1 10.0 1 -0.70711 v
## v.5 10.0 2  0.70711 v
```

Only the *relative* ordering of rows within groups has been retained. Overall, the rows are in a different order. If this is an issue, we can use the **unsplit** function:

```
unsplit(x_groups, x["d"])
##      a b        c d
## 1 10.0 1 -0.70711 v
## 2  1.0 1  1.46357 u
## 3  2.5 4 -0.17823 u
## 4  2.5 3 -0.63478 u
## 5 10.0 2  0.70711 v
## 6  4.0 2 -0.65057 u
```

**Exercise 12.30** *Show how we can perform the above also via the replacement version of* **`split`***.*

**Example 12.31** *(*) Recreating the previous ordering can be done manually, too. It is because the split operation behaves as if we first ordered the data frame with respect to the grouping variable(s) (using a stable sorting algorithm). Here is a transformation of an example data frame split by a combination of two factors:*

```
(x <- `row.names<-`(ToothGrowth[sample(NROW(ToothGrowth), 10), ], NULL))
##     len supp dose
## 1  23.0   OJ  2.0
## 2  23.3   OJ  1.0
## 3  29.4   OJ  2.0
## 4  14.5   OJ  1.0
## 5  11.2   VC  0.5
## 6  20.0   OJ  1.0
## 7  24.5   OJ  2.0
## 8  10.0   OJ  0.5
## 9   9.4   OJ  0.5
## 10  7.0   VC  0.5
(y <- do.call(rbind.data.frame, lapply(
    split(x, x[c("dose", "supp")]),  # two grouping variables
    function(df) {
        df[["len"]] <- df[["len"]] * 100^df[["dose"]] *  # whatever
            ifelse(df[["supp"]] == "OJ", -1, 1)          # do not overthink it
        df
    }
)))
##                len supp dose
## 0.5.OJ.8      -100   OJ  0.5
## 0.5.OJ.9       -94   OJ  0.5
## 1.OJ.2       -2330   OJ  1.0
## 1.OJ.4       -1450   OJ  1.0
## 1.OJ.6       -2000   OJ  1.0
## 2.OJ.1     -230000   OJ  2.0
## 2.OJ.3     -294000   OJ  2.0
## 2.OJ.7     -245000   OJ  2.0
## 0.5.VC.5       112   VC  0.5
## 0.5.VC.10       70   VC  0.5
```

*Section 5.4.4 mentioned that by calling* **`order`***, we can determine the inverse of a given permutation. Hence, we can call:*

```
y[order(order(x[["supp"]], x[["dose"]])), ]  # not: dose, supp
##              len supp dose
## 2.OJ.1    -230000   OJ  2.0
## 1.OJ.2      -2330   OJ  1.0
## 2.OJ.3    -294000   OJ  2.0
## 1.OJ.4      -1450   OJ  1.0
## 0.5.VC.5      112   VC  0.5
## 1.OJ.6      -2000   OJ  1.0
## 2.OJ.7    -245000   OJ  2.0
## 0.5.OJ.8     -100   OJ  0.5
## 0.5.OJ.9      -94   OJ  0.5
## 0.5.VC.10      70   VC  0.5
```

*Additionally, we can manually restore the original* `row.names`*, et voilà.*

### 12.3.9 Metaprogramming-based techniques (*)

Section 9.4.7 mentioned a few functions that provide *convenient* interfaces to some common data frame operations. These include **`transform`**, **`subset`**, **`with`**, and basically every procedure accepting a formula. The popular **`data.table`** and **`dplyr`** packages also belong to this class (Section 12.3.10).

Unfortunately, each method relying on metaprogramming must be studied separately because it is free to interpret the *form* of the passed arguments arbitrarily, without taking into account their *real* meaning. As we are concerned with developing a more universal skill set, we avoid[10] them in this course. They do not go beyond what we have learnt so far.

Withal, they are thought-provoking on their own. Furthermore, they are popular in other users' code. Thus, after all, they deserve the honourable mention.

**Example 12.32** *Consider an example call to the* **`subset`** *function:*

```
subset(iris, Sepal.Length<4.5, -(Sepal.Width:Petal.Width))
##    Sepal.Length Species
## 9           4.4  setosa
## 14          4.3  setosa
## 39          4.4  setosa
## 43          4.4  setosa
```

*Neither the second nor the third argument makes sense as a standalone R expression. We have not defined the named variables used there:*

[10] We are not alone in our calling to refrain from using them. **`help`**`("subset")` (and **`help`**`("transform")` similarly) warns: *This is a convenience function intended for use interactively. For programming, it is better to use the standard subsetting functions like* `` `[` ``*, and in particular the nonstandard evaluation of argument* **`subset`** *can have unanticipated consequences.* The same in **`help`**`("with")`: *For* interactive *use, this is very effective and nice to read. For programming however, i.e., in one's functions, more care is needed, and typically one should refrain from using* **`with`***, as, e.g., variables in data may accidentally override local variables.*

```
Sepal.Length<4.5            # utter nonsense
## Error: ! object 'Sepal.Length' not found
-(Sepal.Width:Petal.Width)  # gibberish
## Error: ! object 'Sepal.Width' not found
```

*Only from* **help("subset")** *we can learn that this tool* assumes *that the expression passed as the second argument plays the role of a row selector. Moreover, the third one is meant to remove all the columns between the two given ones.*

*In our course, we pay attention to developing* transferable *skills. We believe that R is not the only language we will learn during our long and happy lives. It is much more likely that in the next environment, we will become used to writing something of the more basic form:*

```
between <- function(x, from, to) match(from, x):match(to, x)
iris[iris[["Sepal.Length"]]<4.5,
    -between(names(iris), "Sepal.Width", "Petal.Width")]
##    Sepal.Length Species
## 9           4.4  setosa
## 14          4.3  setosa
## 39          4.4  setosa
## 43          4.4  setosa
```

**Example 12.33** *With* **transform***, we can add, modify, and remove columns in a data frame. Existing features can be referred to as if they were ordinary variables:*

```
(mtcars4 <- mtcars[sample(seq_len(NROW(mtcars)), 4), c("hp", "am", "mpg")])
##                     hp am  mpg
## Maserati Bora      335  1 15.0
## Cadillac Fleetwood 205  0 10.4
## Honda Civic         52  1 30.4
## Merc 450SLC        180  0 15.2
transform(mtcars4, log_hp=log(hp), am=2*am-1, hp=NULL, fcon=235/mpg)
##                    am  mpg log_hp    fcon
## Maserati Bora       1 15.0 5.8141 15.6667
## Cadillac Fleetwood -1 10.4 5.3230 22.5962
## Honda Civic         1 30.4 3.9512  7.7303
## Merc 450SLC        -1 15.2 5.1930 15.4605
```

*Similarly,* **attach** *adds any named list to the search path (see* Section 16.2.6*) but it does not support altering their contents. As an alternative,* **within** *may be called:*

```
within(mtcars4, {
    log_hp <- log(hp)
    fcon <- 235/mpg
    am <- factor(am, levels=c(0, 1), labels=c("no", "yes"))
    hp <- NULL
})
##                     am  mpg    fcon log_hp
## Maserati Bora      yes 15.0 15.6667 5.8141
```

```
## Cadillac Fleetwood  no 10.4 22.5962 5.3230
## Honda Civic        yes 30.4  7.7303 3.9512
## Merc 450SLC         no 15.2 15.4605 5.1930
```

*Those who find writing `mtcars4[["name"]]` instead of `name` too exhausting, can save a few keystrokes.*

**Example 12.34** *As mentioned in Section 10.3.4 (see Section 17.6 for more details), formulae are special objects that consist of two unevaluated expressions separated by a tilde, `~`. Functions can support formulae and do what they please with them. However, a popular approach is to allow them to express "something grouped by something else" or "one thing as a function of other things".*

```
do.call(rbind.data.frame, lapply(split(ToothGrowth, ~supp+dose), head, 1))
##         len supp dose
## OJ.0.5 15.2   OJ  0.5
## VC.0.5  4.2   VC  0.5
## OJ.1   19.7   OJ  1.0
## VC.1   16.5   VC  1.0
## OJ.2   25.5   OJ  2.0
## VC.2   23.6   VC  2.0
aggregate(cbind(mpg, log_hp=log(hp))~am:cyl, mtcars, mean)
##   am cyl    mpg log_hp
## 1  0   4 22.900 4.4186
## 2  1   4 28.075 4.3709
## 3  0   6 19.125 4.7447
## 4  1   6 20.567 4.8552
## 5  0   8 15.050 5.2553
## 6  1   8 15.400 5.6950
head(model.frame(mpg+hp~log(hp)+I(1/qsec), mtcars))
##                   mpg + hp log(hp)    I(1/qsec)
## Mazda RX4            131.0  4.7005 0.060753....
## Mazda RX4 Wag        131.0  4.7005 0.058754....
## Datsun 710           115.8  4.5326 0.053734....
## Hornet 4 Drive       131.4  4.7005 0.051440....
## Hornet Sportabout    193.7  5.1648 0.058754....
## Valiant              123.1  4.6540 0.049455....
```

*If these examples seem esoteric, it is because it is precisely the case. We need to consult the corresponding functions' manuals to discover what they do. And, as we do not recommend their use to beginner programmers, we will not explain them here. Don't trip.*

**Exercise 12.35** *In the last example, the peculiar printing of the last column is due to which method's being overloaded?*

In the third part of this book, we will return to these functions for they will serve as an amusing illustration of how to indite our own procedures that rely on metaprogramming techniques.

### 12.3.10 A note on the `dplyr` (`tidyverse`) and `data.table` packages (*)

**`data.table`** and **`dplyr`** are very popular packages that implement common data frame transformations. In particular, the latter is part of an immerse system of interdependent extensions called **`tidyverse`** which became quite invasive over the last few years. They both heavily rely on metaprogramming and introduce entirely new APIs featuring hundreds of functions for the operations we already know well how to perform (the calamity of superabundance).

Still, their users must remember that they will *need* to rely on base functions when the processing of other prominent data structures is required, e.g., of fancy lists and matrices. Base R (and its predecessor, S) has long ago proven to be a versatile tool for rapid prototyping, calling specialised procedures written in C or Java, and wrangling data that *fit into memory*. Even though some operations from the mentioned packages may be much faster for larger datasets, the speed is less often an issue in practice than what most users might think.

For larger problems, techniques for working with batches of data, sampling methods, or aggregating data stored elsewhere are often the way to go, especially when building machine learning models or visualisation[11] is required. Usually, the most recent data will be stored in external, normalised databases, and we will need to join a few tables to fetch something valuable from the perspective of the current task's context.

Thus, we cannot stress enough that, in many situations, SQL, not the other tools, is the most powerful interface to more considerable amounts of data. Learning it will give us the skills we can use later in other programming environments.

---

**Note** Of course, certain functions from **`tidyverse`** and related packages we will find very helpful after all. Quite annoyingly, they tend to return objects of the class `tibble` (`tbl_df`) (e.g., **`haven::read.xpt`** that reads SAS data files). Luckily, they are subclasses of `data.frame`; we can always use **`as.data.frame`** to get our favourite objects back.

---

## 12.4 Exercises

**Exercise 12.36** *Answer the following questions.*

- *What attributes a data frame is equipped with?*
- *If `row.names` is an integer vector, how to access rows labelled 1, 7, and 42?*
- *How to create a data frame that carries a column that is a list of character vectors of different lengths?*

[11] For example, drawing scatter plots of billions of points makes little sense as they may result in unreadable images of large file sizes. The points need to be sampled or summarised somehow (e.g., binned); see Chapter 13.

- *How to create a data frame that includes a matrix column?*
- *How to convert all numeric columns in a data frame to a numeric matrix?*
- *Assuming that `x` is an atomic vector, what is the difference between `as.data.frame(x)`, `as.data.frame(as.list(x))`, `as.data.frame(list(a=x))`, and `data.frame(a=x)`?*

**Exercise 12.37** *Assuming that `x` is a data frame, what is the meaning of/difference between the following:*

- `x["u"]` *vs* `x[["u"]]` *vs* `x[, "u"]`*?*
- `x["u"][1]` *vs* `x[["u"]][1]` *vs* `x[1, "u"]` *vs* `x[1, "u", drop=FALSE]`*?*
- `x[which(x[[1]] > 0), ]` *vs* `x[x[[1]] > 0, ]`*?*
- `x[grep("^foo", names(x))]`*?*

**Exercise 12.38** *We have a data frame with columns named like: `ID` (character), `checked` (logical, possibly with missing values), `category` (factor), `x0`, …, `x9` (ten separate numeric columns), `y0`, …, `y9` (ten separate numeric columns), `coords` (numeric matrix with two columns named `lat` and `long`), and `features` (list of character vectors of different lengths).*

- *How to extract the rows where `checked` is `TRUE`?*
- *How to extract the rows for which `ID` is like three letters and then five digits (e.g., `XYZ12345`)?*
- *How to select all the numeric columns in one go?*
- *How to extract a subset comprised only of the `ID` and `x`-something columns?*
- *How to get rid of all the columns between `x3` and `y7`?*
- *Assuming that the `ID`s are like three letters and then five digits, how to add two columns: `ID3` (the letters) and `ID5` (the five digits)?*
- *How to check whether both `lat` and `long` in `coords` are negative?*
- *How to add the column indicating the number of `features`?*
- *How to extract the rows where `"spam"` is amongst the `features`?*
- *How to convert it to a long data frame with two columns: `ID` and `feature` (individual strings)?*
- *How to change the name of the `ID` column to `id`?*
- *How to make the `y`-foo columns appear before the `x`-bar ones?*
- *How to order the rows with respect to `checked` (`FALSE` first, then `TRUE`) and `ID`s (decreasingly)?*
- *How to remove rows with duplicate IDs?*
- *How to determine how many entries correspond to each `category`?*
- *How to compute the average `lat` and `long` in each `category`?*
- *How to compute the average `lat` and `long` for each `category` and `checked` combined?*

**Exercise 12.39** *Consider the `flights`[12] dataset. Give some ways to select all rows between March and October (regardless of the year).*

**Exercise 12.40** *In this task, you will be working with a version of a dataset on 70k+ Melbourne trees (`urban_forest`[13]).*

1. *Load the downloaded dataset by calling the **`read.csv`** function.*
2. *Fetch the IDs (`CoM.ID`) and trunk diameters (`Diameter.Breast.Height`) of the horse chestnuts with five smallest diameters at breast height. The output data frame must be sorted with respect to `Diameter.Breast.Height`, decreasingly.*
3. *Create a new data frame that gives the number of trees planted in each year.*
4. *Compute the average age (in years, based on `Year.Planted`) of the trees of genera (each genus separately): Eucalyptus, Platanus, Ficus, Acer, and Quercus.*

**Exercise 12.41** *(*) Consider the historic data dumps of Stack Exchange[14] available here[15]. Export these CSV files to an SQLite database. Then, write some R code that corresponds to the following SQL queries. Use **`dbGetQuery`** to verify your results.*

*First:*

```
SELECT
    Users.DisplayName,
    Users.Age,
    Users.Location,
    SUM(Posts.FavoriteCount) AS FavoriteTotal,
    Posts.Title AS MostFavoriteQuestion,
    MAX(Posts.FavoriteCount) AS MostFavoriteQuestionLikes
FROM Posts
JOIN Users ON Users.Id=Posts.OwnerUserId
WHERE Posts.PostTypeId=1
GROUP BY OwnerUserId
ORDER BY FavoriteTotal DESC
LIMIT 10
```

*Second:*

```
SELECT
    Posts.ID,
    Posts.Title,
    Posts2.PositiveAnswerCount
FROM Posts
JOIN (
        SELECT
            Posts.ParentID,
```

[12] https://github.com/gagolews/teaching-data/blob/master/other/flights.csv
[13] https://github.com/gagolews/teaching-data/raw/master/marek/urban_forest.csv.gz
[14] https://travel.stackexchange.com/
[15] https://github.com/gagolews/teaching-data/tree/master/travel_stackexchange_com_2017

```
            COUNT(*) AS PositiveAnswerCount
        FROM Posts
        WHERE Posts.PostTypeID=2 AND Posts.Score>0
        GROUP BY Posts.ParentID
    ) AS Posts2
    ON Posts.ID=Posts2.ParentID
ORDER BY Posts2.PositiveAnswerCount DESC
LIMIT 10
```

*Third:*

```
SELECT
    Posts.Title,
    UpVotesPerYear.Year,
    MAX(UpVotesPerYear.Count) AS Count
FROM (
        SELECT
            PostId,
            COUNT(*) AS Count,
            STRFTIME('%Y', Votes.CreationDate) AS Year
        FROM Votes
        WHERE VoteTypeId=2
        GROUP BY PostId, Year
    ) AS UpVotesPerYear
JOIN Posts ON Posts.Id=UpVotesPerYear.PostId
WHERE Posts.PostTypeId=1
GROUP BY Year
```

*Fourth:*

```
SELECT
    Questions.Id,
    Questions.Title,
    BestAnswers.MaxScore,
    Posts.Score AS AcceptedScore,
    BestAnswers.MaxScore-Posts.Score AS Difference
FROM (
        SELECT Id, ParentId, MAX(Score) AS MaxScore
        FROM Posts
        WHERE PostTypeId==2
        GROUP BY ParentId
    ) AS BestAnswers
JOIN (
        SELECT * FROM Posts
        WHERE PostTypeId==1
    ) AS Questions
    ON Questions.Id=BestAnswers.ParentId
JOIN Posts ON Questions.AcceptedAnswerId=Posts.Id
```

```
WHERE Difference>50
ORDER BY Difference DESC
```

*Fifth:*

```
SELECT
    Posts.Title,
    CmtTotScr.CommentsTotalScore
FROM (
        SELECT
            PostID,
            UserID,
            SUM(Score) AS CommentsTotalScore
        FROM Comments
        GROUP BY PostID, UserID
) AS CmtTotScr
JOIN Posts ON Posts.ID=CmtTotScr.PostID
    AND Posts.OwnerUserId=CmtTotScr.UserID
WHERE Posts.PostTypeId=1
ORDER BY CmtTotScr.CommentsTotalScore DESC
LIMIT 10
```

*Sixth:*

```
SELECT DISTINCT
    Users.Id,
    Users.DisplayName,
    Users.Reputation,
    Users.Age,
    Users.Location
FROM (
        SELECT
            Name, UserID
        FROM Badges
        WHERE Name IN (
            SELECT
                Name
            FROM Badges
            WHERE Class=1
            GROUP BY Name
            HAVING COUNT(*) BETWEEN 2 AND 10
        )
        AND Class=1
    ) AS ValuableBadges
JOIN Users ON ValuableBadges.UserId=Users.Id
```

*Seventh:*

```sql
SELECT
    Posts.Title,
    VotesByAge2.OldVotes
FROM Posts
JOIN (
    SELECT
        PostId,
        MAX(CASE WHEN VoteDate = 'new' THEN Total ELSE 0 END) NewVotes,
        MAX(CASE WHEN VoteDate = 'old' THEN Total ELSE 0 END) OldVotes,
        SUM(Total) AS Votes
    FROM (
        SELECT
            PostId,
            CASE STRFTIME('%Y', CreationDate)
                WHEN '2017' THEN 'new'
                WHEN '2016' THEN 'new'
                ELSE 'old'
                END VoteDate,
            COUNT(*) AS Total
        FROM Votes
        WHERE VoteTypeId=2
        GROUP BY PostId, VoteDate
    ) AS VotesByAge
    GROUP BY VotesByAge.PostId
    HAVING NewVotes=0
) AS VotesByAge2 ON VotesByAge2.PostId=Posts.ID
WHERE Posts.PostTypeId=1
ORDER BY VotesByAge2.OldVotes DESC
LIMIT 10
```

**Exercise 12.42** *(*) Generate a CSV file that stores some random data arranged in a few columns of a size at least two times larger than your available RAM. Then, export the CSV file to an SQLite database. Use file connections (Section 8.3.5) and the* `nrow` *argument to* **`read.table`** *to process it chunk by chunk. Determine whether setting* `colClasses` *in* **`read.table`** *speeds up the reading of large CSV files significantly or not.*

**Exercise 12.43** *(*) Export the whole XML data dump of StackOverflow[16] published at https://archive.org/details/stackexchange (see also https://data.stackexchange.com/) to an SQLite database.*

[16] https://stackoverflow.com/

# 13

# *Graphics*

The R project homepage advertises our free software as an *environment for statistical computing and graphics*. Hence, had we not dealt with the latter use case, our course would have been incomplete.

R is nowadays equipped with two independent (incompatible, yet coexisting) systems for graphics generation; see Figure 13.1.

1. The (historically) newer one, **`grid`** (e.g., [49]), is very flexible but might seem complicated. Some readers might have come across the **`lattice`** [54] and **`ggplot2`** [61, 64] packages before: they are built on top of **`grid`**.

2. On the other hand, its traditional (S-style) counterpart, *base* **`graphics`** (e.g., [7]), is much easier to master. It still gives their users complete control over the drawing process. It is simple, fast, and minimalist, which makes it very attractive from the perspective of this course's philosophy.

This is why we only cover the second system here. Most importantly, *all* figures in this book were generated using **`graphics`** and its dependants. They are *sufficiently* aesthetic, aren't they? Form precedes essence (but see [57, 60]).

## 13.1 Graphics primitives

In **`graphics`**, we do not choose from a superfluity of virtual objects to be placed on an abstract canvas, letting some algorithm decide how and when to delineate them. We *just draw*. We do so by calling functions that plot the following *graphics primitives* (see, e.g., [37, 45]):

- symbols (e.g., pixels, circles, stars) of different shapes and colours,
- line segments of different styles (e.g., solid, dashed, dotted),
- polygons (optionally filled),
- text (using available fonts),
- raster images (bitmaps).

That's it. It will turn out that *all* other shapes (smooth curves, circles) may be easily approximated using the above.

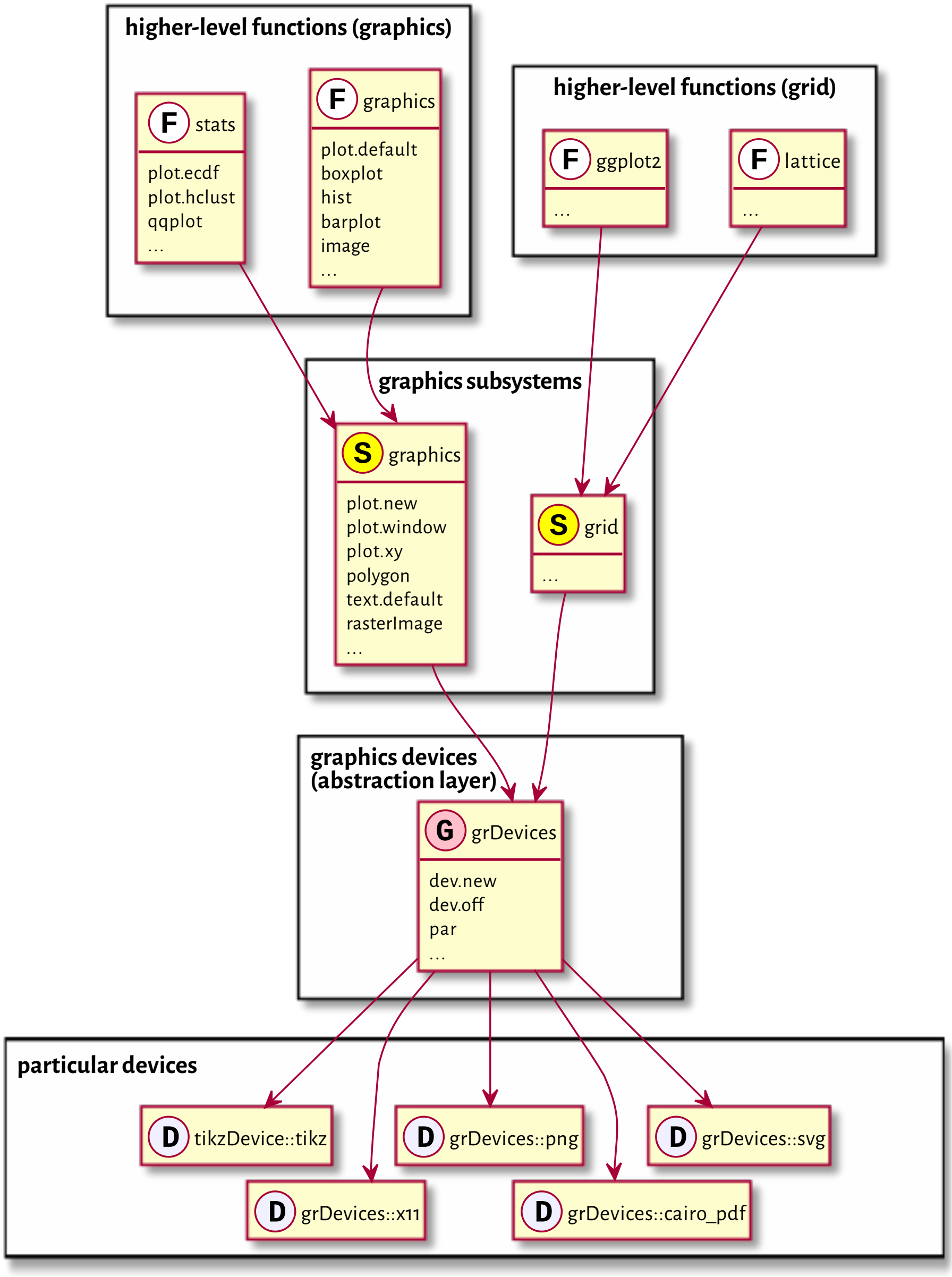

Figure 13.1. Relation between the graphics subsystems.

Of course, in practice, we do not always have to be so low-level. There are many functions that provide the most popular chart types: histograms, bar plots, dendrograms, etc. They will suit our basic needs. We will review them in Section 13.3.

The more primitive routines we discuss next will still be of service for fine-tuning our figures and adding further details. However, if the prefabricated components are not what we are after, we will be able to create any drawing from scratch.

---

**Important** In **graphics**, most of the function calls have immediate effects. Objects are drawn on the active plot one by one, and their state cannot be modified later.

---

**Example 13.1** *Figure 13.2 depicts some graphics primitives, which we plotted using the following program. We will detail the meaning of all the functions in the next sections, but they should be self-explanatory enough for us to be able to find the corresponding shapes in the plot.*

```
par(mar=rep(0.5, 4))  # small plot margins (bottom, left, top, right)
plot.new()  # start a new plot
plot.window(c(0, 6), c(0, 2), asp=1)  # x range: 0-6, y: 0-2; proportional
x <- c(0, 0, NA, 1, 2, 3, 4, 4, 5,    6)
y <- c(0, 2, NA, 2, 1, 2, 2, 1, 0.25, 0)
points(x[-(1:6)], y[-(1:6)])  # symbols
lines(x, y)   # line segments
text(c(0, 6), c(0, 2), c("(0, 0)", "(6, 2)"), col="red")  # two text labels
rasterImage(
    matrix(c(1, 0,  # 2x3 pixel "image"; 0=black, 1=white
             0, 1,
             0, 0), byrow=TRUE, ncol=2),
    5, 0.5, 6, 2,  # position: xleft, ybottom, xright, ytop
    interpolate=FALSE
)
polygon(
    c(4, 5, 5.5,    4),  # x coordinates of the vertices
    c(0, 0,   1, 0.75),  # y coordinates
    lty="dotted",   # border style
    col="#ffff0044"  # fill colour: semi-transparent yellow
)
```

### 13.1.1 Symbols (points)

The **points** function can draw a series of symbols (by default, circles) on the two-dimensional plot region, relative to the user coordinate system. We specify the points' coordinates using the x and y arguments (two vectors of equal lengths; no recycling). Alternatively, we may give a matrix or a data frame with two columns: its first column (regardless of how and if it is named) defines the abscissae, and the second column determines the ordinates.

**points** can draw each point differently if this is what we desire. Thus, it is ideal for making scatter plots, possibly for grouped data (see Figure 13.17 below). The function is vectorised with respect to, amongst others, the `col` (colour; see Section 13.2.1), `cex` (scale, defaults to 1), and `pch` (plotting character or symbol, defaults to 1, i.e., a circle) arguments.

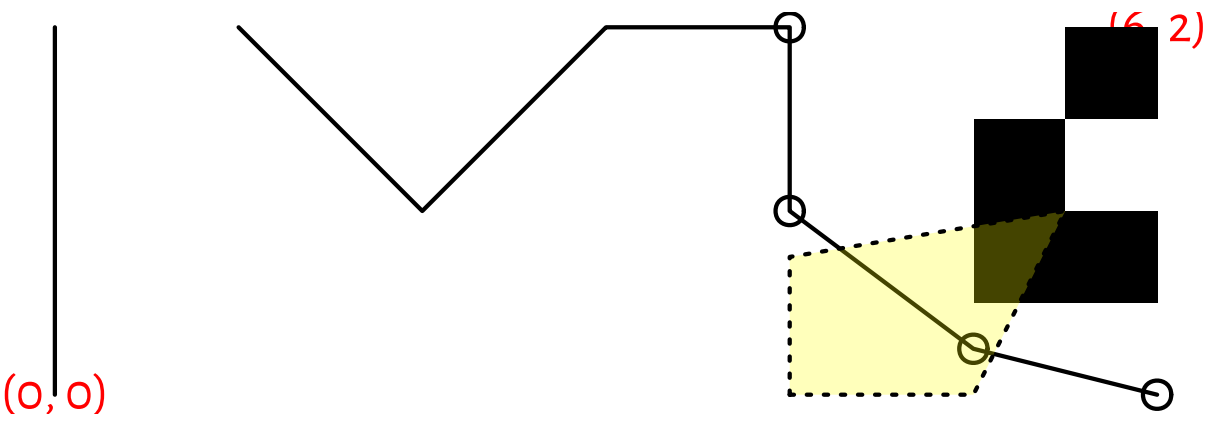

Figure 13.2. Graphics primitives: plotting symbols, line segments, polygons, text labels, and bitmaps. Objects are added one after another, with newer ones drawn *over* the already existing shapes.

**Example 13.2** *Figure 13.3 gives an overview of the plotting symbols available. The most often used ones are:*

- *`NA_integer_` – no symbol,*
- *0, ..., 14 and 15, ..., 18 – unfilled and filled symbols, respectively;*
- *19, ..., 25 – filled symbols with a border of width `lwd`; for codes 21, ..., 25, the fill colour is controlled separately by the `bg` parameter,*
- *`"."` – a tiny point (a "pixel"),*
- *`"a"`, `"1"`, etc. – a single character (not all Unicode characters can be drawn); strings longer than one will be truncated.*

```
par(mar=rep(0.5, 4)); plot.new(); plot.window(c(0.9, 9.1), c(0.9, 4.1))
points(
    cbind(1:9, 1),  # or x=1:9, y=rep(1, 9); bottom row
    col="red",
    pch=c("A", "B", "a", "b", "Spanish Inquisition", "*", "!", ".", "9")
)
xy <- expand.grid(1:9, 4:2)
text(xy, labels=0:(NROW(xy)-1), pos=1, cex=0.89, offset=0.75, col="darkgray")
points(xy, pch=0:(NROW(xy)-1), bg="yellow")
## Warning in plot.xy(xy.coords(x, y), type = type, ...): unimplemented pch
##     value '26'
```

### 13.1.2 Line segments

**`lines`** can draw connected line segments whose mid- and endpoints are given in a similar manner as in the **`points`** function. A series of segments can be interrupted by defining an endpoint whose coordinate is a missing value; compare Figure 13.2.

Actually, **`points`** and **`lines`** are wrappers around the same function, **`plot.xy`**, which we usually do not call directly. Their type arguments determine the object to draw; the only difference is that by default the former uses type="p" whilst the latter relies on type="l" . Changing these to type="b" (both) or type="o" (overplot) will give their combination. Moreover, type="s" and type="S" creates step functions (with post- and pre-increments, respectively), and type="h" draws bar plot-like vertical lines. For an illustration, see Figure 13.4 (implement something similar yourself as an exercise).

Figure 13.3. Plotting characters and symbols (pch).

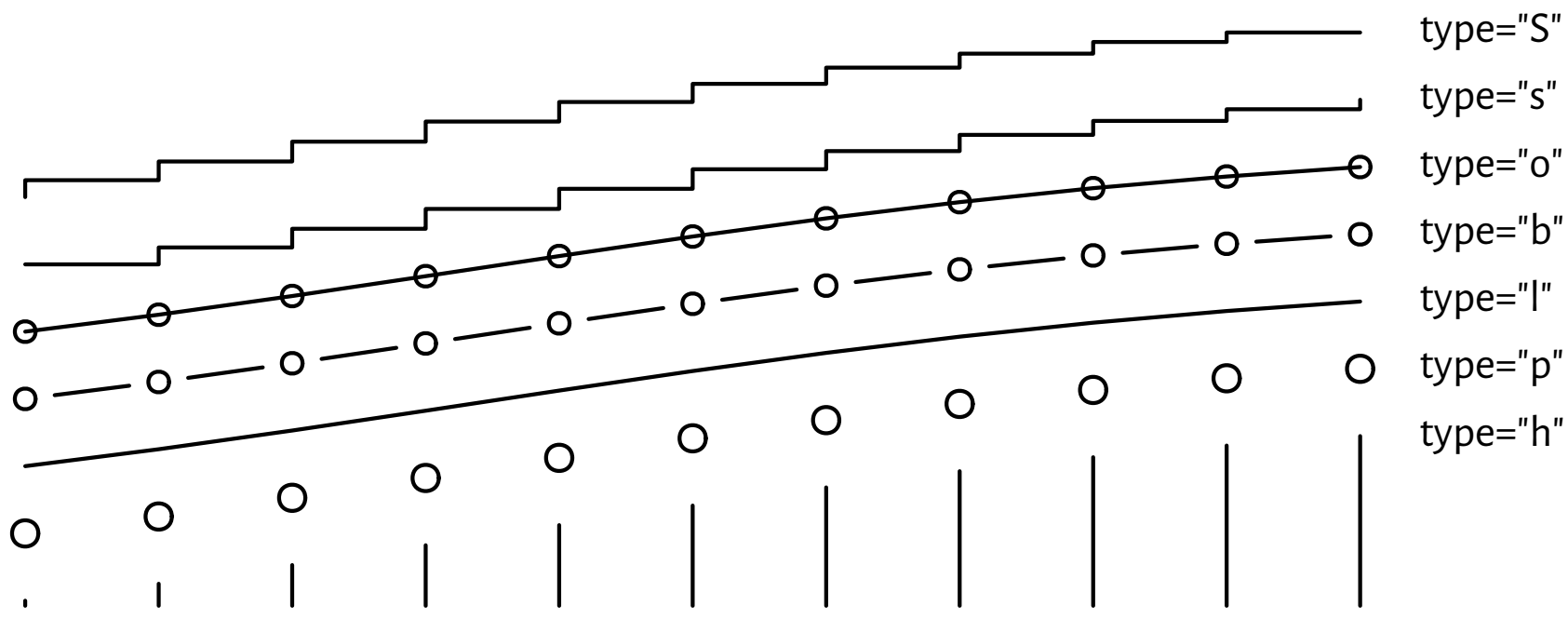

Figure 13.4. Different type argument settings in **lines** or **points**.

The col argument controls the line colour (see Section 13.2.1), and lwd determines the line width (1 by default). Six named line types (lty) are available, which can also be specified via their respective numeric identifiers, lty=1, ..., lty=6; see Figure 13.5 (implementing a similar plot is left as an exercise). Additionally, custom dashes can be defined using strings of up to eight hexadecimal digits. Consecutive digits give the lengths of the dashes and blanks (alternating). For instance, lty="1343" yields a dash of length 1, followed by a space of length 3, then a dash of length 4, followed by a blank of length 3. The whole sequence will be recycled for as long as necessary.

**Example 13.3** *lines can be used for plotting empirical cumulative distribution functions (we will suggest it as an exercise later), regression models (e.g., lines, splines of different degrees), time series, and any other mathematical functions, even when they are smooth and* curvy. *The naked eye cannot tell the difference between a densely sampled piecewise linear approximation of an object and its original version. The code below illustrates this (sad for the high-hearted idealists) truth using the* sine *function; see Figure 13.6.*

"solid" or 1
"dashed", "44", or 2
"dotted", "13", or 3
"dotdash", "1343", or 4
"longdash", "73", or 5
"twodash", "2262", or 6
"5515"
"9515"
"19"
"4484C4"

Figure 13.5. Line types (`lty`).

```
ns <- c(seq(3, 25, by=2), 50, 100)
par(mar=rep(0.5, 4)); plot.new(); plot.window(c(0, length(ns)*pi), c(-1, 1))
for (i in seq_along(ns)) {
    x <- seq((i-1)*pi, i*pi, length.out=ns[i])
    lines(x, sin(x))
    text((i-0.5)*pi, 0, ns[i], cex=0.89)
}
```

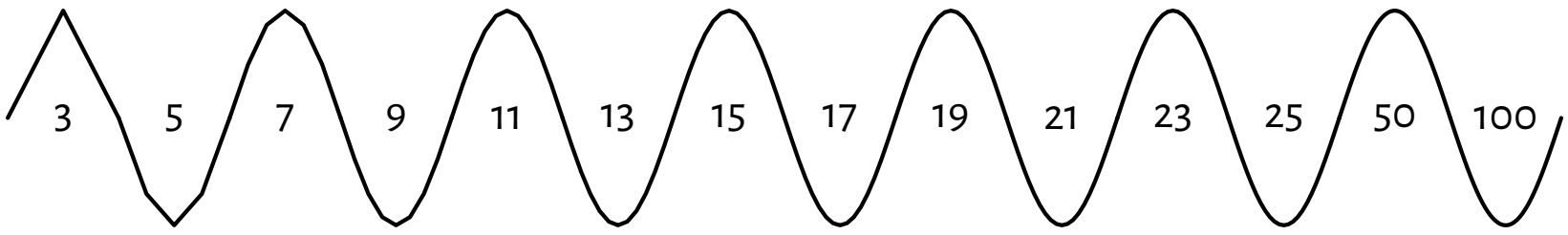

Figure 13.6. The sine function approximated with line segments. Sampling more densely gives the illusion of smoothness.

**Exercise 13.4** *Implement your version of the* **`segments`** *function using a call to* **`lines`**.

**Exercise 13.5** *(*) Implement a simplified version of the* **`arrows`** *function, where the length of edges of the arrowhead is given in user coordinates (and not inches; you will be equipped with skills to achieve this later; see Section 13.2.4). Use the* `ljoin` *and* `lend` *arguments (see* **`help`**`("par")` *for admissible values) to change the line end and join styles (from the default rounded caps).*

### 13.1.3 Polygons

**`polygon`** draws a polygon with a border of specified colour and line type (`border`, `lty`, `lwd`). If the `col` argument is not missing, the polygon is filled (or hatched; cf. the `density` and `angle` arguments).

**Example 13.6** *Let's draw a few regular (equilateral and equiangular) polygons; see Figure 13.7. By increasing the number of sides, we can obtain an approximation to a circle.*

```
regular_poly <- function(x0, y0, r, n=101, ...)
{
```

```
    theta <- seq(0, 2*pi, length.out=n+1)[-1]
    polygon(x0+r*cos(theta), y0+r*sin(theta), ...)
}

par(mar=rep(0.5, 4)); plot.new(); plot.window(c(0, 9.5), c(-1, 1), asp=1)
regular_poly(1, 0, 1, n=3)
regular_poly(3.5, 0, 1, n=7, density=15, angle=45, col="tan", border="red")
regular_poly(6, 0, 1, n=10, density=8, angle=-60, lty=3, lwd=2)
regular_poly(8.5, 0, 1, n=100, border="brown", col="lightyellow")
```

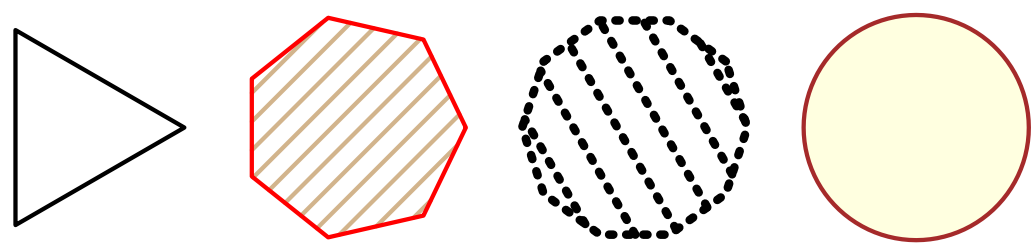

Figure 13.7. Regular polygons drawn using **polygon**.

*Note the asp=1 argument to the **plot.window** function (which we detail below) that guarantees the same scaling of the x- and y-axes. This way, the circle looks like one and not an oval.*

Notice that the last vertex adjoins the first one. Also, if we are absent-minded (or particularly creative), we can produce self-intersecting or otherwise degenerate shapes.

**Exercise 13.7** *Implement your version of the **rect** function using a call to **polygon**.*

### 13.1.4 Text

A call to **text** draws arbitrary strings (newlines and tabs are supported) centred at the specified points. Moreover, by setting the pos argument, the labels may be placed below, to the left of, etc., the pivots. Some further position adjustments are also possible (`adj`, `offset`); see Figure 13.8.

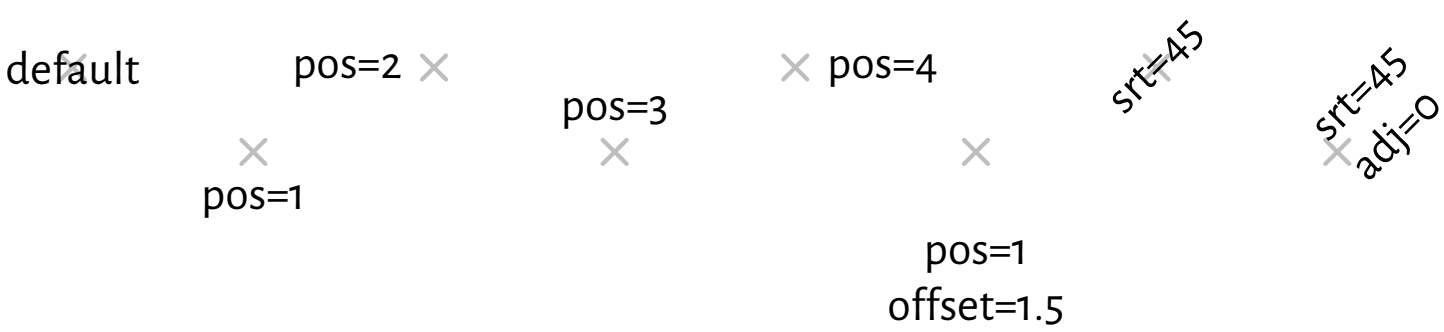

Figure 13.8. The positioning of text with **text** (plotting symbols added for reference).

`col` specifies the colour, `cex` affects the size, and `srt` changes the rotation of the text.

On many graphics devices, we have little but crude control over the font face used: `family` chooses a generic font family (`"sans"`, `"serif"`, `"mono"`), and `font` selects between the normal variant (1), bold (2), italic (3), or bold italic (4). See, however, Section 13.2.6 for some workarounds.

**Note** (*) There is limited support for mathematical symbols and formulae. It relies on some quirky syntax that we enter using unevaluated R expressions (Chapter 15). Still, it should be enough to meet our most basic needs. For instance, passing `quote(beta[i]^j)` as the `labels` argument to `text` will output "$\beta_i^j$". See `help("plotmath")` for more details.

For more sophisticated mathematical typesetting, see the **tikzDevice** graphics device mentioned in Section 13.2.6. It outputs plot specifications that can be rendered by the LaTeX typesetting system.

### 13.1.5 Raster images (bitmaps) (*)

Raster images are encoded in the form of bitmaps, i.e., matrices whose elements represent pixels (see Figure 13.2 for an example). They can be used for drawing heat maps or backgrounds of contour plots; see Section 13.3.4.

**Example 13.8** *Optionally, bilinear interpolation can be applied if the drawing area is larger than the true bitmap size, and we would like to smoothen the colour transitions out. Figure 13.9 presents a very stretched* $4 \times 3$ *pixel image, with and without interpolation.*

```
par(mar=rep(0.5, 4)); plot.new(); plot.window(c(0, 9), c(0, 1))
I <- matrix(nrow=4, byrow=TRUE,
    c(   "red", "yellow",  "white",
      "yellow", "yellow", "orange",
      "yellow", "orange", "orange",
       "white", "orange",    "red")
)
rasterImage(I, 0, 0, 4, 1)  # interpolate=TRUE; left subplot
rasterImage(I, 5, 0, 9, 1, interpolate=FALSE)  # right subplot
```

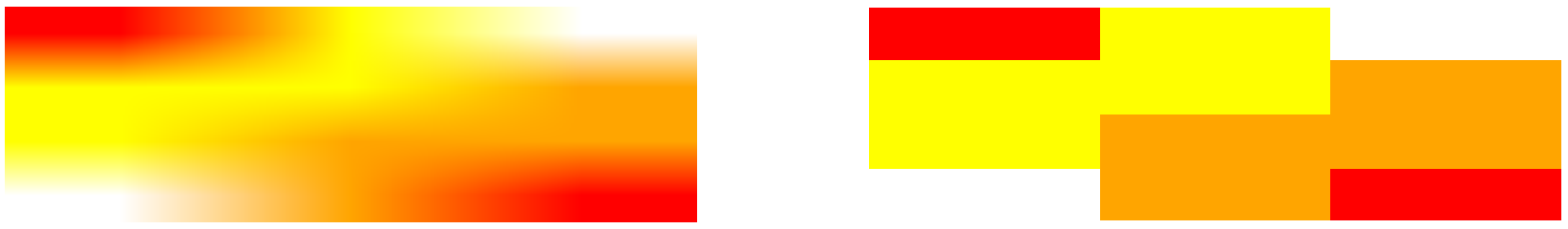

Figure 13.9. Example bitmaps drawn with `rasterImage`, with (left) and without (right) colour interpolation.

## 13.2 Graphics settings

`par` can be used to query and modify (as long as they are not read-only) many graphics options. For instance, we have several parameters related to the current page or device settings, e.g., the plot's margins (see Section 13.2.2) or user coordinates (see Section 13.2.3). The reference list of available parameters is given in **help**("par"). Below we discuss the most noteworthy ones.

Moreover, some functions source[1] the values of their default arguments by querying **par**. This is the case of, e.g., col, pch, lty in the **points** and **lines** function.

**Exercise 13.9** *Study the following pseudocode.*

```
lines(x, y)  # use the default `lty`, i.e., par("lty") == "solid"
old_settings <- par(lty="dashed")  # change setting, save old for reference
lines(x, y)  # use the new default `lty`, i.e., par("lty") == "dashed"
lines(x, y, lty=3)  # use the given `lty`, but only for this call
lines(x, y)  # default lty="dashed" again
par(old_settings)  # restore the previous settings
lines(x, y)  # lty="solid" now
```

### 13.2.1 Colours

Many functions allow for customising colours of the plotted objects or their parts; compare, e.g., col and border arguments to **polygon**, or col and bg to **points**. There are a few ways to specify colours (see the *Colour Specification* section of **help**("par") for more details).

- We can use a "colour name" string, being one of the 657 predefined tags known to the **colours** function:

```
sample(colours(), 8)  # this is just a sample
## [1] "grey23"        "darksalmon"    "tan3"          "violetred4"
## [5] "lightblue1"    "darkorchid3"   "darkseagreen1" "slategray3"
```

- We can pass a "#rrggbb" string, which specifies a position in the RGB colour space: three series of hexadecimal numbers of two digits each, i.e., between $00_{\text{hex}} = 0$ (off) and $\text{FF}_{\text{hex}} = 255$ (full on), giving the intensity of the red, green, and blue channels[2].

  In practice, the **col2rgb** and **rgb** functions can convert between the decimal and hexadecimal representations:

```
C <- c("black", "red", "green", "blue", "cyan", "magenta",
    "yellow", "grey", "lightgrey", "pink")  # example colours
(M <- structure(col2rgb(C), dimnames=list(c("R", "G", "B"), C)))
```

[1] Alas, it is not as straightforward as that. For instance, **polygon** is unaffected by the col setting, **axis** uses col.axis instead, etc. We should always consult the documentation.

[2] From school, we probably know the *subtractive* CMY (cyan, magenta, yellow) model, where we obtain, e.g., a green colour by using blue-ish and yellow crayons (subtracting certain wavelengths from white light). The RGB model, on the other hand, corresponds to the three photoreceptor/cone cells in the retinas of the human eyes. Nonetheless, it is *additive* and, therefore, less intuitive: total darkness emerges when we emit no light, yellow emerges when mixing red and green beams, etc.

(continued from previous page)

```
##   black red green blue cyan magenta yellow grey lightgrey pink
## R     0 255     0    0    0     255    255  190       211  255
## G     0   0   255    0  255       0    255  190       211  192
## B     0   0     0  255  255     255      0  190       211  203
structure(rgb(M[1, ], M[2, ], M[3, ], maxColorValue=255), names=C)
##     black       red     green      blue      cyan   magenta    yellow
## "#000000" "#FF0000" "#00FF00" "#0000FF" "#00FFFF" "#FF00FF" "#FFFF00"
##      grey lightgrey      pink
## "#BEBEBE" "#D3D3D3" "#FFC0CB"
```

- An `"#rrggbbaa"` string is similar, but has the added alpha channel (two additional hexadecimal digits): from $00_{\text{hex}} = 0$ denoting fully transparent, to $\text{FF}_{\text{hex}} = 255$ indicating fully visible (lit) colour; see Figure 13.2 for an example.

  Semi-transparency (translucency) can significantly enhance the expressivity of our data visualisations; see Figure 13.18 and Figure 13.19.

- We can rely on an integer index to select an item from the *current palette* (with recycling), which we can get or set by a call to **palette**. Moreover, `0` identifies the background colour, `par("bg")`.

  Integer colour specifiers are particularly valuable when plotting data in groups defined by factor objects. The underlying integer level codes can be mapped to consecutive colours from any palette; see Figure 13.17 below for an example.

**Example 13.10** *We recommend memorising the colours in the default palette:*

```
palette()  # get current palette
## [1] "black"   "#DF536B" "#61D04F" "#2297E6" "#28E2E5" "#CD0BBC" "#F5C710"
## [8] "gray62"
```

*These are[3], in order: black, red, green, blue, cyan, magenta, yellow, and grey; see[4] Figure 13.10.*

```
k <- length(palette())
par(mar=rep(0.5, 4)); plot.new(); plot.window(c(0, k+1), c(0, 1))
points(1:k, rep(0.5, k), col=1:k, pch=16, cex=3)
text(1:k, 0.5, palette(), pos=rep(c(1, 3), length.out=k), col=1:k, offset=1)
text(1:k, 0.5, 1:k, pos=rep(c(3, 1), length.out=k), col=1:k, offset=1)
```

Choosing usable colours requires talents that most programmers lack. Therefore, we will find ourselves relying on the built-in colour sets. **palette.pals** and **hcl.pals** return the names of the available *discrete* (qualitative) palettes. Then, **palette.colors** and **hcl.colors** (note the American spelling) can generate a given number of colours from a particular named set.

[3] Actually, red-ish, green-ish, etc. The choice is more aesthetic than when pure red, green, etc. was used (before R 4.0.0). It is also expected to be more friendly to people who have colour vision deficiencies. We know that roughly every 1 in 12 men (8%) and 1 in 200 women (0.5%), especially in the red-green or blue-yellow spectrum; see [51] for more details.

[4] The readers of the printed version of this book should know that its online version displays this figure (and all others) in full colour. See you there.

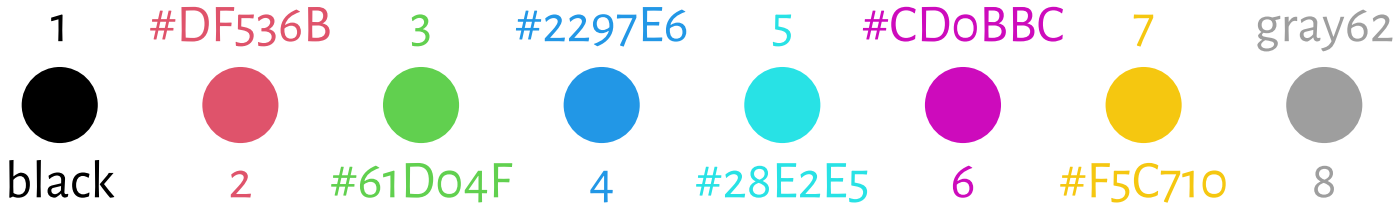

Figure 13.10. The default colour `palette`.

*Continuous* (quantitative) palettes are also available, see `rainbow`, `heat.colors`, `terrain.colors`, `topo.colors`, `cm.colors`, and `gray.colors`. They transition smoothly between predefined pivot colours, e.g., from blue through green to brown (like in a topographic map with elevation colouring). They may be of use, e.g., when drawing contour plots; compare Figure 13.27.

**Exercise 13.11** *Create a demo of the aforementioned palettes in a similar (or nicer) style to that in Figure 13.11.*

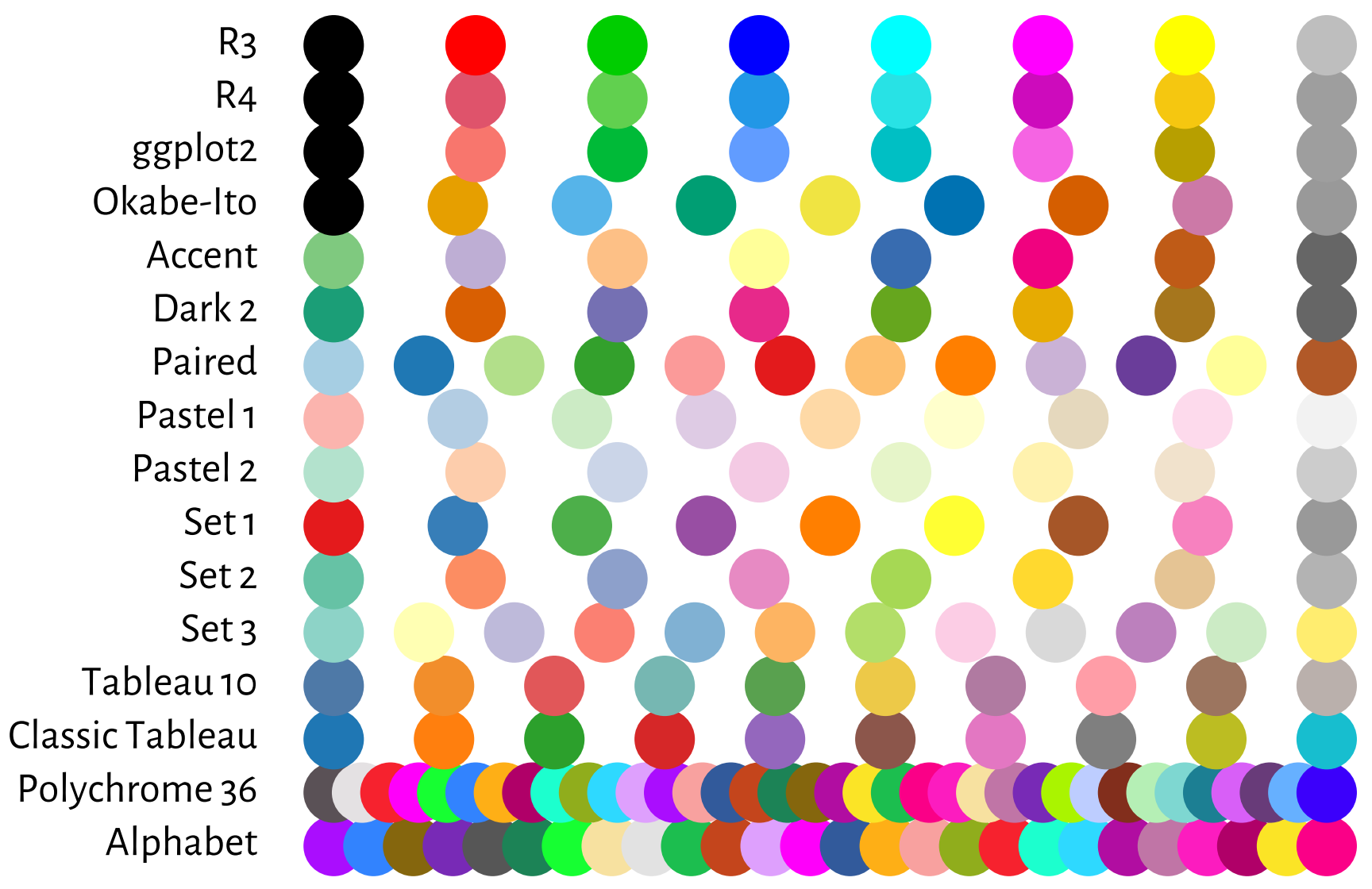

Figure 13.11. Qualitative colour palettes in `palette.pals`; *R4* is the default one, as seen in Figure 13.10.

### 13.2.2 Plot margins and clipping regions

A *device (page) region* represents a single plot window, one raster image file, or a page in a PDF document (see Section 13.2.6 for more information on graphics devices). As we will learn from Section 13.2.5, it is capable of holding many *figures*.

Usually, however, we draw *one figure per page*. In such a case, the device region is divided into the following parts:

1) *outer margins*, which can be set via, e.g., the `oma` graphics parameter (in text lines, based on the height of the default font); by default, they are equal to 0;

2) *figure region*:

   a) *inner (plot) margins*, by default `mar=c(5.1, 4.1, 4.1, 2.1)` text lines (bottom, left, top, right, respectively); this is where we usually emplace the figure title, axes labels, etc.

   b) *plot region*, where we draw graphics primitives positioned relative to the user coordinates.

---

**Note** Typically, all drawings are *clipped* to the plot region, but this can be changed with the `xpd` parameter; see also the more flexible **`clip`** function.

---

**Example 13.12** *Figure 13.12 shows the default page layout. In the code chunk below, note the use of* **`mtext`** *to print a text line in the inner margins,* **`box`** *to draw a rectangle around the plot or figure region,* **`axis`** *to add the two axes (labels and tick marks), and* **`title`** *to print the descriptive labels.*

```
plot.new(); plot.window(c(-2, 2), c(-1, 1))  # whatever
for (i in 1:4) {  # some text lines on the inner margins
    for (j in seq_len(par("mar")[i]))
        mtext(sprintf("Text line %d on inner margin %d", j, i),
            side=i, line=j-1, col="lightgray")
}

title(main="Main", sub="sub", xlab="xlab", ylab="ylab")
box("figure", lty="dashed")  # a box around the figure region
box("plot")  # a box around the plot region
axis(1)  # horizontal axis (bottom)
axis(2)  # vertical axis (left)

rect(-10, -10, 10, 10, col="lightgray")  # rectangle (clipped to plot region)
text(0, 0, "Plot region")
lines(c(-3, 0, 3), c(-2, 2, -2))  # standard clipping (plot region)
lines(c(-3, 0, 3), c(-2, 1.25, -2), xpd=TRUE, lty=3)  # clip to figure region
```

### 13.2.3 User coordinates and axes

**`plot.window`** sets the user coordinates. It accepts the following parameters:

- `xlim`, `ylim` – vectors of length two giving the minimal and maximal ranges on the respective axes; by default, they are extended by 4% in each direction for aesthetic reasons (see, e.g., Figure 13.12) but we can disable this behaviour by setting the `xaxs` and `yaxs` graphics parameters;

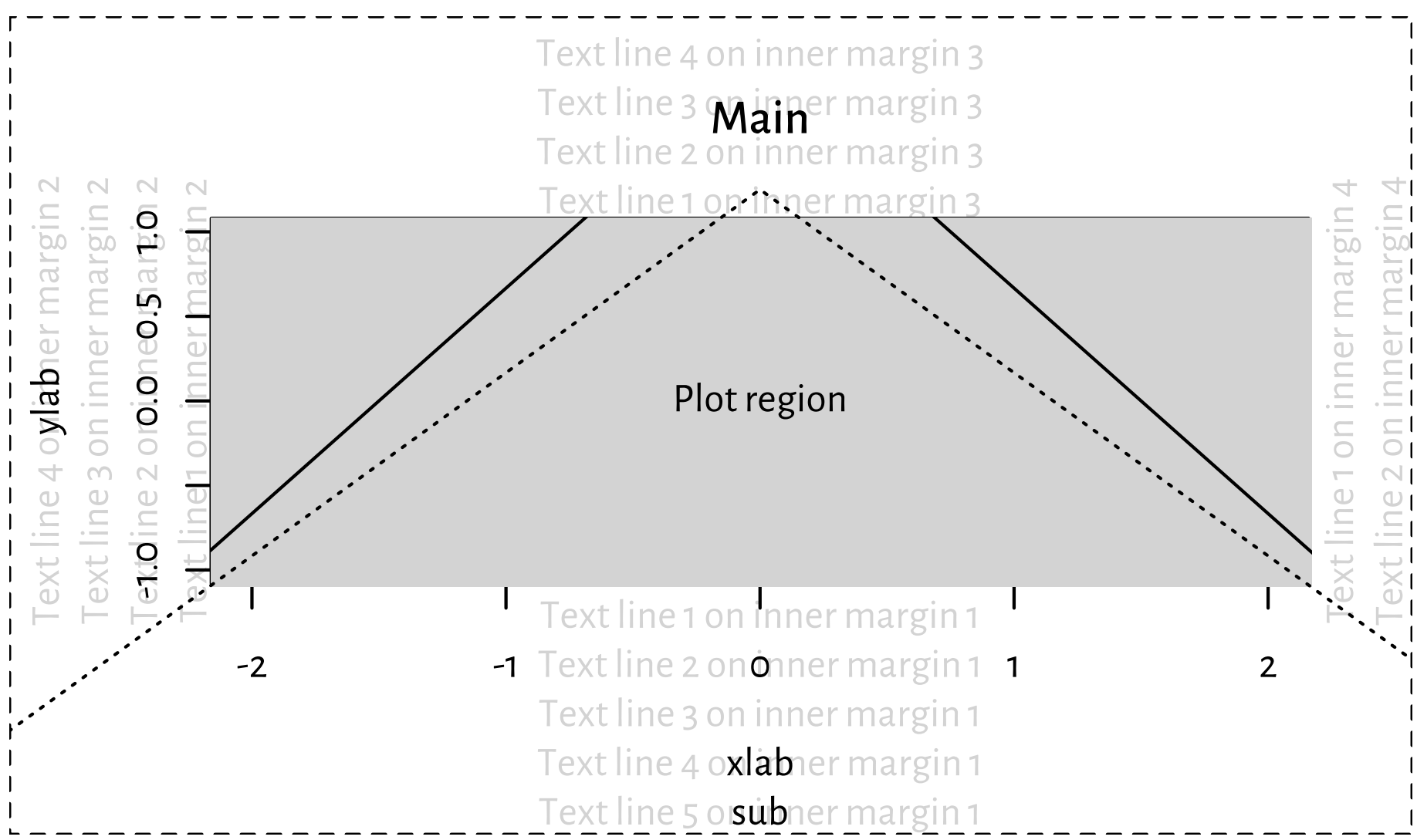

Figure 13.12. Figure layout with default inner and outer margins (`mar=c(5.1, 4.1, 4.1, 2.1)` and `oma=c(0, 0, 0, 0)` text lines, respectively). We see that a lot of space is wasted and hence some tweaking might be necessary to suit our needs better. Note the clipping of the solid line to the grey plot region.

- `asp` – aspect ratio ($y/x$); defaults to `NA`, i.e., no adjustment; use `asp=1` for circles to look like ones, and not ovals;
- `log` – logarithmic scaling on particular axes: `""` (none; default), `"x"`, `"y"`, or `"xy"`.

**Example 13.13** *The graphics parameter `usr` can be used to read (and set) the extremes of the user coordinates in the form $(x_1, x_2, y_1, y_2)$.*

```
plot.new()
plot.window(c(-1, 1), c(1, 1000), log="y", yaxs="i")
par("usr")
## [1] -1.08  1.08  0.00  3.00
```

*Indeed, the x-axis range was extended by 4% in each direction (`xaxs="r"`). We have turned this behaviour off for the y-axis (`yaxs="i"`), which uses the base-10 logarithmic scale. In this case, its actual range is `10^par("usr")[3:4]` because $\log_{10} 1 = 0$ and $\log_{10} 1000 = 3$.*

**Exercise 13.14** *Implement your version of the **`abline`** function using **`lines`**.*

Even though axes (labels and tick marks) can be drawn manually using the aforementioned graphics primitives, it is usually too tedious a work. This is why we tend to rely on the **`axis`** function, which draws the object on one of the plot sides (as usual, 1=bottom, …, 4=right).

Once **`plot.window`** is called, **`axTicks`** can be called to guesstimate the tasteful (*round*)

locations for the tick marks relative to the current plot size. By default, they are based on the xaxp and yaxp graphics parameters, which give the axis ranges and the number of intervals between the tick marks.

```
plot.new(); plot.window(c(-0.9, 1.05), c(1, 11))
par("usr")  # (x1, x2, y1, y2)
## [1] -0.978  1.128  0.600 11.400
par("yaxp")  # (y1, y2, n)
## [1]  2 10  4
axTicks(2)  # left y-axis
## [1]  2  4  6  8 10
par("xaxp")  # (x1, x2, n)
## [1] -0.5  1.0  3.0
axTicks(1)  # bottom x-axis
## [1] -0.5  0.0  0.5  1.0
par(xaxp=c(-0.9, 1.0, 5))  # change
axTicks(1)
## [1] -0.90 -0.52 -0.14  0.24  0.62  1.00
```

`axis` relies on the same algorithm as `axTicks`. Alternatively, we can provide custom tick locations and labels.

**Example 13.15** *Most of the plots in this book use the following graphics settings (except `las=1` to `axis(2)`); see Figure 13.13. Check out `help("par")`, `help("axis")`, etc. and tune them up to suit your needs.*

```
par(mar=c(2.2, 2.2, 1.2, 0.6))
par(tcl=0.25)  # the length of the tick marks (fraction of text line height)
par(mgp=c(1.1, 0.2, 0))  # axis title, axis labels, and axis line location
par(cex.main=1, font.main=2)  # bold, normal size - main in title
par(cex.axis=0.8889)
par(cex.lab=1, font.lab=3)  # bold italic, normal size
plot.new(); plot.window(c(0, 1), c(0, 1))
# a "grid":
rect(par("usr")[1], par("usr")[3], par("usr")[2], par("usr")[4],
    col="#00000010")
abline(v=axTicks(2), col="white", lwd=1.5, lty=1)
abline(h=seq(0, 1, length.out=4), col="white", lwd=1.5, lty=1)
# set up axes:
axis(2, at=seq(0, 1, length.out=4), c("0", "1/3", "2/3", "1"), las=1)
axis(1)
title(xlab="xlab", ylab="ylab", main="main (use sparingly)")
box()
```

### 13.2.4 Plot dimensions (*)

Certain sizes can be read or specified in inches (1" is exactly 25.4 mm):

- `pin` – plot dimensions (width, height),
- `fin` – figure region dimensions,

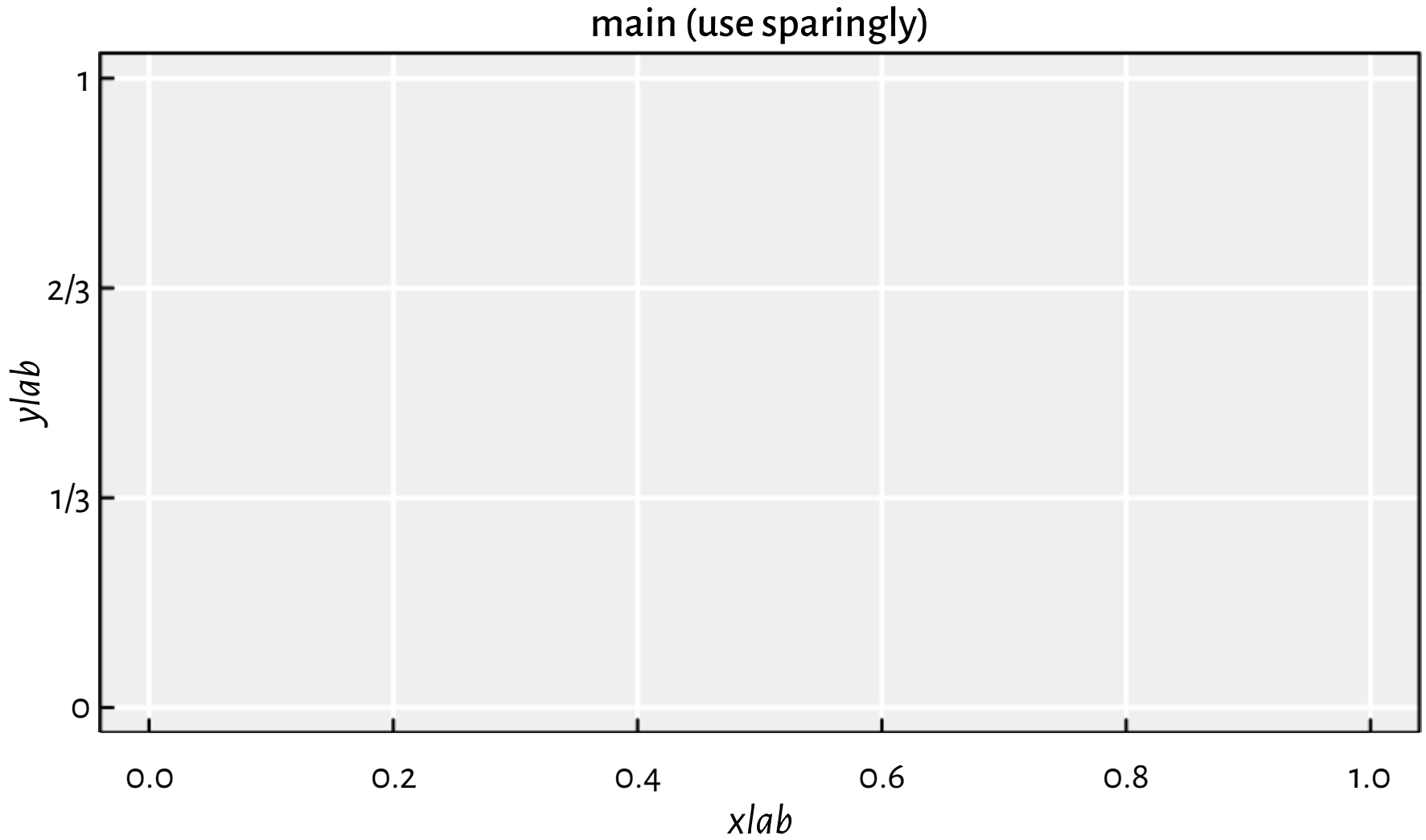

Figure 13.13. Custom axes and other settings.

- `din` – page (device) dimensions,
- `mai` – plot (inner) margin size,
- `omi` – outer margins,
- `cin` – the size of the "default" character (width, height).

If the figure is scaled, the *virtual* inch (the one reported by R) will not match the physical one (e.g., the actual size in the printed version of this book or on the computer screen).

---

**Important** Most objects' positions are specified in *virtual* user coordinates, as given by `usr`. They are automatically mapped to the *physical* device region, taking into account the page size, outer and inner margins, etc.

---

Knowing the above, some scaling can be used to convert between the user and physical sizes (in inches). It is based on the ratios `(usr[2]-usr[1])/pin[1]` and `(usr[4]-usr[3])/pin[2]`; compare the **`xinch`** and **`yinch`** functions.

**Example 13.16** *(*) Figure 13.14 shows how we can pinpoint the edges of the figure and device region in user coordinates.*

**Exercise 13.17** *(*) We cannot use* **`mtext`** *to print text on the right inner margin rotated by 180 degrees compared to what we see in Figure 13.12. Write your version of this function that will allow you to do so. Hint: use* **`text`***, the* `cin` *graphics parameter, and what you can read from Figure 13.14.*

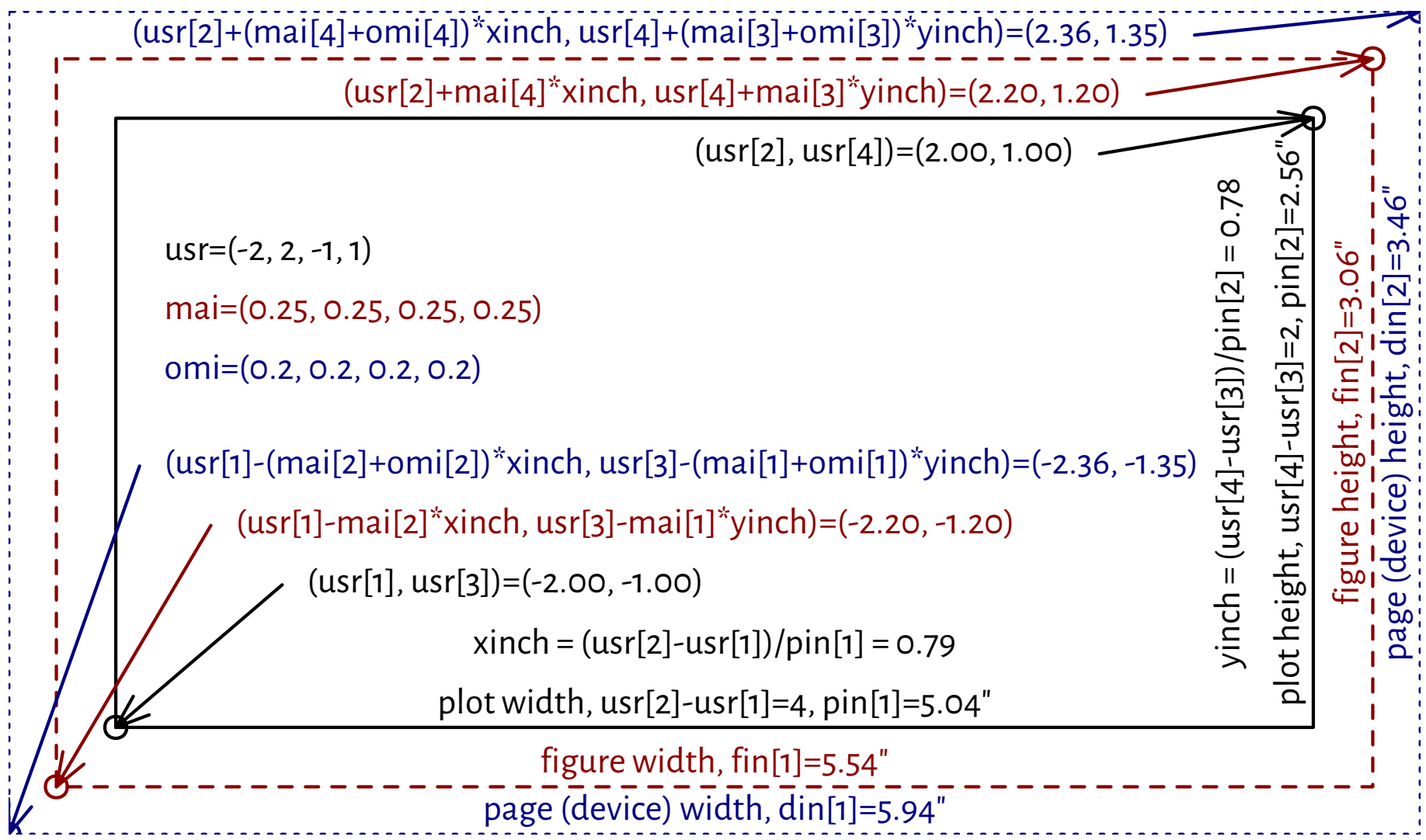

Figure 13.14. User vs device coordinates. Note that the virtual inch does not correspond to the physical one, as some scaling was applied.

### 13.2.5 Many figures on one page (subplots)

It is possible to create many figures on one page. In such a case, each subplot has its own inner margins and plot region.

A call to `par(mfrow=c(nr, nc))` or `par(mfcol=c(nr, nc))` splits the page into a regular grid with `nr` rows and `nc` columns. Each invocation of `plot.new` starts a new figure. Consecutive figures are either placed rowwisely (`mfrow`) or in the column-major order (`mfcol`). Alternatively, any subplot can be activated by referring to the `mfg` parameter.

**Example 13.18** *Figure 13.15 depicts an example page with four figures aligned on a* $2 \times 2$ *grid.*

```
par(oma=rep(1.2, 4))  # outer margins (default 0)
par(mfrow=c(2, 2))  # a 2x2 plot grid

for (i in 1:4) {
    plot.new()
    par(mar=c(3, 3, 2, 2))  # each subplot will have the same inner margins
    plot.window(c(i-1, i+1), c(-1, 1))  # separate user coordinates for each

    text(i, 0, sprintf("Plot region (plot %d)\n(%d, %d)", i,
        par("mfg")[1], par("mfg")[2]))

    box("figure", lty="dashed")  # a box around the figure region
    box("plot")   # a box around the plot region
    axis(1)  # horizontal axis (bottom)
    axis(2)  # vertical axis (left)
```

```
}

box("outer", lty="dotdash")  # a box around the whole page
for (i in 1:4)
    mtext(sprintf("Outer margin %d", i), side=i, outer=TRUE)
```

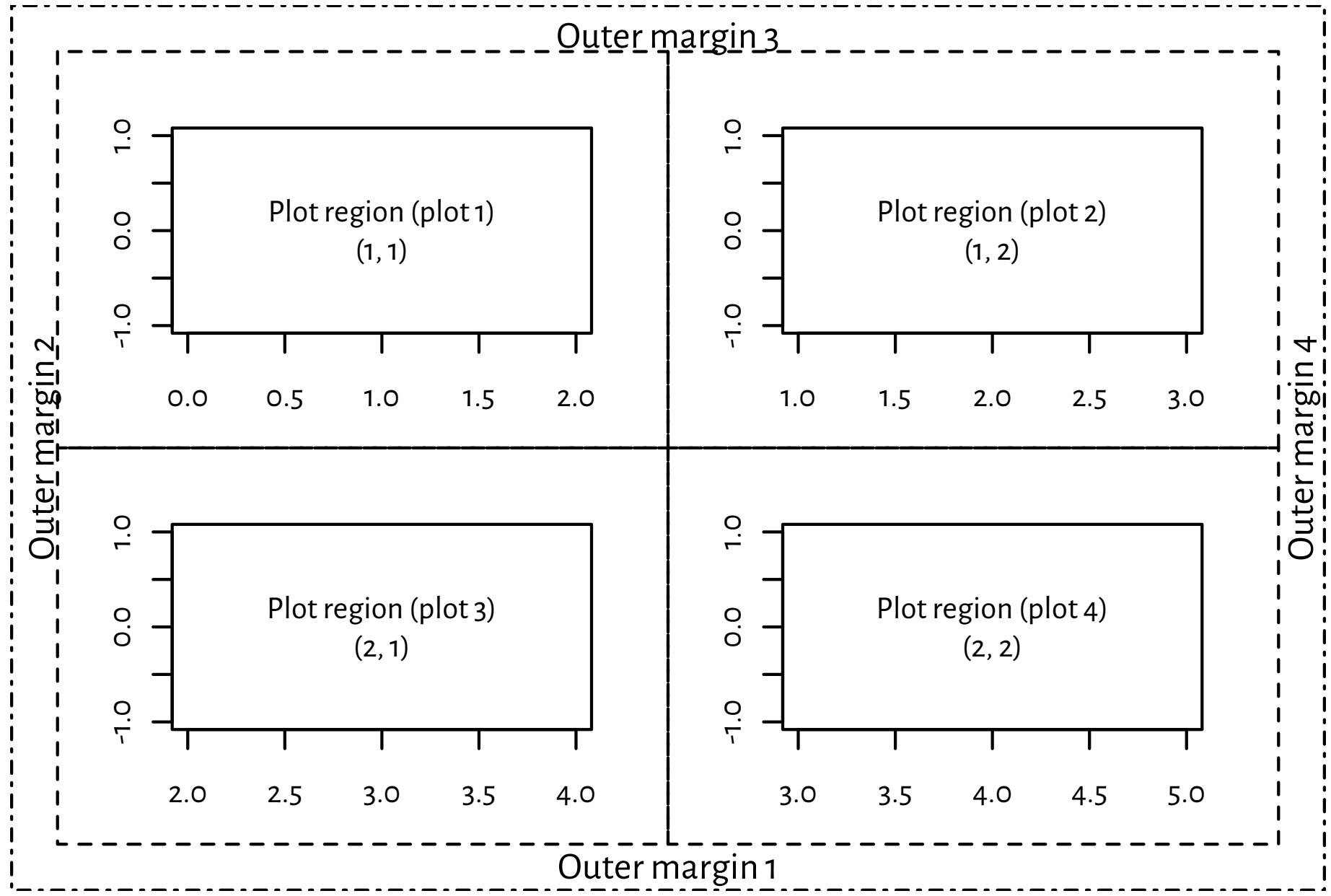

Figure 13.15. A page with four figures created using `par(mfrow=c(2, 2))`.

Thanks to `mfrow` and `mfcol`, we can create, e.g., a scatter plot matrix or different trellis plots. If an irregular grid is required, we can call the slightly more sophisticated **`layout`** function (which is incompatible with `mfrow` and `mfcol`). Examples will follow later; see Figure 13.24 and Figure 13.26. Also, the `fig` parameter (with `new=TRUE` to suppress the creation of a new figure) creates a subplot in an arbitrary rectangular region of the current page.

Certain grid sizes might affect the `mex` and `cex` parameters and hence the default font sizes (amongst others). Refer to the documentation of **`par`** for more details.

### 13.2.6 Graphics devices

Where our plots are *displayed* depends on our development environment (Section 1.2). Users of **`JupyterLab`** see the plots embedded into the current notebook, consumers of **`RStudio`** display them in a dedicated *Plots* pane, working from the console opens a new graphics window (unless we work in a text-only environment), whereas compiling **`utils::Sweave`** or **`knitr`** markup files brings about an image file that will be included in the output document.

In practice, we might be interested in exercising our creative endeavours on different devices. For instance, to draw something in a PDF file, we can call:

```
Cairo::CairoPDF("figure.pdf", width=6, height=3.5)  # open "device"
# ... calls to plotting functions...
dev.off()  # save file, close device
```

Similarly, a call to **CairoPNG** or **CairoSVG** creates a PNG or a SVG file. In both cases, as we rely on the **Cairo** library, we can customise the font family by calling **Cairo::CairoFonts**.

---

**Note** Typically, web browsers can display PNG, JPEG, and SVG files. On the other hand, PDF is a popular choice in printed publications (e.g., articles or books).

It is worth knowing that PNG and JPEG are raster graphics formats, i.e., they store figures as bitmaps (pixel matrices). They are fast to render, but the file sizes might become immense if we want decent image quality (high resolution). Most importantly, they should not be scaled: it is best to display them at their original widths and heights. However, JPEG uses lossy compression. Therefore, it is not a particularly fortunate file format for data visualisations. It does not support transparency either.

On the other hand, SVG and PDF files store vector graphics, where all primitives are described geometrically. This way, the image can be redrawn at any size and is always expected to be aesthetic. Unfortunately, scatter plots with millions of points will result in considerable files size and relatively slow rendition times (but there are tricks to remedy this).

---

Users of TeX should take note of **tikzDevice::tikz**, which creates TikZ files that can be rendered as standalone PDF files or embedded in LaTeX documents (and its variants). It allows for typesetting beautiful equations using the standard `"$...$"` syntax within any R string.

Many other devices are listed in **help**("Devices").

---

**Note** (*) The opened graphics devices form a stack. Calling **dev.off** will return to the last opened device (if any). See **dev.list** and other functions listed in its help page for more information.

Each device has separate graphics parameters. When opening a new device, we start with default settings in place.

Also, **dev.hold** and **dev.flush** can suppress the immediate display of the plotted objects, which might increase the drawing speed on certain interactive devices.

The current plot can be copied to another device (e.g., a PDF file) using **dev.print**.

---

**Exercise 13.19** *(*) Create an animated PNG displaying a large point sliding along the sine curve. Generate a series of video frames like in Figure 13.16. Store each frame in a separate PNG* *file. Then, use* **`ImageMagick`**[5] *(compare Section 7.3.2 or rely on another tool) to combine these files as a single animated PNG.*

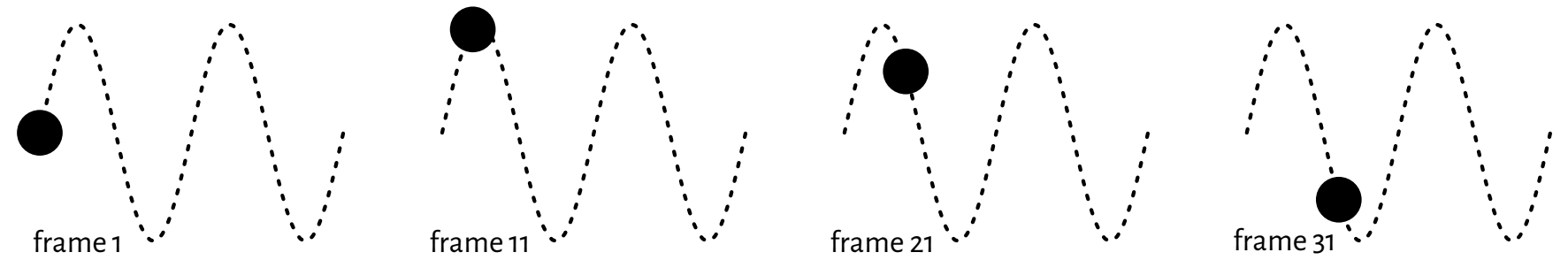

Figure 13.16. Selected frames of an example animation. They can be stored in separate files and then combined as a single animated PNG.

## 13.3 Higher-level functions

Higher-level plotting commands call **`plot.new`**, **`plot.window`**, **`axis`**, **`box`**, **`title`**, etc., and draw graphics primitives on our behalf. They provide ready-to-use implementations of the most common data visualisation tools, e.g., box-and-whisker plots, histograms, pairs plots, etc. Below we review a few of them. We also show how they can be customised or even rewritten from scratch if we are not completely happy with them. They will inspire us to practice lower-level graphics programming.

**Exercise 13.20** *Check out the meaning of the* `ask`*,* `new`*,* `xaxt`*,* `yaxt`*, and* `ann` *graphics parameters and how they affect* **`plot.new`***,* **`axis`***,* **`title`***, and so forth.*

### 13.3.1 Scatter and function plots with `plot.default` and `matplot`

The default method for the S3 generic **`plot`** is a convenient wrapper around **`points`** and **`lines`**.

**Example 13.21** **`plot.default`** *can draw a scatter plot of a set of points in* $\mathbb{R}^2$ *possibly grouped by another categorical variable. From Section 10.3.2 we know that a factor is represented as a vector of small natural numbers. Therefore, its underlying level codes can be used directly as* `col` *or* `pch` *specifiers; see Figure 13.17 for a demonstration. Take note of a call to the* **`legend`** *function.*

```
plot(
    jitter(iris[["Sepal.Length"]]),  # x (it is a numeric vector)
    jitter(iris[["Petal.Width"]]),   # y (it is a numeric vector)
    col=as.numeric(iris[["Species"]]),  # colours (integer codes)
    pch=as.numeric(iris[["Species"]]),  # plotting symbols (integer codes)
    xlab="Sepal length", ylab="Petal width",
    asp=1  # y/x aspect ratio
)
legend(
```

[5] https://imagemagick.org/

```
    "bottomright",
    legend=levels(iris[["Species"]]),
    col=seq_along(levels(iris[["Species"]])),
    pch=seq_along(levels(iris[["Species"]])),
    bg="white"
)
```

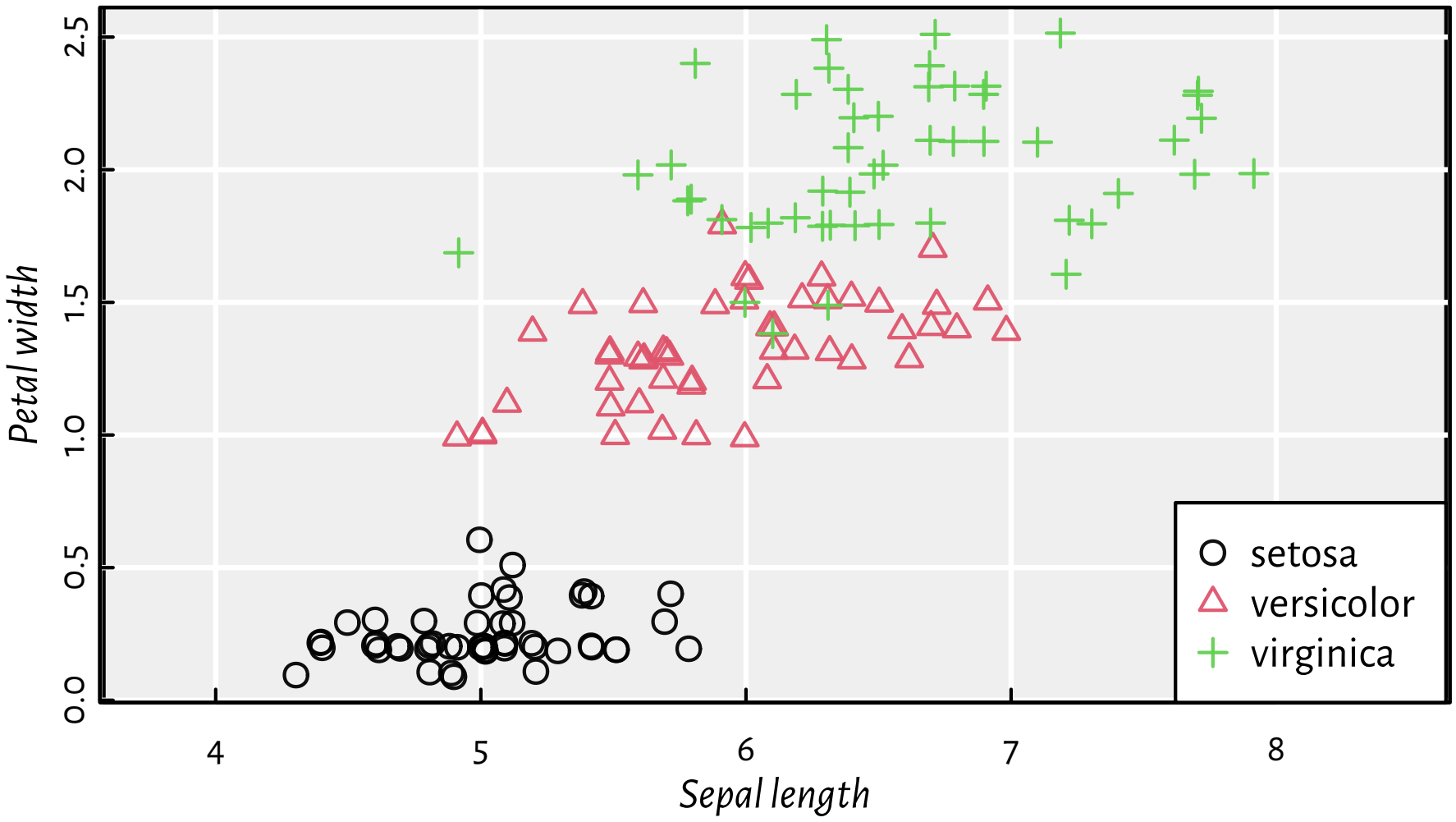

Figure 13.17. `as.numeric` can define different plotting styles for each factor level.

**Exercise 13.22** *Pass* ann=FALSE *and* axes=FALSE *to* **`plot`** *to suppress the addition of axes and labels. Then, draw them manually using the functions discussed in the previous section.*

**Exercise 13.23** *Draw a plot of the* $y = \sin x$ *function using* **`plot`**. *Then, call* **`lines`** *to add* $y = \cos x$. *Later, do the same using a single reference to* **`matplot`**. *Include a legend.*

**Example 13.24** *Semi-transparency may convey additional information. Figure 13.18 shows two scatter plots of adult females' weights vs heights. If the points are fully opaque, we cannot judge the density around them. On the other hand, translucent symbols somewhat imitate the two-dimensional histograms that we will later depict in Figure 13.29.*

```
nhanes <- read.csv(paste0("https://raw.githubusercontent.com/gagolews/",
        "teaching-data/master/marek/nhanes_adult_female_bmx_2020.csv"),
    comment.char="#", col.names=c("weight", "height", "armlen", "leglen",
        "armcirc", "hipcirc", "waistcirc"))
par(mfrow=c(1, 2))
for (col in c("black", "#00000010"))
    plot(nhanes[["height"]], nhanes[["weight"]], col=col,
        pch=16, xlab="Height", ylab="Weight")
```

**Example 13.25** *Figure 13.19 depicts the average monthly temperatures in your next holiday des-*

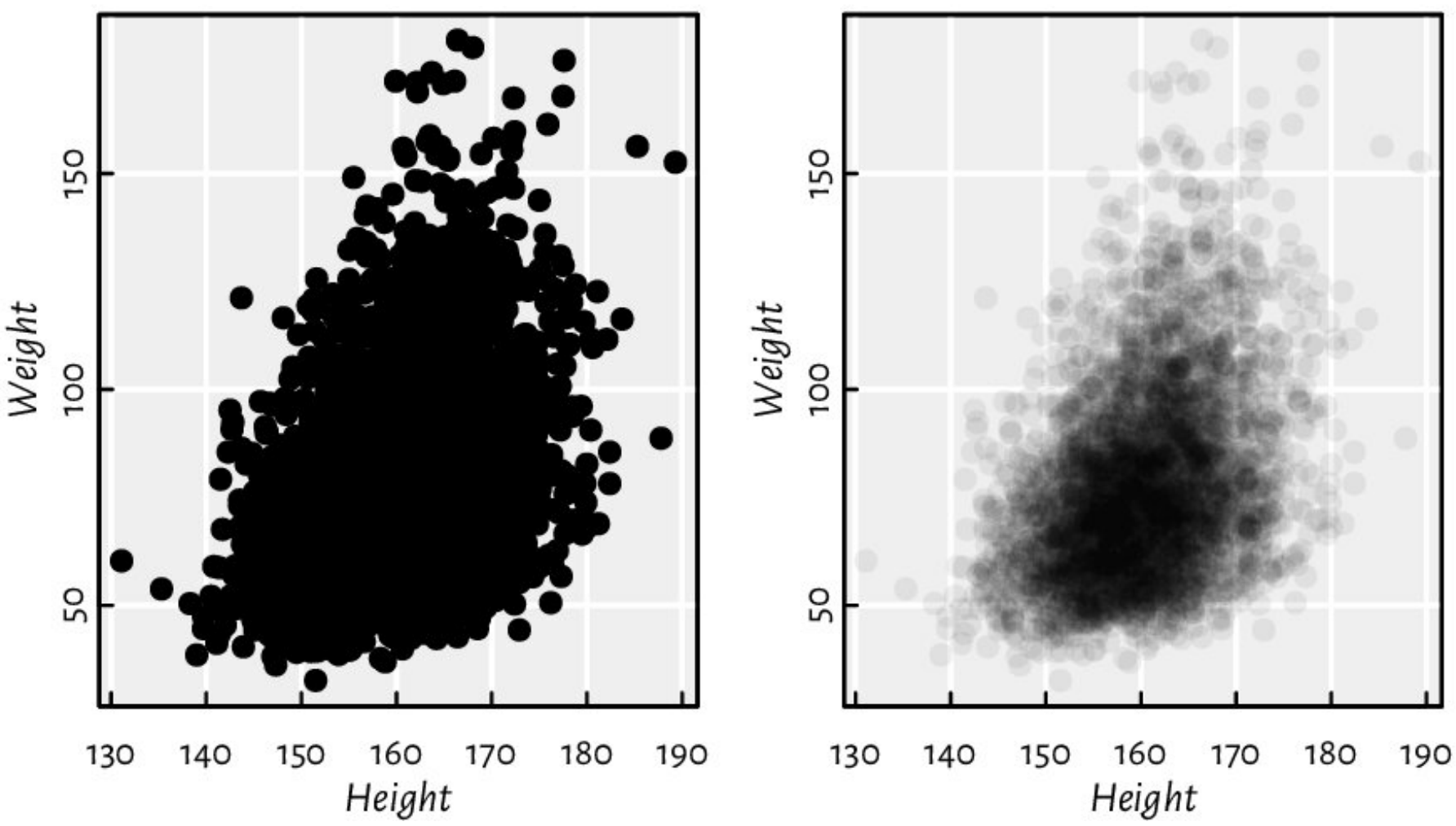

Figure 13.18. Semi-transparent symbols can reflect the points' distribution density.

*tination: Warsaw, Poland (a time series). Note that the translucent ribbon representing the low-high average temperature intervals was added using a call to* **`polygon`**.

```
# Warsaw monthly temperatures; source: https://en.wikipedia.org/wiki/Warsaw
high <- c( 0.6,  1.9,  6.6, 13.6, 19.5, 21.9,
          24.4, 23.9, 18.4, 12.7,  5.9,  1.6)
mean <- c(-1.8, -0.6,  2.8,  8.7, 14.2, 17.0,
          19.2, 18.3, 13.5,  8.5,  3.3, -0.7)
low  <- c(-4.2, -3.6, -0.6,  3.9,  8.9, 11.8,
          13.9, 13.1,  9.1,  4.8,  0.6, -3.0)
matplot(1:12, cbind(high, mean, low), type="o", col=c(2, 1, 4), lty=1,
    xlab="month", ylab="temperature [°C]", xaxt="n", pch=16, cex=0.5)
axis(1, at=1:12, labels=month.abb, line=-0.25, lwd=0, lwd.ticks=1)
polygon(c(1:12, rev(1:12)), c(high, rev(low)), border=NA, col="#ffff0033")
legend("bottom", c("average high", "mean", "average low"),
    lty=1, col=c(2, 1, 4), bg="white")
```

**Example 13.26** *Figure 13.20 depicts a scatter plot similar to Figure 13.18, but now with the points' hue being a function of a third variable.*

```
midpoints <- function(x) 0.5*(x[-1]+x[-length(x)])
z <- nhanes[["waistcirc"]]
breaks <- seq(min(z), max(z), length.out=10)
zf <- cut(z, breaks, include.lowest=TRUE)
col <- hcl.colors(nlevels(zf), "Viridis", alpha=0.5)
layout(matrix(c(1, 2), nrow=1),  # two plots in one page
    widths=c(1, lcm(3)))         # second one is of width "3cm" (scaled)
# first subplot:
```

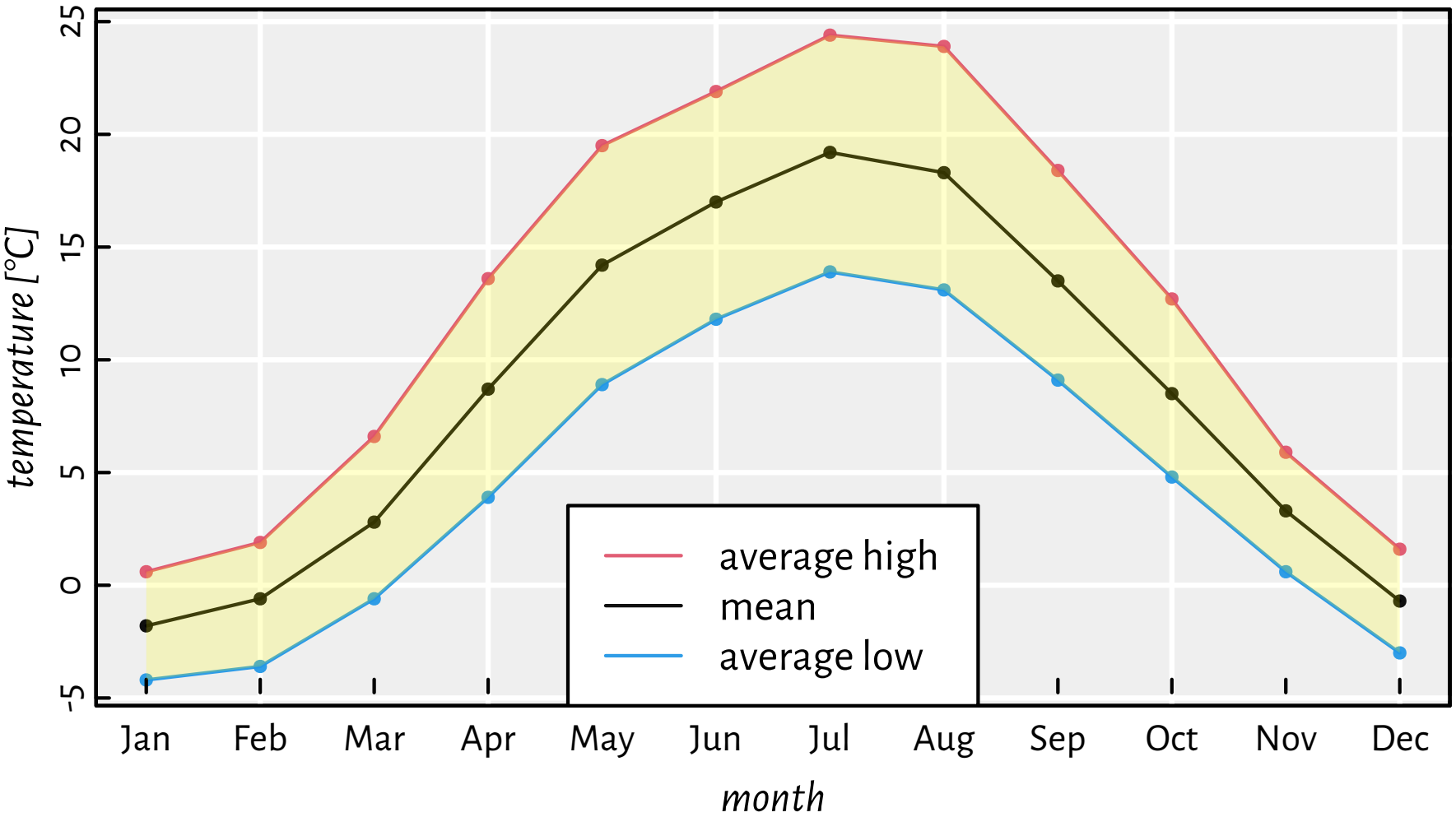

Figure 13.19. Example time series. A semi-transparent ribbon was added by calling **polygon** to highlight the area between the low-high ranges (intervals).

```
plot(nhanes[["height"]], nhanes[["weight"]], col=col[as.numeric(zf)],
    pch=16, xlab="Height", ylab="Weight")
# second subplot:
par(mar=c(2.2, 0.6, 2.2, 0.6))
plot.new(); plot.window(c(0, 1), c(0, nlevels(zf)))
rasterImage(as.matrix(rev(col)), 0, 0, 1, nlevels(zf), interpolate=FALSE)
text(0.5, 1:nlevels(zf)-0.5, sprintf("%3.0f", midpoints(breaks)))
mtext("Waist Ø", side=3)
```

**Exercise 13.27** *Implement your version of* ***pairs****, being the function to draw a scatter plot matrix (a pairs plot).*

**Exercise 13.28** ***ecdf*** *returns an object of the S3 classes* `ecdf` *and* `stepfun`*. There are* ***plot*** *methods overloaded for them. Inspect their source code. Then, inspired by this, create a function to compute and display the empirical cumulative distribution function corresponding to a given numeric vector.*

**Exercise 13.29** ***spline*** *performs cubic spline interpolation, whereas* ***smooth.spline*** *determines a smoothing spline of a given two-dimensional dataset. Plot different splines for* `cars[["dist"]]` *as a function of* `cars[["speed"]]`*. Which of these two functions is more appropriate for depicting this dataset?*

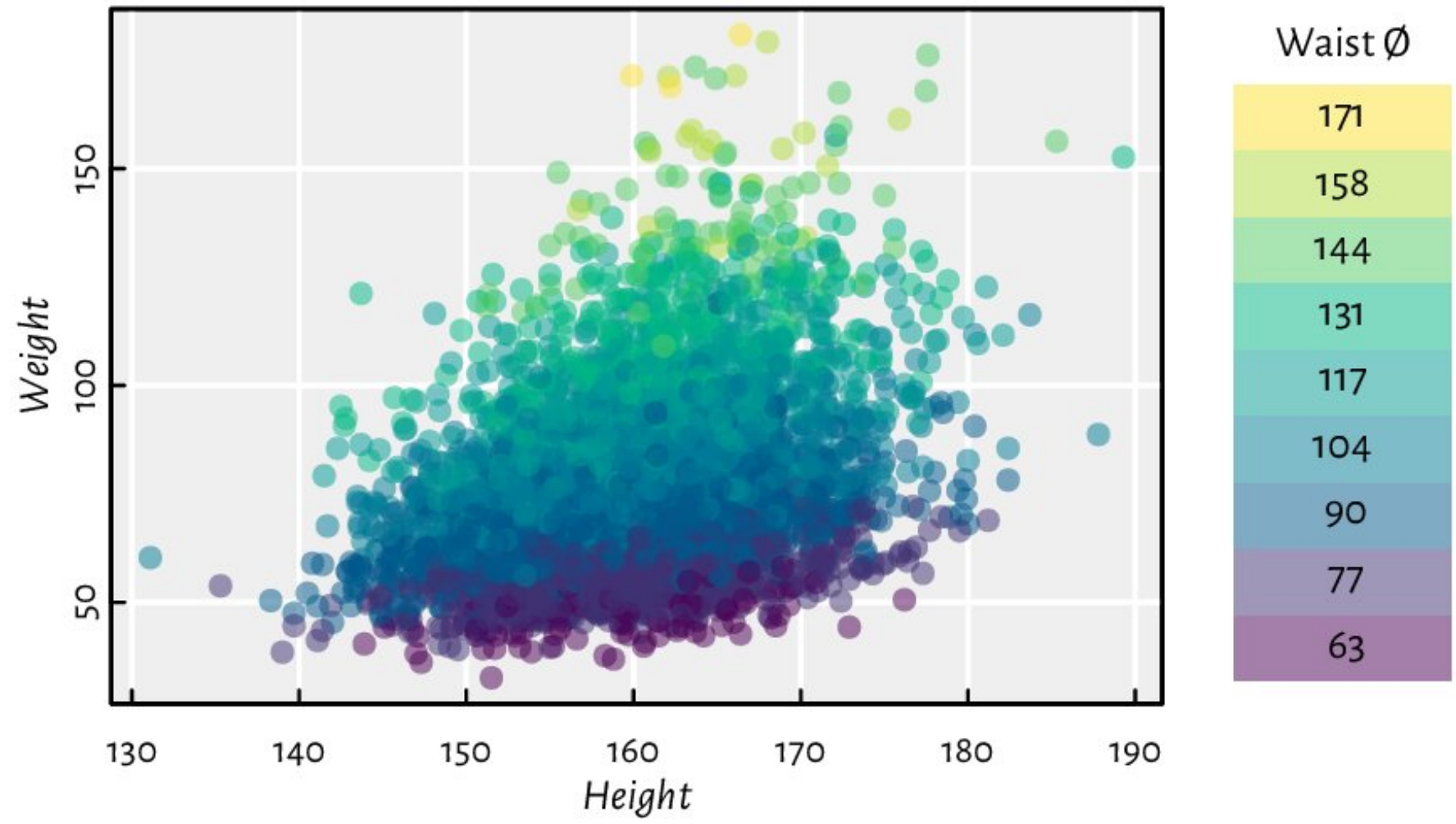

Figure 13.20. A 2D scatter plot with a third variable represented by colours.

### 13.3.2 Bar plots and histograms

A bar plot is drawn using a series of rectangles (i.e., certain polygons) of different heights (or widths, if we request horizontal alignment).

**Example 13.30** *Let's visualise the dataset[6] listing the most frequent causes of medication errors (data are fabricated):*

```
cat_med = c(
    "Unauthorised drug", "Wrong IV rate", "Wrong patient", "Dose missed",
    "Underdose", "Wrong calculation","Wrong route", "Wrong drug",
    "Wrong time", "Technique error", "Duplicated drugs", "Overdose"
)
counts_med = c(1, 4, 53, 92, 7, 16, 27, 76, 83, 3, 9, 59)
```

*A Pareto chart combines a bar plot featuring bars of decreasing heights with a cumulative percentage curve; see Figure 13.21.*

```
o <- order(counts_med)
cato_med <- cat_med[o]
pcto_med <- counts_med[o]/sum(counts_med)*100
cumpcto_med <- rev(cumsum(rev(pcto_med)))
# bar plot of percentages
par(mar=c(2.2, 0.6, 2.2, 6.6))  # wide left margin
midp <- barplot(pcto_med, horiz=TRUE, xlab="%",
    col="white", xlim=c(0, 25), xaxs="r", yaxs="r", yaxt="n",
```

[6] https://www.cec.health.nsw.gov.au/CEC-Academy/quality-improvement-tools/pareto-charts

```
    width=3/4, space=1/3)
text(pcto_med, midp, sprintf("%.1f%%", pcto_med), pos=4, cex=0.89)
axis(4, at=midp, labels=cato_med, las=1)
box()
# cumulative percentage curve in a new coordinate system
par(usr=c(-4, 104, par("usr")[3], par("usr")[4]))  # 0-100 with 4% addition
lines(cumpcto_med, midp, type="o", col=4, pch=18)
axis(3, col=4)
mtext("cumulative %", side=3, line=1.2, col=4)
text(cumpcto_med, midp, sprintf("%.1f%%", cumpcto_med), cex=0.89, col=4,
    pos=c(4, 2)[(cumpcto_med>80)+1], offset=0.5)
```

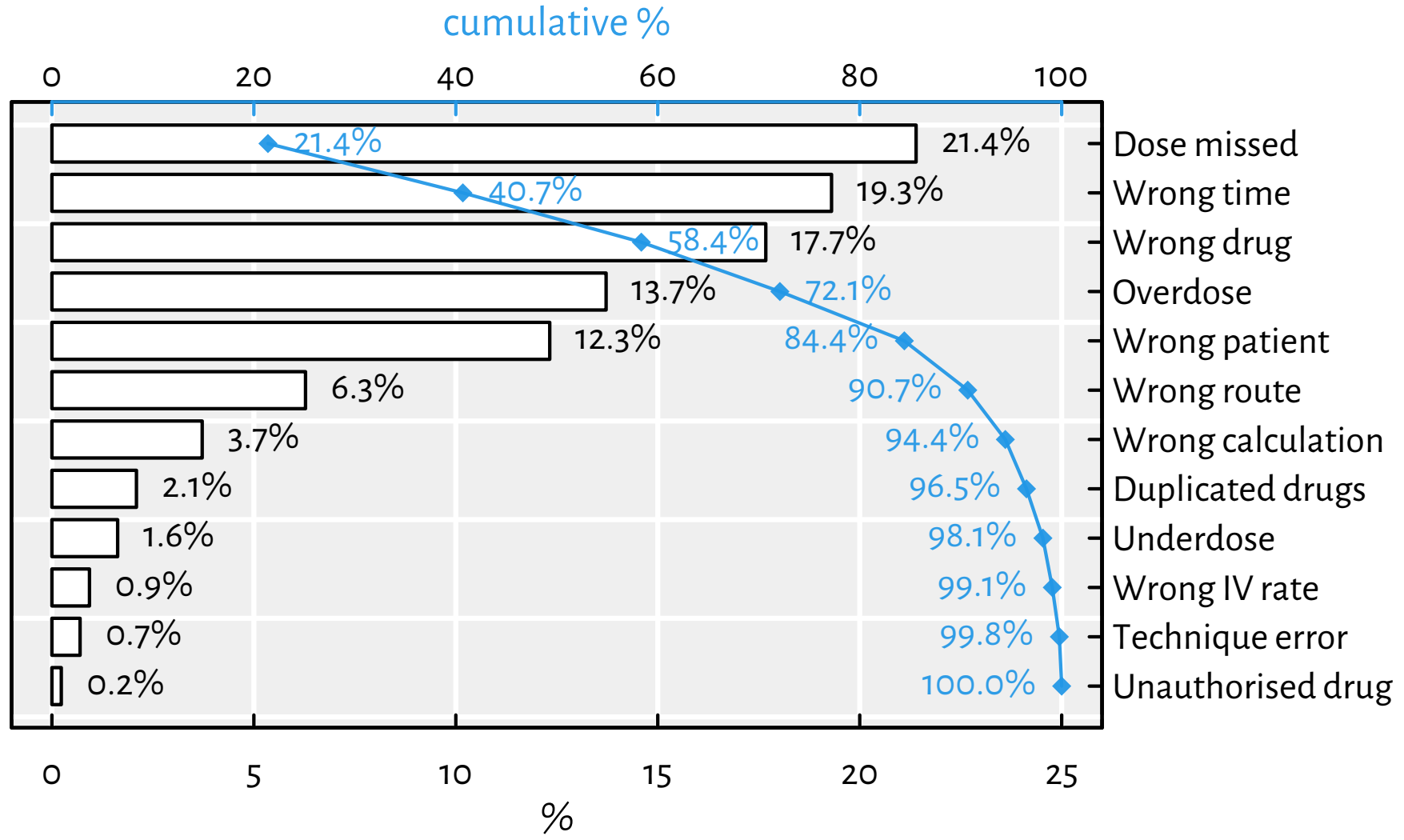

Figure 13.21. An example Pareto chart (a fancy bar plot). Double axes have a general tendency to confuse the reader.

*Note that* **`barplot`** *returned the midpoints of the bars, which we put in good use. By default, it sets the* `xaxs="i"` *graphics parameter and thus does not extend the x-axis range by 4% on both sides. This would not make us happy here, therefore we needed to change it manually.*

**Exercise 13.31** *Draw a bar plot summarising, for each passenger class and sex, the number of adults who did not survive the sinking of the deadliest 1912 cruise; see Figure 13.22 and the* `Titanic` *dataset.*

**Exercise 13.32** *Implement your version of* **`barplot`**, *but where the bars are placed precisely at the positions specified by the user, e.g., allowing the bar midpoints to be consecutive integers.*

We will definitely not cover the (in)famous pie charts in our book. The human brain is not very skilled at judging the relative differences between the areas of geometric objects. Also, they are ugly (pie charts, not geometric objects in general).

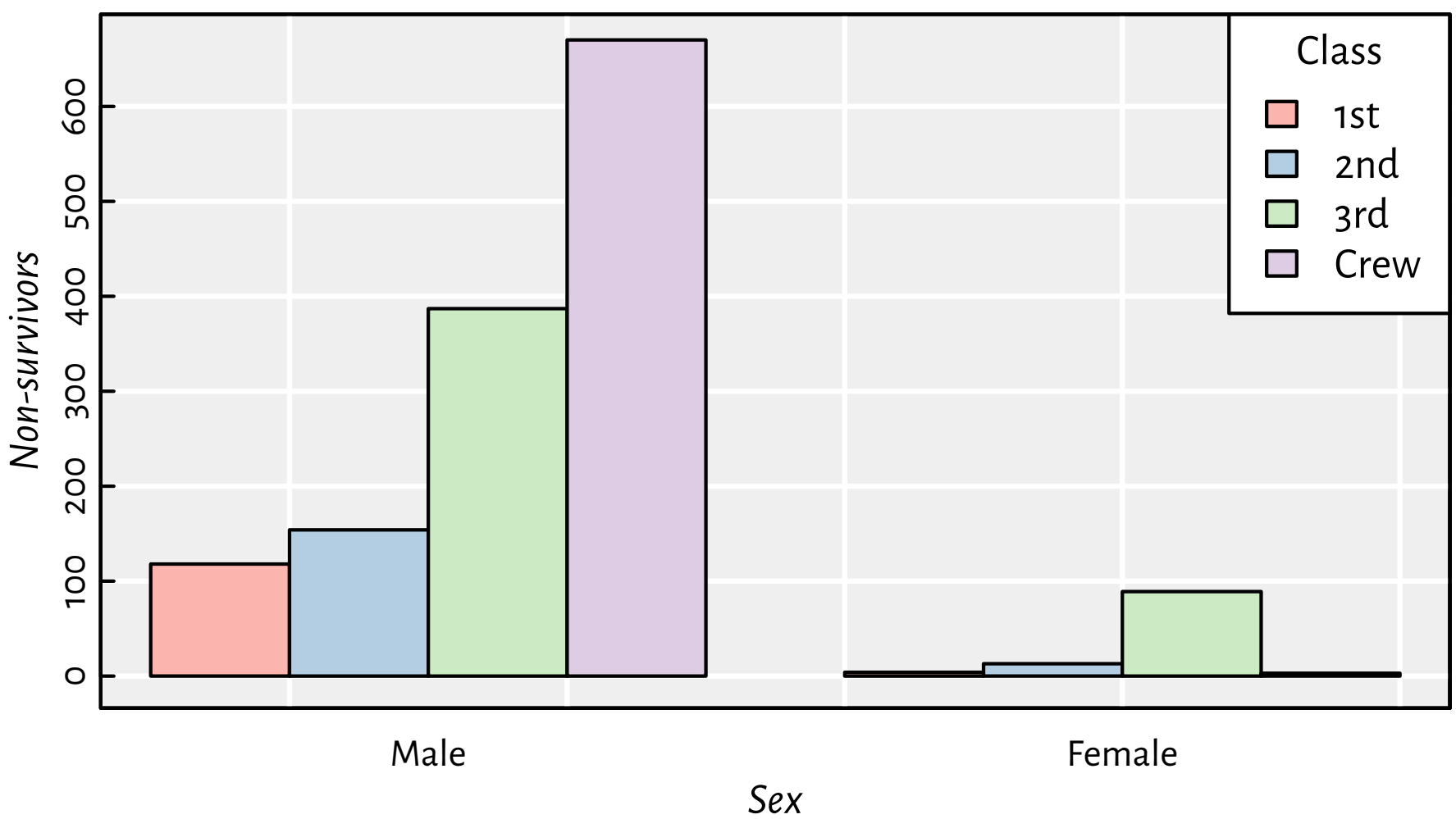

Figure 13.22. An example bar plot representing a two-way contingency table.

Moving on: a histogram is a simple density estimator for continuous data. It can be thought of as a bar plot with bars of heights proportional to the number of observations falling into the corresponding disjoint intervals. Most often, there is no space between the bars to emphasise that the intervals cover the whole data range.

A histogram can be computed and drawn using the high-level function **hist**; see Figure 13.23.

```
par(mfrow=c(1, 2))
for (breaks in list("Sturges", 25)) {
    # Sturges (a heuristic) is the default; any value is merely a suggestion
    hist(iris[["Sepal.Length"]], probability=TRUE, xlab="Sepal length",
        main=NA, breaks=breaks, col="white")
    box()  # oddly, we need to add it manually
}
```

**Exercise 13.33** *Study the source code of* **hist.default**. *Note the invisibly-returned list of the S3 class* `histogram`. *Then, study* **graphics:::plot.histogram**. *Implement similar functions yourself.*

**Exercise 13.34** *Modify your function to draw a scatter plot matrix so that it gives the histograms of the marginal distributions on its diagonal.*

**Example 13.35** *Using* **layout** *mentioned in Section 13.2.5, we can draw a scatter plot with marginal histograms; see Figure 13.24. Note that we split the page into four plots of unequal sizes, but the upper right part of the grid is unused. We use* **hist** *for binning only (*`plot=FALSE`*). Then,* **barplot** *is utilised for drawing as it gives greater control over the process (e.g., supports vertical layout).*

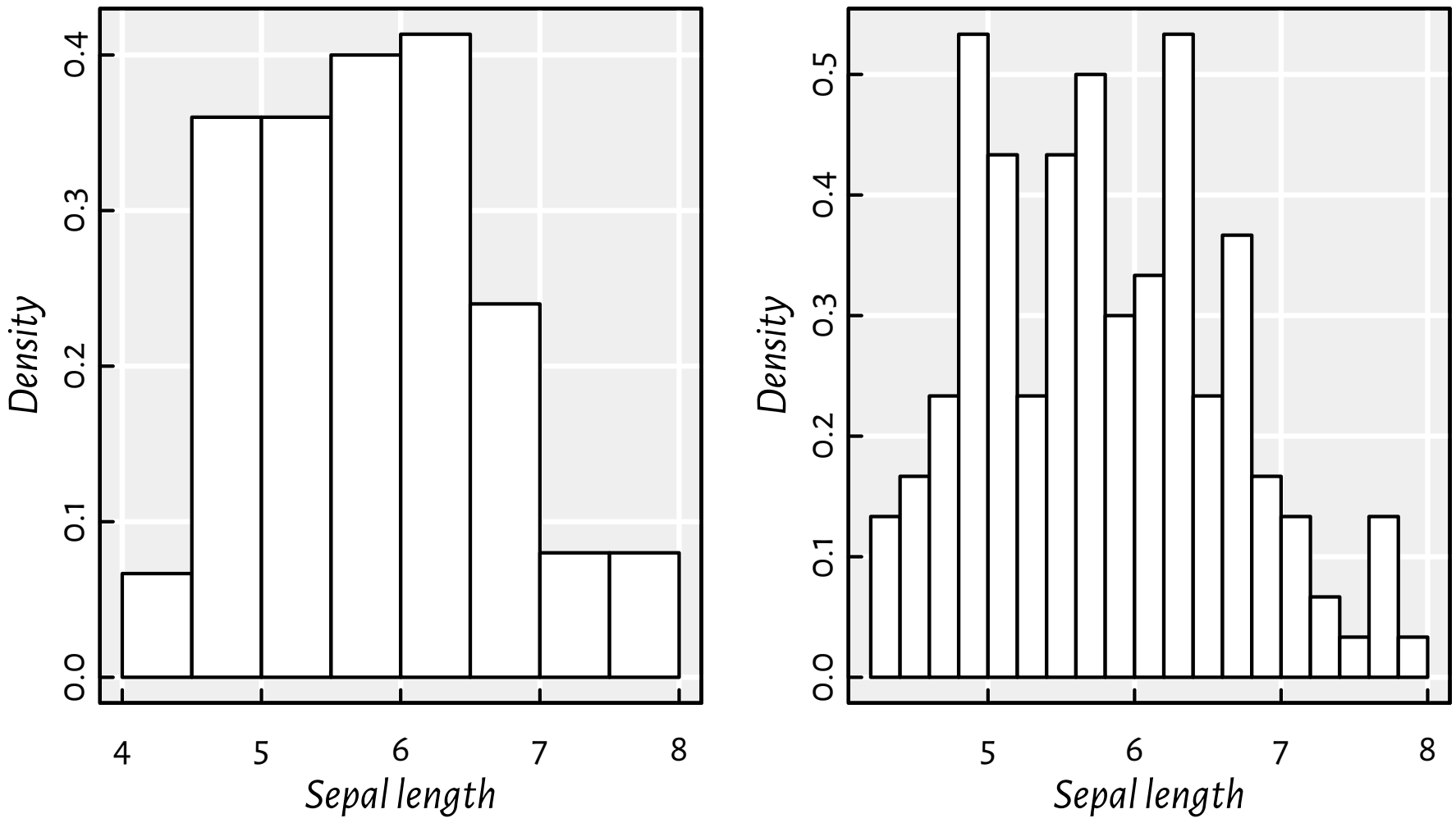

Figure 13.23. Example histograms for the same dataset.

```
layout(matrix(
    c(1, 1, 1, 0,  # the first row: the first plot of width 3 and nothing
      3, 3, 3, 2,  # the third plot (square) and the second (tall) in 3 rows
      3, 3, 3, 2,
      3, 3, 3, 2), nrow=4, byrow=TRUE))
par(mex=1, cex=1)  # the layout function changed this!
x <- jitter(iris[["Sepal.Length"]])
y <- jitter(iris[["Sepal.Width"]])
# the first subplot (top)
par(mar=c(0.2, 2.2, 0.6, 0.2), ann=FALSE)
hx <- hist(x, plot=FALSE, breaks=seq(min(x), max(x), length.out=20))
barplot(hx[["density"]], space=0, axes=FALSE, col="#00000011")
# the second subplot (right)
par(mar=c(2.2, 0.2, 0.2, 0.6), ann=FALSE)
hy <- hist(y, plot=FALSE, breaks=seq(min(y), max(y), length.out=20))
barplot(hy[["density"]], space=0, axes=FALSE, horiz=TRUE, col="#00000011")
# the third subplot (square)
par(mar=c(2.2, 2.2, 0.2, 0.2), ann=TRUE)
plot(x, y, xlab="Sepal length", ylab="Sepal width",
    xlim=range(x), ylim=range(y))  # default xlim, ylim
```

**Example 13.36** *(*) Kernel density estimators (KDEs) are another way to guesstimate the data distribution. The* **`density`** *function, for a given numeric vector, returns a list with, amongst others, the x and y coordinates of the points that we can pass directly to the* **`lines`** *function. Below we depict the KDEs of data split into three groups; see Figure 13.25.*

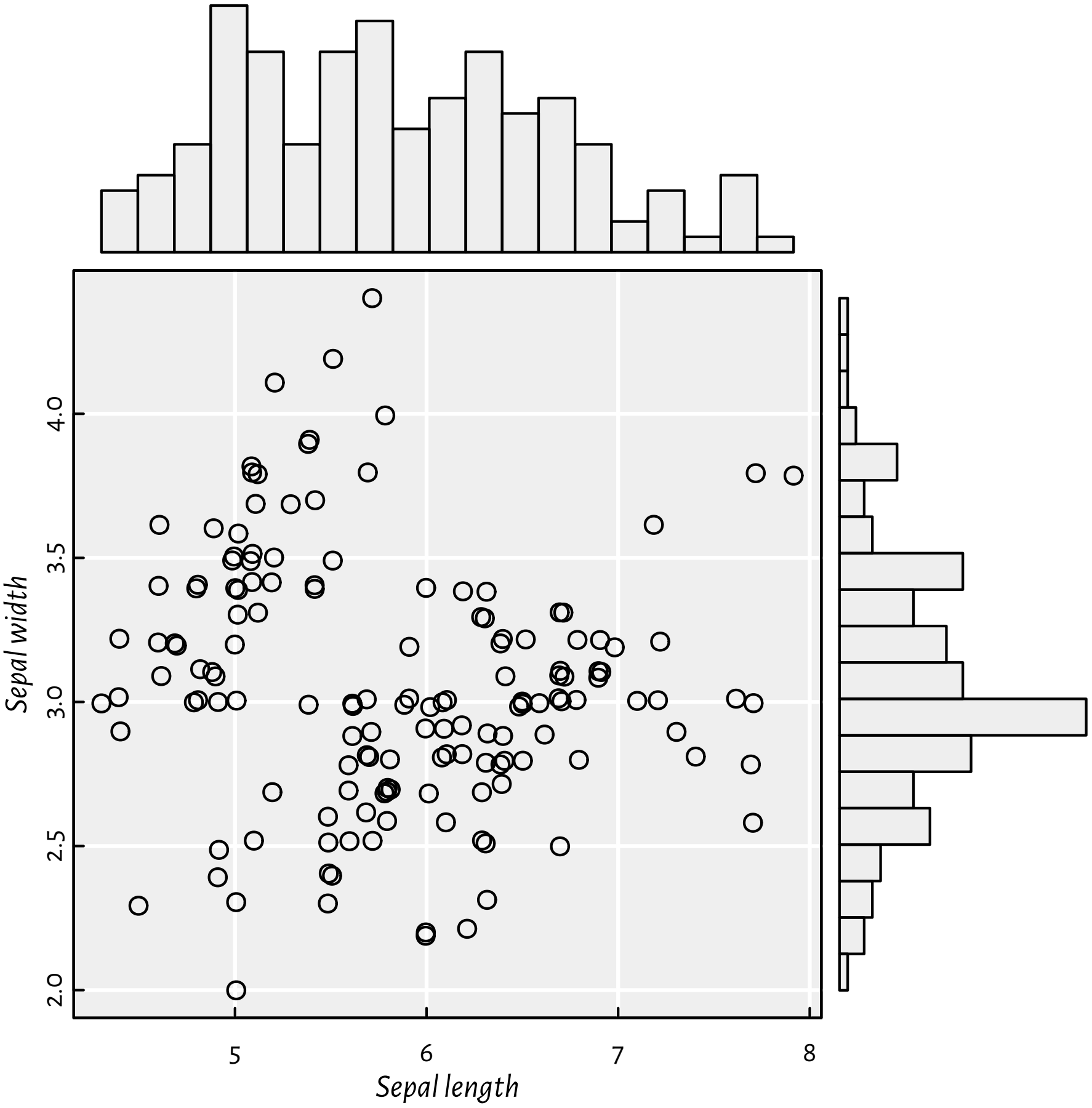

Figure 13.24. A scatter plot with marginal histograms: three (four) plots on one page, but on a nonuniform grid created using **layout**.

```
adjust_transparency <- function(col, alpha)
    rgb(t(col2rgb(col)/255), alpha=alpha)  # alpha in [0, 1]

pal <- adjust_transparency(palette(), 0.2)
kdes <- lapply(split(iris[["Sepal.Length"]], iris[["Species"]]), density)
matplot(sapply(kdes, `[[`, "x"), sapply(kdes, `[[`, "y"),
    type="l", xlab="Sepal length", ylab="density", lwd=1.5)
for (i in seq_along(kdes))
    polygon(kdes[[i]][["x"]], kdes[[i]][["y"]], col=pal[i], border=NA)
legend("topright", legend=levels(iris[["Species"]]), bg="white", lwd=1.5,
    col=seq_along(levels(iris[["Species"]])),
    lty=seq_along(levels(iris[["Species"]])))
```

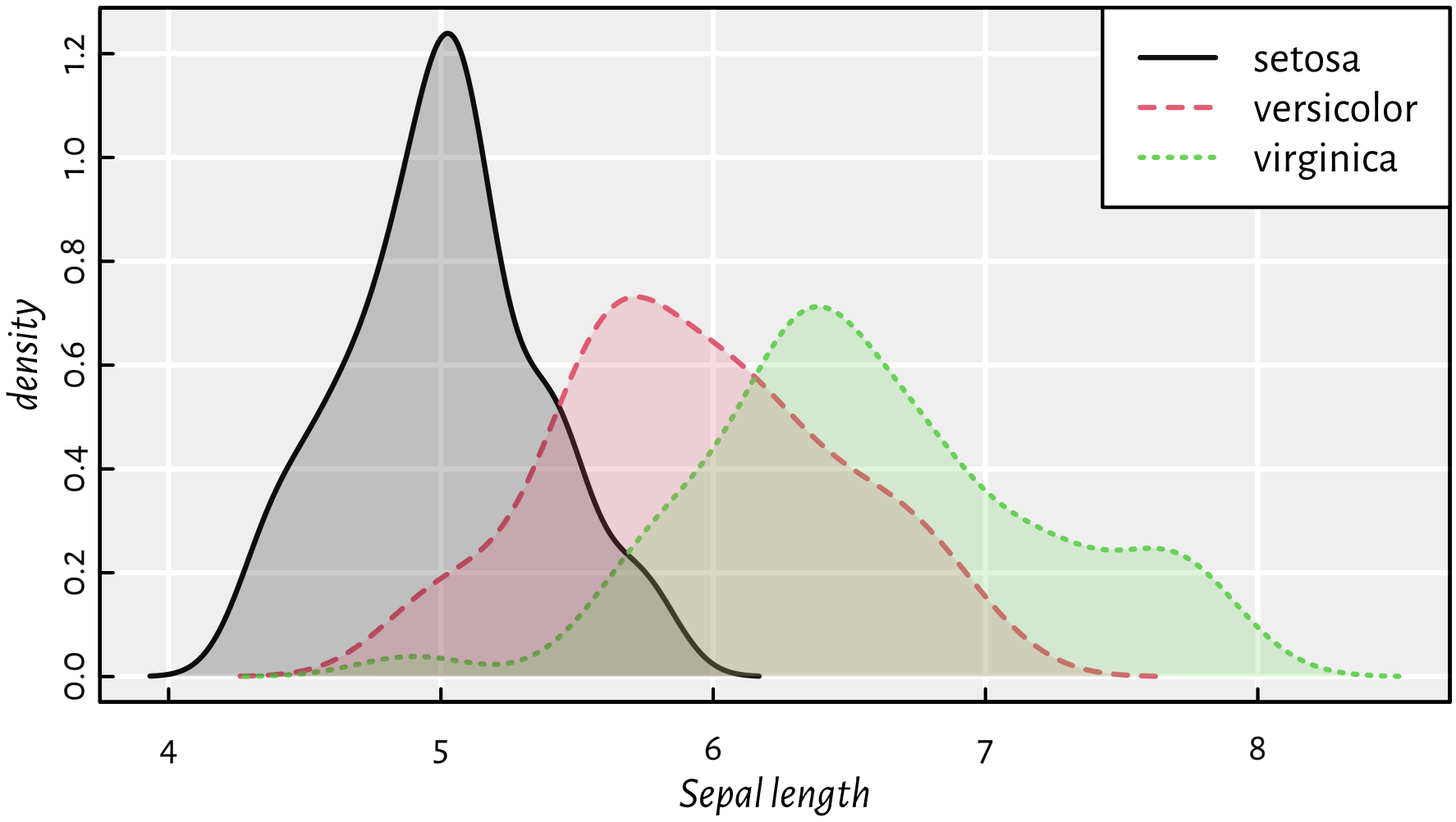

Figure 13.25. Kernel density estimators of sepal length split by species in the `iris` data-set. Note the semi-transparent polygons (again).

**Exercise 13.37** *(*) Implement a function that draws kernel density estimators for a given numeric variable split by a combination of three factor levels; see Figure 13.26 for an example.*

```
grid_kde <- function(data, values, x, y, hue) ...to.do...

tips <- read.csv(paste0("https://raw.githubusercontent.com/gagolews/",
        "teaching-data/master/other/tips.csv"),
    comment.char="#", stringsAsFactors=TRUE)
head(tips, 3)  # preview this example dataset
##   total_bill tip    sex smoker day   time size
## 1      16.99 1.01 Female     No Sun Dinner    2
## 2      10.34 1.66   Male     No Sun Dinner    3
## 3      21.01 3.50   Male     No Sun Dinner    3
grid_kde(tips, values="tip", x="smoker", y="time", hue="sex")
```

### 13.3.3 Box-and-whisker plots

We have already seen a chart generated by the **`boxplot`** function in Figure 5.1. Tinkering with it will give us robust practice, which in turn shall make us perfect.

**Exercise 13.38** *Modify the code generating Figure 5.1 so that:*

1. *same doses are grouped together (more space between different doses added; also, on the x-axis, only* unique *doses are printed),*
2. *different supps have different colours (add a legend explaining them).*

**Exercise 13.39** *Write a function for drawing box plots using graphics primitives.*

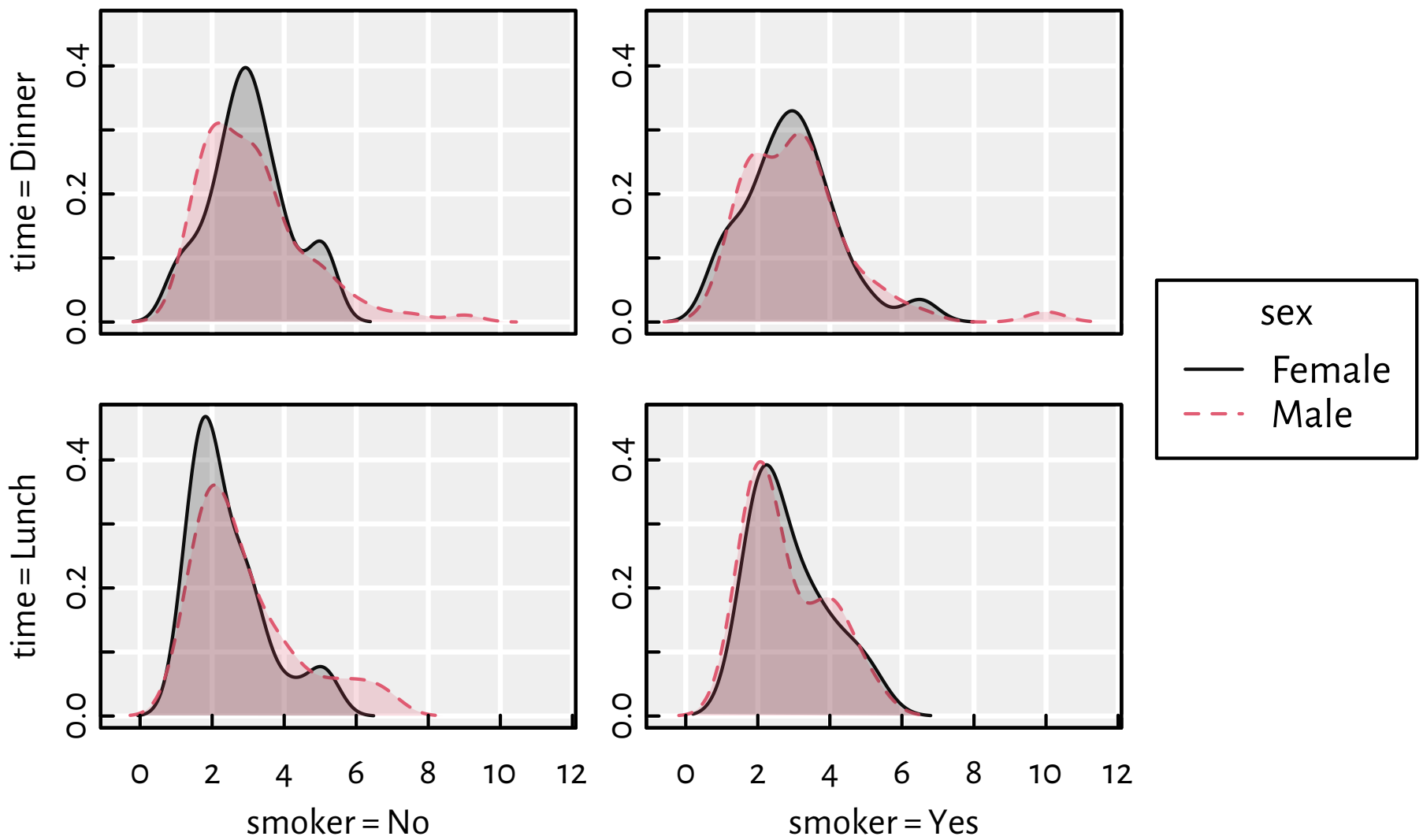

Figure 13.26. An example grid plot (also known as a trellis, panel, conditioning, or lattice plot) with kernel density estimators for a numeric variable (amount of tip in a US restaurant) split by a combination of three factor levels (smoker, time, sex).

**Exercise 13.40** *(*) Write a function for drawing violin plots. They are similar to box plots but use kernel density estimators.*

**Exercise 13.41** *(*) Implement a bag plot, which is a two-dimensional version of a box plot. Use* **`chull`** *to compute the convex hull of a point set.*

### 13.3.4 Contour plots and heat maps

**`image`** is a convenient wrapper around **`rasterImage`**, which can draw contour plots, two-dimensional histograms, heatmaps, etc. In particular, when plotting a function of two variables like $z = f(x, y)$, the magnitude of the $z$ component can be expressed using colour brightness or hue.

**Example 13.42** *Figure 13.27 presents a filled contour plot of Himmelblau's function,* $f(x, y) = (x^2 + y - 11)^2 + (x + y^2 - 7)^2$*, for* $x \in [-5, 5]$ *and* $y \in [-4, 4]$*. A call to* **`contour`** *adds labelled contour lines (which is actually a nontrivial operation).*

```
x <- seq(-5, 5, length.out=250)
y <- seq(-4, 4, length.out=200)
z <- outer(x, y, function(xg, yg) (xg^2 + yg - 11)^2 + (xg + yg^2 - 7)^2)
image(x, y, z, col=grey(seq(1, 0, length.out=16)))
contour(x, y, z, nlevels=16, add=TRUE)
```

In **`image`**, the number of rows in `z` matches the length of `x`, whereas the number of columns is equal to the size of `y`. This might be counterintuitive; when `z` is printed, the image is its 90-degree rotated version.

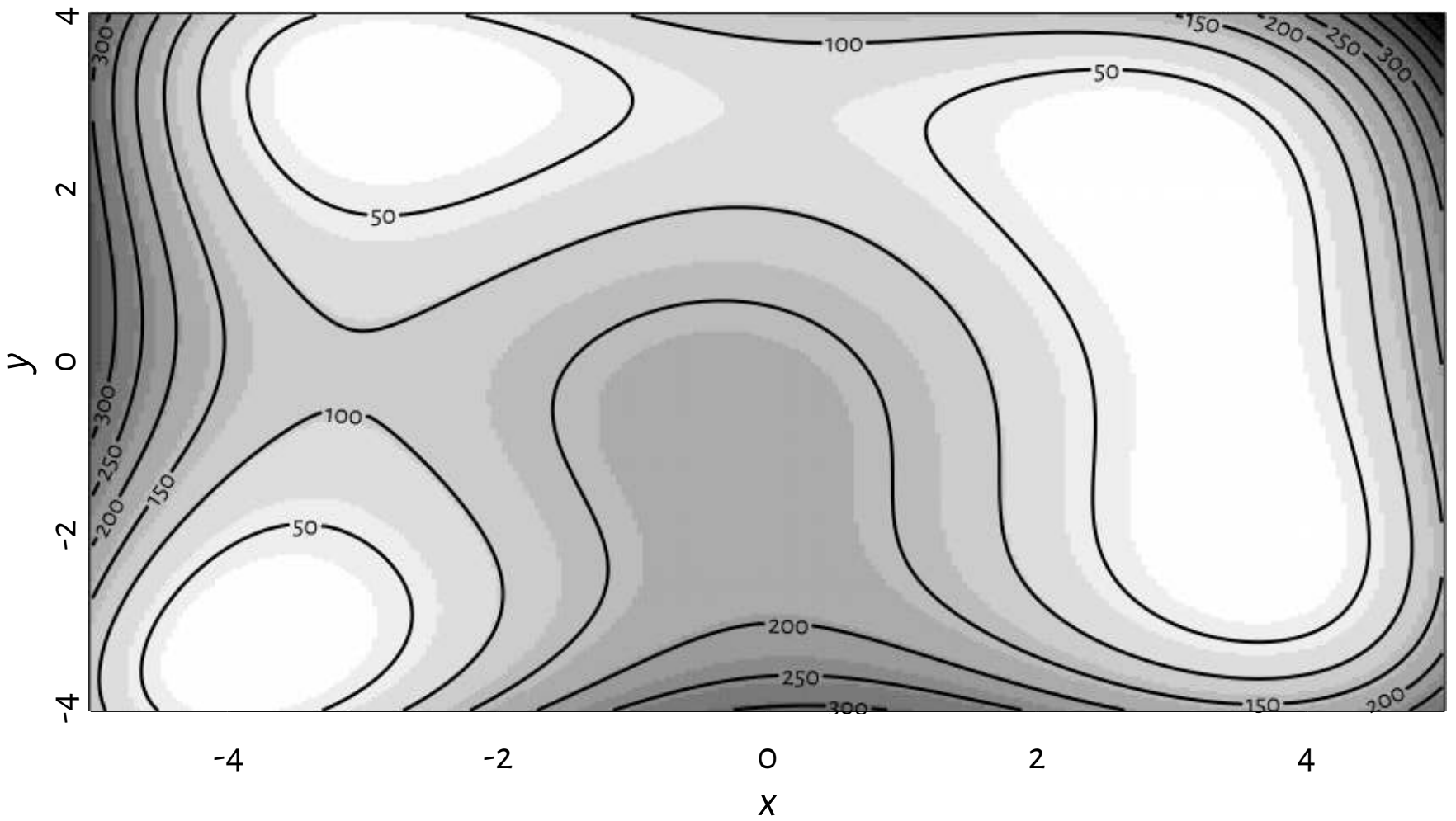

Figure 13.27. A filled contour plot with labelled contour lines.

**Example 13.43** *Figure 13.28 presents an example heatmap depicting Pearson's correlations between all pairs of variables in the* `nhanes` *data frame which we loaded some time ago.*

```
o <- c(6, 5, 1, 7, 4, 2, 3)  # order of rows/cols (by similarity)
R <- cor(nhanes[o, o])
par(mar=c(2.8, 7.6, 1.2, 7.6), ann=FALSE)
image(1:NROW(R), 1:NCOL(R), R,
    ylim=c(NROW(R)+0.5, 0.5),
    zlim=c(-1, 1),
    col=hcl.colors(20, "BluGrn", rev=TRUE),
    xlab=NA, ylab=NA, asp=1, axes=FALSE)
axis(1, at=1:NROW(R), labels=dimnames(R)[[1]], las=2, line=FALSE, tick=FALSE)
axis(2, at=1:NCOL(R), labels=dimnames(R)[[2]], las=1, line=FALSE, tick=FALSE)
text(arrayInd(seq_along(R), dim(R)),
    labels=sprintf("%.2f", R),
    col=c("white", "black")[abs(R<0.8)+1],
    cex=0.89)
```

**Exercise 13.44** *Check out the* **`heatmap`** *function, which uses hierarchical clustering to find an aesthetic reordering of the matrix's items.*

**Example 13.45** *Figure 13.29 depicts a two-dimensional histogram. It approaches the idea of reflecting the points' density differently from the semi-transparent symbols in Figure 13.18.*

```
histogram_2d <- function(x, y, k=25, ...)
{
    breaksx <- seq(min(x), max(x), length.out=k)
    fx <- cut(x, breaksx, include.lowest=TRUE)
```

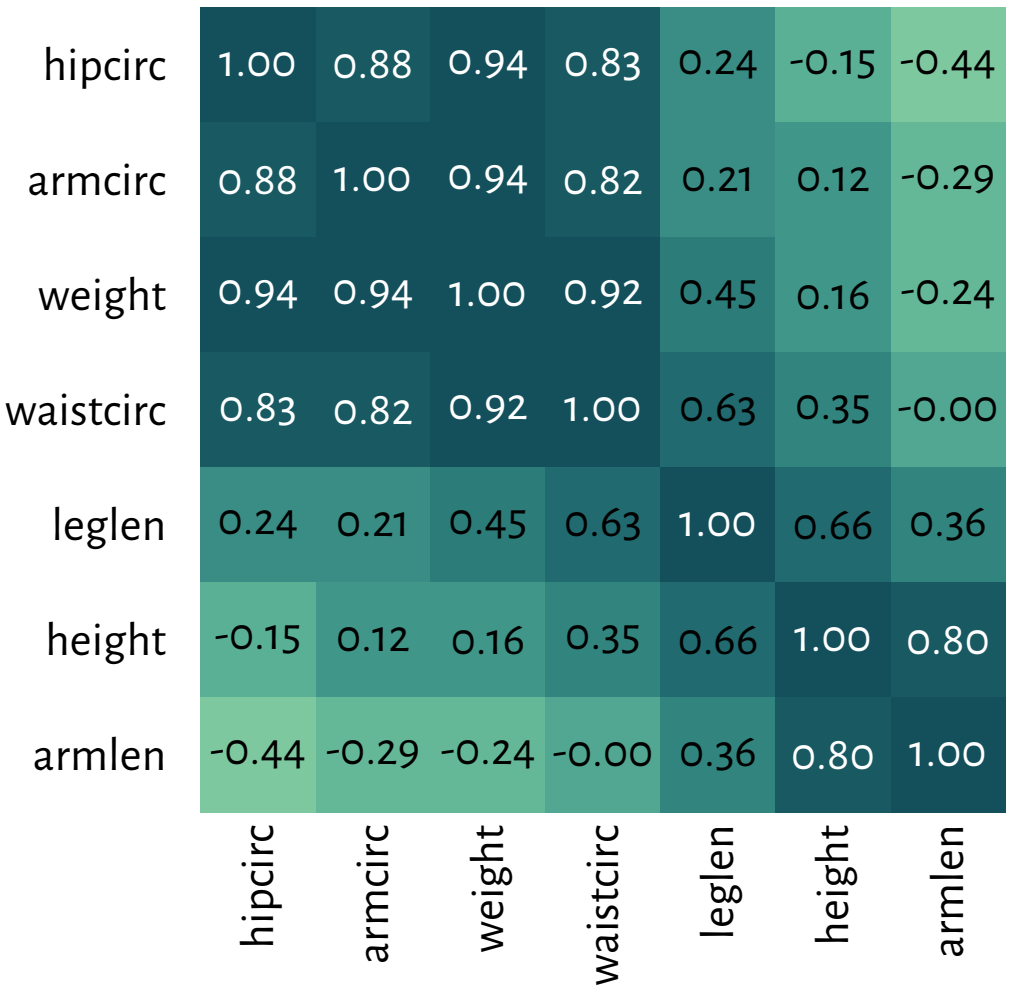

Figure 13.28. A correlation heatmap drawn using `image`.

```
    breaksy <- seq(min(y), max(y), length.out=k)
    fy <- cut(y, breaksy, include.lowest=TRUE)
    C <- table(fx, fy)
    image(midpoints(breaksx), midpoints(breaksy), C,
        xaxs="r", yaxs="r", ...)
}

par(mfrow=c(1, 2))
for (k in c(25, 50))
    histogram_2d(nhanes[["height"]], nhanes[["weight"]], k=k,
        xlab="Height", ylab="Weight",
        col=c("#ffffff00", hcl.colors(25, "Viridis", rev=TRUE))
    )
```

**Exercise 13.46** *(*) Implement some two-dimensional kernel density estimator and plot it using* `contour`.

## 13.4 Exercises

**Exercise 13.47** *Answer the following questions.*

- *Can functions from the* `graphics` *package be used to adjust the plots generated by* `lattice` *and* `ggplot2`*?*

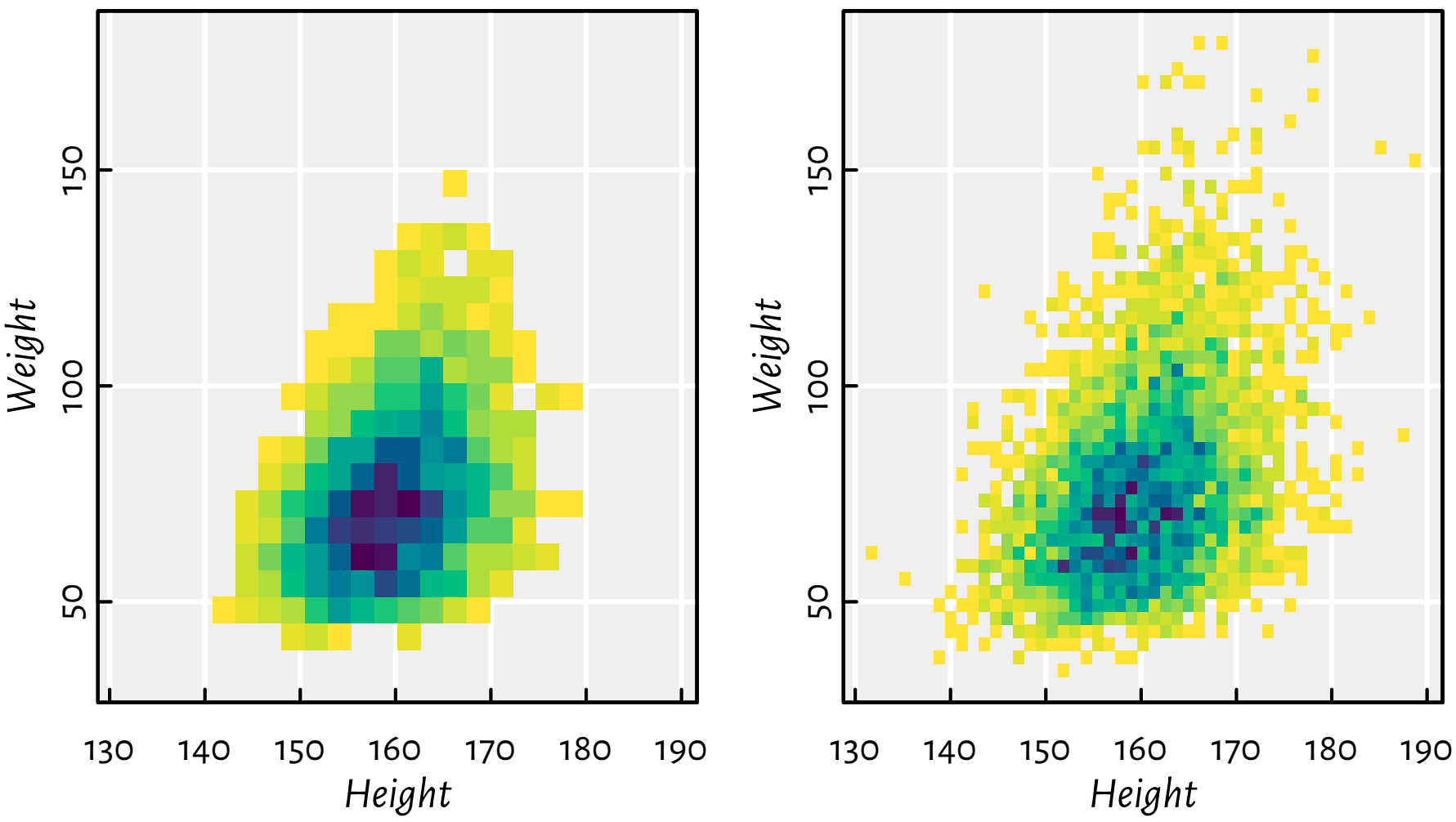

Figure 13.29. Two-dimensional histograms with different numbers of bins, where the bin count is reflected by the colour.

- *What are the most common graphics primitives?*
- *Can all high-level functions be implemented using low-level ones? As an example, discuss the key ingredients used in* **`barplot`**.
- *Some high-level functions discussed in this chapter carry the* `add` *parameter. What is its purpose?*
- *What are the admissible values of* `pch` *and* `lty`*? Also, in the default palette, what is the meaning of colours 1, 2, …, 16? Can their meaning be changed?*
- *Can all graphics parameters be changed?*
- *What is the difference between passing* `xaxt="n"` *to* **`plot.default`** *vs setting it with* **`par`**, *and then calling* **`plot.default`**?
- *Which graphics parameters are set by* **`plot.window`**?
- *What is the meaning of the* `usr` *parameter when using the logarithmic scale on the x-axis?*
- *(*) How to place a plotting symbol exactly 1 centimetre from the top-left corner of the current page (following the page's diagonal)?*
- *Semi-transparent polygons are nice, right?*
- *Can an ellipse be drawn using* **`polygon`**?
- *What happens when we set the graphics parameter* `mfrow=c(2, 2)`?
- *How to export the current plot to a PDF file?*

**Exercise 13.48** *Draw the 2022 BTC-to-USD close rates[7] time series. Then, add the 7- and 30-day moving averages. (*) Also, fit a local polynomial (moving) regression model using the Savitzky–Golay filter (see `loess`).*

**Exercise 13.49** *(*) Draw (from scratch) a candlestick plot for the 2022 BTC-to-USD rates[8].*

**Exercise 13.50** *(*) Create a function to draw a normal quantile-quantile (Q-Q) plot, i.e., for inspecting whether a numeric sample might come from a normal distribution.*

**Exercise 13.51** *(*) Draw a map of the world, where each country is filled with a colour whose brightness or hue is linked to its Gini index of income inequality. You can easily find the data on Wikipedia. Try to find an open dataset that gives the borders of each country as vertices of a polygon (e.g., in the form of a (geo)JSON file).*

**Exercise 13.52** *Next time you see a pleasant data visualisation somewhere, try to reproduce it using base `graphics`.*

For further information on graphics generation in R, see, e.g., Chapter 12 of [59], [49], and [53]. Good introductory textbooks to data visualisation as an art include [57, 60].

In this chapter, we were only interested in *static* graphics, e.g., for use in printed publications or plain websites. Interactive plots that a user might tinker with in a web browser are a different story.

And so the second part of our delightful course is ended.

[7] https://github.com/gagolews/teaching-data/raw/master/marek/btcusd_close_2022.csv
[8] https://github.com/gagolews/teaching-data/raw/master/marek/btcusd_ohlc_2022.csv

# Part III

# Deepest

# 14

# *Interfacing compiled code (**)*

R is an effective *glue* language. It is suitable for composing whole data wrangling pipelines: from data import through processing, analysis, and visualisation to export. It makes using and connecting larger *building blocks* very convenient. R is also a competent tool for developing *usable* implementations of standalone, general-purpose algorithms, especially if they are of *numerical* nature. Nevertheless, for performance reasons, we may consider rewriting computing-intensive tasks in C or C++[1]. Such a move can be beneficial if we need a method that:

- has higher memory or time complexity when programmed using vectorised R functions than its straightforward implementation,
- has an *iterative* or *recursive* nature, e.g., involving unvectorisable *for* or *while* loops,
- relies on complicated dynamic data structures (e.g., hash maps, linked lists, or trees),
- needs methods provided elsewhere and not available in R (e.g., other C or C++ libraries).

In the current chapter, we will demonstrate that R works very well as a user-friendly interface to compiled code.

As the topic is overall very technical, we will only cover the most important rudiments. We will focus on writing or interfacing *portable*[2] function libraries that only rely on *simple*[3] data structures (e.g., arrays of the type `double` and `int`). Thanks to this, we will be able to reuse them in other environments such as Python (e.g., via Cython) or Julia (remember that R is one of *many* languages out there). For those who are interested in specifics, the definitive reference is the *Writing R Extensions* manual [66], but see also Chapter 11 of [11]. Furthermore, R's source code provides many working examples of how to deal with R objects in C.

We assume some knowledge of the C language; see [39]. The reader can skip this

[1] Plain C and C++ are as fast as we can get without applying fancy CPU-specific optimisations or similar hacks. Fortran is also supported but will not be covered in this book because of its smaller popularity. Additionally, certain external packages are gateways to other languages, such as Java.

Nevertheless, D.E. Knuth once said: "The real problem is that programmers have spent far too much time worrying about efficiency in the wrong places and at the wrong times; premature optimisation is the root of all evil (or at least most of it) in programming" [40].

[2] Hence, we are not interested in the overall very convenient **Rcpp** or **cpp11** packages. They define C++ classes that make interacting with R objects more pleasant for some users.

[3] Thus, we will not discuss the ALTREP [56] representation of objects, ways to deal with environments or pairlists, etc.

chapter now and return to it later: the remaining material is not contingent on the current one. Otherwise, from now on, we take for granted that our environment can successfully build a source package with C code, as mentioned in Section 7.3.1. In particular, Win***s and m**OS users should install, respectively, `RTools` and `Xcode`.

**Note** To avoid ambiguity, in the main text, calls to C functions will be denoted by the "`C::`" prefix, e.g., `C::spanish_inquisition()`.

## 14.1 C and C++ code in R

### 14.1.1 Source files for compiled code in R packages

Perhaps the most versatile way to interact with portable C code is via standalone R packages; compare Section 7.3.1 and Section 9.2.2. For the purpose of the current chapter, we created a demo project available at https://github.com/gagolews/cpackagedemo.

**Exercise 14.1** *Inspect the structure of* ***cpackagedemo****. Note that C source files are located in the* `src/` *subdirectory. Build and install the package using* ***install.packages*** *or the "*`R CMD INSTALL`*" command. Then, load the package in R and call* ***my_sum*** *defined there on some numeric vector.*

The package provides an R interface to one C function, `C::my_c_sum`, written in the most portable fashion possible. Its declaration is included in the `src/cfuns.h` file:

```c
#ifndef __CFUNS_H
#define __CFUNS_H
#include <stddef.h>

double my_c_sum(const double* x, size_t n);

#endif
```

The function accepts a pointer to the start of a numeric sequence and its size, which is a standard[4] way of representing an array of `double`s. Its definition is given in `src/cfuns.c`. We see that it is nothing more than a simple sum of all the elements in an array:

```c
#include "cfuns.h"

```

[4] (*) A slightly more sophisticated representation (used, e.g., in GNU GSL and **numpy**) deals with a *sliced* array, where we additionally store the so-called *stride*. Instead of inspecting elements one after another, we advance the iterator by a given step size. This way, we could apply the same function on selected rows of a matrix (if it is in the column-major order).

```
/* computes the sum of all elements in an array x of size n */
double my_c_sum(const double* x, size_t n)
{
    double s = 0.0;
    for (size_t i = 0; i < n; ++i) {
        /* this code does not treat potential missing values specially
           (they are kinds of NaNs); to fix this, add:
        if (ISNA(x[i])) return NA_REAL;  // #include <R.h>   */
        s += x[i];
    }
    return s;
}
```

To make `C::my_c_sum` available in R, we have to introduce a wrapper around it that works with the data structures from the first part of this jolly book. We know that an R function accepts objects of *any* kind as input and yields *anything* as a result. In the next section, we will explain that we get access to R objects via special pointers of the type SEXP (S expressions). Thus, let us declare our R-callable wrapper in `src/rfuns.h`:

```
#ifndef __RFUNS_H
#define __RFUNS_H
#include <R.h>
#include <Rinternals.h>
#include <Rmath.h>

SEXP my_c_sum_wrapper(SEXP x);

#endif
```

The actual definition is included in `src/rfuns.c`:

```
#include "rfuns.h"
#include "cfuns.h"

/* a wrapper around my_c_sum callable from R */
SEXP my_c_sum_wrapper(SEXP x)
{
    double s;

    if (!Rf_isReal(x)) {
        /* the caller is expected to prepare the arguments
           (doing it at the C level is tedious work) */
        Rf_error("`x` should be a vector of the type 'double'");
    }

    s = my_c_sum(REAL(x), (size_t)XLENGTH(x));
```

```
    return Rf_ScalarReal(s);
}
```

The arguments could be, technically speaking, prepared at the C level. For instance, if `x` turned out to be an integer vector, we could have converted it to the `double` one (they are two different types; see Section 6.4.1). Nevertheless, overall, it is very burdensome. It is easier to use pure R code to ensure that the arguments are of the correct form as well as to beautify the outputs. This explains why we only assert the enjoyment of **`C::Rf_isReal`**`(x)`. It guarantees that the **`C::REAL`** and **`C::XLENGTH`** functions correctly return the pointer to the start of the sequence and its length, respectively.

Once **`C::my_c_sum`** is called, we must convert it to an R object so that it can be returned to our environment. Here, it is a newly allocated numeric vector of length one. We did this by calling **`C::Rf_ScalarReal`**.

Although optional (see Section 5.4 of [66]), we will register **`C::my_c_sum_wrapper`** as a callable function explicitly. This way, R will not be struggling to find the specific entry point in the resulting dynamically linked library (DLL). We do this in `src/cpackagedemo.c`:

```
#include <R_ext/Rdynload.h>
#include "rfuns.h"

/* the list of functions available in R via a call to .Call():
   each entry is like {exported_name, fun_pointer, number_of_arguments} */
static const R_CallMethodDef cCallMethods[] = {
    {"my_c_sum_wrapper", (DL_FUNC)&my_c_sum_wrapper, 1},
    {NULL, NULL, 0}  // the end of the list (sentinel)
};

/* registers the list of callable functions */
void R_init_cpackagedemo(DllInfo *dll)
{
    R_registerRoutines(dll, NULL, cCallMethods, NULL, NULL);
    R_useDynamicSymbols(dll, FALSE);
}
```

The function can be invoked from R using **`.Call`**. Here are the contents of `R/my_sum.R`:

```
my_sum <- function(x)
{
    # prepare input data:
    if (!is.double(x))
        x <- as.double(x)

    s <- .Call("my_c_sum_wrapper", x, PACKAGE="cpackagedemo")
```

```
    # some rather random postprocessing:
    attr(s, "what") <- deparse(substitute(x))
    s
}
```

And, finally, here is the package NAMESPACE file responsible for registering the exported R names and indicating the DLL to use:

```
export(my_sum)
useDynLib(cpackagedemo)
```

Once the package is built and installed (e.g., by running "R CMD INSTALL <pkgdir>" in the terminal or calling **install.packages**), we can test it by calling:

```
library("cpackagedemo")
my_sum(runif(100)/100)
## [1] 0.49856
## attr(,"what")
## [1] "runif(100)/100"
```

**Exercise 14.2** *Extend the package by adding a function to compute the index of the greatest element in a numeric vector. Note that C uses 0-based array indexing whereas in R, the first element is at index 1. Compare its run time against* ***which.max*** *using* ***proc.time****.*

### 14.1.2 R CMD SHLIB

The "R CMD SHLIB <files>" shell command compiles one or more source files without the need for turning them into standalone packages; see [66]. Then, **dyn.load** loads the resulting DLL.

**Exercise 14.3** *(*) Compile src/cfuns.c and src/rfuns.c from our demo package using "R CMD SHLIB". Call* ***dyn.load****. Write an R function that uses* ***.Call*** *to invoke* ***C::my_c_sum_wrapper*** *from the second source file.*

The direct SHLIB approach is convenient for learning C programming, including running simple examples. We will thus use it for didactic reasons in this chapter. The inst/examples/csource.R file in our demo package includes the implementation of an R function called **csource**. It compiles a given C source file, and loads the resulting DLL. It also extracts and executes a designated R code chunk preferably defining a function that refers to **.Call**.

Here is an example source file, inst/examples/helloworld.c in the **cpackagedemo** source code repository:

```
// the necessary header files are automatically included by `csource`

SEXP C_hello()
{
    Rprintf("The mill's closed. There's no more work. We're destitute.\n"
        "I'm afraid I've no choice but to sell you all "
        "for scientific experiments.\n");
    return R_NilValue;
}

/* R
# this chunk will be extracted and executed by `csource`.

hello <- function()
    invisible(.Call("C_hello", PACKAGE="helloworld"))

R */
```

Let's compile it and call the aforementioned R function.

```
source("~/R/cpackagedemo/inst/examples/csource.R")  # defines csource
csource("~/R/cpackagedemo/inst/examples/helloworld.c")
hello()
## The mill's closed. There's no more work. We're destitute.
## I'm afraid I've no choice but to sell you all for scientific experiments.
```

**Exercise 14.4** *(*) C++, which can be thought of as a superset of the C language (but the devil is in the detail), is also supported. Change the name of the aforementioned file to* `helloworld2.cpp`*, add* `extern "C"` *before the function declaration, pass* `PACKAGE="helloworld2"` *to* `.Call`*, and run* **`csource`** *on the new file.*

**Exercise 14.5** *(*) Verify that C and C++ source files can coexist in R packages.*

**Example 14.6** *(*) It might be very educative to study the implementation of* **`csource`**. *We should be able to author such functions ourselves now (a few hours' worth of work), let alone read with understanding.*

```
# Compiles a C or C++ source file using R CMD SHLIB,
# loads the resulting DLL, and executes the embedded R code.

csource <- function(
    fname,
    libname=NULL,  # defaults to the base name of `fname` without extension
    shlibargs=character(),
    headers=paste0(
        "#include <R.h>\n",
        "#include <Rinternals.h>\n",
        "#include <Rmath.h>\n"
    ),
    R=file.path(R.home(), "bin/R")
```

```
)
{
    stopifnot(file.exists(fname))
    stopifnot(is.character(shlibargs))
    stopifnot(is.character(headers))
    stopifnot(is.character(R), length(R) == 1)

    if (is.null(libname))
        libname <- regmatches(basename(fname),
            regexpr("[^.]*(?=\\..*)", basename(fname), perl=TRUE))

    stopifnot(is.character(libname), length(libname) == 1)

    # read the source file:
    f <- paste(readLines(fname), collapse="\n")

    # set up output file names:
    tmpdir <- normalizePath(tempdir(), winslash="/")  # tempdir on Win uses \
    dynlib_ext <- .Platform[["dynlib.ext"]]
    libpath <- file.path(tmpdir, sprintf("%s%s", libname, dynlib_ext))
    cfname <- file.path(tmpdir, basename(fname))
    rfname <- sub("\\..*?$", ".R", cfname, perl=TRUE)  # .R extension

    # separate the /* R ... <R code> ... R */ chunk from the source file:
    rpart <- regexec("(?smi)^/\\* R\\s?(.*)R \\*/$", f, perl=TRUE)[[1]]
    rpart_start <- rpart
    rpart_len <- attr(rpart, "match.length")
    if (rpart_start[1] < 0 || rpart_len[1] < 0)
        stop("enclose R code between /* R ... and ... R */")

    rcode <- substr(f, rpart_start[2], rpart_start[2]+rpart_len[2]-1)
    cat(rcode, file=rfname, append=FALSE)

    # write the C/C++ file:
    ccode <- paste(
        headers,
        substr(f, 1, rpart_start[1]-1),
        substr(f, rpart_start[1]+rpart_len[1], nchar(f)),
        collapse="\n"
    )
    cat(ccode, file=cfname, append=FALSE)

    # prepare the "R CMD SHLIB ..." command:
    shlibargs <- c(
        "CMD", "SHLIB",
        sprintf("-o %s", libpath),
        cfname,
        shlibargs
```

```
    )

    # compile and load the DLL, run the extracted R script:
    retval <- FALSE
    oldwd <- setwd(tmpdir)
    tryCatch({
        if (libpath %in% sapply(getLoadedDLLs(), `[[`, "path"))
            dyn.unload(libpath)
        stopifnot(system2(R, shlibargs) == 0)  # 0 == success
        dyn.load(libpath)
        source(rfname)
        retval <- TRUE
    }, error=function(e) {
        cat(as.character(e), file=stderr())
    })
    setwd(oldwd)

    if (!retval) stop("error compiling file or executing R code therein")
    invisible(TRUE)
}
```

## 14.2 Handling basic types

### 14.2.1 SEXPTYPEs

All R objects are stored as instances of the C language structure SEXPREC. Usually, we access them via pointers, which are of the type SEXP (S expression).

A C function referred to via `.Call` takes the very generic SEXPs as input. It outputs another SEXP. Importantly, one of the said structure's fields represents the actual R object type (SEXPTYPE numbers); see Table 14.1 for a selection.

Table 14.1. Basic R types in C.

| SEXPTYPE | Type in R (typeof) | Test in C |
|---|---|---|
| NILSXP | NULL | **Rf_isNull**(x) (*true for R_NilValue only*) |
| RAWSXP | raw | **TYPEOF**(x) == RAWSXP |
| LGLSXP | logical | **Rf_isLogical**(x) |
| INTSXP | integer | **Rf_isInteger**(x) |
| REALSXP | double | **Rf_isReal**(x) |
| CPLXSXP | complex | **Rf_isComplex**(x) |
| STRSXP | character | **Rf_isString**(x) |
| VECSXP | list | **Rf_isVectorList**(x) |
| CHARSXP | char (*scalar string; internal*) | **TYPEOF**(x) == CHARSXP |
| EXTPTRSXP | externalptr (*internal*) | **TYPEOF**(x) == EXTPTRSXP |

**Example 14.7** *To illustrate that any R object is available as a SEXP, consider the* `inst/examples/sexptype.c` *file from* **cpackagedemo**:

```
SEXP C_test_sexptype(SEXP x)
{
    Rprintf("type of x: %s (SEXPTYPE=%d)\n",
        Rf_type2char(TYPEOF(x)),
        (int)TYPEOF(x)
    );
    return R_NilValue;
}

/* R
test_sexptype <- function(x)
    invisible(.Call("C_test_sexptype", x, PACKAGE="sexptype"))
R */
```

*Example calls:*

```
csource("~/R/cpackagedemo/inst/examples/sexptype.c")
test_sexptype(1:10)
## type of x: integer (SEXPTYPE=13)
test_sexptype(NA)
## type of x: logical (SEXPTYPE=10)
test_sexptype("spam")
## type of x: character (SEXPTYPE=16)
```

*We should refer to particular SEXPTYPEs via their descriptive names (constants; e.g., STRSXP), not their numeric identifiers (e.g., 16); see Section 1.1 of [69] for the complete list*[5].

[5] `src/include/Rinternals.h` in R's source code repository; see, e.g., https://svn.r-project.org/R/trunk.

### 14.2.2 Accessing elements in simple atomic vectors

We have already seen an example function that processes a numeric vector; see **`C::my_c_sum_wrapper`** above. Table 14.2 gives other important vector-like `SEXPTYPE`s (atomic and generic), the C types of their elements, and the functions to access the underlying array pointers. It is also worth knowing that as call to **`C::XLENGTH`** returns the length of a given sequence. Let's stress that writing functions that accept `int` and `double` array pointers and their lengths makes them easily reusable in other programming environments. In many data analysis applications, we do not need much more.

Table 14.2. Basic array-like R types and their elements in C.

| SEXPTYPE | Array element type | Pointer access |
|---|---|---|
| `RAWSXP` | `typedef unsigned char Rbyte;` | **`RAW`**`(x)` |
| `LGLSXP` | `int` *(use the FALSE, TRUE, and NA_LOGICAL constants)* | **`LOGICAL`**`(x)` |
| `INTSXP` | `int` | **`INTEGER`**`(x)` |
| `REALSXP` | `double` | **`REAL`**`(x)` |
| `CPLXSXP` | `typedef struct { double r; double i; } Rcomplex;` | **`COMPLEX`**`(x)` |
| `STRSXP` | `SEXP` *(array of SEXPs of the type CHARSXP)* | *(not directly)* |
| `VECSXP` | `SEXP` *(array of SEXPs of any SEXPTYPE)* | *(not directly)* |
| `CHARSXP` | `const char*` *(read-only; trailing 0; check encoding)* | **`CHAR`**`(x)` |

**Important** With raw, logical, integer, floating-point, and complex vectors, we get direct access to data that might be shared amongst many objects (compare Section 16.1.4). `SEXPREC`s are simply passed by pointers (since `SEXP`s are exactly them). We must thus refrain[6] from modifying objects passed as function arguments. Ways to create new vectors, e.g., for storing auxiliary or return values, are discussed below.

**Example 14.8** *Consider* `inst/examples/sharedmem.c`*:*

```
SEXP C_test_sharedmem(SEXP x)
{
    if (!Rf_isReal(x) || XLENGTH(x) == 0)
        Rf_error("`x` should be a non-empty vector of the type 'double'");

    REAL(x)[0] = REAL(x)[0]+1;  // never do it; always make a copy;
            // the underlying array `x` may be shared by many objects

    return R_NilValue;
}

/* R
```

[6] (*) Unless we know what we are doing, e.g., we are certain that we deal with a local variable in an R function that invokes our **`.Call`**.

```
test_sharedmem <- function(x)
    invisible(.Call("C_test_sharedmem", x, PACKAGE="sharedmem"))
R */
```

*Calling the foregoing function on an example vector:*

```
csource("~/R/cpackagedemo/inst/examples/sharedmem.c")
y <- 1
z <- y
test_sharedmem(y)
print(c(y, z))
## [1] 2 2
```

*modifies y and z in place. Hence, to maintain the compatibility with the classic R semantics, we must always make a copy.*

### 14.2.3 Representation of missing values

Most languages do not support the notion of missing values out of the box. Therefore, in R, they have to be emulated. Table 14.3 lists the relevant constants and the conventional ways for testing for missingness.

Table 14.3. Representation of missing values.

| SEXPTYPE | Missing value | Testing |
|---|---|---|
| RAWSXP | *(none)* | *(none)* |
| LGLSXP | `NA_LOGICAL` *(equal to `INT_MIN`)* | `el == NA_LOGICAL` |
| INTSXP | `NA_INTEGER` *(equal to `INT_MIN`)* | `el == NA_INTEGER` |
| REALSXP | `NA_REAL` *(a special NaN)* | `ISNA(el)` |
| CPLXSXP | a pair of `NA_REAL`s | `ISNA(el.r)` |
| STRSXP | `NA_STRING` *(a CHARSXP object)* | `el == NA_STRING` |

In logical and integer vectors, NAs are represented as the smallest 32-bit signed integer. Thus, we need to be careful when performing any operations on these types: testing for missingness must be performed first.

The case of `double`s is slightly less irksome, for a missing value is represented as a special not-a-number. Many arithmetic operations on NaNs return NaNs as well, albeit there is no guarantee[7] that they will be of precisely the same type as `NA_REAL`. Thus, manual testing for missingness is also advised.

[7] (**) Namely, NAs are encoded as un-signalling NaNs `0x7ff00000000007A2` of the type `double` (the lower 32 payload bits are equal to 1954, decimally); see `src/arithmetic.c` in R's source code. The payload propagation is not fully covered by the current IEEE 754 floating point standard; see [23] for discussion. Reliance on such behaviour will thus make our code platform-dependent. R itself sometimes does that; theoretically, this may cause NAs to be converted to (other) NaNs.

**Example 14.9** *The `inst/examples/mean_naomit.c` file defines a function to compute the arithmetic mean of an `int` or a `double` vector:*

```
SEXP C_mean_naomit(SEXP x)
{
    double ret = 0.0;
    size_t k = 0;

    if (Rf_isInteger(x)) {
        const int* xp = INTEGER(x);
        size_t n = XLENGTH(x);
        for (size_t i=0; i<n; ++i)
            if (xp[i] != NA_INTEGER) {  // NOT: ISNA(xp[i])
                ret += (double)xp[i];
                k++;
            }
    }
    else if (Rf_isReal(x)) {
        const double* xp = REAL(x);
        size_t n = XLENGTH(x);
        for (size_t i=0; i<n; ++i)
            if (!ISNA(xp[i])) {  // NOT: xp[i] == NA_REAL
                ret += xp[i];
                k++;
            }
    }
    else
        Rf_error("`x` should be a numeric vector");

    return Rf_ScalarReal((k>0)?(ret/(double)k):NA_REAL);
}

/* R
mean_naomit <- function(x)
{
    if (!is.numeric(x))  # neither integer nor double
        x <- as.numeric(x)  # convert to double (the same as as.double)
    .Call("C_mean_naomit", x, PACKAGE="mean_naomit")
}
R */
```

*There is some inherent code duplication but `int` and `double` are distinct types. Thus, they need to be handled separately (we could have convert them to `double`s at the R level too). Some tests:*

```
csource("~/R/cpackagedemo/inst/examples/mean_naomit.c")
mean_naomit(c(1L, NA_integer_, 3L, NA_integer_, 5L))
## [1] 3
mean_naomit(rep(NA_real_, 10))
## [1] NA
```

**Exercise 14.10** *Implement **`all`** and **`any`** in C. Add the `na.rm` argument.*

### 14.2.4 Memory allocation

To allocate a new vector of length one and set its only element, we can call `C::ScalarLogical`, `C::ScalarInteger`, `C::ScalarReal`, etc. We have already used these functions for returning R "scalars". Vectors of arbitrary lengths can be created using `C::Rf_allocVector(sexptype, size)`. Note that this function does not initialise the elements of logical and numeric sequences (amongst others). They will need to be set manually after creation.

---

**Important** R implements a simple yet effective garbage collector that relies on reference counting. Occasionally[8], memory blocks that can no longer be reached are either freed or marked as reusable.

All allocated vectors must be manually protected from garbage collection. To guard against premature annihilation, R maintains a stack[9] of objects. `C::PROTECT(sexp)` pushes a given object pointer onto the top of the list. `C::UNPROTECT(n)` pops the last *n* elements from it in the last-in-first-out manner. At the end of a `.Call`, R checks if the number of protects matches that of unprotects and generates a warning if there is a stack imbalance.

Protection is *not* needed:

- for arguments to functions referred to by `.Call`, as they are already in use and hence protected;
- for objects assigned as list or character vectors' elements using `C::SET_VECTOR_ELT` and `C::SET_STRING_ELT` (see below); when the container is protected, so are its components;
- when we return the allocated vector to R immediately after creating it (like in `return Rf_ScalarReal(val)` in a C function invoked by `.Call`).

---

**Example 14.11** *Here is a function to compute the square of each element in a numeric vector. Note that the new vector must be protected from garbage collection while data are being prepared.*

```
SEXP C_square1(SEXP x)
{
    // no need to call PROTECT(x), it is already in use
    if (!Rf_isReal(x)) Rf_error("`x` should be of the type 'double'");

    size_t n = XLENGTH(x);
    const double* xp = REAL(x);

    SEXP y = PROTECT(Rf_allocVector(REALSXP, n));  // won't be GC'd
```

[8] A *safe* strategy is to assume that any call to a function from R's API may trigger the memory cleanup. On a side note, we may call the `gc` function in R to enforce rubbish removal. It also reports the current memory usage.

[9] (**) `C::R_PreserveObject` protects an arbitrary SEXP until `C::R_ReleaseObject` is called manually. With this mechanism, objects are not automatically released at the end of a `.Call`.

```
    double* yp = REAL(y);

    for (size_t i=0; i<n; ++i) {
        if (ISNA(xp[i])) yp[i] = xp[i];  // NA_REAL
        else             yp[i] = xp[i]*xp[i];
    }

    UNPROTECT(1);  // pops one object from the protect stack;
        // does not trigger garbage collection, so we can return `y` now
    return y;  // R will retrieve and protect it
}

/* R
square1 <- function(x)
{
    if (!is.double(x)) x <- as.double(x)
    .Call("C_square1", x, PACKAGE="square1")
}
R */
```

As an alternative, in this case, we may use **C::Rf_duplicate**:

```
SEXP C_square2(SEXP x)
{
    if (!Rf_isReal(x)) Rf_error("`x` should be of the type 'double'");

    x = PROTECT(Rf_duplicate(x));  // OK; just replaces the pointer (address)

    size_t n = XLENGTH(x);
    double* xp = REAL(x);
    for (size_t i=0; i<n; ++i)
        if (!ISNA(xp[i])) xp[i] = xp[i]*xp[i];

    UNPROTECT(1);
    return x;
}

/* R
square2 <- function(x)
{
    if (!is.double(x)) x <- as.double(x)
    .Call("C_square2", x, PACKAGE="square2")
}
R */
```

Some tests:

```
csource("~/R/cpackagedemo/inst/examples/square1.c")
```

```
square1(c(-2, -1, 0, 1, 2, 3, 4, NA_real_))
## [1]  4  1  0  1  4  9 16 NA
csource("~/R/cpackagedemo/inst/examples/square2.c")
square2(c(-2, -1, 0, 1, 2, 3, 4, NA_real_))
## [1]  4  1  0  1  4  9 16 NA
```

We can claim auxiliary memory from the heap during a function's runtime using the well-known `C::malloc` (or `new` in C++). We are of course fully responsible for releasing it via `C::free` (or `delete`).

**Example 14.12** *Here is our version of the `which` function.*

```
SEXP C_which1(SEXP x)
{
    if (!Rf_isLogical(x)) Rf_error("`x` should be of the type 'logical'");

    size_t n = XLENGTH(x), i, k;
    const int* xp = LOGICAL(x);

    size_t* d = (size_t*)malloc(n*sizeof(size_t));  // conservative size
    if (!d) Rf_error("memory allocation error");

    for (i=0, k=0; i<n; ++i)
        if (xp[i] != NA_LOGICAL && xp[i])
            d[k++] = i;

    // Rf_allocVector can longjmp, memory leak possible...
    SEXP y = PROTECT(Rf_allocVector(REALSXP, k));
    double* yp = REAL(y);  // yes, the type is double; ready for long vectors
    for (i=0; i<k; ++i)
        yp[i] = (double)d[i]+1;  // R uses 1-based indexing

    free(d);
    UNPROTECT(1);
    return y;
}

/* R
which1 <- function(x)
{
    if (!is.logical(x)) x <- as.logical(x)
    .Call("C_which1", x, PACKAGE="which1")
}
R */
```

*Some tests:*

```r
csource("~/R/cpackagedemo/inst/examples/which1.c")
which1(c(TRUE, FALSE, TRUE, NA, TRUE))
## [1] 1 3 5
```

**Exercise 14.13** *R's* `which` *returns either an* `int` *or a* `double` *vector, depending on the size of the input vector (whether it is shorter than* $2^{31}-1$*). Rewrite the above to take that into account: integer arithmetic is slightly faster.*

---

**Note** (*) R's exception handling uses a long jump[10]. Therefore, when calling `C::Rf_error` (whether directly or not) normal stack unwinding will not occur. This is particularly important when using C++ objects which deallocate memory in their destructors as they might not be invoked whatsoever.

In the preceding example, a call to `C::Rf_allocVector` may trigger a long jump, e.g., if we run out of available memory. In such a case, `d` will not be freed.

Thus, care should be taken to make sure there are no memory leaks. We can sometimes switch to `C::R_alloc(n, size)` which allocates `n*size` bytes. The memory it requests will automatically be garbage-collected at the end of a `.Call`.

Otherwise, we should ensure that blocks relying on manual memory allocation are not mixed with the calls to R API functions. In our `C::which1`, it would be better to determine the desired size of `y` and allocate it before calling `C::malloc`.

---

**Example 14.14** *(*) If we do not like that we are potentially wasting memory in the case of sparse logical vectors, we can rely on dynamically growable arrays. Below is a C++ rewrite of the foregoing function using* `deque` *(double-ended queue) from the language's standard library.*

```cpp
#include <deque>

extern "C" SEXP C_which2(SEXP x)
{
    if (!Rf_isLogical(x)) Rf_error("`x` should be of the type 'logical'");

    size_t n = XLENGTH(x), i, k=0;
    const int* xp = LOGICAL(x);

    // precompute k, Rf_allocVector can do a longjmp
    for (i=0; i<n; ++i) if (xp[i] != NA_LOGICAL && xp[i]) k++;
    SEXP y = PROTECT(Rf_allocVector(REALSXP, k));
    double* yp = REAL(y);  // ready for long vectors

    std::deque<size_t> d;  // allocates memory
    for (i=0; i<n; ++i)
        if (xp[i] != NA_LOGICAL && xp[i])
            d.push_back(i);
```

[10] https://en.wikipedia.org/wiki/Setjmp.h

```
    i=0;
    for (size_t e : d)
        yp[i++] = (double)e+1;  // R uses 1-based indexing

    UNPROTECT(1);
    return y;  // d's destructor will be called automatically
}

/* R
which2 <- function(x)
{
    if (!is.logical(x)) x <- as.logical(x)
    .Call("C_which2", x, PACKAGE="which2")
}
R */
```

*Example calls:*

```
csource("~/R/cpackagedemo/inst/examples/which2.cpp")
x <- (runif(10) > 0.5)
stopifnot(which(x) == which1(x))
stopifnot(which(x) == which2(x))
```

*Alternatively, we could have used* **`C::realloc`** *to extend an initially small buffer created using* **`C::malloc`** *by, say, 50% whenever it is about to overflow.*

### 14.2.5 Lists

For safety reasons[11], we do not get access to the underlying pointers in lists and character vectors. List items can be read by calling **`C::VECTOR_ELT`**`(x, index)` and can be set with **`C::SET_VECTOR_ELT`**`(x, index, newval)`. Note that lists (VECSXPs) are comprised of SEXPs of any type. Hence, after extracting an element, its SEXPTYPE needs to be tested using one of the functions listed in Table 14.1. This can be tiresome.

**Example 14.15** *Here is a rather useless function that fetches the first and the last element in a given numeric vector or a list. However, if the latter case, we apply the function recursively on all its elements.*

```
SEXP C_firstlast(SEXP x)
{
    if (!Rf_isVector(x) || XLENGTH(x) == 0)
        Rf_error("`x` must be a non-empty vector (atomic or generic)");
    else if (Rf_isReal(x)) {
        SEXP y = PROTECT(Rf_allocVector(REALSXP, 2));
```

[11] To get the object reference counting right, **`C::SET_VECTOR_ELT`** needs to unprotect the old element and start protecting the new one.

```
        REAL(y)[0] = REAL(x)[0];             // first
        REAL(y)[1] = REAL(x)[XLENGTH(x)-1];  // last
        UNPROTECT(1);
        return y;
    }
    else if (Rf_isVectorList(x)) {
        SEXP y = PROTECT(Rf_allocVector(VECSXP, 2));
        // VECTOR_ELT(x, i) is PROTECTed by the container;
        // SET_VECTOR_ELT does not trigger GC; no need to call PROTECT
        // on the result of C_firstlast(...) in this context
        SET_VECTOR_ELT(y, 0, C_firstlast(VECTOR_ELT(x, 0)));
        SET_VECTOR_ELT(y, 1, C_firstlast(VECTOR_ELT(x, XLENGTH(x)-1)));
        UNPROTECT(1);
        return y;
    }
    else
        Rf_error("other cases left as an exercise");

    return R_NilValue;  // avoid compiler warning
}

/* R
firstlast <- function(x)
    .Call("C_firstlast", x, PACKAGE="firstlast")
R */
```

*Testing:*

```
csource("~/R/cpackagedemo/inst/examples/firstlast.c")
firstlast(c(1, 2, 3))
## [1] 1 3
firstlast(list(c(1, 2, 3), c(4, 5), 6))
## [[1]]
## [1] 1 3
##
## [[2]]
## [1] 6 6
firstlast(list(c(1, 2, 3), 4, 5, list(6, c(7, 8), c(9, 10, 11))))
## [[1]]
## [1] 1 3
##
## [[2]]
## [[2]][[1]]
## [1] 6 6
##
## [[2]][[2]]
## [1]  9 11
```

**Exercise 14.16** *Implement a C function that returns the longest vector in a given list. Use*

*`C::Rf_isVector` to check whether a given object is an atomic or a generic vector, and hence if `C::XLENGTH` can be called thereon.*

**Exercise 14.17** *Inscribe your version of `unlist`. Consider scanning the input list twice. First, determine the size of the output vector. Second, fill the return object with the un-listed values.*

**Exercise 14.18** *Write a C function that takes a list of numeric vectors of identical lengths. Return their elementwise sum: the first element of the output should be the sum of the first elements in every input vector, and so forth.*

### 14.2.6 Character vectors and individual strings (*)

Character vectors (`STRSXP`s) are similar to `VECSXP`s except that they only carry individual strings which are represented using a separate data type, `CHARSXP`. **`C::STRING_ELT`**`(x, index)` and **`C::SET_STRING_ELT`**`(x, index, newval)` play the role of the element getters and setters.

---

**Important** If we are not interested in text processing but rather in handling *categorical* data or defining grouping variables, we should consider converting character vectors to *factors* before issuing a `.Call`. Comparing small integers is much faster than strings; see below for more details.

---

Because of R's string cache, there are no duplicate strings in the memory. However, this feature could only be guaranteed by making data in `CHARSXP`s *read-only*. We can access the underlying `const char*` pointer by calling **`C::CHAR`**`(s)`. As typical in C, a string is terminated by byte 0.

---

**Note** R strings may be of different encodings; compare Section 6.1.1. For portability and peace of mind, it is best to preprocess the arguments to `.Call` using **`enc2utf8`**, which converts all strings to UTF-8[12].

Despite being the most universal encoding, UTF-8 does not represent each code point using a fixed number of bytes. For instance, computing the string length requires iterating over all its elements. For `CHARSXP`s, **`C::XLENGTH`** returns the number of *bytes*, not including the trailing 0.

It is thus best to leave the processing of strings to the dedicated libraries, e.g., ICU[13] or rely on functions from the **`stringi`** package [27] at the R level.

---

C strings can be converted to `CHARSXP`s by calling **`C::Rf_mkCharCE`**`(stringbuf, CE_UTF8)` or **`C::Rf_mkCharLenCE`**`(stringbuf, buflen, CE_UTF8)`. If we are sure that a string is in ASCII (a subset of UTF-8), we can also call **`C::Rf_mkChar`**`(stringbuf)`.

---

[12] Take care when calling `C::Rprintf`, though. It should only be used to output messages in the *native* encoding, which does not necessarily have to be UTF-8, although this landscape is slowly changing. Sticking to ASCII is a safe choice.

[13] https://icu.unicode.org/

We should never return CHARSXPs as results to R. They are for internal use only. They must be wrapped inside a character vector, e.g., using **C::Rf_ScalarString**.

### 14.2.7 Calling R functions from C (**)

Section 5.11 of [66] discusses ways to call R functions in C. To understand them, we will first need to study the material from the remaining chapters of our book, i.e., environments and the related evaluation model. They can be useful, e.g., when calling optimisation algorithms implemented in C on objective functions written in R.

### 14.2.8 External pointers (**)

For storing arbitrary C pointers, there is a separate basic R type named `externalptr` (SEXPTYPE of EXTPTRSXP); see Section 5.13 of [66] for more details. We can use them to maintain dynamic data structures or resource handlers between calls to R functions. The problem with these is that pointers are passed as… pointers. They can easily break R's pass-by-value-like semantics, where changes to the state of the referenced object will be visible outside the call.

**Example 14.19** *(**) `inst/examples/stack.cpp` provides a C++ implementation of the stack data structure, being a last-in-first-out container of arbitrary R objects:*

```
#include <deque>

class S : public std::deque<SEXP>
{
    public: ~S()
    {   // destructor: release all SEXPs so that they can be GC'd
        while (!this->empty()) {
            SEXP obj = this->front();
            this->pop_front();
            R_ReleaseObject(obj);
        }
    }
};

S* get_stack_pointer(SEXP s, bool check_zero=true)  // internal function
{
    if (TYPEOF(s) != EXTPTRSXP)
        Rf_error("not an external pointer");

    SEXP tag = R_ExternalPtrTag(s);  // our convention, this can be anything
    if (TYPEOF(tag) != CHARSXP || strcmp(CHAR(tag), "stack") != 0)
        Rf_error("not a stack");

    S* sp = (S*)R_ExternalPtrAddr(s);
    if (check_zero && !sp)
        Rf_error("address is 0");
```

```
    return sp;
}

void stack_finaliser(SEXP s)  // internal function
{
    // called during garbage collection
    S* sp = get_stack_pointer(s, false);
    if (sp) {
        delete sp;  // destruct S, release SEXPs
        R_ClearExternalPtr(s);
    }
}

extern "C" SEXP C_stack_create()
{
    S* sp = new S();  // stack pointer
    SEXP s = PROTECT(
        R_MakeExternalPtr((void*)sp, /*tag*/mkChar("stack"), R_NilValue)
    );
    R_RegisterCFinalizerEx(s, stack_finaliser, TRUE);  // auto-called on GC
    UNPROTECT(1);
    return s;
}

extern "C" SEXP C_stack_empty(SEXP s)
{
    S* sp = get_stack_pointer(s);
    return Rf_ScalarLogical(sp->empty());
}

extern "C" SEXP C_stack_push(SEXP s, SEXP obj)
{
    S* sp = get_stack_pointer(s);
    R_PreserveObject(obj);
    sp->push_front(obj);
    return R_NilValue;
}

extern "C" SEXP C_stack_pop(SEXP s)
{
    S* sp = get_stack_pointer(s);
    if (sp->empty())
        Rf_error("stack is empty");
    SEXP obj = sp->front();
    sp->pop_front();
    R_ReleaseObject(obj);
    return obj;
```

```
}

/* R
stack_create <- function()
    .Call("C_stack_create", PACKAGE="stack")

stack_empty <- function(s)
    .Call("C_stack_empty", s, PACKAGE="stack")

stack_push <- function(s, obj)
    .Call("C_stack_push", s, obj, PACKAGE="stack")

stack_pop <- function(s)
    .Call("C_stack_pop", s, PACKAGE="stack")
R */
```

*Note how we preserve R objects from garbage collection. Some tests:*

```
csource("~/R/cpackagedemo/inst/examples/stack.cpp")
s <- stack_create()
print(s)
## <pointer: 0x55b2ec069360>
typeof(s)
## [1] "externalptr"
for (i in c("one", "two", "Spanish Inquisition"))
    stack_push(s, i)
while (!stack_empty(s))
    print(stack_pop(s))
## [1] "Spanish Inquisition"
## [1] "two"
## [1] "one"
```

Note that pointers are not serialisable. They cannot be saved for use in another R session.

## 14.3 Dealing with compound types

### 14.3.1 Reading and setting attributes

From Chapter 10, we know that compound types such as matrices, factors, or data frames are represented using basic data structures. Usually, they are atomic vectors or lists organised in a predefined manner.

`C::Rf_getAttrib(x, attrname)` and `C::Rf_setAttrib(x, attrname, newval)` gets and sets specific attributes of an object `x`. Their second argument, `attrname`, should be a one-element `STRSXP`. For convenience, the `R_ClassSymbol`, `R_DimNamesSymbol`,

R_DimSymbol, R_NamesSymbol, R_RowNamesSymbol, and R_LevelsSymbol constants can be used instead of the STRSXP versions of the "class", "dimnames", "dim", "names", "row.names", and "levels" strings.

**Example 14.20** *Consider a function for testing whether an object is of a given class:*

```
#include <string.h>

SEXP C_isofclass(SEXP x, SEXP class)
{
    if (!Rf_isString(class) && XLENGTH(class) != 1)
        Rf_error("`class` must be a single string");

    if (!Rf_isObject(x))  // is the class attribute set at all?
        return Rf_ScalarLogical(FALSE);

    SEXP xclass = Rf_getAttrib(x, R_ClassSymbol);  // STRSXP (guaranteed)
    const char* c = CHAR(STRING_ELT(class, 0));  // class arg as a C string
    size_t n = XLENGTH(xclass);
    for (size_t i=0; i<n; ++i)
        if (strcmp(CHAR(STRING_ELT(xclass, i)), c) == 0)
            return Rf_ScalarLogical(TRUE);

    return Rf_ScalarLogical(FALSE);
}

/* R
isofclass <- function(x, class)
    .Call("C_isofclass", x, class, PACKAGE="isofclass")
R */
```

*Some tests:*

```
csource("~/R/cpackagedemo/inst/examples/isofclass.c")
isofclass(Sys.time(), "POSIXct")
## [1] TRUE
isofclass(cbind(1:5, 11:15), "matrix")
## [1] FALSE
```

*Note that a matrix has an* implicit *class (reported by the* **class** *function), but its* class *attribute does not have to be set. Hence the negative result.*

**Example 14.21** *Write a function that fetches a particular* named *element in a list.*

### 14.3.2 Factors

Factors (Section 10.3.2) are represented as integer vectors with elements in the set {1, 2, ..., $k$, NA_integer_} for some $k$. They are equipped with the levels attribute, being a character vector of length $k$. Their class attribute is set to factor.

**Example 14.22** *An example implementation of a function to compute the number of occurrences of each factor level is given below.*

```
SEXP C_table1(SEXP x)
{
    if (!Rf_isFactor(x)) Rf_error("`x` is not a 'factor' object");

    size_t n = XLENGTH(x);
    const int* xp = INTEGER(x);  // `x` is INTSXP

    SEXP levels = Rf_getAttrib(x, R_LevelsSymbol);  // `levels` is a STRSXP
    size_t k = XLENGTH(levels);

    SEXP y = PROTECT(Rf_allocVector(REALSXP, k));
    double* yp = REAL(y);
    for (size_t i=0; i<k; ++i)
        yp[i] = 0.0;
    for (size_t j=0; j<n; ++j) {
        if (xp[j] != NA_INTEGER) {
            if (xp[j] < 1 || xp[j] > k)
                Rf_error("malformed factor");  // better safe than sorry
            yp[xp[j]-1] += 1.0;  // levels are 1..k, but C needs 0..k-1
        }
    }

    Rf_setAttrib(y, R_NamesSymbol, levels);  // set names attribute
    UNPROTECT(1);
    return y;
}

/* R
table1 <- function(x)
{
    if (!is.factor(x)) x <- as.factor(x)
    .Call("C_table1", x, PACKAGE="table1")
}
R */
```

*Testing:*

```
csource("~/R/cpackagedemo/inst/examples/table1.c")
table1(c("spam", "bacon", NA, "spam", "eggs", "bacon", "spam", "spam"))
## bacon  eggs  spam
##     2     1     4
```

**Exercise 14.23** *Create a function to compute the most frequently occurring value (mode) in a given factor. Return a character vector. If a mode is ambiguous, return all the possible candidates.*

### 14.3.3 Matrices

Matrices (Chapter 11) are *flat* atomic vectors or lists with the `dim` attribute being an integer vector of length two. The `class` attribute does not have to be set (but the **`class`** function returns `matrix` and `array`). Matrices are so important in data analysis that they have been blessed with a few dedicated functions available at the C level. **`C::Rf_isMatrix`** tests if a given object meets the criteria mentioned above. **`C::Rf_allocMatrix`**`(sexptype, nrows, ncols)` allocates a new matrix.

R relies on the Fortran order of matrix elements, i.e., it uses the column-major alignment. Let `A` be a matrix with $n$ rows and $m$ columns (compare **`C::Rf_nrows`** and **`C::Rf_ncols`**). Then, the element in the $i$-th row and the $j$-th column is at `A[i+n*j]`. The `dimnames` attributes must be handled manually, though.

**Example 14.24** *Here is a function to compute the transpose of a numeric matrix:*

```
SEXP C_transpose(SEXP x)
{
    if (!Rf_isMatrix(x) || !Rf_isReal(x))
        Rf_error("`x` must be a real matrix");

    size_t n = Rf_nrows(x);
    size_t m = Rf_ncols(x);
    const double* xp = REAL(x);

    SEXP y = PROTECT(Rf_allocMatrix(REALSXP, m, n));
    double* yp = REAL(y);
    for (size_t i=0; i<n; ++i)
        for (size_t j=0; j<m; ++j)
            yp[j+m*i] = xp[i+n*j];

    SEXP dimnames = Rf_getAttrib(x, R_DimNamesSymbol);
    if (!Rf_isNull(dimnames)) {
        SEXP tdimnames = PROTECT(Rf_allocVector(VECSXP, 2));
        SET_VECTOR_ELT(tdimnames, 0, VECTOR_ELT(dimnames, 1));
        SET_VECTOR_ELT(tdimnames, 1, VECTOR_ELT(dimnames, 0));
        Rf_setAttrib(y, R_DimNamesSymbol, tdimnames);  // set dimnames
        UNPROTECT(1);
        // dimnames may have the names attribute too (left as an exercise)
    }

    UNPROTECT(1);
    return y;
}

/* R
transpose <- function(x)
{
    if (!is.matrix(x)) x <- as.matrix(x)
    if (!is.double(x)) x[] <- as.double(x)  # preserves attributes
```

```
    .Call("C_transpose", x, PACKAGE="transpose")
}
R */
```

*Testing:*

```
csource("~/R/cpackagedemo/inst/examples/transpose.c")
transpose(cbind(c(1, 2, 3, 4), c(5, 6, 7, 8)))
##      [,1] [,2] [,3] [,4]
## [1,]    1    2    3    4
## [2,]    5    6    7    8
transpose(Titanic[, "Male", "Adult", ])
##     1st 2nd 3rd Crew
## No  118 154 387  670
## Yes  57  14  75  192
```

**Exercise 14.25** *Author a C function named **`table2`** that computes a two-way contingency table.*

### 14.3.4 Data frames

Data frames (Chapter 12) are lists of the S3 class `data.frame` featuring, for some $n$ and $m$, $m$ vectors of identical lengths $n$ or matrices of $n$ rows. The character vectors stored in the `row.names` and `names` attributes give the $n$ row and $m$ column labels.

We process data frames as ordinary lists. However, assuming we only want to process numeric data, we can extract the quantitative columns and put them inside a matrix *at the R level*. If element grouping is required, they can be accompanied by a factor or a list of factor variables. In many applications, this strategy is good enough.

## 14.4 Using existing function libraries

### 14.4.1 Checking for user interrupts

Long computations may lead to R's becoming unresponsive. The user may always request to cancel the evaluation of the current expression by pressing `Ctrl+C`. To allow the processing of the event queue, we should call **`C::R_CheckUserInterrupt()`** occasionally, e.g., in every iteration of a complex `for` loop. Note that R might decide never to return to our function. Thus, we have to prevent memory leaks, e.g., by preferring **`C::R_alloc`** over **`C::malloc`**.

### 14.4.2 Generating pseudorandom numbers

`C::unif_rand` returns a single pseudorandom deviate from the uniform distribution on the unit interval. It is the basis for generating numbers from all other supported distributions (Section 6.7.1 of [66]). It uses the same pseudorandom generator as we described in Section 2.1.5. To read and memorise its seed (the `` `.Random.seed` `` object in the global environment), we have to call `C::GetRNGstate()` and `C::PutRNGstate()` at the beginning and the end of our function, respectively.

**Example 14.26** *Below is a function to generate a pseudorandom bit sequence:*

```
SEXP C_randombits(SEXP n)
{
    if (!Rf_isInteger(n) || XLENGTH(n) != 1)
        Rf_error("`n` should be a single integer");

    int _n = INTEGER(n)[0];
    if (_n == NA_INTEGER || _n < 1)
        Rf_error("incorrect `n`");

    SEXP y = PROTECT(Rf_allocVector(INTSXP, _n));
    int* yp = INTEGER(y);

    GetRNGstate();
    for (int i=0; i<_n; ++i)
        yp[i] = (int)(unif_rand()>0.5);  // not the best way to sample bits
    PutRNGstate();

    UNPROTECT(1);

    return y;
}

/* R
randombits <- function(n)
{
    if (!is.integer(n)) n <- as.integer(n)
    .Call("C_randombits", n, PACKAGE="randombits")
}
R */
```

*Let's play around with it:*

```
csource("~/R/cpackagedemo/inst/examples/randombits.c")
set.seed(123); randombits(10)
##  [1] 0 1 0 1 1 0 1 1 1 0
randombits(10)
##  [1] 1 0 1 1 0 1 0 0 0 1
set.seed(123); randombits(20)
##  [1] 0 1 0 1 1 0 1 1 1 0 1 0 1 1 0 1 0 0 0 1
```

```
set.seed(123); as.integer(runif(20)>0.5)  # it's the same "algorithm"
##  [1] 0 1 0 1 1 0 1 1 1 0 1 0 1 1 0 1 0 0 0 1
```

**Exercise 14.27** *Create a function to compute the most frequently occurring value (mode) in a given factor object. In the case of ambiguity, return a randomly chosen candidate.*

### 14.4.3 Mathematical functions from the R API

Section 6.7 of [66] lists the available statistical distribution functions, mathematical routines and constants, and other numerical utilities.

### 14.4.4 Header files from other R packages (*)

A package may use header files from another package. For this to be possible, it must include the dependency name in the `LinkingTo` field of its `DESCRIPTION` file; see [66] for discussion.

**Exercise 14.28** *The **BH** package on CRAN gives access to Boost, the header-only C++ libraries that define many useful algorithms and data structures. Create an R package that calls **C++::boost::math::gcd** after issuing the `#include <boost/math/common_factor.hpp>` directive.*

### 14.4.5 Specifying compiler and linker flags (**)

We can pass arbitrary flags to the compiler or linker, e.g., to use any library installed on our system. Basic configuration can be tweaked via `Makevars` (or `Makevars.win` on Win***s), e.g., by setting `PKG_CFLAGS` or `PKG_LIBS` variables. For maximum portability across different platforms, which is overall challenging to ensure when we do not wish to exclude Win***s users, we might be required to author custom `configure` (and `configure.win`) scripts. For more information, see [66] which discusses, amongst others, how to use OpenMP[14] in our projects.

## 14.5 Exercises

**Exercise 14.29** *Answer the following questions about the C language API for R.*

- *What are the most common SEXPTYPEs?*

[14] Most R functions are single-threaded by design. It is up to the user to decide whether and how they would like their code to be parallelised. More often than not, computations in the data science domain are naïvely parallelisable (e.g., Monte Carlo simulations, exhaustive grid search, etc.). In such cases, the R package **parallel** might be helpful: it defines parallel versions of **lapply** and **apply**. However, for serious jobs, running *multiple* single-threaded R instances via, e.g., the **slurm** workload manager might be a better idea than starting a process that spawns many child threads.

- *How are missing values represented?*
- *How can we check if an `int` is a missing value? What about a `double`?*
- *How to prevent SEXPs from being garbage-collected?*
- *How are character vectors represented? What is the difference between a CHARSXP and a STRSXP?*
- *Why is it better to handle factor objects rather than raw character vectors if we merely would like to define grouping variables?*
- *How are R matrices represented in C? Does R use the C or Fortran order of matrix elements?*
- *How are R data frames handled in C?*

**Exercise 14.30** *Implement the C versions of the* **`rep`**, **`seq`**, **`rle`**, **`match`**, **`findInterval`**, **`sample`**, **`order`**, **`unique`**, *and* **`split`** *functions.*

**Exercise 14.31** *(*) Read* Writing R Extensions *[66] in its entirety.*

**Exercise 14.32** *(*) Download R's source code from CRAN[15] or its* **`Subversion`**[16] *(SVN) repository. Explore the header files in the `src/include` subdirectory. They are part of the callable API.*

[15] https://stat.ethz.ch/R/daily
[16] https://svn.r-project.org/R/trunk

# 15

# *Unevaluated expressions (*)*

In this and the remaining chapters, we will learn some hocus-pocus that should only be of interest to the advanced-to-be[1] and open-minded R programmers who would like to understand what is going on under our language's bonnet. In particular, we will inspect the mechanisms behind why certain functions act differently from what we would expect them to do if a *standard* evaluation scheme was followed (compare **subset** and **transform** mentioned in Section 12.3.9). Namely, in *normal* programming languages, when we execute something like:

```
plot(x, exp(x))
```

the expression **exp**(x) is evaluated *first*. Then, and only then, its value[2] (in this case: probably a numeric vector) is passed to the **plot** function as an actual parameter. Thus, if x becomes **seq**(0, 10, length.out=1001), the above never means anything else than:

```
plot(c(0.00, 0.01, 0.02, 0.03, ...), c(1.0000, 1.0101, 1.0202, 1.0305, ...))
```

But R was heavily inspired by the S language from which it has taken the notion of lazy arguments (Chapter 17). It is thus equipped with the ability to apply a set of techniques referred to as *metaprogramming* (computing on the language, reflection). With it, we can define functions that do not take their arguments for granted and clearly see the code fragments passed to them. Having access to such *unevaluated expressions*, we can do to them whatever we please: print, modify, evaluate on different data, or ignore whatsoever.

In theory, this enables implementing many *potentially helpful*[3], beginner-friendly features and express certain requests in a more concise manner. For instance, that the y-axis labels in Figure 2.2 could be generated automatically is precisely because **plot** was able to see not only a vector like **c**(1.0000, 1.0101, 1.0202, 1.0305, ...) but also the expression that generated it, **exp**(x).

[1] Remember that this book is supposed to be read from the beginning to the end. Also, if you have not tested yourself against all the 300-odd exercises suggested so far, please do it before proceeding with the material presented here. Only practice makes perfect, and nothing is built in a day. Give yourself time: you can always come back later.

[2] Or a reference/pointer to an object that stores the said value.

[3] The original authors of R (R. Ihaka and R. Gentleman), in [38], mention: "A policy of *lazy arguments* is very useful because it means that, in addition to the value of an argument, its symbolic form can be made available in the function being called. This can be very useful for specifying functions or models in symbolic form."

Nonetheless, as a form of *untamed freedom of expression*[4], metaprogramming has the endless potential to arouse chaos, confusion, and division in the user community. In particular, we can introduce a dialect *within* our language that people outside our circle will not be able to understand. Once it becomes a dominant one, other users will feel excluded.

Cursed be us, for we are about to start eating from the tree of the knowledge of good and evil. But remember: with great power comes great fun (and responsibility).

## 15.1 Expressions at a glance

At the most general level, *expressions* (statements) in a language like R can be classified into two groups:

- *simple expressions*:
  - *constants* (e.g., `3.14`, `2i`, `42L`, `NA_real_`, `Inf`, `NaN`, `NA`, `FALSE`, `TRUE`, `"character string"`, `NULL`, `-1.3e-16`, and `0x123abc`),
  - *names* (symbols, identifiers; e.g., `x`, `iris`, `sum`, `data.frame`, `spam`, `` `+` ``, `` `[<-` ``, and `spanish_inquisition`),
- *compound expressions* – combinations of $n+1$ expressions (simple or compound) of the form:

$$(f, e_1, e_2, \dots, e_n).$$

As we will soon see, compound expressions represent a *call* to $f$ (an *operator*) on a sequence of arguments $e_1, e_2, \dots, e_n$ (*operands*). It is why, equivalently, we denote them by $f(e_1, e_2, \dots, e_n)$.

On the other hand, *names* have no meaning without an explicitly stated context, which we will explore in Chapter 16. Prior to that, we treat them as *meaning-less*. Hence, for the time being, we are only interested in the *syntax* or *grammar* of our language, not the *semantics*. We are abstract in the sense that, in the expression `mean(rates)+2`, which we know from Section 9.3.5 that we can equivalently articulate as `` `+`(mean(rates), 2) ``, neither `mean`, `x`, nor even `` `+` `` has the *usual* sense. Therefore, we should consider it the same as, say, `f(g(x), 2)` or `nobody(expects(spanish_inquisition), 2)`.

[4] In the current author's opinion, R (as a whole, in the sense of *R (GNU S) as a language and an environment*) would be better off if an ordinary programmer was not exposed so much to functions heavily relying on metaprogramming. A healthy user can perfectly manage without (and thus refrain from using) them. The fact that we call them *advanced* will not make us *cool* if we start horsing around with nonstandard evaluation. *Perverse* is perhaps a better label.

## 15.2 Language objects

There are three types of *language objects* in R:

- `name` (`symbol`) represents object names in the sense of *simple expressions: names* in Section 15.1;
- `call` stores unevaluated function calls in the sense of *compound expressions* above;
- `expression`, quite confusingly, represents a *sequence* of simple or compound expressions (constants, names, or calls).

One way to create a simple or compound expression is by *quoting*, where the R interpreter is asked to refrain from evaluating a given command:

```
quote(spam)  # name (symbol)
## spam
quote(f(x))  # call
## f(x)
quote(1+2+3*pi)  # another call
## 1 + 2 + 3 * pi
```

None of the foregoing was executed. In particular, `spam` has no *sense* in the current context (whichever that is). It is not the meaning that we are now after.

Single strings can be converted to names by calling:

```
as.name("spam")
## spam
```

Calls can be built programmatically by invoking:

```
call("sin", pi/2)
## sin(1.5707963267949)
```

Sometimes we would rather have the arguments quoted:

```
call("sin", quote(pi/2))
## sin(pi/2)
call("c", 1, exp(1), quote(exp(1)), pi, quote(pi))
## c(1, 2.71828182845905, exp(1), 3.14159265358979, pi)
```

Objects of the type `expression` can be thought of as list-like sequences that consist of simple or compound expressions.

```
(exprs <- expression(1, spam, mean(x)+2))
## expression(1, spam, mean(x) + 2)
```

All arguments were quoted. We can select or subset the individual statements using the extraction or index operators:

```
exprs[-1]
## expression(spam, mean(x) + 2)
exprs[[3]]
## mean(x) + 2
```

**Exercise 15.1** *Check the type of the object returned by a call to* `c(1, "two", sd, list(3, 4:5), expression(3+3))`.

There is also an option to *parse* a given text fragment or a whole source file:

```
parse(text="mean(x)+2")
## expression(mean(x) + 2)
parse(text="  # two code lines (comments are ignored by the parser)
    x <- runif(5, -1, 1)
    print(mean(x)+2)
")
## expression(x <- runif(5, -1, 1), print(mean(x) + 2))
parse(text="2+")  # syntax error - unfinished business
## Error in `parse()`: ! <text>:2:0: unexpected end of input 1: 2+ ^
```

---

**Important** The **`deparse`** function converts language objects to character vectors, e.g.:

```
deparse(quote(mean(x+2)))
## [1] "mean(x + 2)"
```

This function has the nice side effect of tidying up the code formatting:

```
exprs <- parse(text=
    "`+`(x, 2)->y; if(y>0) print(y**10|>log()) else { y<--y; print(y)}")
```

Let's print them out:

```
for (e in exprs)
    cat(deparse(e), sep="\n")
## y <- x + 2
## if (y > 0) print(log(y^10)) else {
##     y <- -y
##     print(y)
## }
```

---

---

**Note** Calling **`class`** on objects of the three aforementioned types yields `name`, `call`, and `expression`, whereas **`typeof`** returns `symbol`, `language`, and `expression`, respectively.

---

## 15.3 Calls as combinations of expressions

We have mentioned that calls (compound expressions) are combinations of simple or compound expressions of the form $(f, e_1, \ldots, e_n)$. The first expression on the list, denoted above by $f$, plays a special role. This is precisely seen in the following examples:

```
as.call(expression(f, x))
## f(x)
as.call(expression(`+`, 1, x))  # `+`(1, x)
## 1 + x
as.call(expression(`while`, i < 10, i <- i + 1))
## while (i < 10) i <- i + 1
as.call(expression(function(x) x**2, log(exp(1))))
## (function(x) x^2)(log(exp(1)))
as.call(expression(1, x, y, z))  # utter nonsense, but syntactically valid
## 1(x, y, z)
```

Recall from Section 9.3 that operators and language constructs such as `if` and `while` are ordinary functions. Furthermore, keyword arguments to a call result in the underlying sequence's being named:

```
expr <- quote(f(1+2, a=1, b=2))
length(expr)  # three arguments -> length-4 sequence
## [1] 4
names(expr)  # NULL if no arguments are named
## [1] ""  ""  "a" "b"
```

### 15.3.1 Browsing parse trees

Square brackets give us access to the individual expressions constituting an object of the type `call`. For example:

```
expr <- quote(1+x)
expr[[1]]
## `+`
expr[c(1, 3, 2)]
## x + 1
expr[c(2, 3, 1, 3)]
## 1(x, `+`, x)
```

A compound expression was defined recursively: it may consist of other compound expressions. For instance, a statement:

```
expr <- quote(
    while (i < 10) {
        cat("i = ", i, "\n", sep="")
```

```
        i <- i+1
    }
)
```

can be rewritten[5] using the $f(\ldots)$ notation like:

```
quote(
    `while`(
        `<`(i, 10),
        `{`(cat("i = " , i, "\n", sep=""), `<-`(i, `+`(i, 1)))
    )
)
```

We can dig into all the subexpressions using a series of extractions:

```
expr[[2]][[1]]  # expr[[c(2, 1)]]
## `<`
expr[[3]][[3]][[3]]  # expr[[c(3, 3, 3)]]
## i + 1
expr[[3]][[3]][[3]][[1]]  # expr[[c(3, 3, 3, 1)]]
## `+`
```

**Example 15.2** *We can even compose a recursive function to traverse the whole parse tree:*

```
recapply <- function(expr)
{
    if (is.call(expr)) lapply(expr, recapply)
    else expr
}

str(recapply(expr))
## List of 3
```

[5] (*) Equivalently, in the fully parenthesised Polish notation $(f, \ldots)$ (the prefix notation; traditionally used in source code s-expressions in Lisp), we can express it like:

```
# (this is not valid R syntax)
(
    `while`,
    (`<`, i, 10),
    (
        `{`,
        (cat, "i = ", i, "\n", sep=""),
        (
            `<-`,
            i,
            (`+`, i, 1)
        )
    )
)
```

```
##  $ : symbol while
##  $ :List of 3
##   ..$ : symbol <
##   ..$ : symbol i
##   ..$ : num 10
##  $ :List of 3
##   ..$ : symbol {
##   ..$ :List of 5
##   .. ..$    : symbol cat
##   .. ..$    : chr "i = "
##   .. ..$    : symbol i
##   .. ..$    : chr "\n"
##   .. ..$ sep: chr ""
##   ..$ :List of 3
##   .. ..$ : symbol <-
##   .. ..$ : symbol i
##   .. ..$ :List of 3
##   .. .. ..$ : symbol +
##   .. .. ..$ : symbol i
##   .. .. ..$ : num 1
```

### 15.3.2 Manipulating calls

The R language is *homoiconic*: it can treat code as data. This includes the ability to manipulate it on the fly. This is because, just like on lists, we can freely use the replacement versions of `` `[` `` and `` `[[` `` on objects of the type `call`.

```
expr[[2]][[1]] <- as.name("<=")  # was: `<`
expr[[3]] <- quote(i <- i * 2)  # was: {...}
print(expr)
## while (i <= 10) i <- i * 2
```

We are only limited by our imagination. We should spend some time and contemplate how powerful this is, knowing that soon we will become able to evaluate any expression in different contexts.

## 15.4 Inspecting function definition and usage

### 15.4.1 Getting the body and formal arguments

Consider a function:

```
test <- function(x, y=1, z=f(x, y))
    x+y+z  # whatever
```

We know from the first part of this book that calling **print** on a function reveals its source code. But there is more. We can fetch its formal parameters in the form of a named list[6]:

```
formals(test)
## $x
##
##
## $y
## [1] 1
##
## $z
## f(x, y)
```

Expressions corresponding to the default arguments are stored as ordinary list elements (for more details, see Section 17.2).

Furthermore, we can access the function's body:

```
body(test)
## x + y + z
```

It is an object of the now well-known class `call`. Thus, we can customise it as we please:

```
body(test)[[1]] <- as.name("*")  # change `+` to `*`
body(test) <- as.call(list(
    as.name("{"), quote(cat("spam\n")), body(test)
))
print(test)
## function (x, y = 1, z = f(x, y))
## {
##     cat("spam\n")
##     (x + y) * z
## }
```

### 15.4.2 Getting the expression passed as an actual argument

A call to **substitute** reveals the expression passed as a function's argument.

```
test <- function(x) substitute(x)
```

Some examples:

```
test(1)
## [1] 1
test(2+spam)
```

[6] (*) Actually, a special internal datatype called `pairlist`, which is rarely seen at the R level; see [69] and [66] for information on how to deal with them in C. In the current context, seeing pairlists as named lists is perfectly fine.

```
## 2 + spam
test(test(test(!!7)))
## test(test(!!7))
test()  # it is not an error
```

Chapter 17 notes that arguments are evaluated only on demand (lazily): **substitute** triggers no computations. This opens the possibility to author functions that interpret their input whichever way they like; see Section 9.4.7, Section 12.3.9, and Section 17.5 for examples.

**Example 15.3** *library (see Section 7.3.1) specifies the name of the package to be loaded both in the form of a character string and a* name*:*

```
library("gsl")  # preferred
library(gsl)  # discouraged; via as.character(substitute(package))
```

*A user saves two keystrokes at the cost of not being able to prepare the package name programmatically before the call:*

```
which_package <- "gsl"
library(which_package)  # library("which_package")
## Error in `library()`: ! there is no package called 'which_package'
```

*In order to make the above possible, we need to alter the function's character.only argument (which defaults to FALSE):*

```
library(which_package, character.only=TRUE)  # OK
```

**Exercise 15.4** *In many functions, we can see a call like **deparse(substitute(arg))** or **as.character(substitute(arg))**. Study the source code of **plot.default**, **hist.default**, **prop.test**, **wilcox.test.default** and the aforementioned **library**. Explain why they do that. Propose a solution to achieve the same functionality without using reflection techniques.*

### 15.4.3 Checking if an argument is missing

**missing** checks whether an argument was provided:

```
test <- function(x) missing(x)

test(1)
## [1] FALSE
test()
## [1] TRUE
```

**Exercise 15.5** *Study the source code of **sample**, **seq.default**, **plot.default**, **matplot**, and **t.test.default**. Determine the role of a call to **missing**. Would introducing a default argument NULL and testing its value with **is.null** constitute a reasonable alternative?*

### 15.4.4 Determining how a function was called

Even though this somewhat touches on the topics discussed in the two coming chapters, it is worth knowing that **sys.call** can look at the call stack and determine how the current function was invoked. Moreover, **match.call** takes us a step further: it returns a call with argument names matched to a function's list of formal parameters. For instance:

```
test <- function(x, y, ..., a="yes", b="no")
{
    print(sys.call())  # sys.call(0)
    print(match.call())
}

x <- "maybe"
test("spam", "bacon", "eggs", u=("ham"<"jam"), b=x)
## test("spam", "bacon", "eggs", u = ("ham" < "jam"), b = x)
## test(x = "spam", y = "bacon", "eggs", u = ("ham" < "jam"), b = x)
```

In both cases, the results are objects of the type `call`. We know how to manipulate them already.

Another example where we see that we can dig into the call stack much more deeply:

```
f <- function(x)
{
    g <- function(y)
    {
        cat("g:\n")
        print(sys.call(0))
        print(sys.call(-1))  # go back one frame
        y
    }

    cat("f:\n")
    print(sys.call(0))
    g(x+1)
}

f(1)
## f:
## f(1)
## g:
## g(x + 1)
## f(1)
## [1] 2
```

---

**Note** Matching function parameters *to* the passed arguments is done in the following order (see Section 4.3 of [70]):

1. First, keyword arguments with names are matched exactly. Each name is matched at most once.
2. Then, we take the remaining keyword arguments, but with the partial matching of names listed before the ellipsis, `...`. Each match must be unambiguous.
3. Third, we apply the positional matching to the remaining parameters.
4. Last, the ellipsis (if present) consumes all the remaining arguments (named or not).

For instance:

```
test <- function(spam, jasmine, jam, ..., option=NULL)
    print(match.call())
```

Example calls:

```
test(1, 2, 3, 4, option="yes")
## test(spam = 1, jasmine = 2, jam = 3, 4, option = "yes")
test(1, 2, jasmine="no", sp=4, ham=7)
## Warning in test(1, 2, jasmine = "no", sp = 4, ham = 7): partial argument
##     match of 'sp' to 'spam'
## Warning in match.call(definition, call, expand.dots, envir): partial
##     argument match of 'sp' to 'spam'
## test(spam = 4, jasmine = "no", jam = 1, 2, ham = 7)
test(1, 2, ja=7)  # ambiguous match
## Warning in test(1, 2, ja = 7): partial argument match of 'ja' to 'jasmine'
## Error in `test()`: ! argument 3 matches multiple formal arguments
test(o=7)  # partial matching of `option` failed - `option` is after `...`
## test(o = 7)
```

Note again that our environment uses **`options`**`(warnPartialMatchArgs=TRUE)`.

---

**Exercise 15.6** *A function can*[7] *see how it was defined by its maker. Call* ***`sys.function`*** *inside its body to reveal that.*

**Exercise 15.7** *Execute* ***`match.call(sys.function(-1), sys.call(-1))`*** *in the above* ***`g`*** *function.*

## 15.5 Exercises

**Exercise 15.8** *Answer the following questions.*

- *What is a simple expression? What is a compound expression? Give a few examples.*
- *What is the difference between an object of the type `call` and that of the type `expression`?*

[7] Therefore, it is possible to have a function that returns a modified version of itself.

- *What do **formals** and **body** return when called on a function object?*
- *How to test if an argument to a function was given? Provide a use case for such a verification step.*
- *Give a few ways to create an unevaluated call.*
- *What is the purpose of **deparse(substitute(...))**? Give a few examples of functions that use this technique.*
- *What is the difference between **sys.call** and **match.call**?*
- *Why cannot we rely on partial matching in the call **boxplot**(x, horiz=TRUE) and have to write the full argument name like **boxplot**(x, horizontal=TRUE) instead?*

**Exercise 15.9** *Write a function that takes the dot-dot-dot argument. Using **match.call** (amongst others), determine the list of all the expressions passed via `...`. Allow some of them to be named (just like in one of the preceding examples). The solution will be given in Section 17.3.*

**Exercise 15.10** *Write a function **check_if_calls**(f, fun_list) that takes another function **f** as input. Then, it verifies if **f** calls any of the functions (referred to by their names) from a character vector fun_list.*

# 16

# *Environments and evaluation (*)*

In the first part of our book, we discussed the most crucial *basic* object types: numeric, logical, and character vectors, lists (generic vectors), and functions. In this chapter, we introduce another basic type: *environments*. Like lists, they can be classified as recursive data structures; compare the diagram in Figure 17.2.

**Important** Each object of the type `environment` consists of:

- a *frame*[1] (Section 16.1), which stores a set of *bindings* that associate variable names with their corresponding values; it can be thought of as a container of named R objects of any type;
- a reference to an *enclosing environment*[2] (Section 16.2.2), which might be inspected (recursively!) when a requested named variable is not found in the current frame.

Even though we rarely interact with them directly (unless we need a hash table-like data structure with a quick by-name element lookup), they are crucial for the R interpreter itself. Namely, we shall soon see that they form the basis of the *environment model of evaluation*, which governs how expressions are computed (Section 16.2).

## 16.1 Frames: Environments as object containers

To create a new, empty environment, we can call the **`new.env`** function:

```
e1 <- new.env()
typeof(e1)
## [1] "environment"
```

In this section, we treat environments merely as containers for named objects of any kind, i.e., we deal with the *frame* part thereof.

Let's insert a few elements into `e1`:

[1] Not to be confused with a *data frame*, i.e., an object (list) of the S3 class `data.frame`; see Chapter 12.

[2] Some also call it a *parent* environment, but we will not. We will try to follow the nomenclature established in Section 3.2 of [1]. Note that there is a bit of a mess in the R documentation regarding how enclosing environments are referred to.

```
e1[["x"]] <- "x in e1"
e1[["y"]] <- 1:3
e1[["z"]] <- NULL  # unlike in the case of lists, creates a new element
```

The `` `[[` `` operator provides us with a *named list*-like behaviour also in the case of element extraction:

```
e1[["x"]]
## [1] "x in e1"
e1[["spam"]]  # does not exist
## NULL
(e1[["y"]] <- e1[["y"]]*10)  # replace with new content
## [1] 10 20 30
```

### 16.1.1 Printing

Printing an environment leads to an uncanny result:

```
print(e1)  # same with str(e1)
## <environment: 0x5601c5fec3f0>
```

It is the address where `e1` is stored in the computer's memory. It can serve as the environment's unique identifier. As we have said, environments are of rather *internal* interest. Thus, such an esoteric message was perhaps a good design choice; it wards off novices. However, we can easily get the list of objects stored inside the container by calling **names**:

```
names(e1)  # but attr(e1, "names") is not set
## [1] "x" "y" "z"
```

Moreover, **length** gives the number of bindings in the frame:

```
length(e1)
## [1] 3
```

### 16.1.2 Environments vs named lists

Environment frames, in some sense, can be thought of as named lists, but the set of admissible operations is severely restricted. In particular, we cannot extract more than one element at the same time using the index operator:

```
e1[c("x", "y")]  # but see the `mget` function
## Error in `e1[c("x", "y")]`: ! object of type 'environment' is not
##     subsettable
```

nor can we refer to the elements by position:

```
e1[[1]] <- "bad key"
## Error in `e1[[1]] <- "bad key"`: ! wrong args for environment
##     subassignment
```

**Exercise 16.1** *Check if* **`lapply`** *and* **`Map`** *can be applied directly on environments. Also, can we iterate over their elements using a* **`for`** *loop?*

Still, named lists can be converted to environments and vice versa using **`as.list`** and **`as.environment`**.

```
as.list(e1)
## $x
## [1] "x in e1"
##
## $y
## [1] 10 20 30
##
## $z
## NULL
as.environment(list(u=42, whatever="it's not going to be printed anyway"))
## <environment: 0x5601c8ecde10>
as.list(as.environment(list(x=1, y=2, x=3)))  # no duplicates allowed
## $y
## [1] 2
##
## $x
## [1] 3
```

### 16.1.3 Hash maps: Fast element lookup by name

Environment frames are internally implemented using hash tables (hash maps; see, e.g., [15, 42]) with character string keys.

---

**Important** A *hash table* is a data structure that implements a very quick lookup, insertion and deletion of individual elements *by name* (in amortised $O(1)$ time).

---

This comes at a price, including what we have already observed before:

- the elements are not ordered in any particular way: they cannot be referred to via a numeric index;
- all element names must be unique.

---

**Note** A list may be considered a *sequence*, but an environment frame is only, in fact, a *set* (a bag) of *key-value pairs*. In most numerical computing applications, we would rather store, iterate over, and process all the elements *in order*, hence the greater prevalence of the former. Lists still implement the element lookup by name, even though it is slightly slower[3]. However, they are much more universal.

---

**Example 16.2** *A natural use case of manually-created environment frames deals with grouping a series of objects identified by character string keys. Consider a simple pseudocode for counting the number of occurrences of objects in a given container:*

```
for (key in some_container) {
    if (!is.null(counter[["key"]]))
        counter[["key"]] <- counter[["key"]]+1
    else
        counter[["key"]] <- 1
}
```

*Assume that `some_container` is large, e.g., it is generated on the fly by reading a data stream of size $n$. The runtime of the above algorithm will depend on the data structure used. If the `counter` is a list, then, theoretically, the worst-case performance will be $O(n^2)$ (if all keys are unique). On the other hand, for environments, it will be faster by one order of magnitude: down to amortised $O(n)$.*

**Exercise 16.3** *Implement a test function according to the above pseudocode and benchmark the two data structures using **`proc.time`** on example data.*

**Exercise 16.4** *(*) Determine the number of unique text lines in a huge file (assuming that the set of unique text lines fits into memory, but the file itself does not). Also, determine the five most frequently occurring text lines.*

### 16.1.4 Call by value, copy on demand: Not for environments

Given any object `x`, when we issue:

```
y <- x
```

its copy[4] is made so that `y` and `x` are independent. In other words, any change to the state of `x` (or `y`) is not reflected in `y` (or `x`). For instance:

```
x <- list(a=1)
y <- x
y[["a"]] <- y[["a"]]+1
print(y)
## $a
## [1] 2
print(x)  # not affected: `x` and `y` are independent
```

[3] Accessing elements by position (numeric index) in lists takes $O(1)$ time. The worst-case runtime for the element lookup by name is linear with respect to the container size (when the item is not found). Also, inserting new elements at the end takes amortised $O(1)$ time.

[4] Delayed (on demand); see below.

```
## $a
## [1] 1
```

The same happens with arguments that we pass to the functions:

```
mod <- function(y, key)  # it is like: local_y <- passed_argument
{
    y[[key]] <- y[[key]]+1
    y
}

mod(x, "a")[["a"]]  # returns a modified copy of `x`
## [1] 2
x[["a"]]  # not affected
## [1] 1
```

We can thus say that R imitates the *pass-by-value* strategy here.

---

**Important** Environments are the only[5] objects that follow the assign- and pass-by-reference strategies.

---

In other words, if we perform:

```
x <- as.environment(x)
y <- x
```

then the names x and y are bound to the same object in the computer's memory:

```
print(x)
## <environment: 0x5601c9afe068>
print(y)
## <environment: 0x5601c9afe068>
```

Therefore:

```
y[["a"]] <- y[["a"]]+1
print(y[["a"]])
## [1] 2
print(x[["a"]])  # `x` is `y`, `y` is `x`
## [1] 2
```

The same happens when we pass an environment to a function:

[5] We do not count all the tricks we can do at the C language level (Chapter 14). In R, the distinction between pass-by-value and pass-by-reference is slightly more complicated because of the lazy evaluation of arguments (the call-by-need strategy; Chapter 17). We have made an idealisation for didactic purposes.

```
mod(y, "a")[["a"]]  # pass-by-reference (`y` is `x`, remember?)
## [1] 3
x[["a"]]   # `x` has changed
## [1] 3
```

Thus, any changes we make to an environment passed as an argument to a function will be visible *outside* the call. This minimises time and memory use in certain situations.

---

**Note** (*) For efficiency reasons, when we write "`y <- x`", a copy of `x` (unless it is an environment) is created only if it is absolutely necessary.

Here is some benchmarking of the *copy-on-demand* mechanism.

```
n <- 100000000  # like, a lot
```

Creation of a new large numeric vector:

```
t0 <- proc.time();  x <- numeric(n);  proc.time() - t0
##    user  system elapsed
##   0.853   1.993   2.852
```

Creation of a (delayed) copy is instant:

```
t0 <- proc.time();  y <- x;           proc.time() - t0
##    user  system elapsed
##       0       0       0
```

We definitely did not duplicate the `n` data cells.

Copy-on-demand is implemented using some simple *reference counting*; compare Section 14.2.4. We can inspect that `x` and `y` point to the same address in memory by calling:

```
.Internal(inspect(x))  # internal function - do not use it
## @7efba1134010 14 REALSXP g0c7 [REF(2)] (len=100000000, tl=0) 0,0,0,0,...
.Internal(inspect(y))
## @7efba1134010 14 REALSXP g0c7 [REF(2)] (len=100000000, tl=0) 0,0,0,0,...
```

The actual copying is only triggered when we try to modify `x` or `y`. This is when they need to be separated.

```
t0 <- proc.time();  y[1] <- 1;        proc.time() - t0
##    user  system elapsed
##   1.227   1.910   3.142
```

Now `x` and `y` are different objects.

```
.Internal(inspect(x))
## @7efba1134010 14 REALSXP g0c7 [MARK,REF(1)] (len=100000000, tl=0) 0,0,...
```

```
.Internal(inspect(y))
## @7ef9c43ce010 14 REALSXP g0c7 [MARK,REF(1)] (len=1000000000, tl=0) 1,0,...
```

The elapsed time is similar to that needed to create x from scratch. Further modifications will already be quick:

```
t0 <- proc.time();  y[2] <- 2;        proc.time() - t0
##    user  system elapsed
##   0.000   0.001   0.000
```

### 16.1.5 A note on reference classes (**)

In Section 10.5, we briefly mentioned the S4 system for object-orientated programming. We also have access to its variant, called *reference classes*[6], which was first introduced in R version 2.12.0. Reference classes are implemented using S4 classes, with the data part being of the type `environment`. They give a more typical OOP experience, where methods can modify the data they act on in place.

Reference classes are a theoretically interesting concept on its own and may be quite appealing to package developers with C++ or Java background. Nevertheless, in the current author's opinion, such classes are alien citizens of our environment, violating its *functional* nature. Therefore, we will not be discussing them here. A curious reader is referred to `help("ReferenceClasses")` and Chapters 9 and 11 of [12] for more details.

## 16.2 The environment model of evaluation

In Chapter 15, we said that there are three types of expressions: constants (e.g., `1` and `"spam"`), names (e.g., `x`, `` `+` ``, and `spam`), and calls (like `f(x, 1)`).

**Important** Names (symbols) have no meaning by themselves. The *meaning* of a name always depends on the context, which is specified by an environment.

Consider a simple expression that merely consists of the name `x`:

```
expr_x <- quote(x)
```

Let's define two environments that bind the name x to two different constants.

[6] Some call them R5, but we will not.

```
e1 <- as.environment(list(x=1))
e2 <- as.environment(list(x="spam"))
```

---

**Important** An expression is evaluated *within* a specific environment.

---

Let's call **`eval`** on the above.

```
eval(expr_x, envir=e1)  # evaluate `x` within environment e1
## [1] 1
eval(expr_x, envir=e2)  # evaluate the same `x` within environment e2
## [1] "spam"
```

The very same expression has two different meanings, depending on the *context*. This is quite like in the so-called real life: "I'm good" can mean "I don't need anything" but also "My virtues are plentiful". It all depends on who and when is asking, i.e., in which *environment* we *evaluate* the said sentence.

We call this the *environment model of evaluation*, a notion that R authors have borrowed from a Lisp-like language called Scheme[7] (see Section 3.2 of [1] and Section 6 of [70]).

### 16.2.1 Getting the current environment

By default, expressions are evaluated in the *current* environment, which can fetch by calling:

```
sys.frame(sys.nframe())  # get the current environment
## <environment: R_GlobalEnv>
```

We are working *on the R console*. Hence, the current one is the *global* environment (user workspace). We can access it from anywhere by calling **`globalenv`** or referring to the `` `.GlobalEnv` `` object.

**Example 16.5** *Calling any operation, for instance*[8]*:*

```
x <- "spammity spam"
```

*means evaluating it* within the current environment:

```
eval(quote(x <- "spammity spam"), envir=sys.frame(sys.nframe()))
```

*Here, we bound the name x to the string* `"spammity spam"` *in the current environment's frame:*

---

[7] That is why everyone *serious* about R programming should add the *Structure and Interpretation of Computer Programs* [1] to their reading list. Also, R is not the only known marriage between statistics and Lisp-like languages; see also LISP-STAT [55].

[8] For now, let's take for granted that `` `<-` `` is accessible from the current context and denotes the assignment.

```
sys.frame(sys.nframe())[["x"]]  # yes, `x` is in the current environment now
## [1] "spammity spam"
globalenv()[["x"]]  # because the global environment is the current one here
## [1] "spammity spam"
```

*Therefore, when we now refer to* `x` *(from within the current environment):*

```
x  # eval(quote(x), envir=sys.frame(sys.nframe()))
## [1] "spammity spam"
```

*precisely the foregoing named object is fetched.*

**Exercise 16.6** ***save.image*** *saves the current workspace, i.e., the global environment, by default, to the file named* `.Rdata`*. Test this function in combination with* ***load***.

---

**Note** Names starting with a dot are *hidden*. **ls**, a function to fetch all names registered within a given environment, does not list them by default.

```
.test <- "spam"
ls()  # list all names in the current environment, i.e., the global one
## [1] "e1"     "e2"     "expr_x" "mod"    "x"      "y"
```

Compare it with:

```
ls(all.names=TRUE)
## [1] ".Random.seed" ".test"        "e1"           "e2"
## [5] "expr_x"       "mod"          "x"            "y"
```

On a side note, `` `.Random.seed` `` stores the current pseudorandom number generator's seed; compare Section 2.1.5.

---

### 16.2.2 Enclosures, enclosures thereof, etc.

To show that there is much more to the environment model of evaluation than what we have already mentioned, let's try to evaluate an expression featuring two names:

```
e2 <- as.environment(list(x="spam"))  # once again (a reminder)
expr_comp <- quote(x < "eggs")
eval(expr_comp, envir=e2)  # "spam" < "eggs"
## Error in `x < "eggs"`: ! could not find function "<"
```

The meaning of any constant (here, `"spam"`) is context-independent. The environment provided specifies the name `x` but does not define `` `<` ``. Hence the error. Nonetheless, we *feel* that we know the meaning of `` `<` ``. It is a relational operator, obviously, isn't it? To increase the confusion, let's highlight that our experience-grounded intuition is true in the following context:

```
e3 <- new.env()
e3[["x"]] <- "bacon"
eval(expr_comp, envir=e3)  # "bacon" < "eggs"
## [1] TRUE
```

So where does the name `` `<` `` come from? It is neither included in `e2` nor `e3`:

```
e2[["<"]]
## NULL
e3[["<"]]
## NULL
```

Is `` `<` `` hardcoded somewhere? Or is it also dependent on the context? Why is it *visible* when evaluating an expression within `e3` but not in `e2`?

Studying **help**("[[") (see the *Environments* section), we discover that `e3[["<"]]` is equivalent to a call to **get**("<", envir=e3, inherits=FALSE). In **help**("get"), we read that if the `inherits` argument is set to `TRUE` (which is the default in **get**), then *the enclosing frames of the given environment are searched as well*. Continuing the example from the previous subsection:

```
get("<", envir=e2)  # inherits=TRUE
## Error in `get()`: ! object '<' not found
get("<", envir=e3)  # inherits=TRUE
## function (e1, e2)  .Primitive("<")
```

Indeed, we see that `` `<` `` is *reachable* from `e3` but not from `e2`. It means that `e3` *points to* another environment where further information should be sought if the current container is left empty-handed.

---

**Important** The reference (pointer) to the *enclosing environment* is integral to each environment (alongside a *frame* of objects). It can be fetched and set using the **parent.env** function.

---

### 16.2.3 Missing names are sought in enclosing environments

To understand the idea of enclosing environments better, let's create two new environments whose enclosures are explicitly set as follows:

```
(e4 <- new.env(parent=e3))
## <environment: 0x5601c6a16678>
(e5 <- new.env(parent=e4))
## <environment: 0x5601c69b0910>
```

To verify that everything is in order, we can inspect the following:

```
print(e3)  # this is the address of e3
## <environment: 0x5601c99685b8>
parent.env(e4)  # e3 is the enclosing environment of e4
## <environment: 0x5601c99685b8>
parent.env(e5)  # e4 is the enclosing environment of e5
## <environment: 0x5601c6a16678>
```

Also, let's bind two different objects to the name y in e5 and e3.

```
e5[["y"]] <- "spam"
e3[["y"]] <- function() "a function `y` in e3"
```

The current state of matters is depicted in Figure 16.1.

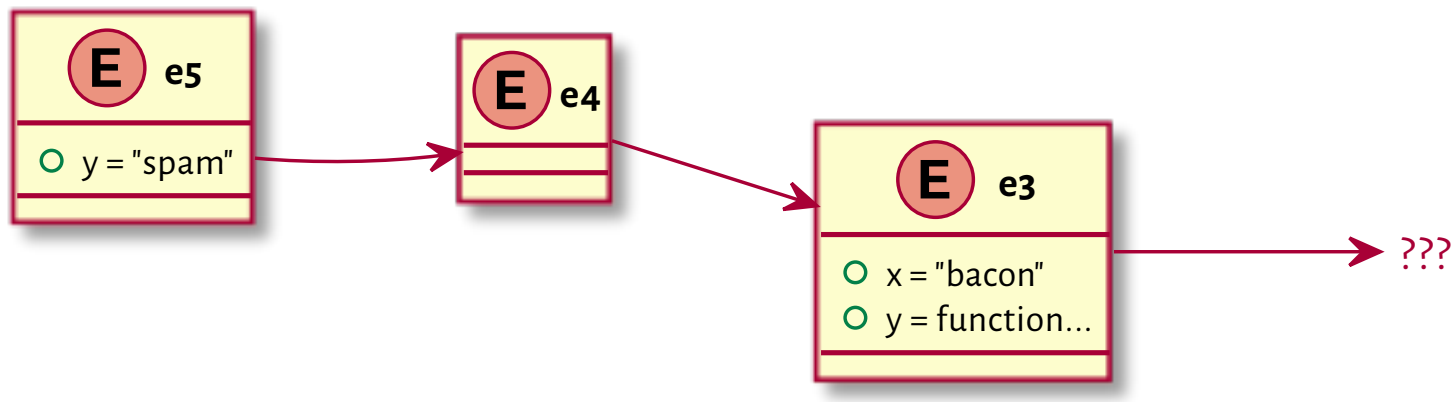

Figure 16.1. Example environments and their enclosures (original setting).

Let's evaluate the name y in the foregoing environments:

```
expr_y <- quote(y)
eval(expr_y, envir=e3)
## function ()
## "a function `y` in e3"
eval(expr_y, envir=e5)
## [1] "spam"
```

No surprises, yet. However, evaluating it in e4, which does not define y, yields:

```
eval(expr_y, envir=e4)
## function ()
## "a function `y` in e3"
```

It returned y from e4's enclosure, e3. Let's play about with the enclosures of e5 and e4 so that we obtain the setting depicted in Figure 16.2:

```
parent.env(e5) <- e3
parent.env(e4) <- e5
```

Evaluating y again in the same e4 nourishes a very different result:

```
eval(expr_y, envir=e4)
## [1] "spam"
```

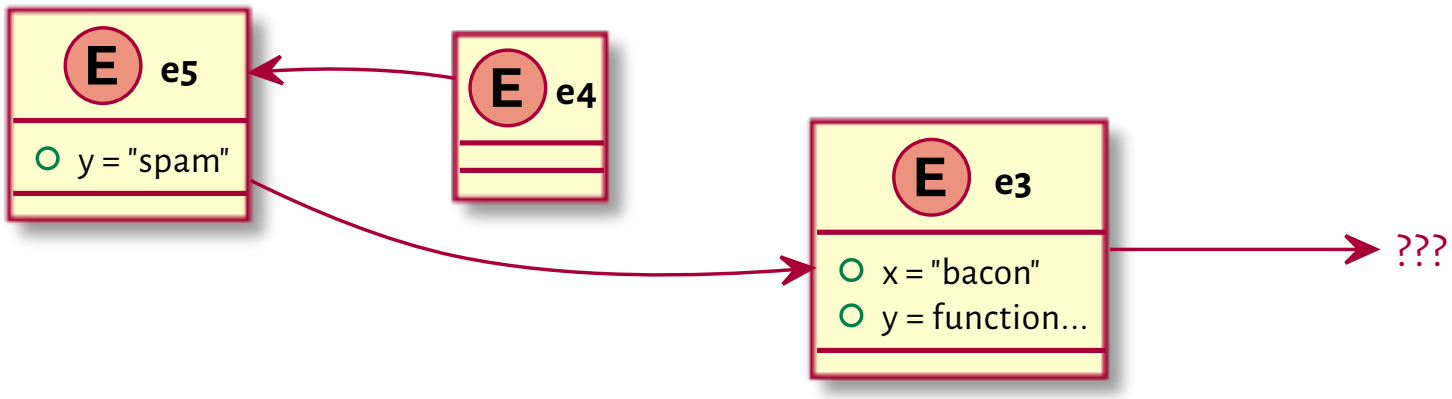

Figure 16.2. Example environments and their enclosures (after the change made).

**Important** Names referred to in an expression but missing in the current environment will be sought in their enclosure(s) *until* successful.

**Note** Here are the functions related to searching within and modifying environments that optionally allow for continuing explorations in their enclosures:

- `inherits=TRUE` by default: **`exists`**, **`get`**,
- `inherits=FALSE` by default: **`assign`**, * **`rm`** (remove).

### 16.2.4 Looking for functions

Interestingly, if a name is used instead of a function to be called, the object sought is always[9] of the mode `function`. Consider an expression similar to the above, but this time including the name y playing a different role:

```
expr_y2 <- quote(y())  # a call to something named `y`
eval(expr_y2, envir=e4)
## [1] "a function `y` in e3"
```

In other words, what we used here was not:

```
get("y", envir=e4)
## [1] "spam"
```

but:

```
get("y", envir=e4, mode="function")
## function ()
## "a function `y` in e3"
```

[9] This is why we can write "`c <- c(1, 2)`" and then still be able to call **`c`** to create another vector.

**Note** `name()`, `"name"()`, and `` `name`() `` are synonymous. However, the first expression is acceptable only if `name` is syntactically valid.

### 16.2.5 Inspecting the search path

Going back to our expression involving a relational operator:

```
expr_comp
## x < "eggs"
```

Why does the following work as expected?

```
eval(expr_comp, envir=e3)  # "bacon" < "eggs"
## [1] TRUE
```

Well, we have gathered all the bits to understand it now. Namely, `` `<` `` is a function that is looked up like:

```
get("<", envir=e3, inherits=TRUE, mode="function")
## function (e1, e2)  .Primitive("<")
```

It is reachable from e3, which means that e3 also has an enclosing environment.

```
parent.env(e3)
## <environment: R_GlobalEnv>
```

This is our global namespace, which was the current environment when e3 was created. Still, we did not define `` `<` `` there. It means that the global environment also has an enclosure.

We can explore the whole *search path* by starting at the global environment and following the enclosures recursively.

```
ecur <- globalenv()  # starting point
repeat {
    cat(paste0(format(ecur), " (", attr(ecur, "name"), ")"))  # pretty-print

    if (exists("<", envir=ecur, inherits=FALSE))  # look for `<`
        cat(strrep(" ", 25), "`<` found here!")
    cat("\n")

    ecur <- parent.env(ecur)  # advance to its enclosure
}
## <environment: R_GlobalEnv> ()
## <environment: 0x5601c6ccda40> (.marekstuff)
## <environment: package:stats> (package:stats)
## <environment: package:graphics> (package:graphics)
```

```
## <environment: package:grDevices> (package:grDevices)
## <environment: package:utils> (package:utils)
## <environment: package:datasets> (package:datasets)
## <environment: package:methods> (package:methods)
## <environment: 0x5601c4817640> (Autoloads)
## <environment: base> ()                          `<` found here!
## <environment: R_EmptyEnv> ()
## Error in `parent.env()`: ! the empty environment has no parent
```

Underneath the global environment, there is a whole list of attached packages:

1. packages attached by the user (**.marekstuff** is used internally in the process of evaluating code in this book),
2. default packages (Section 7.3.1),
3. (**) **Autoloads** (for the promises-to-load R packages; compare **help**("autoload"); it is a technicality we may safely ignore here),
4. the **base** package, which we can access directly by calling **baseenv**; it is where most of the *fundamental* functions from the previous chapters reside,
5. the *empty* environment (**emptyenv**), which is the only one followed by nil (the loop would turn out endless otherwise).

It comes at no surprise that the **`<`** operator has been found in the **base** package.

---

**Note** On a side note, the reason why this operation failed:

```
e2 <- as.environment(list(x="spam"))  # to recall
eval(expr_comp, envir=e2)
## Error in `x < "eggs"`: ! could not find function "<"
```

is because **as.environment** sets the enclosing environment to:

```
parent.env(e2)
## <environment: R_EmptyEnv>
```

See also **list2env** which gives greater control over this (cf. its `parent` argument).

---

### 16.2.6 Attaching to and detaching from the search path

In Section 7.3.1, we mentioned that we can access the objects exported by a package without attaching them to the search path by using the **pkg**::object syntax, which loads the package if necessary. For instance:

```
tools::toTitleCase("`tools` not attached to the search path")
## [1] "`tools` not Attached to the Search Path"
```

However:

```
toTitleCase("nope")
## Error in `toTitleCase()`: ! could not find function "toTitleCase"
```

It did not work because **toTitleCase** is not reachable from the current environment.

Let's inspect the current search path:

```
search()
##  [1] ".GlobalEnv"        ".marekstuff"       "package:stats"
##  [4] "package:graphics"  "package:grDevices" "package:utils"
##  [7] "package:datasets"  "package:methods"   "Autoloads"
## [10] "package:base"
```

Some might find writing "**pkg::**" inconvenient. Thus, we can call **library** to attach the package to the search path immediately below the global environment.

```
library("tools")
```

The search path becomes (see Figure 16.3 for an illustration):

```
search()
##  [1] ".GlobalEnv"        "package:tools"     ".marekstuff"
##  [4] "package:stats"     "package:graphics"  "package:grDevices"
##  [7] "package:utils"     "package:datasets"  "package:methods"
## [10] "Autoloads"         "package:base"
```

Therefore, what follows, now works as expected:

```
toTitleCase("Nobody expects the Spanish Inquisition")
## [1] "Nobody Expects the Spanish Inquisition"
```

We can use **detach**[10] to remove an item from the search path.

```
head(search())  # before detach
## [1] ".GlobalEnv"        "package:tools"     ".marekstuff"
## [4] "package:stats"     "package:graphics"  "package:grDevices"
detach("package:tools")
head(search())  # not there anymore
## [1] ".GlobalEnv"        ".marekstuff"       "package:stats"
## [4] "package:graphics"  "package:grDevices" "package:utils"
```

---

**Note** We can also plug arbitrary environments[11] and named lists into the search path.

[10] Which does not unload the package from memory, though; see **unload** (possibly combined with **library.dynam.unload**).

[11] Or we should rather say, environment frames. When an environment is attached to the search path, it is duplicated so that the changes made to the original environment are not reflected in the copy. Then, its previous enclosure is discarded. After all, we want a series of recursive calls to **parent.env** to form the whole search path.

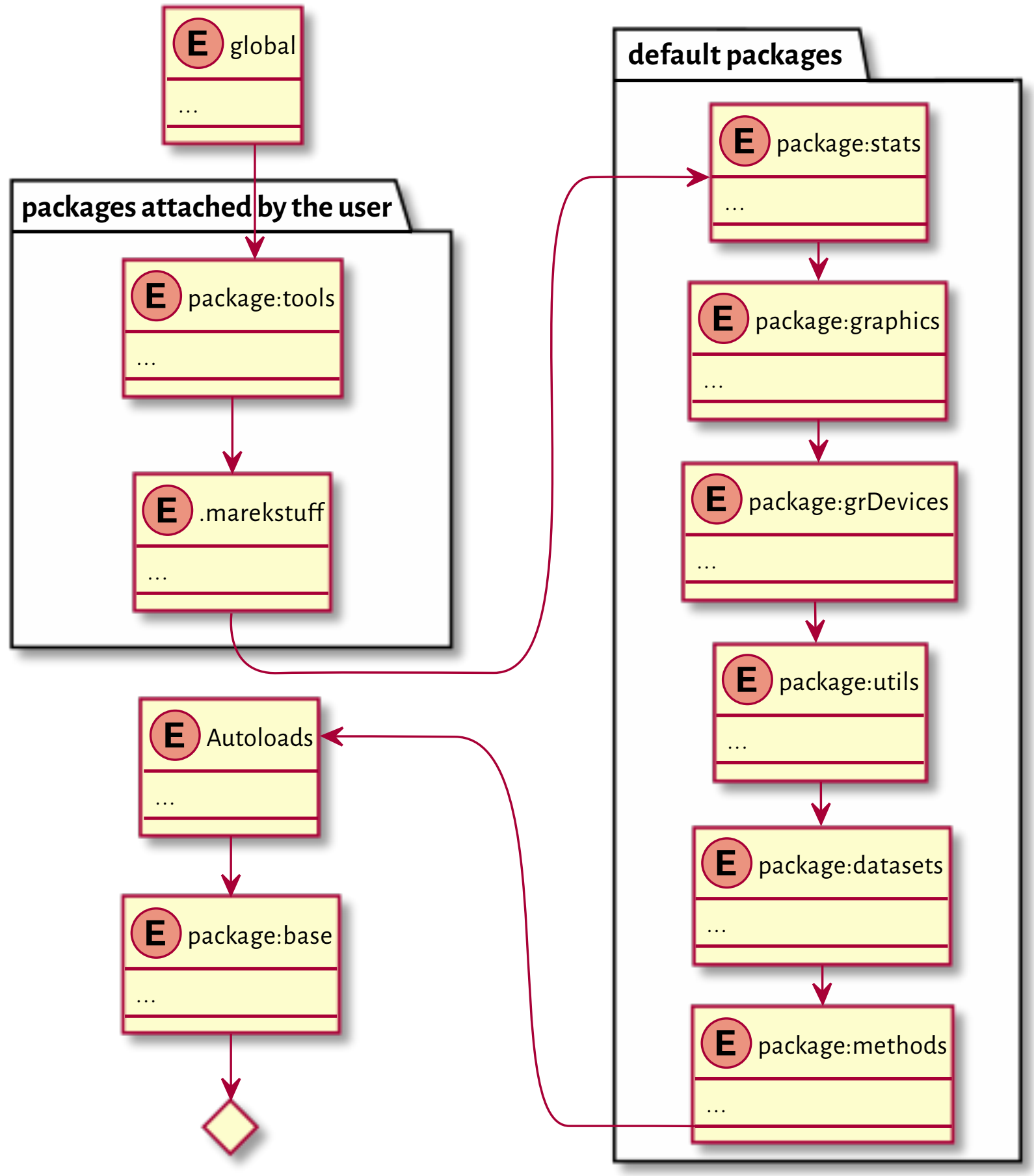

Figure 16.3. The search path after attaching the **tools** package.

Recalling that data frames are built on the latter (Section 12.1.6), some users rely on this technique save a few keystrokes.

```
attach(iris)
head(search(), 3)
## [1] ".GlobalEnv"  "iris"        ".marekstuff"
```

The `iris` list was converted to an environment, and the necessary enclosures were set accordingly:

```
str(parent.env(globalenv()))
## <environment: 0x5601c96e7f88>
##  - attr(*, "name")= chr "iris"
str(parent.env(parent.env(globalenv())))
```

```
## <environment: 0x5601c6ccda40>
##  - attr(*, "name")= chr ".marekstuff"
```

We can now write:

```
head(Petal.Width/Sepal.Width)  # iris[["Petal.Width"]]/iris[["Sepal.Width"]]
## [1] 0.057143 0.066667 0.062500 0.064516 0.055556 0.102564
```

Overall, attaching data frames is discouraged, especially outside the interactive mode. Let's not be too lazy.

```
detach(iris)  # such a relief
```

---

### 16.2.7 Masking (shadowing) objects from down under

An assignment via `` `<-` `` creates a binding in the *current* environment. Therefore, even if the name to bind exists somewhere on the search path, it will not be modified. Instead, a new name will be created.

```
eval(quote("spam" < "eggs"))
## [1] FALSE
```

Here, we rely on `` `<` `` from the base environment. Withal, we can create an object of the same name in the current (global) context:

```
`<` <- function(e1, e2)
{
    warning("This is not the base `<`, mate.")
    NA
}
```

Now we have two different functions of the same name. When we evaluate an expression within the current environment or any of its descendants, the new name *shadows* the base one:

```
"spam" < "eggs"  # evaluate in the global environment
## Warning in "spam" < "eggs": This is not the base `<`, mate.
## [1] NA
eval(quote("spam" < "eggs"), envir=e5)  # its enclosure's enclosure is global
## Warning in "spam" < "eggs": This is not the base `<`, mate.
## [1] NA
```

But we can still call the original function directly:

```
base::`<`("spam", "eggs")
## [1] FALSE
```

It is also reachable from within the current environment's ancestors:

```
eval(quote("spam" < "eggs"), envir=parent.env(globalenv()))
## [1] FALSE
```

Before proceeding any further, we should clean up after ourselves. Otherwise, we will be asking for trouble.

```
rm("<")  # removes `<` from the global environment
```

An attached package may introduce some object names that are also available elsewhere. For instance:

```
library("stringx")
## Attaching package: 'stringx'
## The following objects are masked from 'package:base': casefold, chartr,
##     endsWith, gregexec, gregexpr, grep, grepl, gsub, ISOdate, ISOdatetime,
##     nchar, nzchar, paste, paste0, regexec, regexpr, sprintf, startsWith,
##     strftime, strptime, strrep, strsplit, strtrim, strwrap, sub, substr,
##     substr<-, substring, substring<-, Sys.time, tolower, toupper, trimws,
##     xtfrm, xtfrm.default
```

Therefore, in the current context, we have what follows:

```
toupper("Groß")  # stringx::toupper
## [1] "GROSS"
base::toupper("Groß")
## [1] "GROß"
```

Sometimes[12], we can use **assign**(..., inherits=TRUE) or its synonym, **`<<-`**, to modify the *existing* binding. A new binding is only created if necessary.

---

**Note** Let's attach the `iris` data frame (named list) to the search path again:

```
attach(iris)
Sepal.Length[1] <- 0
```

We did not modify the original `iris` nor its converted-to-an-environment copy that we can find in the search path. Instead, a new vector named `Sepal.Length` was created in the current environment:

[12] We normally cannot modify package namespaces. As we will mention in Section 16.3.5, they are automatically locked.

```
exists("Sepal.Length", envir=globalenv(), inherits=FALSE)  # it is in global
## [1] TRUE
Sepal.Length[1]  # global
## [1] 0
```

We can verify the preceding statement as follows:

```
rm("Sepal.Length")  # removes the one in the global environment
Sepal.Length[1]  # `iris` from the search path
## [1] 5.1
iris[["Sepal.Length"]][1]  # the original `iris`
## [1] 5.1
```

However, we can write:

```
Sepal.Length[1] <<- 0  # uses assign(..., inherits=TRUE)
```

We changed the state of the environment on the search path.

```
exists("Sepal.Length", envir=globalenv(), inherits=FALSE)  # not in global
## [1] FALSE
Sepal.Length[1]  # `iris` from the search path
## [1] 0
```

Yet, the original `iris` object is left untouched. There is no mechanism in place that would *synchronise* the original data frame and its independent copy on the search path.

```
iris[["Sepal.Length"]][1]  # the original `iris`
## [1] 5.1
```

It is best to avoid **`attach`** to avoid confusion.

## 16.3 Closures

So far, we have only covered the rules of evaluating *standalone* R expressions. In this section, we look at what happens *inside* the invoked functions.

### 16.3.1 Local environment

When we call a function, a new temporary environment is created. It is where all argument values[13] and local variables are emplaced. This environment is the current one while the function is being evaluated. After the call, it ceases to exist, and we return to the previous environment from the call stack.

[13] Function arguments are initially unevaluated; see Chapter 17.

Consider the following function:

```
test <- function(x)
{
    print(ls())  # list object names in the current environment
    y <- x^2  # creates a new variable
    print(sys.frame(sys.nframe()))  # get the ID of the current environment
    str(as.list(sys.frame(sys.nframe())))  # display its contents
}
```

First call:

```
test(2)
## [1] "x"
## <environment: 0x5601c9b59ed0>
## List of 2
##  $ y: num 4
##  $ x: num 2
```

Second call:

```
test(3)
## [1] "x"
## <environment: 0x5601c9e61190>
## List of 2
##  $ y: num 9
##  $ x: num 3
```

Each time, the current environment is different. This is why we do not see the variable y at the start of the second call. It is a brilliantly simple implementation of the storage for local variables.

### 16.3.2 Lexical scope and function closures

We were able to access the **print** function (amongst others) in the preceding example. This should make us wonder what the enclosing environment of that local environment is.

```
print_enclosure <- function()
    print(parent.env(sys.frame(sys.nframe())))

print_enclosure()
## <environment: R_GlobalEnv>
```

It is the global environment. Let's invoke the same function from another one:

```
call_print_enclosure <- function()
    print_enclosure()
```

```
call_print_enclosure()
## <environment: R_GlobalEnv>
```

It is the global environment again. If R used the so-called *dynamic scoping*, we would see the local environment of the function that invoked the one above. If this was true, we would have access to the caller's local variables from within the callee. But this is not the case.

---

**Important** Objects of the type `closure`, i.e., user-defined[14] functions, consist of three components:

- a list of formal arguments (compare **formals** in Section 15.4.1);
- an expression (see **body** in Section 15.4.1);
- a reference to the *associated environment* where the function might store data for further use (see **environment**).

By default, the associated environment is set to the current environment where the function was created.

A local environment created during a function's call has this associated environment as its closure.

Due to this, we say that R has *lexical (static) scope*.

---

Thence, in the foregoing example, we have:

```
environment(print_enclosure)  # print the associated environment
## <environment: R_GlobalEnv>
```

**Example 16.7** *Consider a function that prints out x defined outside of its scope:*

```
test <- function() print(x)
```

*Now:*

```
x <- "x in global"
test()
## [1] "x in global"
```

*It printed out x from the* user *workspace as it is precisely the environment associated with the function. However, setting the associated environment to another one that also happens to define x will give a different result:*

[14] There are two other types of functions: a `special` is an internal function that does not necessarily evaluate its arguments (e.g., **switch**, **if**, or **quote**; compare also Chapter 17), whereas a `builtin` always evaluates its actual parameters, e.g., **sum**.

```
environment(test) <- e3  # defined some time ago
test()
## [1] "bacon"
```

**Example 16.8** *Consider the following:*

```
test <- function()
{
    cat(sprintf("test: current env: %s\n", format(sys.frame(sys.nframe()))))

    subtest <- function()
    {
        e <- sys.frame(sys.nframe())
        cat(sprintf("subtest: enclosing env: %s\n", format(parent.env(e))))
        cat(sprintf("x = %s\n", x))
    }

    x <- "spam"
    subtest()
    environment(subtest) <- globalenv()
    subtest()
}

x <- "bacon"
test()
## test: current env: <environment: 0x5601c87db698>
## subtest: enclosing env: <environment: 0x5601c87db698>
## x = spam
## subtest: enclosing env: <environment: R_GlobalEnv>
## x = bacon
```

*Here is what happened.*

1. *A call to **`test`** creates a local function **`subtest`**, whose associated environment is set to the local frame of the current call. It is precisely the current environment where **`subtest`** was created (because R has lexical scope).*
2. *The above explains why **`subtest`** can access the local variable x inside its maker.*
3. *Then we change the environment associated with **`subtest`** to the global one.*
4. *In the next call to **`subtest`**, unsurprisingly, we gain access to x in the user workspace.*

---

**Note** In lexical (static) scoping, which variables a function refers to can be deduced by reading the function's body only and not how it is called in other contexts. This is the theory. Nevertheless, the fact that we can freely modify the associated environment anywhere can complicate the program analysis greatly.

If we find the rules of lexical scoping complicated, we should refrain from referring to objects outside of the current scope ("global" or "non-local" variables") except for the functions defined as top-level ones or imported from external packages. It is what we have been doing most of the time anyway.

### 16.3.3 Application: Function factories

As closures are functions with associated environments, and the role of environments is to store information, we can consider closures = functions + data. We have already seen that in Section 9.4.3, where we mentioned **approxfun**. To recall:

```
x <- seq(0, 1, length.out=11)
f1 <- approxfun(x, x^2)
print(f1)
## function (v)
## .approxfun(x, y, v, method, yleft, yright, f, na.rm)
## <environment: 0x5601c8a6d7a8>
```

The variables x, y, etc., that **f1**'s source code refers to, are stored in its associated environment:

```
ls(envir=environment(f1))
## [1] "f"      "method" "na.rm"  "x"      "y"      "yleft"  "yright"
```

**Important** Routines that return functions whose non-local variables are memorised in their associated environments are referred to as *function factories*.

**Example 16.9** *Consider a function factory:*

```
gen_power <- function(p)
    function(x) x^p  # p references a non-local variable
```

*A call to* **gen_power** *creates a local environment that defines one variable,* p*, where the argument's value is stored. Then, we create a function whose associated environment (remember that R uses lexical scoping) is that local one. It is where the reference to the non-local* p *in its body will be resolved. This new function is returned by* **gen_power** *to the caller. Normally, the local environment would be destroyed, but it is still used after the call. Thus, it will not be garbage-collected.*

*Example calls:*

```
(square <- gen_power(2))
## function (x)
## x^p
## <environment: 0x5601c91a4e10>
(cube <- gen_power(3))
## function (x)
## x^p
## <environment: 0x5601c9004740>
```

```
square(2)
## [1] 4
cube(2)
## [1] 8
```

*The underlying environment can, of course, be modified:*

```
assign("p", 7, envir=environment(cube))
cube(2)  # so much for the cube
## [1] 128
```

**Example 16.10** *`Negate` is another example of a function factory. The function it returns stores `f` passed as an argument.*

```
notall <- Negate(all)
notall(c(TRUE, TRUE, FALSE))
## [1] TRUE
```

*Study its source code:*

```
print(Negate)
## function (f)
## {
##     f <- match.fun(f)
##     function(...) !f(...)
## }
## <environment: namespace:base>
```

**Example 16.11** *In [38], the following example is given:*

```
account <- function(total)
    list(
        balance  = function() total,
        deposit  = function(amount) total <<- total+amount,
        withdraw = function(amount) total <<- total-amount
    )

Robert <- account(1000)
Ross <- account(500)
Robert$deposit(100)
Ross$withdraw(150)
Robert$balance()
## [1] 1100
Ross$balance()
## [1] 350
```

*We can now fully understand why this code does what it does. The return list consists of three functions whose enclosing environment is the same. `account` somewhat resembles the definition*

*of a class with three methods and one data field. No wonder why reference classes (Section 16.1.5) were introduced at some point: they are based on the same concept.*

**Exercise 16.12** *Write a function factory named* **gen_counter** *which implements a simple counter that is increased by one on each call thereto.*

```
gen_counter <- function() ...to.do...
c1 <- gen_counter()
c2 <- gen_counter()
c(c1(), c1(), c2(), c1(), c2())
## [1] 1 2 1 3 2
```

*Moreover, compose a function that resets a given counter to zero.*

```
reset_counter <- function(counter_fun) ...to.do...
reset_counter(c1)
c1()
## [1] 1
```

### 16.3.4 Accessing the calling environment

We know that the environment associated with a function is not necessarily the same as the environment from which the function was called, sometimes confusingly referred to as the *parent frame*.

R maintains a whole *frame stack*. The global environment is assigned the number 0. Each call to a function increases the stack by one frame, whereas returning from a call decreases the counter. To get the current frame number, we call **sys.nframe**. This is why **sys.frame(sys.nframe())** returns the current environment.

We can fetch the calling environment by referring to **parent.frame()** or **sys.frame(sys.parent())**, amongst others[15]. Thanks to **parent.frame**, we may evaluate arbitrary expressions in (on behalf of) the calling environment. Normally, we should never be doing that. However, a few functions rely on this feature, hence our avid interest in this possibility.

### 16.3.5 Package namespaces (*)

An R package pkg defines two environments:

- namespace:pkg is where all objects are defined (functions, vectors, etc.); it is the enclosing environment of all closures in the package;
- package:pkg contains selected[16] objects from namespace:pkg that can be accessed by the user; it can be attached to the search path.

[15] In **help**("sys.parent"), we read that the parent frame number, as returned by **sys.parent**(), is not necessarily equal to **sys.nframe**()-1. It is certainly true if we are at the top (global) level.

[16] Exported using the export or exportPattern directive in the package's NAMESPACE file; see Section 1 of [66].

As an illustration, we will use the example package discussed in Section 7.3.1.

```
library("rpackagedemo")  # https://github.com/gagolews/rpackagedemo/
## Loading required package: tools
```

Here is its `DESCRIPTION` file:

```
Package: rpackagedemo
Type: Package
Title: Just a Demo R Package
Version: 1.0.2
Date: 1970-01-01
Author: Anonymous Llama
Maintainer: Unnamed Kangaroo <roo@inthebush.au>
Description: Provides a function named bamboo(), just give it a shot.
License: GPL (>= 2)
Imports: stringx
Depends: tools
```

The `Import` and `Depends` fields specify which packages (apart from **base**) ours depends on. As we can see above, all items in the latter list are attached to the search path on a call to **`library`**.

The `NAMESPACE` file specifies the names imported from other packages and those that are expected to be visible to the user:

```
importFrom(stringx, sprintf)
importFrom(tools, toTitleCase)
S3method(print, koala)
S3method(print, kangaroo, .a_hidden_method_to_print_a_roo)
export(bamboo)
```

Thus, our package exports one object, a function named **`bamboo`** (we will discuss the S3 methods in the next section). It is included in the `package:rpackagedemo` environment attached to the search path:

```
ls(envir=as.environment("package:rpackagedemo")) # ls("package:rpackagedemo")
## [1] "bamboo"
```

Let's give it a shot:

```
bamboo("spanish inquisition")  # rpackagedemo::bamboo
## G'day, Spanish Inquisition!
```

We did not expect this at all, nor that its source code looks like:

```
print(bamboo)
## function (x = "world")
## cat(prepare_message(toTitleCase(x)))
## <environment: namespace:rpackagedemo>
```

We see a call to **toTitleCase** (most likely from **tools**, and this is indeed the case). Also, **prepare_message** is invoked but it is not listed in the package's imports (see the NAMESPACE file). We definitely cannot access it directly:

```
prepare_message
## Error: ! object 'prepare_message' not found
```

It is the package's *internal* function, which is included in the namespace:rpackagedemo environment.

```
(e <- environment(rpackagedemo::bamboo))  # or getNamespace("rpackagedemo")
## <environment: namespace:rpackagedemo>
ls(envir=e)
## [1] "bamboo"          "prepare_message" "print.koala"
```

We can fetch it via the `:::` operator:

```
print(rpackagedemo:::prepare_message)
## function (x)
## sprintf("G'day, %s!\n", x)
## <environment: namespace:rpackagedemo>
```

All functions defined in a package have the corresponding namespace as their associated environment. As a consequence, **bamboo** can refer to **prepare_message** directly.

It will be educative to inspect the enclosure of namespace:rpackagedemo:

```
(e <- parent.env(e))
## <environment: 0x5601c8310910>
## attr(,"name")
## [1] "imports:rpackagedemo"
ls(envir=e)
## [1] "sprintf"     "toTitleCase"
```

It is the environment carrying the bindings to all the imported objects. This is why our package can also refer to **stringx::sprintf** and **tools::toTitleCase**. Its enclosure is the *namespace* of the **base** package (not to be confused with package:base):

```
(e <- parent.env(e))
## <environment: namespace:base>
```

The next enclosure is, interestingly, the global environment:

```
(e <- parent.env(e))
## <environment: R_GlobalEnv>
```

Then, of course, the whole search path follows; see Figure 16.4 for an illustration.

---

**Note** (**) All environments related to packages are locked, which means that we can-not change any bindings inside their frames; compare **help**("lockEnvironment"). In the extremely rare event of our needing to *patch* an existing function within an already loaded package, we can call **unlockBinding** followed by **assign** to change its definition.

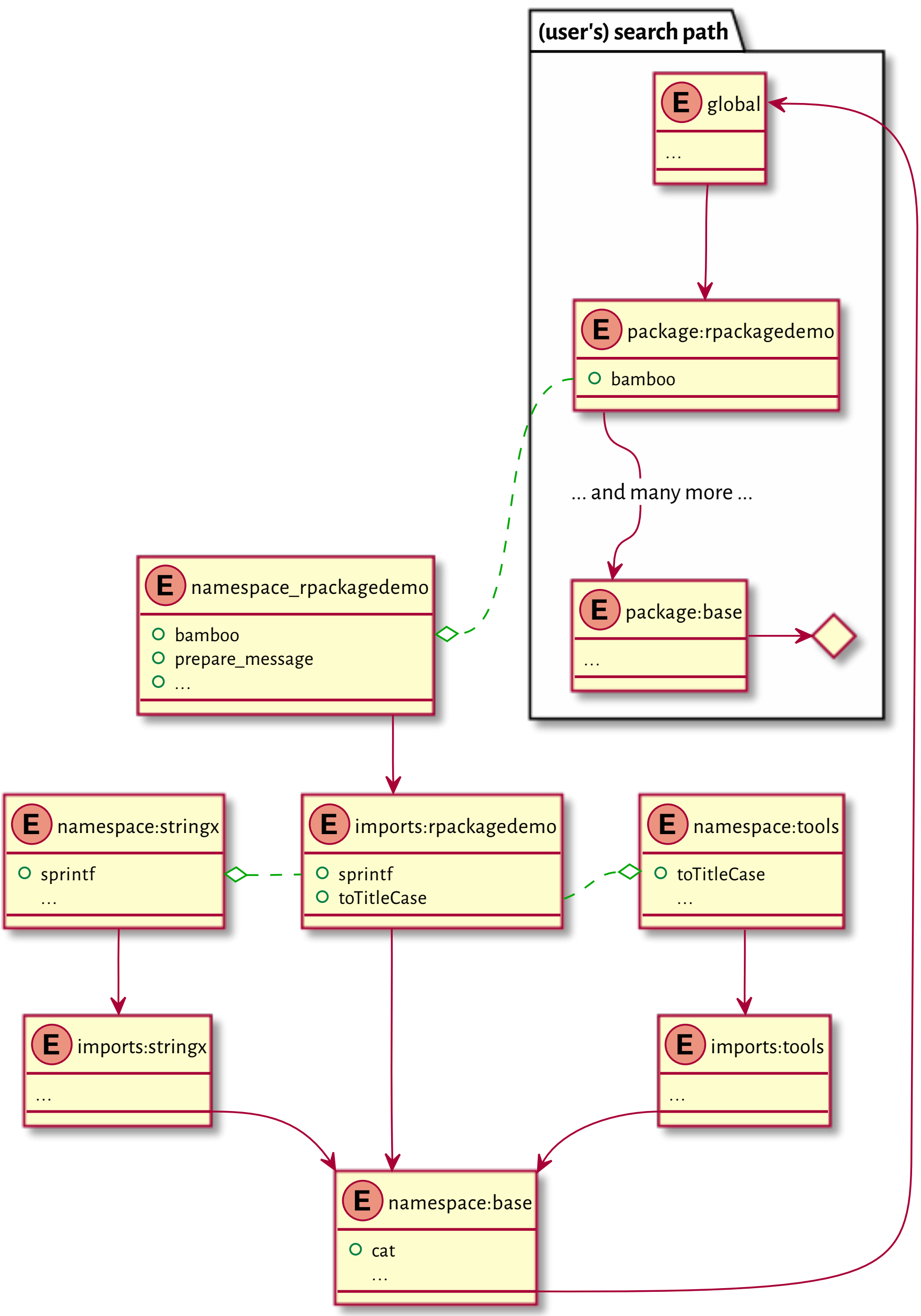

Figure 16.4. A search path for an example package. Dashed lines represent environments associated with closures, whereas solid lines denote enclosing environments. References to objects within each package are resolved inside their respective namespaces.

```
new_message <- function (x) sprintf("Nobody expects %s!\n", x)
e <- getNamespace("rpackagedemo")
environment(new_message) <- e  # set enclosing environment (very important!)
unlockBinding("prepare_message", e)
assign("prepare_message", new_message, e)
rm("new_message")
bamboo("the spanish inquisition")
## Nobody expects The Spanish Inquisition!
```

R is indeed a quite hackable language (except in the cases where it is not).

---

**Exercise 16.13** *(**) A function or a package might register certain functions (hooks) to be called on various events, e.g., attaching a package to the search patch; see* ***help("setHook")*** *and* ***help(".onAttach")****.*

1. *Inspect the source code of* ***plot.new*** *and notice a reference to a hook named "before.plot.new". Try setting such a hook yourself (e.g., one that changes some of the graphics parameters discussed in Section 13.2) and see what happens on each call to a plotting function.*
2. *Define the* ***.onLoad****,* ***.onAttach****,* ***.onUnload****, and* ***.onDetach*** *functions in your own R package and take note of when they are invoked.*

**Exercise 16.14** *(**) For the purpose of this book, we have registered a custom "before.plot.new" hook that sets our favourite graphics parameters that we listed in Section 13.2.3. Moreover, to obtain a white grid on a grey background, e.g., in Figure 13.13, we modified* ***plot.window*** *slightly. Apply similar hacks to the* ***graphics*** *package so that its outputs suit your taste better.*

### 16.3.6 S3 method lookup by UseMethod (*)

Inspecting the NAMESPACE file in **rpackagedemo**, we see that the package defines two **print** methods for objects of the classes koala and kangaroo. As the package is still attached to the search path, we can access these methods via a call to the corresponding generic:

```
print(structure("Tiny Teddy", class="koala"))
## This is a cute koala, Tiny Teddy
print(structure("Moike", class="kangaroo"))
## This is a very naughty kangaroo, Moike
```

The package does not make the definitions of these S3 methods available to the user, at least not directly. It is not the first time when we have experienced such an obscuration. In the first case, the method is simply hidden in the package namespace because it was not marked for exportation in the NAMESPACE file. However, it is still available under the expected name:

```
rpackagedemo:::print.koala
## function (x, ...)
## cat(sprintf("This is a cute koala, %s\n", x))
## <environment: namespace:rpackagedemo>
```

In the second case, the method appears under a very different identifier:

```
rpackagedemo:::.a_hidden_method_to_print_a_roo
## function (x, ...)
## cat(sprintf("This is a very naughty kangaroo, %s\n", x))
## <environment: namespace:rpackagedemo>
```

Since the base **UseMethod** is still able to find them, we suspect that there must be a global register of all S3 methods. And this is indeed the case. We can use **getS3method** to get access to what is available via **UseMethod**:

```
getS3method("print", "kangaroo")
## function (x, ...)
## cat(sprintf("This is a very naughty kangaroo, %s\n", x))
## <environment: namespace:rpackagedemo>
```

---

**Important** Overall, the search for methods is performed in two places:

1. in the environment where the generic is called (the current environment); this is why defining **print.kangaroo** in the current scope will use this method instead of the one from the package:

   ```
   print.kangaroo <- function(x, ...) cat("Nobody expects", x, "\n")
   print(structure("the Spanish Inquisition", class="kangaroo"))
   ## Nobody expects the Spanish Inquisition
   ```

2. in the internal S3 methods table (registration database).

See **help**("UseMethod") for more details. Also, recall that in Section 10.2.3, we said that **UseMethod** is not the only way to perform method dispatching. There are also internal generics and group generic functions.

---

**Exercise 16.15** *(*) Study the source code of* ***getS3method****. Note the reference to the* `base::`.__S3MethodsTable__.`` *object which is for R's internal use (we ought not to tinker with it directly). Moreover, study the* ***.S3method*** *function with which we can define new S3 methods not necessarily following the* ***generic.classname*** *convention.*

## 16.4 Exercises

**Exercise 16.16** *Asking too many questions is not very charismatic, but challenge yourself by finding the answer to the following.*

- *What is the role of a frame in an environment?*
- *What is the role of an enclosing environment? How to read it or set it?*
- *What is the difference between a named list and an environment?*
- *What functions and operators work on named lists but cannot be applied on environments?*
- *What do we mean by saying that environments are not passed by value to R functions?*
- *What do we mean by saying that objects are sometimes copied on demand?*
- *What happens if a name listed in an expression to be evaluated is not found in the current environment?*
- *How and what kind of objects can we attach to the search path?*
- *What happens if we have two identical object names on the search path?*
- *What do we mean by saying that package namespaces are locked when loaded?*
- *What is the current environment when we evaluate an expression "on the console"?*
- *What is the difference between* **`` `<-` ``** *and* **`` `<<-` ``***?*
- *Do packages have their own search paths?*
- *What is a function closure?*
- *What is the difference between the dynamic and the lexical scope?*
- *When evaluating a function, how is the enclosure of the current (local) environment determined? Is it the same as the calling environment? How to get it/them programmatically?*
- *How and why function factories work?*
- *(*) What is the difference between the* `package:pkg` *and* `namespace:pkg` *environments?*
- *How do we fetch the definition of an S3 method that does not seem to be available directly via the standard accessor* **`generic.classname`***?*
- *(*)* **`base::print.data.frame`** *calls* **`base::format.data.frame`** *(directly). Will the introduction of* **`print.data.frame`** *in the current environment affect how data frames are printed?*
- *(*) On the other hand,* **`base::format.data.frame`** *calls the generic* **`base::format`** *on all the input data frame's columns. Will the overloading of the particular methods affect how data frames are printed?*

**Exercise 16.17** *Calling:*

```
pkg <- available.packages()
pkg[, "Package"]  # a list of the names of available packages
pkg[, "Depends"]  # dependencies
```

*gives the list of available packages and their dependencies. Convert the dependency lists to a list of character vectors (preferably using regular expressions; see Section 6.2.4).*

*Then, generate a list of reverse dependencies: what packages depend on each given package.*

*Use an object of the type `environment` (a hash table) to map the package names to numeric IDs (indexes). It will significantly speed up the whole process (compare it to a named list-based implementation).*

**Exercise 16.18** *According to [70], compare also Section 9.3.6, a call to:*

```
add(x, f(x)) <<- v
```

*translates to:*

```
`*tmp*` <- get(x, envir=parent.env(), inherits=TRUE)
x <<- `add<-`(`*tmp*`, f(x), v)  # note: not f(`*tmp*`)
rm(`*tmp*`)
```

*Given:*

```
`add<-` <- function(x, where=TRUE, value)
{
    x[where] <- x[where] + value
    x  # the modified object that will replace the original one
}

y <- 1:5
f <- function() { y <- -(1:5); add(y, y==-3) <<- 1000; y }
```

*explain why we get the following results:*

```
f()
## [1] -1 -2 -3 -4 -5
print(y)
## [1]    1    2 1003    4    5
```

# 17

# *Lazy evaluation (**)*

The ability to create, store, and manipulate unevaluated expressions so that they can be computed later is not particularly special. Many languages enjoy such *metaprogramming* (computing *on* the language, reflection) capabilities, e.g., Lisp, Scheme, Wolfram, Julia, amongst many others. However, R inherited from its predecessor, the S language, a variation of lazy[1] (nonstrict, noneager, delayed) evaluation of function arguments. They are only computed when their *values* are first needed. As we can take the expressions used to generate them (via **`substitute`**; see Section 15.4.2), we shall see that we can ignore their meaning in the original (caller's) context and compute them in a very different one.

## 17.1 Evaluation of function arguments

We know that calls such as `` `if`(test, ifyes, ifno) ``, `` `||`(mustbe, maybe) ``, or `` `&&`(mustbe, maybe) `` do not have to evaluate all their arguments.

```
{cat(" first "); FALSE} && {cat(" second "); FALSE}
##  first
## [1] FALSE
{cat(" first "); TRUE } && {cat(" Spanish Inquisition "); FALSE}
##  first  Spanish Inquisition
## [1] FALSE
```

We can compose such functions ourselves. For instance:

```
test <- function(a, b, c) a + c  # b is not used
test({cat("spam\n"); 1}, {cat("eggs\n"); 10}, {cat("salt\n"); 100})
## spam
## salt
## [1] 101
```

The second argument was not referred to in the function's body. Therefore, it was not evaluated (no printing of *eggs* occurred).

[1] *Call-by-need* but without the memoisation of results generated by expressions which is available, e.g., in Haskell. In other words, in an expression like `c(f(x), f(x))`, the call `f(x)` will still be performed twice.

**Example 17.1** *Study the following very carefully.*

```
test <- function(a, b, c)
{
    cat("Arguments passed to `test` (expressions): \n")
    cat("a = ", deparse(substitute(a)), "\n")
    cat("b = ", deparse(substitute(b)), "\n")
    cat("c = ", deparse(substitute(c)), "\n")

    subtest <- function(x, y, z)
    {
        cat("Arguments passed to `subtest` (expressions): \n")
        cat("x = ", deparse(substitute(x)), "\n")
        cat("y = ", deparse(substitute(y)), "\n")
        cat("z = ", deparse(substitute(z)), "\n")
        cat("Using x and z... ")
        retval <- x + z  # does not refer to `y`
        cat("Cheers!\n")
        retval
    }

    cat("Using c... ")
    c  # force evaluation; we do not even have to be particularly creative

    subtest(a, ~!~b*2 := headache ->> ha@x$y, c*10)  # no evaluation yet!
}

environment(test) <- new.env()  # to spice things up

test(
    {testx <- "goulash"; cat("spam\n"); 1},
    {testy <- "kabanos"; cat("eggs\n"); MeAn(egGs+whatever&!!weird[stuff])},
    {testx <- "kransky"; cat("salt\n"); 100}
)
## Arguments passed to `test` (expressions):
## a =  {     testx <- "goulash"     cat("spam\n")     1 }
## b =  {     testy <- "kabanos"     cat("eggs\n")     MeAn(egGs + whatever …
## c =  {     testx <- "kransky"     cat("salt\n")     100 }
## Using c... salt
## Arguments passed to `subtest` (expressions):
## x =  a
## y =  `:=`(~!~b * 2, ha@x$y <<- headache)
## z =  c * 10
## Using x and z... spam
## Cheers!
## [1] 1001
print(testx)
## [1] "goulash"
print(testy)
## Error: ! object 'testy' not found
```

*On a side note, the `~` (formula) operator will be discussed in Section 17.6. Furthermore, the `:=` operator was used in an ancient version of R for assignments. The parser still recognises it, yet now it has no associated meaning.*

---

**Important** We note what follows.

- Either the evaluation of an argument does not happen, or is triggered only once (in which case the result is cached).

  This is why, in our example, *salt* was printed once.

- Evaluation is *delayed* until the very first request for the underlying value. We call it *lazy evaluation*.

  It can be delayed forever; *eggs* is never printed and `testy` is undefined.

- Evaluation takes place in the calling environment (parent frame).

  `testx` is equal to *goulash* after all.

- Merely passing arguments further to another function *usually* does not trigger the evaluation.

  We wrote *usually* because functions of the type `builtin` (e.g., **`c`**, **`list`**, **`sum`**, `` `+` ``, `` `&` ``, and `` `:` ``) always evaluate the arguments. There is no lazy evaluation in the case of the arguments passed to group generics; see **`help`**`("groupGeneric")` and Section 10.2.6. Furthermore, replacement functions' `values` arguments (Section 9.3.6) are computed eagerly.

- Fetching the expression passed as an argument using **`substitute`** (Section 15.4.2) or checking if an argument was provided with **`missing`** (Section 15.4.3) does not trigger the evaluation.

  We see *spam* printed much later.

---

**Exercise 17.2** *Study the source code of **`system.time`** and notice the use of delayed evaluation to measure the duration of the execution of a given expression. Note that **`on.exit`** (Section 17.4) reacts to possible exceptions.*

**Example 17.3** *The role of **`substitute`** is broader than just getting the expression passed as an argument. We can actually replace each occurrence of every name from a given dictionary (a named list or an environment). For instance:*

```
test <- function(x)
{
    subtest <- function(y)
    {
        ex <- substitute(x, env=parent.frame())  # substitute(x) is just `x`
        ey <- substitute(y)
        cat("ex =", deparse(ex), "\n")
        cat("ey =", deparse(ey), "\n")
```

```
        # not: eval(substitute(ey, list(x=ex)))
        eval(as.call(list(substitute, ey, list(x=ex))))
    }

    subtest(spam(!x[x](x)))
}

test(eels@hovercraft)
## ex = eels@hovercraft
## ey = spam(!x[x](x))
## spam(!eels@hovercraft[eels@hovercraft](eels@hovercraft))
```

*We fetched the expression passed as the `x` argument to the calling function. Then, we replaced every occurrence of `x` in the expression `ey`. On a side note, as **`substitute`** does not evaluate its first argument, if we called **`substitute`**`(ey, ...)` in the last expression of **`subtest`**, we would treat `ey` as a quoted name.*

**Exercise 17.4** *Study the source code of **`replicate`**:*

```
print(replicate)
## function (n, expr, simplify = "array")
## sapply(integer(n), eval.parent(substitute(function(...) expr)),
##     simplify = simplify)
## <environment: namespace:base>
```

*It creates a function that evaluates `expr` inside its local environment, which is new every time. Note that **`eval.parent`**`(expr)` is a shorthand for **`eval`**`(expr, `**`parent.frame`**`())`.*

---

**Note** (*) Internally, lazy evaluation of arguments is implemented using the so-called *promises*, compare [70], which consist of:

- an expression (which we can access by calling **`substitute`**);
- an environment where the expression is to be evaluated (once this happens, it is set to `NULL`);
- a cached value (computed on demand, once).

This interface is not really visible from within R, but see **`help`**`("delayedAssign")`.

---

**Exercise 17.5** *Inspect the definition of **`match.fun`**. Why is it called by, e.g., **`apply`**, **`Map`**, or **`outer`**? Note that it uses **`eval.parent(substitute(substitute`**`(FUN)))` to fetch the expression representing the argument passed by the calling function (but it is probably very rarely needed there). Compare:*

```
test <- function(x)
{
    subtest <- function(y)
```

```
    {
        # NOT: substitute(y)
        # NOT: eval.parent(substitute(y))
        eval.parent(substitute(substitute(y)))
    }

    subtest(x*3)
}

test(1+2)
## (1 + 2) * 3
```

**Exercise 17.6** (*) *Implement your version of the* **bquote** *function.*

## 17.2 Evaluation of default arguments

As we know from Section 9.4.4, default arguments are special expressions specified in a function's parameter list.

**Important** When a function's body requires the value of an argument that the caller did not provide, the default expression will be evaluated *in the current (local) environment* of the function.

It is thus different from the case of normally passed arguments, which are interpreted in the context of the calling environment.

**Example 17.7** *Study the following very carefully.*

```
x <- "banana"

test <- function(y={cat("spam\n"); x})
{
    cat(deparse(substitute(y)), "\n")
    cat("bacon\n")
    x <- "rotten potatoes"
    cat(y, y, "\n")
}

test({cat("spam\n"); x})
## {     cat("spam\n")     x }
## bacon
## spam
## banana banana
```

*As usual, the evaluation is triggered only once, where it was explicitly requested, and only when needed.* y *was bound to the value of* x *from the* calling *environment* (banana *in the global one*).

```
test()
## {     cat("spam\n")     x }
## bacon
## spam
## rotten potatoes rotten potatoes
```

*The expression for the default* y *was evaluated in the* local *environment. It happened after the creation of the local* x.

**Example 17.8** *Consider the following example from [38]:*

```
sumsq <- function(y, about=mean(y), na.rm=FALSE)
{
    if (na.rm)
        y <- y[!is.na(y)]
    sum((y - about)^2)
}

y <- c(1, NA_real_, NA_real_, 2)
sumsq(y, na.rm=TRUE)
## [1] 0.5
```

*In the case where we rely on the default argument, the computation of the mean may take into account the request for missing value removal. Still, the following will not work as* intended*:*

```
sumsq(y, mean(y), na.rm=TRUE)  # we should rather pass mean(y, na.rm=TRUE)
## [1] NA
```

*However, as the idea of lazy evaluation of arguments is alien to most programmers (especially those coming from different languages), it might be better to rewrite the above using a call to* **missing** (Section 15.4.3)*:*

```
sumsq <- function(y, about, na.rm=FALSE)
{
    if (na.rm)
        y <- y[!is.na(y)]
    if (missing(about))
        about <- mean(y)
    sum((y - about)^2)
}

sumsq(y, na.rm=TRUE)
## [1] 0.5
```

*or better even:*

```
sumsq <- function(y, about=NULL, na.rm=FALSE)
{
    if (na.rm)
        y <- y[!is.na(y)]
    if (is.null(about))
        about <- mean
    sum((y - about(y))^2)
}

sumsq(y, na.rm=TRUE)
## [1] 0.5
```

**Exercise 17.9** *The default arguments to* **`do.call`**, **`list2env`**, *and* **`new.env`** *are set to* **`parent.frame`**. *What does that mean?*

**Exercise 17.10** *Study the source code of the* **`local`** *function:*

```
print(local)
## function (expr, envir = new.env())
## eval.parent(substitute(eval(quote(expr), envir)))
## <environment: namespace:base>
```

## 17.3 Ellipsis revisited

If our function has the dot-dot-dot parameter, `` `...` ``, whatever we pass through it is packed into a pairlist of promise expressions. Thus, we can relish the benefits of lazy evaluation. In particular, we can redirect all `` `...` ``-fed arguments to another call as-is.

```
test <- function(...)
{
    subtest <- function(x, ...)
    {
        cat("x   = "); str(x)
        cat("... = "); str(list(...))
    }

    subtest(...)
}

test({cat("eggs! "); 1}, {cat("spam! "); 2}, z={cat("rice! "); 3})
## x   = eggs!  num 1
## ... = spam! rice! List of 2
##  $  : num 2
##  $ z: num 3
```

**Exercise 17.11** *In the documentation of* **`lapply`**, *we read that this function is called like* **`lap-`**

*`ply(X, FUN, ...)`, where `` `...` `` are* optional arguments to `FUN`. *Verify that whatever is passed via the ellipsis is evaluated only once and not on each application of `FUN` on the elements of X.*

**Example 17.12** *We know from Chapter 13 that many high-level graphics functions rely on multiple calls to more primitive routines that allow for setting a variety of parameters (e.g., via `par`). A common scenario is for a high-level function to pass* all *the arguments down. Each underlying procedure can then decide by itself which items it is interested in.*

```
test <- function(...)
{
    subtest1 <- function(..., a=1) c(a=a)
    subtest2 <- function(..., b=2) c(b=b)
    subtest3 <- function(..., c=3) c(c=c)

    c(subtest1(...), subtest2(...), subtest3(...))
}

test(a="A", b="B", d="D")
##   a   b   c
## "A" "B" "3"
```

*Here, for instance, `subtest1` only consumes the value of a and ignores all other arguments whatsoever. `plot.default` (amongst others) relies on such a design pattern.*

`` `...length` `` fetches the number of items passed via the ellipsis, `` `...names` `` retrieves their names (in the case they are given as keyword arguments), and `` `...elt`(i) `` gives the *value* of the $i$-th element. Furthermore, `` `..1` ``, `` `..2` ``, and so forth are synonymous with `` `...elt`(1) ``, `` `...elt`(2) ``, etc.

```
test <- function(...)
{
    cat("length:", ...length(), "\n")
    cat("names: ", paste(...names(), collapse=", "), "\n")
    for (i in seq_len(...length()))
        cat(i, ":", ...elt(i), "\n")
    print(substitute(...elt(i)))
}

test(u={cat("honey! "); "a"}, {cat("gravy! "); "b"}, w={cat("bacon! "); "c"})
## length: 3
## names:  u, , w
## honey! 1 : a
## gravy! 2 : b
## bacon! 3 : c
## ...elt(3L)
```

Note that `` `...elt`(i) `` triggers the evaluation of the respective argument. Unfortunately, we cannot use `substitute` to fetch the underlying expression. Instead, we can rely on `match.call` discussed in Section 15.4.4:

```
test <- function(a, b, ..., z=1)
{
    e <- match.call()[-1]
    as.list(e[names(e) %notin% names(formals(sys.function()))])
}

str(test(1+1, 2+2, 3+3, 4+4, a=2, z=8, w=4))
## List of 4
##  $  : language 2 + 2
##  $  : language 3 + 3
##  $  : language 4 + 4
##  $ w: num 4
```

**Note** Objects passed via `...`, even if they are specified as keyword arguments, cannot be referred to by their name as if they were local variables:

```
test <- function(...) zzz
test(zzz=3)
## Error in `test()`: ! object 'zzz' not found
```

In other words, no assignment in the local environment is triggered.

**Exercise 17.13** *Implement your version of the **`switch`** function.*

**Exercise 17.14** *Write your version of the **`stopifnot`** function.*

## 17.4 on.exit (*)

**`on.exit`** registers an expression to be evaluated at the very end of a call, regardless of whether the function exited due to an error or not. It might be used to reset the temporarily modified graphics parameters (see **`par`**) and system options (**`options`**) or clean up the allocated resources (e.g., close all open file connections). For instance:

```
test <- function(reset=FALSE, error=FALSE)
{
    on.exit(cat("eggs\n"))
    on.exit(cat("bacon\n"))  # replace
    on.exit(cat("spam\n"), add=TRUE)  # add

    cat("roti canai\n")

    if (reset)
        on.exit()  # cancels all (replace by nothing)
```

```
    if (error)
        stop("aaarrgh!")

    cat("end\n")
    "return value"
}

test()
## roti canai
## end
## bacon
## spam
## [1] "return value"
test(reset=TRUE)
## roti canai
## end
## [1] "return value"
test(error=TRUE)
## roti canai
## Error in `test()`: ! aaarrgh!
```

We can always manage without **on.exit**, e.g., by applying exception handling techniques; see Section 8.2.

**Exercise 17.15** *In the definition of* ***scan****, notice the call to:*

```
on.exit(close(file))
```

*Is its purpose to close the file on exit?*

**Exercise 17.16** *Why does* ***graphics::barplot.default*** *call the following expressions?*

```
dev.hold()
opar <- if (horiz) par(xaxs="i", xpd=xpd) else par(yaxs="i", xpd=xpd)
on.exit({
    dev.flush()
    par(opar)
})
```

## 17.5 Metaprogramming and laziness in action: Examples (*)

Due to lazy evaluation, we can define functions that permit any random yet syntactically valid gibberish to be fed as their arguments. Nothing but basic decency stops us from interpreting them in any way we want. Each such function can become a micro-verse (a microlanguage?) by itself. This will surely confuse[2] our users, as they will have to analyse every procedure's behaviour separately.

In this section, we extend on our notes from Section 9.4.7 and Section 12.3.9. We look at a few functions relying on metaprogramming and laziness, mostly because studying them is a good exercise. It can help extend our programming skills and deepen our understanding of the concepts discussed in this part of the book. By no means is it an invitation to use them in practice. Nevertheless, R's computing on the language capabilities might interest some advanced programmers (e.g., package developers).

### 17.5.1 match.arg

**match.arg** was mentioned in Section 9.4.7. When called normally, it matches a string against a set of possible choices, similarly to **pmatch**:

```
choices <- c("spam", "bacon", "eggs")
match.arg("spam", choices)
## [1] "spam"
match.arg("s", choices)  # partial matching
## [1] "spam"
match.arg("eggplant", choices)  # no match
## Error in `match.arg()`: ! 'arg' should be one of "spam", "bacon", "eggs"
match.arg(choices, choices)  # match first
## [1] "spam"
```

However, skipping the second argument, this function will fetch the choices from the default argument of the function it is enclosed in!

```
test <- function(x=c("spam", "bacon", "eggs"))
    match.arg(x)

test("spam")
## [1] "spam"
test("s")
## [1] "spam"
test("eggplant")
## Error in `match.arg()`: ! 'arg' should be one of "spam", "bacon", "eggs"
test()
## [1] "spam"
```

**Exercise 17.17** *Inspect the source code of* **`stats::binom.test`***, which looks like:*

```
function(..., alternative = c("two.sided", "less", "greater"))
{
    # ...
    alternative <- match.arg(alternative)
```

[2] Novices are prone to overgeneralising when they learn new material that they are still far from comfortable with. Such exceptions go against this natural coping strategy of theirs.

```
    # ...
}
```

*Read the description of the `alternative` argument in the documentation.*

**Exercise 17.18** *Study the source code of **`match.arg`**. In particular, notice the following fragment:*

```
if (missing(choices)) {
    formal.args <- formals(sys.function(sysP <- sys.parent()))
    choices <- eval(
        formal.args[[as.character(substitute(arg))]],
        envir=sys.frame(sysP)
    )
}
```

### 17.5.2 curve

The **`curve`** function can be called, e.g., like:

```
curve(sin(1/x^2), 1/pi, 3, 1001, lty=2)
```

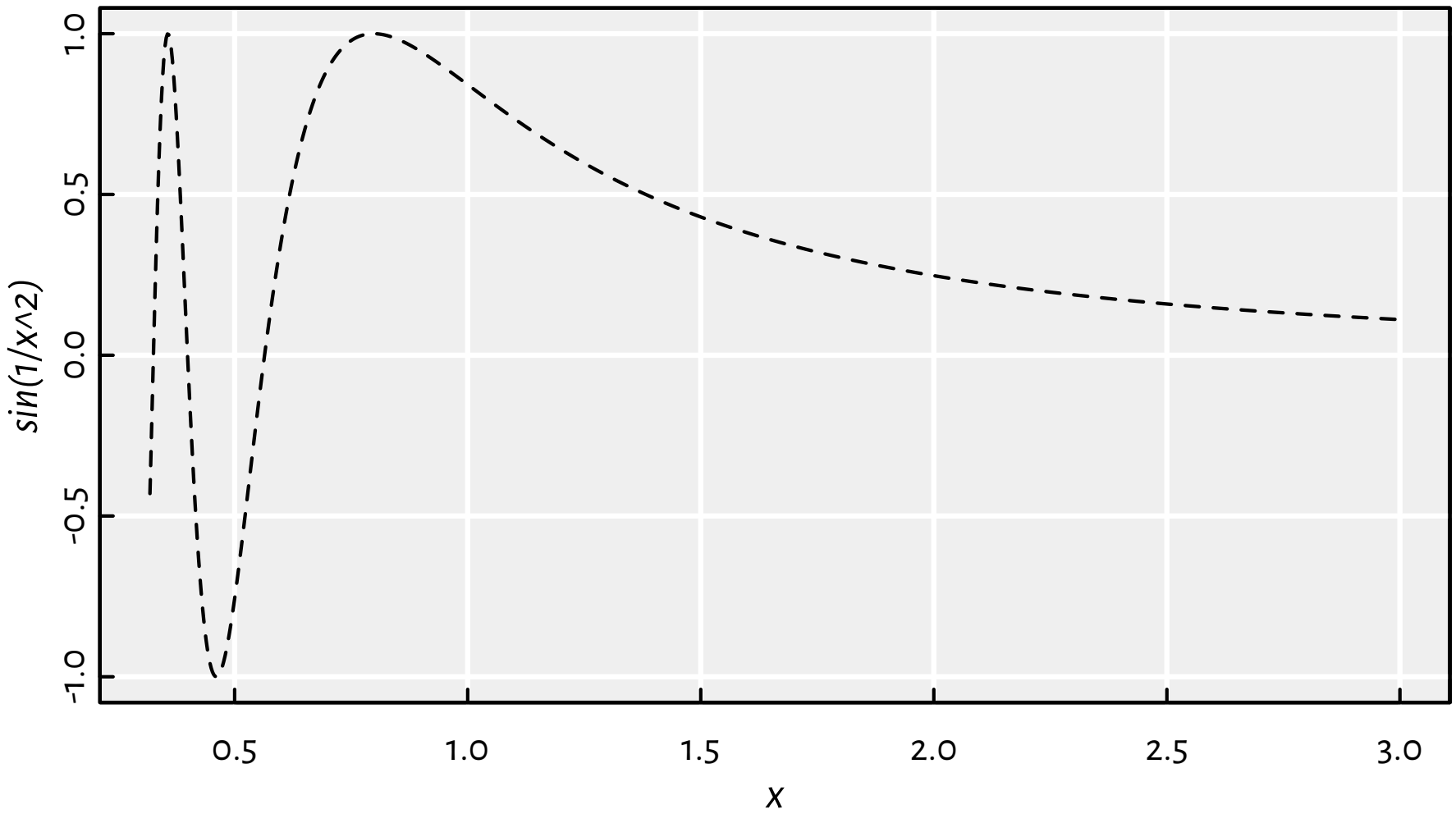

Figure 17.1. An example plot generated by calling **`curve`**.

It results in Figure 17.1. Wait a minute… We did not define `x` as a sequence ranging between about 0.3 and 3!

**Exercise 17.19** *Study the source code of **`curve`**. Take note of the following code fragment:*

```
function(expr, from=NULL, to=NULL, n=101, xlab="x", type="l", ...)
{
    # ...
    expr <- substitute(expr)
    ylab <- deparse(expr)
    x <- seq.int(from, to, length.out=n)
    ll <- list(x=x)
    y <- eval(expr, envir=ll, enclos=parent.frame())
    plot(x=x, y=y, type=type, xlab=xlab, ylab=ylab, ...)
    # ...
}
```

### 17.5.3 `with` and `within`

Environments and named lists (and hence data frames) are similar (Section 16.1.2). Due to this, the `envir` argument to **`eval`** can be set to either. Therefore, for instance:

```
eval(quote(head(Sepal.Length)), envir=iris)
## [1] 5.1 4.9 4.7 4.6 5.0 5.4
```

It evaluates the given expression in something like **`list2env`**`(iris, parent=`**`parent.frame`**`())`. Thus, even though `Sepal.Length` is not a standalone variable, it is treated as one *inside* the `iris` data frame.

Moreover, the enclosure is set to the calling frame. Hence, we can successfully refer to the **`head`** function located somewhere on the search path. This is somewhat similar to **`attach`** (Section 16.2.6) but without modifying the search path.

The **`with`** function does exactly the above:

```
print(with.default)
## function (data, expr, ...)
## eval(substitute(expr), data, enclos = parent.frame())
## <environment: namespace:base>
```

Example use:

```
with(iris, {
    mean(Sepal.Length)  # `Sepal.Length` is in `iris`
})
## [1] 5.8433
```

As we evaluate it in the local (temporary) environment, we cannot modify the existing columns of the data frame this way. However, the **`within`** function includes a way to detect and apply any changes made.

```
within(iris, {
    Sepal.Length <- Sepal.Length/1000
    Spam <- "yum!"
```

```
}) -> iris2
head(iris2, 3)
##   Sepal.Length Sepal.Width Petal.Length Petal.Width Species Spam
## 1       0.0051         3.5          1.4         0.2  setosa yum!
## 2       0.0049         3.0          1.4         0.2  setosa yum!
## 3       0.0047         3.2          1.3         0.2  setosa yum!
```

**Exercise 17.20** *Study the source code of* **`within`***:*

```
print(within.data.frame)
## function (data, expr, ...)
## {
##     parent <- parent.frame()
##     e <- evalq(environment(), data, parent)
##     eval(substitute(expr), e)
##     l <- as.list(e, all.names = TRUE)
##     l <- l[!vapply(l, is.null, NA, USE.NAMES = FALSE)]
##     nl <- names(l)
##     del <- setdiff(names(data), nl)
##     data[nl] <- l
##     data[del] <- NULL
##     data
## }
## <environment: namespace:base>
```

*Note that* **`evalq`**`(expr, ...)` *is equivalent to* **`eval`**`(`**`quote`**`(expr), ...)`*. Also,* **`vapply`**`(X, FUN, NA, ...)` *is like a call to* **`sapply`***, but it guarantees that the result is a logical vector.*

### 17.5.4 `transform`

We can call **`transform`** to modify/add columns in a data frame using vectorised functions. For instance:

```
head(transform(mtcars, log_hp=log(hp), am=2*am-1, hp=NULL), 3)
##                mpg cyl disp drat    wt  qsec vs am gear carb log_hp
## Mazda RX4     21.0   6  160 3.90 2.620 16.46  0  1    4    4 4.7005
## Mazda RX4 Wag 21.0   6  160 3.90 2.875 17.02  0  1    4    4 4.7005
## Datsun 710    22.8   4  108 3.85 2.320 18.61  1  1    4    1 4.5326
```

If we suspect that this function evaluates all expressions passed as `` `...` `` *within* the data frame, we are brilliantly right. Furthermore, there must be a mechanism to detect newly created variables so that new columns can be added.

**Exercise 17.21** *Study the source code of* **`transform`***:*

```
print(transform.data.frame)
## function (`_data`, ...)
## {
```

```
##     e <- eval(substitute(list(...)), `_data`, parent.frame())
##     tags <- names(e)
##     inx <- match(tags, names(`_data`))
##     matched <- !is.na(inx)
##     if (any(matched)) {
##         `_data`[inx[matched]] <- e[matched]
##         `_data` <- data.frame(`_data`, check.names = FALSE)
##     }
##     if (!all(matched)) {
##         args <- e[!matched]
##         args[["check.names"]] <- FALSE
##         do.call("data.frame", c(list(`_data`), args))
##     }
##     else `_data`
## }
## <environment: namespace:base>
```

*In particular, note that* `e` *is a named list.*

### 17.5.5 subset

The **`subset`** function selects rows and columns of a data frame that meet certain criteria. For instance:

```
subset(airquality, Temp>95 | Temp<57, -(Month:Day))
##     Ozone Solar.R Wind Temp
## 5      NA      NA 14.3   56
## 120    76     203  9.7   97
## 122    84     237  6.3   96
```

The second argument, the row selector, must definitely be evaluated *within* the data frame. We expect it to reduce itself to a logical vector which can then be passed to the index operator.

The "select all columns except those between the given ones" part can be implemented by assigning each column a consecutive integer and then treating them as numeric indexes.

**Exercise 17.22** *Study the source code of* **`subset`***:*

```
print(subset.data.frame)
## function (x, subset, select, drop = FALSE, ...)
## {
##     chkDots(...)
##     r <- if (missing(subset))
##         rep_len(TRUE, nrow(x))
##     else {
##         e <- substitute(subset)
```

```
##         r <- eval(e, x, parent.frame())
##         if (!is.logical(r))
##             stop("'subset' must be logical")
##         r & !is.na(r)
##     }
##     vars <- if (missing(select))
##         rep_len(TRUE, ncol(x))
##     else {
##         nl <- as.list(seq_along(x))
##         names(nl) <- names(x)
##         eval(substitute(select), nl, parent.frame())
##     }
##     x[r, vars, drop = drop]
## }
## <environment: namespace:base>
```

### 17.5.6 Forward pipe operator

Section 10.4 mentioned the pipe operator, `` `|>` ``. We can compose its simplified version manually:

```
`%>%` <- function(e1, e2)
{
    e2 <- as.list(substitute(e2))
    e2 <- as.call(c(e2[[1]], substitute(e1), e2[-1]))
    eval(e2, envir=parent.frame())
}
```

This function imputes `e1` as the first argument in a call `e2` and then evaluates the new expression.

Example calls:

```
x <- c(1, NA_real_, 2, 3, NA_real_, 5)
x %>% mean  # mean(x)
## [1] NA
x %>% `-`(1)  # x-1
## [1]  0 NA  1  2 NA  4
x %>% na.omit %>% mean  # mean(na.omit(x))
## [1] 2.75
x %>% mean(na.rm=TRUE)  # mean(x, na.rm=TRUE)
## [1] 2.75
```

Moreover, we can memorise the value of `e1` so that it can be referred to in the expression on the right side of the operator. This comes at a cost of forcing the evaluation of the left-hand side argument and thus losing the potential benefits of laziness, including access to the generating expression.

```
`%.>%` <- function(e1, e2)
{
    env <- list2env(list(.=e1), parent=parent.frame())
    e2 <- as.list(substitute(e2))
    e2 <- as.call(c(e2[[1]], quote(.), e2[-1]))
    eval(e2, envir=env)
}
```

This way, we can refer to the value of the left side multiple times in a single call. For instance:

```
runif(5) %.>% `[`(.>0.5)    # x[x>0.5] with x=runif(5)
## [1] 0.78831 0.88302 0.94047
```

This is crazy, I know. I made this. Your author. One more then:

```
# x[x >= 0.5 & x <= 0.9] <- NA_real_ with x=round(runif(5), 2):
runif(5) %.>% round(2) %.>% `[<-`(.>=0.5 & .<=0.9, value=NA_real_)
## [1] 0.29   NA 0.41   NA 0.94
```

I cannot wait for someone to put this operator into a new R package (it is a brilliant idea, by the way, isn't it?) and then confuse thousands of users ("What is this thing?").

### 17.5.7 Other ideas (**)

Why stop ourselves here? We can create way more invasive functions that read the local variables in the calling functions (unless they are primitive; in R, there are often exceptions to general rules…). Here is an operator which helps select a range of columns in a data frame between two given labels:

```
`%:%` <- function(e1, e2)
{
    # get the `x` argument in the caller (hoping its `[`)
    x <- get("x", envir=sys.frame(sys.nframe()-1))
    n <- names(x)
    from <- pmatch(substitute(e1), n)
    to <- pmatch(substitute(e2), n)
    from:to
}

head(iris[, Sepal.W%:%Petal.W])
##   Sepal.Width Petal.Length Petal.Width
## 1         3.5          1.4         0.2
## 2         3.0          1.4         0.2
## 3         3.2          1.3         0.2
## 4         3.1          1.5         0.2
## 5         3.6          1.4         0.2
## 6         3.9          1.7         0.4
```

This operator relies on the assumption that it is called in the expression passed as an argument to a non-primitive function which also takes a named vector x as an actual parameter. So ugly, but saves a few keystrokes. We will not be using it because it is not good for us.

**Exercise 17.23** *Make the foregoing more foolproof:*

- *if `%:%` is used outside of `[` or `[<-`, raise a polite error,*
- *permit x to be a matrix (is it possible?),*
- *prepare better for the case of less expected inputs.*

**Exercise 17.24** *Modify the definition of the aforementioned operator so that both:*

```
iris[, -Sepal.W%:%Petal.W]
iris[, -(Sepal.W%:%Petal.W)]
```

*mean "select everything except".*

**Exercise 17.25** *Define `%:%` for data frames so that:*

- *x[%:%3, ] means "select the first three rows",*
- *x[3%:%, ] means "select from the third to the end",*
- *x[-3%:%, ] means "select from the third last to the end",*
- *x[%:%-10, ] means "select all but the last nine".*

*You can go one step further and redefine `[` entirely to support such kinds of indexers.*

The ceiling is the limit. Please, do not use it in production.

## 17.6 Processing formulae, `~` (*)

Formulae were introduced to S in the early 1990s [14]. Their original raison d'être was to specify *statistical models*; compare Section 10.3.4. From the language perspective, they are merely unevaluated calls to the `~` (tilde) operator. When creating them, we do not even have to apply **quote** explicitly. For instance:

```
f <- (y ~ x1 + x2)  # or: `~`(y, x1+x2)
mode(f)
## [1] "call"
class(f)
## [1] "formula"
```

Hence, formulae are compound objects in the sense given in Chapter 10. Usually, they are equipped with an additional attribute:

```
attr(f, ".Environment")  # environment active when the formula was created
## <environment: R_GlobalEnv>
```

**Exercise 17.26** *Write a function that generates a list of formulae of the form "*`y ~ x1+x2+...+xk`*", for all possible combinations* `x1`*,* `x2`*, ...,* `xk` *(of any cardinality) of elements in a given set of* `xs`*. For instance:*

```
formula_allcomb <- function(y, xs, env=parent.frame()) ...to.do...
str(formula_allcomb("len", c("supp", "dose")))
## List of 3
##  $ :Class 'formula'  language len ~ supp + dose
##   .. ..- attr(*, ".Environment")=<environment: R_GlobalEnv>
##  $ :Class 'formula'  language len ~ dose
##   .. ..- attr(*, ".Environment")=<environment: R_GlobalEnv>
##  $ :Class 'formula'  language len ~ supp
##   .. ..- attr(*, ".Environment")=<environment: R_GlobalEnv>
str(formula_allcomb(
    "y",
    c("x1", "x2", "x3"),
    env=NULL
))
## List of 7
##  $ :Class 'formula'  language y ~ x1 + x2 + x3
##  $ :Class 'formula'  language y ~ x2 + x3
##  $ :Class 'formula'  language y ~ x1 + x3
##  $ :Class 'formula'  language y ~ x3
##  $ :Class 'formula'  language y ~ x1 + x2
##  $ :Class 'formula'  language y ~ x2
##  $ :Class 'formula'  language y ~ x1
```

As they are unevaluated calls, functions can assign any fantastic meaning to formulae. We cannot really do anything about this freedom of expression. However, many functions, especially in the **stats** and **graphics** packages, rely on a call to **model.frame** and related routines. Thanks to this, we can at least find a few behavioural patterns. In particular, `help("formula")` lists the typical meanings of operators that can be used in a formula.

**Example 17.27** *Here are a few examples (executing these expressions is left as an exercise).*

- *Draw a box plot for* `iris[["Sepal.Width"]]` *split by* `iris[["Species"]]`*:*

  ```
  boxplot(Sepal.Width~Species, data=iris)
  ```

- *Draw a box plot for* `ToothGrowth[["len"]]` *split by a combination of levels in* `ToothGrowth[["supp"]]` *and* `ToothGrowth[["dose"]]`*:*

  ```
  boxplot(len~supp:dose, data=ToothGrowth)
  ```

- *Split the given data frame by a combination of values in two specified columns therein:*

```
split(ToothGrowth, ~supp:dose)
```

- *Order a data frame with respect to one or more columns:*

```
sort_by(mtcars, ~list(am, -mpg))
```

- *Fit a linear regression model of the form $y = a+bx$, where $y$ is* `iris[["Sepal.Length"]]` *and $x$ is* `iris[["Petal.Length"]]`*:*

```
lm(Sepal.Length~Petal.Length, data=iris)
```

- *Fit a linear regression model without the intercept term of the form $z = ax + by$, where $z$ is* `iris[["Sepal.Length"]]`*, $x$ is* `iris[["Petal.Length"]]`*, and $y$ is* `iris[["Sepal.Width"]]`*:*

```
lm(Sepal.Length~Petal.Length+Sepal.Width+0, data=iris)
```

- *Fit a linear regression model of the form $z = a+bx+cy+dxy$, where $z$ is* `iris[["Sepal.Length"]]+e` *(with e fetched from the associated environment), and $x$ and $y$ are like above:*

```
e <- rnorm(length(iris[["Sepal.Length"]]), 0, 0.05)
lm(I(Sepal.Length+e)~Petal.Length*Sepal.Width, data=iris)
```

- *Draw scatter plots of* `warpbreaks[["breaks"]]` *vs their indexes for data grouped by a combination of* `warpbreaks[["wool"]]` *and* `warpbreaks[["tension"]]`*:*

```
Index <- seq_len(NROW(warpbreaks))
coplot(breaks ~ Index | wool * tension, data=warpbreaks)
```

From the perspective of this book, which focuses on more universal aspects of the R language, formulae are not interesting enough to describe them in any more detail. However, the tender-hearted reader is now equipped with all the necessary knowledge to solve the following very educative exercises.

**Exercise 17.28** *Study the source code of* **`graphics:::boxplot.formula`**, **`stats::lm`**, *and* **`stats:::t.test.formula`** *and notice how they prepare and process the calls to* **`model.frame`**, **`model.matrix`**, **`model.response`**, **`model.weights`**, *etc. Note that their main aim is to prepare data to be passed to* **`boxplot.default`**, **`lm.fit`** *(it is just a function with such a name, not an S3 method), and* **`t.test.default`**.

**Exercise 17.29** *Write a function similar to* **`curve`**, *but one that lets us specify the function to plot using a formula.*

## 17.7 Exercises

**Exercise 17.30** *Answer the following questions.*

- *What is the role of promises?*
- *Why do we generally discourage the use of functions relying on metaprogramming?*
- *How are default arguments evaluated?*
- *Is there anything special about formulae from the language perspective?*
- *We said that R evaluates function arguments lazily. Does it mean that "`y[c(length(y)+1, length(y)+1, length(y)+1)] <- list(1, 2, 3)`" extends a list y by three elements? Or are there cases where evaluation is eager?*

**Exercise 17.31** *Why the two following calls yield different results?*

```
test <- function(x, y=deparse(substitute(x)), force_first=FALSE)
{
    if (force_first) y  # just force the evaluation of `y` here
    x <- x**2
    print(y)
}

test(1:5)
## [1] "c(1, 4, 9, 16, 25)"
test(1:5, force_first=TRUE)
## [1] "1:5"
```

## 17.8 Outro

Recall our first approximation to the classification of R data types that we presented in the *Preface*. To summarise what we have covered in this book, let's contemplate Figure 17.2, which gives a much broader picture.

If we omitted something, it was most likely on purpose: either we can now study it on our own easily, it is not really worth our attention, or it violates our *minimalist* design principles that we set forth in the *Preface*.

Now that we have reached the end of this course, we might be interested in reading:

- *R Language Definition* [70],
- *R Internals* [69],
- *Writing R Extensions* [66],
- R's source code available at https://cran.r-project.org/src/base.

What is more, the `NEWS` files available at https://cran.r-project.org/doc/manuals/r-release will keep us updated with fresh features, bug fixes, and newly deprecated functionality; see also the **news** function.

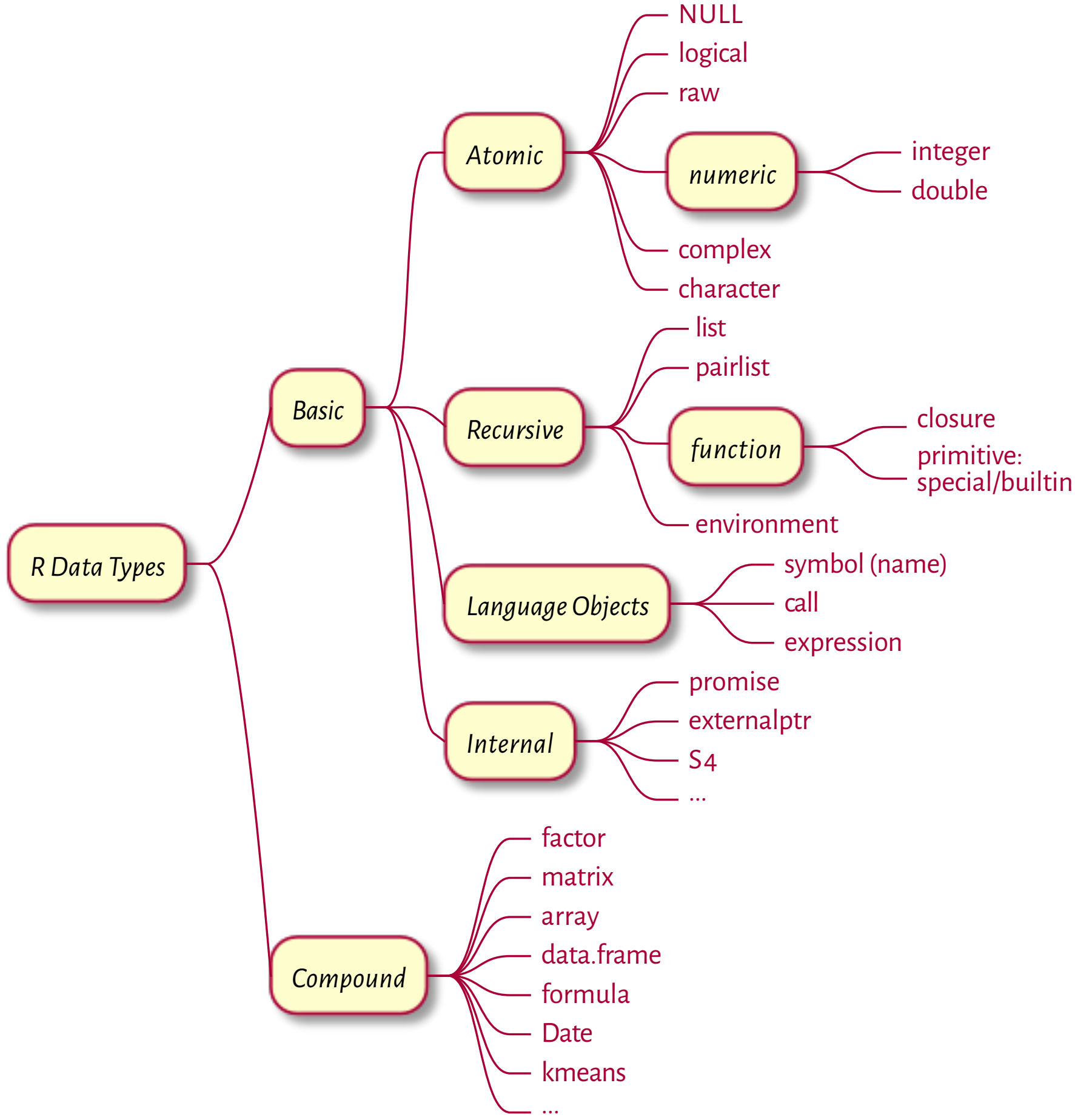

Figure 17.2. R data types.

Please spread the news about this book. Also, check out another open-access work by yours truly, *Minimalist Data Wrangling with Python*[3] [28]. Thank you.

Good luck with your further projects!

[3] https://datawranglingpy.gagolewski.com/

# Changelog

**Important** Any bug/typo reports/fixes[4] are appreciated. The most up-to-date version of this book can be found at https://deepr.gagolewski.com/.

Below is the list of the most noteworthy changes:

- **2026-07-30** (v1.0.2):
  - Updated to R 4.6.1.
  - Sections on the QR and SVD decompositions of matrices (Section 11.4) and permutations (Section 5.4.4) were expanded.
  - Minor extensions and bug fixes.
- **2024-08-27** (v1.0.1):
  - Updated to R 4.4.1.
  - Minor extensions and bug fixes.
- **2023-06-28** (v1.0.0):
  - Final proofreading and copyediting.
  - Minor extensions.
- **2023-05-19** (v0.9.0):
  - Chapter on interfacing compiled code drafted.
  - Minor extensions.
- **2023-04-27** (v0.2.1):
  - Chapter on graphics drafted.
- **2023-04-09** (v0.2.0):
  - New HTML theme (with light and dark modes).
  - Chapter on unevaluated expressions drafted.
  - Chapter on environments and evaluation drafted.

[4] https://github.com/gagolews/deepr

- Chapter on lazy evaluation drafted.

- **2022-12-29** (v0.1.12):

  - The first public release at https://deepr.gagolewski.com/.

  - Chapters 1–12 (basic and compound types, functions, control flow, etc.) drafted.

  - Preface drafted.

  - ISBN 978-0-6455719-2-9 reserved.

  - Cover.